\documentclass[12pt]{iopart}
\usepackage{cite}
\usepackage{iopams}
\usepackage{epsfig}
\usepackage{epsf}
\usepackage{iopams}
\usepackage{graphicx}
\usepackage{psfrag}
\usepackage[T1]{fontenc}
\usepackage[latin1]{inputenc}
\usepackage{color}
\usepackage{amssymb}
\usepackage{bm}
\usepackage{latexsym}
\usepackage{subfigure}
\usepackage{needspace}
\usepackage{pslatex}
\usepackage{txfonts}
\usepackage{caption2}

\def\be {\begin{equation}}
\def\ee {\end{equation}}
\def\bea {\begin{eqnarray}}
\def\eea {\end{eqnarray}}

\def\eq {E_q}

\newcommand{\bef}{\begin{figure}}
\newcommand{\eef}{\end{figure}}

\newcommand{\del}{\partial}

\def\noi{\noindent}
\def\beq{\begin{equation}}   \def\eeq{\end{equation}}

\def\beeq{\begin{eqnarray}} \def\eeeq{\end{eqnarray}}

\newenvironment{narrow}[2]{%
\begin{list}{}{%
\setlength{\topsep}{0pt}%
\setlength{\leftmargin}{#1}%
\setlength{\rightmargin}{#2}%
\setlength{\listparindent}{\parindent}%
\setlength{\itemindent}{\parindent}%
\setlength{\parsep}{\parskip}}%
\item[]}{\end{list}}

\newcommand{\pts}{\langle{p_T}^2\rangle}
\newcommand{\kts}{\langle{k_T}^2\rangle}

\newcommand{\ave}[1]{\left\langle #1 \right\rangle}

\newcommand{\GeV}{\hbox{GeV}}
\newcommand{\fm}{\hbox{fm}}

\def\eq#1{{(\ref{#1})}}
\def\fig#1{{Fig.~\ref{#1}}}

\begin{document}

\newcommand{\A}{{\rm A}}
\newcommand{\rpa}{R_{_{p\A}}}
\newcommand{\xt}{x_{_\perp}}
\newcommand{\ptrans}{p_{_\perp}}
\newcommand{\rag}{R^{^\A}_{_G}}
\newcommand{\rafd}{R^{^\A}_{_{F_2}}}
\newcommand{\dd}{{\rm d}}
\newcommand{\X}{{\rm X}\,}
\newcommand{\sqrtsnn}{\sqrt{s_{_{\mathrm{NN}}}}}
\newcommand{\rapprox}{R^{^{\rm approx}}}
\newcommand{\rppb}{R_{_{p {\rm Pb}}}}
\newcommand{\rpbpb}{R_{_{{\rm Pb}{\rm Pb}}}}
\newcommand{\ud}{\mathrm{d}}
\newcommand{\red}  { \textcolor{red}}
\newcommand{\blue}  { \textcolor{blue}}
\newcommand{\nn}{\nonumber}
\newcommand{\D}{{\cal D}}
\newcommand{\Pom}{{\hspace{ -0.1em}I\hspace{-0.25em}P}}
\newcommand{\vphot}{{\gamma^*}}
\newcommand{\Mcc}{{M_{c\bar{c}}}}
\newcommand{\QQ}{{Q \bar{Q}}}
\newcommand{\cc}{{c\bar{c}}}
\newcommand{\sqrts}{\sqrt{s}}
\newcommand{\Et}{E_{\rm T}}                    % E_T
\newcommand{\as}{\alpha_s}
\newcommand{\BUp}{''Bottom-Up'' }
\newcommand{\B}{{\rm A}}
\newcommand{\raa}{R_{_{\A\A}}}

\providecommand{\jpsi}{J/\psi}
\providecommand{\psip}{\psi^{'}}
\providecommand{\ups}{\Upsilon}
\providecommand{\upsp}{\Upsilon^{'}}
\providecommand{\upspp}{\Upsilon^{''}}
\providecommand{\mean}[1]{\ensuremath{\left<#1\right>}}
\providecommand{\qqbar}{Q\overline{Q}}
\providecommand{\ccbar}{c\overline{c}}
\providecommand{\bbbar}{b\overline{b}}
\providecommand{\dNdeta}{dN_{\rm ch}/d\eta|_{\eta=0}}
\providecommand{\PYTHIA}{{\sc pythia }}
\providecommand{\HERWIG}{{\sc herwig }}

\renewcommand \thesubfigure{{\bf Figure \arabic{subfigure}}}
\renewcommand{\lsim}{~{\buildrel < \over {_\sim}}~}
\newcommand{\gsim}{~{\buildrel > \over {_\sim}}~}
\newcommand{\sqrtsNN}{\sqrt{s_{\scriptscriptstyle{{\rm NN}}}}}
\newcommand{\av}[1]{\left\langle #1 \right\rangle}
\newcommand{\eV}{\mathrm{eV}}
\newcommand{\kev}{\mathrm{keV}}
\newcommand{\mev}{\mathrm{MeV}}
\newcommand{\gev}{\mathrm{GeV}}
\newcommand{\tev}{\mathrm{TeV}}
\newcommand{\mm}{\mathrm{mm}}
\newcommand{\cm}{\mathrm{cm}}
\newcommand{\m}{\mathrm{m}}
\newcommand{\mum}{\mathrm{\mu m}}
\newcommand{\s}{\mathrm{s}}
\newcommand{\mrad}{\mathrm{mrad}}
\newcommand{\mb}{\mathrm{mb}}
\newcommand{\de}{{\rm d}}
\newcommand{\PbPb}{\mbox{Pb--Pb}}
\newcommand{\pPb}{\mbox{p--Pb}}
\newcommand{\Pbp}{\mbox{Pb--p}}
\newcommand{\NN}{\mbox{nucleon--nucleon}}
\newcommand{\pp}{\mbox{proton--proton}}
\newcommand{\pA}{\mbox{proton--nucleus}}
\renewcommand{\AA}{\mbox{nucleus--nucleus}}
\newcommand{\RAA}{R_{\rm AA}}
\newcommand{\RCP}{R_{\rm CP}}
\renewcommand{\pt}{p_{\rm t}}
\renewcommand{\d}{{\rm d}}
\newcommand{\dNdy}{{\rm d}N_{\rm ch}/{\rm d}y}

\def\lsim{\raise0.3ex\hbox{$<$\kern-0.75em\raise-1.1ex\hbox{$\sim$}}}
\def\gsim{\raise0.3ex\hbox{$>$\kern-0.75em\raise-1.1ex\hbox{$\sim$}}}
\def\imb{\fam\imbf\tenimbf}
\def\Tdec{{T_{\rm dec}}}
\def\psat{{p_{\rm sat}}}
\def\LQCD{{\Lambda_{\rm QCD}}}
\def\lq{{\lambda_{\rm q}}}
\def\lg{{\lambda_{\rm g}}}
\def\pt{p_{_T}}
\def\kt{k_{_T}}
\def\xt{x_{_T}}
\def\mf{M_{_F}}
\def\beq{\begin{equation}}
\def\eeq{\end{equation}}
\def\bea{\begin{eqnarray}}
\def\eea{\end{eqnarray}}
\def\noi{\noindent}
\def\a{\alpha}
\def\ab{\bar{\alpha}}
\def\eq#1{{Eq.~(\ref{#1})}}
\def\fig#1{{Fig.~\ref{#1}}}
\def\ud{\underline}
\def\nn{\nonumber}
\def\del{\partial}
\def\Eq#1{Eq.~(\ref{#1})}

 \newcommand\la{\langle}
 \newcommand\beqn{\begin{eqnarray}}
 \newcommand\eeqn{\end{eqnarray}}

\def\mb{\,\mbox{mb}}
\def\mub{\,\mu\mbox{b}}
\def\fm{\,\mbox{fm}}
\def\GeV{\,\mbox{GeV}}
\def\TeV{\,\mbox{TeV}}
\def\eps{\varepsilon}
\def\inf{\int_{-\infty}^{\infty}}
\def\Jpsi{J\!/\!\psi}
\def\Pom{{\bf I\!P}}
\def\Reg{{\bf I\!R}}
\def\lsim{\mathrel{\rlap{\lower4pt\hbox{\hskip1pt$\sim$}}
    \raise1pt\hbox{$<$}}}         %less than or approx. symbol
\def\gsim{\mathrel{\rlap{\lower4pt\hbox{\hskip1pt$\sim$}}
 \raise1pt\hbox{$>$}}}         %greater than or approx. sy
\def\lessim{\lower.5ex\hbox{$\; \buildrel < \over \sim \;$}}

\newcommand{\bfig}{\begin{figure}}
\newcommand{\efig}{\end{figure}}
\newcommand{\infinitessimal}{\mathrm{d}}
\newcommand{\infinitesmal}{\mathrm{d}}
\newcommand{\infinitesimal}{\mathrm{d}}
\newcommand{\intd}{\mathrm{d}}

%define new commands for common words such as RAA, \raa
%also define them for inside of math equations, as \eqnraa
%finally, define them without a space at the end so that the spacing works correctly when punctuation follows the symbol, as \raacomma

\newcommand{\raah}{$R_{AA}$ }
\newcommand{\raacomma}{$R_{AA}$}
\newcommand{\raaphi}{$R_{AA}(\phi)$ }
\newcommand{\raaphicomma}{$R_{AA}(\phi)$}
\newcommand{\raaphipt}{$R_{AA}(\phi;\,\eqnpt)$ }
\newcommand{\raaphiptcomma}{$R_{AA}(\phi;\,\eqnpt)$} 
\newcommand{\raapt}{$R_{AA}(\eqnpt)$ } 
\newcommand{\raaptcomma}{$R_{AA}(\eqnpt)$} 
\newcommand{\eqnraapt}{R_{AA}(\eqnpt)} 
\newcommand{\raaq}{$R_{AA}^q$ }
\newcommand{\raaqcomma}{$R_{AA}^q$}
\newcommand{\eqnraaq}{R_{AA}^q}
\newcommand{\raaqphi}{$R_{AA}^q(\phi)$ }
\newcommand{\raaqphicomma}{$R_{AA}^q(\phi)$} 
\newcommand{\eqnraaqphi}{R_{AA}^q(\phi)}
\newcommand{\raaqphipt}{$R_{AA}^q(\phi;\,\eqnpt)$ }
\newcommand{\raaqphiptcomma}{$R_{AA}^q(\phi;\,\eqnpt)$} 
\newcommand{\eqnraaqphipt}{R_{AA}^q(\phi;\,\eqnpt)} 
\newcommand{\raaqpt}{$R_{AA}^q(\eqnpt)$ } 
\newcommand{\raaqptcomma}{$R_{AA}^q(\eqnpt)$} 
\newcommand{\eqnraaqpt}{R_{AA}^q(\eqnpt)}
\newcommand{\mq}{$m_q$ }
\newcommand{\mqcomma}{$m_q$}
\newcommand{\eqnmq}{m_q}
\newcommand{\muq}{$\mu_q$ }
\newcommand{\muqcomma}{$\mu_q$}
\newcommand{\eqnmuq}{\mu_q}
 
\newcommand{\raaQ}{$R_{AA}^Q$ }
\newcommand{\raaQcomma}{$R_{AA}^Q$}
\newcommand{\eqnraaQ}{R_{AA}^Q}
\newcommand{\raaQphi}{$R_{AA}^Q(\phi)$ }
\newcommand{\raaQphicomma}{$R_{AA}^Q(\phi)$} 
\newcommand{\eqnraaQphi}{R_{AA}^Q(\phi)}
\newcommand{\raaQphipt}{$R_{AA}^Q(\phi;\,\eqnpt)$ }
\newcommand{\raaQphiptcomma}{$R_{AA}^Q(\phi;\,\eqnpt)$} 
\newcommand{\eqnraaQphipt}{R_{AA}^Q(\phi;\,\eqnpt)} 
\newcommand{\raaQpt}{$R_{AA}^Q(\eqnpt)$ } 
\newcommand{\raaQptcomma}{$R_{AA}^Q(\eqnpt)$} 
\newcommand{\eqnraaQpt}{R_{AA}^Q(\eqnpt)}
\newcommand{\mQ}{$m_Q$ }
\newcommand{\mQcomma}{$m_Q$}
\newcommand{\eqnmQ}{m_Q}
\newcommand{\muQ}{$\mu_Q$ }
\newcommand{\muQcomma}{$\mu_Q$}
\newcommand{\eqnmuQ}{\mu_Q}

\newcommand{\raac}{$R_{AA}^c$ }
\newcommand{\raaccomma}{$R_{AA}^c$}
\newcommand{\eqnraac}{R_{AA}^c}
\newcommand{\raacphi}{$R_{AA}^c(\phi)$ }
\newcommand{\raacphicomma}{$R_{AA}^c(\phi)$} 
\newcommand{\eqnraacphi}{R_{AA}^c(\phi)}
\newcommand{\raacphipt}{$R_{AA}^c(\phi;\,\eqnpt)$ }
\newcommand{\raacphiptcomma}{$R_{AA}^c(\phi;\,\eqnpt)$} 
\newcommand{\eqnraacphipt}{R_{AA}^c(\phi;\,\eqnpt)} 
\newcommand{\raacpt}{$R_{AA}^c(\eqnpt)$ } 
\newcommand{\raacptcomma}{$R_{AA}^c(\eqnpt)$} 
\newcommand{\eqnraacpt}{R_{AA}^c(\eqnpt)} 

\newcommand{\raab}{$R_{AA}^b$ }
\newcommand{\raabcomma}{$R_{AA}^b$}
\newcommand{\eqnraab}{R_{AA}^b}
\newcommand{\raabphi}{$R_{AA}^b(\phi)$ }
\newcommand{\raabphicomma}{$R_{AA}^b(\phi)$} 
\newcommand{\eqnraabphi}{R_{AA}^b(\phi)}
\newcommand{\raabphipt}{$R_{AA}^b(\phi;\,\eqnpt)$ }
\newcommand{\raabphiptcomma}{$R_{AA}^b(\phi;\,\eqnpt)$} 
\newcommand{\eqnraabphipt}{R_{AA}^b(\phi;\,\eqnpt)} 
\newcommand{\raabpt}{$R_{AA}^b(\eqnpt)$ } 
\newcommand{\raabptcomma}{$R_{AA}^b(\eqnpt)$} 
\newcommand{\eqnraabpt}{R_{AA}^b(\eqnpt)}

\newcommand{\cbratio}{$\eqnraacpt/\eqnraabpt$ }
\newcommand{\cbratiocomma}{$\eqnraacpt/\eqnraabpt$}
\newcommand{\eqncbratio}{\eqnraacpt/\eqnraabpt}

\newcommand{\raag}{$R_{AA}^g$ }
\newcommand{\raagcomma}{$R_{AA}^g$}
\newcommand{\eqnraag}{R_{AA}^g}
\newcommand{\raagphi}{$R_{AA}^g(\phi)$ }
\newcommand{\raagphicomma}{$R_{AA}^g(\phi)$} 
\newcommand{\eqnraagphi}{R_{AA}^g(\phi)}
\newcommand{\raagphipt}{$R_{AA}^g(\phi;\,\eqnpt)$ }
\newcommand{\raagphiptcomma}{$R_{AA}^g(\phi;\,\eqnpt)$} 

\newcommand{\eqnraagphipt}{R_{AA}^g(\phi;\,\eqnpt)} 
\newcommand{\raagpt}{$R_{AA}^g(\eqnpt)$ } 
\newcommand{\raagptcomma}{$R_{AA}^g(\eqnpt)$} 
\newcommand{\eqnraagpt}{R_{AA}^g(\eqnpt)} 

\newcommand{\RAAcomma}{\raacomma}
\newcommand{\RAAphi}{\raaphi}
\newcommand{\RAAphicomma}{\raaphicomma}
\newcommand{\RAAphipt}{\raaphipt}
\newcommand{\RAAphiptcomma}{\raaphiptcomma}
\newcommand{\raapi}{$R_{AA}^\pi$ }
\newcommand{\raae}{$R_{AA}^{e^-}$ }
\newcommand{\raapicomma}{$R_{AA}^\pi$}
\newcommand{\raaecomma}{$R_{AA}^{e^-}$}
\newcommand{\eqnraapi}{R_{AA}^\pi}
\newcommand{\eqnraae}{R_{AA}^{e^-}}

\newcommand{\vtwo}{$v_2$ }
\newcommand{\vtwocomma}{$v_2$}
\newcommand{\vtwopt}{$v_2(\eqnpt)$ }
\newcommand{\vtwoptcomma}{$v_2(\eqnpt)$}
\newcommand{\eqnraa}{R_{AA}}
\newcommand{\eqnraaphi}{R_{AA}(\phi)}
\newcommand{\eqnraaphipt}{R_{AA}(\phi;\,\eqnpt)} 
\newcommand{\eqnRAA}{\eqnraa}
\newcommand{\eqnRAAphi}{\eqnraaphi}
\newcommand{\eqnRAAphipt}{\eqnraaphipt}
\newcommand{\eqnvtwo}{v_2}
\newcommand{\vtwovsraa}{\vtwo vs.~\raa}
\newcommand{\vtwovsraacomma}{\vtwo vs.~\raacomma}
\newcommand{\eqnvtwopt}{v_2(\eqnpt)}

\newcommand{\ppcomma}{$p+p$}
\newcommand{\dau}{$d+Au$ }
\newcommand{\daucomma}{$d+Au$}
\newcommand{\auau}{$Au+Au$ }
\newcommand{\auaucomma}{$Au+Au$}
\newcommand{\aplusa}{$A+A$ }
\newcommand{\aplusacomma}{$A+A$}
\newcommand{\cucu}{$Cu+Cu$ }
\newcommand{\cucucomma}{$Cu+Cu$}

\newcommand{\rhopart}{$\rho_{\textrm{\footnotesize{part}}}$ }
\newcommand{\rhopartcomma}{$\rho_{\textrm{\footnotesize{part}}}$}
\newcommand{\eqnrhopart}{\rho_{\textrm{\footnotesize{part}}}}
\newcommand{\npart}{$N_{\textrm{\footnotesize{part}}}$ }
\newcommand{\npartcomma}{$N__{\textrm{\footnotesize{part}}}$}
\newcommand{\eqnnpart}{N_{\textrm{\footnotesize{part}}}}
\newcommand{\taa}{$T_{AA}$ }
\newcommand{\taacomma}{$T_{AA}$}
\newcommand{\eqntaa}{T_{AA}}
\newcommand{\rhocoll}{$\rho_{\textrm{\footnotesize{coll}}}$ }
\newcommand{\rhocollcomma}{$\rho_{\textrm{\footnotesize{coll}}}$}
\newcommand{\eqnrhocoll}{\rho_{\textrm{\footnotesize{coll}}}}
\newcommand{\ncoll}{$N_{\textrm{\footnotesize{coll}}}$ }
\newcommand{\ncollcomma}{$N_{\textrm{\footnotesize{coll}}}$}
\newcommand{\eqnncoll}{N_{\textrm{\footnotesize{coll}}}}
\newcommand{\dndy}{$\frac{dN_g}{dy}$ }
\newcommand{\dndycomma}{$\frac{dN_g}{dy}$}
\newcommand{\eqndndy}{\frac{dN_g}{dy}}
\newcommand{\eqndndyabs}{\frac{dN_g^{abs}}{dy}}
\newcommand{\eqndndyrad}{\frac{dN_g^{rad}}{dy}}
\newcommand{\dnslashdy}{$dN_g/dy$ }
\newcommand{\dnslashdycomma}{$dN_g/dy$}
\newcommand{\eqndnslashdy}{dN_g/dy}
\newcommand{\dngdy}{$\frac{dN_g}{dy}$ }
\newcommand{\dngdycomma}{$\frac{dN_g}{dy}$}
\newcommand{\eqndngdy}{\frac{dN_g}{dy}}
\newcommand{\eqndngdyabs}{\frac{dN_g^{abs}}{dy}}
\newcommand{\eqndngdyrad}{\frac{dN_g^{rad}}{dy}}
\newcommand{\dngslashdy}{$dN_g/dy$ }
\newcommand{\dngslashdycomma}{$dN_g/dy$}
\newcommand{\eqndngslashdy}{dN_g/dy}
\newcommand{\alphas}{$\as$ }
\newcommand{\alphascomma}{$\as$}
\newcommand{\eqnalphas}{\as}

\renewcommand{\pt}{$p_T$ }
\newcommand{\pT}{\pt}
\newcommand{\ptcomma}{$p_T$}
\newcommand{\pTcomma}{\ptcomma}
\newcommand{\eqnpt}{p_T}
\newcommand{\ptf}{$p_T^f$ }
\newcommand{\ptfcomma}{$p_T^f$}
\newcommand{\eqnptf}{p_T^f}
\newcommand{\pti}{$p_T^i$ }
\newcommand{\pticomma}{$p_T^i$}
\newcommand{\eqnpti}{p_T^i}
\newcommand{\lowpt}{low-\pt}
\newcommand{\lowptcomma}{low-\ptcomma}
\newcommand{\midpt}{mid-\pt}
\newcommand{\midptcomma}{mid-\ptcomma}
\newcommand{\intermediatept}{intermediate-\pt}
\newcommand{\intermediateptcomma}{intermediate-\ptcomma}
\newcommand{\highpt}{high-\pt}
\newcommand{\highptcomma}{high-\ptcomma}
\newcommand{\Aperp}{$A_\perp$ }
\newcommand{\Aperpcomma}{$A_\perp$}
\newcommand{\eqnAperp}{A_\perp}
\newcommand{\rperp}{$r_\perp$ }
\newcommand{\rperpcomma}{$r_\perp$}
\newcommand{\eqnrperp}{r_\perp}
\newcommand{\eqnrperpHS}{r_{\perp,HS}}
\newcommand{\eqnrperpWS}{r_{\perp,WS}}
\newcommand{\Rperp}{$R_\perp$ }
\newcommand{\Rperpcomma}{$R_\perp$}
\newcommand{\eqnRperp}{R_\perp}

\newcommand{\pizero}{$\pi^0$ }
\newcommand{\eqnpizero}{\pi^0}
\newcommand{\pizerocomma}{$\pi^0$}

\newcommand{\qhat}{$\hat{q}$ }
\newcommand{\qhatcomma}{$\hat{q}$}
\newcommand{\eqnqhat}{\hat{q}}

\newcommand{\gym}{$g_{SYM}$ }
\newcommand{\gymcomma}{$g_{SYM}$}
\newcommand{\eqngym}{g_{SYM}}
\newcommand{\gsym}{\gym}
\newcommand{\gsymcomma}{\gymcomma}
\newcommand{\eqngsym}{\eqngym}
\newcommand{\gs}{$g_{s}$ }
\newcommand{\gscomma}{$g_{s}$}
\newcommand{\eqngs}{g_{s}}
\newcommand{\asym}{$\alpha_{SYM}$ }
\newcommand{\asymcomma}{$\alpha_{SYM}$}
\newcommand{\eqnasym}{\alpha_{SYM}}
\newcommand{\alphasym}{\asym}
\newcommand{\alphasymcomma}{\asymcomma}
\newcommand{\eqnalphasym}{\eqnalsym}

\newcommand{\stronglycoupled}{strongly-coupled }
\newcommand{\weaklycoupled}{weakly-coupled }
\newcommand{\stronglycoupledcomma}{strongly-coupled}
\newcommand{\weaklycoupledcomma}{weakly-coupled}
\newcommand{\ads}{AdS/CFT }
\newcommand{\asd}{\ads}
\newcommand{\adscomma}{AdS/CFT}
\newcommand{\asdcomma}{\adscomma}
\newcommand{\thooft}{'t Hooft }
\newcommand{\thooftcomma}{'t Hooft}

\newcommand{\infinity}{\infty}

\newcommand{\rightleftarrow}{\leftrightarrow}

%Normal commands for Eq., Fig., etc.
%\newcommand{\eq}[1]{Eq.~(\ref{#1})}
%\newcommand{\eqn}[1]{Eq.~(\ref{#1})}
%\newcommand{\fig}[1]{Fig.~\ref{#1}}
\newcommand{\figtwo}[2]{Figs.~\ref{#1}, \ref{#2}}
\newcommand{\tab}[1]{Table \ref{#1}}

%Commands for Eq., Fig., etc. for IOP journals
%\newcommand{\eq}[1]{Eq.~(\ref{#1})}
%\newcommand{\eqn}[1]{\ref{#1}}
%\newcommand{\fig}[1]{Figure \ref{#1}}
%\newcommand{\figtwo}[2]{Figures \ref{#1}, \ref{#2}}
%\newcommand{\tab}[1]{Table \ref{#1}}

%\newcommand{\captionsize}{\small}

\title[Heavy Ion Collisions at the LHC - Last Call for Predictions]{Heavy Ion Collisions at the LHC - Last Call for Predictions}

\author{
N Armesto$^1$, N Borghini$^2$, S Jeon$^3$ and U A Wiedemann$^4$ (editors)
}

\author{
S~Abreu$^{5}$,
V Akkelin$^{6}$,
J Alam$^{7}$,
J L Albacete$^{8}$,
A Andronic$^{9}$,
D Antonov$^{10}$
% ...
(authors)\footnote{The contributors on this author list have contributed only to those subsections of the
  report, which they cosign with their name. Only those have collaborated together,
  whose names appear together in the header of a given subsection.}}
%Journal of Physics G
%recognizes that not all names appearing in the list above are co-authors.
%Only the names appearing together in each subsection should be
%regarded as co-authors.}}

\address{
$^1$ Departamento de F{\'\i}sica de Part{\'\i}culas and IGFAE,
Universidade de Santiago de Compostela, 15782 Santiago de Compostela,
Spain\\
$^2$ Fakult{\"a}t f{\"u}r Physik, Universit{\"a}t Bielefeld,
D-33501 Bielefeld, Germany\\
$^3$ Department of Physics, McGill University, Montr\'eal,
  Canada H3A 2T8\\
$^4$ CERN, PH Department, TH Division,
1211 Geneva 23, Switzerland\\
$^{5}$ Instituto Superior T\'ecnico/CENTRA, Av. Rovisco Pais, P-1049-001 Lisboa, Portugal\\
$^{6}$
Bogolyubov Institute for Theoretical Physics, Metrolohichna str. 14-b, 03680 Kiev-143, Ukraine\\
$^{7}$
Variable Energy Cyclotron Centre, 1/AF Bidhan Nagar, Kolkata 700 064, India\\
$^{8}$ Department of Physics, The Ohio State University, 191
  W. Woodruff Avenue, OH-43210, Columbus, USA\\
$^{9}$ Gesellschaft f\"ur Schwerionenforschung, GSI,
D-64291 Darmstadt, Germany\\
$^{10}$ Institut f\"ur Theoretische Physik, Universit\"at Heidelberg,
Philosophenweg 19, D-69120 Heidelberg, Germany\\
% ...
}

\begin{abstract} This writeup is a compilation of the predictions for the
forthcoming Heavy Ion Program at the Large Hadron Collider, as presented at the
CERN Theory Institute 'Heavy Ion Collisions at the LHC - Last Call for
Predictions', held from May 14th to June 10th 2007.
\end{abstract}

\maketitle

\tableofcontents

\markboth{Heavy Ion Collisions at the LHC - Last Call for Predictions}{Heavy Ion Collisions at the LHC - Last Call for Predictions}

\section*{Preface}
In August 2006, the CERN Theory Unit announced to restructure its visitor program
and to create a "CERN Theory Institute", where 1-3 month long specific programs
can take place. The first such Institute was held from 14 May to 10 June 2007,
focussing on "Heavy Ion Collisions at the LHC - Last Call for Predictions". It
brought together close to 100 scientists working on the theory of ultra-relativistic
heavy ion collisions.
The aim of this workshop was to review and document the status of expectations and
predictions for the heavy ion program at the Large Hadron Collider
LHC before its start. LHC will explore
heavy ion collisions at $\sim 30$ times higher center of mass energy than explored
previously at  the Relativistic Heavy Ion Collider RHIC. So, on the one hand, the charge
of this workshop provided a natural forum for the exchange of the most recent ideas, and
allowed to monitor how the understanding of heavy ion collisions has evolved in recent
years with the data from RHIC, and with the preparation of the LHC experimental program.
On the other hand, the workshop aimed at a documentation which helps to distinguish
pre- from post-dictions. An analogous documentation of the "Last Call for Predictions"
\cite{Bass:1999zq} was prepared
prior to the start of the heavy-ion program at the Relativistic Heavy Ion Collider RHIC, and
it proved useful in the subsequent discussion and interpretation of RHIC data.
The present write-up is the documentation of predictions for the LHC heavy ion
program, received or presented during the CERN TH Institute. The set-up of the
CERN TH Institute allowed us to aim for the wide-most coverage of predictions.
There were more than 100 presentations and discussions during the workshop.
Moreover, those unable to attend could still participate by submitting predictions in
written form during the workshop. This followed the spirit that everybody interested
in making a prediction had the right to be heard.
%We received 3 contributions in this
%latter way.

To arrive at a concise document, we required that each prediction should be
summarized on at most two pages, and that predictions should be presented,
whenever possible, in figures which display measurable quantities. Full model descriptions
were not accepted - the authors were encouraged to indicate the relevant references for the
interested reader. Participants had the possibility to submit multiple contributions on
different topics, but it was part of the subsequent editing process to ensure that predictions
on neighboring topics were merged wherever possible.
The contributions summarized here are organized in several sections,
 - though some of them contain material related with more than one section -,
roughly by going
from low transverse
momentum to high transverse momentum and from abundant to rare measurements.
In the low transverse momentum regime, we start with predictions on multiplicity distributions,
azimuthal asymmetries in particle production and hadronic flavor observables, followed by
correlation and fluctuation measurements. The contributions on hard probes at the LHC
start with predictions for single inclusive high transverse momentum spectra,
and jets, followed by heavy quark and
quarkonium measurements, leptonic probes and photons.
A final section "Others" encompasses those predictions which do not fall naturally
within one of the above-mentioned categories, or discuss the more speculative phenomena
that may be explored at the LHC.

We would like to end this Preface by thanking the TH Unit at CERN for
its generous support of this workshop. Special thanks go to Elena Gianolio,
Michelle Mazerand, Nanie Perrin and Jeanne Rostant, whose help and
patience was invaluable.

%We would like to end this Preface by warmly thanking the TH Division at CERN, particularly
%its head
%Luis \'Alvarez-Gaum\'e, Elena Gianolio,
%and the secretaries Michelle Mazerand, Nanie Perrin and Jeanne
%Rostant, for their encouragment, unvaluable help and patience without which this program
%could have hardly been carried on.

N\'estor Armesto

Nicolas Borghini

Sangyong Jeon

Urs Achim Wiedemann

\section{Multiplicities and multiplicity distributions}
\label{sec:mult}

\subsection{Multiplicity distributions in rapidity for Pb-Pb and $p$-Pb central collisions from a simple model}
\label{s:Abreu}

{\em S. Abreu, J. Dias de Deus and J. G. Milhano}
\vskip 0.5cm

The simple model~\cite{DiasdeDeus:2007wb} for the distribution of rapidity 
extended objects (longitudinal glasma colour fields or coloured strings) 
created in a heavy ion collision combines the generation of lower centre-of-mass
rapidity objects from higher rapidity ones with asymptotic saturation in the 
form of the well known logistic equation for population dynamics 
\begin{equation}
\label{eq:Abreu-eq1}
\frac{\partial \rho}{\partial (-\Delta)} = \frac{1}{\delta}(\rho - A \rho^2)\, ,
\end{equation}
where $\rho \equiv \rho(\Delta, Y)$ is the particle density, $Y$ is the beam 
rapidity, and $\Delta \equiv |y| - Y$. 
The $Y$-dependent limiting value of $\rho$ is determined by the saturation 
condition $\partial_{(-\Delta)} \rho = 0 \longrightarrow\rho_Y = 1/A$, while the 
separation between the low density (positive curvature) and high density 
(negative curvature) regions is given by 
${\partial^2}_{(-\Delta)} \rho\big |_{\Delta_0} = 0\longrightarrow \rho_0 \equiv \rho( \Delta_0, Y) =  {\rho_Y}/{2}$. 
Integrating (\ref{eq:Abreu-eq1}) we get
\begin{equation}
\label{eq:Abreu-eq2}
\rho( \Delta, Y) = \frac{\rho_Y}{{\rm e}^{\frac{\Delta - \Delta_0}{\delta}}+1}\, .
\end{equation}
In the String Percolation Model~\cite{DiasdeDeus:2007wb} the particle density 
is proportional, once the colour reduction factor is taken into account, to the 
average number of participants $\rho\propto N_A$; 
the normalized particle density at mid-rapidity is related to the gluon 
distribution at small Bjorken-$x$, $\rho\propto {\rm e}^{\lambda\, Y}$; and the 
dense-dilute separation scale decreases, from energy conservation, linearly 
with $Y$, $\Delta_0=-\alpha Y$ with $0<\alpha <1$. 
Rewriting (\ref{eq:Abreu-eq2}) in rapidity
\begin{equation}
\label{eq:Abreu-eq3}
\rho \equiv  \frac{{\rm d}N}{{\rm d}y} =  
\frac{N_A \cdot {\rm e}^{\lambda Y}}{{\rm e}^{\frac{|y|- (1-\alpha) Y}{\delta}} +1}\, .
\end{equation}
The values $\lambda=0.247$, $\alpha=0.269$ and $\delta=0.67$ for the parameters 
in the solution (\ref{eq:Abreu-eq3}) are fixed by an overall 
fit~\cite{Brogueira:2006nz} of Au-Au RHIC data~\cite{Back:2004je}.

\begin{figure}[h] 
   \centering
   \includegraphics[angle=0,width=10cm]{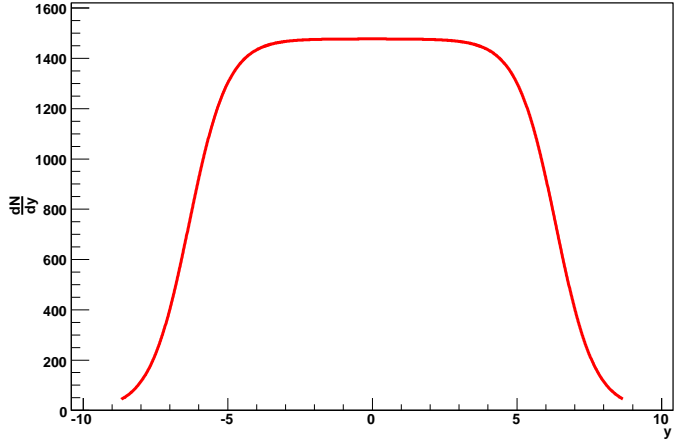} 
   \caption{$\frac{{\rm d}N}{{\rm d}y}$ from (\ref{eq:Abreu-eq3}) for Pb-Pb 
  (0-10\% central) collisions at $\sqrtsnn=5.5$~TeV with $N_A=173.3$}
   \label{fig:Abreu-fig1}
\end{figure}

In Fig.~\ref{fig:Abreu-fig1} we show the predicted multiplicity 
distribution for the 10\% most central Pb-Pb collisions at $\sqrtsnn=5.5$~TeV 
with $N_A=173.3$ taken from the Glauber calculation in~\cite{Kharzeev:2004if}.

\begin{figure}[h] 
\centering
\includegraphics[angle=0,width=10cm]{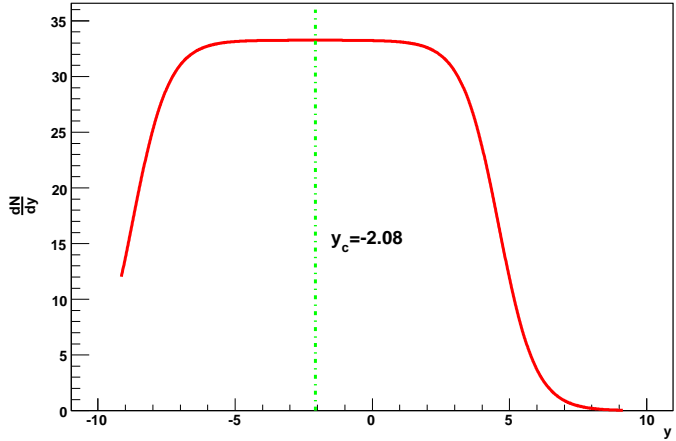} 
\caption{$\frac{{\rm d}N}{{\rm d}y}$ from asymmetric version of 
  (\ref{eq:Abreu-eq3})~\cite{DiasdeDeus:2007wb} for $p$-Pb (0-20\% central) 
  collisions at $\sqrtsnn=8.8$~TeV with $N_{\rm part}=13.07$.}
\label{fig:Abreu-fig2}
\end{figure}

In Fig.~\ref{fig:Abreu-fig2} we show the predicted multiplicity 
distribution for the 20\% most central $p$-Pb collisions at $\sqrtsnn=8.8$~TeV 
with $N_{part}=13.07$ also from~\cite{Kharzeev:2004if}. 
In this case the solution (\ref{eq:Abreu-eq3}) have been modified to account 
for the asymmetric geometry and the shift of the centre of mass of the system 
relatively to the laboratory centre of mass~\cite{DiasdeDeus:2007wb}. 
The resulting rapidity shift $y_c = -2.08$ is marked in Fig. \ref{fig:Abreu-fig2}.

\subsection{Multiplicities
  in Pb-Pb central collisions at the LHC from running coupling
  evolution and RHIC data} 
\label{albacete}
{\it J. L. Albacete}

{\small
Predictions for the pseudorapidity density of charged particles
produced in Pb-Pb
central collisions at $\sqrt{s_{NN}}=5.5$ TeV are presented. Particle
production in such collisions is computed in the framework of
$k_t$-factorization, using running coupling non-linear evolution to
determine the transverse momentum and rapidity dependence of the
nuclear unintegrated gluon distributions.
}
\vskip 0.5cm

Predictions for the pseudorapidity density of charged particles produced
in Pb-Pb central collisions at $\sqrt{s_{NN}}=5.5$ TeV
presented in \cite{Albacete:2007sm} are summarized. Primary gluon production in
such collisions can be computed perturbatively in the framework of
$k_t$-factorization. Under the additional assumption of local parton-hadron
duality, the rapidity density of produced charged particles in nucleus-nucleus 
collisions at energy $\sqrt{s}$ and impact parameter $b$ is given by,
\cite{Kharzeev:2001gp}:
\begin{center}
\beq
\!\!\!\!\!\!\!\!\!\!\!\!\!\!\!\!\!\!\!\!\!\!\!\!\!\!\!
\frac{dN}{dy\, d^2b}=C\frac{4\pi N_c}{N_c^2-1}\int^{p_{kin}}\frac{d^2p_t}{p_t^2}\int^{p_t}\,d^2k_t\, \alpha_s(Q)\,\varphi\left(x_1,\frac{\vert\ud{k_t}+\ud{p_t}\vert}{2}\right) \varphi\left(x_2,\frac{\vert\ud{k_t}-\ud{p_t}\vert}{2}\right),
\label{ktfact}
\eeq
\end{center}
where $p_t$ and $y$ are the transverse momentum and rapidity of the
produced particle, $x_{1,2}=(p_t/\sqrt{s})\,e^{\pm y}$ and
$Q\!=\!0.5\max\left\{\vert p_t\pm k_t\vert \right\}$. The lack of
impact parameter integration in this calculation and the gluon to
charged hadron
ratio are accounted for by the constant $C$, which sets the normalization.  
The nuclear unintegrated gluon distributions (u.g.d.), $\varphi(x,k)$,
entering \eq{ktfact} are 
taken from numerical solutions of the Balitsky-Kovchegov evolution
equation including running coupling corrections, \cite{Albacete:2007yr}:
\beq
\qquad\qquad\frac{\partial N(Y,r)}{\partial
  Y}=\mathcal{R}[N(Y,r)]-\mathcal{S}[N(Y,r)]
\label{evol}
\eeq
Explicit expressions for the {\it running}, $\mathcal{R}[N]$, and {\it
  subtraction}, $\mathcal{S}[N]$, functionals in the r.h.s. of
\eq{evol} can be found in \cite{Albacete:2007yr}. The nuclear u.g.d.
are given by the Fourier transform of the dipole scattering amplitude
evolved according to \eq{evol}, $\varphi(Y,k)=\int \frac{d^2r}{2\pi
  r^2}\,e^{i\,\ud{k}\cdot\ud{r}}\,\mathcal{N}(Y,r)$, with
$Y=\ln(0.05/x)+\Delta Y_{ev}$, where $\Delta Y_{ev}$ is a free
parameter.
Large-$x$ effects have been included by replacing $\varphi(x,k)\rightarrow
\varphi(x,k)(1-x)^4$. 
The initial condition for the evolution is taken from the
McLerran-Venugopalan model \cite{McLerran:1993ni}, which is believed
to provide a good 
description of nuclear distribution functions at moderate energies:
\beq
N^{MV}(Y=0,r)=1-\exp\left\{-\frac{r^2Q_0^2}{4}\ln\left(\frac{1}{r\Lambda}+e\right)\right\}, 
\eeq
where $Q_0$ is the initial saturation scale and $\Lambda\!=\!0.2$ GeV.
In order to compare \eq{ktfact} with
experimental data it is
necessary to correct the difference between rapidity, $y$, and the
experimentally measured pseudorapidity, $\eta$. This is managed by
introducing an effective hadron mass, $m_{eff}$. The variable
transformation, $y(\eta,p_t,m_{eff})$, and its corresponding jacobian
are given by Eqs.(25-26) in 
\cite{Kharzeev:2001gp}. Corrections to the kinematics due to
the hadron mass are also considered by replacing $p_t\rightarrow
m_t=\left(p_t^2+m_{eff}^2\right)^{1/2}$ in the evaluation of
$x_{1,2}$. This replacement affects the predictions for the LHC by less than a
$5\%$, \cite{Albacete:2007sm}.

\begin{figure}[ht]
 \begin{center}
 \includegraphics[height=10cm]{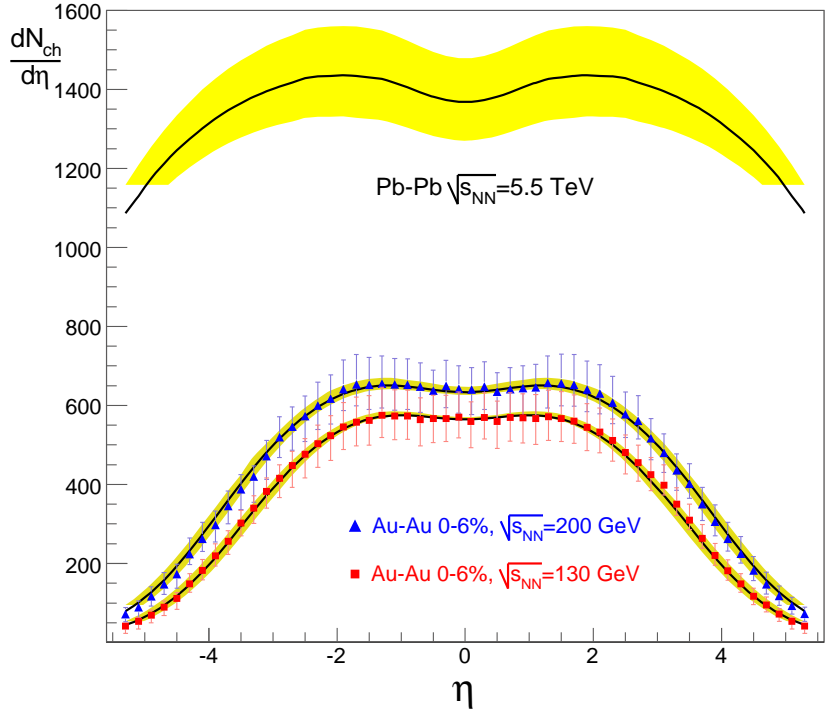}
 \vskip -0.5cm
 \caption{Multiplicity densities for Au-Au central collisions at RHIC
   (experimental data taken from \cite{Back:2004je}), and prediction
   for Pb-Pb central 
   collisions at $\sqrt{s_{NN}}\!=\!5.5$ TeV. The best fits to
   data (solid lines) are obtained with $Q_0\!=\!1$
   GeV, $\Delta Y_{ev}\!=\!1$ and $m_{eff}\!=\!0.25$ GeV. The upper limit of
   the error bands correspond  to $\Delta Y_{ev}\!=\!3$ and $Q_0\!=\!0.75$ GeV,
   and the lower limit to $\Delta Y_{ev}\!=\!0.5$ and $Q_0\!=\!1.25$ 
   GeV, with $m_{eff}\!=\!0.25$ GeV in both cases. } 
 \label{pred}
 \end{center}
 \end{figure}

The results for the pseudorapidity density of charged particles in
central Au-Au collisions at
$\sqrt{s_{NN}}\!=\!130$, $200$ and $5500$ GeV are shown in \fig{pred}. A
remarkably good description 
of RHIC data is obtained with $Q_0\!=\!0.75\div 1.25$ GeV, $\Delta
Y_{ev}\!\lesssim\!3$ and $m_{eff}\!=\!0.2\div0.3$ GeV. Assuming no
difference between Au 
and Pb nuclei, the extrapolation of the fits to RHIC data yields
the following band:
$\frac{dN^{Pb-Pb}_{ch}}{d\eta}(\sqrt{s_{NN}}\!=\!5.5\,\mbox{TeV})\vert_{\eta=0}\approx 
1290\div 1480$ for central Pb-Pb collisions at the LHC. The central
value of our predictions
$\frac{dN^{Pb-Pb}_{ch}}{d\eta}(\sqrt{s_{NN}}\!=\!5.5\,\mbox{TeV})\vert_{\eta=0}\approx 
1390$ corresponds to the best fits to RHIC data.
%\maketitle
 %%%%%%%%%%%%%%%%%%%%%%%%%%%%%
\vskip -1cm
%%%%%%%%%%%%%%%%%%%%%%%%%%%%

\subsection{Identified hadron spectra in Pb-Pb at $\sqrtsnn$ = 5.5 TeV: hydrodynamics+pQCD predictions}
\label{denterria}
 
{\it F. Arleo, D. d'Enterria and D. Peressounko}

{\small
  The single inclusive charged hadron $p_T$ spectra in Pb-Pb collisions at the LHC, 
  predicted by a combined hydrodynamics+perturbative QCD (pQCD) approach are presented.
}
\vskip 0.5cm

We present predictions for the inclusive transverse momentum distributions of pions, kaons and 
(anti)protons produced at mid-rapidity in Pb-Pb collisions at $\sqrtsnn$ = 5.5 TeV based on 
hydrodynamics+pQCD calculations. The bulk of the spectra ($p_T\lesssim$ 5 GeV/$c$) in central 
Pb-Pb at the LHC is computed with a hydrodynamical model -- successfully tested at RHIC~\cite{d'Enterria:2005vz} --
%the identified charged hadron spectra in Au-Au collisions at RHIC energies. %The basic input parameter of the hydro evolution, t
using an initial entropy density extrapolated empirically from the hadron multiplicities 
measured at RHIC: $\dNdeta/(0.5\,N_{\rm part})\approx 0.75\ln(\sqrtsnn/1.5)$~\cite{Adler:2004zn}. %In addition, a
Above $p_T\approx$ 3 GeV/$c$, additional hadron production from (mini)jet fragmentation 
is computed from collinearly factorized pQCD cross sections at next-to-leading-order (NLO)
accuracy~\cite{Aurenche:1999nz}. We use recent parton distribution functions (PDF)~\cite{Pumplin:2002vw} 
and fragmentation functions (FF)~\cite{AKK1}, modified respectively
to account for initial-state shadowing~\cite{deFlorian:2003qf} 
and final-state parton energy loss~\cite{Arleo:2007bg}.\\
%in the produced medium, appropiately scaling the quenching 
%weights from RHIC to LHC energies.\\

We use cylindrically symmetric boost-invariant 2+1-D relativistic hydrodynamics, fixing the initial
conditions for Pb-Pb at $b$ = 0 fm and employing a simple Glauber prescription to obtain the 
corresponding source profiles at all other centralities~\cite{d'Enterria:2005vz}. The initial source is 
assumed to be formed at a time $\tau_0 = 1/Q_s\approx$ 0.1 fm/$c$, with an initial entropy density 
of $s_0$~=~1120~fm$^{-3}$ (i.e. %corresponding to an energy density of 
$\varepsilon_o\propto s_0^{4/3}\approx$ 650~GeV/fm$^3$) so as to reproduce the 
expected final hadron multiplicity $\dNdeta\approx$ 1300 at the LHC~\cite{Adler:2004zn}.
We follow the evolution of the system by solving the equations of ideal hydrodynamics 
including the current conservation for net-baryon number (the system is almost baryon-free, 
$\mu_B\approx$ 5 MeV). For temperatures above (below) $T_{\rm crit}\approx$ 170 MeV 
the system is described with a QGP (hadron gas) equation of state (EoS).
The QGP EoS -- obtained from a parametrization to recent lattice QCD results -- is Maxwell
%~\cite{dde_peress} -- is Maxwell
connected to the hadron resonance gas phase assuming a first-order phase transition.
%For the QGP phase, we assume two different 
%Equation of States (EoS): (i) an ideal massless gas of 2+1 quarks and gluons ($g_{\rm QGP}=42.5$)
%with bag constant $B$ = 0.38 GeV/fm$^3$, and (ii) a $P(\varepsilon,T)$ parametrization of the 
%lattice results  %h, and a hadron resonance gas of all states with $m<2$ GeV at
%$T<T_c$. Taking $B^{1/4}=239$ MeV leads to 1st-order transition with $T_c=165$ MeV.
As done for RHIC energies, we chemically freeze-out the system (i.e. fix the hadron ratios) at $T_{\rm crit}$.
%explicitly conserving particle numbers by introducing individual (temperature-dependent)
%chemical potentials for each hadron.
Final state hadron spectra are obtained with the Cooper-Frye prescription %on a decoupling surface 
at $T_{\rm fo}\approx$ 120 MeV followed by decays of unstable resonances using the known 
branching ratios. Details can be found at~\cite{d'Enterria:2005vz}.\\

Our NLO pQCD predictions are obtained with the code of ref.~\cite{Aurenche:1999nz} 
with %renormalization, factorization and fragmentation 
all scales set to $\mu=p_{T}$. Pb-Pb yields are obtained scaling the NLO 
cross-sections by the number of incoherent nucleon-nucleon collisions for each centrality class given by
a Glauber model ($N_{\rm coll}$ = 1670, 12.9 for 0-10\%-central and 60-90\%-peripheral). 
Nuclear (isospin and shadowing) corrections of the CTEQ6.5M PDFs~\cite{Pumplin:2002vw} 
are introduced using the NLO nDSg %gluon shadowing
parametrization~\cite{deFlorian:2003qf}. Final-state energy loss in the hot and dense
medium is accounted for by modifying the AKK FFs~\cite{AKK1} with BDMPS
quenching weights as described in~\cite{Arleo:2007bg}. The BDMPS medium-induced 
gluon spectrum depends on a single scale $\omega_c=\mean{\hat{q}}\,L^2$, 
related to the transport coefficient and length of the medium. We use 
$\omega_c\approx$ 50 GeV, from the expected energy dependence of the 
quenching parameter and the measured $\omega_c\approx$ 20 GeV at 
RHIC~\cite{Arleo:2007bg}. The inclusive hadron spectra in central Pb-Pb 
are suppressed by up to a factor $\sim$10 (2), $R_{PbPb}\approx$~0.1 (0.5), at $p_T$ = 10 (100) GeV/$c$.\\
%compared to $N_{\rm coll}$-scaled p-p collisions.\\

\begin{figure}[htbp]
\begin{centering}
%\resizebox*{5cm}{!}{\includegraphics{dNdpT_hadrons_PbPb_5500GeV_latt_tau01_eps0_650GeV.eps}}
\hspace{6mm}
\includegraphics[width=0.45\textwidth]{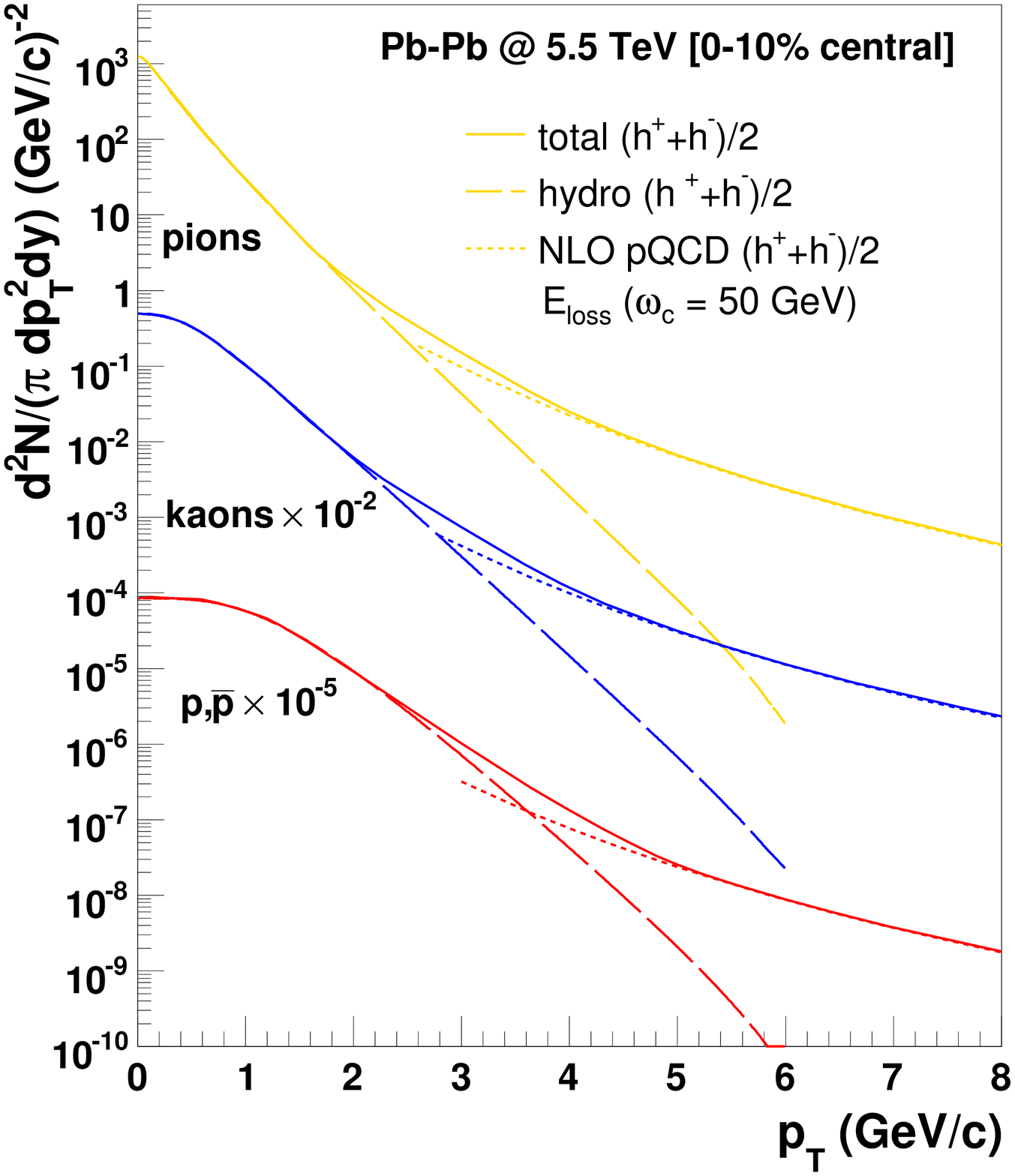}
\hspace{6mm}
\includegraphics[width=0.45\textwidth]{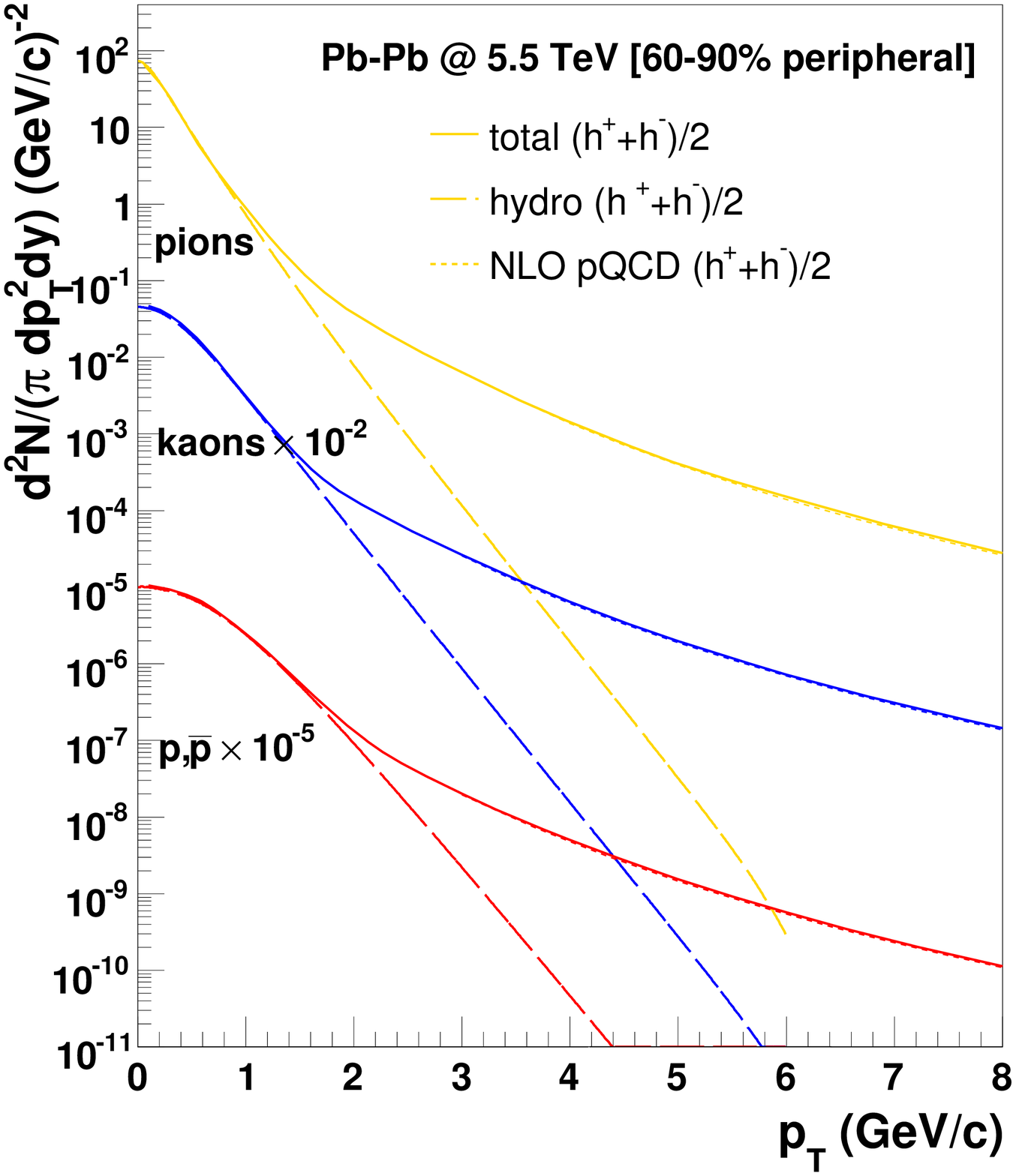}
\end{centering}
\vspace{-1cm}
\caption{\label{fig:spectra}Spectra at y=0 for $\pi^{\pm}$,$K^{\pm}$, $p,\bar{p}$ in 0-10\% central (left)
and 60-70\% peripheral (right) Pb-Pb at $\sqrtsnn$ = 5.5 TeV, obtained with hydrodynamics + (quenched) pQCD calculations.}
\end{figure}

%Our prediction for the identified hadron spectra in central (peripheral) Pb-Pb collisions at 5.5 TeV 
%is shown in the left (right) plot of Fig.~\ref{fig:spectra}. The hydrodynamical contribution 
Our predictions for the identified hadron spectra in Pb-Pb collisions at 5.5 TeV are shown 
in Figure~\ref{fig:spectra}. The hydrodynamical contribution dominates over the (quenched) pQCD 
one up to $p_T\approx$ 4 (1.5) GeV/$c$ in central (peripheral) Pb-Pb. As expected, the hydro-pQCD 
$p_T$ crossing point increases with the hadron mass. In the absence of recombination effects 
(not included here), bulk protons may be boosted up to $p_T\approx$ 5 GeV/$c$ in central Pb-Pb at the LHC.

%%%%%%%%%%%%%%%%%%%%%%%%%%%%%%%%%%%%%%%%%%%%%%%%%%%%%%%%%%%%%%%
\vspace{-2mm}

\subsection{Multiplicities at the LHC in a geometric scaling model}

%{\it N\'estor Armesto, Carlos A.~Salgado and Urs Achim Wiedemann}
{\it N.~Armesto, C.~A.~Salgado and U.~A.~Wiedemann}

{\small
We present predictions for charged multiplicities at mid-rapidity in PbPb collisions, as well as transverse momentum distributions at different pseudorapidities in pPb collisions,  at LHC energies. We use geometric scaling as found in lepton-proton and lepton-nucleus scattering, to determine the evolution of multiplicities with energy, pseudorapidity and centrality. The only additional free parameter required to obtain the multiplicities is fixed from RHIC data.
}
\vskip 0.5cm

Geometric scaling - the phenomenological finding that virtual photon-hadron cross sections in lepton-proton \cite{Stasto:2000er} and lepton-nucleus \cite{Armesto:2004ud,Albacete:2005ef} collisions, are functions of a single variable which encodes all dependences on Bjorken-$x$, virtuality $Q^2$ and nuclear size $A$ - is usually considered as one of most important evidences in favor of saturation physics at work \cite{Albacete:2005ef} . In the scaling variable $\tau_A=Q^2/Q_{s,A}^2(x)$ the quantity $Q_{s,A}$, the saturation momentum, shows a behavior with energy or Bjorken-$x$ determined by lepton-proton data, while  the dependence on $A$ is fixed by lepton-nucleus data \cite{Armesto:2004ud}:
\begin{equation}
Q_{s,A}^2(x) \propto x^{-\lambda} A^{1/(3\delta)}, \ \ \lambda=0.288,\ \ \delta=0.79\pm 0.02.
\label{eq1.salgado}
\end{equation}

\begin{figure}[h]
\begin{minipage}{0.5\textwidth}
\begin{center}
\includegraphics[width=7.cm]{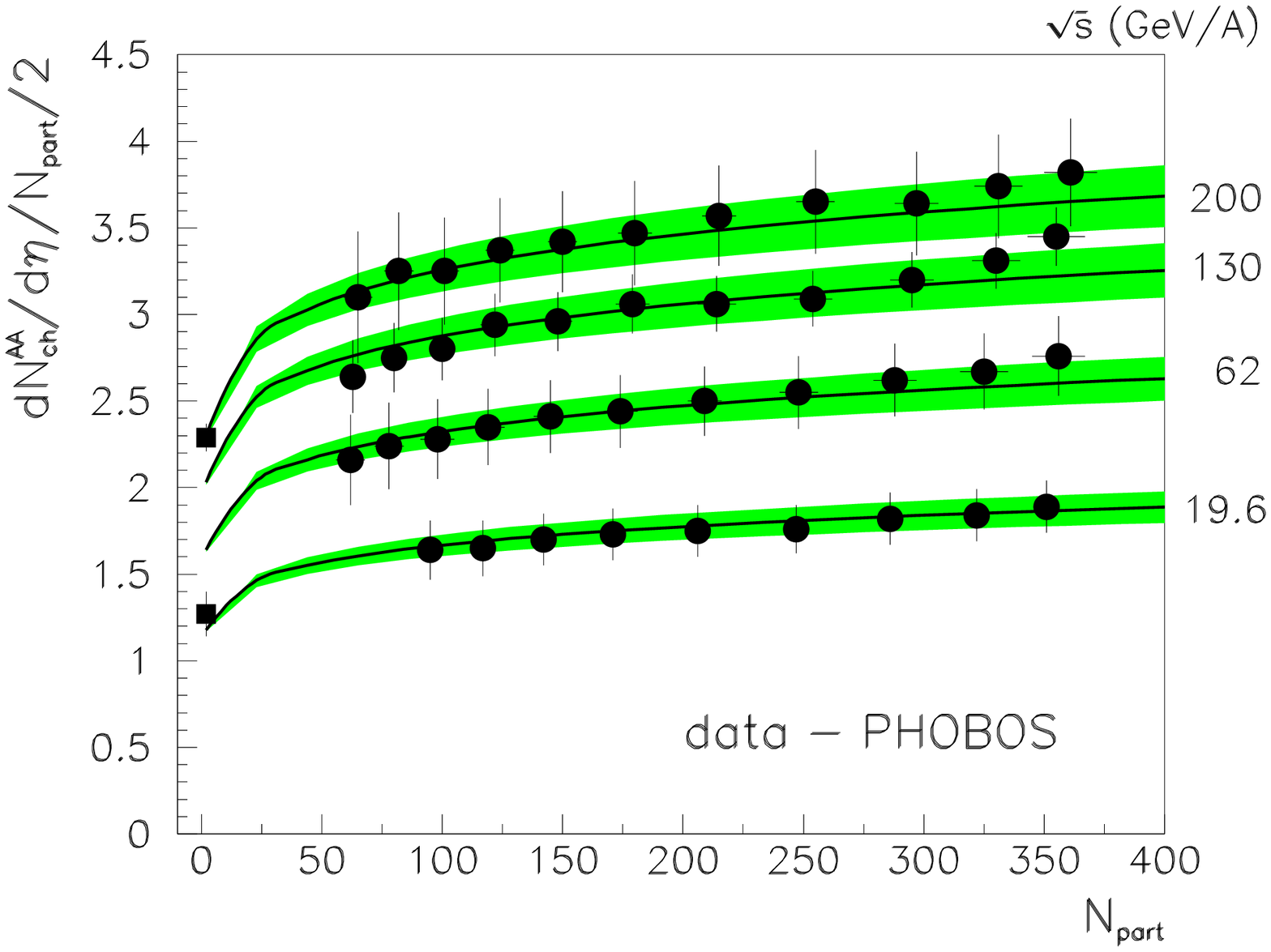}
\end{center}
\end{minipage}
\hskip -0.5cm
\begin{minipage}{0.5\textwidth}
\begin{center}
\includegraphics[width=7.cm]{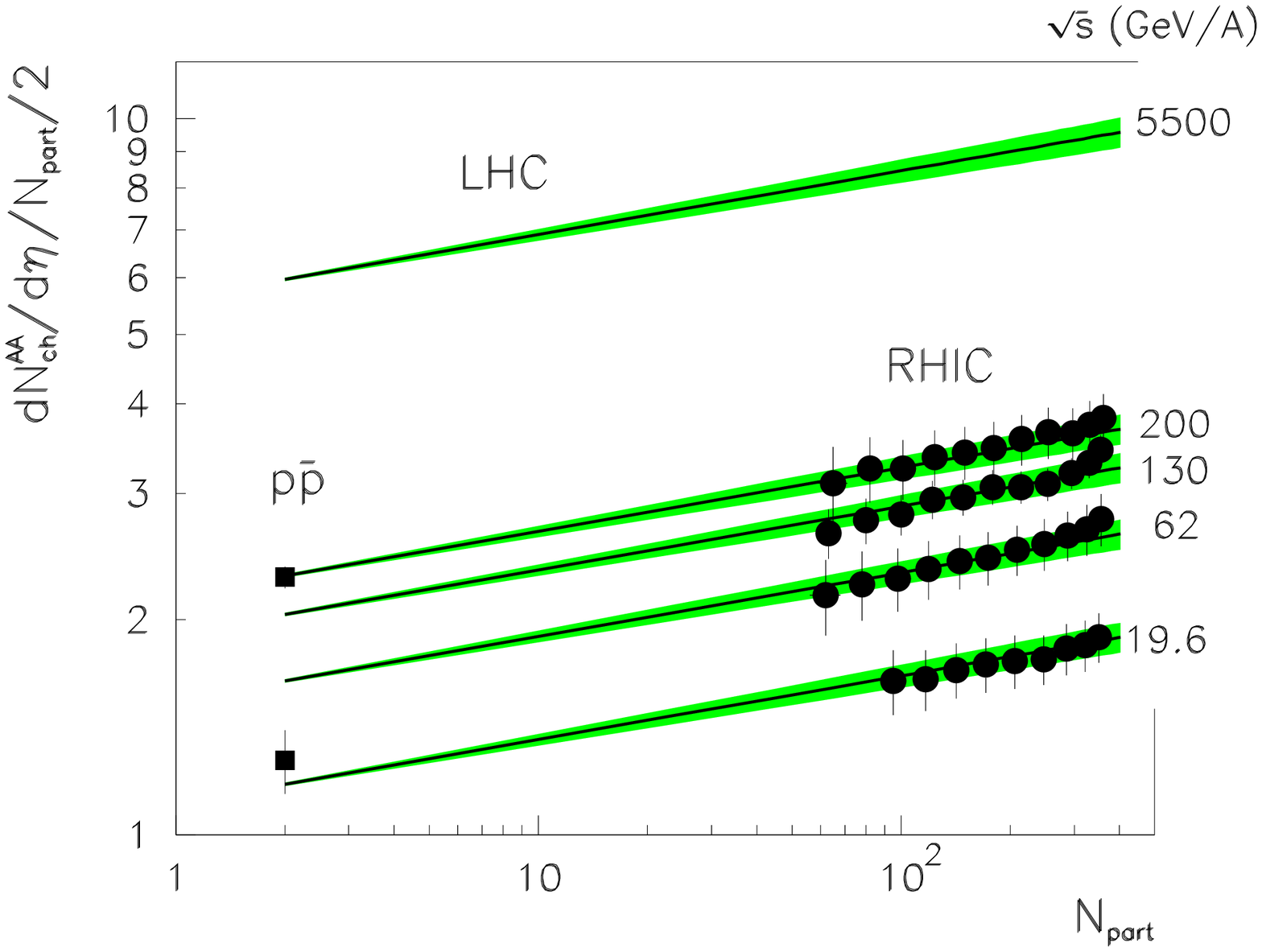}
\end{center}
\end{minipage}
\caption{Charged multiplicity at mid-pseudorapidity per participant pair, for four RHIC energies and for LHC energies, from Eq. (\ref{eqmult}). The band shows the uncertainty coming from $\delta$, see Eq. (\ref{eq1.salgado}).}
\label{fig:raa}
\end{figure}

To compute particle production, we assume that geometric scaling holds for the distributions assigned to the projectile and target. Without invoking factorization, dimensional analysis allows us to factor
out the geometrical information. Then, the multiplicity at central pseudorapidity can be written in the form \cite{Armesto:2004ud} (with $N_{\rm part} \propto A$)
\begin{equation}
\frac{2}{N_{\rm part}}
\frac{dN^{AA}}{d\eta}\Bigg\vert_{\eta\sim 0}=N_0\,\left(s/\rm{GeV}^2\right)^{\lambda/2}
N_{\rm part}^{\frac{1-\delta}{3\delta}}\, .
\label{eqmult}
\end{equation}
The only new parameter is $N_0$, a normalization constant which takes into account the parton-hadron conversion and the change from mid-rapidity to mid-pseudorapidity. Once fixed for a set of data ($N_0=0.47$), this formula has predictive power and establishes a factorization of the energy and centrality dependences in agreement with data. 
In Fig. \ref{fig:raa} we show the results of Eq. (\ref{eqmult}) compared to RHIC data (including those of intermediate energies \cite{Back:2005hs}) and our prediction for the LHC, where our numbers for $dN^{AA}/d\eta\vert_{\eta\sim 0}$ are $1550\div 1760$ for $N_{\rm part}=350$ and $1670 \div 1900$ for $N_{\rm part}=375$, with the range in the predictions reflecting the uncertainty coming from $\delta$, see Eq. (\ref{eq1.salgado}). We note that these values are based on a $\sqrt{s}$-powerlaw dependence
in Eq. (2) and can be discriminated clearly from a log-extrapolation of RHIC data.

%figure---------------------------------------------
%figure---------------------------------------------

\begin{figure}
\begin{center}
\includegraphics[width=0.47\textwidth]{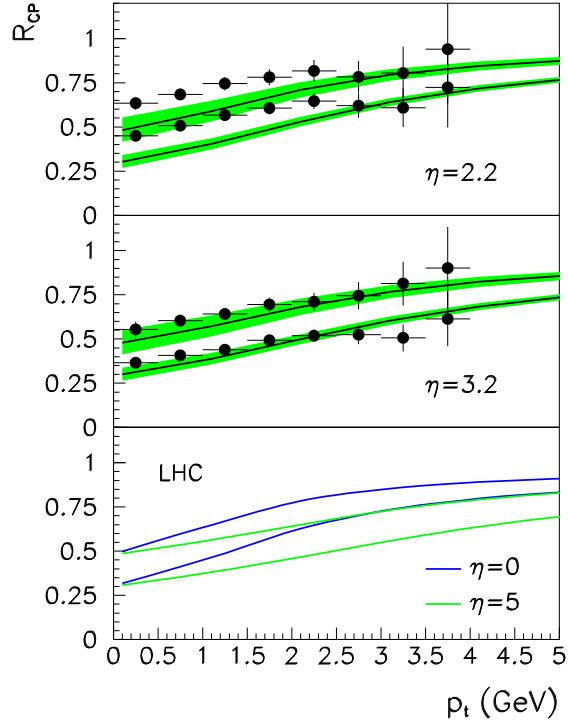}
\end{center}
\caption{$R_{CP}$ versus $p_t$, Eq. (\ref{eqratpt}), in dAu collisions at RHIC compared to experimental data (upper and middle plots), and in pPb collisions at the LHC (lower plot), for different pseudorapidities and centrality classes: central to peripheral (lower curves) and semicentral to peripheral (upper curves). The bands reflect the uncertainty in the definition of the centrality class.}
\label{fig:kinem}
\end{figure}

A more model-dependent application of this formalism \cite{Armesto:2004ud} concerns particle production in hadron-nucleus collisions at forward rapidities or large energies. Assuming factorization, geometric scaling and a steeply falling parton distribution in the proton or deuteron, one gets for one-particle distributions in two centrality classes $c_1$ and $c_2$
\begin{equation}
   \frac{dN^{\rm dAu}_{c_1}}{N_{\rm coll_1}d\eta d^2p_t}\left/
   \frac{dN^{\rm dAu}_{c_2}  }{N_{\rm coll_2}d\eta d^2p_t}\right.
   \approx
   \frac{N_{\rm coll_2}\phi_A(p_t/Q_{\rm sat_1})}
    {N_{\rm coll_1}\phi_A(p_t/Q_{\rm sat_2})}
   \approx
   \frac{N_{\rm coll_2}\Phi(\tau_1)}{N_{\rm coll_1}\Phi(\tau_2)}\, ,
   \label{eqratpt}
\end{equation}
where $\Phi$ is the geometric scaling function in lepton-hadron collisions.
In this way, particle ratios in hadron-nucleus collisions provide a check of parton densities in the nucleus through geometric scaling.
In Fig. \ref{fig:kinem} we show the results compared to RHIC data and our predictions for the LHC. The definition of the centrality classes is $N_{\rm coll_1}=13.6\pm0.3$ (central), $7.9\pm0.4$ (semicentral) and
$N_{\rm coll_2}=3.3\pm0.4$ (peripheral). The suppression for mid-pseudorapidities at the LHC turns out to be as large as that for forward pseudorapidities at RHIC.

%figure---------------------------------------------
%figure---------------------------------------------

\subsection{Multiplicity and cold-nuclear matter effects from Glauber-Gribov theory}

{\it I. C. Arsene, L. Bravina, A. B. Kaidalov, K. Tywoniuk and
E. Zabrodin}

{\small
We present predictions for nuclear modification factor in proton-lead
collisions at LHC energy 5.5 TeV from Glauber-Gribov theory of
nuclear shadowing. We
have also made predictions for baseline cold-matter nuclear effects in
lead-lead collisions at the same energy.
}

\subsubsection{Introduction}
The system formed in nucleus-nucles (AA)
collisions at LHC will provide further insight into the
dynamics of the deconfined state of nuclear matter. There are also 
interesting effects anticipated for the initial state of the incoming
nuclei related 
to shadowing of nuclear parton distributions and the space-time
picture of the interaction. These should be studied in the more
``clean'' environment of a proton-nucleus collision. The initial-state effects
constitute a baseline for calculation of the density of particles at
all rapidities and affect therefore also high-$p_\bot$ particle
suppression and jet quenching, as well as the total multiplicity.

Both soft and relatively high-$p_\bot$, $p_\bot < 10$ GeV/c, particle
production in pA at LHC energies probe the low-$x$ 
gluon distribution of the target nucleus at moderate scales, $Q^2
\sim p_\bot^2$, and is therefore mainly influenced by nuclear shadowing. In
the Glauber-Gribov theory \cite{Gribov:1968jf}, shadowing at low-$x$ is
related to diffractive structure functions of the nucleon, which are
studied at HERA. The space-time picture 
of the interaction is altered from a longitudinally ordered
rescattering at low energies, to a coherent interaction of the
constituents of the incoming wave-functions at high energy.
Shadowing affects both soft and hard processes. Calculation of gluon
shadowing was performed in our recent paper \cite{Tywoniuk:2007xy},
where Gribov
approach for the calculation of nuclear structure functions was used.
The Schwimmer model was used to account for higher-order
rescatterings. The gluon diffractive distributions are 
taken from the most recent experimental parameterizations \cite{Aktas:2006hx}.

\subsubsection{Particle production at LHC}
\begin{figure}[t!]
  \begin{center}
    \includegraphics[scale=.5]{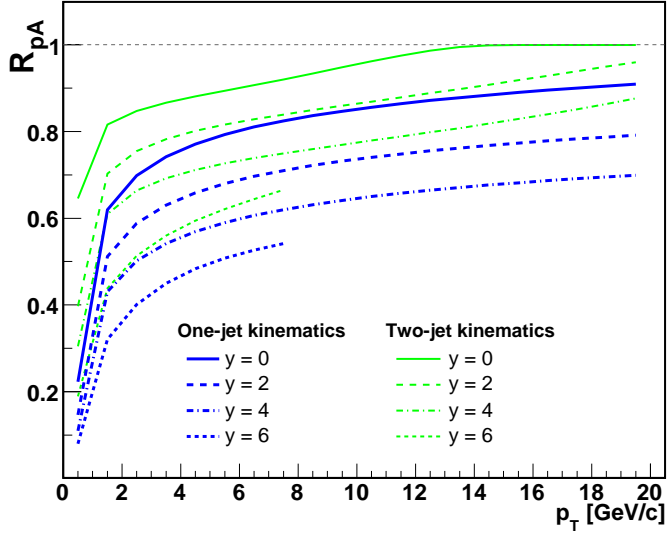}
  \end{center}
  \caption{Shadowing as a function of transverse for p+Pb collisions
    at $\sqrt{s}$ = 5.5 TeV.}
  \label{fig:R_pA}
\end{figure}

Shadowing will lead to a suppression both at mid- and forward rapidities in
p+Pb collisions at $\sqrt{s} = 5.5$ TeV as seen in
Fig.~\ref{fig:R_pA}. We have plotted the curves for two
distinct kinematical scenarios of 
particle production; one-jet kinematics which may be well motivated for
particle production at $p_\bot < 2$ GeV/c and two-jet kinematics that
apply for high-$p_\bot$ particle producion. The uncertainty in the
curves is due to uncertainty in the parameterization of gluon
diffractive distribution functions. Cronin effect is not included in
the curves of Fig.~\ref{fig:R_pA}. We estimate it to be a 10\% effect
at these energies.

\begin{figure}[t!]
  \begin{minipage}[t]{1.\linewidth}
    \begin{center}
      \includegraphics[scale=.5]{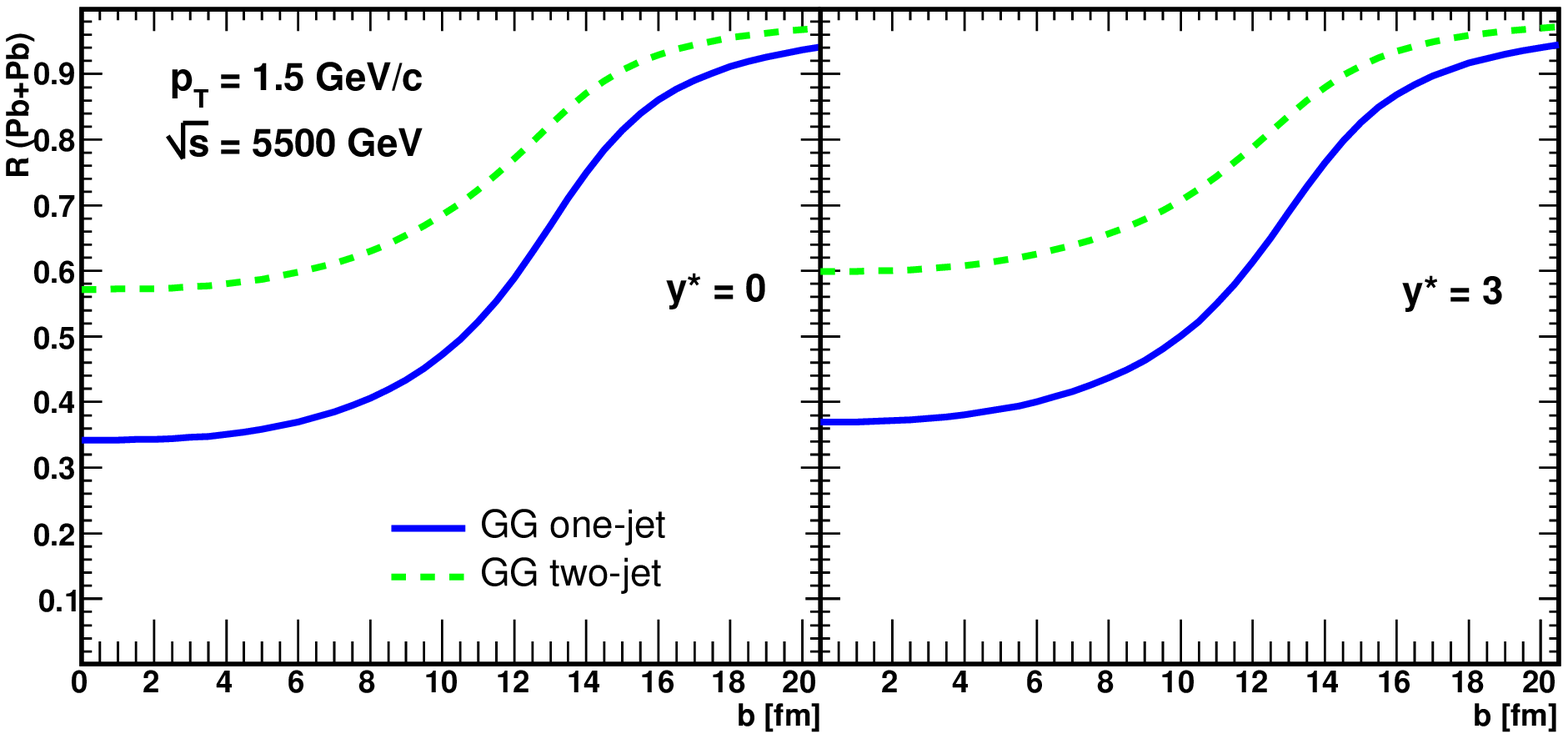}\\
      \includegraphics[scale=.4]{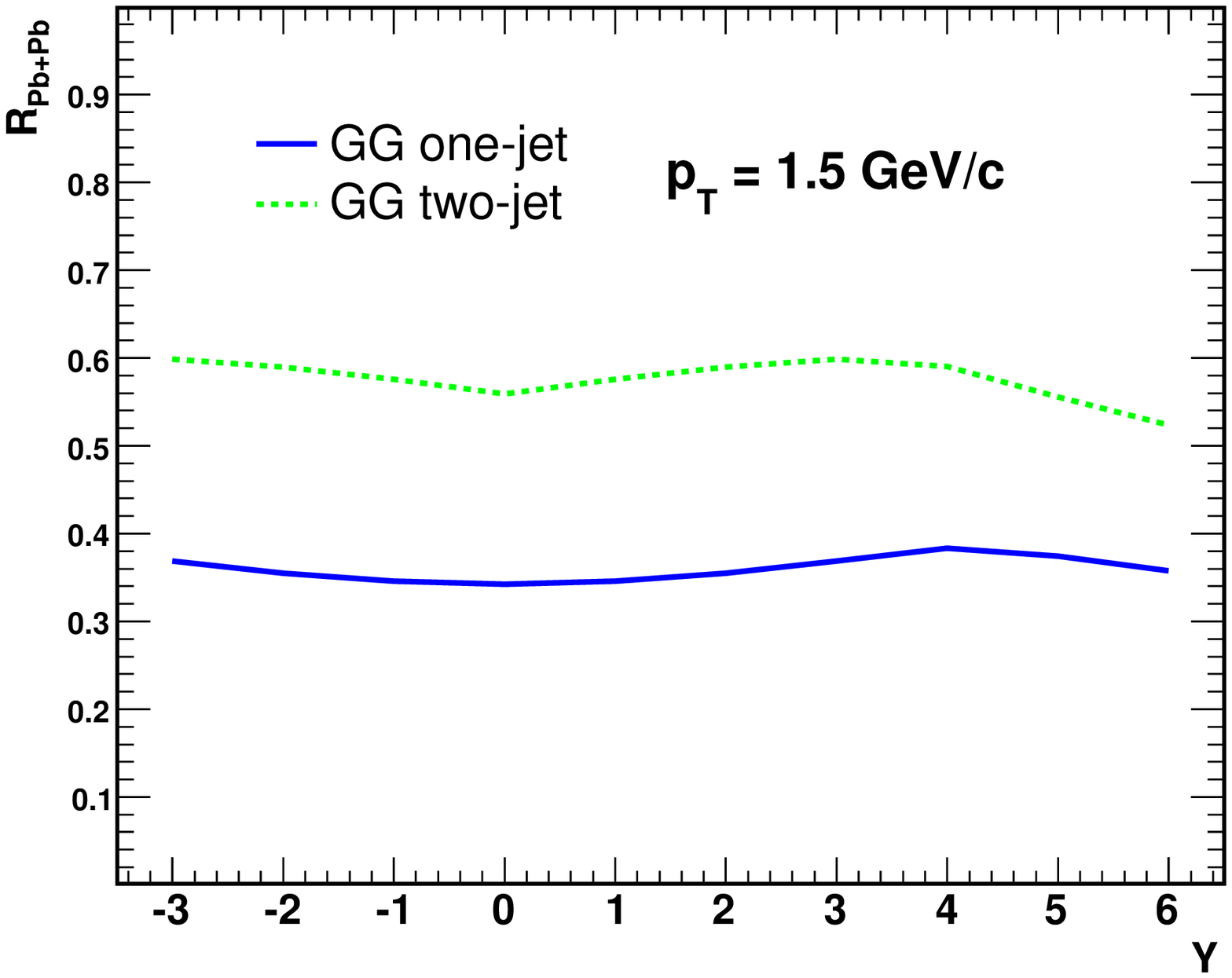}
    \end{center}
  \end{minipage}
  \caption{Shadowing as a function of centrality (top) and rapidity
    (bottom) for Pb+Pb collisions
    at $\sqrt{s}$ = 5.5 TeV.}
  \label{fig:ColdNucl2}
\end{figure}

In Fig.~\ref{fig:ColdNucl2} we present the suppresion due to
cold-nuclear effects in Pb+Pb collisions at $\sqrt{s} = 5.5$ TeV as a
function of centrality (top) and rapidity (bottom). Also here we
present the results for two kinematics.


\subsection{Stopping Power from SPS to LHC energies.}

{\it V.~Topor Pop, J.~Barrette, C.~Gale, S.~Jeon and
M.~Gyulassy}

{\small
We investigate the energy dependence of hadron production and 
of stopping power based on HIJING/B\=B v2.0 model calculations.
Pseudorapidity spectra and $p_T$ distributions 
for produced charged particles as well as net baryons (per pair of 
partcipants) and their
rapidity loss are compared to data at RHIC and predictions for
LHC energies are discussed.
}
\vskip 0.5cm

In previous papers \cite{Pop:2005uz} we studied 
the possible role of topological baryon junctions \cite{Kharzeev:1996sq}
\cite{Vance:1998vh},
and the effects of strong color field (SCF) 
in nucleus-nucleus collisions at RHIC energies.
In the framework of HIJING/B\=B v2.0 model,
the new algorithm for junction anti-junction J\=J loops provide a possible 
explanation for baryon/meson anomaly. 
The SCF effects as implemented within our model  
gives a better description of this anomaly. 
At LHC energies, due to higher initial energy density    
(or temperature) we expect an increase of 
the mean value of the string tension ($\kappa$) \cite{Pop:these}.

The day 1 measurements at the LHC will include 
results on multiplicity distributions 
with important consequences for our understanding 
of matter produced in the collisions \cite{Armesto:2000xh},
\cite{Wiedemann:2007ss}. 
From our model calculations one expects 
dN$^{\rm ch}_{\rm PbPb}$/d$\eta$ $\approx$ 3500 at $\eta =0$
in central (0-5 \%) Pb +Pb collision. 
This correspond to $\approx$ 17.5 produced charged hadrons 
per participant pair. 
These values are higher than those obtained 
by requiring that both limiting fragmentation and the trapezoidal shape
of the pseudo-rapidity distribution persist at the LHC
\cite{Wiedemann:2007ss}.
Our model predicts a characteristic violations of the apparently universal
trend, seen up to maximum RHIC energy. In contrast saturation models 
\cite{Kharzeev:2004if} offer a justification 
for the predicted very weak $\sqrt{s_{\rm NN}}$ dependence 
of event multiplicity.

%%%%%%%%%\section{ Multi}

\begin{figure}[h]
\vspace*{-0.2cm}
\begin{center}
\includegraphics*[width=11.8cm,height=5.9cm]{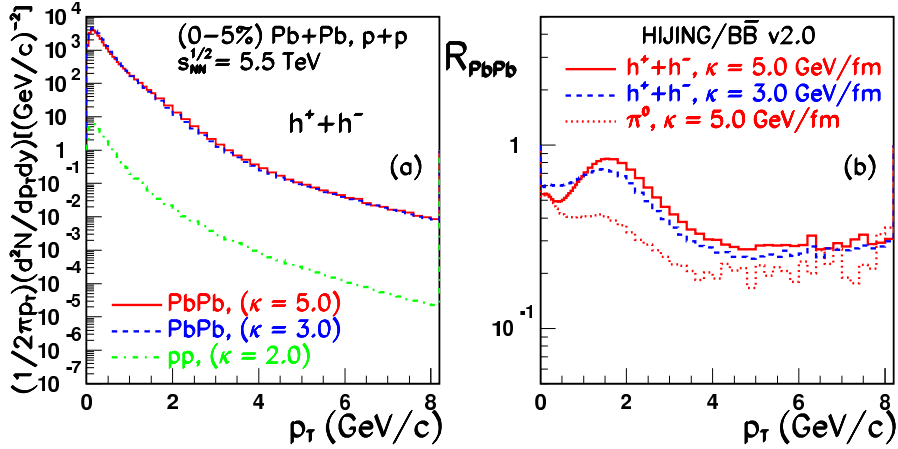}
\end{center}
\vspace*{-0.3cm}
\caption{Left: HIJING/B\=B v2.0 predictions
for $p_T$ spectra at mid-rapidity of total inclusive charged hadrons
for central (0-5\%) Pb+Pb and $p+p$ collisions. Right: 
Predicted nuclear modification 
factors for charged hadrons and for neutral pions.}
\label{chpbpb}
\end{figure}

Figure \ref{chpbpb} presents predictions for $p_T$ spectra at midrapidity
and NMF $R_{\rm PbPb}^{\rm ch}$ of total inclusive charged hadrons
for central (0-5\%) Pb+Pb and $p+p$ collisions at $\sqrt{s_{\rm NN}}$ = 
5.5 TeV. The predicted NMF $R_{\rm PbPb}^{\pi^0}$ of neutral pions 
is also presented.
From our model calculations we conclude that baryon/meson anomaly, will
persist at the LHC with a slight increase for increasing strength
of the chromoelectric field ($\kappa = e_{eff} E$). 
A somewhat higher sensitivity to $\kappa$ is obtained for
NMF of identified particles \cite{Pop:these}. 

\begin{figure}[h]
\vspace*{-0.2cm}
\begin{center}
\includegraphics*[width=11.8cm,height=5.9cm]{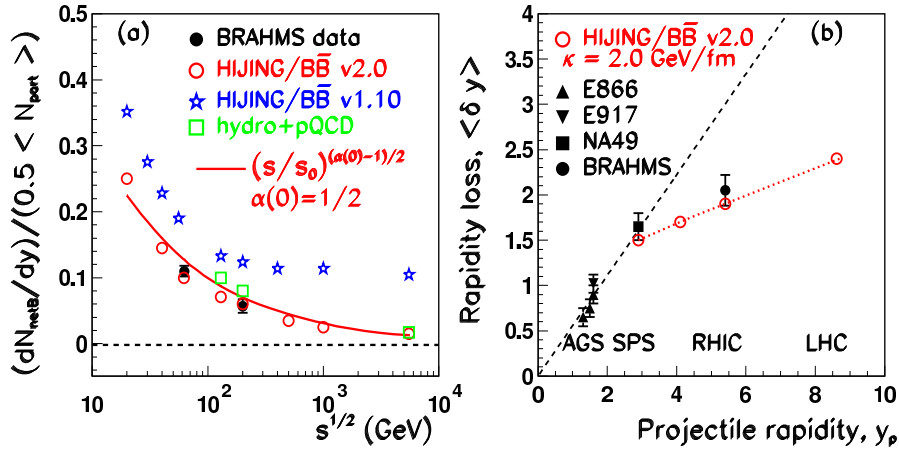}
\end{center}
\vspace*{-0.3cm}
\caption{Left: HIJING/B\=B v2.0 predictions for net-baryon 
(per participant pair) at mid-rapidity as function of $\sqrt{s_{\rm NN}}$. 
Right: Average rapidity loss versus beam rapidity. 
The data and dashed line extrapolation are from ref. \cite{Videbaek:2005qn} 
and from BRAHMS \cite{Bearden:2003hx}.}
\label{stlhc}
\end{figure}

%Because baryon number is conserved, the measured baryon distribution 
%retains information about energy loss and allows the degree 
%of nuclear stopping to be determined. 
The net-baryon rapidity distribution measured at RHIC 
is both qualitatively and quantitatively
very different from those at lower energies indicating that a significantly 
different system is formed near mid rapidity \cite{Videbaek:2005qn}.
Fig. \ref{stlhc} (left panel) presents the energy dependence  of net-baryon at 
mid-rapidity per participant pair. 
Shown are the results for central (0-5\%) Au+Au collisions, 
which indicate a net decrease with increasing energy.
This picture, corroborated with an increase of the ratio $\bar{p}/p$ to
$\approx$ 1 suggests that the reaction at the LHC is more transparent
in contrast to the situation at lower energy.
For central (0-5\%) Pb+Pb collisions 
at $\sqrt{s_{\rm NN}}$ = 5.5 TeV, our prediction 
for net-baryon per participant pair is  
$\approx$ 0.065 with $N_{\rm part}$ = 398, assuming $\kappa$ = 5 GeV/fm.
Similar values (open squares) are obtained within
pQCD+hydro model \cite{Eskola:2005ue}. However, this model 
predicts (Fig. 15 from ref. \cite{Eskola:2005ue}) much steeper slopes of 
charged hadron $p_T$ spectra.

In our model the main mechanisms for baryon production are 
quark di-quark ($q-{\rm qq}$) strings fragmentation and
J\=J loops in which baryons are produced approximatively in pairs. 
The energy dependence is $\propto \, (s/s_0)^{-1/4 +\Delta/2}$ 
similar with those predicted in ref. \cite{Kharzeev:1996sq} (eq. 11)
with the assumption that J\=J is a dominant mechanisms. 
This dependence is obtained if we choose for the parameters:
$s_0$ = 1 GeV$^2$ the usual parameter  of Regge theory,
$\alpha(0)=1/2$ the reggeon ($M_0^J$) intercept and
$\alpha_{P}(0) = 1 + \Delta$ (where $\Delta \approx 0.01$) for the
pomeron intercept. 
If confirmed, the measurements at LHC energies will
help us to determine better these values.
In contrast, results from HIJING/B\=B v1.10 model 
\cite{Vance:1998vh} (star symbol) 
give a slow energy dependence with a higher pomeron intercept 
$\alpha_{P}(0) = 1 + 0.08$ and over-estimate the stopping 
in the entire energy region.

Baryon conservation in the reactions can be used to predict rapidity 
loss and the energy loss per baryon.
The results are illustrated in Fig.~\ref{stlhc} (right panel)
for average rapidity loss 
$<\delta y>$ defined as in ref. \cite{Pop:2005uz}.
The predicted values for RHIC and LHC energies, 
clearly depart from the linear extrapolation
for constant relative rapidity loss \cite{Videbaek:2005qn}, which 
seems to be valid  only at lower energies 
($\sqrt{s_{\rm NN}}\, \leq \,$ 20 GeV).

%%%%%%%\section{Summary}

%\myack
%{
%This work was partly supported by the Natural Sciences and Engineering 
%Research Council of Canada and by the U. S. DOE 
%under Contract No. DE-AC03-76SF00098 and DE-FG02-93ER-40764.
%One of us (MG), gratefully acknowledges partial 
%support also from FIAS and GSI, Germany.
%}

\subsection{Investigating the extended geometric
  scaling region at LHC with polarized and unpolarized final states}

%{\it Dani\"el Boer, Andre Utermann, Erik Wessels}
{\it D.~Boer, A.~Utermann and E.~Wessels}

{\small
We present predictions for charged hadron production and $\Lambda$
polarization in $p\,$-$p$ and $p\,$-$Pb$ collisions at the LHC using
the saturation inspired DHJ model for the dipole cross section in the
extended geometric scaling region.
}
\vskip 0.5cm

At high energy, scattering of a particle off a nucleus can
be described in terms of a colour dipole scattering off small-$x$
partons, predominantly gluons, in the nucleus. At very high energy
(small $x$), the dipole amplitude starts to evolve nonlinearly with
$x$, leading to saturation of the density of these small-$x$ gluons.
The scale associated with this nonlinearity, the saturation scale
$Q_s(x)$, grows exponentially with $\log(1/x)$.

The nonlinear evolution of the dipole amplitude is expected to be
characterized by geometric scaling, which means that the dipole
amplitude depends only on the combination $r_t^2Q^2_s(x)$, instead of
on $r_t^2$ and $x$ independently. Moreover, the scaling behaviour is
expected to hold approximately in the so-called extended geometric
scaling (EGS) region between $Q_s^2(x)$ and $Q^2_{gs}(x)\sim
Q_s^4(x)/\Lambda^2$.

The small-$x$ DIS data from HERA, which show geometric scaling, were
successfully described by the GBW model \cite{Golec-Biernat:1998js}.
To describe the RHIC data on hadron production in $d\,$-$Au$ in the
EGS region a modification of the GBW model was proposed by Dumitru,
Hayashigaki and Jalilian-Marian (DHJ), incorporating scaling
violations in terms of a function $\gamma$\footnote{We note that at
  central rapidities we cannot reproduce exactly the results of
  \cite{Dumitru:2005kb} for large $p_t$. Therefore, a modification of
  the model may be needed to describe all RHIC
  data.} \cite{Dumitru:2005kb}. This DHJ model also describes $p\,$-$p$
data at forward rapidities \cite{Boer:2006rj}.

\subsubsection{DHJ model prediction for charged hadron production}

Using the DHJ model we can make a prediction for the $p_t$-spectrum of
charged hadron production in both $p\,$-$Pb$ and $p\,$-$p$ collisions
at the LHC, at respectively $\sqrt{s}=8.8$ TeV and $\sqrt{s}=14$ TeV.
Figure \ref{fig.wessels}a shows the minimum bias invariant yield for an
observed hadron rapidity of $y_h=2$ in the centre of mass frame, which
for $1$ GeV $\lesssim p_t\lesssim 10$ GeV predominantly probes the EGS
region.  We note that at this rapidity the result is not sensitive to
details of the DHJ model in the saturation region $r_t^2>1/Q^2_s$.
Further, from \cite{Dumitru:2005kb} we expect that $p_t$-independent
$K$-factors are needed to fix the normalization. We conclude that the
LHC data on hadron production in both $p\,$-$Pb$ and $p\,$-$p$
collisions will provide valuable data to further study the dipole
scattering amplitude near the onset of saturation, particularly the
behaviour of the function $\gamma$, which is discussed in e.g.\
\cite{Boer:2007wf}.

\subsubsection{DHJ model prediction for $\Lambda$ polarization}

\begin{figure}[htb]
\includegraphics*[height=53mm]{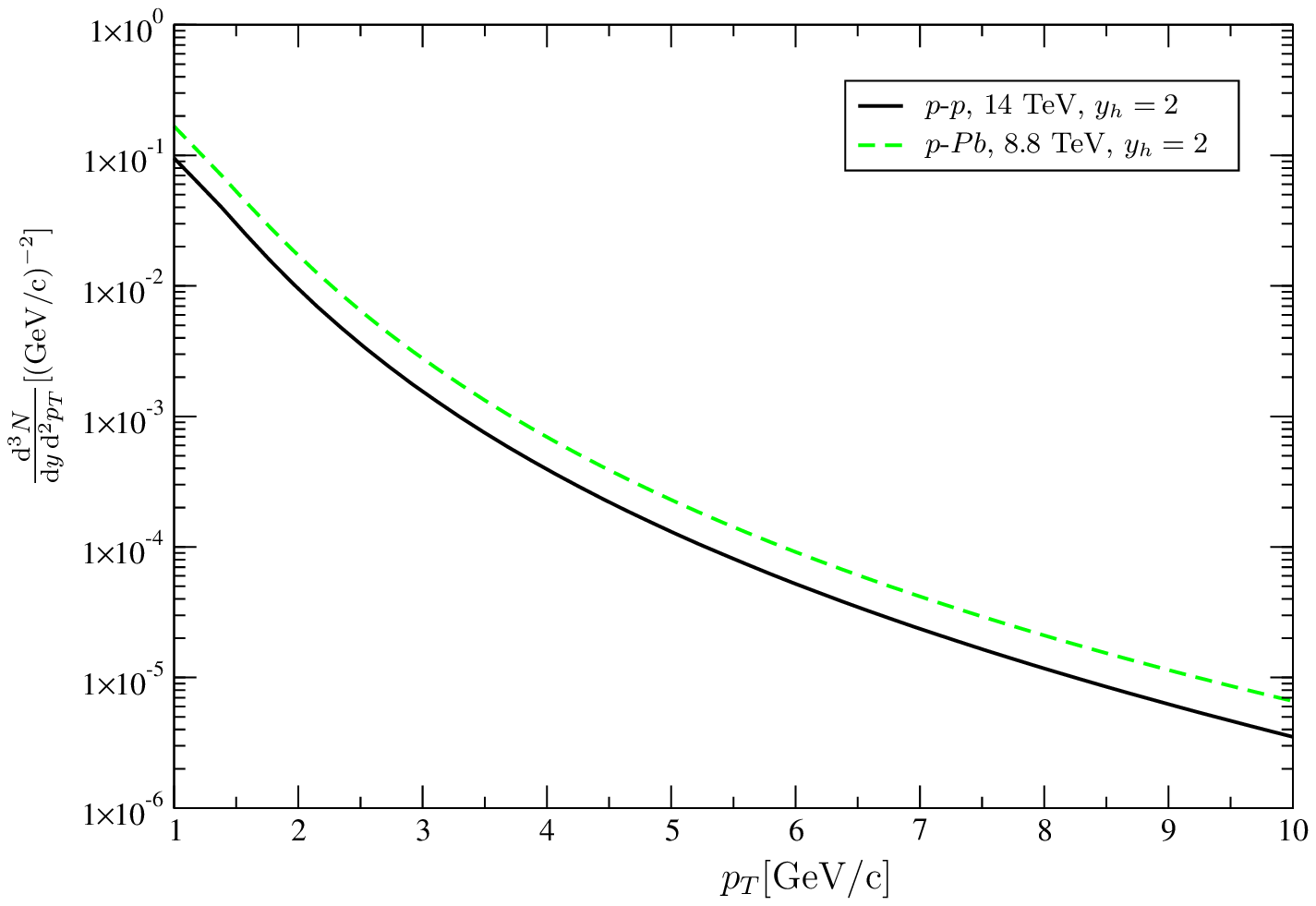}
\includegraphics*[height=53mm]{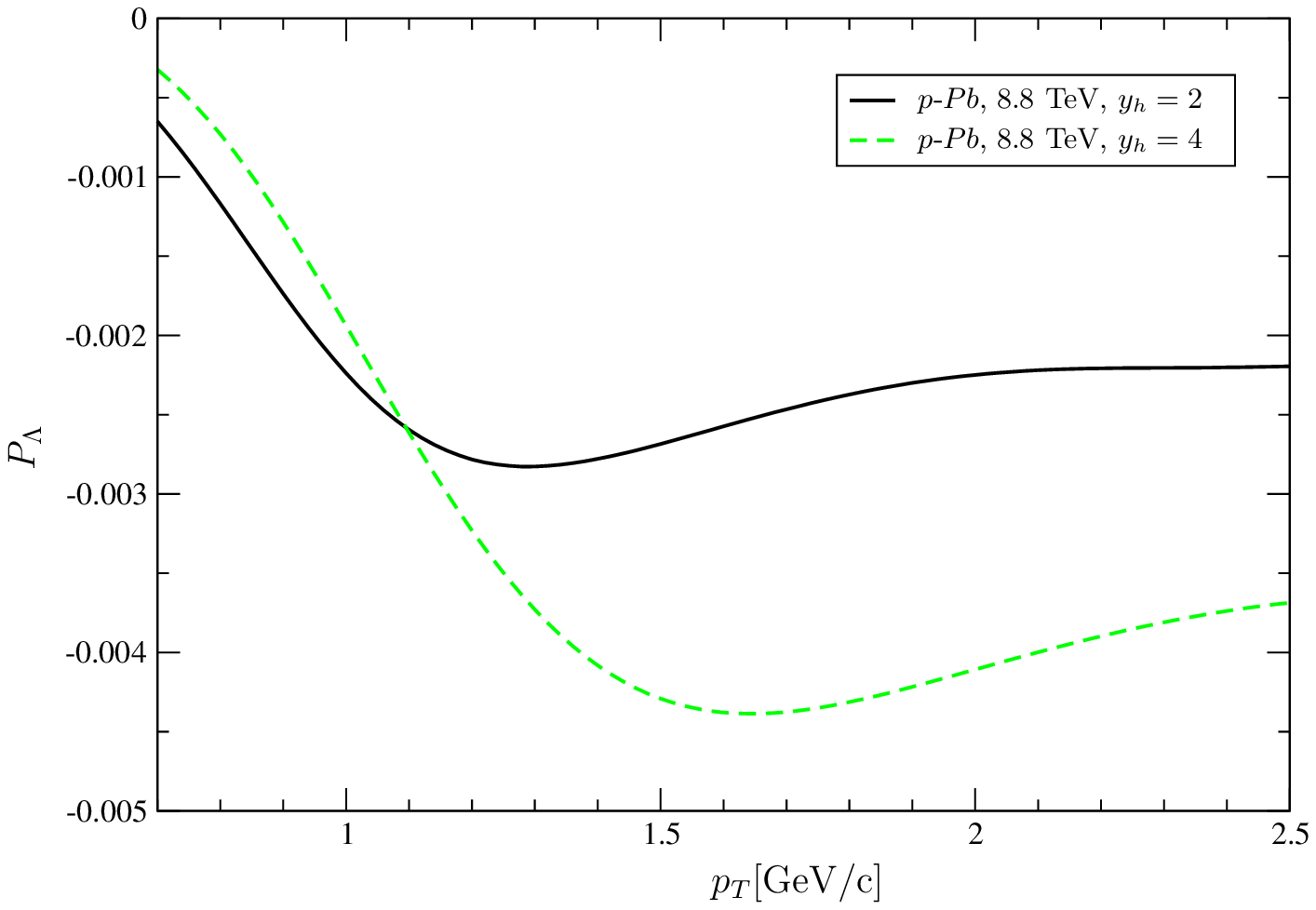}
\caption{\label{fig.wessels}a.\ Charged hadron production. b.\ $\Lambda$
  polarization. In both plots, $A_{\rm eff}=20$, and parton
  distributions and fragmentation functions of \cite{Dumitru:2005kb}
  and \cite{Boer:2002ij} were used.}
\end{figure}

Another interesting small-$x$ observable is the polarization of
$\Lambda$ hyperons produced in $p\,$-$A$ collisions, $P_\Lambda$. This
polarization, oriented transversely to the production plane, was shown
to essentially probe the derivative of the dipole scattering
amplitude, hence displaying a peak around $Q_s$ when described in the
McLerran-Venugopalan model \cite{Boer:2002ij}.  If this feature
persists when $x$-evolution of the dipole scattering amplitude is
taken into account, $P_\Lambda$ would be a valuable probe of
saturation effects. Using the DHJ model for the $x$-evolution of the
scattering amplitude, we find that $P_\Lambda$ displays similar
behaviour as in the MV model. This is depicted for fixed $\Lambda$
rapidities of 2 and 4 in figure \ref{fig.wessels}b. The position of the peak
scales with the average value of the saturation scale $\langle
Q_s(x)\rangle$. In the plotted region, the peak is located roughly at
$\langle Q_s(x)\rangle/2$.

The figure also shows that, like in the MV model, in the DHJ model
$|P_\Lambda|$ scales approximately linearly with $x_F$, which means
that at the LHC it is very small due to $\sqrt{s}$ being very large:
rapidities around 6 are required for $P_\Lambda$ to be on the $1\%$
level, although there is a considerable model uncertainty in the
normalization.

We conclude that the polarization of $\Lambda$ particles in $p\,$-$Pb$
collisions is an interesting probe of $\langle Q_s(x)\rangle$, but is
probably of measurable size only at very forward rapidities.

%\myack
%We thank Adrian Dumitru and Jamal Jalilian-Marian for helpful
%discussions.

\subsection{Inclusive distributions at the LHC as predicted from the DPMJET-III 
  model with chain fusion}
\label{s:Bopp3}

{\em F. Bopp, R. Engel, J. Ranft and S. Roesler}

{\small
DPMJET-III with chain fusion is used to calculate inclusive distributions of 
Pb-Pb collisions at LHC energies.
We present rapidity distributions as well as scaled multiplicities at 
mid-rapidity as function of the collision energy and the number of participants.
}
\vskip 0.5cm

Monte Carlo codes based on the two--component Dual Parton Model (soft hadronic 
chains and hard hadronic collisions) are available since 10--20 years:
The present codes are PHOJET for $hh$ and $\gamma h$ collisions~\cite{%
  Engel:1994vs,Engel:1995yd} and DPMJET-III based on PHOJET for $h\A$ and $\A\A$
collisions~\cite{Roesler:2000he}.
To apply DPMJET--III to central collisions of heavy nuclei, the percolation and 
fusion of the hadronic chains had to be implemented~\cite{Ranft:2003xxx}.

\begin{figure}
\centerline{\includegraphics[width=14cm]{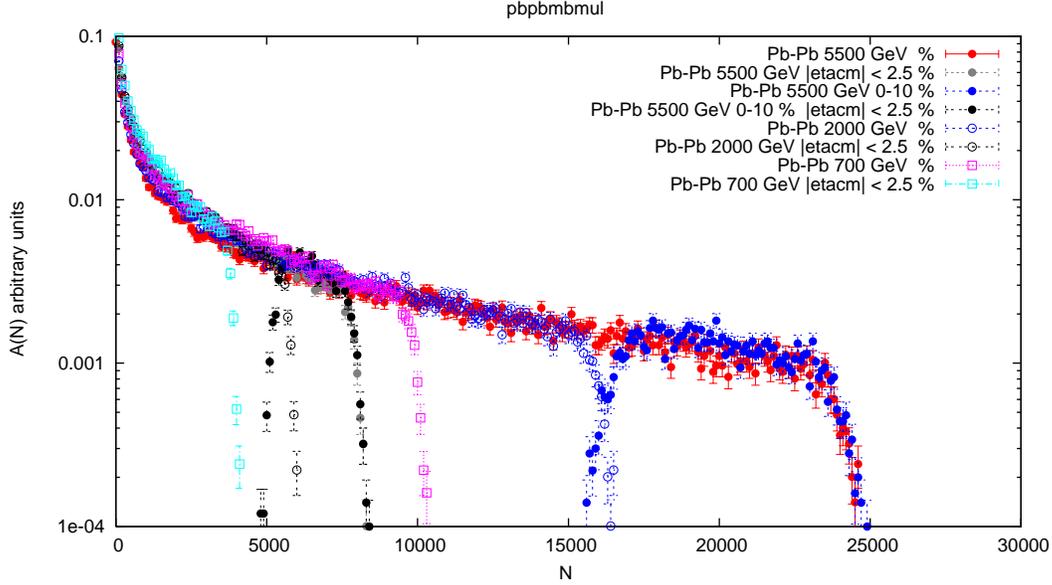}}
\caption{Multiplicity distributions in minimum bias and 0-10\% central 
  collisions in Pb-Pb collisions in the full $\eta_{\rm cm}$ range and for 
  $|\eta_{\rm cm}| \leq 2.5$ (from DPMJET-III).}
\label{fig:Bopp3-fig1}
\end{figure}

\begin{figure}
\includegraphics[width=7cm]{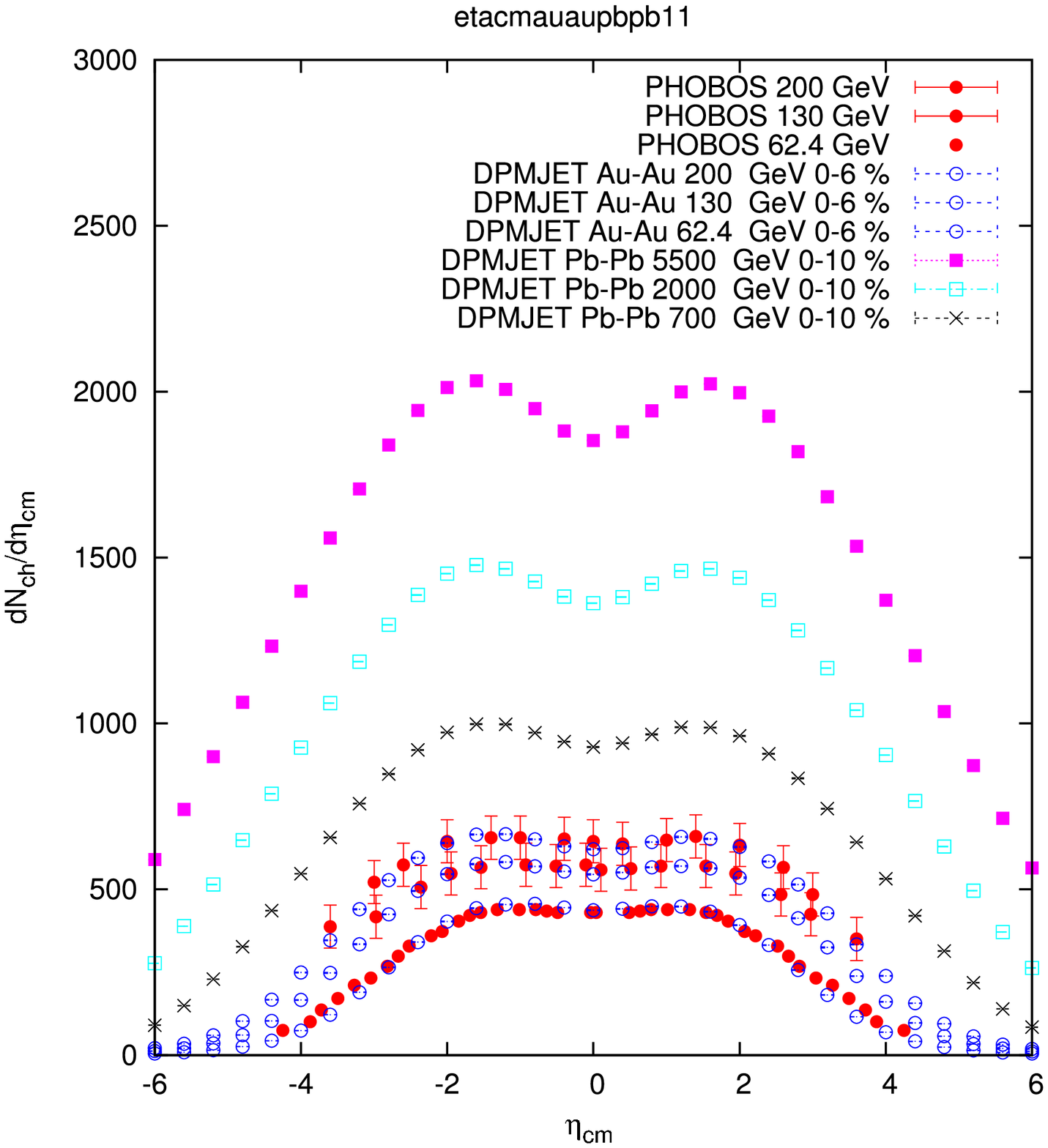}\hfill
\includegraphics[width=7cm]{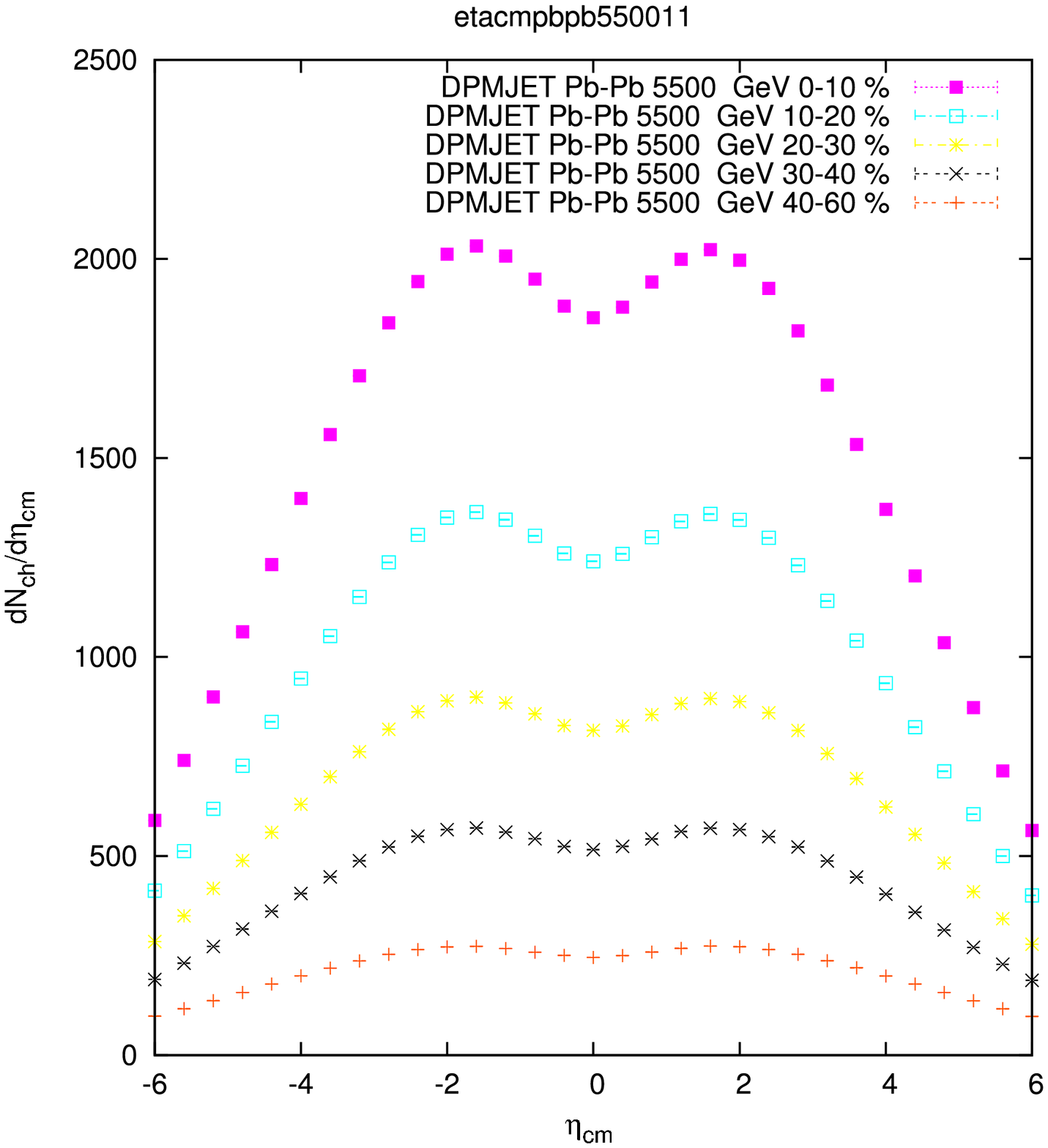}\\
\caption{(left) Central RHIC and LHC collisions. 
  (right) LHC Pb-Pb collisions from DPMJET-III.} 
\label{fig:Bopp3-fig2}
\end{figure}

In figures~\ref{fig:Bopp3-fig1} and \ref{fig:Bopp3-fig2} we apply this model to 
minimum bias and central collisions of heavy nuclei at the LHC and at RHIC. 
We find an excellent agreement to RHIC data on inclusive distributions.

\begin{figure}
\includegraphics[width=7cm]{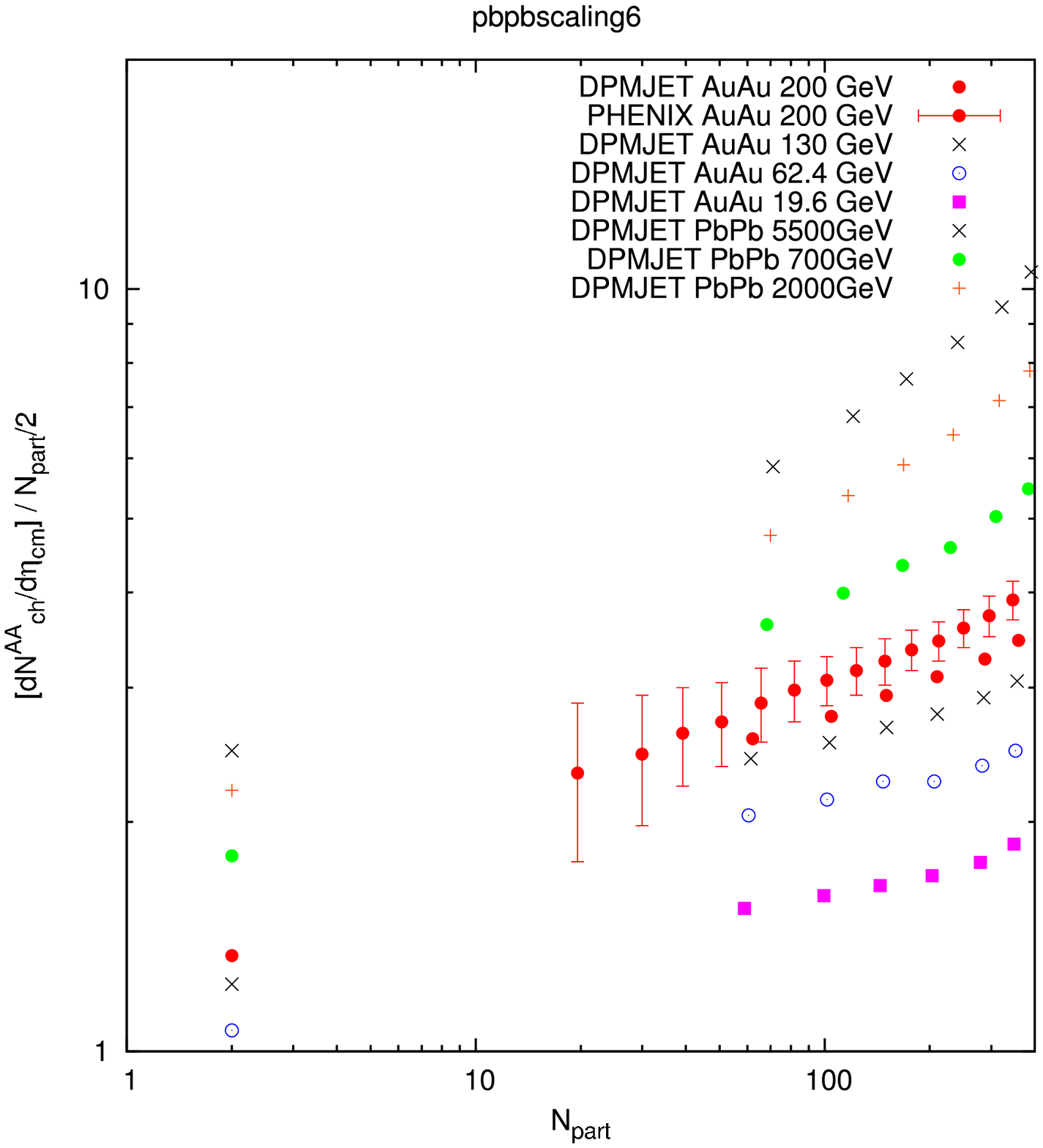}\hfill
\includegraphics[width=7cm]{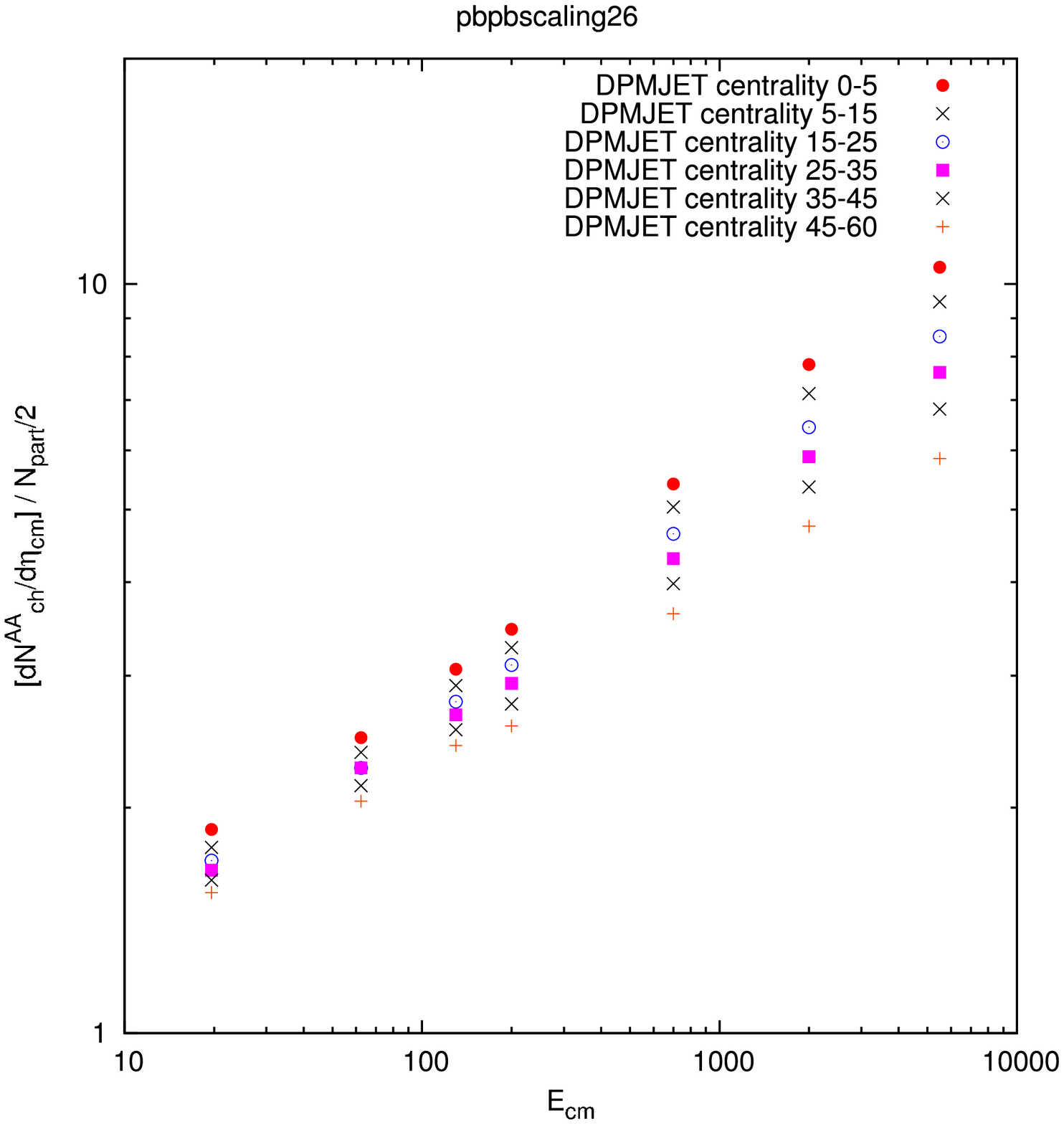}\\
\caption{$\frac{{\rm d}N}{{\rm d}\eta_{\rm cm}}/\frac{N_{\rm part}}{2}$ (left) over 
  $N_{\rm part}$ (right) over $E_{\rm cm}$, Pb-Pb and Au-Au collisions.}
\label{fig:Bopp3-fig3}
\end{figure}

The behaviour of the inclusive hadron production becomes particular simple if 
we plot it in the form 
$\frac{{\rm d}N}{{\rm d}\eta_{\rm cm}}/\frac{N_{\rm part}}{2}$. 
$N_{\rm part}$ is the number of participants in the $\A\A$ collisions. 
In figure~\ref{fig:Bopp3-fig3} we plot this quantity as function of 
$N_{\rm part}$ and as function of $E_{\rm cm}$, in both plots we find a rather 
simple behaviour.  

\begin{figure}
\includegraphics[width=7cm]{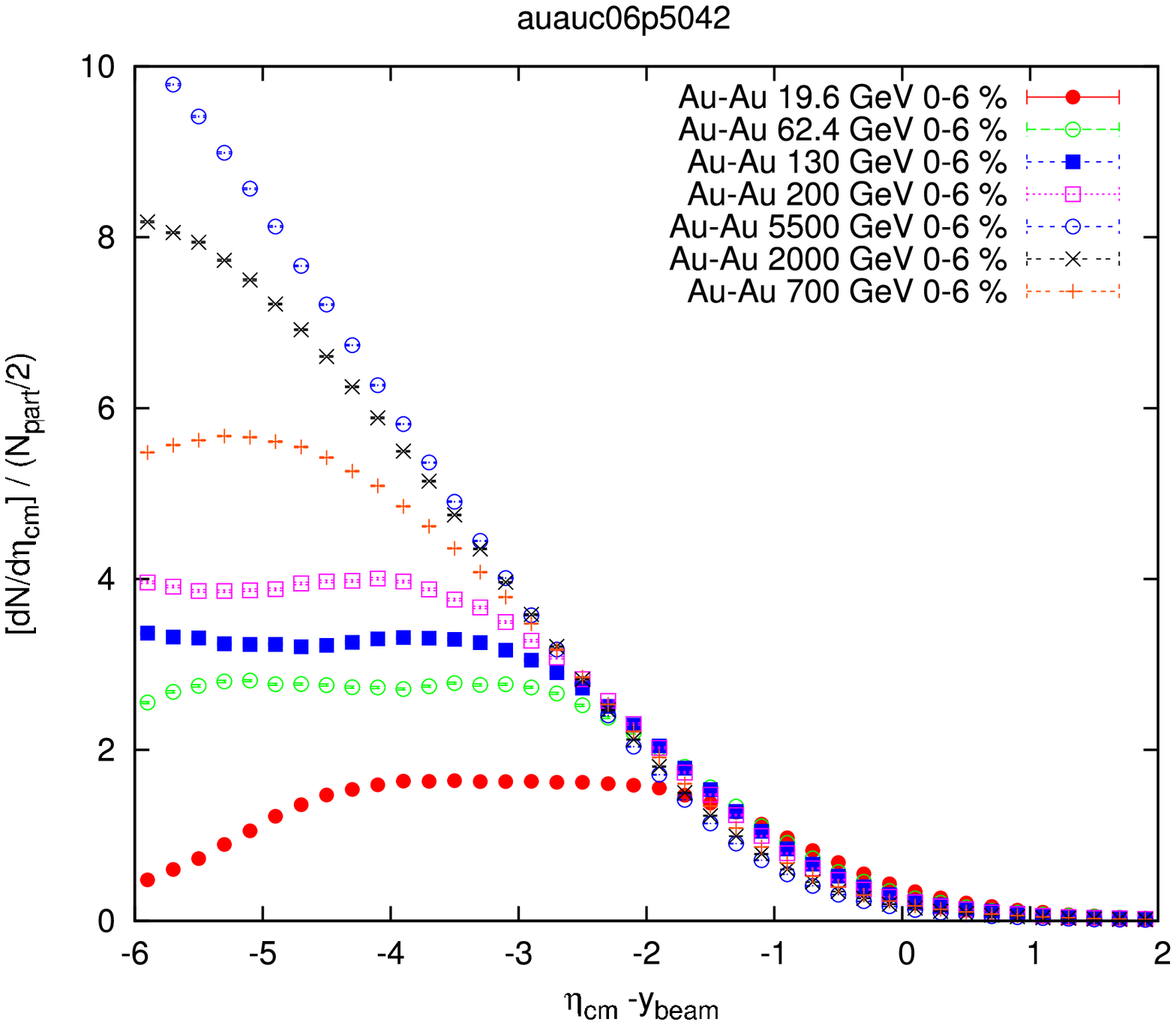}\hfill
\includegraphics[width=7cm]{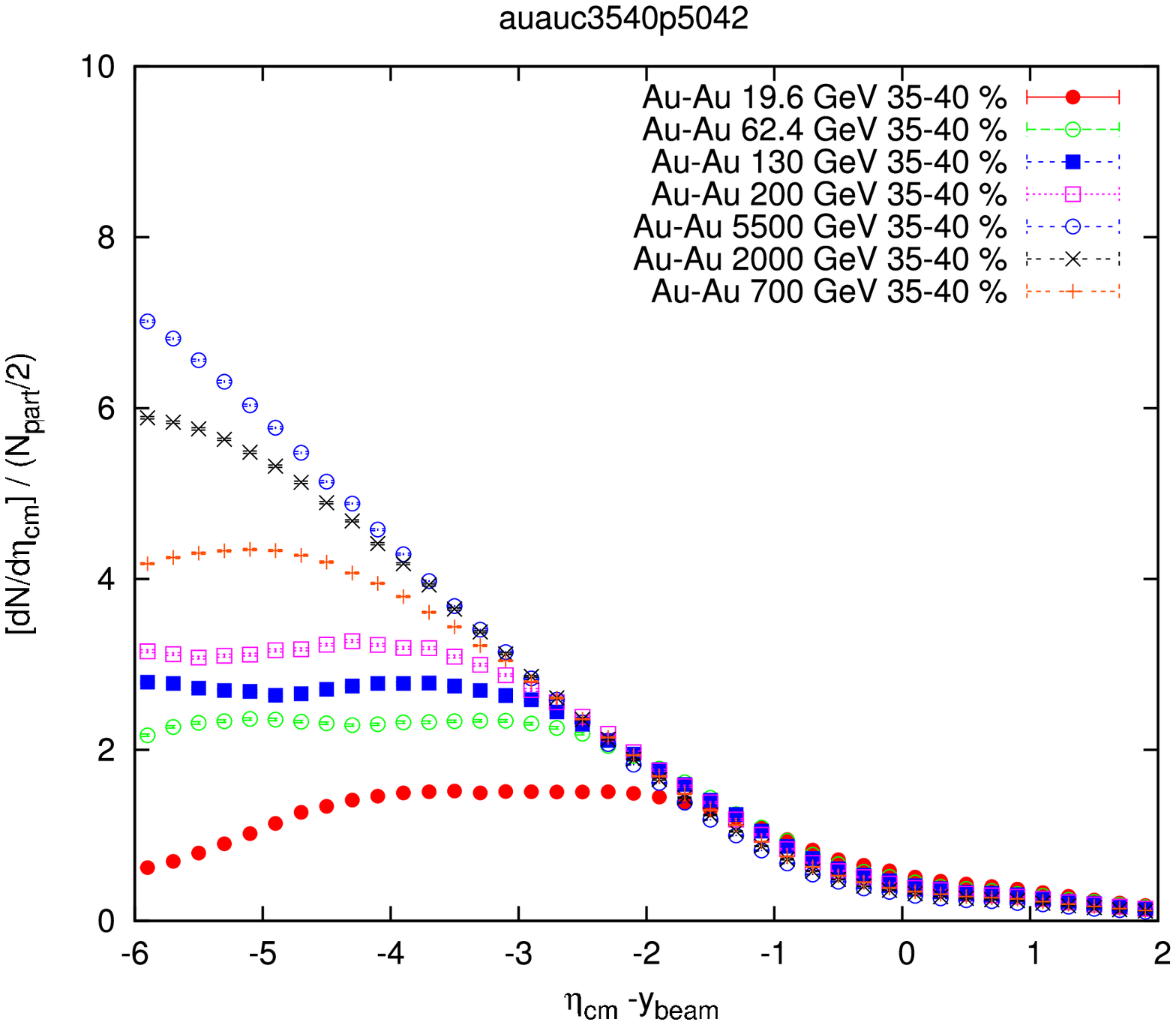}\\
\caption{$\frac{{\rm d}N}{{\rm d}\eta_{\rm cm}}/\frac{N_{\rm part}}{2}$ in Au-Au
  collisions over $\eta_{cm} - y_{beam}$ (left) central, (right) less central.} 
\label{fig:Bopp3-fig4}
\end{figure}

The limiting fragmentation hypothesis was proposed in 1969 by 
Benecke \etal~\cite{Benecke:1969sh}.
If we apply it to nuclear collisions we have to plot 
$\frac{{\rm d}N}{{\rm d}\eta_{\rm cm}}/\frac{N_{\rm part}}{2}$ as function of 
$\eta_{\rm cm}-y_{\rm beam}$. 
In figure~\ref{fig:Bopp3-fig4} we plot central and less central Au-Au collisions
at RHIC and LHC energies in this form. 
We find that DPMJET-III shows in the fragmentation region only small deviations 
from limiting fragmentation.

\subsection{Some ``predictions'' for PbPb and pp at LHC, based on the
extrapolation of data at lower energies}
\label{busza}

{\it W. Busza}
\vskip 0.5cm

The global characteristics of multiparticle production in pp, pA, AA
and even $e^+e^-$ collisions, over the entire energy range studied to
date, show remarkably similar trends.  Furthermore it is a fair
characterization of the data to say that the data appears simpler than
current explanations of it.  These trends allow us to
``predict'', with high precision, several important results which will
be seen in pp and PbPb collisions at LHC.

Such predictions are valuable
from a practical point of view.  More important, if they
turn out to be correct, and the trends seen to date are
not some accidental consequence of averaging over many species and
momenta, the observed trends must be telling us something profound
about how QCD (most likely how the
vacuum) determines particle production.  At a minimum, if the current
belief is correct that the intermediate state between the instant of collision
and final free-streaming of the produced particles is very different
in  $e^+e^-$ and AA colisions, or for that matter, in pp collisions,
AA collisions below SPS energies and AA collisions at the top RHIC
energy, the global characteristics of multiparticle productions must
be insensitive to the intermediate state.  One consequence is that no
successful prediction of any selected set of global properties
can be used as evidence that a particular model correctly describes the
intermediate state.

On the other hand, if these ``predictions" turn out to be false, it
will be a strong indicator of the onset of some new physics at LHC.

So what are these universal simple trends?

We find \cite{Lots,Busza:2004mc}, as a first approximation, that

\begin{figure}[h] %  figure placement: here, top, bottom, or page
  \centering
  \subfigure[Extrapolation of midrapidity particle density per
   participant pair, for central PbPb collisions at $\sqrt{s_{NN}}$ =
   5.5TeV.  The data are from a PHOBOS compilation
   \cite{Back:2004je,WhitePaper}. The predicted value for $\frac{dN_ch}{d\eta}/
   (N_{part}/2)= 6.2\pm0.4 $, which for $N_{part}$ = 386 (top 3 \%)
   corresponds to $dN_{ch}/d\eta=1200\pm 100$.]{
   \includegraphics[width=2.9in]{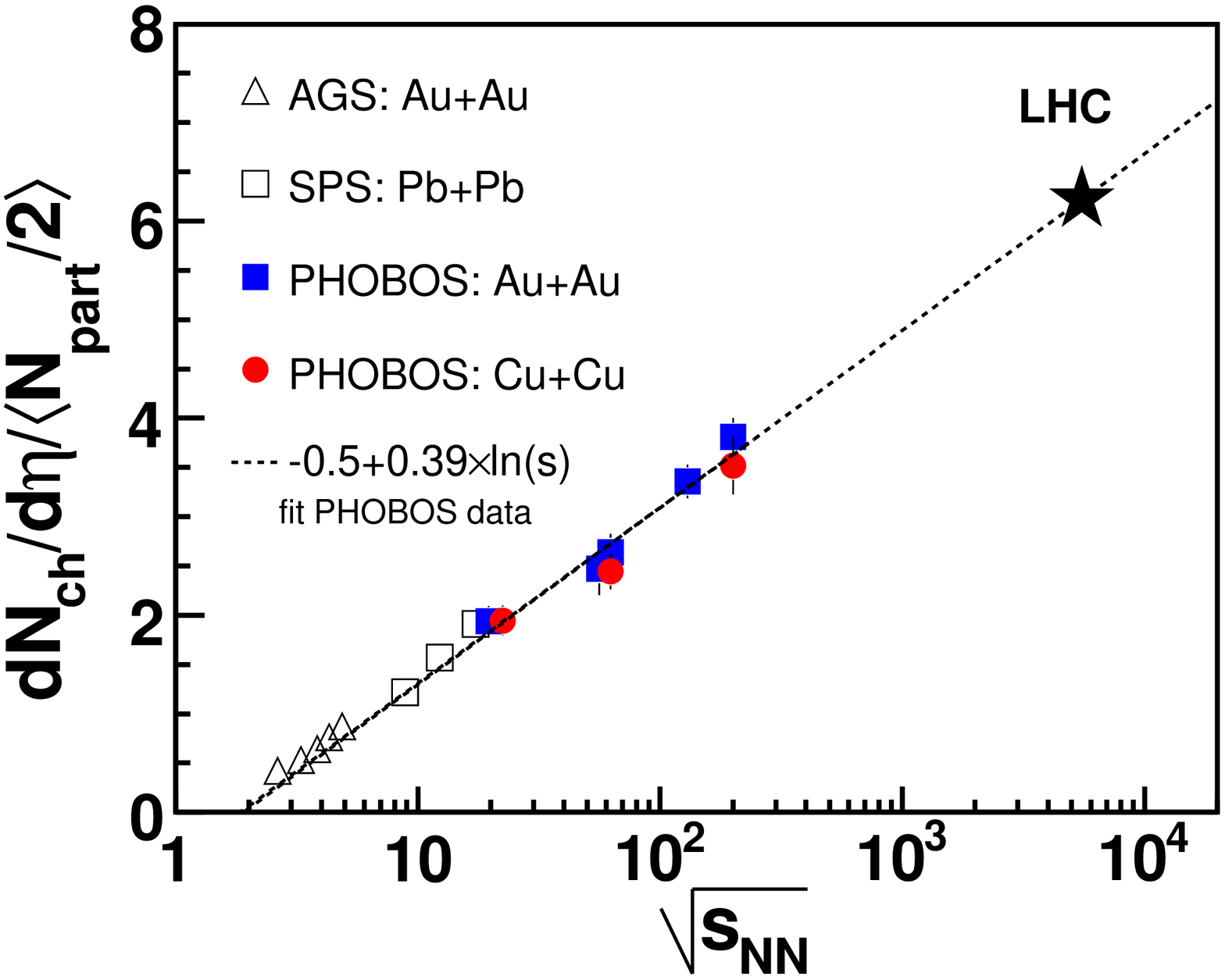}
   \label{fig1busza}}
 \hspace{0.05in}
 \subfigure[Extrapolated central ($N_{part}$ = 360, 0-10\% centrality)
   PbPb pseudorapidity distribution at $\sqrt{s_{NN}}$ =
   5.5TeV. PHOBOS AuAu data \cite{Back:2004je,WhitePaper}, longitudinal scaling with a linear
   ``limiting curve", and the observed $ln \sqrt{s_{NN}}$ energy
   dependence of the mid rapidity density were used in the
   extrapolation.]{
   \includegraphics[width=2.9in]{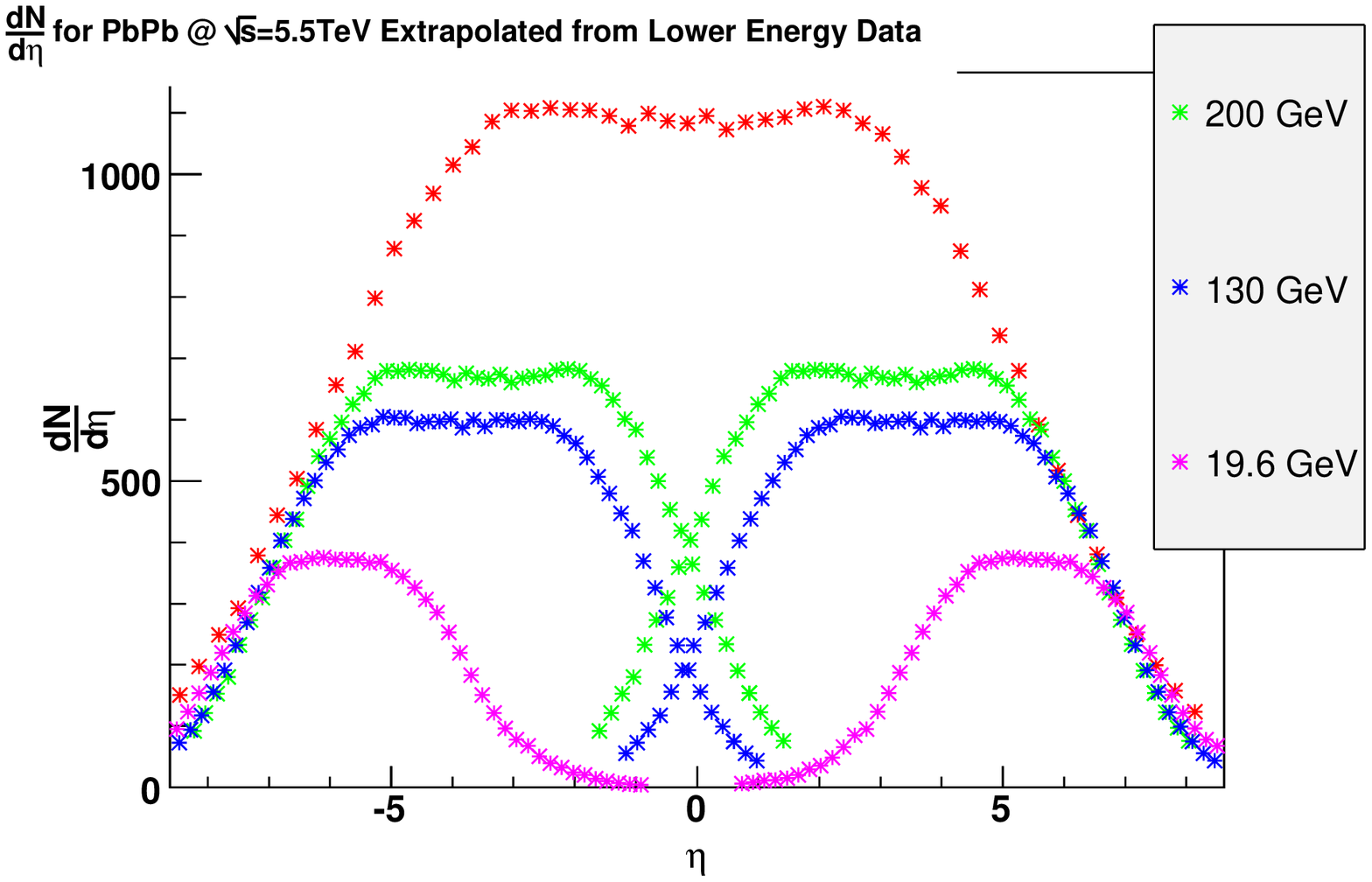}
  \label{fig2} }
\\
%\end{figure}
%\begin{figure}[h]
%  \centering

\subfigure[Predictions of PbPb LHC pseudorapidity distributions for
  different centrality collisions.  For each extrapolated curve,
  PHOBOS AuAu or CuCu data \cite{Back:2004je,WhitePaper}, with $N_{part}$  closest
  to the required value, were taken and normalized to the quoted $N_{part}$
  value.  The vertical and horizontal scales were then re-scaled by
  $ln \sqrt{s_{NN}}$.  This is equivalent to trends (iii) and (iv), provided
  that $dn/d\eta \rightarrow 0$ as $\eta \rightarrow beam rapidity$. 
  It must be emphasized that each curve is an independent
  extrapolation, however each extrapolation relies on the same method.]{
   \includegraphics[width=2.9in]{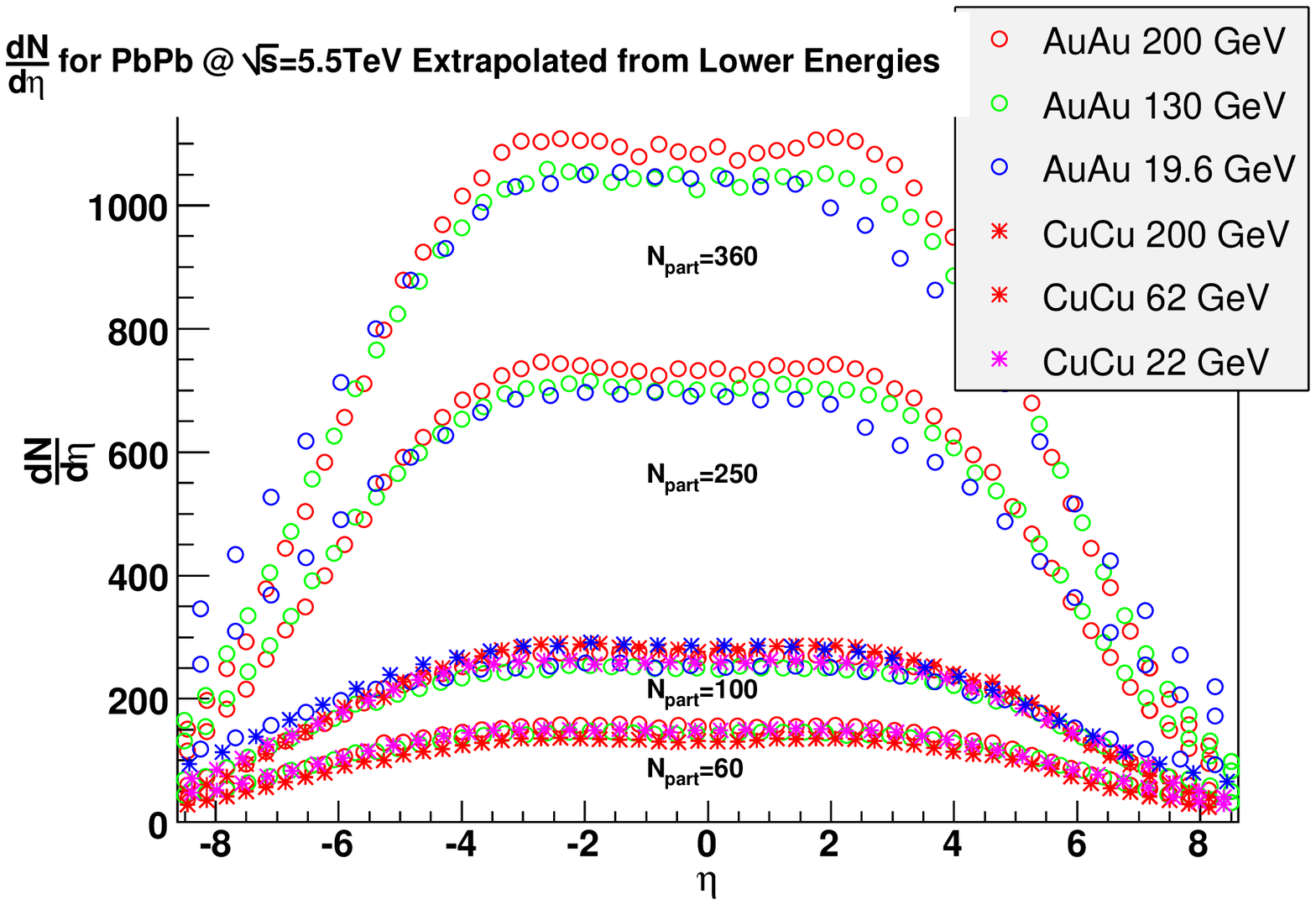}
  \label{fig3} }
\hspace{0.05in}
\subfigure[Total charged particle multiplicity per participant pair
   plotted as a function of $ln^2 \sqrt{s}$ (with $\sqrt{s}$ in GeV)
   for various colliding systems.  The data are taken from the
   compilation in Busza \cite{Lots}.  The extrapolation of the AA data to
   $\sqrt{s_{NN}}$=5.5 TeV predicts $15000\pm1000$ charges particles at
   LHC for $N_{part} = 386$ (top 3\%). Extrapolation of the non-single
   diffractive (NSD) pp data to
   $\sqrt{s_{NN}}$ = 14 TeV predicts $70\pm8$ charged particles at LHC.
   The $ln^2 \sqrt{s}$ extrapolation for the total multiplicity
   is a consequence of the extrapolation procedure described in the
   caption of fig 3.]{
   \includegraphics[width=2.9in]{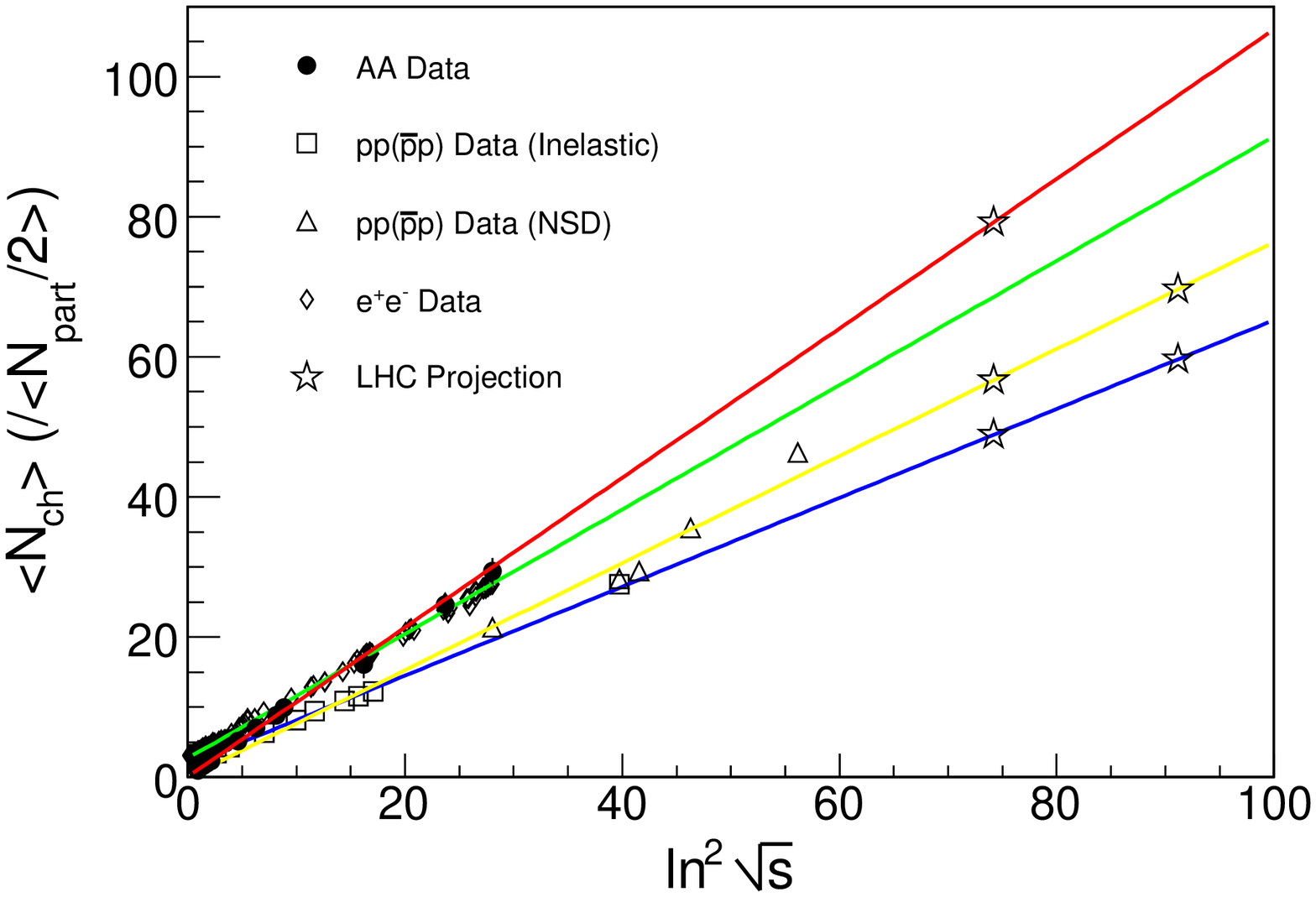}
 \label{fig4} }
\caption{}
\label{largebusza1}
\end{figure}

\begin{figure}[h]
  \centering
\subfigure[Predicted PbPb pseudorapidity distributions of charged
   particles at $\sqrt{s_{NN}}$=5.5 TeV for two centralities, 0-6\%
   and 35-45\%.  These predicted curves are obtained by extrapolating
   the corresponding centrality 200GeV PHOBOS AuAu data
   \cite{Back:2004je,WhitePaper} using the
   same procedure as in fig. \ref{largebusza1}.3.]{
   \includegraphics[width=2.9in]{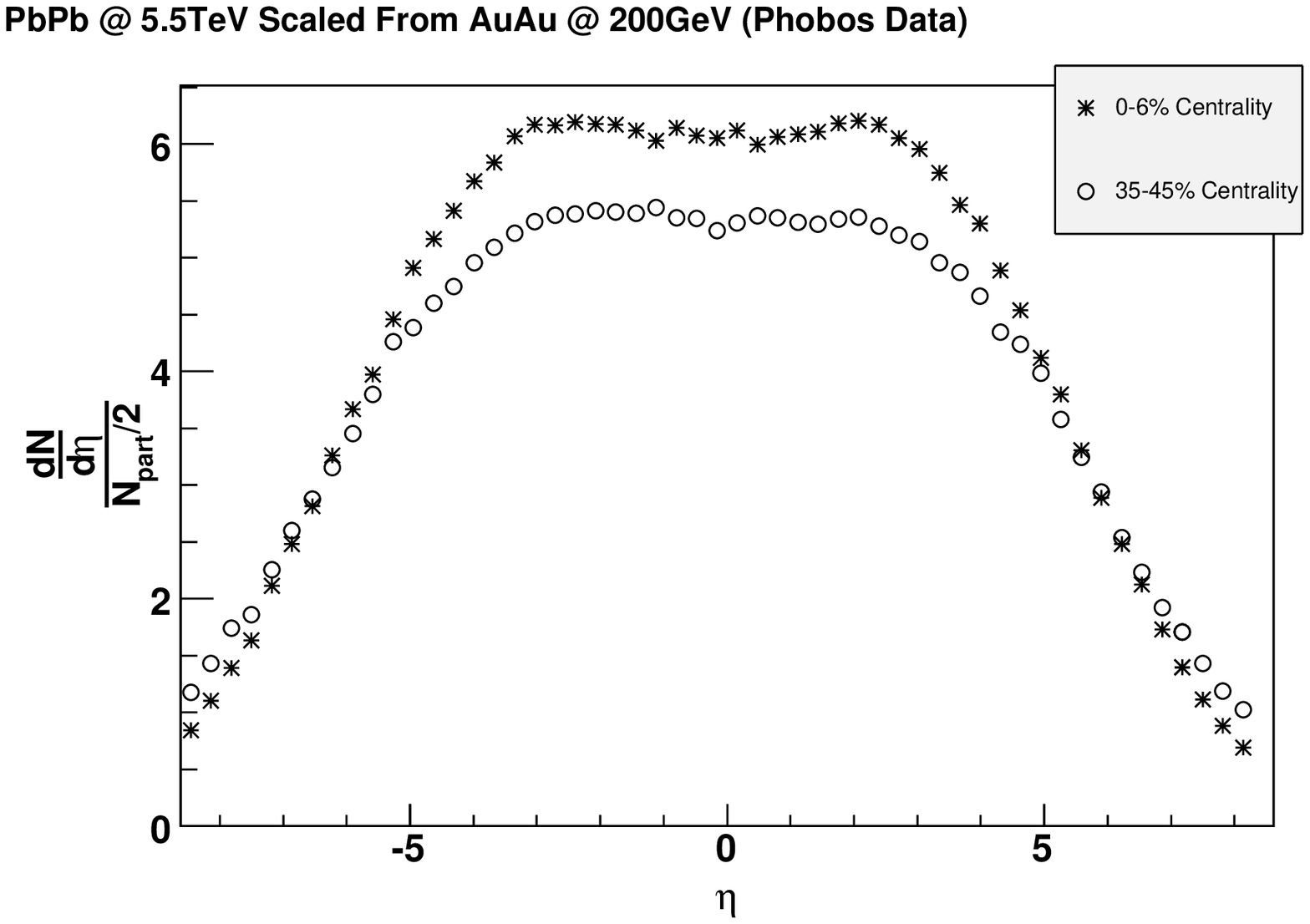}
  \label{fig5} }
\hspace{0.05in}
\subfigure[Predicted $N_{part}$ dependence of the mid rapidity
   charged particle density per participant pair for PbPb at
   $\sqrt{s_{NN}}$ = 5.5 TeV.  These results are obtained from the
   PHOBOS  200 GeV AuAu data \cite{Back:2004je,WhitePaper} scaled by $ln \sqrt{s_{NN}}$ (see
   trends (i) and (iv) in the text).]{
   \includegraphics[width=2.9in]{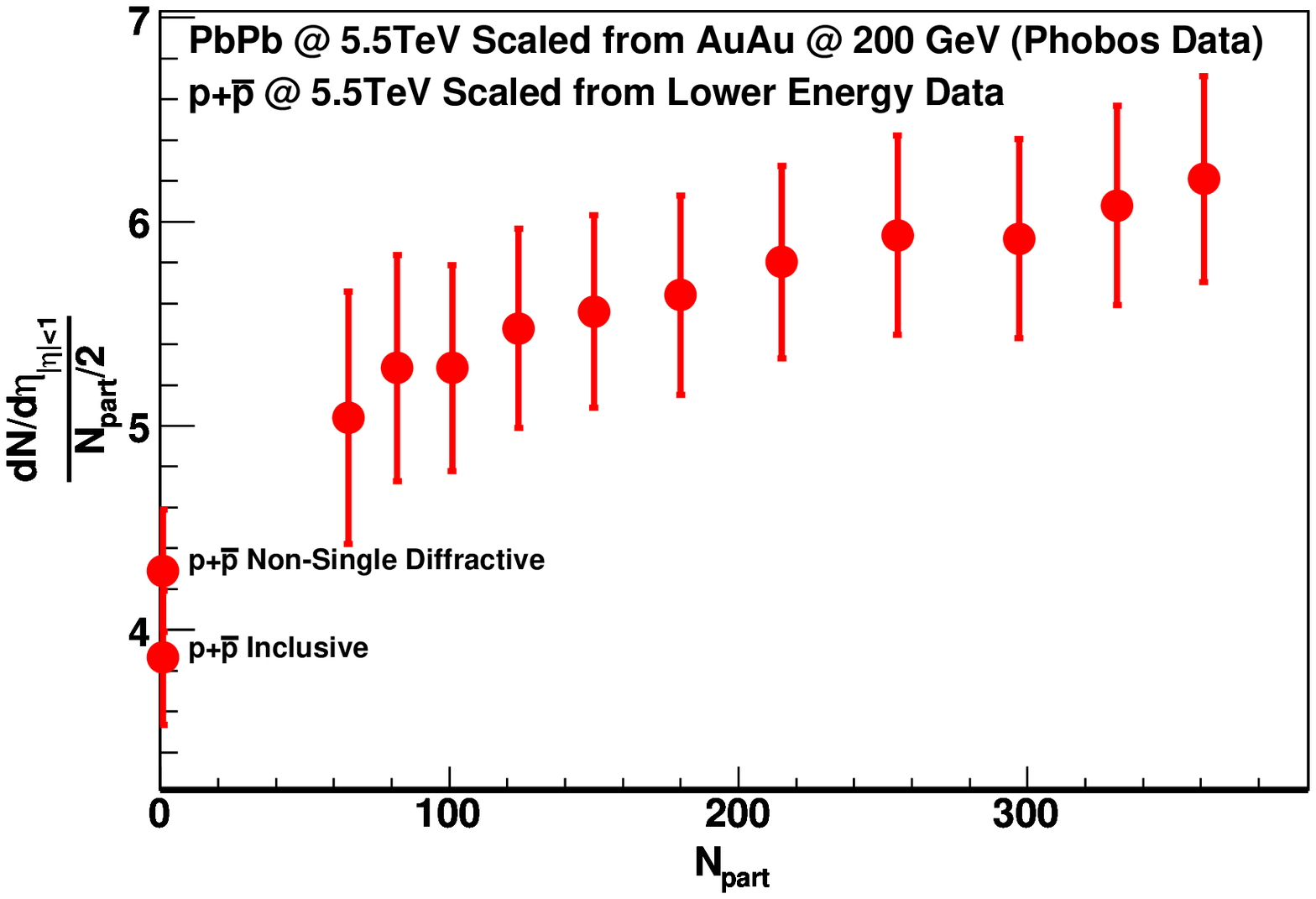}
  \label{fig6} }
\\
%\end{figure}
%\begin{figure}[h]
%  \centering
\subfigure[Predicted pA pseudorapidity distribution for
   $\sqrt{s_{NN}}$ = 5.5 TeV for $N_{part}$ = 3.4.  Each curve is an
   independent extrapolation of lower energy p- emulsion data using
   the same procedure as in Fig. \ref{largebusza1}.3.  The p-emulsion data are from ref
   \cite{Otterlund}, and covers the energy range $\sqrt{s_{NN}}$ =
   11.3 GeV to 38.8 GeV (ie. proton beams with momentum 67 GeV/c to
   800 GeV/c)]{
   \includegraphics[width=2.9in]{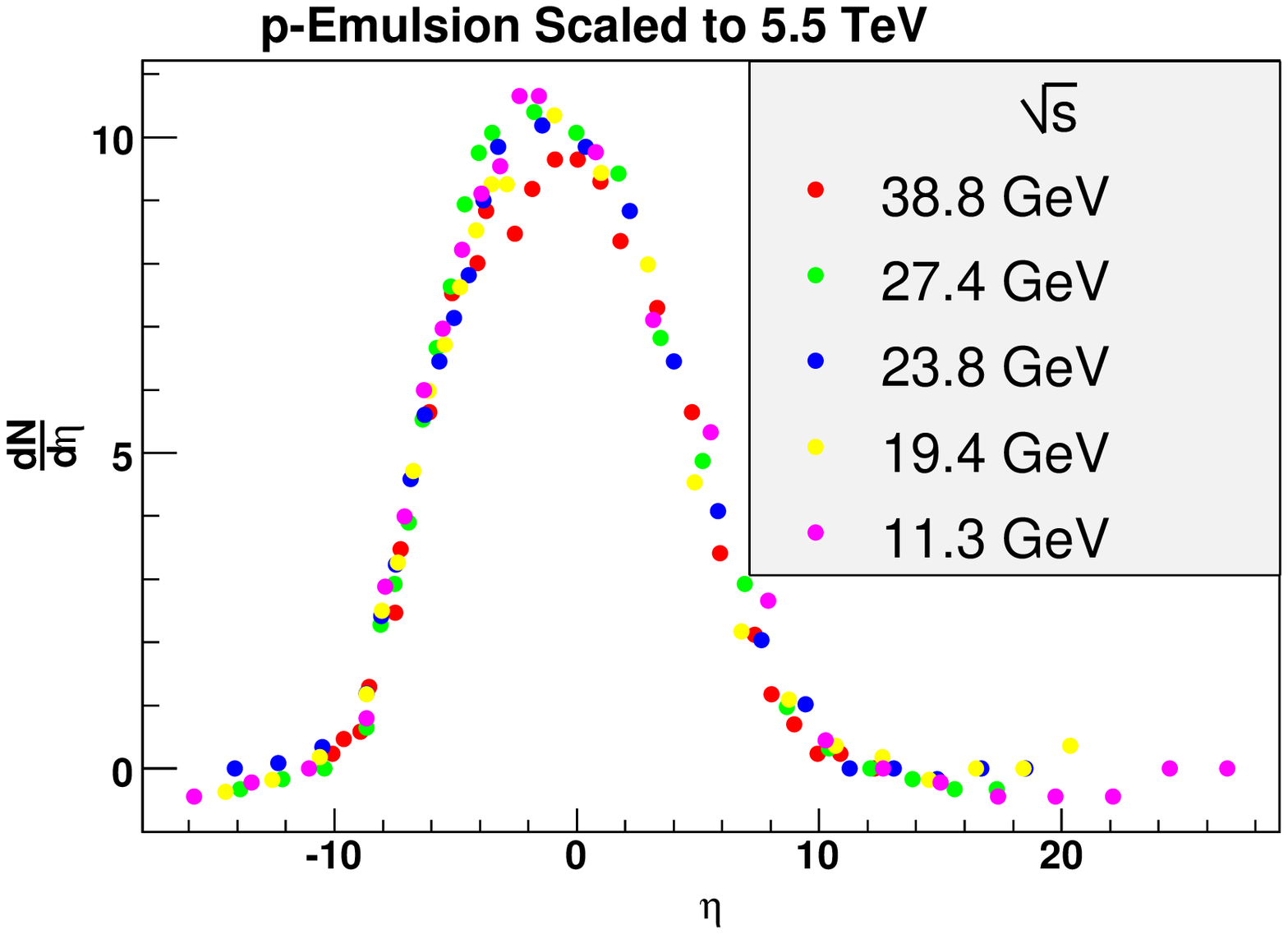}
  \label{fig7}}
\hspace{0.05in}
\subfigure[Predicted pp Non-Single-Diffractive (NSD) pseudorapidity
   distributions at $\sqrt{s_{NN}}$ = 14 TeV.  Each curve is an
   independent extrapolation of lower energy $p\bar{p}$ data \cite{BBBack}, using
   the same procedure as in Fig. \ref{largebusza1}.3.  The scatter of points at central
   rapidities most likely reflects systematic errors in these difficult
   measurements.]{
   \includegraphics[width=2.9in]{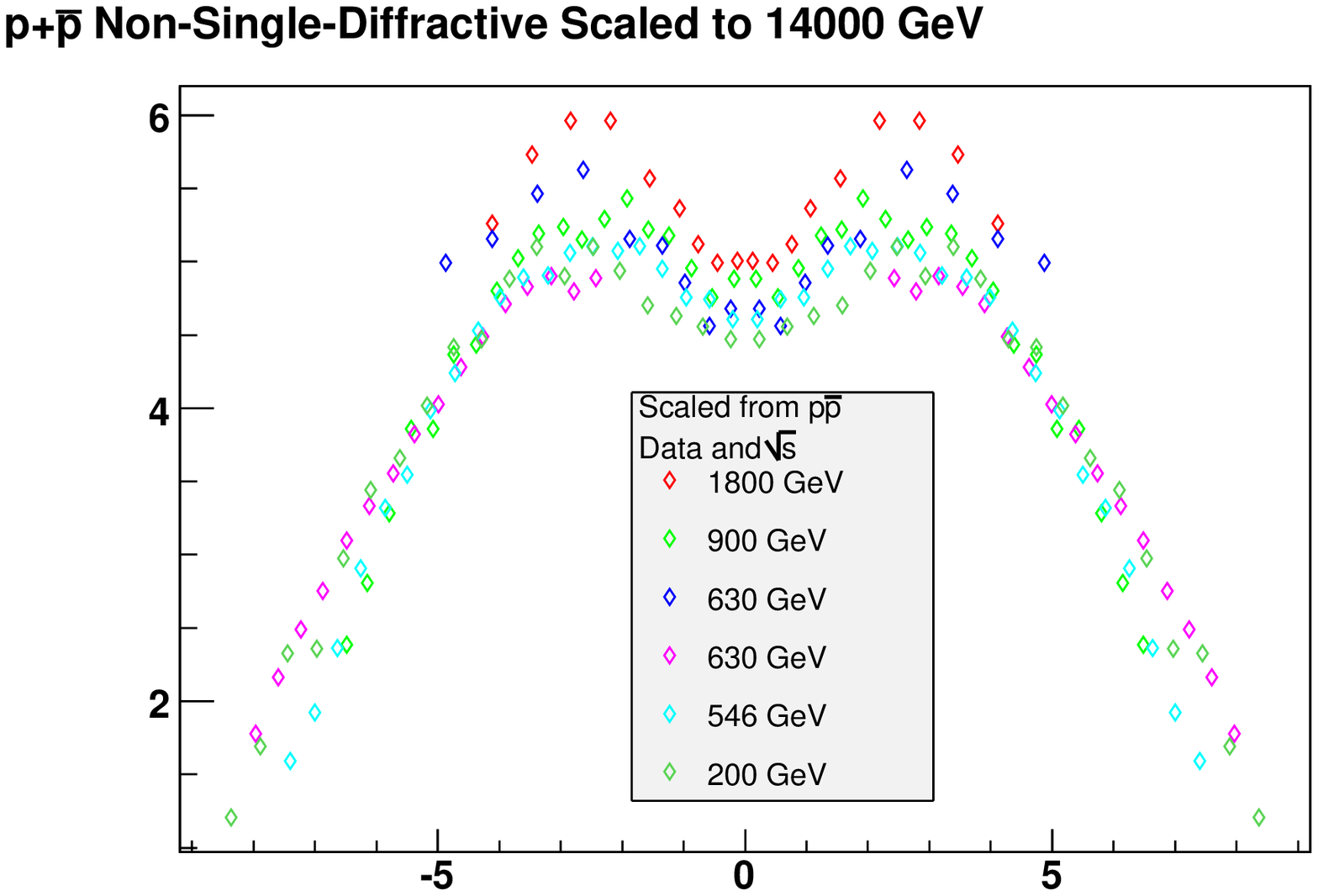}
  \label{fig8} }
\caption{}
\label{largebusza2}
\end{figure}

\begin{figure}[h]
  \centering
\subfigure[Predicted pseudorapidity dependance of the elliptic flow parameter $\nu_2$ for the 40\% most central collisions of PbPb at $\sqrt{s_{NN}}$ = 5.5 TeV.  An extrapolation procedure similar to that in fig. \ref{largebusza1}.2 was used, with input data from PHOBOS \cite{BBBack}.]{
   \includegraphics[width=2.9in]{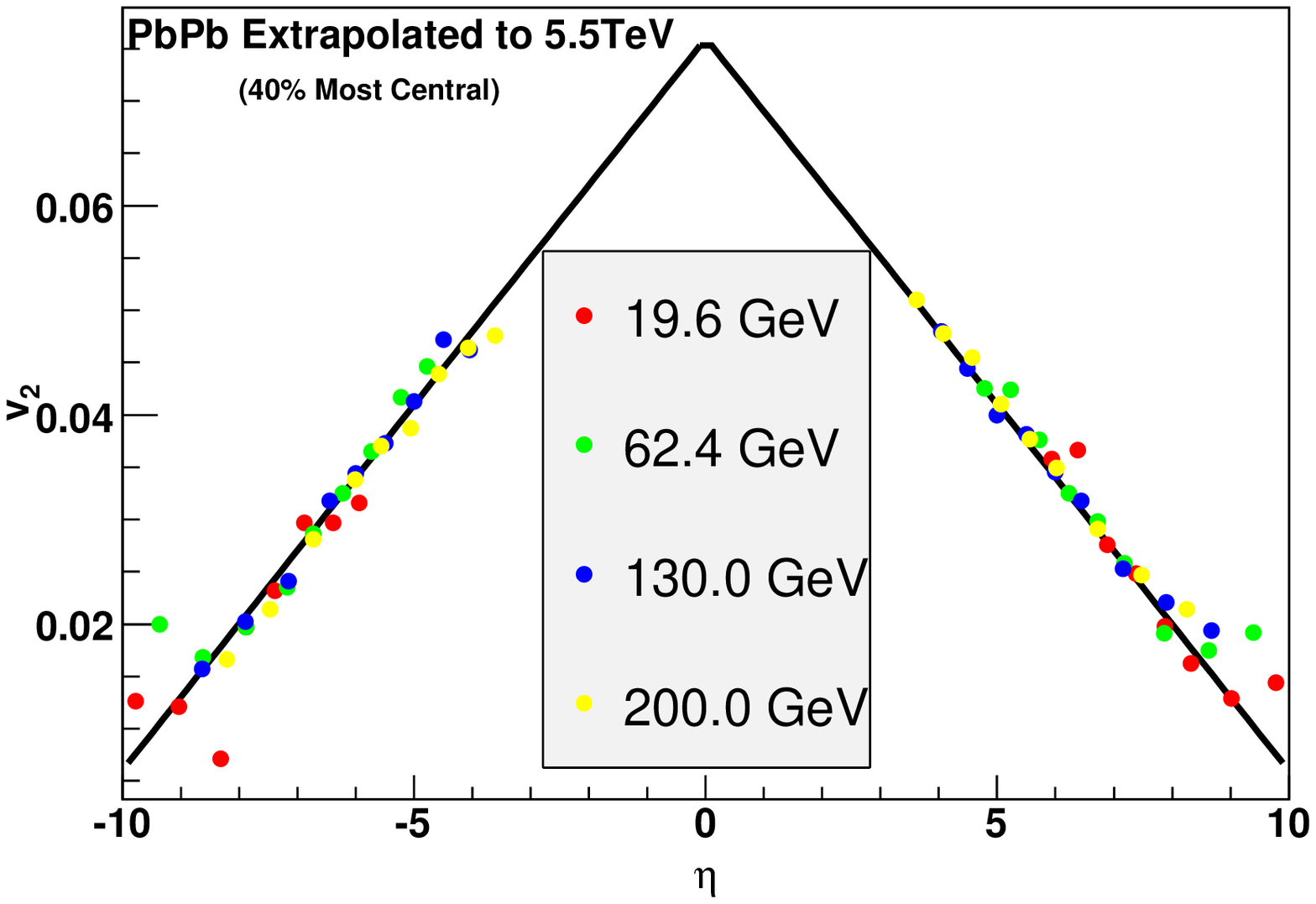}
  \label{fig9} }
\hspace{0.05in}
\subfigure[Extrapolation of the elliptic flow parameter $\nu_2$ for
  the 40\% most central collisions in PbPb at $\sqrt{s_{NN}}$ = 5.5
  TeV.  The data is a compilation in \cite{Alt}.]{
   \includegraphics[width=2.9in]{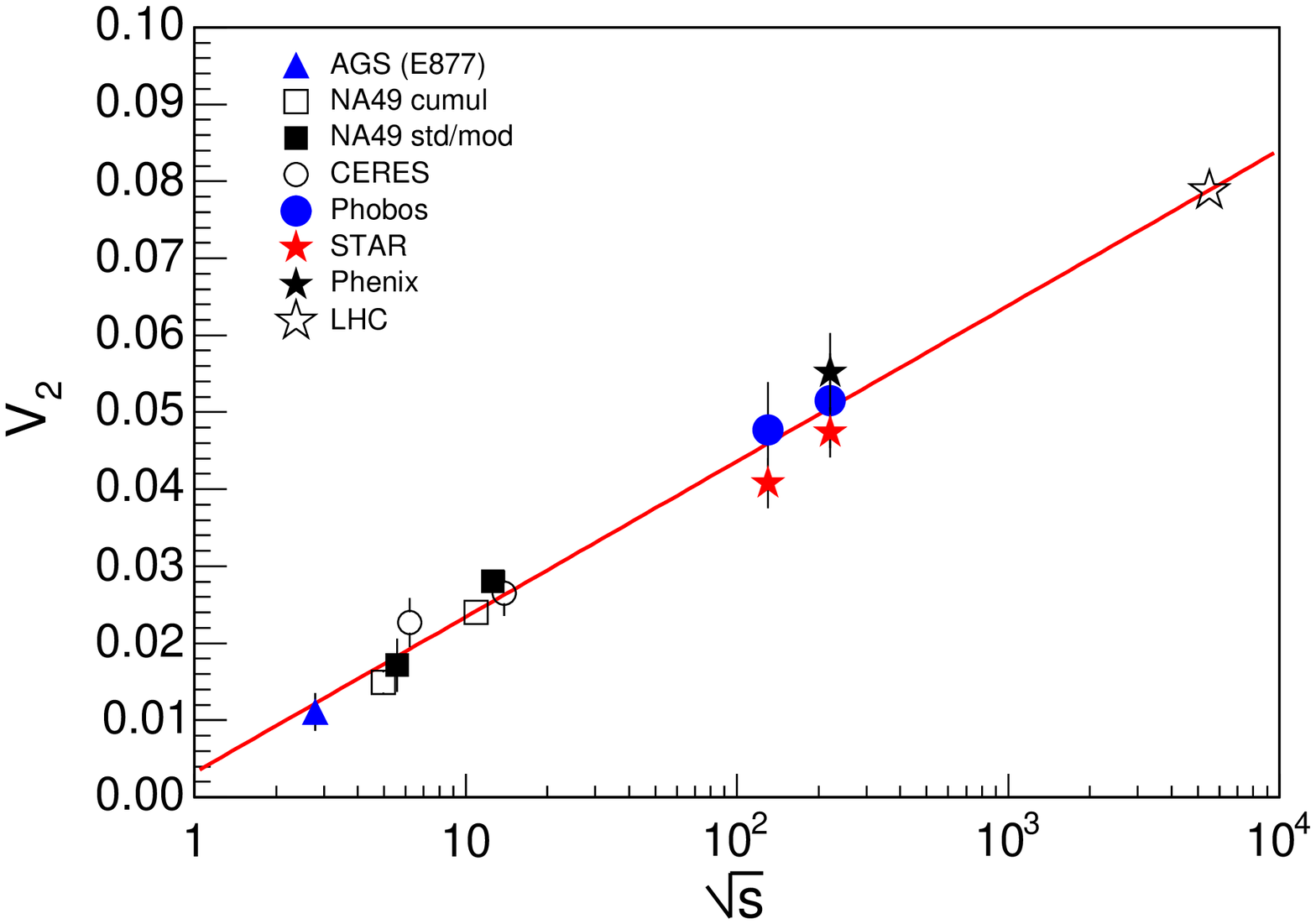}
\label{fig10} }
\caption{}
\label{largebusza3}
\end{figure}

\begin{enumerate}
\item The global distributions of charged particles factorize into an
      energy dependent part and a geometry, or incident system,
      dependent part.
      
    \item At a given energy, in pA and AA the distributions do depend
          in detail on the colliding systems or geometry (eg. impact
          parameter).  However the total number of produced particles
          is simply proportional to the total number of participants
          $N_{part}$ (or wounded nucleons, in the language of Bialas and
          Czyz). [Note: there is a systematic difference in the
          constant of proportionality in pA and AA, that can be
          attributed to the leading particle effect in pp collisions]
          
        \item The total charged particle density $dN/d\eta$ (where
              $\eta$ is the pseudorapidity), and the directed and
              elliptic flow parameters $\nu_1$ and $\nu_2$ satisfy
              extended longitudinal scaling.
              Furthermore, over most of its range the ``limiting
              curve" is linear.
              
            \item The mid-rapidity (in the cm system) particle density
                  $\frac{dN}{d\eta} |_{y=0}$ and the elliptic flow
                  parameter $\nu_2$ both increase linearly with
                  $ln\sqrt{s}$.  It is not clear if this is the origin
                  or consequence of item (iii) above. [Note: for elliptic
                  flow, (iii) and (iv) are directly related only if we
                  postulate that at all energies there is a
                  ``pedestal" in the value of $\nu_2$, i.e. there is a
                  part of the source of flow that is independent of energy]
                  
\end{enumerate}
In the figures \ref{largebusza1}, \ref{largebusza2} and
\ref{largebusza3} we use the above observed trends at
lower energies to ``predict" LHC results.  A more detailed version of
this work will be submitted to Acta Physica Polonica.  
\\


\subsection{Multiplicities 
and $J/\psi$ suppression at LHC energies}

{\it A. Capella and E. G. Ferreiro}

{\small
We present our predictions on multiplicities and $J/\psi$ suppression at LHC energies.
Our results take into account shadowing effects in the initial state and final state interactions 
with the hot medium. We obtain 1800 charged particles at LHC and 
the $J/\psi$ suppression increases by a factor 5 to 6 compared to RHIC. 
}

\subsubsection{Multiplicities with shadowing corrections}

\begin{center}
\begin{figure*}
\begin{minipage}[t]{80mm}
\vskip 1cm
\epsfxsize=7.5cm
\epsfysize=8.5cm
\centerline{\epsfbox{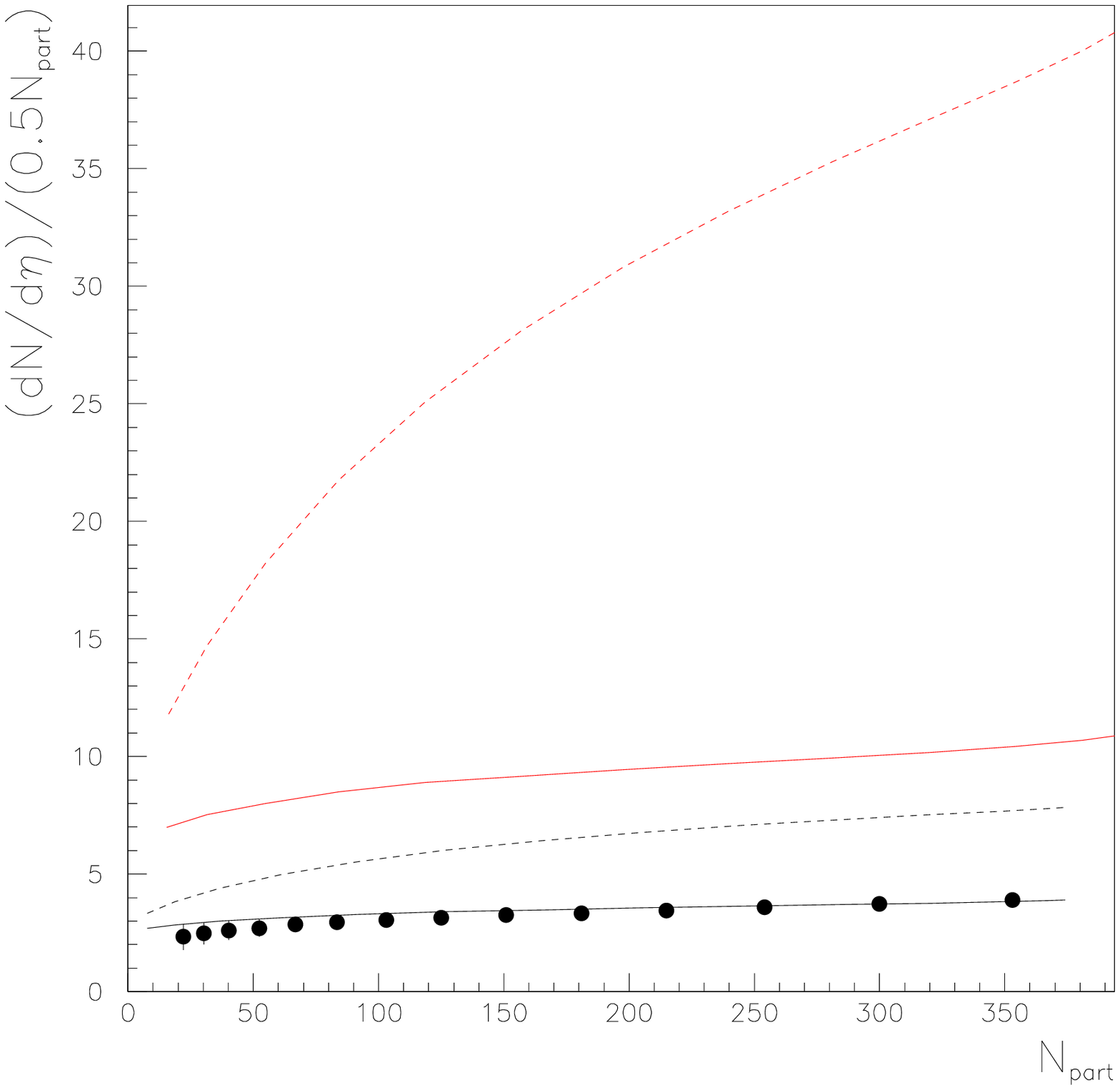}}
\vskip -2.3cm
\caption{
Multiplicities of charged particles with (solid lines) and without (dashed lines)
shadowing corrections at RHIC and LHC.}
\label{ferreirojpsif1}
\end{minipage}
\hspace{\fill}
\begin{minipage}[t]{80mm}
\vskip 1.4cm
\epsfxsize=7.5cm
\epsfysize=8.5cm
\centerline{\epsffile[4 4 500 650]{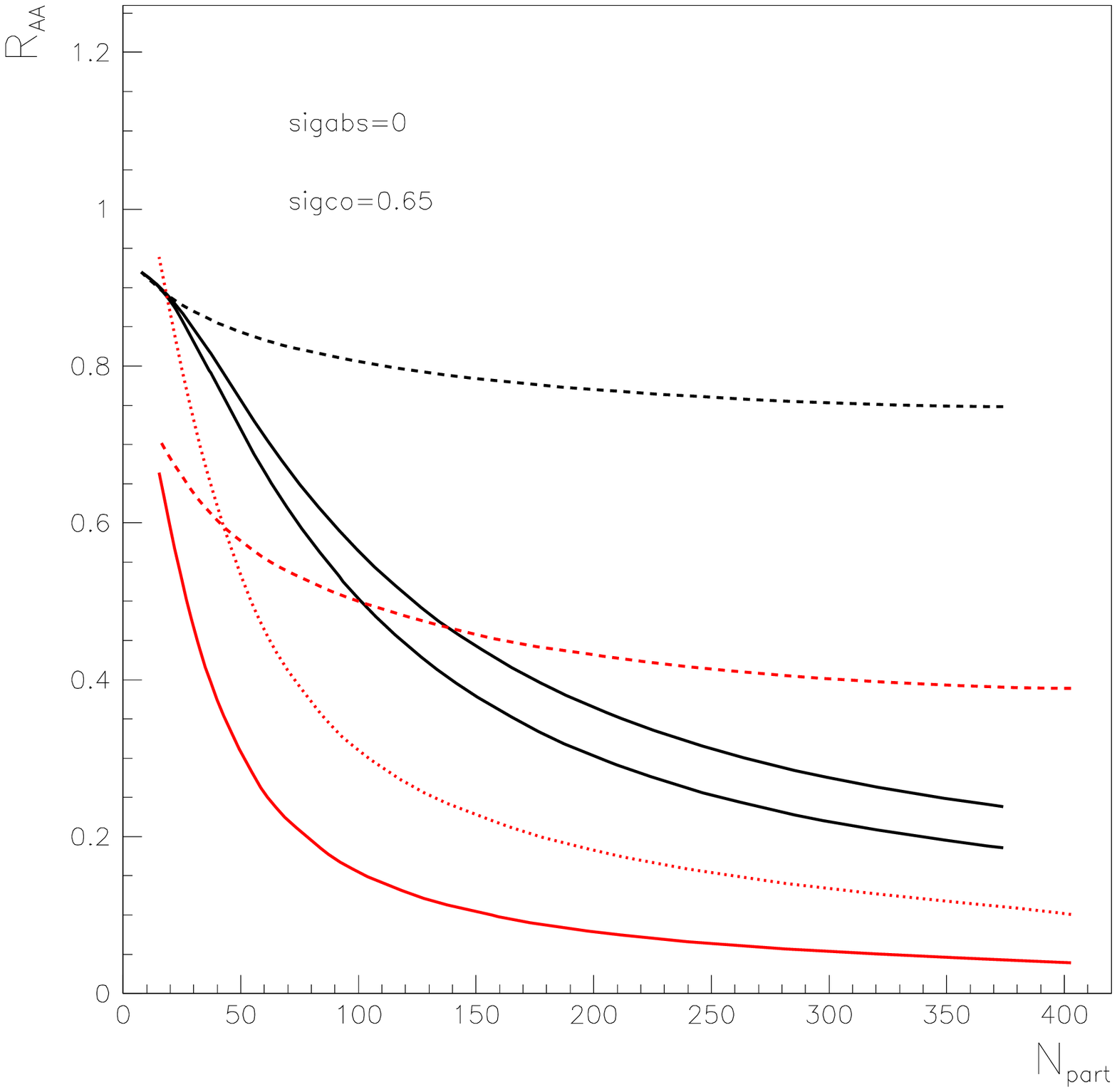}}
\vskip -2.7cm
\caption{$J/\psi$ production at RHIC and LHC. 
Dashed: $J/\psi$ shadowing, pointed: comovers suppression, continuous: total suppression.}
\label{ferreirojpsif2}
\end{minipage}
\vskip -0.3cm
\end{figure*}
\end{center}

%Multiplicities are usually considered as the addition of two contributions: one proportional to 
%number of
%participant
%nucleons $A$ %(valence-like contribution) 
%and a second one proportional to the number of 
%inelastic nucleon-nucleon collisions $A^{4/3}$, dominant at
%asymptotic energies.
%In order to get the right multiplicities at
%RHIC, d
At high energy, different mechanisms in the initial state {\it -shadowing-}, that lower the 
total multiplicity, have to be taken into account. 
%have been included to lower this second contribution.
The shadowing makes the
nuclear structure functions in nuclei different from the superposition of
those of their constituents nucleons. Its effect 
increases with decreasing $x$ and decreases with increasing
$Q^2$.
We have included a dynamical, non linear  of shadowing \cite{Capella:2005cn}, controlled by triple pomeron diagrams.
It is determined in terms of the diffractive cross sections. 
Our results for charged particles multiplicities at RHIC and LHC energies are presented in Fig. \ref{ferreirojpsif1}.
In absence of shadowing we obtain a maximal multiplicity, $dN_{AA}/dy = A^{4/3}$. With shadowing 
corrections the multiplicity behaves as $dN_{AA}/dy = A^{\alpha}$, with $\alpha=1.13$ at RHIC
and $\alpha=1.1$ at LHC.

\subsubsection{$J/\psi$ suppression}

%The $J/\psi$ production in proton-nucleus collisions
%is suppressed with
%respect to the characteristic $A^1$ scaling of lepton pair production.
%This
% suppression is generally
%interpreted as a
%result of the multiple scattering of a pre-resonance $c\overline{c}$
%with the nucleons of the nucleus: {\it nuclear absorption}.
An anomalous 
$J/\psi$ suppression 
-that clearly exceeds the one
expected from nuclear absorption-
has been found in $PbPb$ collisions at SPS.
Such a phenomenon was predicted by Matsui and Satz as a consequence
of
deconfinement in a dense medium.
It can also be described as a result
of final state interaction of the $c\overline{c}$ pair with the dense
medium produced in the collision: {\it comovers
interaction} \cite{Capella:2000zp}.
Here we present our results for the ratio of the $J/\psi$
yield over the average
number of binary nucleon-nucleon collisions at RHIC and LHC energies:
\beq
R_{AB}^{J/\psi}(b) = {dN_{AB}^{J/\psi}(b)/dy  \over n(b)} =
{dN_{pp}^{J/\psi}  \over dy}  {\int d^2s\ \sigma_{AB}(b)\ n(b,s) \
S^{abs}(b,s)\ S^{co}(b,s) \over \int d^2s\ \sigma_{AB}(b)\ n(b,s)}\ .
\eeq
$S^{abs}$ refers to the survival probability due to nuclear absorption and
$S^{co}$ is the survival probability due to the medium interactions.
The data on
$dAu$ collisions at RHIC favorize a small $\sigma_{abs}=0$ mb, so $S^{abs}=1$
\cite{Capella:2006mb}.
The interaction of a particle or a parton with the medium
is described by the
gain and loss differential equations which govern the final state
interactions:
\beq
\tau {d\rho^{J/\psi}(b,s,y) \over d \tau} = - \sigma_{co}\
\rho^{J/\psi}(b,s,y)\ \rho^{medium}(b,s,y)\ ,
\eeq
where $\rho^{J/\psi}$ and
$\rho^{co}=\rho^{medium}$ are the densities
of $J/\psi$ and comovers. 
We neglect a gain term resulting
from the recombination of $c$-$\overline{c}$ into $J/\psi$.
Our equations have to be integrated
between initial time $\tau_0$ and freeze-out time $\tau_f$. We use the inverse proportionality between proper time and densities,
$\tau_f/\tau_0 = \rho(b,s,y)/\rho_{pp}(y)$.
Our densities
 can be either hadrons or partons, so $\sigma_{co}$ represents an  effective cross-section
averaged over the interaction time.
We obtain the survival probability
$S_{co}(b,s)$
of the $J/\psi$ due to the medium interaction:
\begin{equation}
S^{co}(b,s) \equiv \frac{N^{J/\psi (final)}(b,s,y)}{N^{J/\psi
(initial)}(b,s,y)} \nonumber \\
= \exp \left [ - \sigma_{co}\ \rho^{co}(b,s,y) \ell n \left
(\frac{\rho^{co}(b,s,y)}{\rho_{pp}(0)} \right ) \right ] \ .
\end{equation}
The shadowing produces a decrease of the medium density. Because of this, the 
shadowing corrections on comovers increase
the $J/\psi$ survival probability $S^{co}$. On the other side, the shadowing corrections on ${J/\psi}$ 
decrease the $J/\psi$ yield.
Our results for RHIC and LHC are presented in Fig. \ref{ferreirojpsif2}.
We use the same
value of the comovers cross-section, $\sigma_{co} = 0.65$~mb that we have used at SPS energies.
We neglect the nuclear absorption.
%, $\sigma_{abs} = 0$~mb. 
The 
shadowing is introduced in both the comovers and the $J/\psi$ yields.

\subsection{Heavy ion collisions at LHC in a Multiphase Transport Model}
\label{koampt}

{\it L.-W. Chen, C. M. Ko, B.-A. Li,
Z.-W. Lin and B.-W. Zhang}
\vskip 0.5cm

The AMPT model \cite{Lin:2004en} is a hybrid model that uses the
\textrm{HIJING} model \cite{Wang:1991ht} to generate the initial
conditions, the ZPC \cite{Zhang:1997ej} for modeling the partonic
scatterings, and the ART model \cite{Li:1995pr} for treating
hadronic scatterings. In the default version \cite{Zhang:1999bd},
the initial conditions are strings and minijets from the HIJING
model and particle production is based on the Lund string
fragmentation, while in the string melting version
\cite{Lin:2001zk}, the initial conditions are valence quarks and
antiquarks from hadrons produced in the HIJING model and
hadronization is described by a coordinate-space coalescence model.
Using the AMPT model, we predict in the following the hadron
rapidity and transverse momentum distributions, the elliptic flows
of both light and heavy hadrons, the two-pion and two-kaon
correlation functions in Pb+Pb collisions at center-of-mass energy
of $\sqrt{s_{NN}}=5.5$ TeV at LHC \cite{ko}.

\begin{figure}[ht]
\centerline{
\includegraphics[width=2.4in,height=2.4in,angle=0]{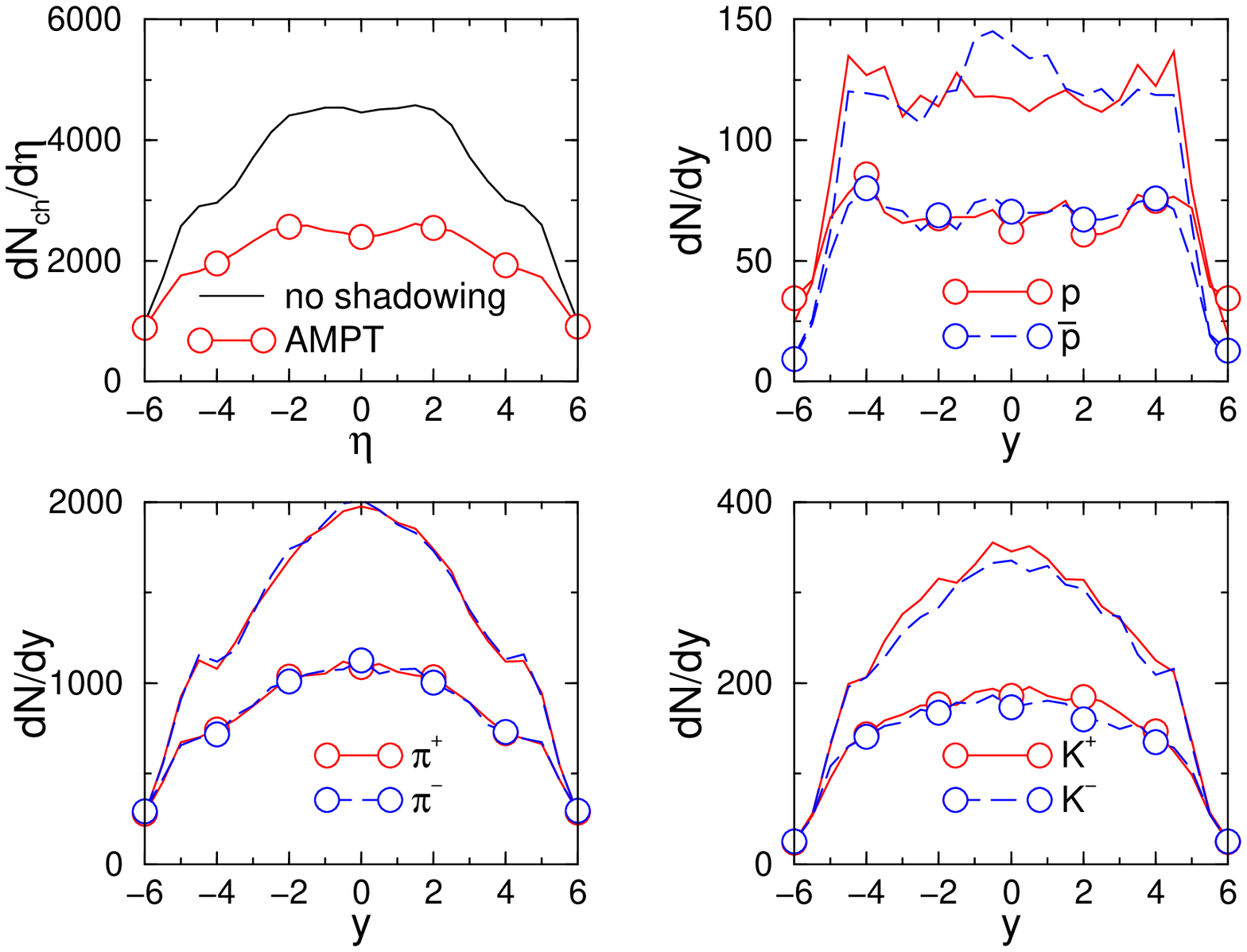}
\hspace{0.7cm}
\includegraphics[width=2.6in,height=2.6in,angle=0]{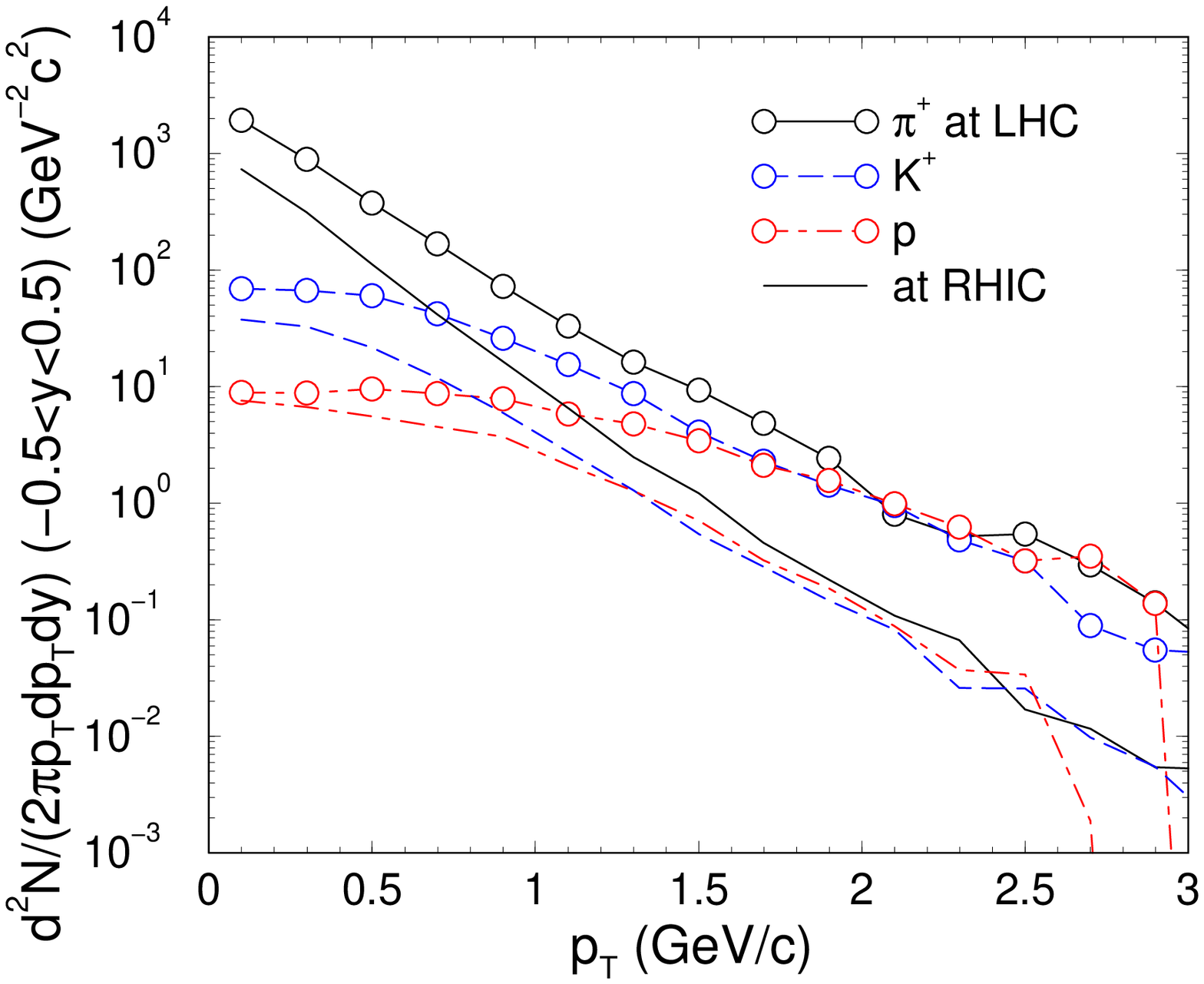}}
\caption{Left window: Pseudorapidity distributions of charged
hadrons and rapidity distributions of identified hadrons in central
($b\leq 3$ fm) Pb+Pb collisions at $\sqrt {s_{NN}}=5.5$ TeV from the
default AMPT model with (lines with circles) and without (solid
lines) nuclear shadowing. Right window: Transverse momentum spectra
of identified midrapidity hadrons from same collisions as well as
central Au+Au collisions at $\sqrt {s_{NN}}=200$ GeV.} \label{ypt}
\end{figure}

Shown in the left window of Fig.~\ref{ypt} are the charged hadron
pseudorapidity distribution and the rapidity distributions of
identified hadrons obtained with (lines with circles) and without
(solid lines) nuclear shadowing of nucleon parton distribution
functions. Compared to results from the AMPT model for RHIC
\cite{Lin:2000cx}, the distributions at LHC are significantly wider
and higher. For mid-pseudorapidity charged hadrons, the distribution
shows a clear plateau structure with a value of about 4500 and 2500,
respectively, without and with nuclear shadowing. The latter is more
than a factor of three higher than that at RHIC. The transverse
momentum spectra of identified midrapidity hadrons are shown in the
right window of Fig.~\ref{ypt} by lines with circles. The inverse
slope parameters, particularly for kaons and protons with transverse
momenta below 0.5 GeV/$c$ and 1 GeV/$c$, respectively, are larger
than those at RHIC (solid lines) as a result of stronger final-state
rescatterings at LHC than at RHIC. Similar to that observed at RHIC,
the proton spectrum is below that of pions at low transverse
momenta, but they become comparable at about 2 GeV/$c$.

\begin{figure}[htb]
\centerline{
\includegraphics[width=2in,height=2.5in,angle=0]{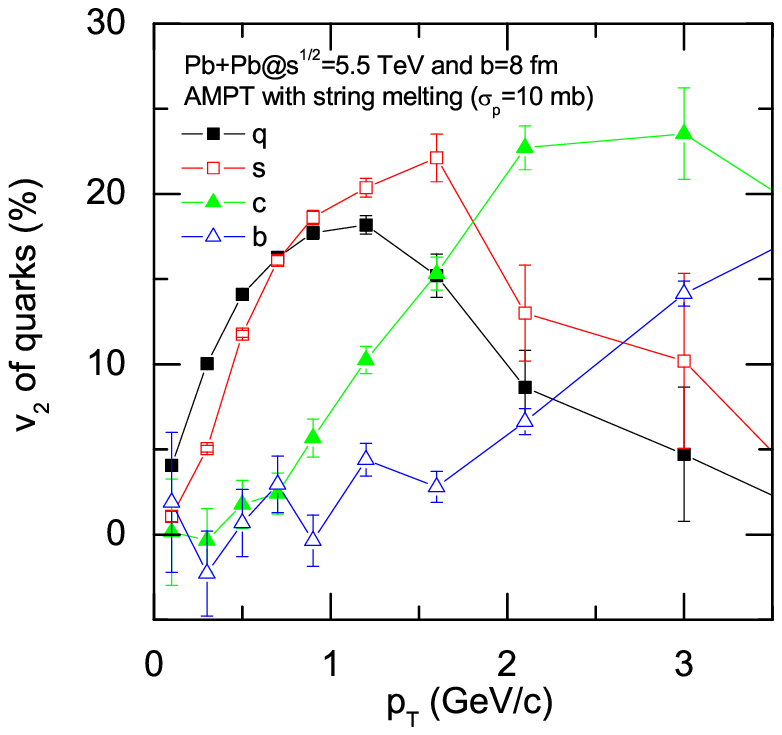}
\includegraphics[width=2in,height=2.5in,angle=0]{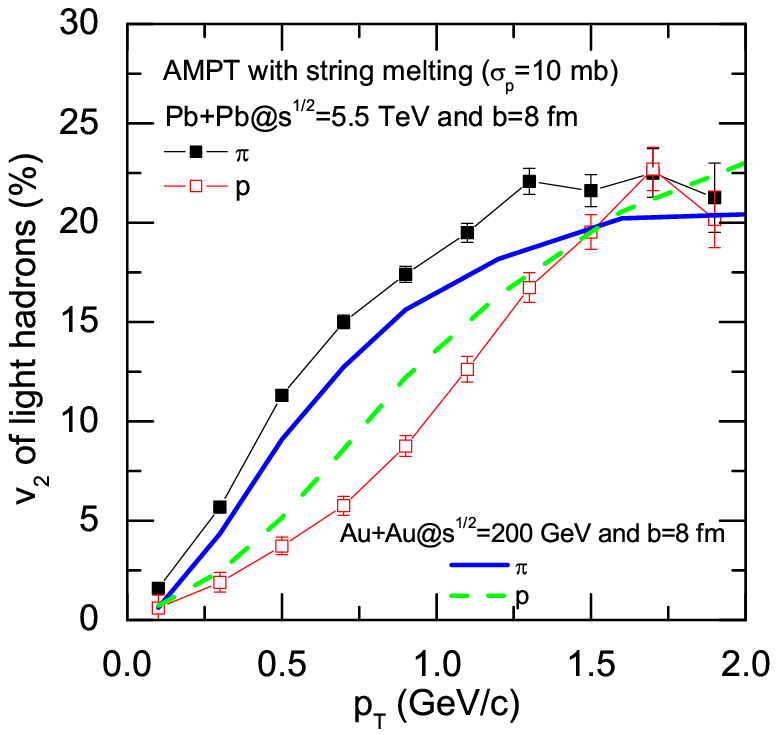}
\includegraphics[width=2in,height=2.5in,angle=0]{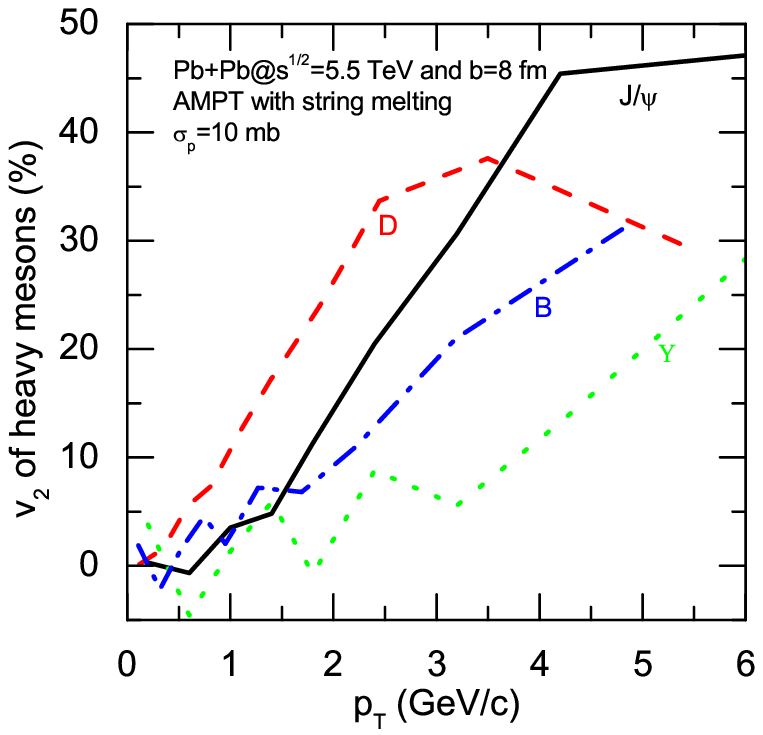}}
\caption{Elliptic flows of quarks (left window), light hadrons
(middle window), and heavy mesons (right window) in Pb+Pb collisions
at $\sqrt{s_{NN}}=5.5$ TeV and $b = 8$ fm from the AMPT model with
string melting and a parton scattering cross section of 10 mb.}
\label{v2}
\end{figure}

Hadron elliptic flows based on a parton scattering cross section of
10 mb, which is needed to describe observed hadron elliptic flows at
RHIC \cite{Lin:2001zk}, are shown in Fig.~\ref{v2}. The left window
gives the elliptic flows of light and heavy quarks as functions of
their transverse momenta, and they display the expected mass
ordering at low transverse momenta, i.e., the elliptic flow is
smaller for quarks with larger masses. At larger transverse momenta,
the elliptic flows of heavy quarks become, however, larger than
those of light and strange quarks, which peak at around 1-1.5 GeV/c.
The elliptic flows of pions and protons at LHC are shown in the
middle window of Fig.~\ref{v2}. Compared to corresponding ones at
RHIC for Au+Au collisions at $\sqrt{s_{NN}}=200$ GeV and same impact
parameter shown in the figure, the elliptic flow of pions at LHC is
larger while that of protons is smaller. As at RHIC
\cite{Zhang:2005ni}, elliptic flows of heavy mesons are estimated
from those of quarks using the quark coalescence or recombination
model \cite{Lin:2003jy,Greco:2003vf}
%via $v_{2,M}(p_T)\approx
%v_{2,Q}((m_Q/m_M)p_T)+v_{2,q}((m_q/m_M)p_T)$, where $v_{2,Q}$ and
%$v_{2,q}$ are elliptic flows of heavy and light quarks,
%respectively, while $m_M$, $m_Q$, and $m_q$ are, respectively,
%masses of heavy meson, heavy quark, and light quark. Results using
%$m_q = 300 $ MeV, $m_c = 1.5 $ GeV and  $m_b = 4.8 $ GeV
and are shown in the right window of Fig.~\ref{v2}. While elliptic
flows of heavy mesons are dominated by those of heavy quarks,
particularly for bottomed mesons, those of heavy mesons with hidden
charm or bottom, i.e., quarkonia $J/\psi$ and $\Upsilon$ consisting
of a heavy quark and its antiquark, at transverse momentum $p_T$ are
simply twice those of their constituent heavy quarks at $p_T/2$.

\begin{figure}[ht]
\centerline{
\includegraphics[width=2.7in,height=2.7in,angle=0]{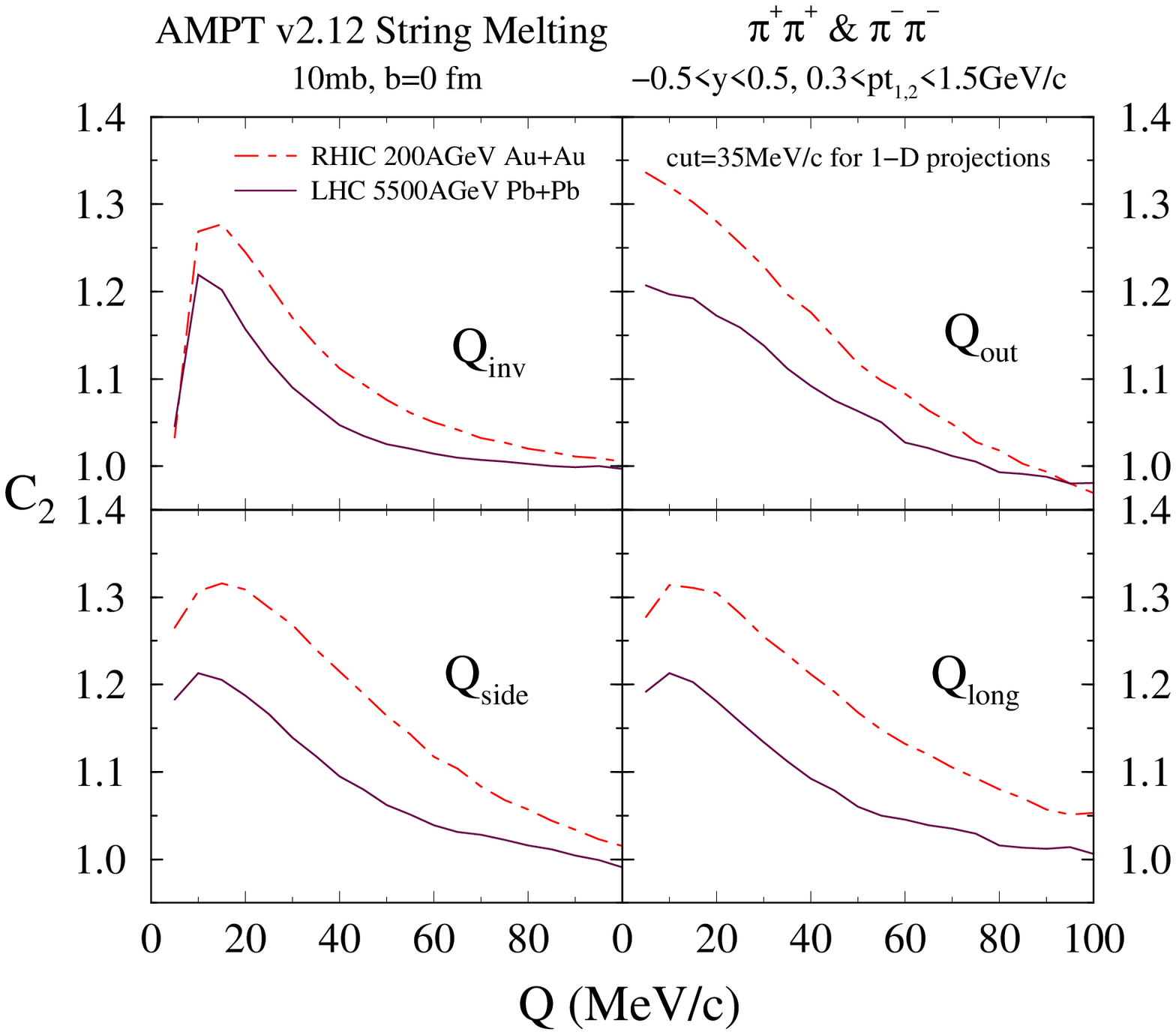}
\hspace{-0.5cm}
\includegraphics[width=2.7in,height=2.7in,angle=0]{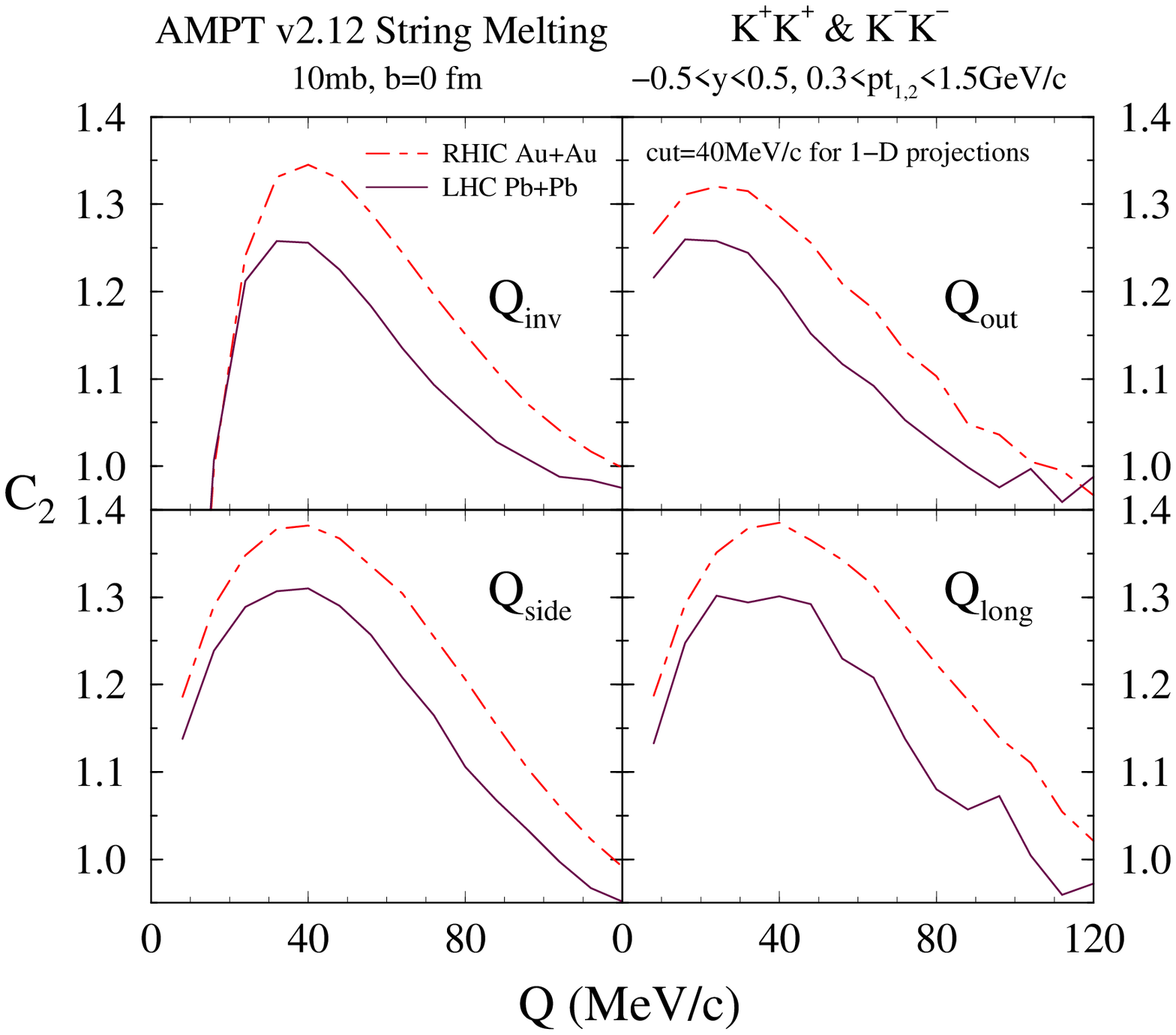}}
\caption{Correlation functions for midrapidity charged pions (left
windows) and kaons (right window) with $300<p_{\rm T}<1500$ MeV$/c$
from the AMPT model with string melting and a parton cross section
of 10 mb for central ($b=0$ fm) Pb+Pb collisions at
$\sqrt{s_{NN}}=5.5$ TeV (solid lines) and Au+Au collisions at
$\sqrt{s_{NN}}=200$ GeV (dashed lines).} \label{hbt}
\end{figure}

From the positions and momenta of pions or kaons at freeze out,
their correlation functions in the longitudinally comoving frame can
be calculated using the program Correlation After Burner
\cite{Pratt:1994uf} to take into account their final-state strong and
Coulomb interactions. Shown in the left and right windows of
Fig.~\ref{hbt} are, respectively, one-dimensional projections of the
correlation functions of midrapidity ($-0.5<y<0.5$) charged pions
and kaons with transverse momentum $300<p_{\rm T}<1500$ MeV$/c$ and
their comparison with corresponding ones for central Au+Au
collisions at $\sqrt{s_{NN}}=200$ GeV at RHIC, which have been shown
to reproduce reasonably measured ones for pions \cite{Lin:2002gc}.
The correlation functions at LHC are seen to be narrower than at
RHIC.

\begin{table}[th]
\caption{Radii from Gaussian fit to correlation functions.}
\medskip
\begin{tabular}{cccccc}\hline\hline
\hfil & $R_{\rm out}({\rm fm})$ & $R_{\rm side}({\rm fm})$ & $R_{\rm
long}({\rm fm})$ & $\lambda$ & $R_{\rm out}/R_{\rm side}$
\\ \hline
 RHIC ($\pi$)& 3.60 & 3.52 & 3.23 & 0.50 & 1.02 \\ \hline
 LHC ($\pi$) & 4.23 & 4.70 & 4.86 & 0.43 & 0.90 \\ \hline
 RHIC (K)& 2.95 & 2.79 & 2.62 & 0.94 & 1.06 \\ \hline
 LHC (K) & 3.56 & 3.20 & 3.16 & 0.89 & 1.11 \\ \hline
\end{tabular}
\label{radii}
\end{table}

Fitting the correlation functions by the Gaussian function $C_2({\bf
Q},{\bf K})=1+\lambda\exp(-\sum_{i}R_{ii}^2(K)Q_i^2)$, where ${\bf
K}$ is the average momentum of two mesons. Extracted radii of the
emission source are shown in Table I. Predicted source radii at LHC
are larger than those at RHIC, consistent with the narrower
correlation functions at LHC than at RHIC. In both collisions, radii
of the emission source for pions are larger than those for kaons.
The smaller lambda parameter for pions than for kaons is due to the
large halo in the pion emission source from decays of omega mesons.
Also, the emission source is non-Gaussian and shifted in the
direction of pion or kaon transverse momentum.

\subsection{Multiplicity distributions and percolation of strings}
\label{s:Pajares}

{\em J. Dias de Deus and C. Pajares}

{\small
In the framework of percolations of strings the rapidity distributions for 
central $\A\A$ collisions are shown for SPS, RHIC and LHC energies. 
The obtained value for LHC is lower than the one predicted for the rest of 
models but larger than the linear energy extrapolation from SPS and RHIC.
}
\vskip 0.5cm

Multiparticle production is currently described in terms of color strings
stretched between the partons of the projectile and the target these color 
strings may be viewed as small areas in the transverse space $\pi r^2_0$, 
$r_0\simeq 0.2-0.3$~fm, filled with color field created by the colliding 
partons. 
Particles are produced via emision of $q\bar{q}$ pairs in this field.
With growing energy and/or atomic number of colliding nuclei, the number of 
strings grows, and they start to overlap, forming clusters, very much similar 
to disks in the two dimensional percolation theory. 
At a certain critical density a macroscopical cluster appears that marks the 
percolation phase transition~\cite{Armesto:1996kt}. 
A cluster behaves as a single string with a higher color field $\vec{Q}_n$ 
corresponding to the vectorial sum of the color changes of each individual 
$\vec{Q}_1$ string.
The resulting color field covers the area $S_n$ of the cluster. 
As $\vec{Q}_n=\sum \vec{Q}_1$, and the individual string colors may be oriented 
in an arbitrary manner respective to one another, the average 
$\vec{Q}_{1i}\cdot \vec{Q}_{1j}$ is zero and $\vec{Q}_n^2=n\vec{Q}_1^2$.

In this way, the multiplicity $\mu_n$ and the average $p^2_T$ of particles
$<p^2_T>_n$ produced by a cluster of $n$ strings, are given by
\begin{equation}
\mu_n=\sqrt{\frac{nS_n}{S_1}\mu_1}\ ;\  
\langle p_T^2\rangle_n=\sqrt{\frac{nS_1}{S_n}}\langle p^2_T\rangle_1
\label{eq:Pajares-eq1}
\end{equation}
where $\mu_1$ and $\langle p^2_T\rangle_1$ are the mean multiplicity and the 
mean transverse momentum of particles produced by a simple string with a 
transverse area $S_1=\pi r^2_0$.

Equation~(\ref{eq:Pajares-eq1}) is the main tool of our calculations. 
In order to compute the multiplicities we generate strings according to the 
quark gluon string model and using a Monte Carlo code. 
Each string is produced at an identified impact parameter. 
From this, knowing the transverse area of each string, we identify all the 
clusters formed is each collision and subsequently compute for each of them the 
rapidity multiplicity spectrum.
\begin{figure}[htb]
\centerline{\includegraphics[width=9 cm]{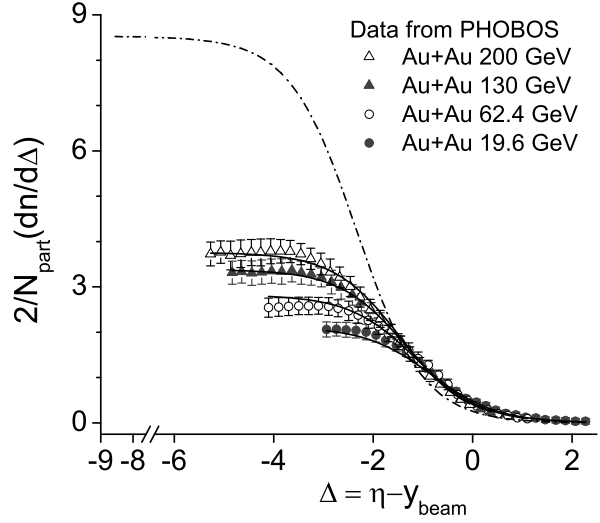}}
\caption{LHC prediction, together with RHIC data and results.}
\label{fig:Pajares-fig1}
\end{figure}

In figure \ref{fig:Pajares-fig1} is shown the results, (see 
reference~\cite{Brogueira:2006nz} for details) for central Au-Au collisions at 
different energies, including the curve for $\sqrtsnn=5.5$~TeV. 
The value at midrapidity 8.5 is similar to other computations in the same 
framework (7.3~\cite{DiasdeDeus:2000gf}, 8.6~\cite{Pajares:2005kk}). 
This strong reduction of the multiplicities relative to simple multicollision 
models, due to the interaction of strings, was anticipated 12 years 
ago~\cite{vanEijndhoven:1995}. 
Nowdays models have incorporated effects, like strong shadowing or triple
pomeron couplings to suppress their original values. 
However, our value is smaller than the one obtained by most of the other 
existing models. 
Only, the extrapolation of the observed geometrical scaling in $lN$ to $\A\A$ 
given a close value: 9.5. 
The linear log of energy extrapolation of the SPS and RHIC values gives a lower 
value of around 6.5.

At SPS and RHIC has been observed an aproximated limiting fragmentation scaling,
which is well reproduced in our approach. 
A clear breaking of this scaling is predicted at LHC.

\subsection{Shear Viscosity to Entropy within a Parton Cascade }
\label{greiner}

{\it A. El, C. Greiner and Z. Xu}

{\small
The shear viscosity is calculated  
by means of the perturbative kinetic partonic cascade BAMPS with CGC initial conditons
for various saturation momentum scale $Q_s$. $\eta /s \approx $ 0.15 stays approximately
constant when going from RHIC to LHC.
}
\vskip 0.5cm

The measured momentum anisotropy parameter $v_2$
at RHIC energy can be well understood 
if the expanding quark-gluon matter is assumed to be described by ideal hydrodynamics.
This suggests that a strongly interacting and locally thermalized state of matter 
has been created which behaves almost like a perfect fluid. 
Since the initial situation of the quark-gluon system is far from thermal equilibrium, it is 
important to understand how and which microscopic partonic interactions can 
thermalize the system within a short timescale and can be responsible
as well for its (nearly) ideal hydrodynamical behaviour.
Furthermore one would like to know  the transport properties of the QGP,
most prominently the shear viscosity. 

A kinetic parton cascade (BAMPS) \cite{Xu:2004mz, Xu:2007aa} has been developed with 
strictly perturbative QCD inspired processes including for the first time inelastic (''Bremsstrahlung'') collisions $gg \leftrightarrow ggg $. The multiparticle back reation channel is treated fully consistently by respecting detailed balance within the same 
algorithm. In \cite{Xu:2007aa} it is demonstrated that the inelastic processes 
dominate the total transport collision rate and thus contribute much stronger 
to momentum isotropization then elastic ones. Within a default setting
of minijet initial conditions, the 
overall build up of elliptic flow $v_2$ can be reasonably described \cite{Xu:2005wv}
(a more dedicated study is presently undertaken \cite{Xu-next}).

\begin{figure}
\hskip -0.5cm
\begin{minipage}[]{50mm}
\epsfxsize=5.0 cm
\epsfysize=5.0 cm
\epsfbox{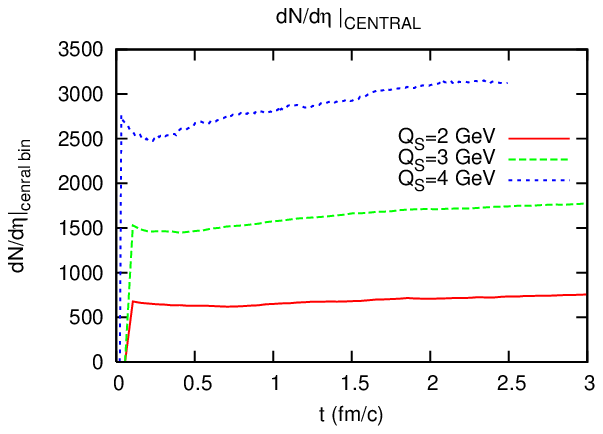}
%\vskip -0.35cm
%\caption{$\frac{dN}{d\eta}$(central rapidity bin), $\alpha_s=0.3$}
\end{minipage}
\begin{minipage}[]{50mm}
\epsfxsize=5.0 cm
\epsfysize=5.0 cm
\epsfbox{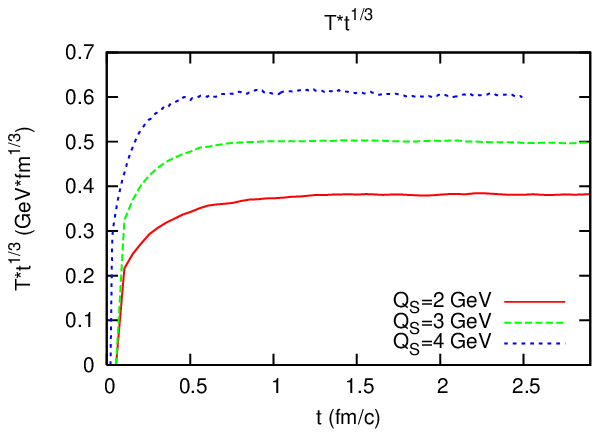}
%\vskip -0.35cm
%\caption{Effective temperature, $\alpha_s=0.3$}
%\label{Tt}
\end{minipage}
\begin{minipage}[]{50mm}
\epsfxsize=5.0cm
\epsfysize=5.0cm
\epsfbox{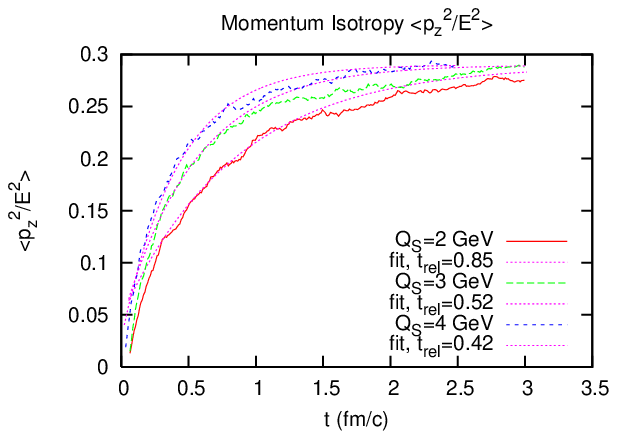}
%\vskip -0.35cm
%\caption{Momentum isotropy, $\alpha_s=0.3$ }

%\label{momis}
\end{minipage}
\caption{Time evolution of $\frac{dN}{d\eta}$ (left), of the effective temperature (middle) and
of the momentum anisotropy (right).}
\label{greinerfig1}
\vskip -0.2cm
\end{figure}

One can thus expect to see thermalization of a QGP on a short time scale less than 1 $fm/c$ 
for LHC relevant initial conditions as can be seen in the evolution in time of 
the temperature and the momentum isotropy depicted in Fig. \ref{greinerfig1}. 
We apply Bjorken expanding geometry in one dimension. 
For the initial condition a simple Color Glass Condensate (CGC) 
gluon distribution is assumed: 
The initial partons are described by the boost-invariant 
form of the distribution function
$f(x,p)|_{z=0}=\frac{c}{\alpha_s \, N_c}\, \frac{1}{\tau_f}\, \delta(p_z)\, \Theta(Q_s^2-p_T^2)$
at a characteristic time $\tau_0 =c/(\alpha_s N_c Q_s)$.

Due to $3\rightarrow 2$ collisions the particle number first decreases (see Fig. \ref{greinerfig1}) 
\cite{El:2006xj}.
This is in contrast to the idealistic ''Bottom-Up'' scenario of thermalization,
where an ongoing particle production in the soft sector ($p_T < \alpha_s Q_S$) is predicted
with a strong increase in the total particle number. The present calculation show that
the particle number roughly stays constant. For the above simple CGC 
parametrization
$Q_s=2$ GeV corresponds to RHIC energy whereas $Q_s \approx 3-4$ GeV is expected for LHC.

\begin{figure} 
\hskip -0.0cm
\epsfxsize=8.0cm
\epsfysize=4.0cm
\centerline{\epsfbox{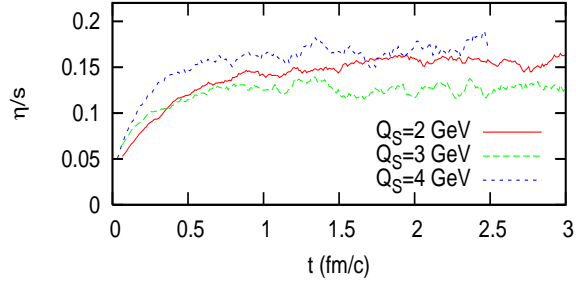}}
\caption{Ratio of the shear viscosity to entropy density ($\alpha_s=0.3$).  }
\label{etaovers}
\vskip -0.2cm
\end{figure}

For all energies a nearly ideal hydrodynamical behavior is observed after $0.5~fm/c$ 
(middle Fig. \ref{greinerfig1}). 
The thermalization time lies in the same range when looking at the momentum isotropy.
It is of crucial importance to extract out of these simulations the transport
properties of QCD matter to quantify the dissipative properties of the fluid.
Using standard dissipative hydrodynamics in expanding geometry shear viscosity and ratio $\frac{\eta}{s}$ can be calculated \cite{El:2006xj}: $\eta=\frac{\tau}{4}\left(T_{xx}+T_{yy}-2\cdot T_{zz}\right)$ and 
$s=4n-n\cdot ln\left(\lambda\right)$, where $\lambda $ denotes the gluon fugacity. 
As depicted in Fig. \ref{etaovers}, the value $\frac{\eta}{s}\approx 0.15$ proves to be a universal number
within the BAMPS simulations, being nearly independent of $Q_S$.
This is in line also with full 3-dim calculations employing minijets and Glauber geometry
for the initial condition \cite{Xu-next}.  $\frac{\eta}{s}$ basically only depends on the 
employed coupling strength $\alpha_s$ (taken to be 0.3 as default setting).
Hence, within BAMPS, we do not expect any change in the shear viscosity ratio
$\eta /s $ when going from RHIC to LHC.

\subsection{Hadron multiplicities, $p_T$ spectra and net-baryon number in central 
  Pb+Pb collisions at the LHC}
\label{eskola1}
{\it K. J. Eskola, H. Honkanen, H. Niemi, P. V. Ruuskanen and S. S. R\"as\"anen}
\vskip 0.5cm
 
We summarize here our recent LHC predictions~\cite{Eskola:2005ue}, obtained in 
the framework of perturbative QCD (pQCD)+saturation+hydrodynamics (EKRT model 
for brief)~\cite{Eskola:1999fc}. This model has successfully predicted~\cite{%
  Eskola:1999fc,Eskola:2001bf} the charged particle multiplicities in central 
Au+Au collisions at different $\sqrtsnn$, and it also describes the low-$p_T$ 
spectra of pions and kaons at RHIC quite well~\cite{Eskola:2005ue,%
  Eskola:2002wx}. 

Primary parton production in the EKRT model is computed from collinearly 
factorized pQCD cross sections~\cite{Eskola:1988yh} by extending the calculation
towards smaller $p_T$ until the abundant gluon production vertices overlap and 
gluon fusions~\cite{Gribov:1984tu} saturate the number of produced partons 
(gluons). The saturation scale is determined as $p_0=p_{\rm sat}$ from a 
saturation condition~\cite{Eskola:1999fc}
$N_{\A\A}(p_0, \sqrt s)\cdot \pi/p_0^2  = c \cdot \pi R_{\A}^2$,
where $N_{\A\A}(p_0, \sqrt s)$ is the average number of partons produced at 
$|y|\le 0.5$ and $p_T\ge p_0$. With a constant $c=1$ the framework is closed. 
For central Pb+Pb collisions at the LHC $p_{\rm sat}\approx 2$~GeV. We obtain the 
initial conditions for the cylindrically symmetric boost invariant 
(2+1)-dimensional hydrodynamical description by converting the computed 
transverse energy $E_T(p_{\rm sat})$ and net-baryon number $N_B(p_{\rm sat})$ into 
densities $\epsilon(r,\tau_0)$ and $n_B(r,\tau_0)$ using binary collision 
profiles and formation time $\tau_0=1/p_{\rm sat}$.

Assuming a fast thermalization at $\tau_0$, and zero initial transverse fluid
velocity, we proceed by solving the standard equations of ideal hydrodynamics 
including the current conservation equation for net-baryon number. In the 
Equation of State we assume an ideal gas of gluons and massless quarks 
($N_f=3$), the QGP, with a bag constant $B$ at $T>T_c$, and a hadron resonance 
gas of all states with $m<2$~GeV at $T<T_c$. Taking $B^{1/4}=239$~MeV leads to 
first-order transition with $T_c=165$ MeV. Final state hadron spectra are 
obtained with Cooper-Frye procedure on a decoupling surface at $T_{\rm dec}$ 
followed by strong and electromagnetic 2- and 3-body decays of unstable states 
using the known branching ratios. Extensive comparison~\cite{Eskola:2005ue,%
  Eskola:2002wx} with RHIC data suggests a single decoupling temperature 
$T_{\rm dec}=150$~MeV which is also used to calculate the predictions for the LHC.
For details, see~\cite{Eskola:2005ue}.

Our predictions~\cite{Eskola:2005ue} for the LHC multiplicities, transverse 
energies and net-baryon number at $y=\eta=0$ for 5\% most central Pb+Pb 
collisions at $\sqrtsnn=5.5$~TeV are summarized in table \ref{eskola-tab1}. 
Note that the predicted charged particle multiplicity 
${\rm d}N_{\rm ch}/{\rm d}\eta$ is 2570,
i.e. only a third of the initial ALICE design value (see 
also~\cite{Eskola:2001bf}). Whereas the multiplicity of initially produced 
partons and observable hadrons are close to each other, the transverse energy is
reduced by a factor as large as 3.4 in the evolution from initial state to final
hadrons. Due to this reduction the very high initial temperature, 
$T_0\gtrsim 1$~GeV, possibly observable through the emission of photons, need 
not lead to contradiction between predicted and observed $E_T$.
\begin{table}[hbt]
  \caption{ }
\lineup
\begin{tabular}{@{}*{13}{l}}
\br                              
$\frac{{\rm d}N}{{\rm d}y}^{\rm tot}$&$\frac{{\rm d}N}{{\rm d}\eta}^{\rm tot}$&$\frac{{\rm d}N}{{\rm d}y}^{\rm ch}$& %%@
$\frac{{\rm d}N}{{\rm d}\eta}^{\rm ch}$&$\frac{{\rm d}N}{{\rm d}y}^{B}$&$\frac{{\rm d}E}{{\rm d}y}^{T}$&$\frac{{\rm d}E}{{\rm d}\eta}^{T}$& %%@
$\frac{{\rm d}N}{{\rm d}y}^{\pi\pm}$&$\frac{{\rm d}N}{{\rm d}y}^{\pi0}$
&$\frac{{\rm d}N}{{\rm d}y}^{K\pm}$&$\frac{{\rm d}N}{{\rm d}y}^{p}$&$\frac{{\rm d}N}{{\rm d}y}^{\bar p}$&$p/\bar p$
\cr 
\mr
{\small 4730} 					& {\small 4240} 						& {\small 2850}					%%@
&\textbf{{\small 2570}}					&
{\small 3.11}					& {\small 4070}						& {\small 3710}
& {\small 1120} & {\small 1240} & {\small 214} & {\small 70.8} & {\small 69.6} & {\small 0.98}
\cr
\br
\end{tabular}
\label{eskola-tab1}
\end{table}

\begin{figure}[h]
 \begin{flushleft}
   \vspace{-1.4cm}
  \includegraphics*[width=7.5cm]{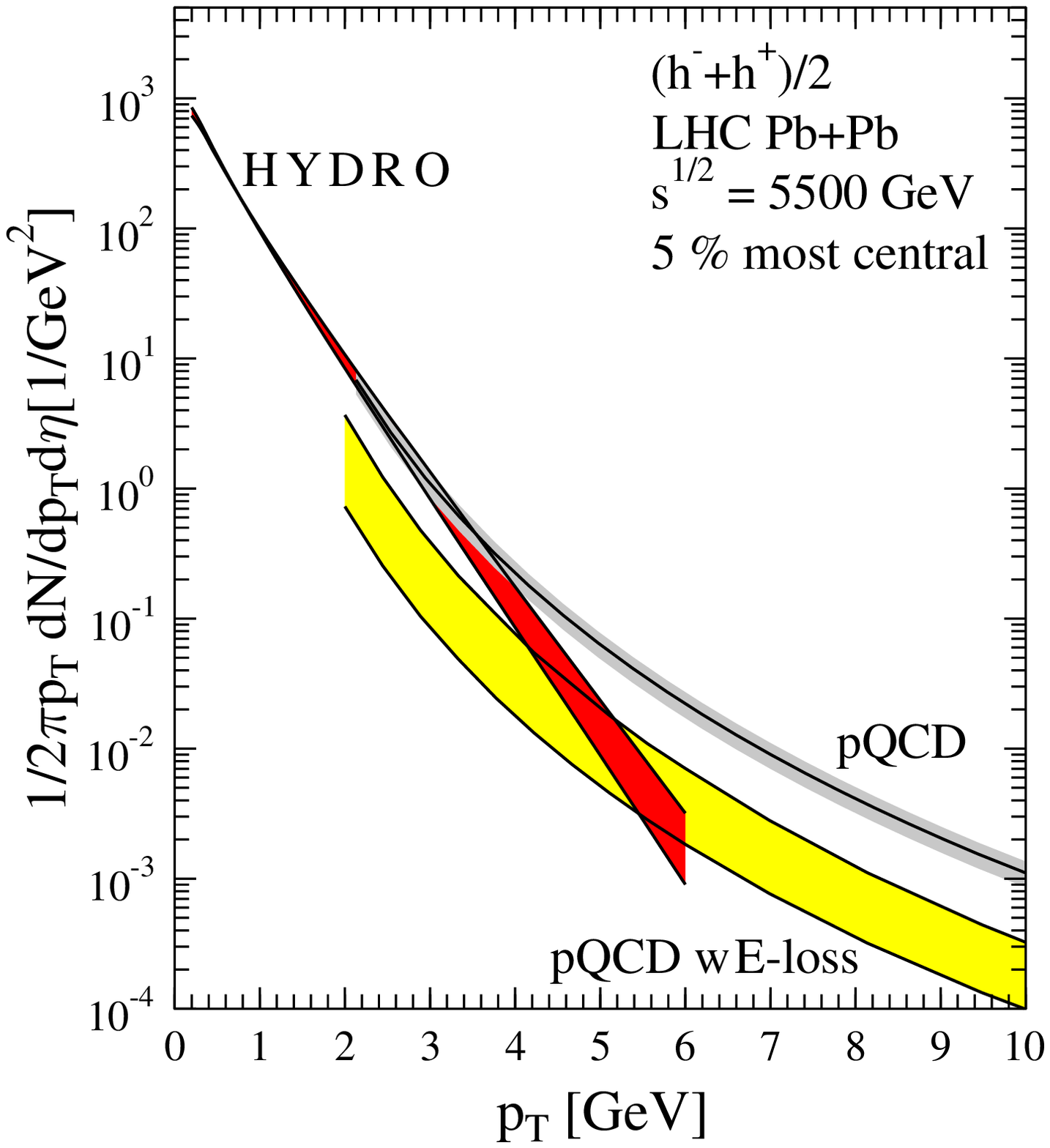}
\end{flushleft}
\begin{flushright}
	\vspace{-9.5cm}
  \includegraphics*[width=7.5cm]{Eskola1-fig1.eps}
   \vspace{-1cm}
 \end{flushright}
  \caption{ }
  \protect{\label{fig1eska}}
\end{figure}

Our prediction for the charged hadron $p_T$ spectrum is the lower limit of the 
red band (HYDRO, the width corresponding to $T_{\rm dec}=120\dots 150$~MeV) in 
the l.h.s. of figure \ref{fig1eska} \cite{Eskola:2005ue}. The 
corresponding $p_T$ distributions of $\pi^+$ and $K^+$ are shown in the r.h.s. 
of the figure (solid lines). The pQCD reference spectra, obtained by folding the
LO pQCD cross sections with the nuclear PDFs and fragmentation functions (KKP) 
and accounting for the NLO contributions with a $\sqrtsnn$-dependent $K$-factor 
from~\cite{Eskola:2002kv}, are also shown (pQCD) on the r.h.s. The yellow bands 
(pQCD+E-loss) show the results with parton energy losses included as 
in~\cite{Eskola:2004cr}. We thus predict the applicability region of 
hydrodynamics at the LHC to be $p_T\lesssim 4\dots 5$~GeV, i.e. a wider region 
than at RHIC.  

\subsection{Melting the Color Glass Condensate at the LHC}

{\it
H. Fujii, F. Gelis, A. Stasto and R. Venugopalan}

{\small
The charged particle multiplicity in central
  AA collisions  and the production of heavy flavors
  in pA collisions at the LHC is predicted in the CGC framework.
  }
%\maketitle

\subsubsection{Introduction}
In the  Color Glass Condensate (CGC) framework,  fast (large $x$) partons are described as frozen light cone color sources while the soft
(small $x$) partons are described as gauge fields. The distribution of
the fast color sources and their evolution with rapidity is described by the JIMWLK evolution
equation; it is well approximated for large nuclei by the
Balitsky-Kovchegov (BK) equation. When two hadrons collide,  a time dependent color field is 
produced  that eventually decays into gluons~\cite{Gelis:2006dv}.  When the projectile is dilute (e.g.,AA collisions at forward rapidity or  pA collisions), $k_\perp$ factorization holds for gluon production, thereby simplifying computations. For quark production, $k_\perp$ factorization breaks down and is recovered only for large invariant masses and momenta.

\subsubsection{Particle multiplicity in central AA collisions}
The $k_\perp$ factorized cross-sections are convolutions over ``dipole" scattering amplitudes in the projectile and target. Initial conditions for the BK evolution of these are specified at an initial $x=x_0$ (chosen here to be  $x_0\approx 10^{-2}$). In this work \cite{Gelis:2006tb}, we consider two  initial conditions, based respectively on the McLerran-Venugopalan (MV) model or on the Golec-Biernat--Wusthoff (GBW) model.We adjust the free parameters  to reproduce the limiting fragmentation curves measured at RHIC from$\sqrt{s}=20$~GeV to $\sqrt{s}=200$~GeV. The value of $\alpha_s$ in the {\sl fixed coupling} BK equation is tuned to obtain the observed rate of growth of the saturation scale.The rapidity distribution $dN/dy$ is converted into the
pseudo-rapidity distribution $dN/d\eta$ by asuming the produced particles have $m\sim
200$~MeV.
\begin{figure}[htbp]
\begin{center}
\resizebox*{5cm}{!}{\includegraphics{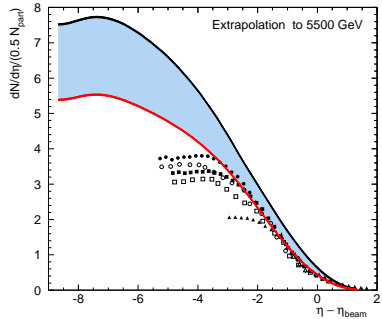}}
\end{center}
\caption{\label{fig:dndeta}Number of charged particles per unit of
  pseudo-rapidity at the LHC energy.}
\end{figure}
A prediction for AA collisions at the LHC is obtained by
changing $\sqrt{s}$ to $5.5$~GeV. From  Fig. \ref{fig:dndeta}, we can infer $dN_{\rm
  ch}/d\eta\big|_{\eta=0}=1000-1400$; the two endpoints correspond to GBW and MV initial conditions respectively.

\subsubsection{Heavy quark production in pA collisions}
\begin{figure}[htbp]
\begin{center}
\resizebox*{5cm}{!}{\includegraphics{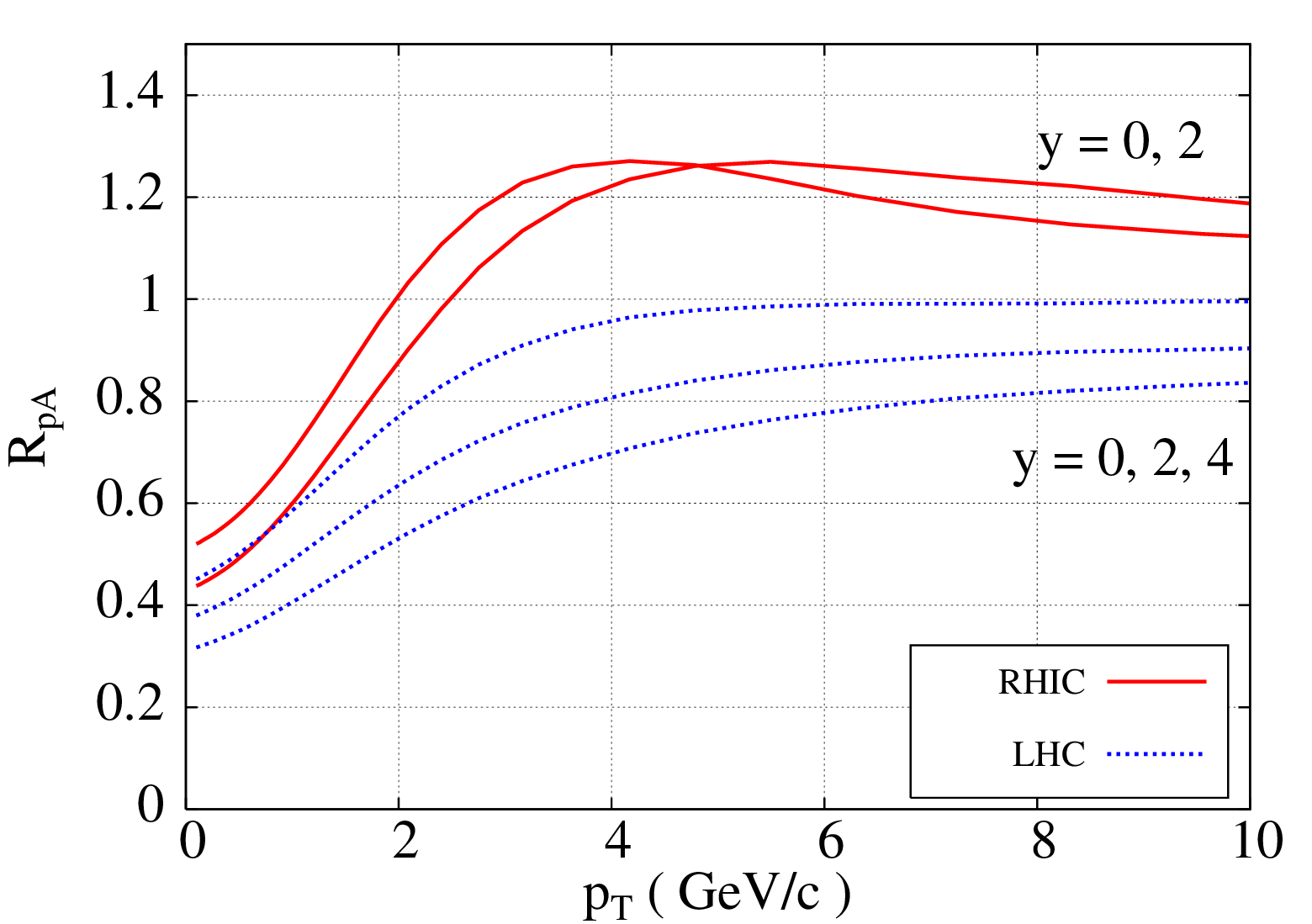}}
\hskip 5mm
\resizebox*{5cm}{!}{\includegraphics{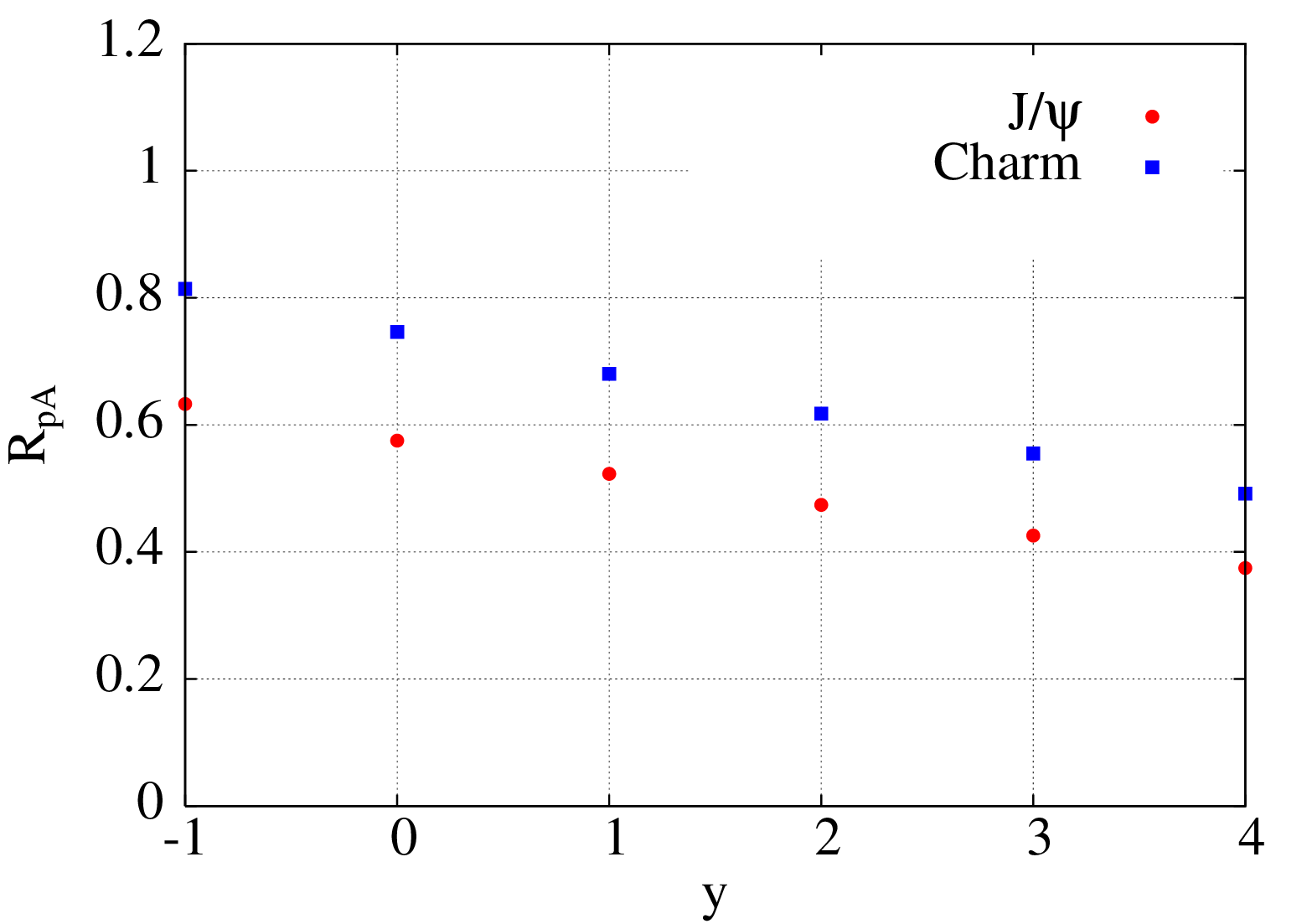}}
\end{center}
\caption{\label{fig:rpa}Left: nuclear modification factor for $D$
  mesons as a function of $p_\perp$. Right: the same ratio as a
  function of rapidity, for $D$ mesons and for $J/\psi$.}
\end{figure}

The cross-section for the
production of a pair of heavy quarks \cite{Gelis:2006cr} is the simplest process for which  $k_\perp$-factorization breaks down\cite{Fujii:2005vj} in pA collisions. This is due to the sensitivity of the cross-section to 3- and 4-point correlations in the
nucleus. Integrating out the antiquark and convoluting with a
fragmentation function, one obtains the cross-section for open heavy
flavor production, e.g., $D$ mesons. Alternatively, one can use the
Color Evaporation Model to obtain the cross-section for
quarkonia bound states.  The
nuclear modification ratio is displayed in figure \ref{fig:rpa}. The main
difference at the LHC compared to RHIC energy is that this ratio is smaller than unity already at mid rapidity, and decreases further towards the proton fragmentation region.

\subsection{$R_{\rm pA}$ ratio: total shadowing due to running coupling}
%\protect{\label{iancu}}

{\it E. Iancu and D. N. Triantafyllopoulos}

{\small 
We predict that the $\rpa$ ratio at the most forward rapidities to be measured 
at LHC should be strongly suppressed, close to ``total shadowing'' 
($\rpa\simeq \A^{-1/3}$), as a consequence of running coupling effects in the 
nonlinear QCD evolution.
}
\vskip 0.5cm

\begin{figure}[h]
\centerline{\includegraphics[width=7.5cm]{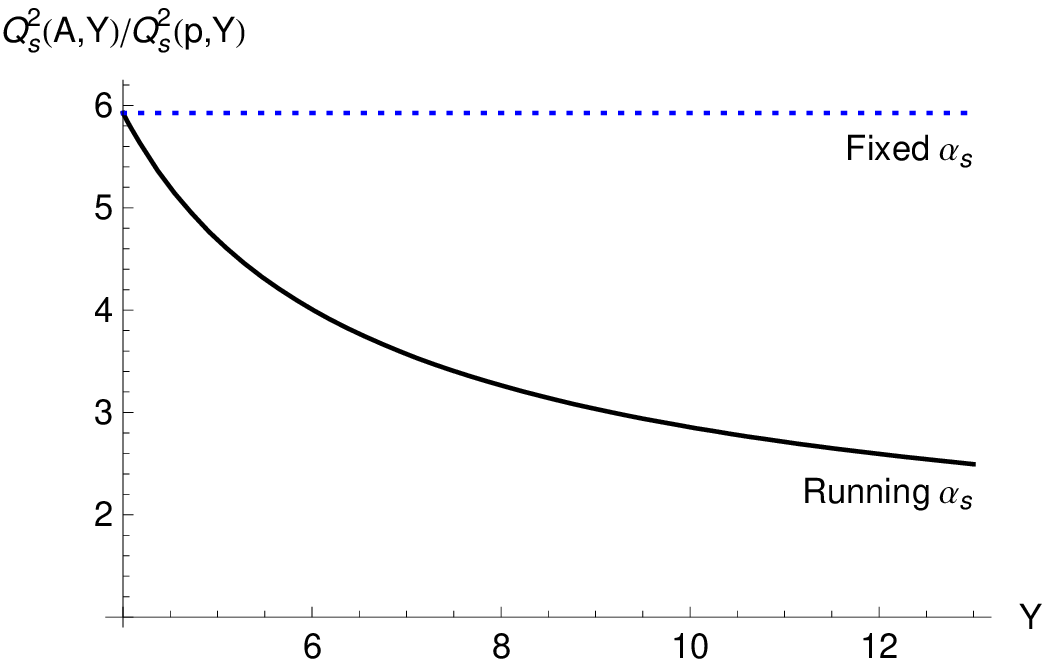}\hspace{0.5cm}
    \includegraphics[width=7.5cm]{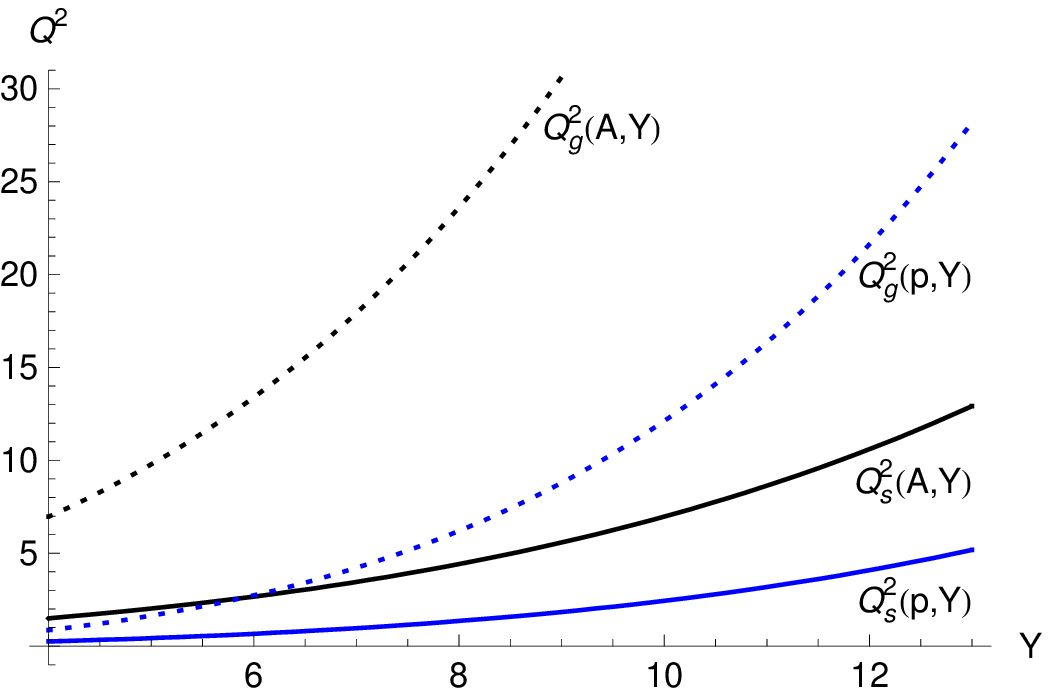}}
    \caption{Left: The ratio of the saturation momenta.
    ($Y=12$ corresponds to a pseudorapidity
    $\eta=6$ for the produced particles). Right: Geometric scaling windows.}
%\vspace{-0.5cm}
\protect{\label{iancufig1}}
\end{figure}

We present predictions for the nuclear modification factor, or ``$\rpa$ ratio'',
at forward pseudorapidities ($\eta>0$) and relatively large transverse momenta 
($\ptrans$) for the produced particles, in the kinematical range to be 
accessible at LHC. These predictions are based on a previous, systematic, study 
of the $\rpa$ ratio within the Color Glass Condensate formalism with running
coupling~\cite{Iancu:2004bx}. The ratio can be approximated by
\begin{equation}
\protect{\label{eq:iancu-rpadef}}
\rpa \simeq \frac{1}{\A^{1/3}} \frac{\Phi_{\A}(Y,\ptrans)}{\Phi_p(Y,\ptrans)},
\end{equation}
where $Y=\eta + \ln\sqrt{s}/\ptrans$ and $\Phi(Y,\ptrans)$ is the unintegrated 
gluon distribution of the corresponding target hadron at fixed impact parameter.
When the energy increases one expects more and more momentum modes of this 
distribution to saturate to a value of order $1/\alpha_s$, and the corresponding
saturation momentum reads
\begin{equation}
\label{eq:iancu-Qsat}
Q_s^2(Y) = \Lambda^2 \exp{\sqrt{B (Y-Y_0) + \ln^2\frac{Q_s^2(Y_0)}{\Lambda^2}}},
\end{equation}
with $\Lambda=0.2\mbox{ GeV}$, $B=2.25$ and $Y_0=4$. The initial condition for 
the nucleus and the proton are taken as $Q_s^2(\A,Y_0) = 1.5\mbox{ GeV}^2$ and 
$Q_s^2(p,Y_0) = 0.25\mbox{ GeV}^2$ respectively, so that 
$Q_s^2(\A,Y_0) = \A^{1/3} Q_s^2(p,Y_0)$ for $\A=208$. The functional form of this 
expression is motivated by the solution to the nonlinear QCD evolution equations
with running coupling~\cite{Mueller:2002zm,Mueller:2003bz}, while the actual 
values of the numbers $B$ and $Y_0$ have been chosen in such a way to agree with
the HERA/RHIC phenomenology. As shown in Fig.~\ref{iancufig1}, with 
increasing $Y$ the two saturation momenta approach to each other and clearly for
sufficiently large $Y$, a nucleus will not be more dense than a 
proton~\cite{Mueller:2003bz}.

For momenta $\ptrans$ larger than $Q_s$, the gluon distribution satisfies 
geometrical scaling~\cite{Iancu:2002tr,Mueller:2002zm}, i.e.\ it is a function 
of only the combined variable $\ptrans/Q_s(Y)$:
\begin{equation}
\label{eq:iancu-phi}
\Phi(p_{\bot},Y) \propto \left[\frac{Q_s^2(Y)}{p_{\bot}^2} \right]^{\gamma} 
  \left( \ln\frac{p_{\bot}^2}{Q_s^2(Y)} + c \right),
\end{equation}
with $\gamma=0.63$ and $c=\mathcal{O}(1)$. This holds within the scaling window 
$Q_s \lesssim \ptrans \lesssim Q_g$, where 
$\ln Q_g^2(Y)/Q_s^2(Y) \sim [\ln Q_s^2(Y)/\Lambda^2]^{1/3}$ and for large $Y$ this 
is proportional to $Y^{1/6}$. The geometrical scaling lines for a proton and a 
nucleus are shown in Fig.~\ref{iancufig1}. Note that, since $Q_g$ is 
increasing much faster than $Q_s$, a {\em common scaling window\/} exists, at 
$Q_s(\A,Y) \lesssim \ptrans \lesssim Q_g(p,Y)$ (and for sufficiently large $Y$),
where the gluon distributions for both the nucleus and the proton are described 
by Eq.~(\ref{eq:iancu-phi}).

\begin{figure}[h]
\centerline{\includegraphics[width=7.5cm]{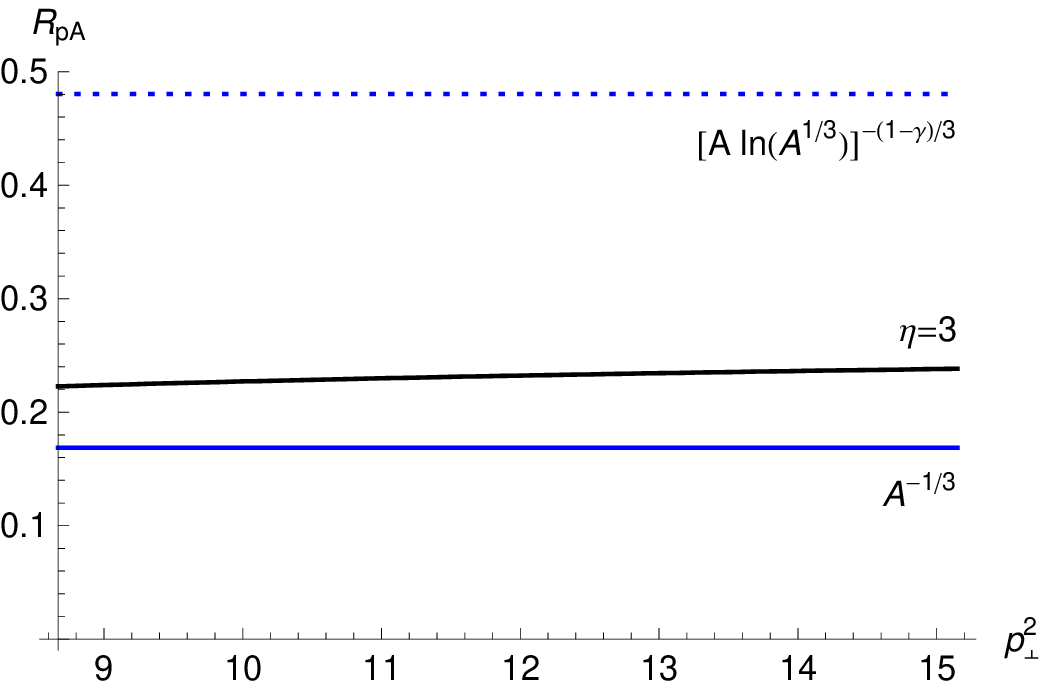}\hspace{0.5cm}
    \includegraphics[width=7.5cm]{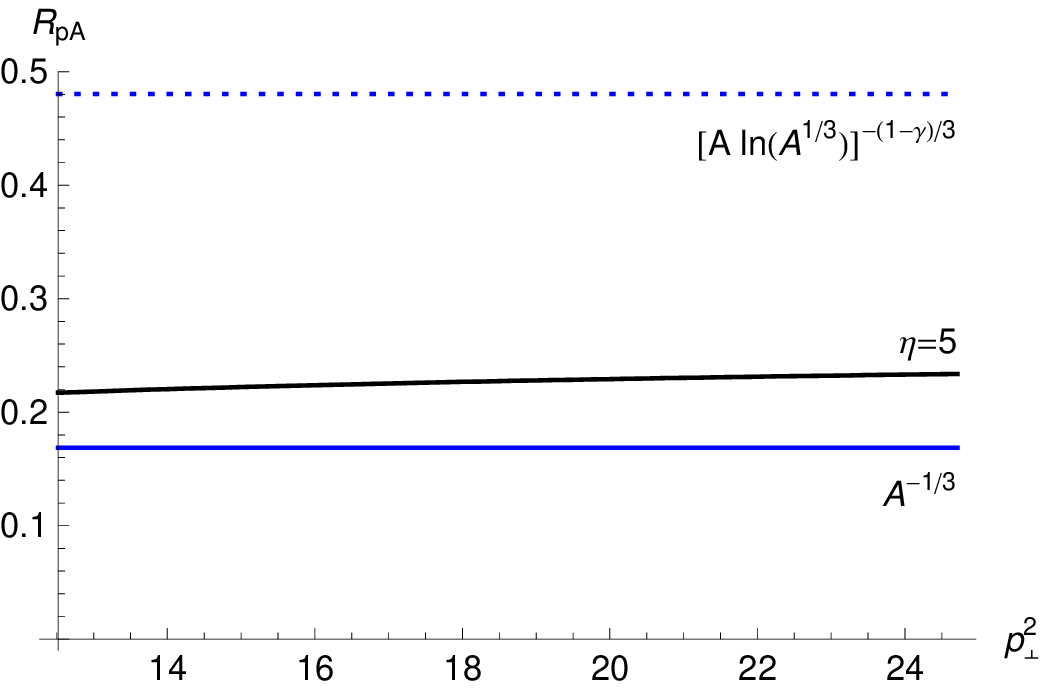}}
    \caption{The ratio $\rpa$ as a function of $\ptrans^2$ at 
       $\sqrt{s}=8.8\,$TeV.}
%\vspace{-0.4cm}
\protect{\label{iancufig2}}
\end{figure}

Within this window, it is straightforward to calculate the $\rpa$ ratio. This is
shown in Fig.~\ref{iancufig2} for two values of pseudorapidity. The upper, 
dotted, line is the asymptotic prediction of a fixed-coupling scenario, in which
the ratio ${Q_s^2(\A,Y)}/{Q_s^2(p,Y)}={\rm const.}=\A^{1/3}$, while the lowest, 
straight, curve is the line of total shadowing $\rpa=1/\A^{1/3}$. Our prediction 
with running coupling is the line in between and it is very close to total 
shadowing. This is clearly a consequence of the fact that the proton and the 
nuclear saturation momenta approach each other with increasing energy.

Note finally that in the present analysis we have neglected the effects of 
particle number fluctuations (or ``Pomeron loops''). This is appropriate since 
Pomeron loops effects are suppressed by the running of the 
coupling~\cite{Dumitru:2007ew}, and thus can be indeed ignored at all energies 
of phenomenological interest (in particular, at LHC).

\def\be{\begin{eqnarray}}
\def\ee{\end{eqnarray}}
\def\del{\partial}
\def\Eq#1{Eq.~(\ref{#1})}

\subsection{LHC ${\rm d}N_{\rm ch}/{\rm d}\eta$ and $N_{\rm ch}$ from Universal Behaviors}
\label{s:Jeon1}
{\it S. Jeon, V. Topor Pop and M. Bleicher}

{\small
RHIC ${\rm d}N_{\rm ch}/{\rm d}\eta$ contains {\em two\/} universal curves, one 
for limiting fragmentation and one for the transition region. By extrapolating,
we predict ${\rm d}N_{\rm ch}/{\rm d}\eta$ and $N_{\rm ch}/N_{\rm part}$ at the LHC 
energy.
}
\vskip 0.5cm

\begin{figure}[thb]
\vspace{0.6cm}
\begin{center}
\includegraphics[width=0.9\textwidth,height=6cm]{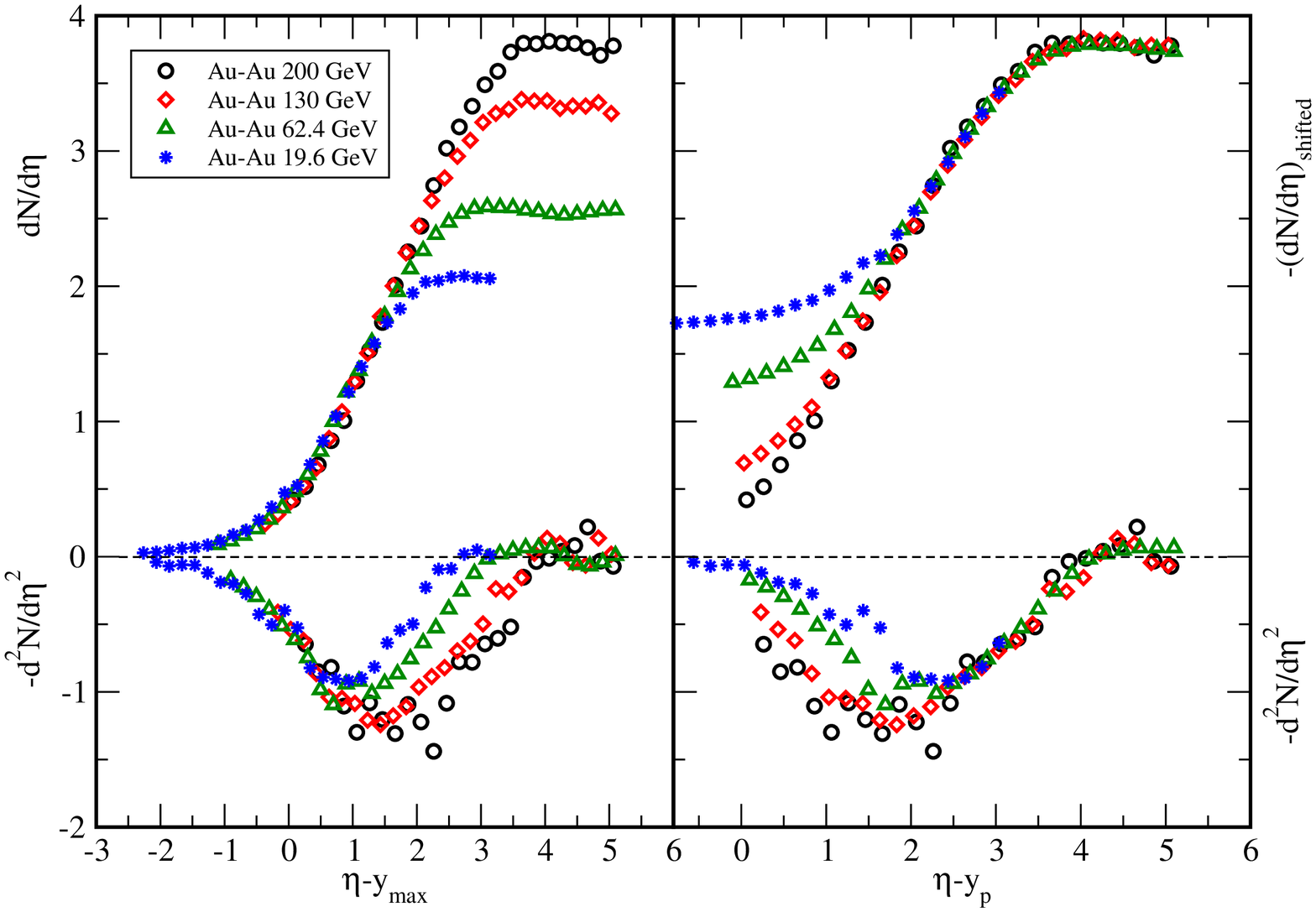}
\end{center}
\caption{Evidence of two universal curves in RHIC ${\rm d}N/{\rm d}\eta$ data.
The slope ${\rm d}^2N/{\rm d}\eta^2$ is inverted for visibility. In the left 
panel, $y_{\rm max} \approx \ln(\sqrt{s}/m_N)$ are matched whereas in the right 
panel, the shoulders of ${\rm d}N/{\rm d}\eta$ are matched.}
\label{fig:Jeon1-fig1}
\end{figure}

Data from RHIC at all energies clearly show limiting fragmentation 
phenomena~\cite{Back:2002wb} for very forward and very backward rapidities. 
In reference~\cite{Jeon:2003nx}, we have shown that in the RHIC 
${\rm d}N/{\rm d}\eta$ (normalized to the number of colliding nucleon pairs) 
spectra at various energies, there are in fact {\em two\/} universal curves.  
This fact is not readily visible if one compares the ${\rm d}N/{\rm d}\eta$ from
different energies directly. It is, however, clearly visible in the slope
${\rm d}^2N/{\rm d}\eta^2$ as shown in the left panel in 
figure~\ref{fig:Jeon1-fig1}
In this panel, we have plotted ${\rm d}N/{\rm d}\eta$ per participant pair from 
the PHOBOS collaboration for $\sqrtsnn = 19.6, 62.4, 130, 200\mbox{ GeV}$ as a
function of $\eta-y_{\rm max}$ with the corresponding ${\rm d}^2N/{\rm d}\eta^2$.

Even though the curves all look similar in ${\rm d}N/{\rm d}\eta$, it is rather 
obvious in ${\rm d}^2N/{\rm d}\eta^2$ that the true universal behavior is 
maintained only up to about $50\%$ of the maximum height. More interestingly, 
there emerges another universal behavior beyond that point as shown in the
right panel in Fig.1. In this panel, we have shifted ${\rm d}N/{\rm d}\eta$ 
vertically and horizontally to match the shoulder. The {\em common\/} straight 
line in ${\rm d}^2N/{\rm d}\eta^2$ in this region clearly show that the shoulder
region in ${\rm d}N/{\rm d}\eta$ is a quadratic function of $\eta$. Moreover 
the {\em curvature\/} of the quadratic function is independent of the colliding 
energy. The universality of these two curves also implies that 
$({\rm d}N/{\rm d}\eta)_{\eta=0}$ will at most grow like $\ln^2(\sqrtsnn/m_N)$ and
the total number of produced particle $N_{\rm ch}$ can at most grow like 
$\ln^3(\sqrtsnn/m_N)$.

\begin{figure}[thb]
\raisebox{3cm}{
 \begin{tabular}{|c|c|c|}
 \hline
 &  $({\rm d}N/{\rm d}\eta)_0$ & $N_{\rm total}$ \\
 \hline
 Param I  & 6.9 & 87\\
 \hline
 Param II  & 6.5 & 83\\
 \hline
 K \& L  &  10.7 & 110\\
 \hline
 HIJING w/ $p_0=3.5\,{\rm GeV}$ & 21.4 & 160\\
 \hline
 HIJING w/ $p_0=5.0\,{\rm GeV}$ & 13.6 & 110\\
 \hline
 \end{tabular}
}
\includegraphics[height=6cm]{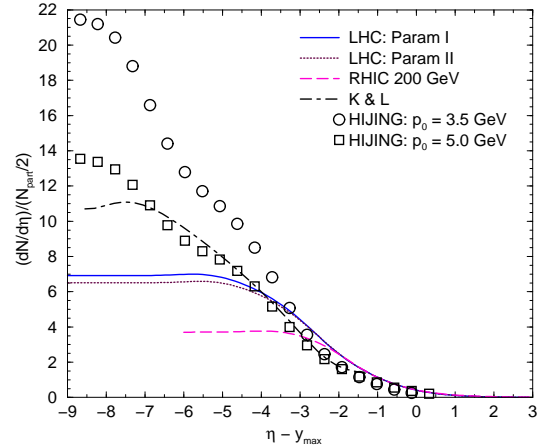}
\caption{Predictions for 6\% central Au-Au collisions at LHC. Curves and rows 
  labeled Param I \& II are our predictions. For comparison, HIJING predictions 
  with two different minijet parameters and Kharzeev and Levin formula~\cite{%
  Kharzeev:2001gp} extrapolated to LHC are also shown.}
\label{fig:Jeon1-fig2}
\end{figure}

Parameterizing the ${\rm d}^2N/{\rm d}\eta^2$ with simple functions in two 
slightly different ways (for details see reference~\cite{Jeon:2003nx}, we can 
easily extrapolate to the LHC energy as shown in figure~\ref{fig:Jeon1-fig2}.
Our prediction is slightly higher than purely linear extrapolation carried out 
by W.~Busza in reference~\cite{Busza:2004mc}.

\subsection{Hadron multiplicities at the LHC}
{\em D. Kharzeev, E. M. Levin and M. Nardi}

{\small
We present the predictions for hadron multiplicities in $pp$, $p\A$ and $\A\A$ 
collisions at the LHC based on our approach to the Color Glass Condensate.
}
\vskip 0.5cm

\begin{figure}[ht]
  \begin{tabular}{cc} 
      \includegraphics[width=6.3cm]{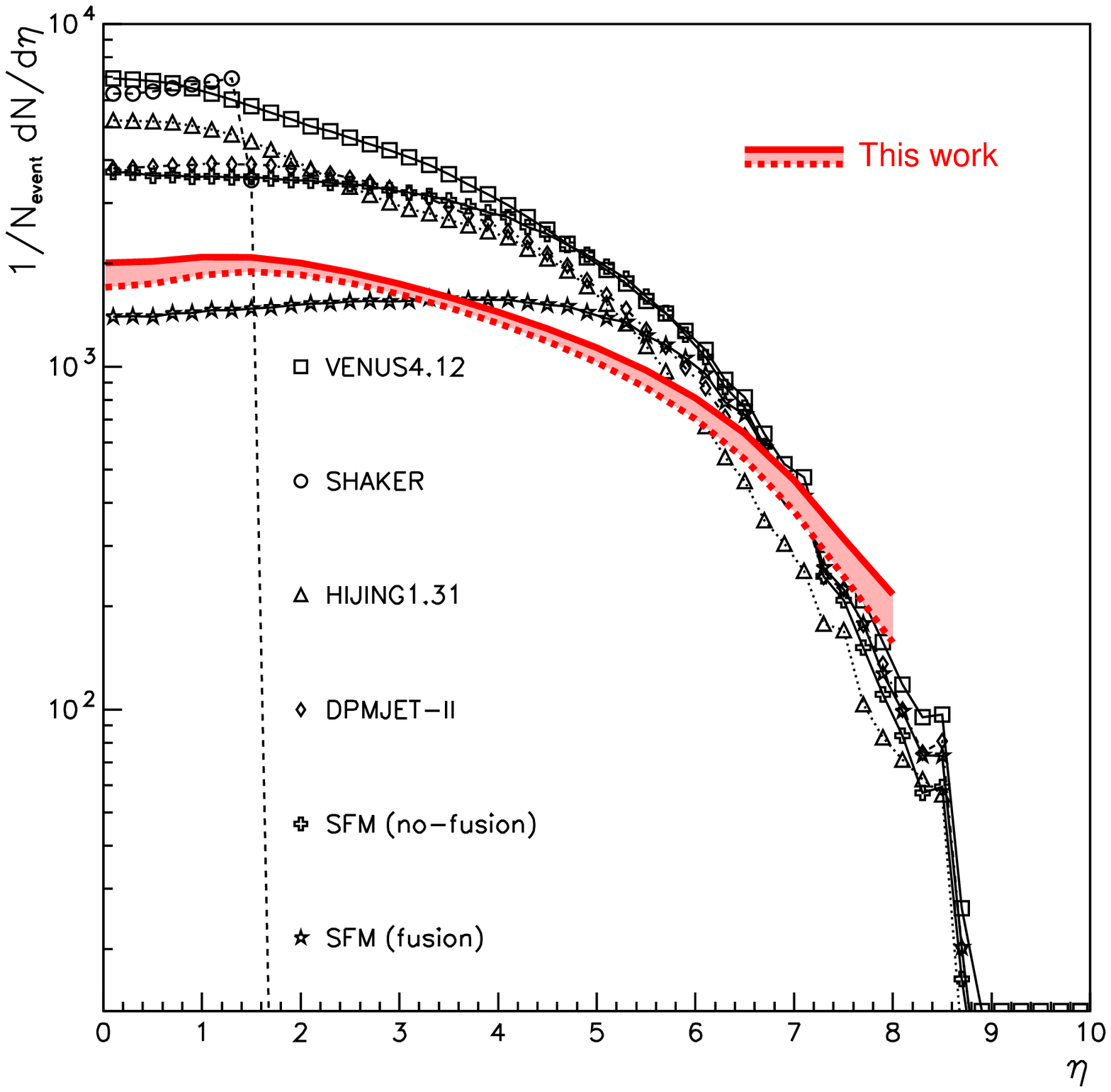} &
      \includegraphics[width=6.3cm]{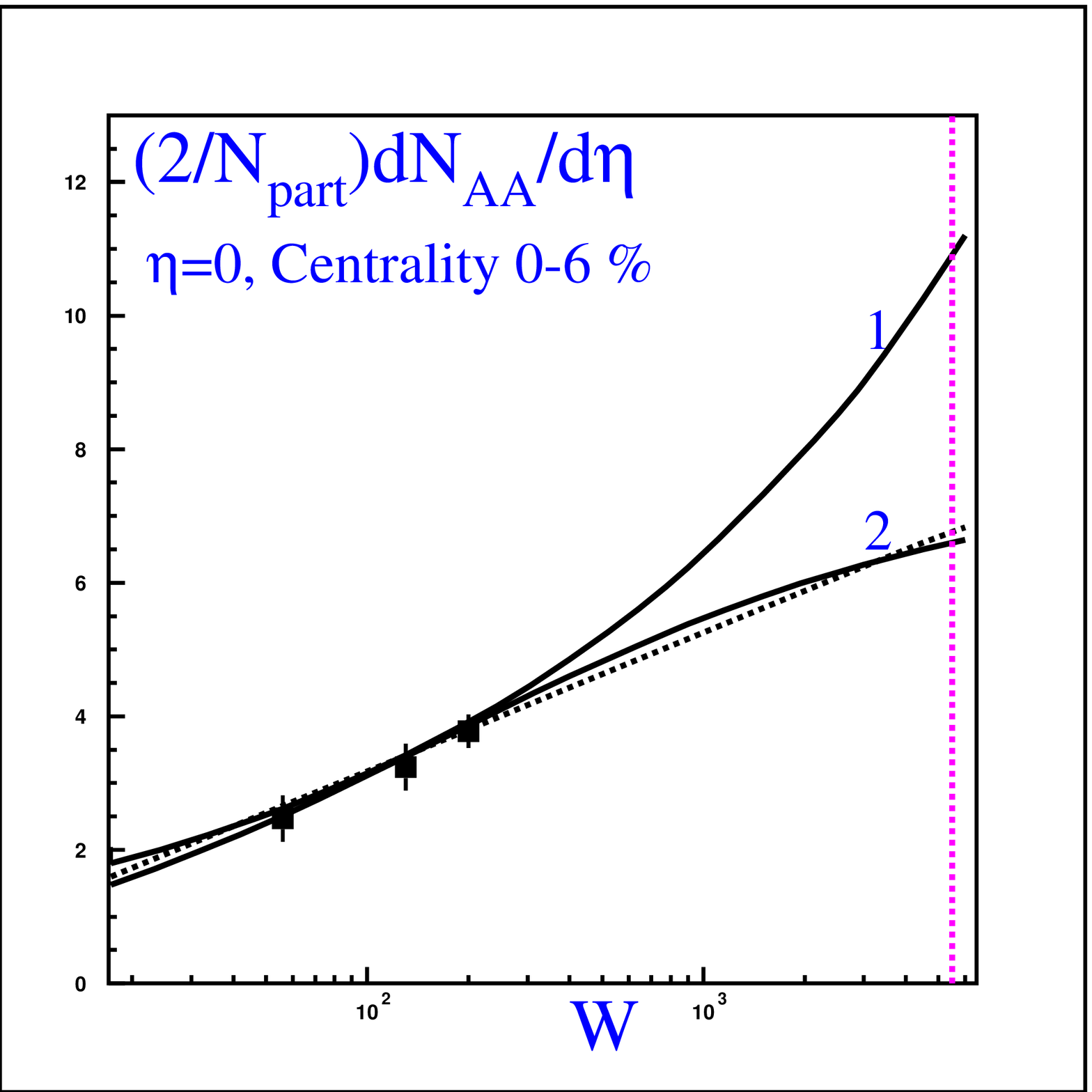}\\
         (a) & (b) 
  \end{tabular}
\caption{(a) Charged hadron multiplicity in Pb-Pb collisions as a function of 
  pseudo-rapidity at $\sqrtsnn = 5.5$~TeV; also shown are predictions from other
  approaches (from~\cite{Kharzeev:2004if}); (b) Energy dependence of charged 
  hadron multiplicity per participant pair in central $\A\A$ collisions for
  different approaches to parton evolution (curves 1 and 2); also shown is the 
  logarithmic fit, dashed curve (from~\cite{{Kharzeev:2007??}}). }
\label{fig:Kharzeev-fig}
\end{figure}

We expect that at LHC energies, the dynamics of soft and semi-hard interactions 
will be dominated by parton saturation. 
In this short note we summarize our results for hadron multiplicities basing on 
the approach that we have developed and tested at RHIC energies in recent 
years~\cite{Kharzeev:2000ph,Kharzeev:2001gp,Kharzeev:2001yq,Kharzeev:2002ei}; 
a detailed description of our predictions for the LHC energies can be found in 
reference~\cite{Kharzeev:2004if}. 
In addition, we will briefly discuss the properties of non-linear evolution at 
high energies, and their implications; details will be presented 
elsewhere~\cite{Kharzeev:2007??}.  
Our approach is based on the description of initial wave functions of colliding 
hadrons and nuclei as sheets of Color Glass Condensate. 
We use a corresponding ansatz for the unintegrated parton distributions, and 
compute the inclusive cross sections of parton production using 
$k_{\perp}$-factorization.  
The hadronization is implemented through the local parton-hadron duality -- 
namely, we assume that the transformation of partons to hadrons is a soft 
process which does not change significantly the angular (and thus 
pseudo-rapidity) distribution  of the produced particles. 
Because of these assumptions, we do not expect our results be accurate for the 
transverse momentum distributions in $\A\A$ collisions, but hope that our 
calculations (see figure~\ref{fig:Kharzeev-fig}a) will apply to the total 
multiplicities. 

While our approach has been extensively tested at RHIC,  an extrapolation of 
our calculations to the LHC energies requires a good theoretical control over 
the rapidity dependence of the saturation momentum $Q_s(y)$.  
The non-linear parton evolution in QCD is a topic of vigorous theoretical 
investigations at present. 
Recently, we have investigated the role of longitudinal color fields in parton 
evolution at small $x$, and found that they lead to the following dependence of 
the saturation momentum on rapidity~\cite{Kharzeev:2007??}:
\begin{equation} 
\label{eq:Kharzeev1}
Q^2_s(Y)=\frac{Q^2_s(Y=Y_0)\,\exp\left( \frac{2\,\alpha_S}{\pi}(Y-Y_0) \right)}
{1+B\,Q^2_s(Y=Y_0)\left(\exp\left(\frac{2\,\alpha_S}{\pi}(Y-Y_0)\right)-1\right)},
\end{equation}
where $B=1/(32\,\pi^2) \,\,(\pi\,R^2_A/{\alpha_S})$; 
$R_A$ is the area of the nucleus, and $\alpha_S$ is the strong coupling 
constant. 
At moderate energies, equation~(\ref{eq:Kharzeev1}) describes an exponential 
growth of the saturation momentum with rapidity; when extrapolated to the LHC 
energy this results in the corresponding growth of hadron multiplicity, see 
curve "1" in figure~\ref{fig:Kharzeev-fig}b. 
At high energies, equation~(\ref{eq:Kharzeev1}) predicts substantial slowing 
down of the evolution, which results in the decrease of hadron multiplicity as 
shown in figure~\ref{fig:Kharzeev-fig}b by the curve "2". 
In both cases, the growth of multiplicity is much slower than predicted in the 
conventional "soft plus hard" models, see figure~\ref{fig:Kharzeev-fig}. 
We thus expect that the LHC data on hadron multiplicities will greatly advance 
the understanding of QCD in the strong color field regime.

\subsection{CGC at LHC}
\label{s:Kopeliovich1}

{\em B. Kopeliovich and I. Schmidt}

{\small
Data strongly indicate the localization of glue in hadrons within small spots. 
This leads to a small transverse overlap of gluons in nuclei, i.e. to weak CGC 
effects. 
We predict a weak Cronin effect for LHC, not considerably altered by gluon 
shadowing.
}
\vskip 0.5cm

There are many experimental evidences for the localization of the glue in 
hadrons within spots of small size, $r_0\approx 0.3$~fm~\cite{%
  Kopeliovich:2006bm,Kopeliovich:2004ex}. 
Correspondingly, the mean transverse momentum of gluons in the proton should be 
rather large, about $700$~MeV/$c$.
One of the manifestation of this phenomenon is a weak Cronin enhancement for 
gluons. 
Indeed, the Cronin effect is a result of the interplay between the primordial 
transverse momentum, $\langle k_T^2\rangle$, of the incoming parton and the 
additional momentum, $\Delta\langle k_T^2\rangle$, gained in the nucleus 
(broadening). 
The relative significance of the latter controls the magnitude of the Cronin 
enhancement. Apparently, the larger the original $\langle k_T^2\rangle$ is, the 
weaker is the Cronin effect. 
The $p_T$-slope of the cross section also matters: the steeper it is, the 
stronger is the nuclear enhancement.

\begin{figure}[htbp]
\centerline{\includegraphics[width=8cm]{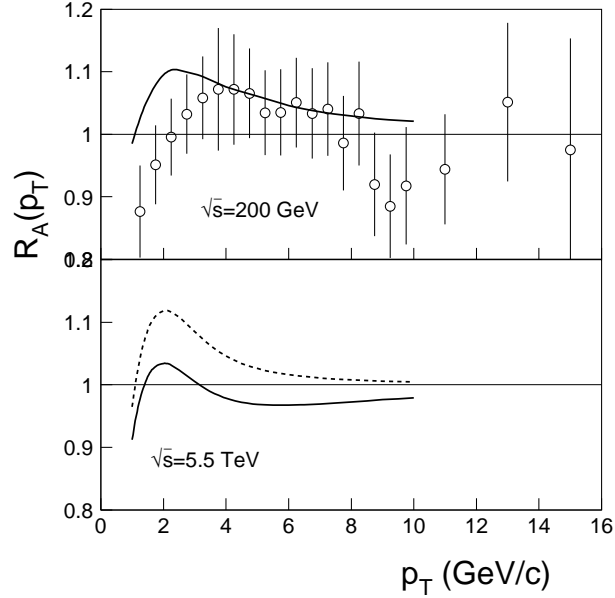}}
\caption{Nucleus-to-proton ratio for pion production versus $p_T$. 
  Dashed and solid curves correspond to calculations without or with gluon
  shadowing~\cite{Kopeliovich:2002yh}.}
\label{fig:Kopeliovich1-fig1}
\end{figure}

Although a rather strong Cronin effect was observed in fixed target experiments,
the production of high-$p_T$ hadrons is dominated by scattering of valence 
quarks~\cite{Kopeliovich:2002yh}. 
One can access the gluons only at sufficiently high energies. 
Relying on the above consideration, a very weak Cronin enhancement was 
predicted in~\cite{Kopeliovich:2002yh} at $\sqrtsnn=200$~GeV, as is depicted in
figure~\ref{fig:Kopeliovich1-fig1}. 
A several times stronger effect was predicted in~\cite{Wang:1998ww}\footnote{%
  The extremely strong gluon shadowing implemented into the HIJING model is 
  ruled out by the recent NLO analysis~\cite{deFlorian:2003qf} of DIS data.}, 
and a suppression, rather than enhancement, was the expectation of the color 
glass condensate (CGC) model~\cite{Kharzeev:2002pc}. 
The latest data from the PHENIX experiment at RHIC support the prediction 
of~\cite{Kopeliovich:2002yh}.

At LHC energies one can access quite small values of Bjorken $x$, such that the 
lifetime of gluonic fluctuations, $t_c\approx 0.05/(xm_N)$~\cite{%
  Kopeliovich:2000ra}, becomes longer than the nuclear size.
Then one might expect coherence effects, in particular pronounced signatures of 
CGC. 
However, the longitudinal overlap of gluons is not sufficient, since they also 
have to overlap in impact parameter, which is something problematic for small 
gluonic spots.
The mean number of gluons overlapping with a given one in a heavy nucleus is, 
$\langle n\rangle=\frac{3\pi}{4}r_0^2\langle T_A\rangle=\pi r_0^2\rho_AR_A=0.3$, 
and such a small overlap results in a quite weak CGC and gluon shadowing. 
The latter is confirmed by the NLO analysis of nuclear structure functions 
performed in~\cite{deFlorian:2003qf}. 
Missing this important observation, one could easily overestimate both the CGC 
and gluon shadowing.

Thus, we expect that the effects of CGC, both the Cronin enhancement and 
shadowing suppression, to be rather weak at the LHC, and nearly compensating 
each other. 
Therefore in this case the nucleus-to proton ratio is expected to approach 
unity from below at high $p_T$.

\subsection{Fluctuation Effects on $R_{pA}$ at High Energy}

%{\it Misha~Kozlov, ~Arif~I.~Shoshi ~and ~Bo-Wen Xiao}
{\it M.~Kozlov, A.~I.~Shoshi ~and ~B.-W. Xiao}
  
  {\small
  We discuss a new physical phenomenon for $R_{pA}$ in the fixed coupling case, 
  the total gluon shadowing, which arises due to the effect of gluon number
  fluctuations.
  }
\vskip 0.5cm

We study the ratio of the unintegrated gluon distribution of a nucleus
$h_A(k_{\perp},Y)$ over the unintegrated gluon distribution of a proton
$h_p(k_{\perp},Y)$ scaled up by $A^{1/3}$
\begin{equation}
R_{pA} = \frac{h_{A}\left( k_{\perp },Y\right) }{A^{\frac{1}{3}}\ h_{p}\left(
    k_{\perp },Y\right) } \ .
\label{R_pA}
\end{equation}
This ratio is a measure of the number of particles produced in a
proton-nucleus collision versus the number of particles in proton-proton
collisions times the number of collisions. The transverse
momentum of gluons is denoted by $k_{\perp}$ and the rapidity variable by $Y$.

\begin{figure}[htb]
\setlength{\unitlength}{1.cm}
\par
\begin{center}
\epsfig{file=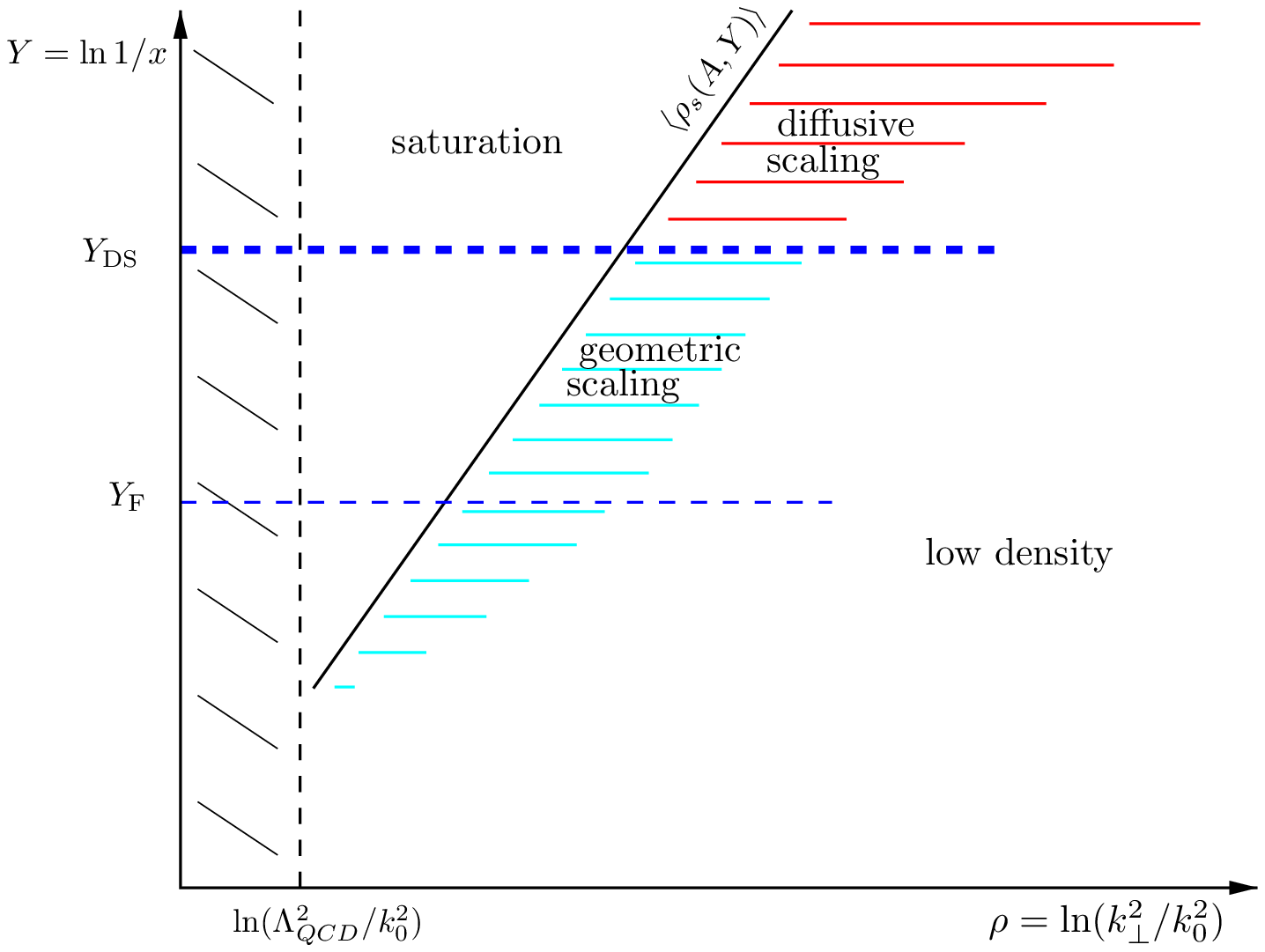, width=7.5cm} \hfill
\epsfig{file=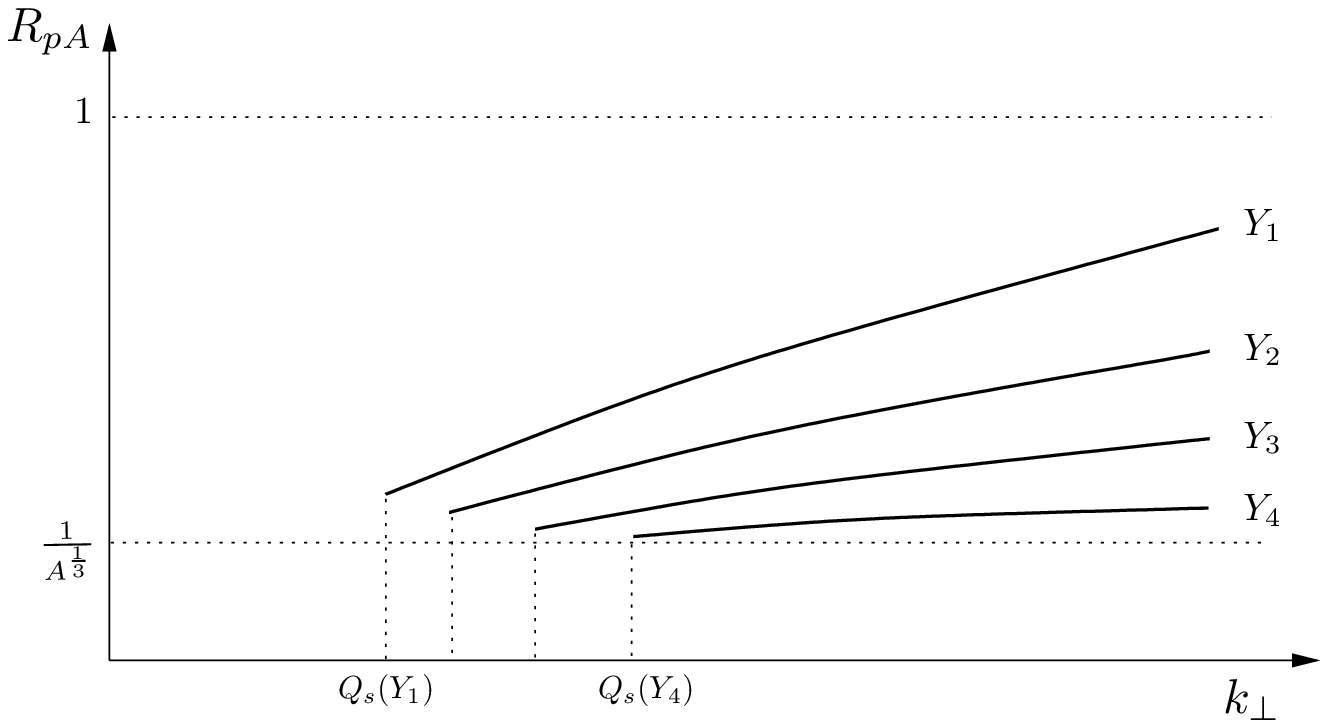, width=8cm}
\end{center}
\caption{(a) Phase diagram of a highly evolved nucleus/proton. (b) $R_{p,A}$ versus
  $k_{\perp}$ at different rapidities $Y_4 \gg Y_3\gg Y_2\gg Y_1$.} 
\label{had_wf}
\end{figure}

In the geometric scaling region shown in Fig.~\ref{had_wf}a the small-$x$
physics is reasonably described by the BK-equation which emerges in the mean
field approximation. Using the BK-equation one finds in the geometric scaling
regime in the fixed coupling case that the shape of the unintegrated gluon
distribution of the nucleus and proton as a function of $k_{\perp}$ is
preserved with increasing $Y$, because of the geometric scaling behaviour
$h_{p,A}(k_{\perp},Y)=h_{p,A}(k_{\perp}^2/Q^2_s(Y))$, and therefore the
leading contribution to the ratio $R_{pA}$ is $k_{\perp}$ and $Y$ independent,
scaling with the atomic number $A$ as $R_{pA} =1/A^{1/3(1-\gamma_{_0})}$, where
$\gamma_{_0}=0.6275$~\cite{Mueller:2003bz}. This means that gluons inside the
nucleus and proton are somewhat shadowed since $h_A/h_p = A^{\gamma_{_0}/3}$ lies
between total ($h_A/h_p=1$) and zero ($h_A/h_p=A^{1/3}$) gluon shadowing. The
{\em partial gluon shadowing} comes from the anomalous behaviour of the
unintegrated gluon distributions which stems from the BFKL evolution.

We have recently shown~\cite{Kozlov:2006qw} that the behaviour of $R_{pA}$ as
a function of $k_{\perp}$ and $Y$ in the fixed coupling case is completely
changed because of the effects of gluon number fluctuations or Pomeron loops
at high rapidity. According to~\cite{Iancu:2004es} the influence of
fluctuations on the unintegrated gluon distribution is as follows: Starting
with an intial gluon distribution of the nucleus/proton at zero rapidity, the
stochastic evolution generates an ensamble of distributions at rapidity $Y$,
where the individual distributions seen by a probe typically have different
saturation momenta and correspond to different events in an experiment. To
include gluon number fluctuations one has to average over all individual
events, $h^{fluc.}_{p,A}(k_{\perp},Y) = \langle h_{p,A}(k_{\perp},Y) \rangle$,
with $h_{p,A}(k_{\perp},Y)$ the distribution for a single event. The main
consequence of fluctuations is the replacement of the geometric scaling by a
new scaling, the diffusive scaling~\cite{Mueller:2004se,Iancu:2004es},
$\langle h_{p,A}(k_{\perp},Y) \rangle = h_{p,A}\left(\ln(k^2_{\perp}/\langle
  Q_s(Y)\rangle^2))/[D Y]\right)$. The diffusive scaling, see
Fig.~\ref{had_wf}a, sets in when the dispersion of the different events is
large, $\sigma^2 = \langle \rho_s(Y)^2 \rangle- \langle \rho_s(Y) \rangle^2 =
D Y \gg 1$, i.e., $Y \gg Y_{DS} =1/D$, where $\rho_s(Y) = \ln(Q^2_s(Y)/k_0^2)$
and $D$ is the diffusion coefficient, and is valid in the region $\sigma \ll
\ln(k^2_{\perp}/\langle Q_s(Y)\rangle^2) \ll \gamma_{_0}\, \sigma^2$. The new
scaling means that the shape of the unintegrated gluon distribution of the
nucleus/proton becomes flatter and flatter with increasing rapidity $Y$, in
contrast to the preserved shape in the geometric scaling regime. This is the
reason why the ratio in the diffusive scaling regime~\cite{Kozlov:2006qw}
\begin{equation}
R_{pA}(k_{\perp},Y) \simeq \frac{1}{A^{\frac{1}{3}\left(1-\frac{\ln A^{1/3}}{2\sigma^{2}}\right)}} \
      \left[\frac{k_{\perp }^{2}}{\langle Q_{s}(A,Y)\rangle^2}\right]^{\frac{\ln
      A^{1/3}}{\sigma^{2}}} \ 
\label{eq:A_R_DS}
\end{equation}
yields {\em total gluon shadowing}, $R_{pA} = 1/A^{1/3}$, at asymptotic
rapidity $Y$ (at fixed $A$). This result is universal since it does not depend
on the initial conditions. Moreover the slope of $R_{pA}$ as a function of
$k_{\perp}$ descreases with increasing $Y$. The qualitative behaviour of
$R_{pA}$ at fixed $\alpha_s$ due to fluctuation effects is shown in
Fig.~\ref{had_wf}b.

The above effects of fluctuations on $R_{pA}$ are valid in the fixed coupling
case and at very large energy. It isn't clear yet whether the energy at LHC is
high enough for them to become important. Moreover, in the case where
fluctuation effects are neglected but the coupling is allowed to run, a
similar behaviour for $R_{pA}$ is obtained~\cite{Iancu:2004bx}, including the
total gluon shadowing. It remains for the future to be clarified how important
fluctuation or running coupling effects are at given energy windows, e.g.,
at LHC energy.

\subsection{Particle Production at the LHC: Predictions from EPOS}

{\it S. Porteboeuf, T. Pierog and K. Werner}

{\small
We present EPOS predictions for proton-proton scattering and for lead-lead 
collisions at different centralities at LHC energies. 
We focus on soft physics and show particle spectra of identified
particles and some results on elliptical flow.
We claim that collective affects are already quite important in proton-proton
scattering.
}
\vskip 0.5cm

EPOS is a consistent quantum mechanical multiple scattering approach
based on partons and strings, where cross sections
and the particle production are calculated consistently, taking into
account energy conservation in both cases~\cite{Drescher:2000ha}.
A special feature is a careful treatment of projectile and target
remnants.

Nuclear effects related to Cronin transverse
momentum broadening, parton saturation, and screening have been introduced
into EPOS~\cite{Werner:2005jf}. 

Furthermore, high density effects leading
to collective behavior in heavy ion collisions are also taken into
account ("plasma core")~\cite{Werner:2007bf}.

\begin{figure}
\includegraphics[%
  scale=0.62, angle=270]{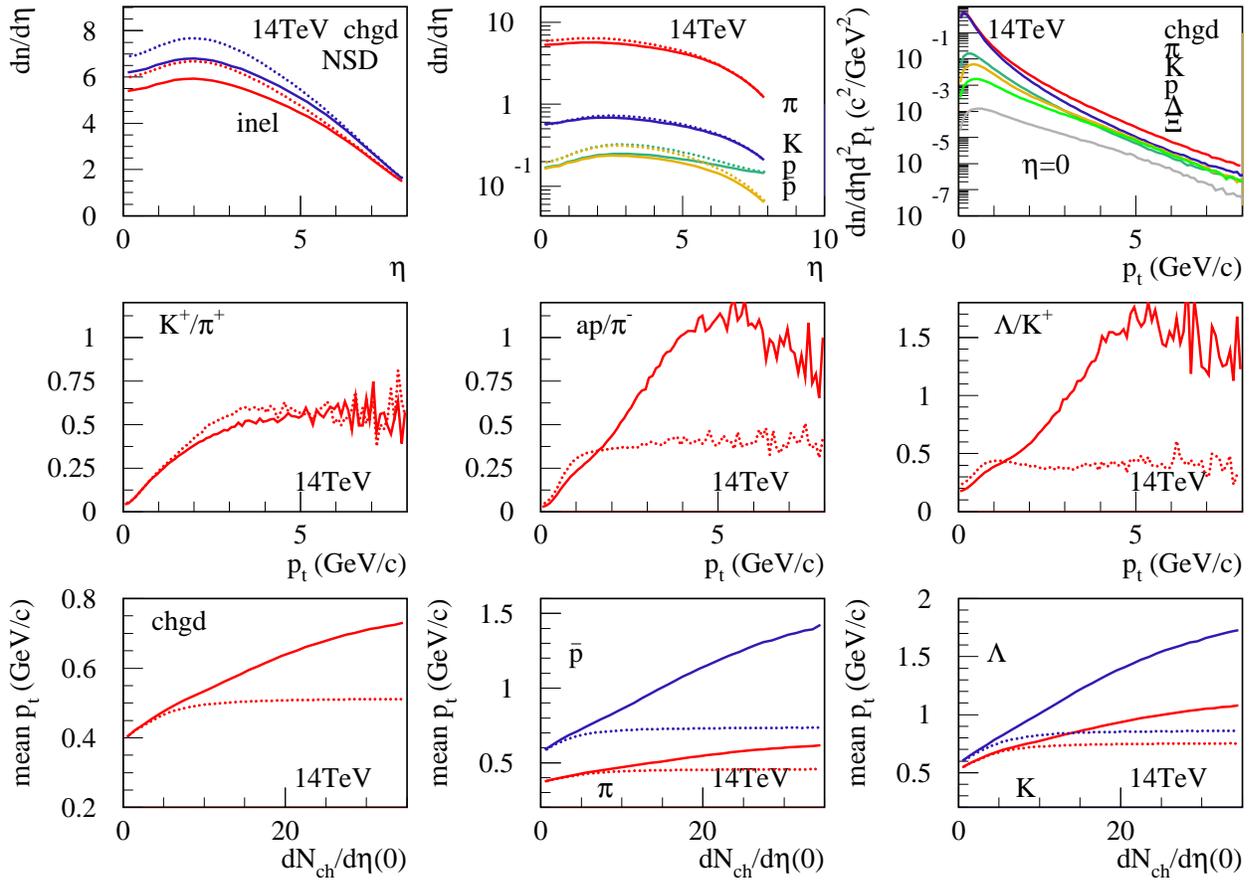} 
\caption{Proton-proton scattering at 14~TeV: pseudorapidity distributions of charged 
particles (upper row left), and of different
identified hadrons (upper row middle), as well as tranverse momentum spectra of different
identified hadrons at $\eta=0$ (upper row right),transverse momentum dependence of particle 
ratios at $\eta=0$ (middle row),the average transverse momentum  of
charged particles and of different identified hadrons at $\eta=0$ (lower row).
 The full lines refer to the "mini-plasma option", the dotted ones refer to
the "conventional option (mini-plasma turned off)". \label{fig1wernernew}}
\end{figure}

We first show in fig. \ref{fig1wernernew} pseudorapidity and transverse 
momentum spectra of charged  particles  and of different identified hadrons, as well
as some particle ratios, in proton-proton scattering
at 14~TeV.
As for heavy ions, the default version of EPOS considers 
also in proton-proton scattering the formation of a core (dense area), with a
hydrodynamical collective expansion. Whereas such "mini-plasma cores" are negligible
in proton-proton scattering at RHIC, they play an important role at the LHC, which
can be seen from the difference between the full curves (full EPOS, including 
"mini-plasma") and the dotted curves ("mini-plasma option turned off"). 
The effect is even more
drastic when we investigate the multiplicity dependence of particle production, see
fig. \ref{fig1wernernew}.

\begin{figure}
\includegraphics[%
  scale=0.60, angle=270]{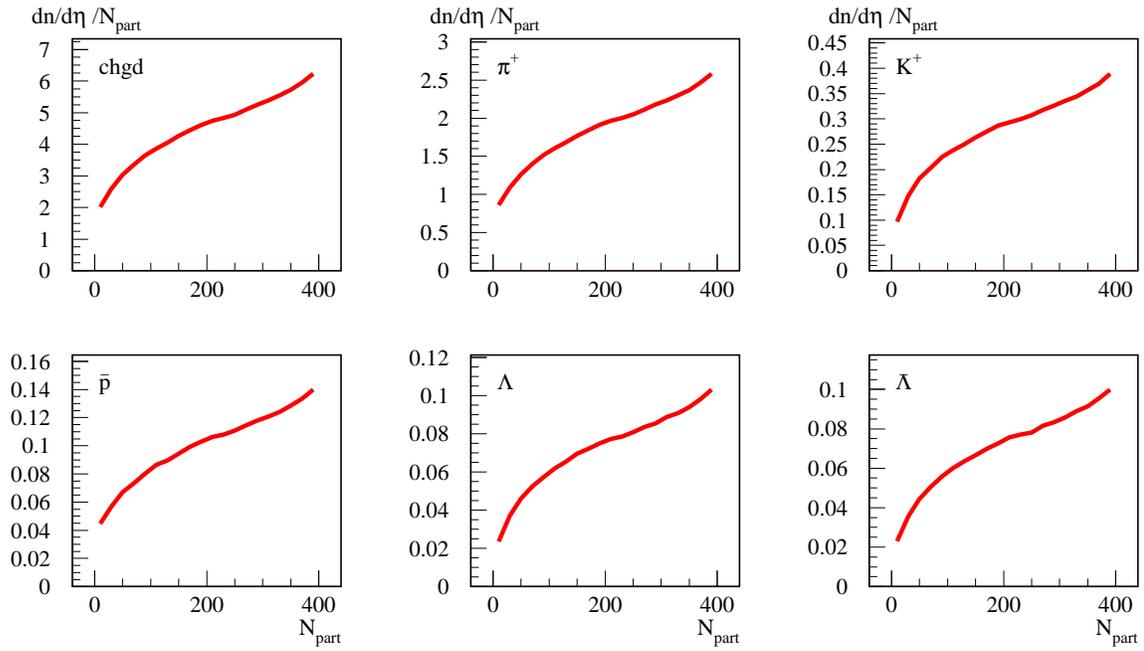} 
\caption{Lead-lead collisions at 5.5~TeV: centrality dependence of particle 
yields (central pseudorapidity density per participant), 
for charged particles and different identified hadrons.
\label{fig4wernernew}}
\end{figure}

In the following, we investigate lead-lead collisions at 5.5~TeV. In fig. \ref{fig4wernernew},
we plot the centrality dependence of particle 
yields for charged particles and different identified hadrons. We observe an increase
by roughly 2.5 for pions, and a bigger factor for the heavier particles.  

\begin{figure}
\includegraphics[%
  scale=0.60, angle=270]{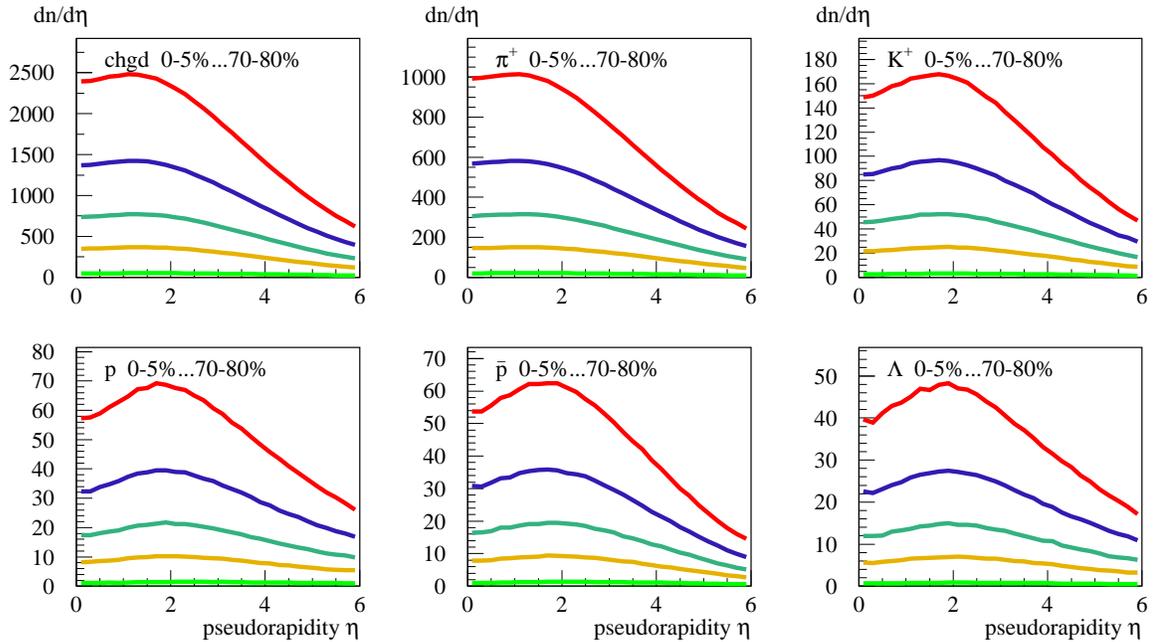} 
\caption{Lead-lead collisions at 5.5~TeV: pseudorapidity distributions of charged 
particles  and of different identified hadrons, at different centralities. For each
plot, from top to bottom: 0-5\%, 10-20\%, 25-35\%, 40-50\% 70-80\%.  \label{fig5wernernew}}
\end{figure}

In fig. \ref{fig5wernernew}, we show pseudorapidity spectra, for different particles, at
different centralities. The pseudorapidity density of charged particles 
at $\eta=0$ is around 2500, for central collisions.

\begin{figure}
\includegraphics[%
  scale=0.60, angle=270]{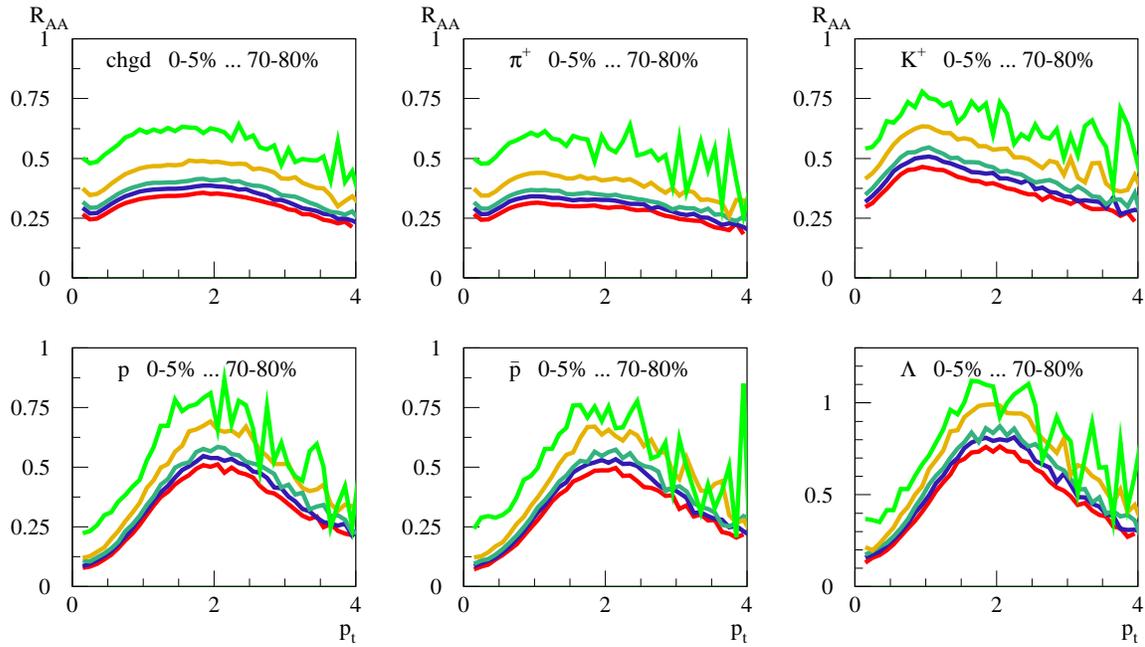} 
\caption{Lead-lead collisions at 5.5~TeV: 
the nuclear modification factor $R_{\rm AA}$  at $\eta=0$ of charged 
particles  and of different identified hadrons, at different centralities. 
For each plot, from top to bottom: 
70-80\%, 40-50\%, 25-35\%, 10-20\%, 0-5\%.
\label{fig7wernernew}}
\end{figure}

In fig. \ref{fig7wernernew}, we show nuclear modification factors $R_{\rm AA}$ 
(ratios with
respect to proton-proton, divided by $N_{\rm coll}$), considering  charged 
particles  and different identified hadrons. The peak structure of the baryon
results is related to the concave form of the baryon spectra from 
the radially flowing core in PbPb collisions. All curves are well below one,
indicating strong screening effects.

\begin{figure}
\includegraphics[%
  scale=0.60, angle=270]{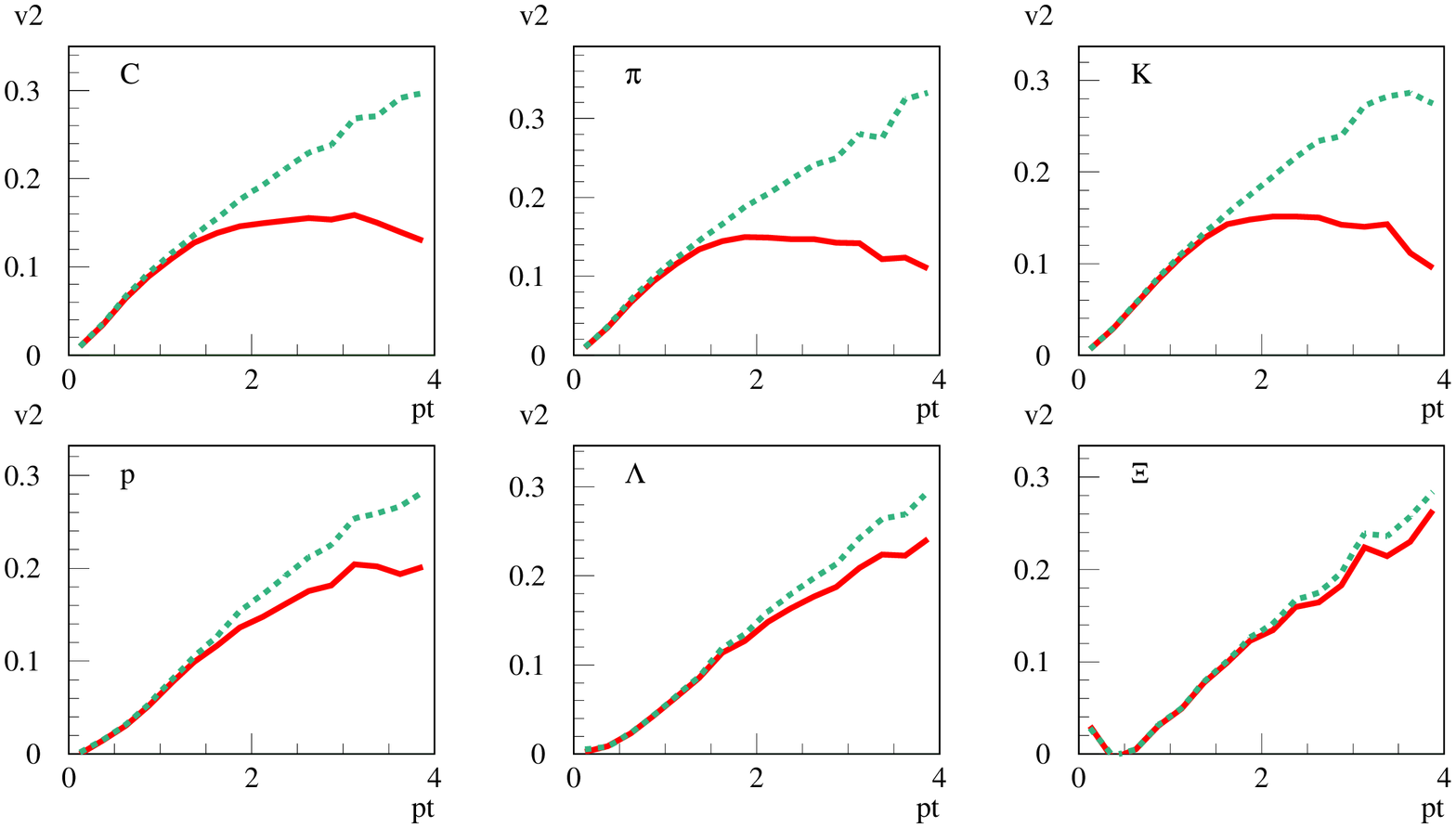} 
\caption{Lead-lead collisions at 5.5~TeV: 
the transverse momentum dependence of the elliptical flow at $\eta=0$ of charged 
particles  and of different identified hadrons, for minimum bias collisions. 
The full line is the full calculation, the dashed one only the core contribution. 
\label{fig8wernernew}}
\end{figure}

In fig. \ref{fig8wernernew}, we finally show the transverse momentum dependence of the
elliptical flow. The full line is the full calculation, the dashed one only the core
contribution. The big difference between the two is due to the fact that high $p_t$
jets are allowed to freely leave the core (no jet quenching). 


\subsection{Forward hadron production in high energy pA collisions}

{\it K.~L.~Tuchin}

{\small
We present a calculation of $\pi$, $D$ and $B$ production at RHIC and 
LHC energies based upon the KKT model of gluon saturation. 
}
\vskip 0.5cm

\begin{figure}[ht]
  \begin{tabular}{cc} 
      \includegraphics[width=6.5cm]{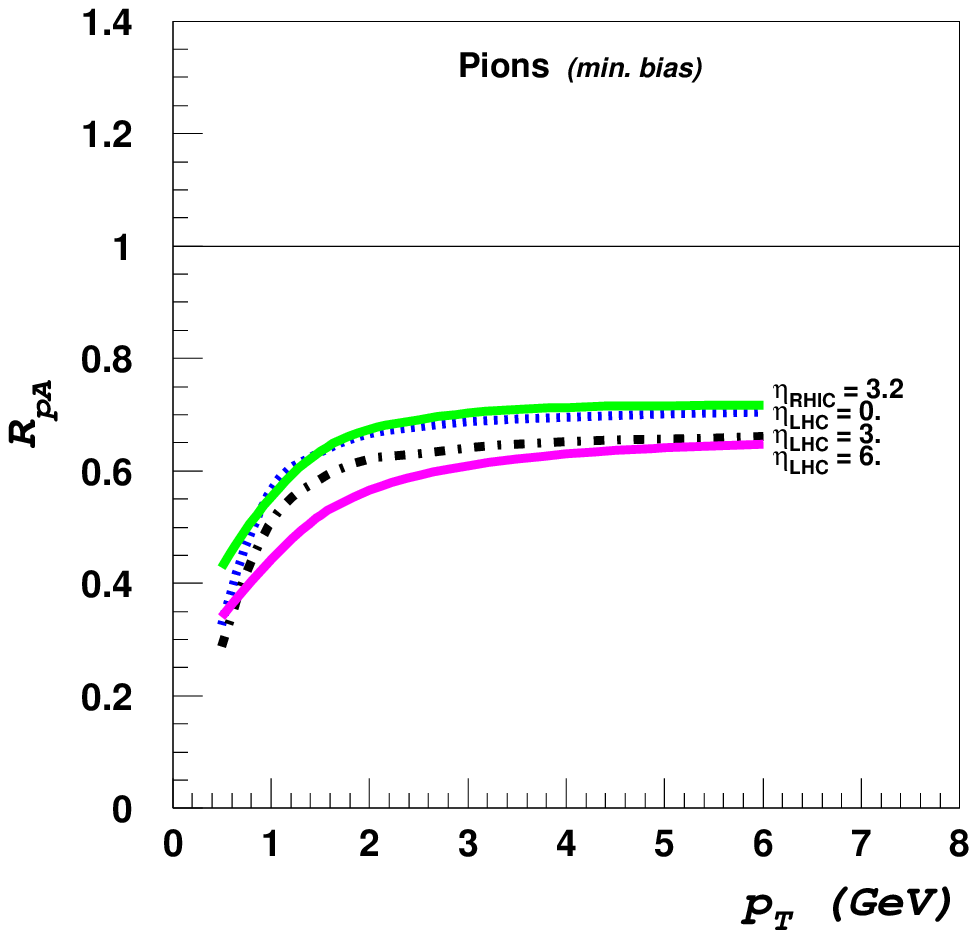} &
        \includegraphics[height=6.5cm]{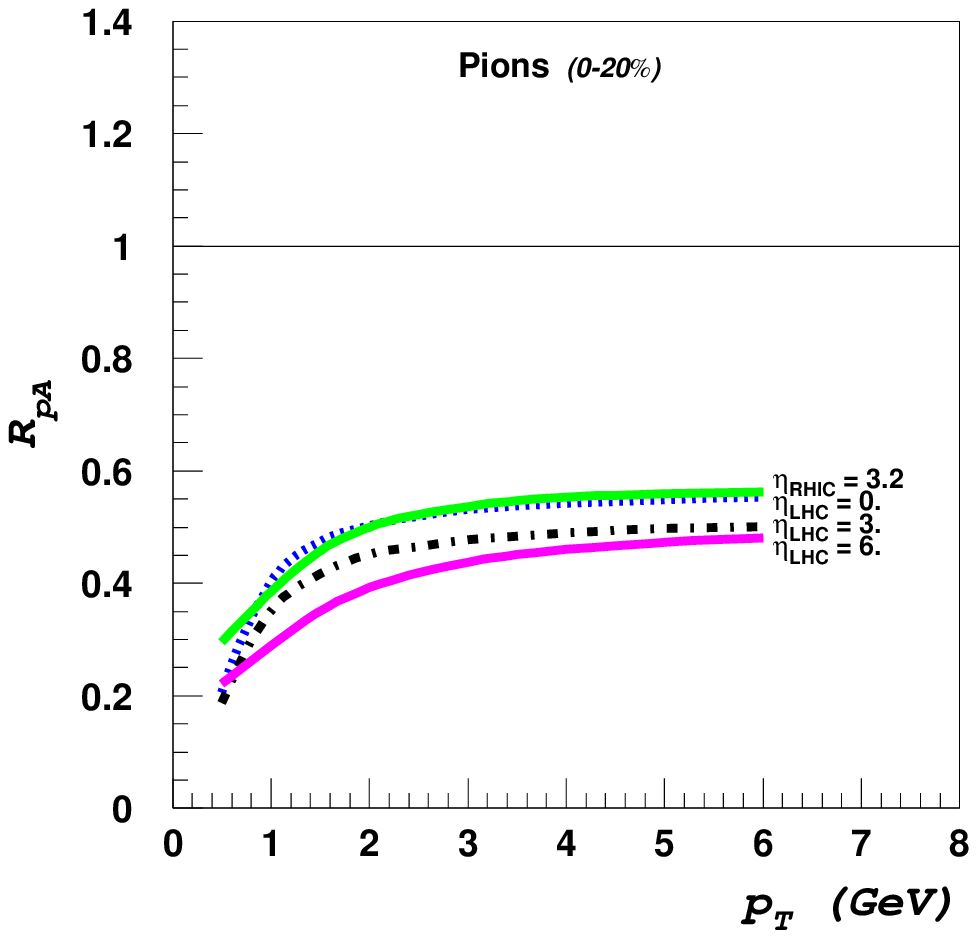}%\\
 %        (a) & (b) 
\end{tabular}
\caption{Nuclear modification factor for pion production at RHIC and LHC.}
\label{fig1_tuchin}
\end{figure}

In this proceedings we present a calculation of forward hadron production in
pA collisions at RHIC and LHC. The theoretical framework for inclusive gluon
production including the effect of gluon saturation was set up in
Ref.~\cite{Kharzeev:2003wz}. It has been successfully applied to study the
inclusive light hadron production  at RHIC \cite{Kharzeev:2004yx}. Since the
KKT model of Ref.~\cite{Kharzeev:2004yx} works so well at RHIC we decided to
extend it to the LHC kinematical region. Doing so we explicitly neglect a
possible effect of gluon saturation in a proton which is perhaps a good
approximation for the nuclear modification factor. The results of
calculation of inclusive pion production are shown in \fig{fig1_tuchin}. 
%%%%
%%%%%

\begin{figure}[h]
  \begin{tabular}{cc} 
      \includegraphics[width=6.5cm]{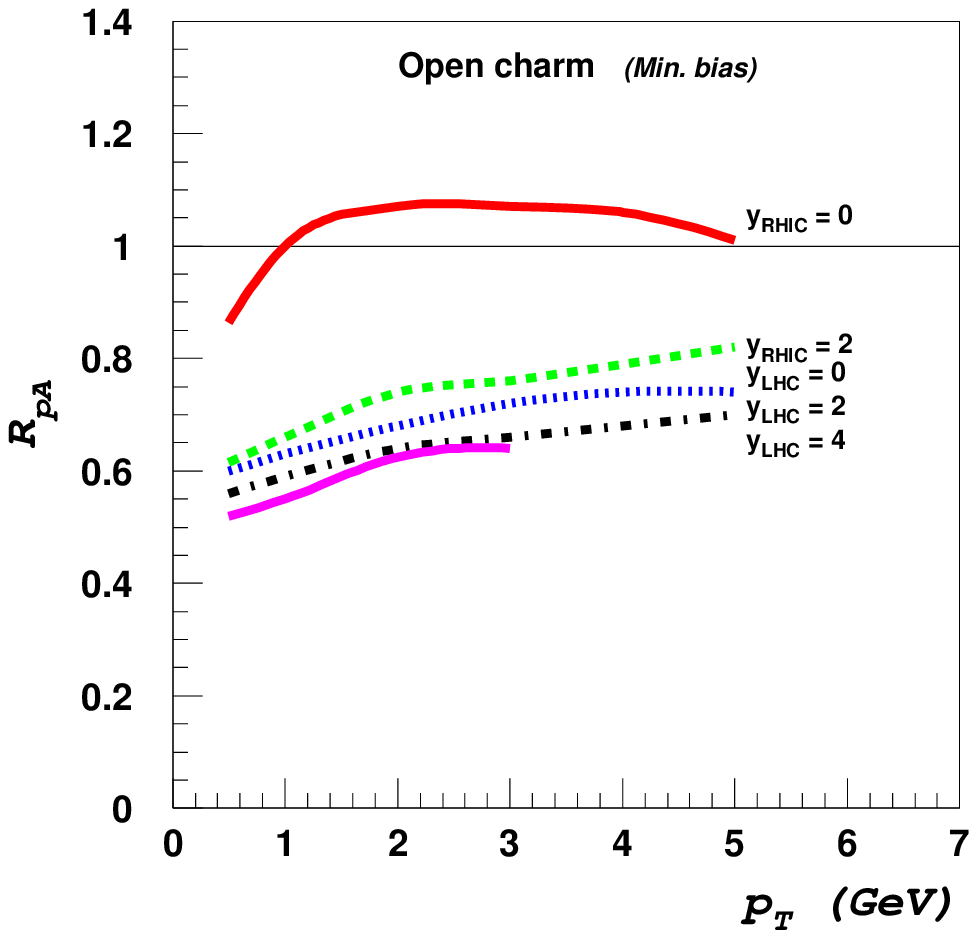} &
        \includegraphics[height=6.5cm]{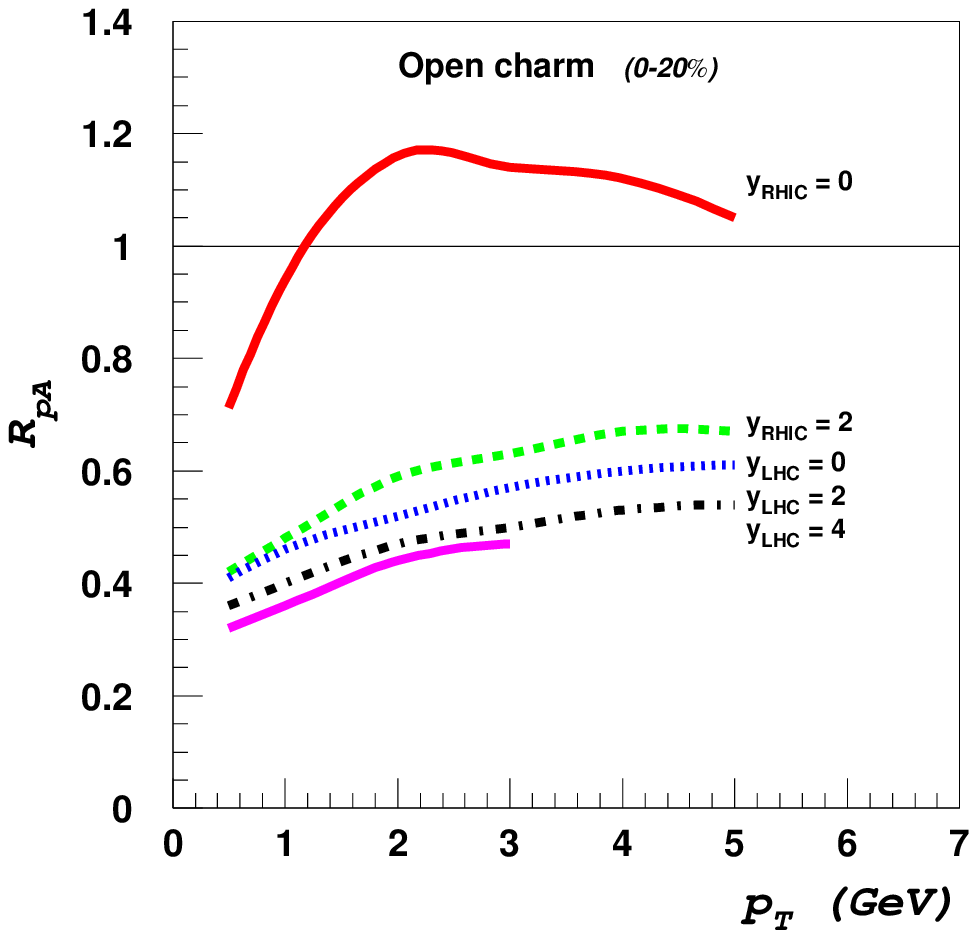}%\\
 %        (a) & (b) 
\end{tabular}
\caption{Nuclear modification factor for open charm production at RHIC and LHC.}
\label{fig2tuchin}
\end{figure}

\begin{figure}[ht]
  \begin{tabular}{cc} 
      \includegraphics[width=6.5cm]{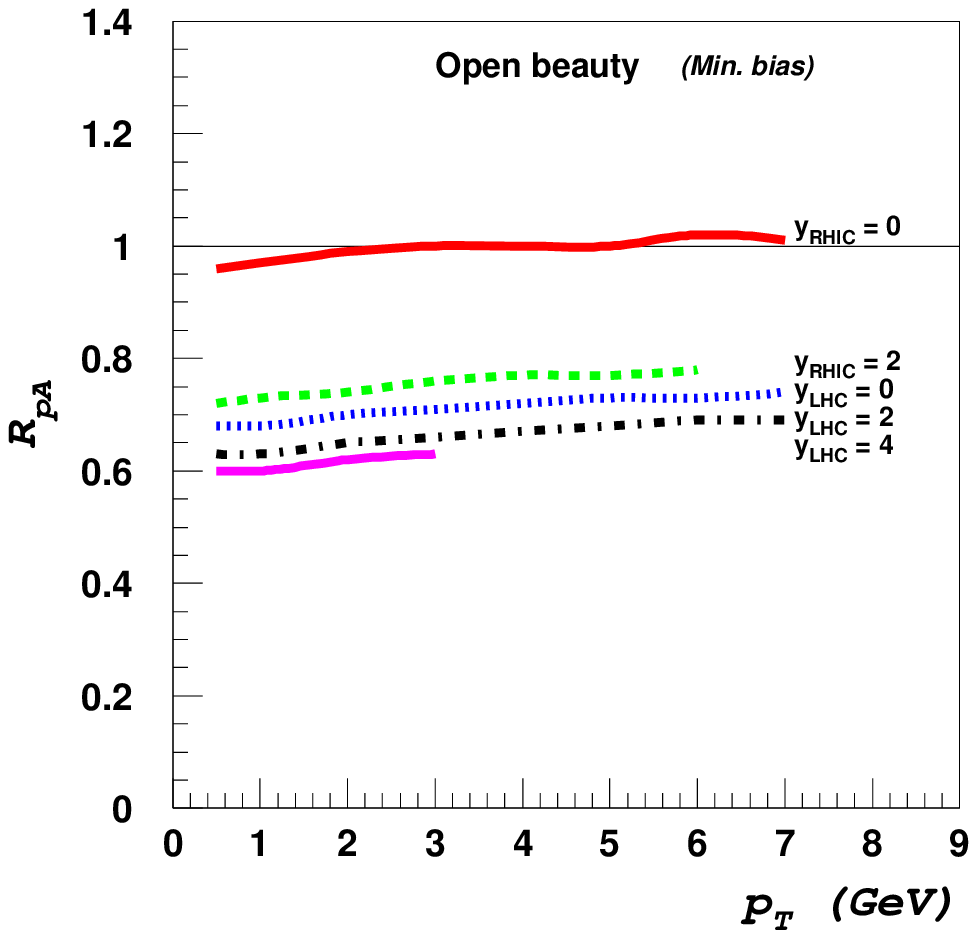} &
        \includegraphics[height=6.5cm]{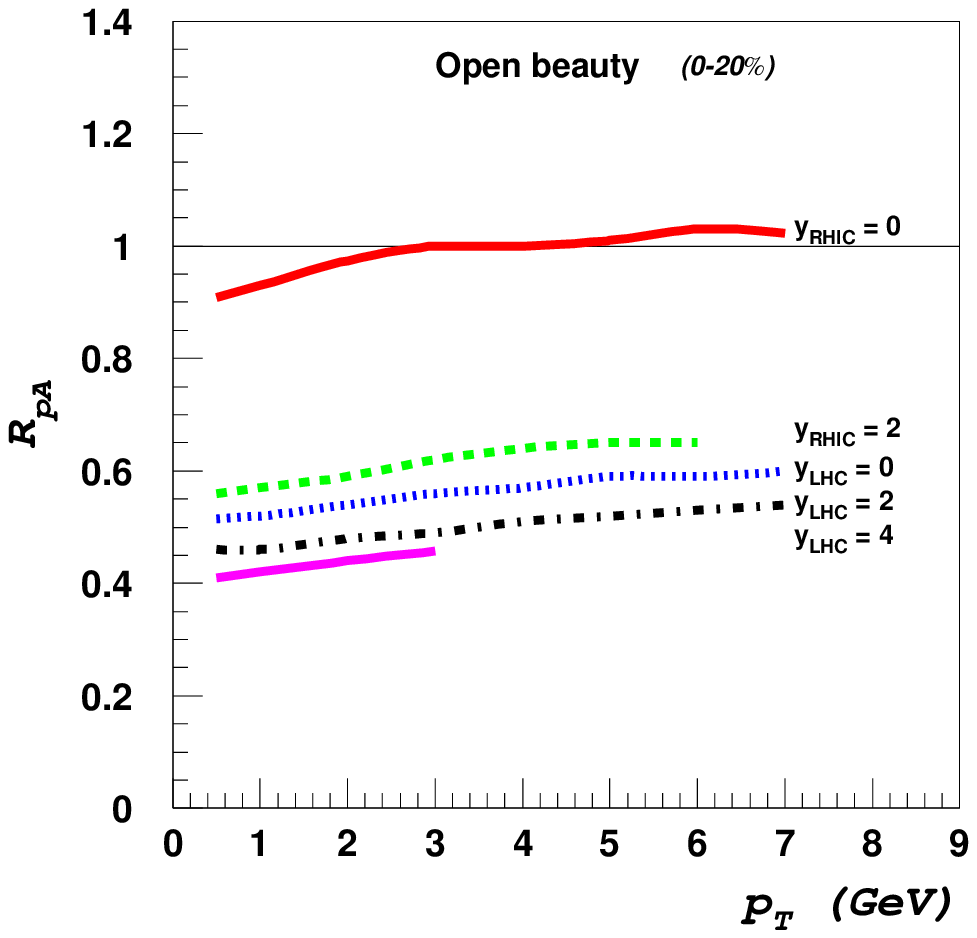}%\\
 %        (a) & (b) 
\end{tabular}
\caption{Nuclear modification factor for open beauty production at RHIC and LHC. Note, that the calculations of \cite{Tuchin:2004rb} break down at $y=0$ at RHIC ($x$ is not small enough); the corresponding result (solid line) is shown for completeness.}
\label{fig3tuchin}
\end{figure}

Production of heavy quarks at small $x$ is also affected by gluon saturation in a way similar to that of gluons \cite{Kharzeev:2003sk}. The main difference, however, is that the effect of gluon saturation is postponed to higher energies/rapidities  for heavier quarks as compared to lighter quarks and gluons. This is because the relevant $x$ is proportional to $m_\bot=(m^2+k_\bot^2)^{1/2}$ and hence is higher for heavier quarks at the same values of $\sqrt{s},y,k_\bot$. In  \fig{fig2tuchin}  and \fig{fig3tuchin} the nuclear modification factors for open charm and beauty are shown. The calculations are  based upon the theoretical result of Ref.~\cite{Tuchin:2004rb} and the KKT model \cite{Kharzeev:2003wz}.

%%%%
%%%%%

%%%%
%%%%%

\begin{figure}[ht]
      \includegraphics[width=6.5cm]{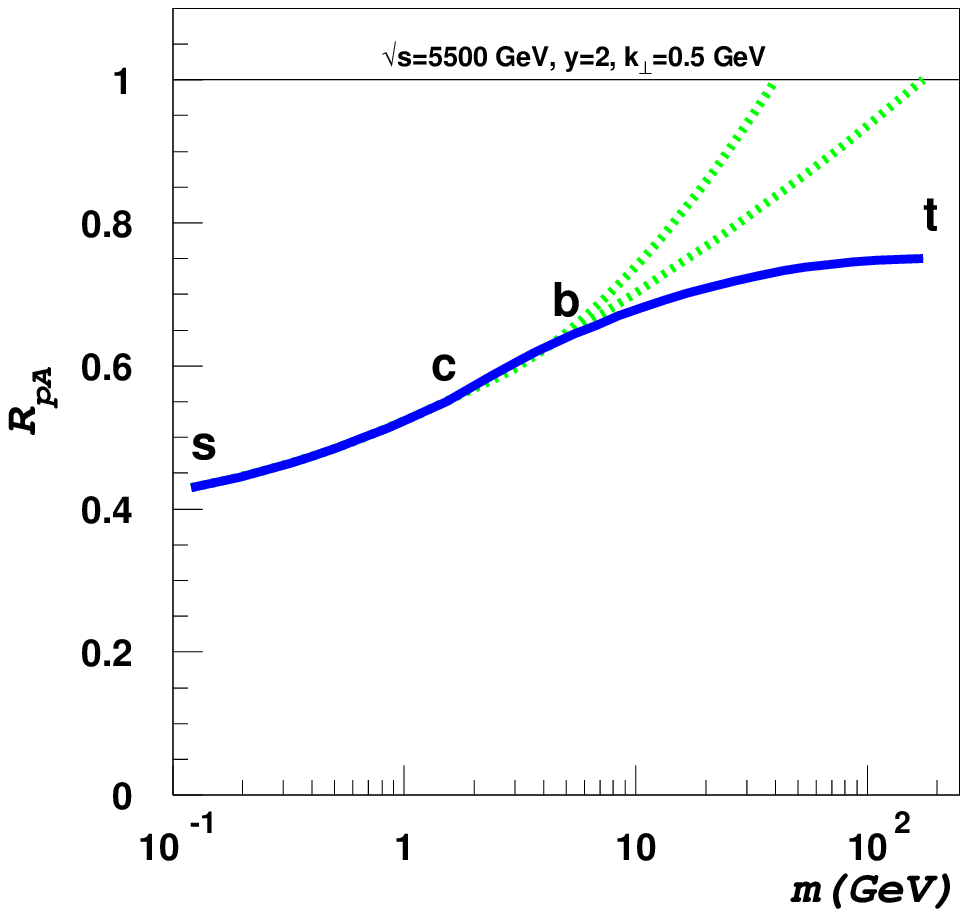} 
\caption{Dependence of the nuclear modification factor on quark mass. Solid line  is $R_{pA}$ for quarks. Geometric scaling is expected to break down at $m_\bot\sim Q_\mathrm{geom}\simeq Q_s^2/\Lambda$, and therefore $R_{pA}$ is anticipated to deviate from the solid line towards unity. Dotted lines illustrate a possible behavior of $R_{pA}$. }
\label{fig4tuchin}
\end{figure}

If the nuclear modification factor is measured to as high transverse mass as possible, we can observe transition from the geometric scaling (described by the KKT model) to the collinear factorization regime. This is shown in \fig{fig4tuchin}. Had the geometric scaling held for all $m_\bot$ and  $x<0.01$, the nuclear modification factor would have been described by the solid line. However, one expect the breakdown of the geometric scaling as illustrated by the dotted lines. 
%%%%
%%%%%

A more detailed description of the theoretical approach to the heavy quark production as well as discussion of the obtained results will be provided in a forthcoming publication.

%\paragraph{Acknowledgements}
%I am grateful to Javier Albacete for showing me the results of his
%calculations of open heavy quark production; his results are in a
%qualitative agreement with \fig{fig2tuchin} and \fig{fig3tuchin}. I would like to thank
%RIKEN, BNL and the U.S. Department of Energy (Contract No.
%DE-AC02-98CH10886) for providing the facilities essential for the completion
%of this work.

\subsection
{Rapidity distributions  at LHC in the Relativistic Diffusion Model}

{\it G.~Wolschin}

{\small
Stopping and particle production in heavy-ion collisions at LHC energies
are investigated in a Relativistic Diffusion Model (RDM).
Using three sources for particle production, the energy-
and centrality dependence of rapidity distributions of net protons, 
and pseudorapidity spectra of charged hadrons in heavy systems 
are studied from SPS to LHC energies.
The transport coefficients are extrapolated from Au + Au 
at RHIC energies ($\sqrt{s_{NN}}$=19.6 - 200 GeV) to Pb + Pb at LHC 
energies of $\sqrt{s_{NN}}$= 5.52 TeV. Rapidity distributions for net protons,
and pseudorapidity spectra for produced charged particles are 
calculated at LHC energies.
}
\vskip 0.5cm

\begin{figure}[htb]
\begin{minipage}[t]{80mm}
%\framebox[79mm]{\rule[-26mm]{0mm}{52mm}}
{\rule[-26mm]{0mm}{52mm}\includegraphics[width=70mm]{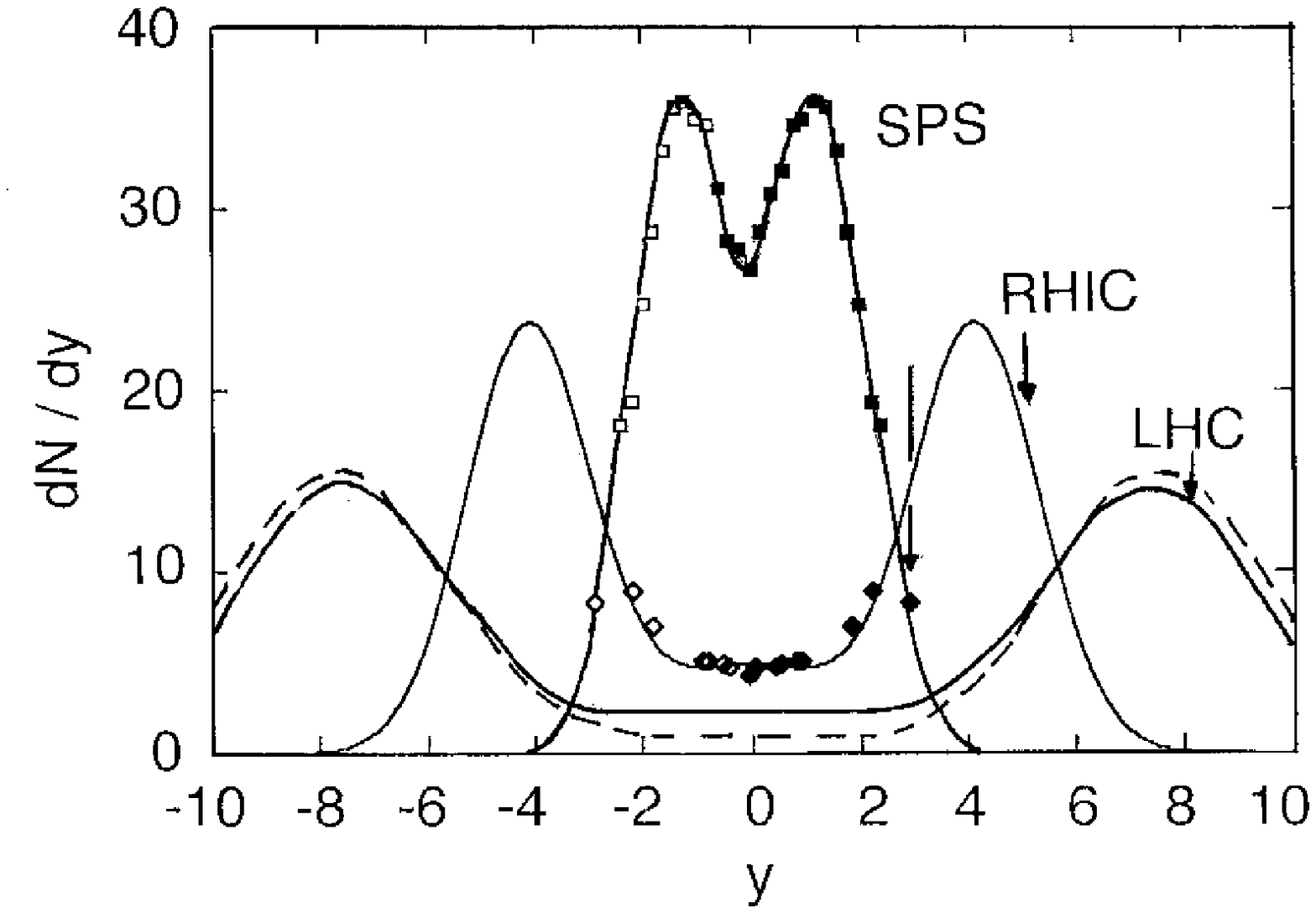}}
%{\rule{0mm}{52mm}\includegraphics[width=80mm]{wolschin_1/lhc1.pdf}}
\vskip -2.5cm
\caption{Net-proton rapidity spectra \cite{Wolschin:2007zz} in the Relativistic Diffusion Model
(RDM), solid curves: Transition from the double-humped shape at SPS 
energies of $\sqrt{s_{NN}}$ = 17.3 GeV to a broad midrapidity valley
%in the three-sources model 
at RHIC (200 GeV) and LHC (5.52 TeV).}
%The percentage of particles in the midrapidity source 
%is indicated.}
\label{fig1.wolschin}
\end{minipage}
%
%\hspace{\fill}
\hspace{4mm}
\begin{minipage}[t]{75mm}
%\framebox[74mm]{\rule[-26mm]{0mm}{52mm}}
{\rule[-26mm]{0mm}{52mm}\includegraphics[width=70mm]{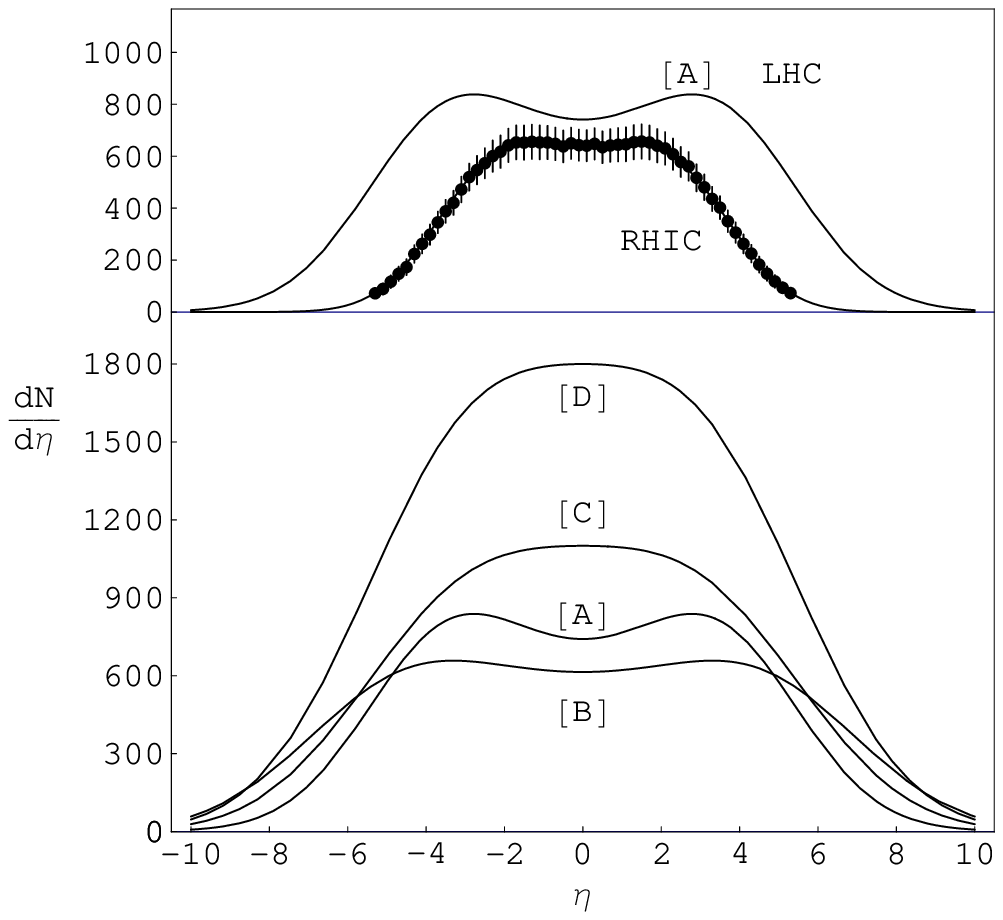}}
%{\rule{0mm}{52mm}\includegraphics[width=70mm]{wolschin_1/lhc2.pdf}}
\vskip -2.5cm
\caption{Produced charged hadrons 
for central Au + Au collisions at RHIC compared with 200 A GeV
PHOBOS data, and diffusion-model 
extrapolation to Pb + Pb at LHC energies of 5520 GeV. See 
%with a charged-particle content in the central source of 50{\%}.
 \cite{Kuiper:2006si} for 
%transport parameters
%, total number of 
%charged particles, and 
curves [A] to [D] at LHC energies.}
\label{fig4.wolschin}
\end{minipage}
\end{figure}

%\subsubsection{Relativistic Diffusion Model at LHC}
Net-proton and charged-hadron distributions
in collisions of heavy systems have been calculated
in a three-sources Relativistic Diffusion Model (RDM) for
multiparticle interactions from SPS to LHC energies.
Analytical results for the rapidity
distribution of net protons in central collisions, and produced
charged hadrons are found to be in 
good agreement with the available data (Figs. \ref{fig1.wolschin}, \ref{fig4.wolschin}) at RHIC. 

An extrapolation of the transport coefficients for net protons, 
and produced hadrons
to Pb + Pb at LHC 
energies of $\sqrt{s_{NN}}$= 5.52 TeV has been 
performed in \cite{Wolschin:2007zz,Kuiper:2006si}, and the corresponding
rapidity distributions have been calculated as shown in Figs. \ref{fig1.wolschin}, \ref{fig4.wolschin}.

The net-proton result for LHC is shown for particle contents of
7 {\%} and 14 {\%} in the central source, respectively \cite{Wolschin:2007zz}.
Kinematical constraints
will modify the result at large values of the rapidity y.
For produced particles, the curves (A) - (D) in Fig. \ref{fig4.wolschin}
are discussed in \cite{Kuiper:2006si}.
The essential parameters relaxation time, diffusion
coefficients or widths of the distribution
functions of the three sources, and number of particles 
in the local equilibrium source 
will have to be adjusted to the ALICE data.
%I thank Rolf Kuiper for his participating in the RDM-calculations.

\section{Azimuthal asymmetries}
\label{sec:v2}

\subsection{Transverse momentum spectra and elliptic flow: Hydrodynamics with 
QCD-based equations of state}
\label{bluhm}

{\it M. Bluhm, B. K\"ampfer and U. Heinz}

{\small
  We present a family of equations of state 
  within a quasiparticle model adjusted to lattice QCD 
  and study the impact on azimuthal flow anisotropies and transverse momentum spectra 
  within hydrodynamic simulations for heavy-ion collisions at energies relevant for LHC. 
}

\subsubsection{Introduction}

\begin{figure}[h]
  \hspace{0.3cm}
  \includegraphics[scale=0.51]{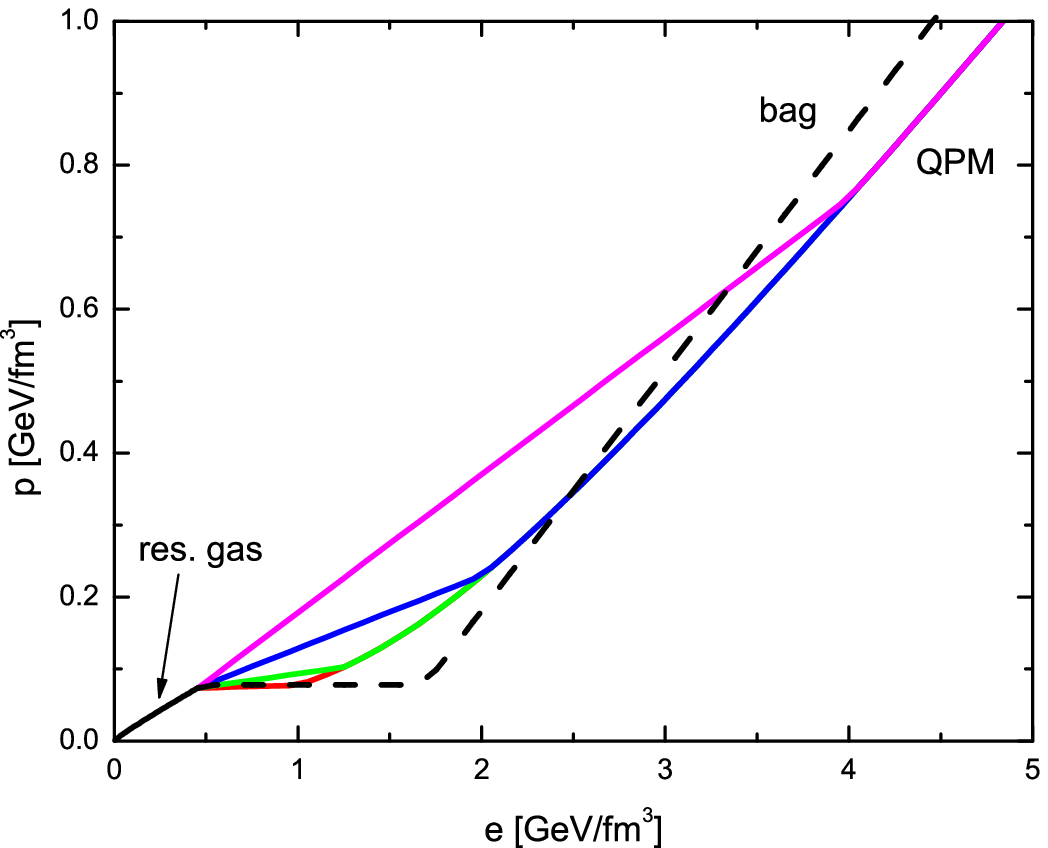}
  \hspace{0.7cm}
  \includegraphics[scale=0.515]{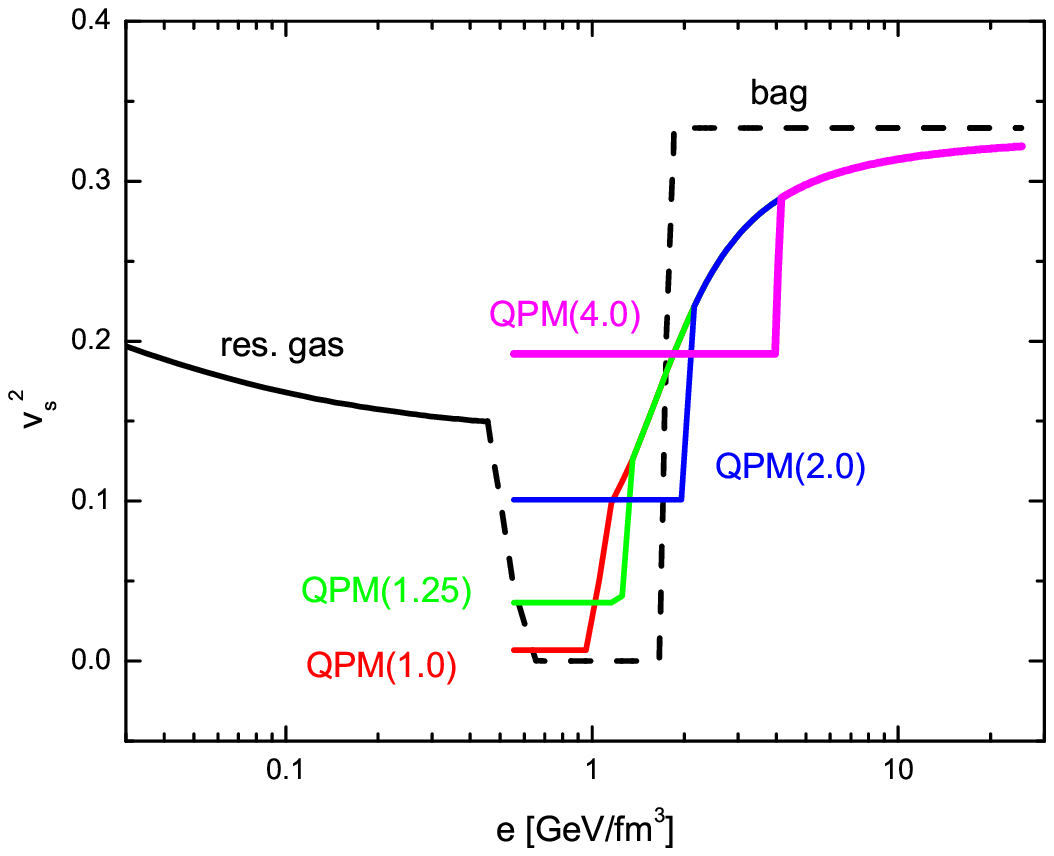}
%  \vspace{4mm}
    \caption{Left panel: Family of EoS's 
    	$p(e)$ labelled in the following as 
	QPM($e_m$) with $e_m=$ 4.0, 2.0, 1.25, 1.0 
	GeV/fm$^3$ (solid curves) combining QPM adjusted to lattice 
	data~\cite{Karsch:2003zq,Karsch:2003vd} 
	and hadron resonance gas at matching point $e_m$. For 
	comparison the bag model EoS (dashed line) is shown. Right panel: corresponding $v_s^2$.}
\label{fig1bluhm}
\end{figure}

The equation of state (EoS) represents the heart of hydrodynamic simulations for ultra-relativistic 
heavy-ion collisions. Here, we present a realistic EoS for QCD matter delivered by our 
quasiparticle model (QPM) faithfully reproducing lattice QCD results. 
The approach is based 
on~\cite{Peshier:1994zf,Peshier:1995ty,Peshier:1999ww,Peshier:2002ww,Bluhm:2006yh} 
adjusted to the pressure $p$ and energy density $e$ of $N_f=2+1$ quark 
flavors~\cite{Karsch:2003zq,Karsch:2003vd}. As the QPM EoS does not 
automatically fit to the hadron resonance gas EoS in the confinement region, we construct a family of EoS's by an interpolation between the hadron resonance gas at 
$e_1=$ 0.45 GeV/fm$^3$ and the QPM at flexible $e_m$ (cf.~\cite{Bluhm:2007nu} for details). 
In this way, the influence of details in the transition region on hydrodynamic flow can 
be studied, since for $e<e_1$ and $e>e_m$ the EoS is uniquely given by the resonance gas 
and the QCD-based QPM, respectively. In Figure \ref{fig1bluhm}, 
we exhibit the EoS family in the form $p=p(e)$ and the corresponding speed of sound $v_s^2=\partial p / \partial e$. For LHC, baryon density effects are negligible. 

\subsubsection{Predictions for heavy-ion collisions at LHC}

\begin{figure}[h]
\begin{center}
  \includegraphics[scale=0.515]{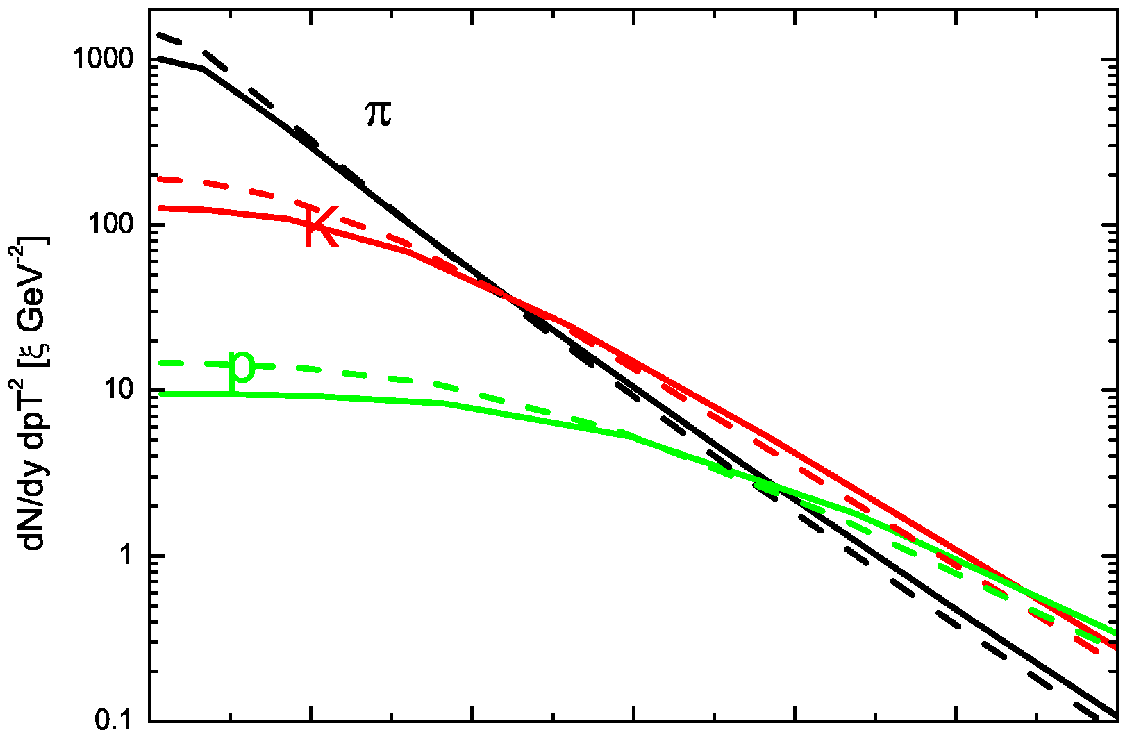}
%  \hskip -1.2cm
   \includegraphics[scale=0.515]{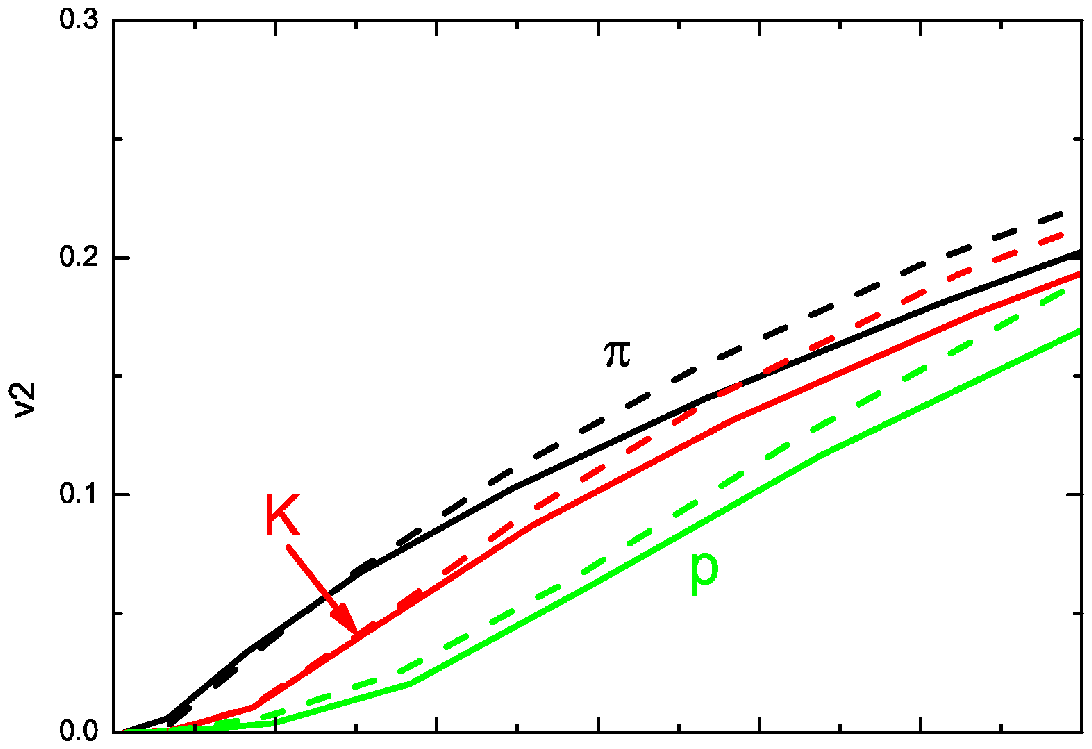}
  \includegraphics[scale=0.518]{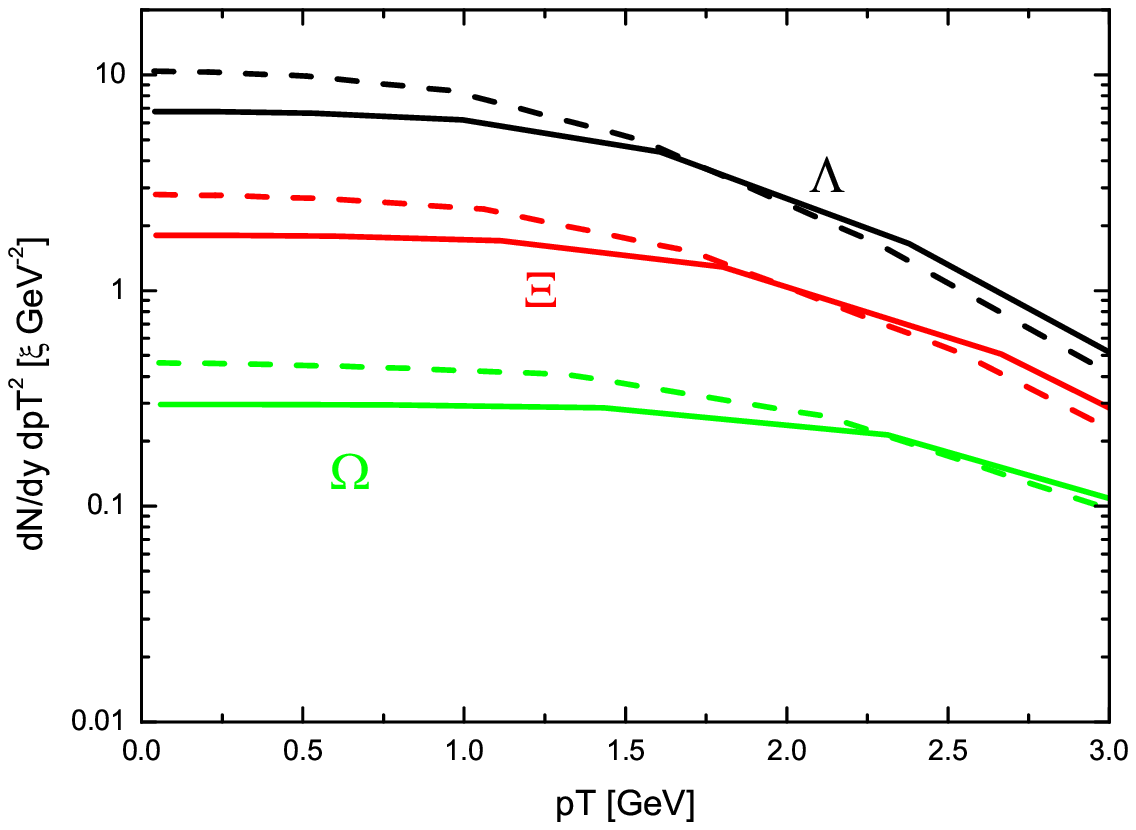}
%  \hfill
  \includegraphics[scale=0.515]{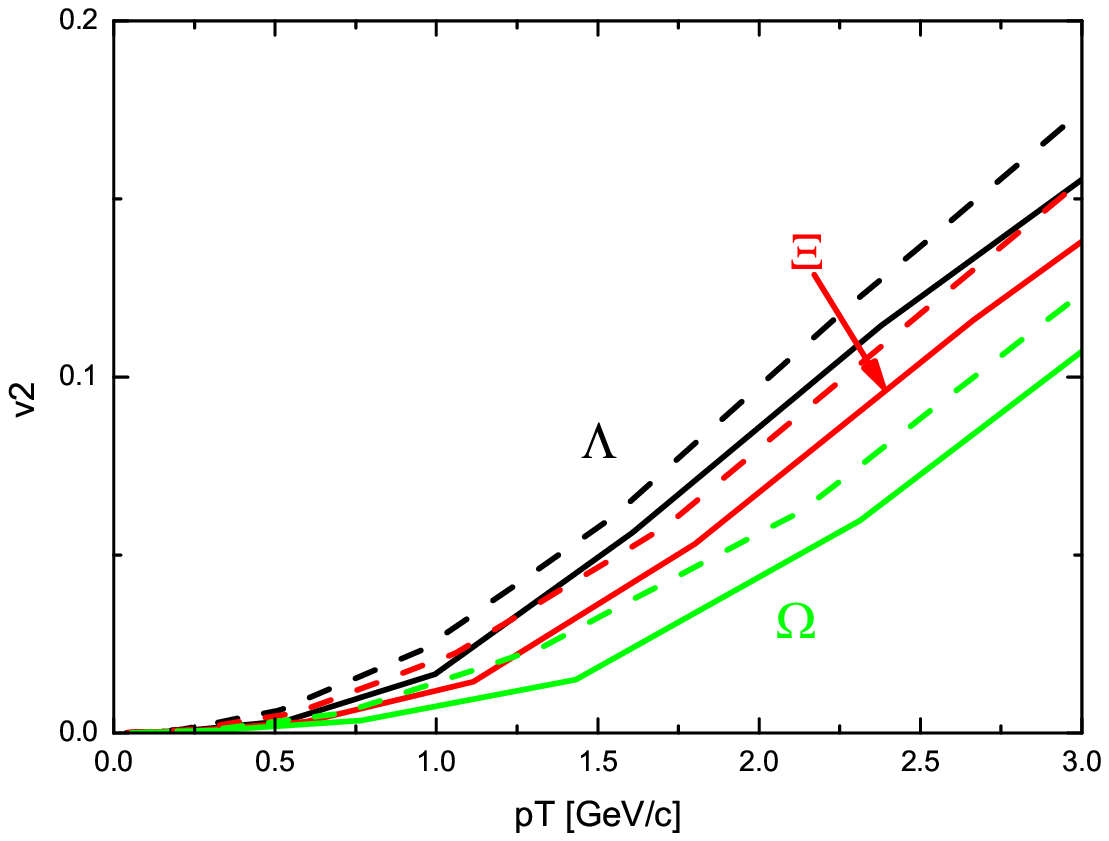}
\end{center}
  \vspace{-7mm}
  \caption{Transverse momentum spectra (left panels) and azimuthal anisotropy (right panels) for 
    directly emitted pions, kaons and protons (upper row) and strange baryons (lower row). Solid 
    and dashed curves are for EoS QPM(4.0) and the bag model EoS, respectively.}
\label{fig2bluhm}
\end{figure}

We concentrate on two extreme EoS's, QPM(4.0) and the bag model EoS being similar 
to QPM(1.0). We calculate transverse momentum spectra and elliptic flow $v_2(p_T)$ using 
the relativistic 
hydrodynamic program package~\cite{Kolb:1999it,Kolb:2000sd} 
with initial conditions for Pb$+$Pb 
collisions at impact parameter $b=5.2$ fm. For the further initial parameters 
required by the program we conservatively guess $s_0=$ 330 fm$^{-3}$, $n_0=$ 0.4 fm$^{-3}$ and 
$\tau_0=$ 0.6 fm/c for initial entropy density, baryon density and time. 
Within the QPM these translate into $e_0=$ 127 GeV/fm$^3$, 
$p_0=$ 42 GeV/fm$^3$ and $T_0=$ 515 MeV. The freeze-out temperature is set $T_{f.o.}=$ 100 MeV. 
In Figure \ref{fig2bluhm}, we exhibit our results at midrapidity for various primordial hadron species. 
Striking is the strong radial flow as evident from the flat $p_T$-spectra and a noticeably 
smaller $v_2(p_T)$ than at RHIC in particular at low $p_T$~\cite{Bluhm:2007nu}. 
Details of the Eos in the transition region as mapped out by our family are still visible. 

\subsection{The centrality dependence of elliptic flow at LHC}
\label{s:Ollitrault}

{\em H.-J. Drescher, A. Dumitru 
and J.-Y. Ollitrault}

{\small 
We present predictions for the centrality dependence of elliptic flow at 
mid-rapidity in Pb-Pb collisions at the LHC. 
}
\vskip 0.5cm

The centrality and system-size dependence of elliptic flow ($v_2$) provides 
direct information on the thermalization of the matter created in the collision.
Ideal (non-viscous) hydrodynamics predicts that $v_2$ scales like the 
eccentricity, $\varepsilon$, of the initial distribution of matter in the 
transverse plane.  
Our predictions are based on this eccentricity scaling, together with a simple 
parameterization of deviations from hydrodynamics~\cite{Drescher:2007cd}: 
\begin{equation}
\label{eq:Ollitrault-eq1}
v_2=\frac{h\varepsilon}{1+K/0.7}, 
\end{equation}
where the scale factor $h$ is independent of system size and centrality, but 
may depend on the collision energy. 
The Knudsen number $K$ can be expressed as
\[
\frac{1}{K}=\frac{\sigma}{S}\frac{{\rm d}N}{{\rm d}y}\frac{1}{\sqrt{3}}.
\]
It vanishes in the hydrodynamic limit.
${\rm d}N/{\rm d}y$ is the total (charged + neutral) multiplicity per unit 
rapidity, $S$ is the transverse overlap area between the two nuclei, and 
$\sigma$ is an effective (transport) partonic cross section. 

The model has two free parameters, the ``hydrodynamic limit'' $h$, and the 
partonic cross section $\sigma$. 
The other quantities, $\varepsilon$, $S$, ${\rm d}N/{\rm d}y$, must be obtained 
from a model for the initial condition. 
Here, we choose the Color Glass Condensate (CGC) approach, including the effect 
of fluctuations in the positions of participant nucleons, which increase
$\varepsilon$~\cite{Drescher:2006ca}. 
The model provides a perfect fit to RHIC data for Au-Au and Cu-Cu collisions 
with $h=0.22$ and $\sigma=5.5$~mb~\cite{Drescher:2007cd}.

\begin{figure}
\centerline{\includegraphics*[width=0.7\linewidth]{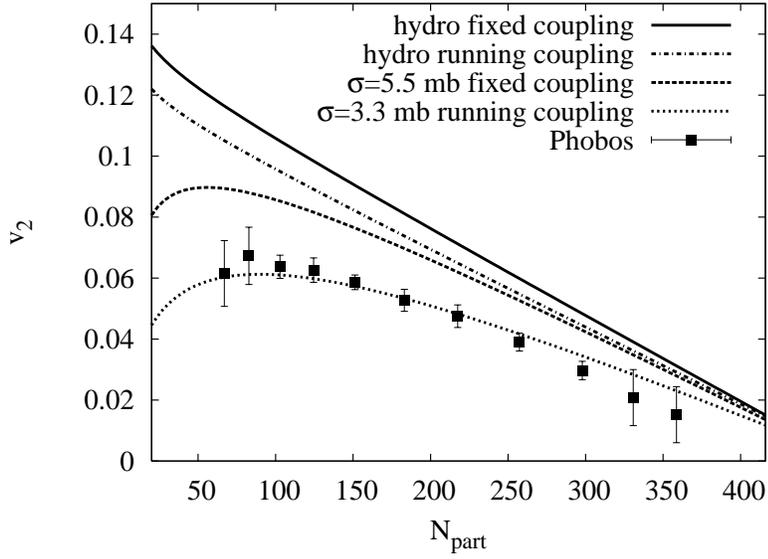}}
\caption{$v_2$ as a function of $N_{\rm part}$ at mid-rapidity for Pb-Pb 
  collisions at LHC ($\sqrtsnn=5.5$~TeV). 
  solid- and dash-dotted lines: $\varepsilon$ scaling ($K=0$ in 
  (\ref{eq:Ollitrault-eq1})); dashed- and dotted lines: incl.\ incomplete
  thermalization, with two values of the partonic cross section.
  Squares: PHOBOS data for Au-Au collisions at RHIC~\cite{Back:2004mh}. 
  The vertical scale is arbitrary (see text).  
\label{fig:Ollitrault-fig1}}
\end{figure}

We now briefly discuss the extrapolation to LHC. 
The hydrodynamic limit $h$ is likely to increase from RHIC to LHC, as the QGP 
phase will last longer; however, we do not have a quantitative prediction for 
$h$. 
We predict only the centrality dependence of $v_2$, not its absolute value. 
Figure~\ref{fig:Ollitrault-fig1} is drawn with $h=0.22$. 

The second parameter is $\sigma$, which parameterizes deviations from ideal 
hydrodynamics, i.e., viscous effects. 
We consider two possibilities: 1) $\sigma=5.5$~mb at LHC, as at RHIC.  
2) $\sigma\sim1/T^2$ (on dimensional grounds, assuming that no non-perturbative 
scales arise), where the temperature $T\sim ({\rm d}N/{\rm d}y)^{1/3}$.  
This gives the value 3.3~mb in figure~\ref{fig:Ollitrault-fig1}.

The remaining quantities ($S$, ${\rm d}N/{\rm d}y$ and $\varepsilon$) are 
obtained by extrapolating the CGC from RHIC to LHC, either with fixed-coupling 
(fc) or running-coupling (rc) evolution of the saturation scale $Q_s$. 
The multiplicity per participant increases by a factor of~3 (resp.\ 2.4) with 
fc (resp. rc).
The eccentricity $\varepsilon$ is 10\% larger with fc (solid curve in 
figure~\ref{fig:Ollitrault-fig1}) than with rc (dash-dotted curve) evolution. 
Deviations from hydrodynamics  (the $K$-dependent factor in 
equation~(\ref{eq:Ollitrault-eq1})) are somewhat smaller than at RHIC:
$v_2$ is 90\% (resp. 80\%) of the hydrodynamic limit for central collisions if 
$\sigma=5.5$~mb (resp. 3.3~mb). 
Our predictions lie between the dashed and dotted curves, up to an overall 
normalization factor. 
The maximum value of $v_2$ occurs for $N_{\rm part}$ between 60
($\sigma\approx$ const.) and 80 ($\sigma\sim1/T^2$).

Elliptic flow will be a first-day observable at LHC. 
Both its absolute magnitude and its centrality dependence are sensitive probes 
of initial conditions, and will help to improve our understanding of 
high-density QCD.

\subsection{Elliptic flow from pQCD+saturation+hydro model}
\label{s:Eskola2}
{\it K. J. Eskola, H. Niemi and P. V. Ruuskanen}
\vskip 0.5cm

We have previously predicted multiplicities and transverse momentum spectra for 
the most central LHC Pb+Pb collisions at $\sqrtsnn=5.5$~TeV using pQCD + 
saturation + hydro (EKRT model)~\cite{Eskola:2005ue,Eskola:1999fc}. We now 
extend these calculations for non-central collisions and predict low-$p_{T}$ 
elliptic flow. Our model is in good agreement with RHIC data for central 
collisions, and we show that our extension of the model is also in good 
agreement with minimum bias $v_2$ data from RHIC Au+Au collisions at 
$\sqrtsnn=200$~GeV.

We obtain the primary partonic transverse energy production and the formation 
time in central \A\A collisions from the EKRT model~\cite{Eskola:1999fc}. With 
the assumption of immediate thermalization we can use these to estimate the 
initial state for hydrodynamic evolution. For centrality dependence we consider 
here two limits which correspond to models eWN and eBC in~\cite{Kolb:2001qz}, 
where the profile and normalization are obtained from optical Glauber model, 
once the parameters in central collisions are fixed. In the eWN (eBC) model the 
energy density profile and normalization are proportional to the density and the
number of wounded nucleons (binary collisions), respectively. These energy 
density profiles are used to initialize boost invariant hydro code with 
transverse expansion. We use the bag model equation of state with massless 
gluons and quarks ($N_f =3$), and hadronic phase with all hadronic states up to 
a mass $2$~GeV included. Phase transition temperature is fixed to 165 MeV. 
Decoupling is calculated using standard Cooper-Frye formula, and all decays of 
unstable hadronic states are included. Freeze-out temperature is fixed from RHIC
$p_{T}$ spectra for the most central collisions and is $150$~MeV for binary
collision profile~\cite{Eskola:2005ue} and $140$~MeV for wounded nucleon 
profile. The same freeze-out temperatures are used at the LHC. Both 
initializations give a good description of the low-$p_{T}$ spectra for different
centralities at RHIC.

\begin{figure}
\hspace{-1.0cm}
\begin{minipage}[b]{0.6\linewidth}
\centering
\hspace*{+0.1cm}
\includegraphics[width=8.2cm]{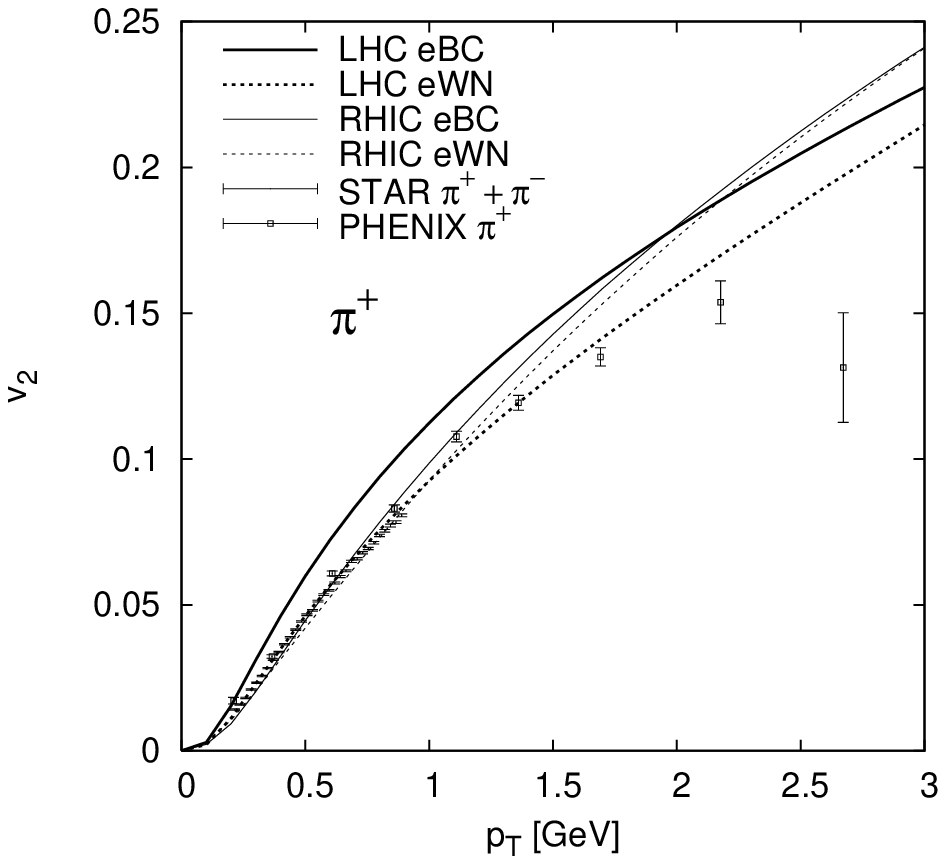}
\end{minipage}
\begin{minipage}[b]{0.6\linewidth}
\centering
\hspace*{-2.8cm}
\includegraphics[width=8.2cm]{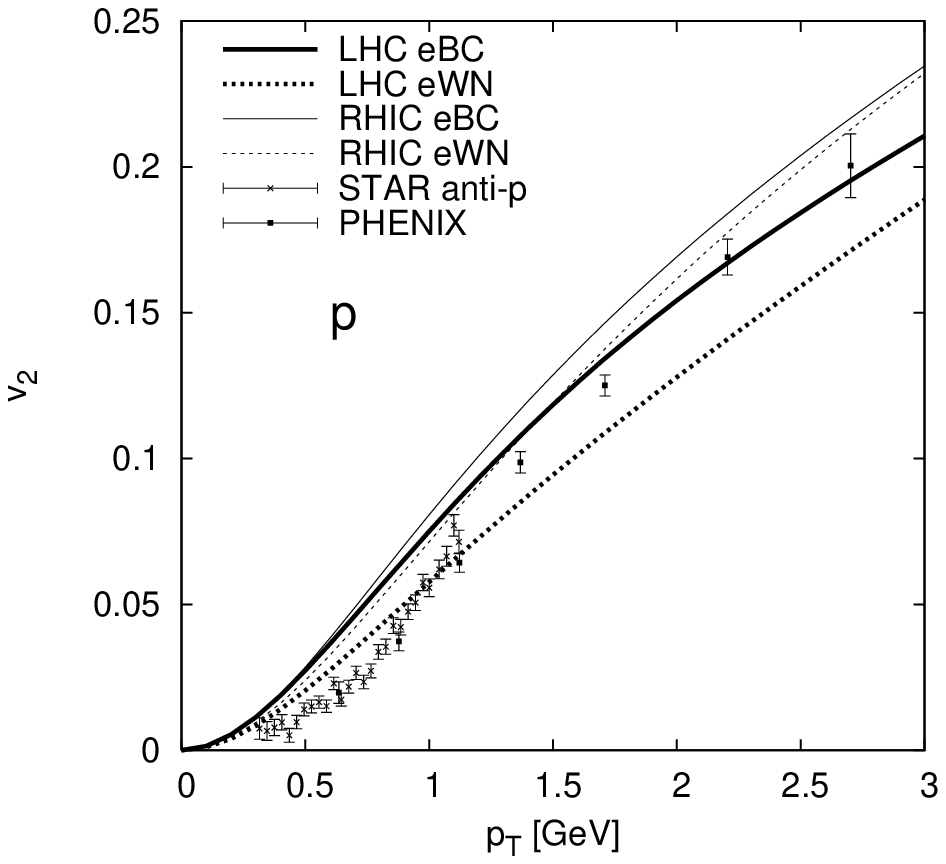}
\end{minipage}
\vspace{-0.8cm}\caption{ }
\label{fig:Eskola2-fig1}
\end{figure}
%%%%%%%%%%%%%%%%%%%%% FIGURE %%%%%%%%%%%%%%%%%%%%%%%%%%%%%%%%

The left panel of figure~\ref{fig:Eskola2-fig1} shows our calculations for 
$p_{T}$ dependence of minimum bias $v_{2}$ for positive pions. RHIC results 
are compared with STAR~\cite{Adams:2004bi} and PHENIX~\cite{Adler:2003kt} data. 
Our minimum bias centrality selection ($0-80$\%) corresponds to the one used by 
STAR collaboration. Solid lines are calculations with the eBC model and dashed 
lines are from the eWN model. Thin lines are our results for RHIC Au+Au 
collisions at $\sqrtsnn=200$~GeV and thick lines show our predictions for the 
LHC Pb+Pb collisions at $\sqrtsnn=5.5$~TeV. Largest uncertainty in $v_{2}$ 
calculations for pions comes here from the initial transverse profile of the 
energy density. Sensitivity to initial time and freeze-out temperature is much 
weaker. In general the eWN profile leads to weaker elliptic flow than the eBC 
profile. At the LHC the lifetime of the QGP phase is longer, which results in 
stronger flow asymmetry than at RHIC. On the other hand the magnitude of 
transverse flow is also larger, which decreases the $v_{2}$ value at fixed 
$p_{T}$. The net effect is that, for a given profile, $v_{2}$ of low-$p_{T}$ pions
is larger at the LHC than at RHIC. Since jet production at the LHC starts to 
dominate over the hydrodynamic spectra at larger $p_{T}$ than at 
RHIC~\cite{Eskola:2005ue}, we expect that the hydrodynamic calculations should 
cover a larger $p_{T}$ range at the LHC. Thus we predict that the minimum bias 
$v_{2}$ of pions at fixed $p_{T}$ is larger at the LHC than at RHIC, and can 
reach values as high as $0.2$.

Our model clearly overshoots the proton $v_{2}$ data from STAR~\cite{%
  Adams:2004bi} and PHENIX~\cite{Adler:2003kt}. A more detailed treatment of 
the hadron gas dynamics and freeze-out is needed to describe both the proton 
spectra and elliptic flow simultaneously. However, we can still predict the 
{\it change\/} in the behaviour of $v_{2}$ of protons when going from RHIC to 
the LHC. This is shown in the r.h.s. of figure~\ref{fig:Eskola2-fig1}. Although 
the flow asymmetry increases at the LHC, for more massive particles like protons
the overall increase in the magnitude of radial flow is more important than for 
light pions. This results in smaller $v_{2}$ at the LHC than at RHIC in the 
whole $p_{T}$ range for protons. Even if $v_{2}$ at fixed $p_{T}$ is smaller at 
the LHC, $p_{T}$-integrated $v_{2}$ is always larger at the LHC for all 
particles, due to the increase in the relative weight at larger $p_{T}$'s.

\subsection{From RHIC to LHC: 
       Elliptic and radial flow effects on hadron spectra}
\label{heinzv2}

{\it G. Kestin and U. Heinz}
\vskip 0.5cm

%\section[Elliptic and radial flow from RHIC to LHC]
%        {From RHIC to LHC: Elliptic and radial flow effects on hadron 
%         spectra\footnote[7]{$\!\!\!$Authors: Gregory Kestin and Ulrich 
%         Heinz, Ohio State University, Columbus, Ohio, USA.\\ 
%         Work supported in part by U.S. DOE grant DE-FG02-01ER41190 and 
%         NSF grant PHY-0354916.}}

Using (2+1)-d ideal hydrodynamics \cite{Kolb:2000sdd}, we computed 
the evolution from AGS to LHC energies of the $p_T$-spectra and elliptic 
flow at midrapidity for several hadrons~\cite{KH07}. While ideal fluid 
dynamics begins to break down below RHIC energies, due to viscous effects 
in the late hadronic stage which persist even at RHIC \cite{Hirano:2005xf}, 
its validity is expected to improve at the LHC where the elliptic flow 
saturates in the quark-gluon plasma (QGP) stage, and effects from late 
hadronic viscosity become negligible \cite{Hirano:2007xd}. Early QGP 
viscous effects seem small at RHIC 
\cite{Hirano:2005xf,Kolb:2003dz}, and recent results from Lattice 
QCD indicate little change of its specific shear viscosity $\eta/s$ 
from RHIC to LHC \cite{Meyer:2007ic}. The following {\em ideal fluid} 
dynamical predictions for soft ($p_T \lesssim 2{-}3$\,GeV/$c$) hadron 
production in $(A{\approx}200){+}(A{\approx}200)$ collisions at the LHC 
should thus be robust.

\begin{figure}[ht]
  \begin{center}
  \includegraphics[bb=60 66 554 740,width=6.8cm,height=0.8\linewidth,%
                   angle=270,clip=]{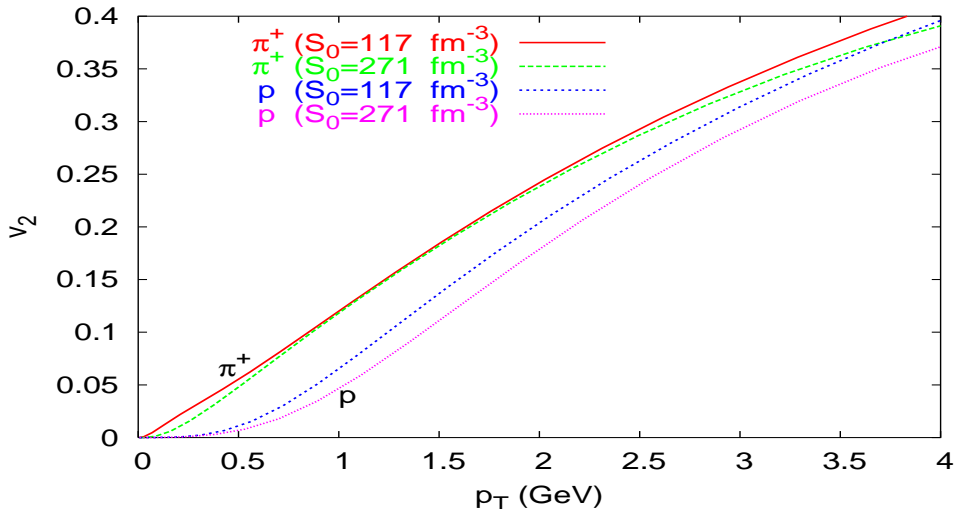}
  \end{center}
  \vspace*{-4mm}
   \caption{\label{v2pip}(Color online)
   Pion and proton elliptic flow as function of $p_T$ for $b{\,=\,}7$\,fm 
   Au+Au collisions at RHIC ($s_0{=}117$\,fm$^{-3}$) and LHC 
   ($s_0{=}271$\,fm$^{-3}$).
   } 
\end{figure}

\begin{figure}[t]
  \begin{center}
  \includegraphics[bb=37 129 580 758,width=\linewidth,height=13cm,clip=]%
                  {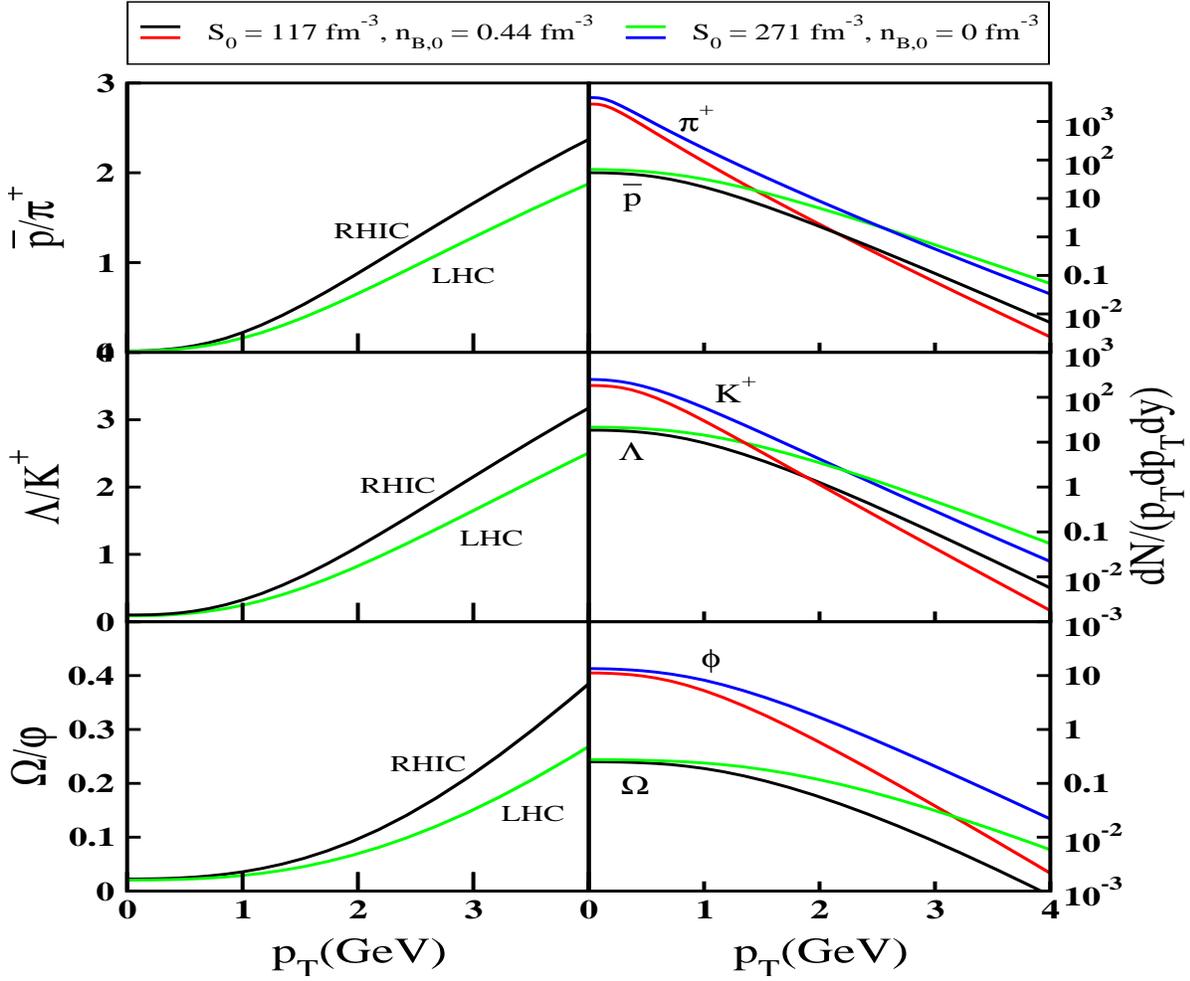}
  \vspace*{-8mm}
  \end{center}
  \caption{\label{spectra}(Color online)
   Normalized $p_T$-spectra (right) and $p_T$-dependent
   particle ratios (left) for ($\bar p,\pi^+$), ($\Lambda,K^+$),
   and ($\Omega,\phi$) in central Au+Au collisions at RHIC and LHC.
   Hadron yields are assumed to freeze out at 
   $T_\mathrm{c}{\,=\,}164$\,MeV.\\[-5mm]
  }
\end{figure}

For Au+Au at RHIC we use standard initial ($s_0{\,=\,}117/\mathrm{fm}^3$,
$n_{B0}{\,=\,}0.44/\mathrm{fm}^3$ at $\tau_0{\,=\,}0.6$\,fm/$c$, 
corresponding to $dN_\mathrm{ch}/dy(y{=}b{=}0){\,=\,}680$) and final 
conditions ($e_\mathrm{f}{\,=\,}75$ MeV/fm$^3$, $T_\mathrm{f}{\,=\,}100$\,MeV) 
\cite{Kolb:2000sdd,Kolb:2003dz}. For the LHC we assume 
$dN_\mathrm{ch}/dy(y{=}b{=}0){\,=\,}1200$ (the lower end of the 
predicted range), using $s_0{\,=\,}271/\mathrm{fm}^3$ and $n_{B0}{\,=\,}0$ at 
$\tau_0{\,=\,}0.45$ fm/$c$, keeping the product $T_0\tau_0$ and $T_\mathrm{f}$ 
unchanged. Predictions for other multiplicities, for interpolation to the 
actually measured LHC value, can be found in \cite{KH07}.\\
%
%%%%%%%%%%%%%%%%%%%%%%%%%%%%%%%%%%%%%%%%%%%%%%%%%%%%%%%%%%%%%%%%%%%%%%%%
\noindent\textbf{\emph{1.~Elliptic flow of pions and protons:}}
%%%%%%%%%%%%%%%%%%%%%%%%%%%%%%%%%%%%%%%%%%%%%%%%%%%%%%%%%%%%%%%%%%%%%%%%
%
\Fref{v2pip} shows the pion and proton elliptic flow at RHIC and LHC.
%
%%%%%%%%%%%%%%%%%%%%%%%%%%%%%%% Fig. 1 %%%%%%%%%%%%%%%%%%%%%%%%%%%%%%%
%%%%%%%%%%%%%%%%%%%%%%%%%%%%%%%%%%%%%%%%%%%%%%%%%%%%%%%%%%%%%%%%%%%%%%
%
While the total ($p_T$-integrated) pion elliptic flow increases from RHIC 
to LHC by about 25\% \cite{Hirano:2007xd}, very little of this increase 
($\sim5\%$) is of ideal fluid dynamical origin, most of it stemming from the 
{\em disappearance} of late hadronic viscous effects between RHIC and 
LHC. At fixed $p_T$, \Fref{v2pip} shows a {\em decrease} of $v_2$, 
reflecting a shift of the momentum anisotropy to larger $p_T$ by increased 
radial flow, which flattens the LHC $p_T$-spectra, affecting the heavier 
protons more than the lighter pions (\Fref{spectra}, right column). 
These radial flow effects on $v_2(p_T)$ are very small for pions but 
clearly visible for protons.\\
%
%%%%%%%%%%%%%%%%%%%%%%%%%%%%%%%%%%%%%%%%%%%%%%%%%%%%%%%%%%%%%%%%%%%%%%%%
\noindent\textbf{\emph{2.~$\bm{p_T}$-dependence of hadron ratios:}}
%%%%%%%%%%%%%%%%%%%%%%%%%%%%%%%%%%%%%%%%%%%%%%%%%%%%%%%%%%%%%%%%%%%%%%%%
%
Hydrodynamic flow, which leads to flatter $p_T$-spectra for heavy
than light particles, is a key contributor to the observed strong rise 
%
%%%%%%%%%%%%%%%%%%%%%%%%%% Fig.2 %%%%%%%%%%%%%%%%%%%%%%%%%%%%%%%%%%%%%%%%%% 
%%%%%%%%%%%%%%%%%%%%%%%%%%%%%%%%%%%%%%%%%%%%%%%%%%%%%%%%%%%%%%%%%%%%%%%%%%%
%
of the $\bar p/\pi$ and $\Lambda/K$ ratios at low $p_T$ at RHIC 
\cite{Kolb:2003dz}. \Fref{spectra} shows that this rise is slower 
at LHC than at RHIC (left column) since {\em all\,} spectra are flatter at 
LHC due to increased radial flow (right column) while their asymptotic 
ratios at $p_T{\,\to\,}\infty$ (given by their fugacity and spin 
degeneracy ratios \cite{Kolb:2003dz}) remain similar.
%
%%%%%%%%%%%%%%%%%%%%%%%%%%%%%%%%%%%%%%%%%%%%%%%%%%%%%%%%%%%%%%%%%%%%%%%%%%

\subsection{Differential elliptic flow prediction at the LHC from parton transport}
\label{s:Molnar}

{\em D. Moln\'ar}
\vskip 0.5cm

\noindent {\bf Introduction.} 
General physics arguments and calculations for a class of conformal field 
theories suggest~\cite{Danielewicz:1984ww,Kovtun:2004de} that quantum effects 
impose a lower bound on transport coefficients. 
For example, the shear viscosity to entropy density ratio is above a small 
value $\eta/s\gtrsim 0.1$ (``most perfect fluid'' limit). 
Dissipative effects can therefore never vanish in a finite, expanding system.
On the other hand, ideal (nondissipative) hydrodynamic modelling of Au+Au 
collisions at RHIC ($\sqrtsnn\sim 100$~GeV) is rather successful, leading many 
to {\em postulate\/} that the hot and dense QCD matter created is in fact such 
a ``most perfect fluid'' (at least during the early stages of the RHIC 
evolution). 
We predict here how differential elliptic flow $v_2(p_T)$ changes from RHIC to 
LHC collision energies (Pb+Pb at $\sqrtsnn=5.5$~TeV), {\em if the quark-gluon 
  system created at both RHIC and the LHC has a ``minimal'' shear viscosity 
  $\eta/s = 1/(4\pi)$\/}.

\smallskip
\noindent {\bf Covariant transport theory} is a nonequilibrium framework with 
two main advantages: 
i) it has a hydrodynamic limit (i.e., capable of thermalization); and 
ii) it is {\em always} causal and stable.
In contrast, hydrodynamics (whether ideal, Navier-Stokes, or second-order 
Israel-Stewart theory~\cite{Israel:1979wp}) shows instabilities and acausal 
behavior in certain, potentially large, regions of the hydrodynamic ``phase 
space''.

We consider here Lorentz-covariant, on-shell Boltzmann transport theory, with a 
$2\to 2$ rate~\cite{Molnar:2001ux,Zhang:1997ej}
\[
p_1^\mu \partial_\mu f_1 = 
S(x, {\vec p}_1)  + \frac{1}{\pi} \int_2\int_3\int_4
(f_3 f_4 - f_1 f_2) W_{12\to 34} \ \delta^4(p_1{+}p_2{-}p_3{-}p_4)
\]
The integrals are shorthands for $\int_i\equiv\int{\rm d}^3p_i/(2E_i)$.
For dilute systems, $f$ is the phase space distribution of quasi-particles, 
while the transition probability $W = s (s-4m^2){\rm d}\sigma/{\rm d}t$ is 
given by the scattering matrix element.
Our interest here, on the other hand, is to study the theory {\em near its 
  hydrodynamic (local equilibrium) limit\/}.

Near local equilibrium, the transport evolution can be characterized via 
transport coefficients of shear and bulk viscosities ($\eta,\zeta$) and heat 
conductivity ($\lambda$) that are determined by the differential cross section.
For the massless dynamics ($\epsilon = 3p$ equation of state) considered here 
$\eta \approx 0.8 T/\sigma_{\rm tr}$, $\zeta = 0$, and 
$\lambda \approx 1.3/\sigma_{\rm tr}$, 
$\tau_\pi \approx 1.2 \lambda_{\rm tr}$~\cite{deGroot:1980,Israel:1979wp}
($\sigma_{\rm tr}$ and $\lambda_{\rm tr}$ are the {\em transport\/} cross section 
and mean free path, respectively).

\smallskip
\noindent {\bf Minimal viscosity and elliptic flow.}
Finite cross sections lead to dissipative effects that reduce elliptic 
flow~\cite{Molnar:2004yh,Zhang:1999rs}.
For a system near thermal and chemical equilibrium undergoing longitudinal 
Bjorken expansion, $T \sim\tau^{-1/3}$, $s\approx 4n\sim T^3$, and thus 
$\eta/s = {\rm const}$ requires a growing $\sigma_{\rm tr} \sim \tau^{2/3}$.
With $2\to 2$ processes chemical equilibrium is broken, therefore 
$\sigma_{\rm tr}$ also depends on the density through $\mu/T \sim \ln n$ (because 
$s = 4(n-\mu/T)$).
We ignore this weak logarithm and take 
$\sigma_{\rm tr}(\tau) = \sigma_{0,\rm tr} (\tau/0.1\mbox{ fm})^{2/3}$ with 
$\sigma_{0,\rm tr}$ large enough to ensure that most of the system is at, or 
below, the viscosity bound (thus we somewhat {\em underestimate\/} viscous 
effects, i.e., overestimate $v_2(p_T)$).

For $\A\A$ at $b=8$~fm impact parameter we use the class of initial conditions 
in~\cite{Molnar:2001ux} that has three parameters: 
parton density ${\rm d}N/{\rm d}\eta$, formation time $\tau_0$, and effective
temperature $T_0$ that sets the momentum scale. 
Because of scalings of the transport solutions~\cite{Molnar:2001ux}, 
$v_2(p_T/T_0)$ only depends on two combinations
$\sigma_{\rm tr}\,{\rm d}N/{\rm d}\eta\sim A_\perp \tau_0 / \lambda_{\rm tr}$ and 
$\tau_0$. 
This may look worrisome because ${\rm d}N/{\rm d}\eta$ at the LHC is uncertain 
by at least a factor of two. 
However, the ``minimal viscosity'' requirement $T\,\lambda_{\rm tr} \approx 0.5$ 
{\em fixes\/} $\sigma_{\rm tr}{\rm d}N/{\rm d}\eta$ (e.g., with 
${\rm d}N/{\rm d}\eta(b{=}0) = 1000$ at RHIC, $\sigma_{0,\rm tr} \approx 2.7$~mb),
while on dimensional grounds $\tau_0 \sim 1/T_0$.

This means that the {\em main difference between LHC and RHIC is in the typical
  momentum scale $T_0$\/} (gold and lead nuclei are basically identical), and 
therefore to good approximation {\em one expects the simple scaling 
$v_2^{\rm LHC}(p_T) \approx v_2^{\rm RHIC}(p_T T_0^{\rm LHC}/T_0^{\rm RHIC})$\/}.
From gluon saturation physics we estimate 
$r\equiv T_0^{\rm LHC}/T_0^{\rm RHIC} \approx 1.3-1.5$ at $b=8$~fm via 
Gribov-Levin-Ryshkin formula as applied in~\cite{Adil:2005bb} (we take 
$T_{\rm eff} \sim Q_s \sim \sqrt{\langle p_T^2\rangle}$).

\begin{figure}[htpb]
\centerline{\includegraphics[height=5.2cm]{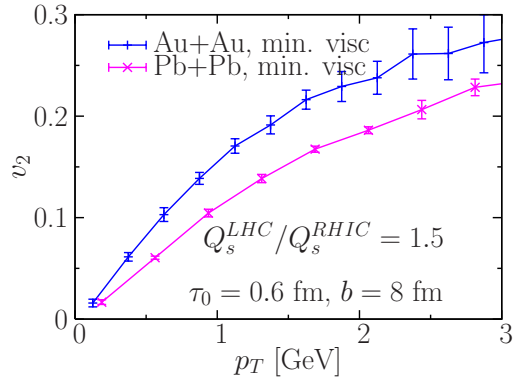}}
\caption{Differential elliptic flow at RHIC and the LHC, assuming a ``minimally 
  viscous'' quark-gluon system $\eta/s = 1/(4\pi)$ at both energies.}
\label{fig:Molnar-fig1}
\end{figure}

As depicted in figure~\ref{fig:Molnar-fig1}, at a given $p_T$ the scaling 
predicts a striking {\em reduction\/} of $v_2(p_T)$ at the LHC relative to RHIC.
This is the opposite of both ideal hydrodynamic expectations and what was seen 
going from SPS to RHIC (where $v_2(p_T)$ increased slightly with energy). 
Experimental determination of the scaling factor 
$r\equiv Q_s^{\rm LHC}/Q_s^{\rm RHIC}$ would provide a further test of gluon 
saturation models.

We note that higher momenta at the LHC would also imply somewhat earlier 
thermalization $\tau_0 \sim 1/T_0$. 
This is expected to prolong longitudinal Bjorken cooling at the LHC, changing 
the scale factor in $v_2(p_T)$ from $r$ towards 
$r^{1-1/3} = r^{2/3} \approx 1.2-1.3$.

\section{Hadronic flavor observables}
\label{sec:flavor}

\subsection{Thermal model predictions of hadron ratios}
\label{andronic_therm}

{\it A. Andronic, P. Braun-Munzinger and J. Stachel}

{\small
We present predictions of the thermal model for hadron ratios in central 
Pb+Pb collisions at LHC.
}
\vskip 0.5cm

Based on the latest analysis within the thermal model of the hadron yields 
in central nucleus-nucleus collisions \cite{Andronic:2005yp}, the
expected values at LHC for the chemical freeze-out temperature and 
baryochemical potential are $T$=161$\pm$4 MeV and 
$\mu_b$=0.8$^{+1.2}_{-0.6}$ MeV, respectively.
For these values, the thermal model predictions for hadron yield ratios 
in central Pb+Pb collisions at LHC are shown in Table~\ref{ttab_aa_t1}.
We have assumed no contribution of weak decays to the yield of 
pions, kaons and protons.

\begin{table}[hbt]
\caption{Predictions of the thermal model for hadron ratios in central 
Pb+Pb collisions at LHC. The numbers in parantheses represent the error
in the last digit(s) of the calculated ratios.\\}
\label{ttab_aa_t1}
\begin{tabular}{cccccc}
$\pi^-/\pi^+$ & $K^-/K^+$ & $\bar{p}/p$ & $\bar{\Lambda}/\Lambda$ & 
$\bar{\Xi}/\Xi$ & $\bar{\Omega}/\Omega$ \\ \br
1.001(0) & 0.993(4) & 0.948$^{-0.013}_{+0.008}$ & 0.997$^{-0.011}_{+0.004}$ & 
1.005$^{-0.007}_{+0.001}$ & 1.013(4) \\ \br \\
$p/\pi^+$ & $K^+/\pi^+$ & $K^-/\pi^-$ & $\Lambda/\pi^-$ & 
$\Xi^-/\pi^-$ & $\Omega^-/\pi^-$ \\ \br
0.074(6) & 0.180(0) & 0.179(1) & 0.040(4) & 0.0058(6) & 0.00101(15) \\ \br \\
\end{tabular}
\end{table}

The antiparticle/particle ratios are all very close to unity, with the
exception of the ratio $\bar{p}/p$, reflecting the expected small, 
but nonzero, $\mu_b$ value. The errors are determined by the errors of 
$\mu_b$ in case of antiparticle/particle ratios and by the errors of
$T$ for all other ratios. 

\begin{table}[hbt]
\caption{Predictions for the relative abundance of resonances at
chemical freeze-out.
\\}
\label{tab_aa_t2}
\begin{tabular}{ccccccc}
$\phi/K^-$ & $K^{*0}/K^0_S$ & $\Delta^{++}/p$ & $\Sigma(1385)^+/\Lambda$ & 
$\Lambda^*/\Lambda$ & $\Xi(1530)^0/\Xi^-$ \\ \br
0.137(5) & 0.318(9) & 0.216(2) & 0.140(2) & 0.075(3) & 0.396(7)\\ \br
\end{tabular}
\end{table}

Assuming that the yield of resonances is fixed at chemical freeze-out,
we show in Table~\ref{tab_aa_t2} predictions for the relative yield of 
various resonance species.
We emphasize that the above hypothesis needs to be checked at LHC, 
in view of the data at RHIC \cite{Adams:2006yu}, which may indicate
rescattering and regeneration of resonances after chemical freeze-out.

\subsection{(Multi)Strangeness Production in Pb+Pb collisions at LHC. 
HIJING/B\=B v2.0 predictions.}

{\it V.~Topor Pop, J.~Barrette, C.~Gale, S.~Jeon and M.~Gyulassy}

{\small
Strangeness and multi-strangeness particles production
can be used to explore the initial transient field fluctuations 
in heavy ion collisions.  
We emphasize the role played by Junction anti-Junction (J\=J) loops and 
strong color electric fields (SCF) in these collisions. 
Transient field fluctuations of SCF  
on the baryon production in central (0-5 \%) Pb+Pb collisions at 
$\sqrt{s_{\rm NN}}$ = 5.5 TeV will be discussed  in the framework of
HIJING/B\=B v2.0 model, looking in particular to the predicted 
evolution of nuclear modification factors ($R_{\rm AA}$) 
from RHIC to LHC energies.  
Our results indicate the importance of a good description of the baseline
elementary $p+p$ collisions at this energy.
}
\vskip 0.5cm

In previous publications \cite{Topor-Pop:2007hb} we studied 
the possible role of topological baryon junctions \cite{Kharzeev:1996sq},
and the effects of strong color field (SCF)  
in nucleus-nucleus collisions at RHIC energies.
We have shown that the dynamics of the 
production process can deviate
considerably from that based on Schwinger-like estimates for 
homogeneous and constant color fields.
An increase of the string tension from $\kappa_0$= 1 GeV/fm, 
to {\em in medium mean values} 
of 1.5-2.0 GeV/fm and 2.0-3.0 GeV/fm, for d+Au and Au+Au 
respectively, results in a consistent description of the 
observed nuclear modification factors (NMF) $R_{\rm AA}$ 
in both reactions and point to the relevance of fluctuations 
on transient color fields. 
The model provides also an explanation of the baryon/meson anomaly,
and is an alternative dynamical description 
of the data to recombination models \cite{Fries:2003kq}.

Strangeness enhancement \cite{Rafelski:1982pu}, 
strong baryon transport, 
and increase of intrinsic transverse momenta $k_T$ \cite{Soff:2004yc} 
are all expected consequences of SCF.
These are modeled  in our microscopic models as
an increase of the effective  string tension that controls the
quark-anti-quark ({\it q}$\bar{q}$) and 
diquark - anti-diquark (qq$\overline{\rm {qq}}$) pair creation rates
and the strangeness suppression factors.
A reduction of the strange ($s$) quark mass from $M_s$=350 MeV to the current
quark mass of approximately $m_s$=150 MeV, gives a
strangeness suppression factor $\gamma_s^1 \approx$ 0.70.
A similar value of $\gamma_s^1$ (0.69) is obtained by increasing   
the string tension from $\kappa_0$=1.0 GeV/fm to $\kappa$=3.0 GeV/fm
\cite{Topor-Pop:2007hb}.
Howeover, if we consider that Schwinger tunneling could explain
the thermal character of hadron spectra we can define an apparent 
temperature as function of the average value of string tension ($ <\kappa>$),
$T=\sqrt{3<\kappa>/4\pi}$ \cite{Castorina:2007eb}.
The predictions at LHC for initial energy density and temperature are 
$\epsilon_{\rm LHC} \approx$ 200 GeV/fm$^3$ and 
$T_{\rm LHC} \approx $ 500 MeV, respectively \cite{Muller:2006ru}.
Both values would lead in the framework of our model to an estimated increase 
of the average value of string tension 
to $\kappa \, \approx \,$ 5.0 GeV/fm at LHC energy.

%%%%%%%%%%%%%%%%%%%%%%%%%%%%

%%%%%%%%%\section{(Multi)Strangeness Production in Pb+Pb collisions at LHC.}

\begin{figure}[h]
\vspace*{-0.2cm}
\begin{center}
\includegraphics*[width=12.0cm,height=6.0cm]{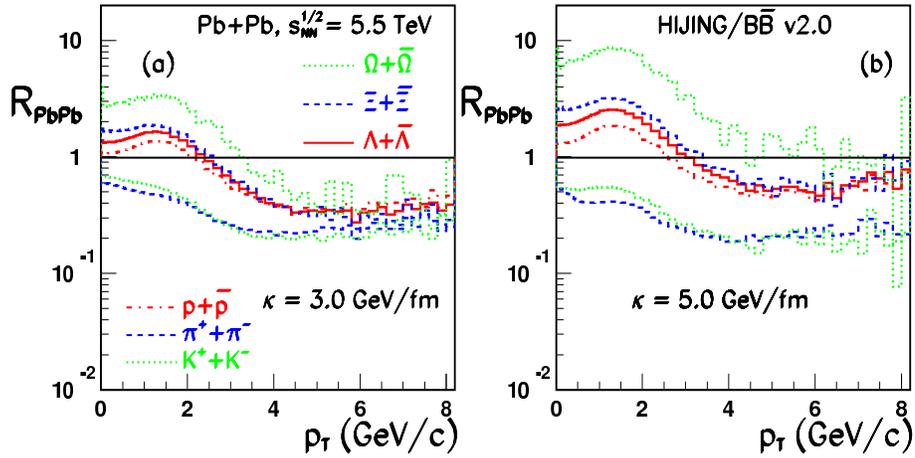}
\end{center}
\vspace*{-0.3cm}
\caption{HIJING/B\=B v2.0 predictions including SCF effects 
for NMF of identified particles. The results for proton and lambda particles
are for inclusive measurements.} 
\label{rpbpb}
\end{figure}

The {\it p}+{\it p} cross sections serve as a baseline
reference to calculate NMF for $A+A$ collisions ({\it R}$_{\rm {AA}}$).
In $p+p$ collisions high baryon/mesons ratio (i.e. close to unity) 
at intermediate $p_T$ were reported at $\sqrt{s_{\rm NN}}$= 1.8 TeV 
\cite{Acosta:2005prd}. These data could be fitted assuming a string tension 
$\kappa$=2.0 GeV/fm. This value is used in our calculations at  
$\sqrt{s_{\rm NN}}$= 5.5 TeV. This stresses the need for a reference 
$p+p$ measurements at LHC energies.

The predictions for NMF {\it R}$_{\rm {PbPb}}$
of identified particles at the LHC energy are presented in Fig.~\ref{rpbpb}
for two values of the string tension.
From our model we conclude that baryon/meson anomaly, will persist at the LHC 
with a slight increase for increasing strength of 
the chromoelectric field.
The NMF {\it R}$_{\rm {PbPb}}$ also 
exhibit an ordering with strangeness content at low and intermediate $p_T$.
The increase of the yield being higher for multi-strange hyperons 
than for (non)strange hyperons 
({\it R}$_{\rm {PbPb}}$($\Omega$) $>$ {\it R}$_{\rm {PbPb}}$($\Xi$)
$>$ {\it R}$_{\rm {PbPb}}$($\Lambda$) 
$>$ {\it R}$_{\rm {PbPb}}$($p$) ). 
At high $p_T > 4 GeV/c $ for $\kappa$=3.0 GeV/fm,   
a suppression independent of flavours is predicted due to quench effects.  
In contrast, this independence seems to happen 
at $p_T > $ 8 GeV/c for $\kappa$=5.0 GeV/fm.   

\begin{figure}[h]
\begin{center}
\vspace*{-0.3cm}
\includegraphics*[width=12.0cm,height=6.0cm]{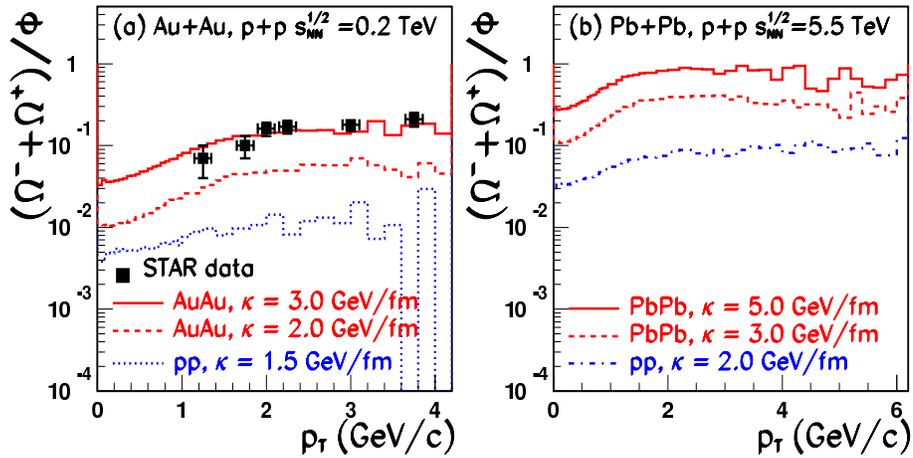}
\end{center}
\vspace*{-0.3cm}
\caption{Predictions of HIJING/B\=B v2.0 for  
the $(\Omega^++\Omega^-)/\Phi$ ratio as function of $p_T$ 
for RHIC (left panel) and LHC (right panel) energies. 
The experimental data are from STAR \cite{Ma:2006pn}.}
\label{omphi}
\end{figure}

As expected, a higher sensitivity to SCF effects on the $p_T$ dependence 
of multi-strange particle yield ratio is predicted. As an example, 
 Fig. \ref{omphi} presents our results for the ratio 
($\Omega^{-} + \Omega^{+}$)/$ \Phi $ in central (0-5\%) Pb+Pb collisions
and  $p+p$ collisions.
The results and data at RHIC top energy are also included (left panel).

%%%%%%%\section{Summary}

The mechanisms of (multi)strange particles production 
is very sensitive to the early phase of nuclear collisions, 
when fluctuation in 
the color field strength are highest. Their mid-rapidity yield favors 
a large value of the average string tension as shown at RHIC 
and we expect similar dynamical effects at LHC energy.
The precision of these predictions depens on our knowledge of  
initial conditions, parton distribution functions at low Bjorken-$x$,
the values of the scale parameter $p_0$, constituent 
and current (di)quark masses, energy loss for gluon and quark jets.

\subsection{Antibaryon to Baryon
Production Ratios
 in Pb-Pb and p-p collision at LHC energies
 of the DPMJET-III Monte Carlo}
\label{bopp}

{\it F. Bopp, R. Engel,  J. Ranft    and
S. Roesler}

{\small
A sizable component of stopped baryons is predicted for $pp$ and
$PbPb$ collisions at LHC. Based on an analysis of RHIC data within
framework of our multichain Monte Carlo DPMJET-III the LHC predictions
are presented. 
}
\vskip 0.5cm

This addendum to Ranft's talk about the main
DPMJET III prediction addresses baryon stopping. The interest is a
component without leading quarks. Where the flavor decomposition is
not determined by final state interactions the valence-quarkless component
can be enhanced by considering net strange baryons.

In models, in which soft gluons can arbitrarily
arrange colors, a configuration can appear in which the baryonic charge
ends up moved to the center. The actual transport is just an effect
of the orientation of the color-compensation during the soft hadronisation.
Various other ideas about fast baryon stopping exist but to have it
caused by such an {}``initial'' process is an attractive option.

The {}``Dual-Topological'' phenomenology
of such baryon transport processes was developed 30 years ago\cite{venez77}.
Critical are various baryonium-exchange intercepts which were  estimated
at that time. Some ambiguity remains until today for the quarkless
com\-po\-nent (also called ``string junction" exchange denoted as $\{SJ\}$)  
and a confirmation of the flat net-baryon distribution indicated
by RHIC data at LHC would be helpful.

Nowadays it is postulated that at very high
energy hadronic scattering can be understood as extrapolation of BFKL
Pomeron exchanges\cite{lipatov99} and their condensates in the minimum
bias region. BFKL Pomerons are described by ladders of dispersion
graphs, in which soft effects are included using effective gluons.
In principle these soft effects include the color compensating mechanism
usually modelled as two strings neutralizing triplet colors. A necessary
ingredient in this approach are \emph{Odderons} exchanges with Pomeron-like
intercepts and with presumably much smaller couplings. As these Odderons
can produce a baryon exchange of the type discussed above, a small
rather flat net baryon component is expected.

Experimentally, the first indication for a flat component
came from never finalized preliminary ZEUS data at HERA. As RHIC runs
$pp$ or \emph{heavy ions} instead of $p\bar{p}$ this question could
be addressed much better than before and the data seem to require
a flat contribution. In a factorizing Quark-Gluon-String model calculation\cite{Bopp:2006dv}
the best fit  to RHIC BRAHMS $pp$ data at $\sqrt{{s}}=200$
GeV required diquarks with a probability of $\epsilon=0.024$ to involve
a quarkless baryonium-exchange with an intercept $\alpha_{\{ SJ\}}=0.9$.

To obtain such a  quarkless baryonium-exchange
in the microscopic generator DPMJET III\cite{dpmjet2} a new
string interaction  reshuffling the initial strings was introduced.
It introduces an exchange with a conservative intercept of $\alpha_{\{ SJ\}}=0.5$.
With this baryonium addition good fits were obtained for various baryon
ratios in $p-p$  and $d-Au$ RHIC and $\pi-p$ FERMILAB
processes\cite{Bopp:2005cr}.
There are of course a number of more conventional baryon transport
mechanisms implemented in the model. As the string interaction requires
multiple Pomeron exchanges the new mechanism is actually only a $10$\%
effect at $pp$  RHIC. It is, however, important for heavy
ion scattering or at LHC energies.

\begin{flushleft}\begin{minipage}[b]{6cm}\begin{flushleft}For \emph{}\(pp\) LHC the DPMJET III prediction for
the pseudo rapidity of \(p\), \(\bar{p}\), and \(p-\bar{p}\)
is shown in the Figure on the right. The new baryon stopping is now 
a 40\% effect\emph{.} Of course, with the effective
intercept of \(0.5\) the present implementation of the baryon stopping
is a rather conservative estimate. For an intercept of \(1.0\) the
value at \(\eta=0\) would roughly correspond to the present value of
\(\eta=4\)\end{flushleft}\end{minipage}\vspace*{2cm} 
\includegraphics[scale=0.50]{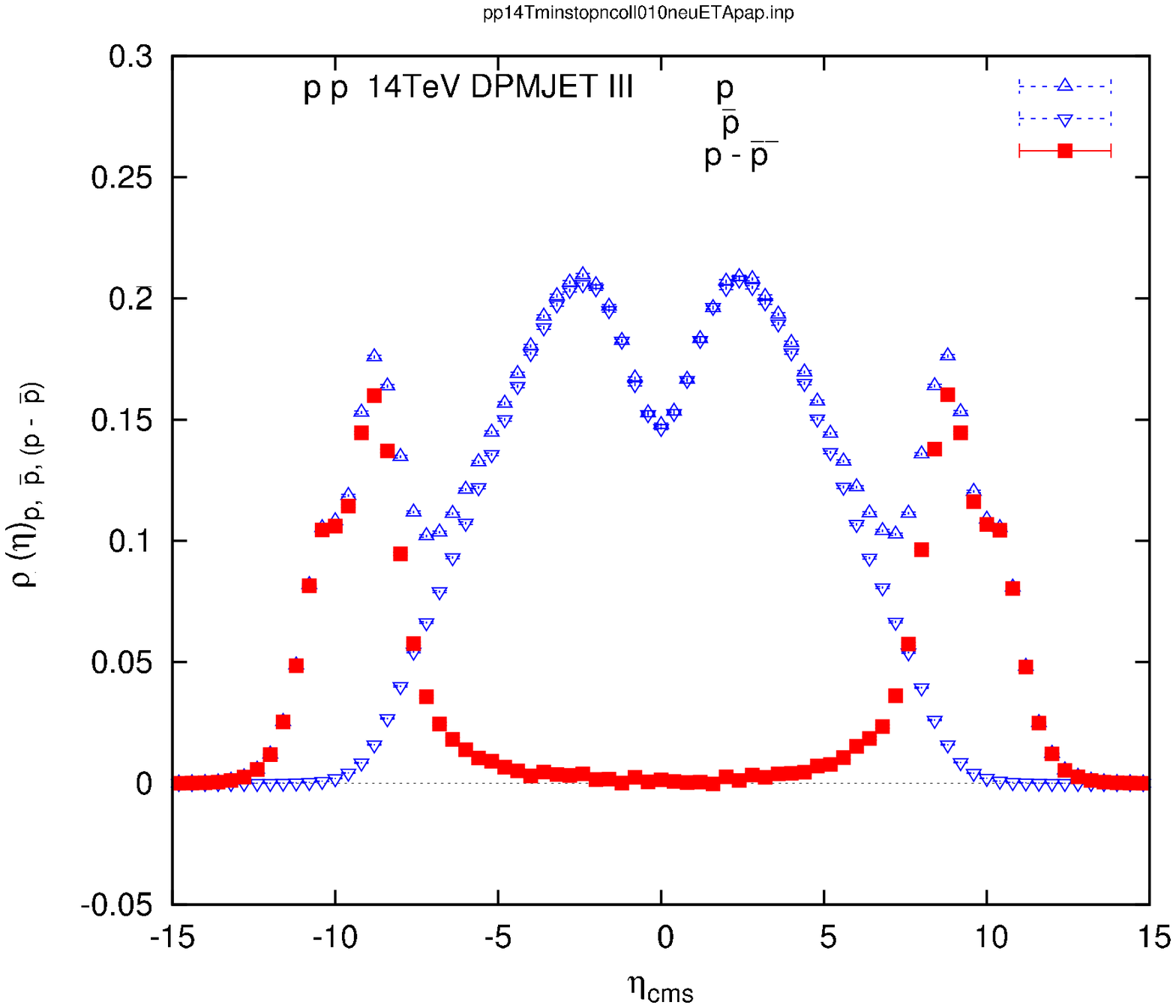}\end{flushleft}\vspace*{-1.5cm}

\noindent
\begin{minipage}[b]{11cm}
\begin{flushleft} \includegraphics[scale=0.50]{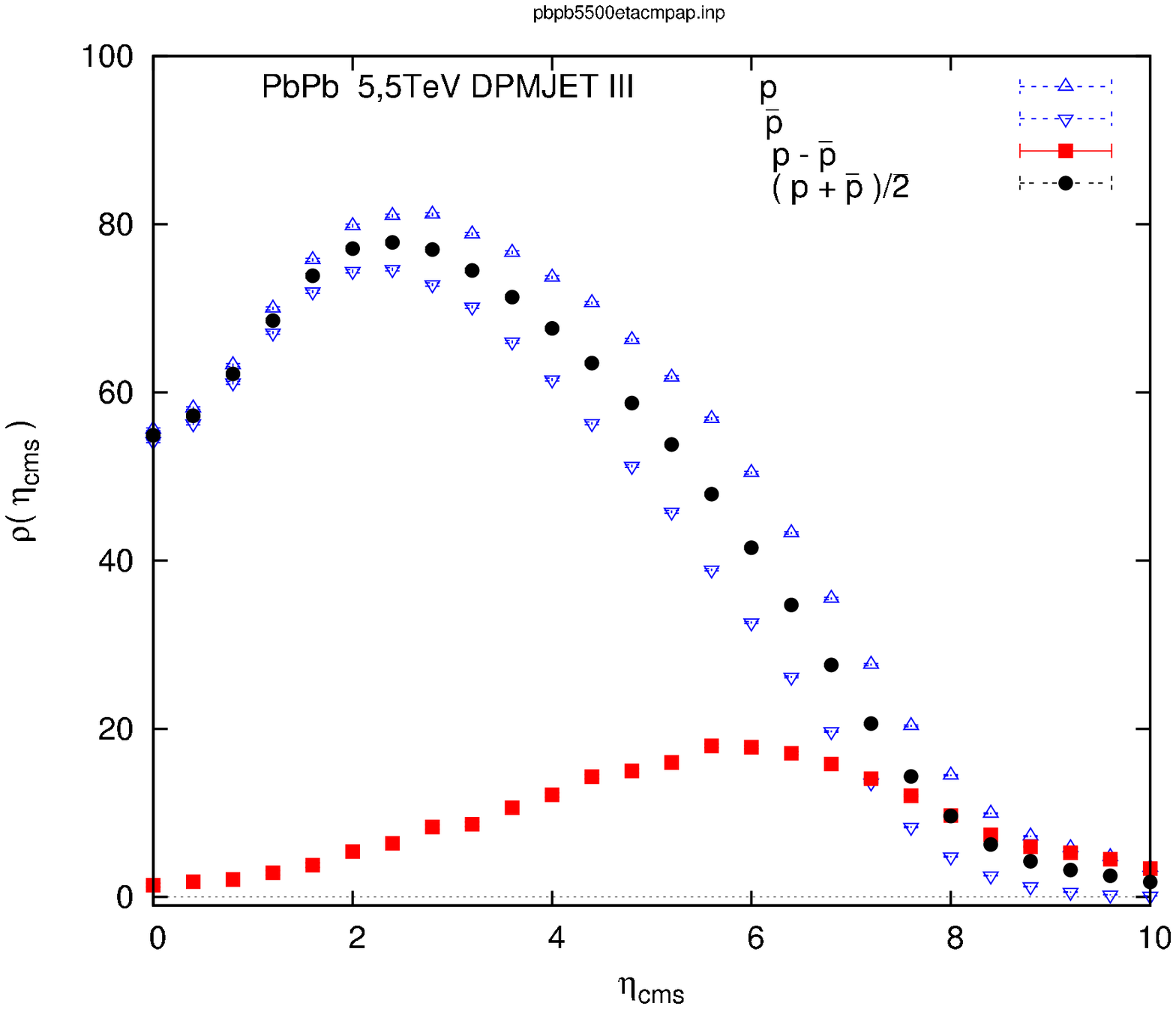}
\end{flushleft} 

\end{minipage}    \begin{minipage}[b]{4cm}
We now turn to DPMJET III prediction for central $PbPb$ LHC. For
the most central $10\%$ of the heavy ion events the pseudorapidity
proton and $\Lambda$ distributions are given in figures below. The
$PbPb$ results are preliminary, as the model is not well tested
in this region.\vspace{-1.5cm}\end{minipage}
\vspace{-0.2cm}

\noindent
\begin{minipage}[b]{11cm}
   \begin{flushleft}
\includegraphics[scale=0.50]{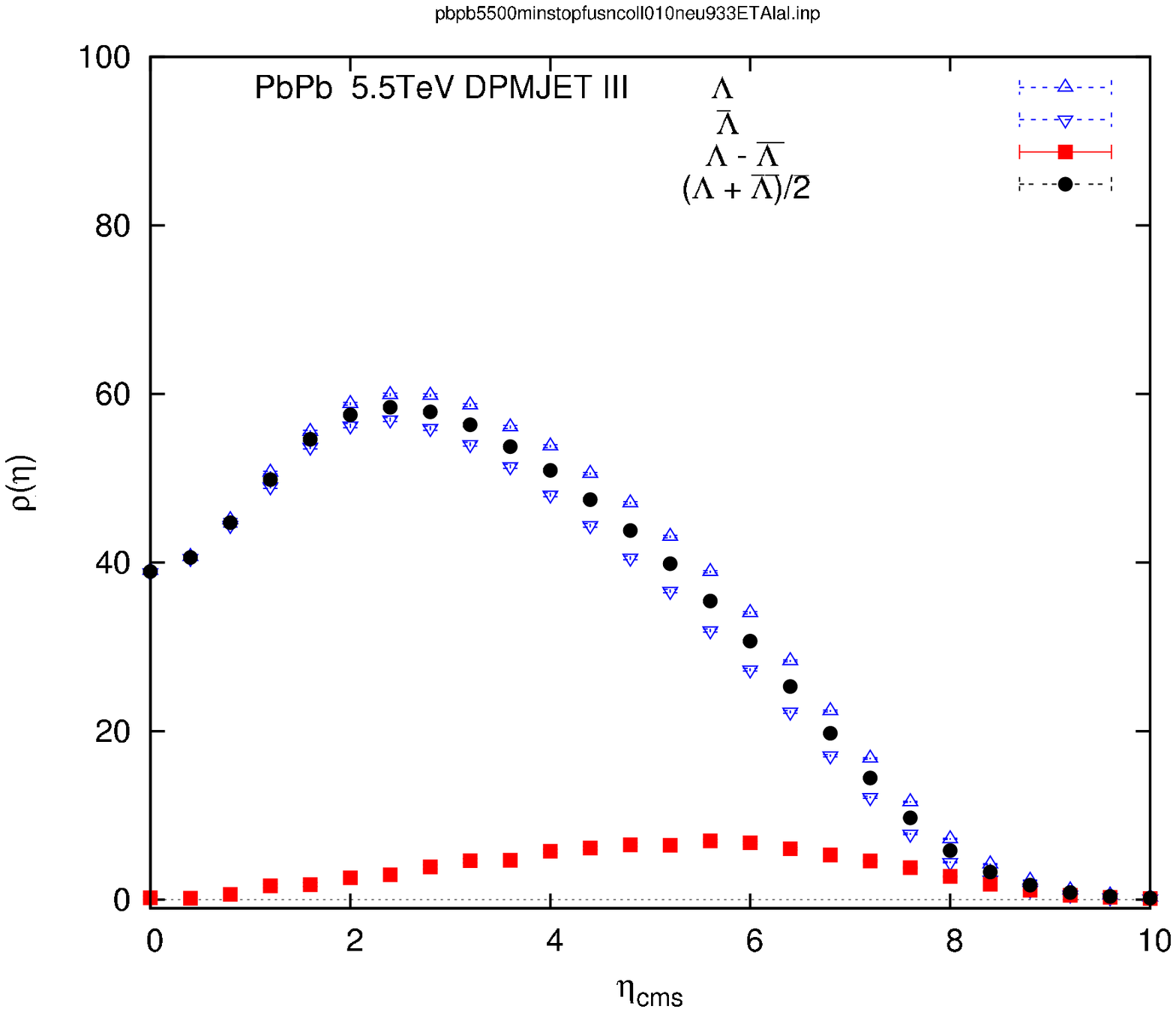}\end{flushleft}

\end{minipage}   \begin{minipage}[b]{4cm}$  $ \end{minipage}

\subsection{Statistical model predictions for $pp$ and Pb-Pb collisions at LHC}
\label{s:Kraus}

{\em I. Kraus, J.~Cleymans, H.~Oeschler, K.~Redlich and S.~Wheaton}

{\small
Predictions for particle production at LHC are discussed in the context of the 
statistical model. 
Moreover, the capability of particle ratios to determine the freeze-out point 
experimentally is studied, and the best suited ratios are specified.
Finally, canonical suppression in p-p collisions at LHC energies is discussed 
in a cluster framework. 
Measurements with $pp$ collisions will allow us to estimate the strangeness 
correlation volume and to study its evolution over a large range of incident 
energies.
}
\vskip 0.5cm

Particle production in heavy-ion collisions is, over a wide energy range, 
consistent with the assumption that hadrons originate from a thermal source 
with a given temperature $T$ and a given baryon chemical potential $\mu_B$. 
In the framework of the statistical model, we exploit the feature that the 
freeze-out points appear on a common curve in the $T-\mu_B$ plane. 
The parameterization of this curve, taken from reference~\cite{Cleymans:2005xv},
is used to extrapolate to the LHC energy of $\sqrtsnn=5.5$~TeV: 
$T\approx 170$~MeV, $\mu_B\approx 1$~MeV.

\begin{figure}[htb]
\begin{minipage}[b]{0.29\linewidth}
\centering
\includegraphics*[width=\linewidth]{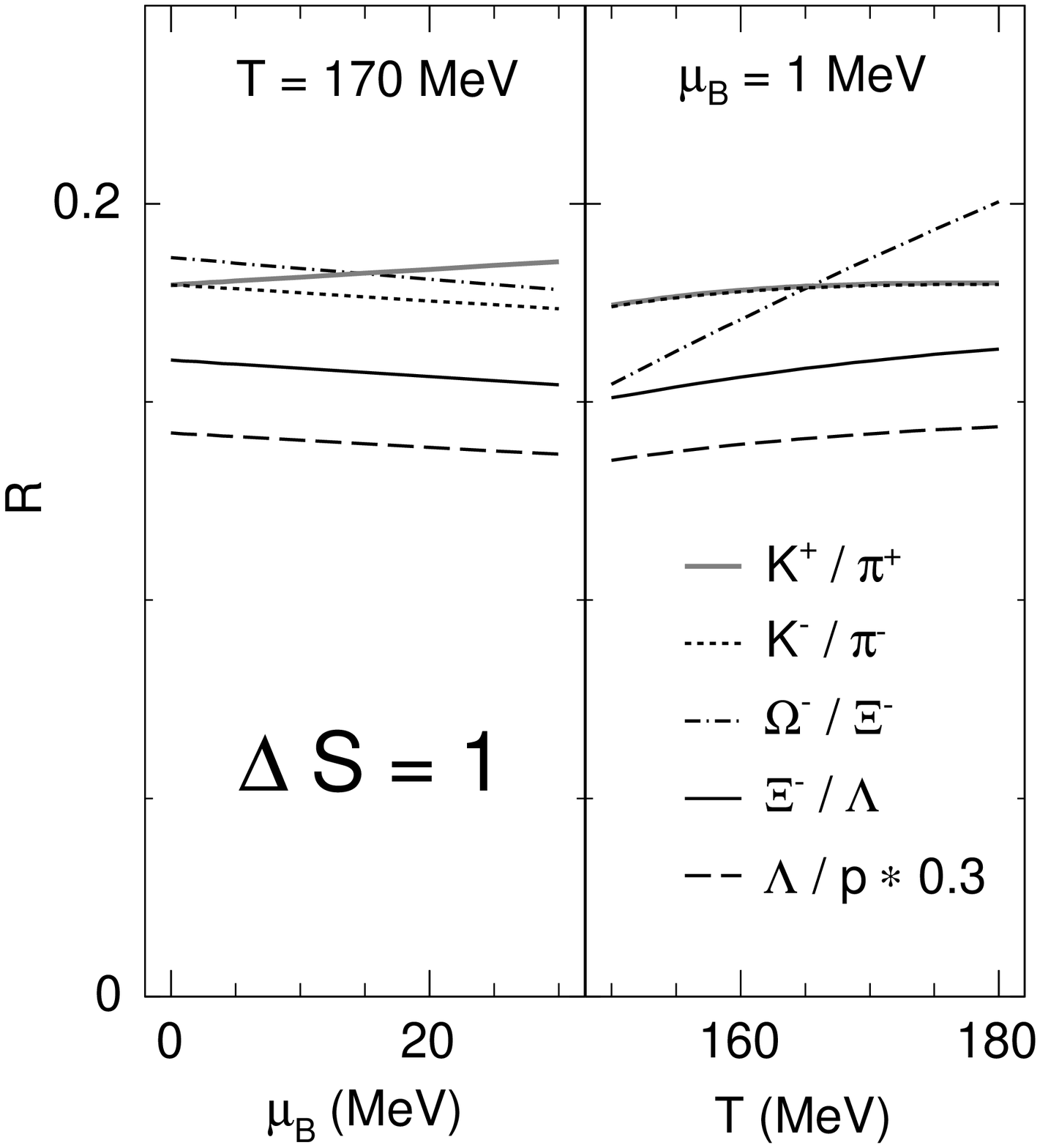}
\end{minipage}\hfill
\begin{minipage}[b]{0.36\linewidth}
\centering
\includegraphics[width=\linewidth]{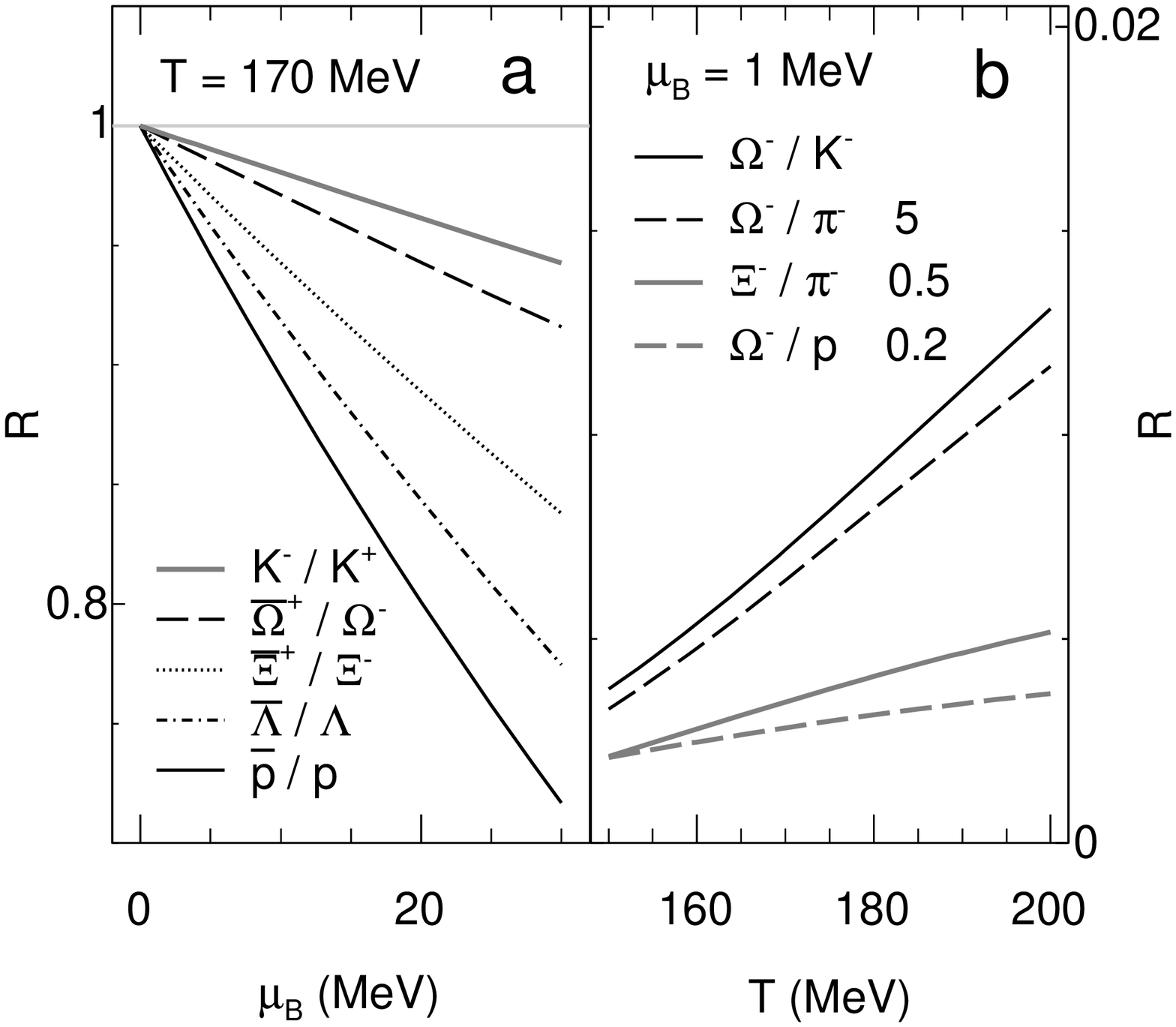}
\end{minipage}
\begin{minipage}[b]{0.34\linewidth}
\centering
\includegraphics*[width=\linewidth]{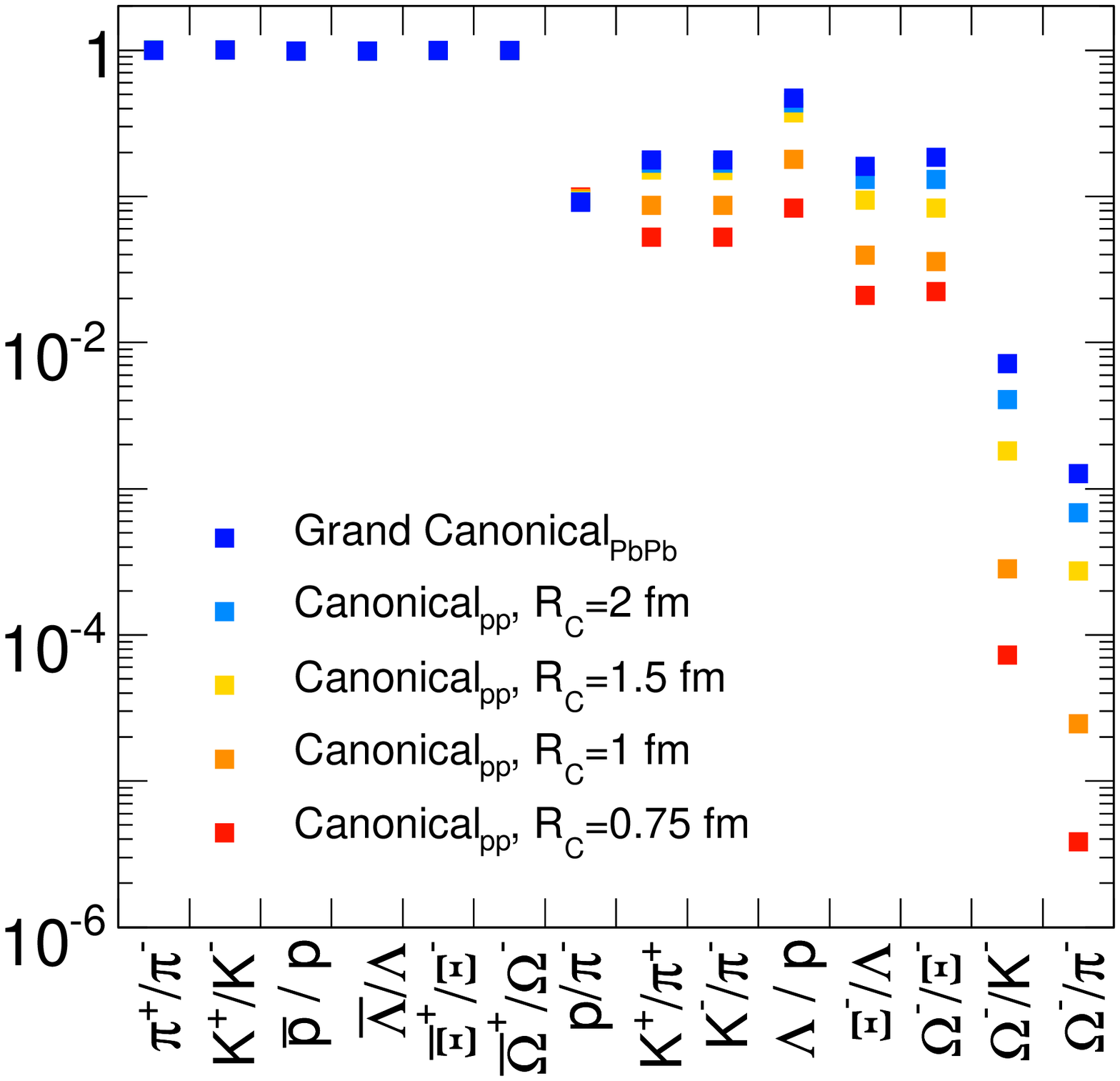}
\end{minipage}\hfill
\caption{\label{fig:Kraus-fig1}
  Left: Ratios $R$ of particles with unequal strangeness content as a function 
  of $\mu_B$ for $T=170$~MeV (left) and as a function of $T$ for $\mu_B=1$~MeV 
  (right).\\
  Middle: Antiparticle/particle ratios $R$ as a function of $\mu_B$ for 
  $T=170$~MeV (left) (the horizontal line at 1 is meant to guide the eye).
  Particle ratios $R$ involving hyperons as a function of  $T$ for $\mu_B=1$~MeV
  (right).\\
  Right: Ratios $R$ of particles in the grand-canonical ensemble and with 
  suppressed strange-particle phase-space in different canonical volumes 
  indicated by the spherical radius $R_C$, calculated at $\mu_B=1$~MeV and 
  $T=170$~MeV.} 
\end{figure}

For the given thermal conditions, particle ratios in central Pb-Pb collisions 
were calculated; numerical values are given in reference~\cite{Cleymans:2006xj}.
As soon as experimental results become available, the extrapolation can be 
cross-checked with particle ratios that exhibit a large sensitivity to the 
thermal parameters. 
The ratios shown in figure~\ref{fig:Kraus-fig1} (left) hardly vary over a broad 
range of $T$ and $\mu_B$. 
This feature can be used to investigate the validity of the statistical model 
at LHC: 
Especially the prediction for the $K/\pi$ ratio is limited to a narrow range. 
It would be hard to reconcile experimental results outside of this band with 
the statistical model.

Antiparticle over particle ratios, on the other hand, strongly depend on $\mu_B$
(figure~\ref{fig:Kraus-fig1}, middle panel). 
Most of all, the $\bar{p}/p$ ratio almost directly translates to the baryon 
chemical potential, since the $T$ dependence is very weak. 
Better suited for the temperature determination are ratios with large mass 
differences, i.e. $\Omega/\pi$ and $\Omega/K$, which increase in the studied 
range by 25\% per 10~MeV change in $T$. 
The astonishing similarity between $K$ and $\pi$ in this respect is caused by 
the huge contribution of 75\% from resonance decays to pions for the given 
thermal conditions~\cite{Kraus:2006yb}.

In collisions of smaller systems, the strange-particle phase-space exhibits a 
suppression beyond the expected canonical suppression. 
A modification of the statistical model is proposed in references~\cite{%
  Kraus:2005ad,Kraus:2007hf}, which is based on the assumption that strangeness 
conservation is maintained in correlated sub-volumes of the fireball. 
The size of these clusters, which could be smaller than the volume defined by 
all hadrons, was estimated from relative strangeness production in collisions 
of small systems at top SPS and RHIC energy. 
The radius $R_C$ of a spherical cluster is of the order of 1~-~2~fm and shows 
only a weak energy dependence.
Additionally it is not clear at which stage of the interaction the strangeness 
abundance is formed. 
Possibly the early, dense phase is crucial, so the cluster size should be the 
same at RHIC and LHC, or, on the contrary, the total number of particles at the 
late stage of hadronisation is relevant; thus $R_C$ should increase as the 
multiplicity will increase with colliding energy.

Instead of precise predictions as shown for Pb-Pb collisions, the correlation 
volume will be extracted from measurement. 
As displayed in figure~\ref{fig:Kraus-fig1} (right), especially the $\Omega/\pi$
ratio varies over orders of magnitude in a reasonable range of the correlation 
length. 
This allows for a good estimate of the cluster size which will give us more 
insight into the mechanism of strangeness production.

\subsection{Universal behavior of baryons and mesons' transverse momentum 
  distributions in the framework of percolation of strings}
\label{s:Cunqueiro}

{\em L. Cunqueiro, J. Dias de Deus, E. G. Ferreiro and C. Pajares}
\vskip 0.5cm

The clustering of color sources~\cite{Armesto:1996kt} reduces the average 
multiplicity and enhances the average $\langle p_T\rangle$ of an event in a 
factor $F(\eta)$ with respect to those resulting from pure superposition of 
strings:
\begin{equation}
\langle\mu\rangle=N_sF(\eta) \langle\mu\rangle_{1}, \;
\langle p_T^2\rangle = \langle p_T^2\rangle_1/F(\eta) 
\label{eq:Cunqueiro-eq1} 
\end{equation}
where $N_s$ is the number of strings and 
$F(\eta)=\sqrt{(1-{\rm e}^{-\eta})/\eta}$ is a function of the density of strings
$\eta$~\cite{Braun:2000hd}.
The invariant cross section can be written as a superposition of the transverse 
momentum distributions of each cluster, $f(x,p_T)$ (Schwinger formula for the 
decay of a cluster), weighted with the distribution of the different tension of 
the clusters, $W(x)$ ($W(x)$ is the gamma function whose width is proportional 
to $1/k$ where $k$ is a determined function of $\eta$ related to the measured 
dynamical transverse momentum and multiplicity fluctuations)~\cite{%
  DiasdeDeus:2003ei,Pajares:2005kk,Ferreiro:2003dw,Cunqueiro:2005hx}:
\begin{equation}
\fl
\frac{{\rm d}N}{{\rm d}p_T^2{\rm d}y}=
\int_0^\infty {\rm d}x\, W(x) f(p_{T},x)=
\frac{{\rm d}N}{{\rm d}y}\frac{k-1}{k}\frac{1}{\langle p_T^2\rangle_{1i}}F(\eta)
\frac{1}{(1+\frac{F(\eta)p_T^2}{k\langle p_T^2\rangle_{1i}})^{k}}.
\label{eq:Cunqueiro-eq2} 
\end{equation}

\begin{figure}
\begin{minipage}[t]{0.49\linewidth}
\centerline{\includegraphics[width=7cm,height=6cm]{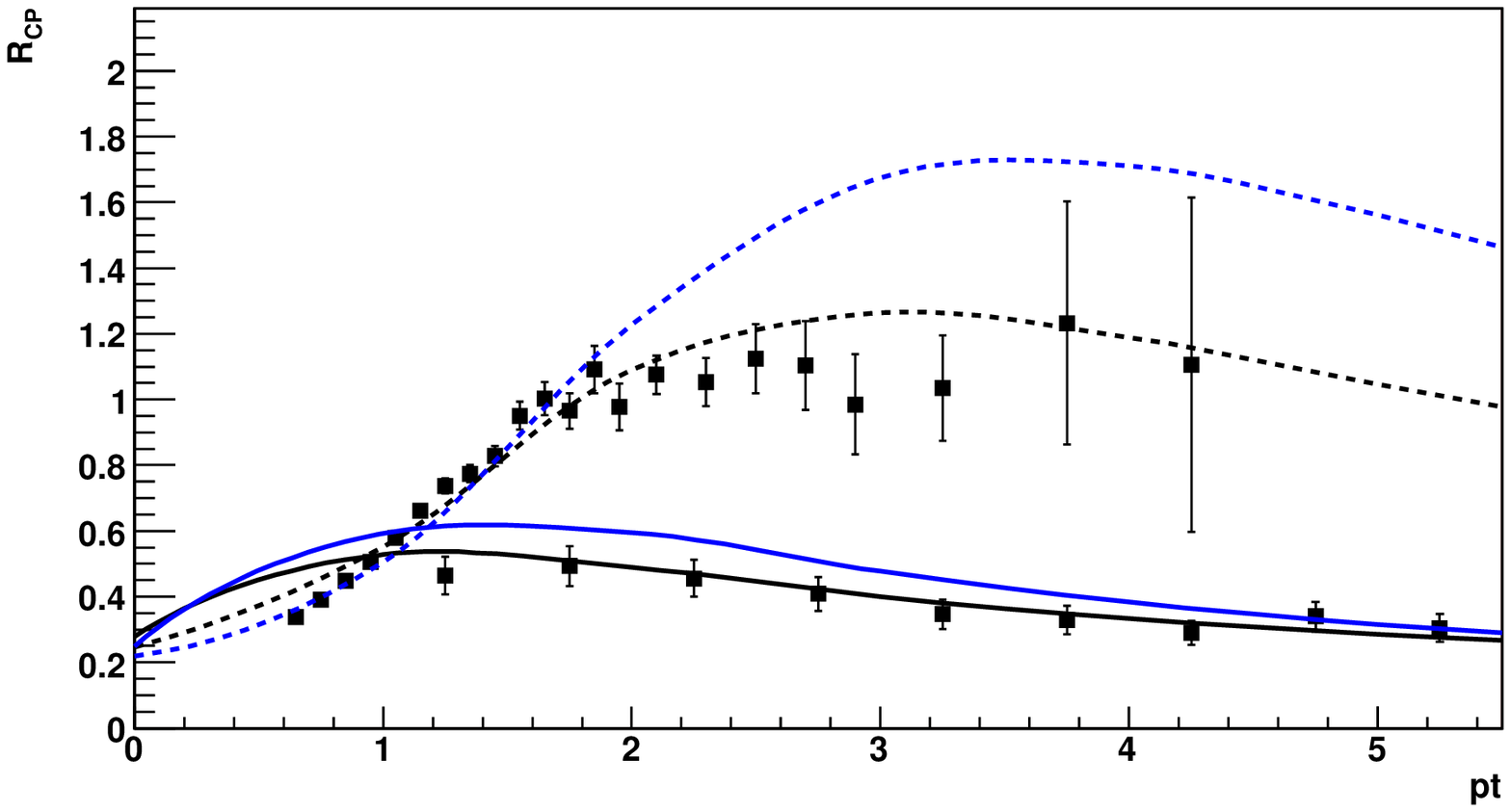}}
\end{minipage} 
\hfill
\begin{minipage}[t]{0.49\linewidth}
\centerline{\includegraphics[width=7cm,height=6cm]{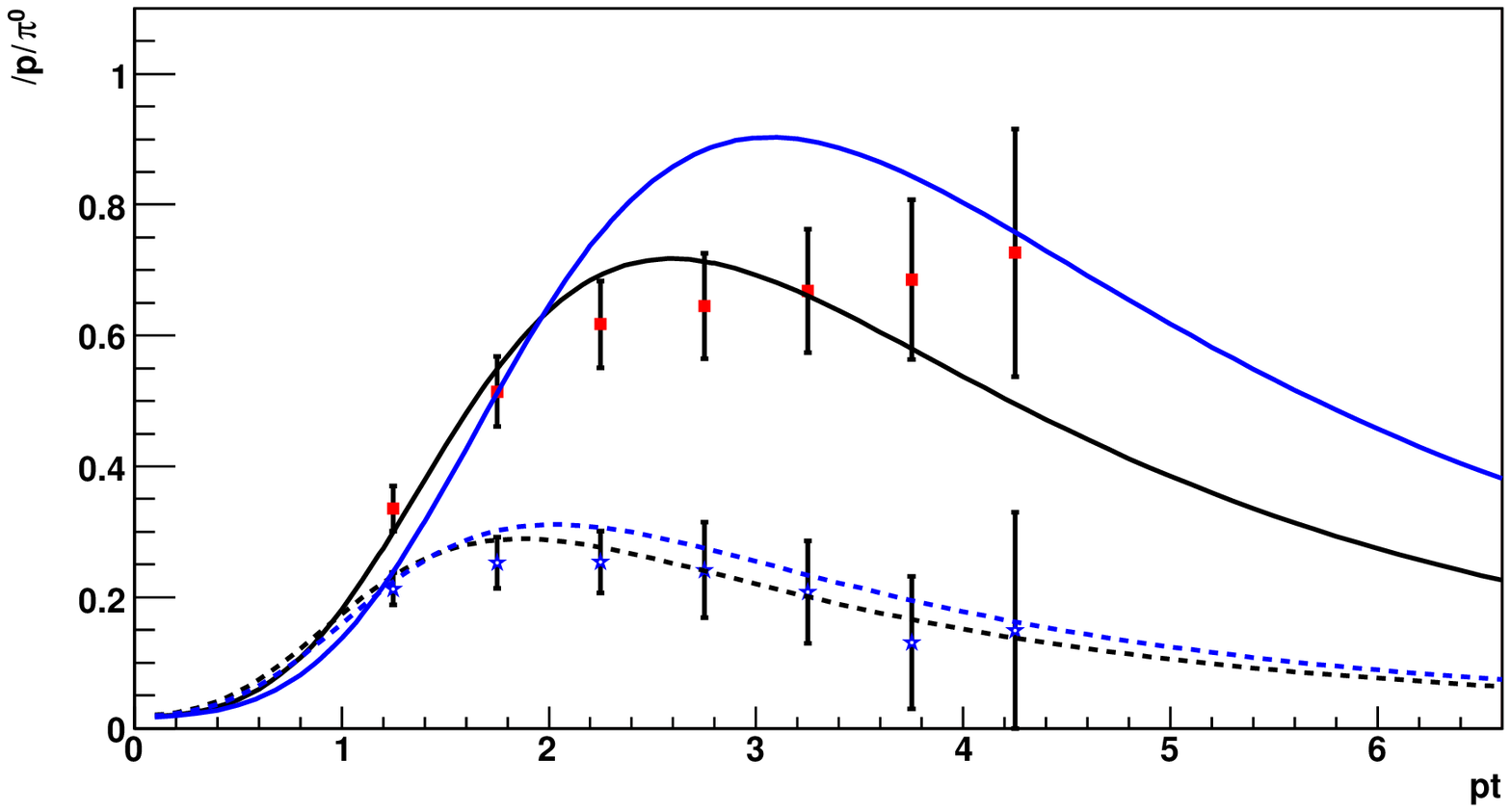}}
\end{minipage} 
\caption{Left: $R_{CP}$ for neutral pions (solid) and antiprotons (dashed). 
  Right: ${\bar p}$ to $\pi^0$ ratio for the centrality bins 0-10\% (solid) and 
  60-92\% (dashed). RHIC results in black and LHC predictions in blue.}
\label{fig:Cunqueiro-fig1}
\end{figure}

\begin{figure}
\begin{minipage}[t]{0.49\linewidth}
\centerline{\includegraphics[width=7cm,height=6cm]{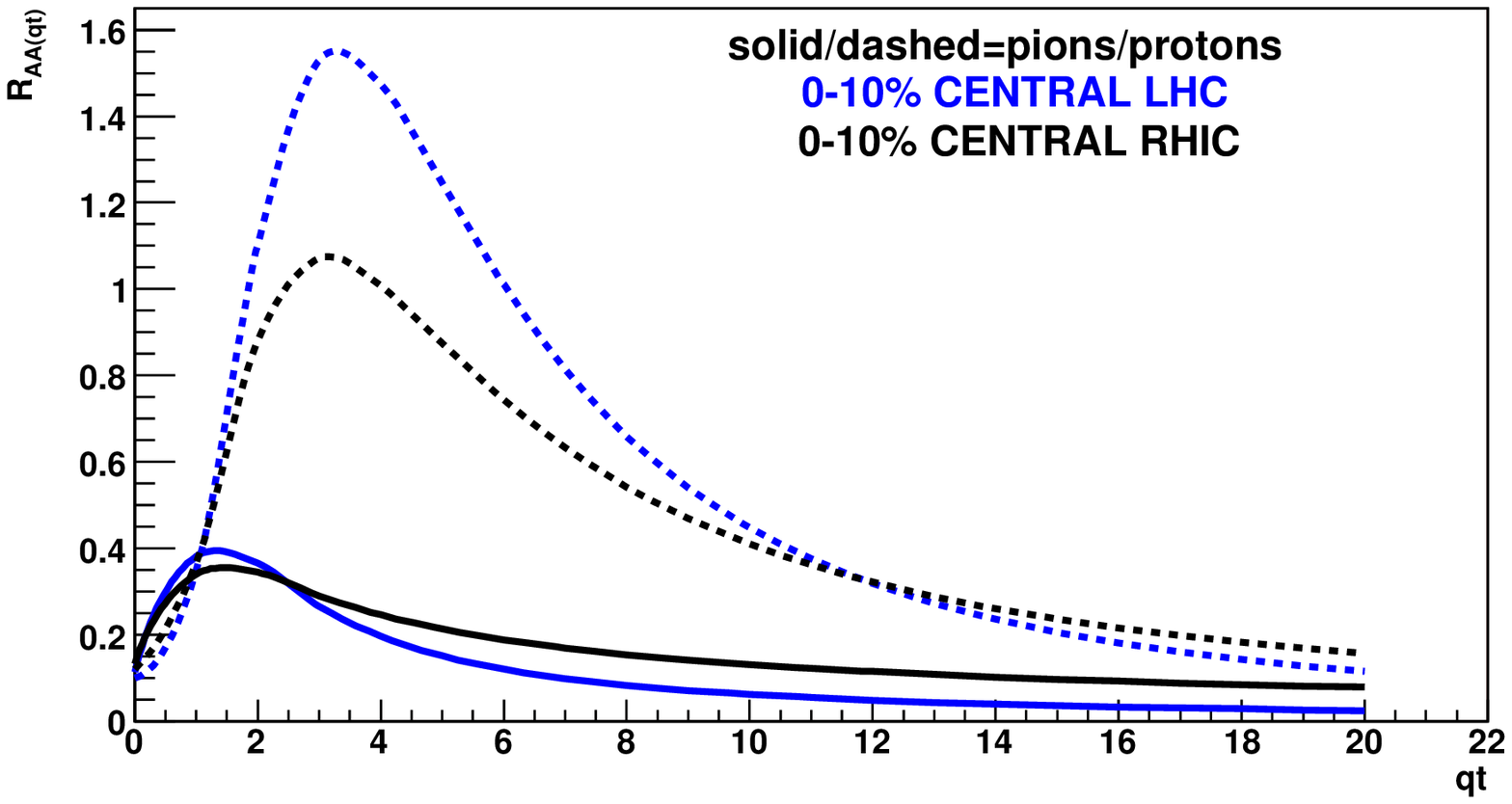}}
\end{minipage} 
\hfill
\begin{minipage}[t]{0.49\linewidth}
\centerline{\includegraphics[width=7cm,height=6cm]{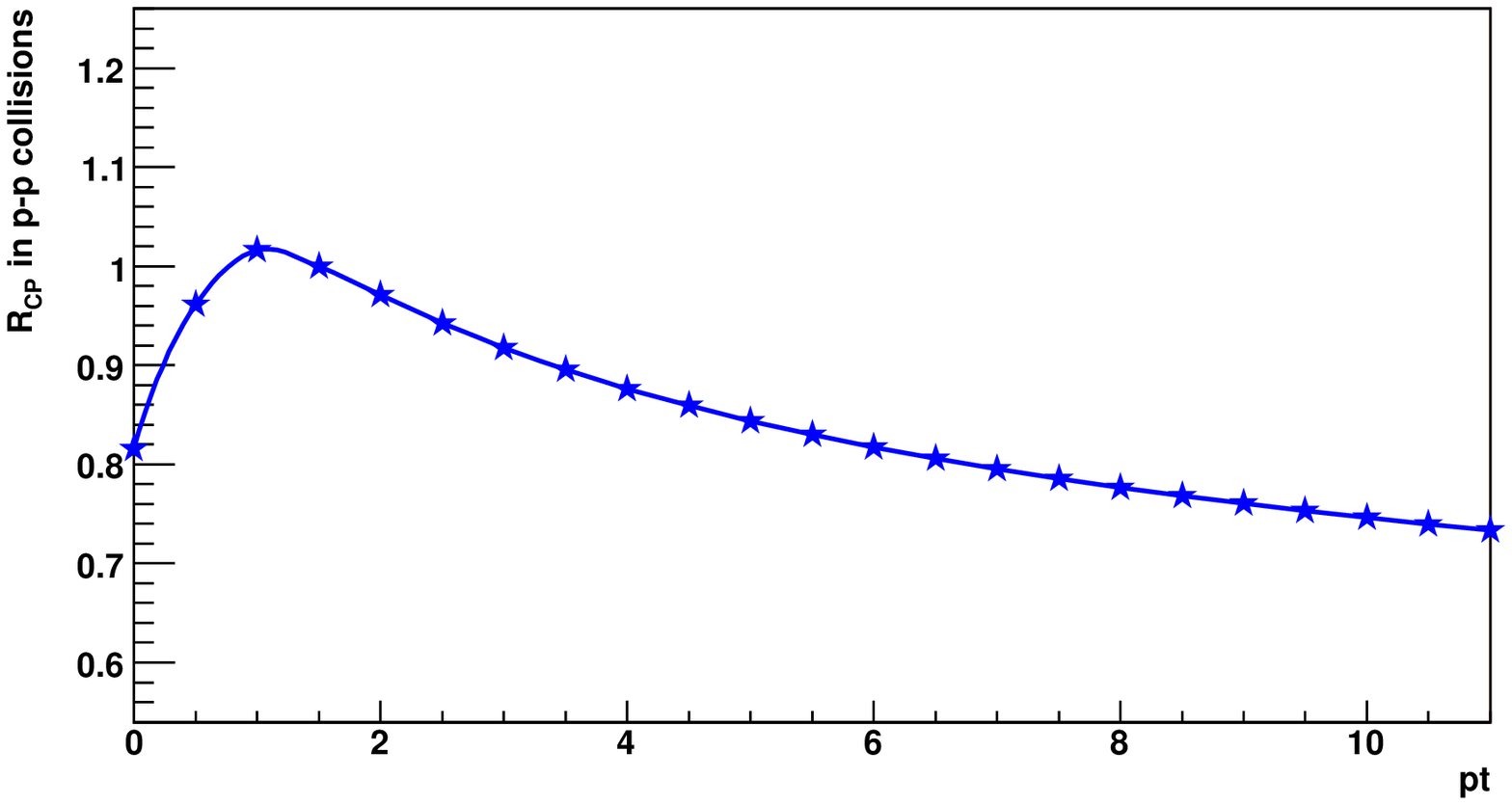}}
\end{minipage} 
\caption{Left: Nuclear modification factor for $\pi^0$ (solid) and ${\bar p}$ 
  (dashed) for 0-10\% central events, RHIC results in black and LHC predictions 
  in blue. Right: $R_{CP}$ for pions in $pp$ collisions at LHC.}
\label{fig:Cunqueiro-fig2}
\end{figure}

For (anti)baryons, equation (\ref{eq:Cunqueiro-eq1}) must be changed to 
$\langle\mu_{\bar B}\rangle=N_s^{1+\alpha}F(\eta_{\bar B})\langle\mu_{1\bar B}\rangle$
to take into account that baryons are enhanced over mesons in the fragmentation 
of a high density cluster. 
The parameter $\alpha=0.09$ is fixed from the experimental dependence of 
${\bar p}/\pi$ on $N_{\rm part}$.
The (anti)baryons probe  higher densities than mesons, $\eta_B=N_s^{\alpha}\eta$. 
On the other hand, from the constituent counting rules applied to the high 
$p_T$ behavior we deduce that for baryons $k_B=k(\eta_B)+1$.
In figure~\ref{fig:Cunqueiro-fig1}, we show the ratios $R_{CP}$ and 
${\bar p}/\pi^0$ defined as usual, compared to RHIC experimental data for pions 
and antiprotons together with the LHC predictions. 
In figure~\ref{fig:Cunqueiro-fig2} left we show the nuclear modification factor 
$R_{\A\A}$ for pions and protons for central collisions at RHIC. 
LHC predictions are also shown.
We note that $pp$ collisions at LHC energies will reach enough string density
for nuclear-like effects to occur. 
In this respect, in figure~\ref{fig:Cunqueiro-fig2} right, we show the ratio 
$R_{CP}$ for $pp\to \pi X $ as a function of $p_T$, where the denominator is 
given by the minimum bias inclusive cross section and the numerator is the 
inclusive cross section corresponding to events with twice multiplicity than 
minimum bias.
According to our formula (\ref{eq:Cunqueiro-eq2}) a suppression at large $p_T$
occurs.


\subsection{Bulk hadron(ratio)s at the LHC-ions}

{\it J. Rafelski  and J. Letessier}

{\small
The expected LHC-heavy ion yields of strange 
and non-strange hadrons, mesons and baryons, are evaluated within the 
statistical hadronization model.
}
%\pacs{12.38.Mh,24.85.+p,25.75.Ld,25.75.-q}
%\maketitle
%\section{Introduction}
\vskip 0.5cm

This summary of our recent work on bulk hadronization in LHC-ion 
interactions is based on methods and ideas presented in \cite{Rafelski:2005jc}, with 
the present update using the results
obtained for strangeness production in \cite{Letessier:2006wn}. This presentation
is more specific regarding the yields in order to allow ``first-day" understanding of the 
mechanisms of hadronization dynamics of the deconfined quark--gluon plasma phase formed in 
most central $\sqrt{s_{\rm NN}}=5520$ GeV Pb--Pb reactions at the LHC. 

Our detailed results rely on SHARE-2.2 suite of programs \cite{Torrieri:2006xi}, which 
have been extensively tested, with several typos, and errors corrected compared to earlier
releases SHARE-1.x  \cite{Torrieri:2004zz}, and SHARE-2.1 . An important feature of the 
SHARE suite of programs is that one can obtain the particle multiplicities
for any consistent {\it mixed} set of    extensive/intensive bulk matter parameters and/or particle yields. 
What `consistency'  means can be understood considering the variables in the Gibbs-Duham relation:
\begin{equation}
PV+E=TS+\sum_a\mu_a N_a, \quad \mu_a =\sum_{i\in a} T\ln\gamma_i + b_i\mu_B+s_i\mu_S\,,
\end{equation}
where the extensive   $V$ (volume), $E\equiv V\epsilon $ (energy),  
$S\equiv V\sigma$ (entropy), $N_i\equiv V\rho_i$ (particle number)
appears along with intensive $P$  (pressure), $T$  (temperature),
 and $\mu_i$ (particle chemical potential).  Aside of the above strict 
constraint, other qualitative constraints arise and thus, in our approach, we allow
for a deviation from prescribed parameter values within a margin of a few percent, to be
chosen in a quasi-fit procedure in order to alleviate inconsistencies in the choices made.

Considering the limited central rapidity experimental coverage, we refer instead of the
total volume $V$ to the range associated with the central rapidity $dV/dy$, 
thus $dS/dy=(dS/dV) (dV/dy)$ is  the entropy (multiplicity) yield per unit of rapidity. 
One can show that $dS/dy$ is conserved in the hydrodynamic expansion of the bulk matter, thus the
final observed entropy (multiplicity) content per unit of rapidity is the outcome of the initial
state entropy production.  

The soft hadron production at LHC-Ion relies on the following  input:

{\bf  I}  The entropy content: $dS/dy\equiv $\,hadron multiplicity --- this is normalizer
of all particle yields for which the predictions most widely vary. The straight line 
extrapolation as function of $\ln \sqrt{s_{\rm NN}}$ implies an increase of $dS/dy$ by only 
a factor 1.65 from RHIC top energy reach $\sqrt{s_{\rm NN}}=200$ GeV to the LHC-ion top energy of 
$\sqrt{s_{\rm NN}}=5520$ GeV. The charged particle yield per unit rapidity is expected, in this case, at
about $h_{\rm ch}=1150$. Since this extrapolation is done based on PHOBOS multiplicity, only partial
K$_s$ weak interaction decay is allowed for. We will also state the corresponding  $h_{\rm ch}^{\rm vis}$ which
is computed assuming acceptance of weak decays akin to the STAR detector.
The entropy content determines up to about 15\% the energy content 
$dE/dy\simeq T_h dS/dy$ which
thus increases, in essence, by the same factor. We note that model differences in hadronization 
temperature $T_h$ which are in the range of up to 20\% impact accordingly the thermal energy content.  

However, one can wonder if the factor 1.65 correctly accounts for the 28-fold reaction
energy increase between RHIC and LHC-ion. The widening  of the particle production 
rapidity window accounts for much of the collision energy increase.  How this widening 
occurs i.e the strength of stopping, determines the central rapidity energy deposition.
We thus consider   in the
second example the case with 3.4-fold increase in entropy/multiplicity   content per unit of rapidity.
This value is fine-tuned such
that  the visible charged hadron yield is identical to the TPC-visible 
charged hadron multiplicity yield   in the chemical equilibrium model,
where the hadronization volume was set  to be $V=6200\,{\rm fm}^3$ (our 3rd table entry). This 
allows to compare  the yields of both models normalized to same hadron yield.

{\bf   II}  The strangeness content $ds/dy=d\bar s/dy$ and/or $(ds/dy)/(dS/dy)=s/S$.  The production 
of strangeness has been evaluated within pQCD, for a given  entropy content.  
The final strangeness yield does not depend 
in a significant way on how the parton entropy content is implemented in the early reaction 
times where thermal distributions are reached (e.g., high $T$, low chemical abundances, 
low $T$, high chemical abundances). This  is so, since strangeness, being a relatively strongly interacting probe, does not convey
a detailed information about the early $\tau < 2$\,fm/c times of the heavy ion  collision. 
For the case of a greater (3.4-times increased)   entropy/multiplicity content, 
the pQCD computation suggests  $s/S\simeq 0.037$ yield, which is 10--15\% above QGP chemical equilibrium, 
the lower entropy variant (extrapolated factor 1.65 increase in multiplicity)  implies for QGP-strangeness 
a small excess above chemical equilibrium,   we will use  $s/S\simeq 0.034$.  For the third
case, the hadron chemical equilibrium, the ratio $s/S=0.025$ results. Thus, strangeness enhancement, where 
it is not washed out  by a lower hadronization temperature, is the salient feature of the non-equilibrium hadronization 
picture we have developed and present here. 

{\bf III} The net  baryon stopping ${d(b-\bar b)/ dy}$ is  unknown, and will be difficult to 
measure.  An extrapolation of the energy per baryon retained per unit of rapidity yields
$E/b\simeq 412\pm20 $\,GeV at LHC. This value is  consistent with the here considered two cases,
when, as an example, we fix $\lambda_q=1.0056$ which  determines  
the baryon and hyperon chemical potentials $\mu_{\rm B}$ and $ \mu_{\rm S}$. We note, in passing, that 
in all the cases considered here, we find  for the baryon asymmetry at LHC $(b-\bar b)/(b+\bar b)\simeq 0.015$,  
which is  6--7 orders of magnitude  larger compared to the conditions prevailing in the early universe.

There are constrains  which  we use to fully determine the 
system properties:\\
1)   For the chemical non-equilibrium hadronization we will use $T_h=140$\,MeV while for
chemical equilibrium we adopt  $T_h=162$\,MeV. Both values are taken from  the study of highest RHIC 
energies. The lower $T$ arises due to supercooling expansion, 
leading to sudden hadronization \cite{Rafelski:2000by},
and thus, we also impose a bias for $E/TS>1$. \\
2)  Strangeness balance $\langle s\rangle = \langle \bar s\rangle$ in the central unit of rapidity.\\
3)  Net charge per net baryon ratio  $Q/b=0.4$ (value in colliding nuclei) is
 implemented. Since the net baryon number is rather small,
the charge asymmetry is for all purposes invisible, the purpose of this exercise is 
to assure  physical consistency and to fix the isospin asymmetry statistical parameter $\lambda_3$.
 
Our results are presented in detail in the table. We note that the
total charged hadron multiplicity will be   a first-day observable at LHC and hence
much of the uncertainty we have in discussing the absolute hadron yields will disappear. 
When comparing 
hadronization models at fixed total hadron yield one sees clear differences in yield pattern:\\
a) Multi-strange hadron yields are, in general, greatly enhanced in our non-equilibrium approach
as compared to yields  assuming chemical equilibrium hadronization, yet single strange yields are often 
similar, since the differences in hadronization (temperature) conditions compensate for the 
strangeness yield enhancement;\\ 
b) The yields of non-strange resonances are, in general, significantly greater in the chemical equilibrium model, 
 due to the higher hadronization temperature.\\ 
c) This suppression is compensated in resonances with single and partial multi-strange content ($\eta,\eta'$).\\
The above differences, already seen at RHIC, are much more striking at LHC, since  the specific strangeness
per entropy yield enhancement is by factor 1.5.  Even the visible K$^+/\pi^+_{\rm vis}$ ratio is increased 
from the RHIC level, to K$^+/\pi^+_{\rm vis}\simeq0.17$ --0.18,  however this enhancement effect 
is much better visible  once weak decays have been 
vetoed in the pion yield, in which case, we predict K$^+/\pi^+\simeq 0.21$.
While the yield of nucleons may be difficult to determine, the measurement of baryon 
resonances  such as $\Delta(1230)$ could help considerably in the characterization of the
baryon yield. 

%\begin{center}
\hspace*{-0.9cm}\begin{tabular}{|c| c | c | c |  }
\hline
$T$[MeV]        &$140^*$& $140^*$& $162^*$    \\
$dV/dy$[ fm$^3$]        &2036  &4187  &6200$^*$    \\
$dS/dy$          &7517  & 15262  &  18021   \\
$dh_{\rm ch}/dy$ &$1150^*$    &$2351 $      &$2430 $     \\
$dh_{\rm ch}^{\rm vis}/dy$&$1351 $  &$2797^* $      &$ 2797$    \\
\hline
$1000\cdot (\lambda_{\rm q,\, s}-1)$     &$5.6^*, \,2.1 $   &$ 5.6^*, \,2.1$    &$ 5.6^*, \,2.0$       \\
$\mu_{\rm B,\, S}$[MeV]     &$2.4, \,0.5$   &$2.3,\, 0.5$    &$2.7,\, 0.6$       \\
$ \gamma_{q,\,s}$      &$1.62,\, 2.42$  &$1.6^*, \,2.6$   &$1^*,\,1^* $    \\
%$\gamma_s $      &2.42  &2.6  &1*    \\
\hline
$s/S$              &$0.034^*$ &$0.037^*$  &$0.025$    \\
$E/b$              &$420^*$ &428 &408  \\
$E/TS$              &1.02 & 1.05& 0.86 \\
$P/E$              & 0.165& 0.164&0.162  \\
$E/V$[MeV/fm$^3$ ]            &530 &538 &400\\
$P$[MeV]   &87&88&65\\
\hline
%$dN/dy$ & & & \\
$p$   &$25/45 $  &$49/95$    &$66/104 $     \\
$ b-\bar b $  &2.6&5.3&6.1 \\ 
$(b+\bar b)/h^-$  &0.335&0.345&0.363 \\ 
$0.1\cdot\pi^\pm$            &49/67  &99/126  &103/126    \\
${\rm K}^\pm$        &94   &207   &175       \\
$\phi $            &14  &33    &23       \\                                %<<<<<<<
$\Lambda$          &19/28 &41/62    &37/50         \\
$\Xi^-$             &4  &9.5   &5.8   \\                     %<<<<<<<<<<<<<<<<<<<<
$\Omega^-$         &0.82 &2.08   &0.98     \\    %<<<<<<<<<<<<<<<<<<<
\hline
$\Delta^{0},\,\Delta^{++} $      
                 &4.7   &9.3   &13.7        \\
$K^*_0(892)$ & 22&48&52 \\
$\eta$ &62&136&127 \\
$\eta'$ & 5.2 &11.8&11.5\\
$\rho$ &36&73 &113\\
$\omega$ &32& 64& 104\\
$f_0$ & 2.7&5.5 & 9.7\\
\hline
K$^+ /\pi^+_{\rm vis}$   & 0.165 &0.176  &  0.148 \\
$\Xi^-/\Lambda_{\rm vis}$ &0.145 & 0.153& 0.116\\
$\Lambda(1520)/\Lambda_{\rm vis}$ &0.043 &0.042 &0.060 \\
$\Xi(1530)^0/\Xi^-$ &0.33 &0.33 &0.36 \\
$\phi/{\rm K}^+$ &0.15 & 0.16&0.13 \\
$K^*_0(892)/K^-$ & 0.236&0.234 &0.301 \\
\hline
\end{tabular}%\vspace*{0.1cm}
%\end{center}
\hspace*{0.5cm}\parbox[t]{6cm}{ \vskip -10.3cm Table: \small LHC predictions, two variants of 
our non-equilibrium hadronization model are shown on left, 
the chemical equilibrium model results are stated for comparison in the right column.  
To obtain results n the first column, we considered an  overall hadron yield
chosen to increase at central rapidity by factor 1.65 compared to PHOBOS results (star `*'  indicates a fixed input value). The chemical equilibrium 
model shown on right is matched in the middle column by assuming a TPC-visible charged hadron yield to be the same, 2797.  These  characteristic
properties along with the entropy content, and chemical conditions at hadronization, are stated in the two top sections 
of the table. In the third section, we show bulk properties
at hadronization, with specific strangeness content prescribed as arising in pQCD computation \cite{Letessier:2006wn}, 
except for the equilibrium model in which case the specific yield $s/S$ is a consequence of the equilibrium assumption. 
One  notes for the equilibrium model that
the energy density and pressure at hadronization  is smaller, which agrees with the greater volume of hadronization
required to obtain the same hadron yield. This is due to particle density scaling roughly 
with $\gamma_q^2 T^3$, the change in  $\gamma_q$ outweighs that in $T$. 
When  we present the hadron yields, we give (separated by slash) the ranges 
with/without weak decays for protons $p$, $\pi$  and $\Lambda$. Clearly the 
properties  of the detector will  impact the uncorrected
%}\\[-0.1cm]
%\noindent {\small
yields. We also note that,
while baryon density in rapidity can vary depending on dynamics of the reaction, the specific
total baryon yield, compared to that of mesons, remains nearly constant and model independent. The difference  
to the equilibrium model is most pronounced in the multi-strange hadron $\Xi,\ \omega$ and $\phi$ yields.
The ratios or resonances with the stable decay product are shown in
the bottom section of the table.}

\section{Correlations at low transverse momentum}
\label{sec:correl}

\subsection{Pion spectra and HBT radii at RHIC and LHC}

{\it Yu. M. Sinyukov, S. V. Akkelin and Iu. A. Karpenko }

{\small
We describe RHIC pion data in central A+A collisions and make
predictions for LHC based on hydro-kinetic model, describing
continuous 4D particle emission, and initial conditions taken from
Color Glass Condensate (CGC) model.
}
\vskip 0.5cm

%\section{Hydro-kinetic approach to A+A collisions for RHIC and LHC energies}
Hydro-kinetic approach to heavy ion collisions proposed in Ref.
\cite{Sinyukov:2002if} accounts for  continuous particle emission from 4D
volume of hydrodynamically expanding fireball as well as back
reaction of the emission  on the fluid dynamics. The approach is
based on the generalized relaxation time approximation  for
relativistic finite expanding systems,
\begin{eqnarray}
\frac{p^{\mu }}{p_{0}}\frac{\partial f(x,p)}{\partial x^{\mu }} =
- \frac{f(x,p)-f^{l.eq.}(x,p)}{\tau_{rel}(x,p)}, \label{boltz-1}
\end{eqnarray}
where $f(x,p)$ is phase-space distribution function (DF),
$f^{(l.eq.)}(x,p)$ is  local equilibrium distribution and
$\tau_{rel}(x,p)$ is relaxation time, $\tau_{rel} (x,p)$ as well
as $f^{l.eq}$ are functional of hydrodynamic variables. Complete
algorithm described in detail in Ref. \cite{AkkSin2} includes:
solution of equations of ideal hydro; calculation of a non local
equilibrium DF and emission function  in the first approximation;
solution of equations for ideal hydro with non-zero
right-hand-side that accounts for conservation laws at the
particle emission  during expansion; calculation of "improved" DF
and emission function; evaluation  of spectra and Bose-Einstein
correlations. Here we present our results for the pion momentum
spectra and interferometry  radii calculated for RHIC and LHC
energies in the first approximation of the hydro-kinetic approach.

\begin{figure}
\begin{narrow}{-0.0in}{-0.0in}
\centering
\begin{minipage}[c]{1.0\textwidth}
 \includegraphics[width=3.2in]{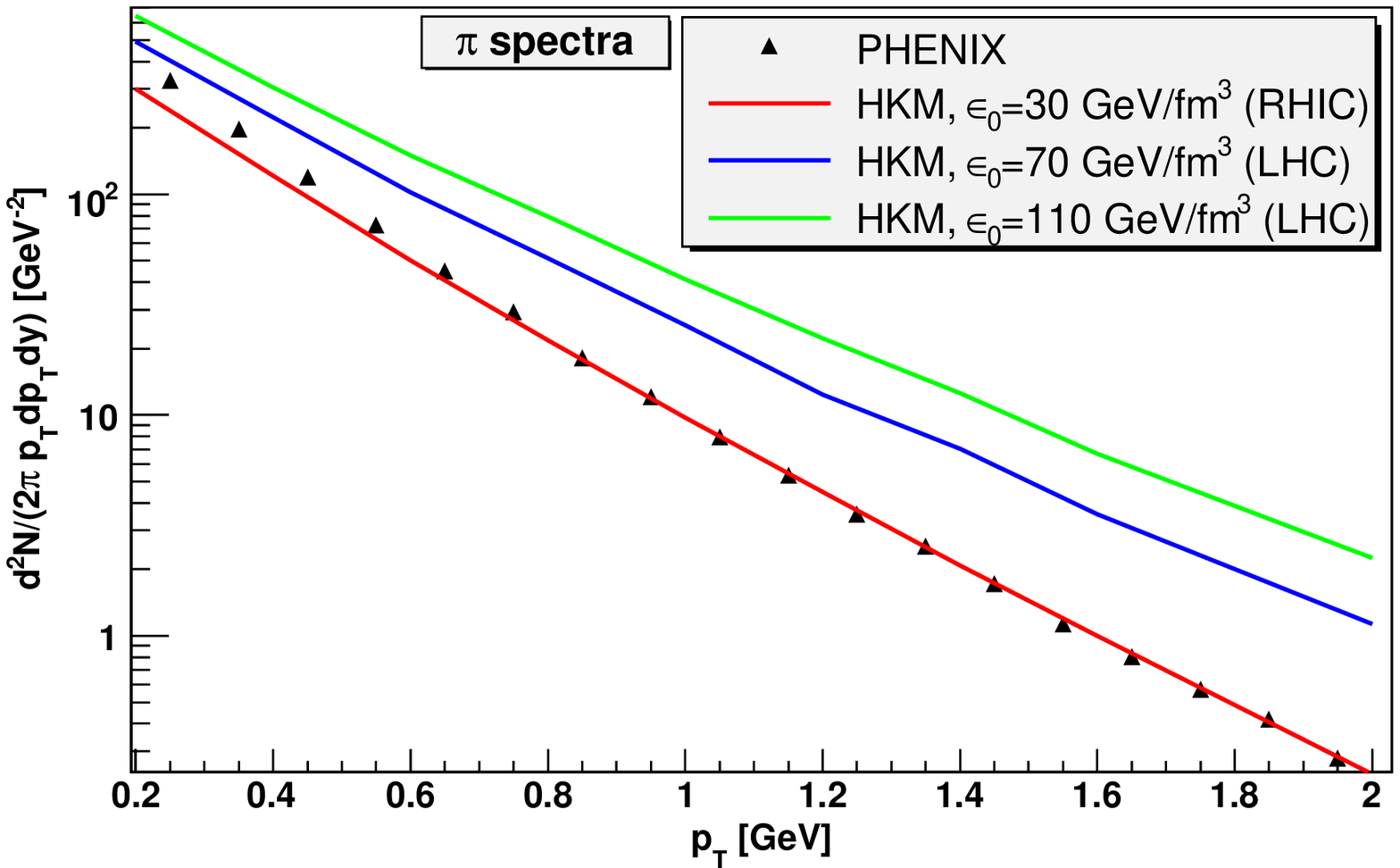}
\hspace{-0.3in}
\includegraphics[width=3.2in]{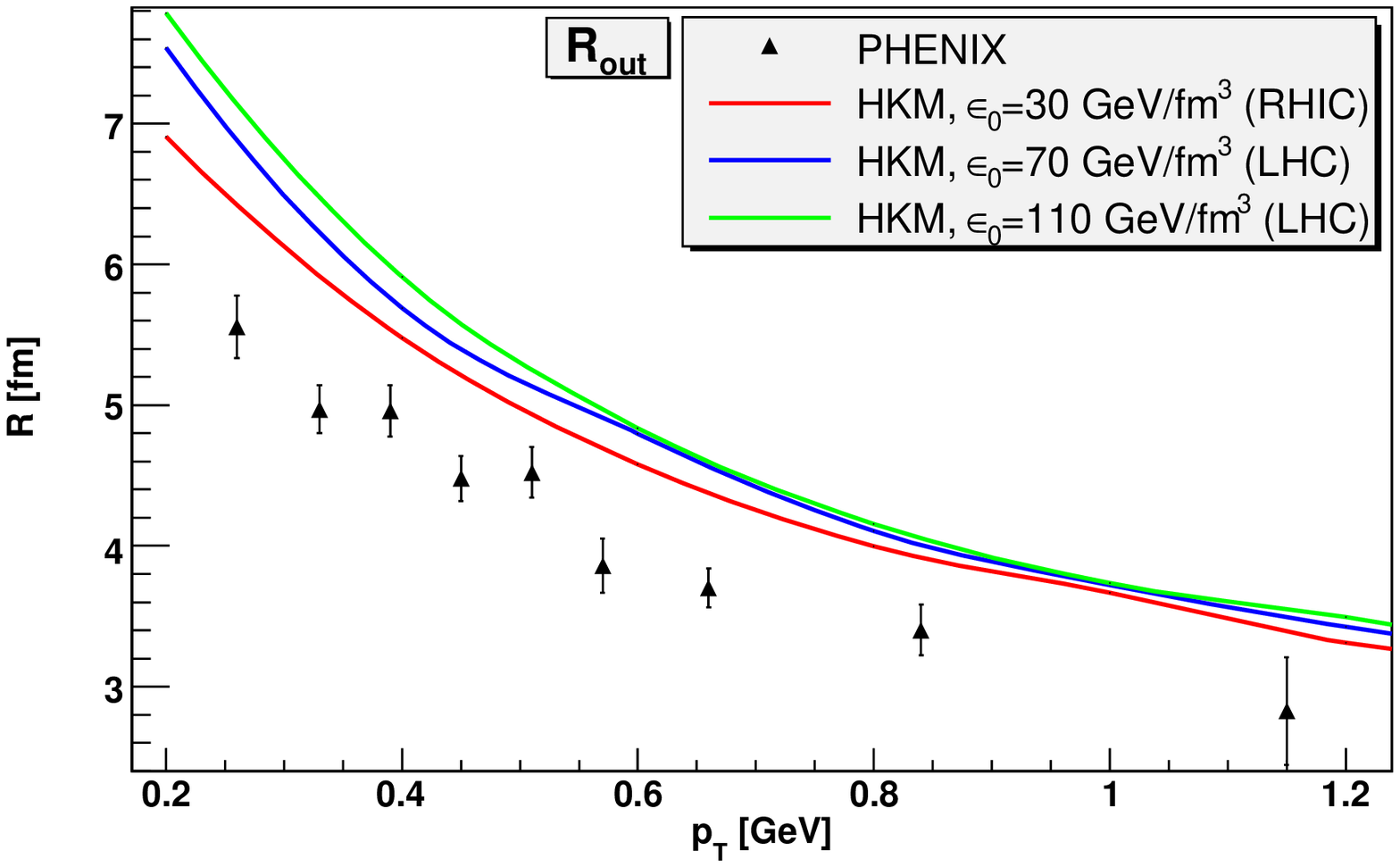}
\end{minipage}

\begin{minipage}[c]{1.0\textwidth}
\hspace{0.0in}
\includegraphics[width=3.2in]{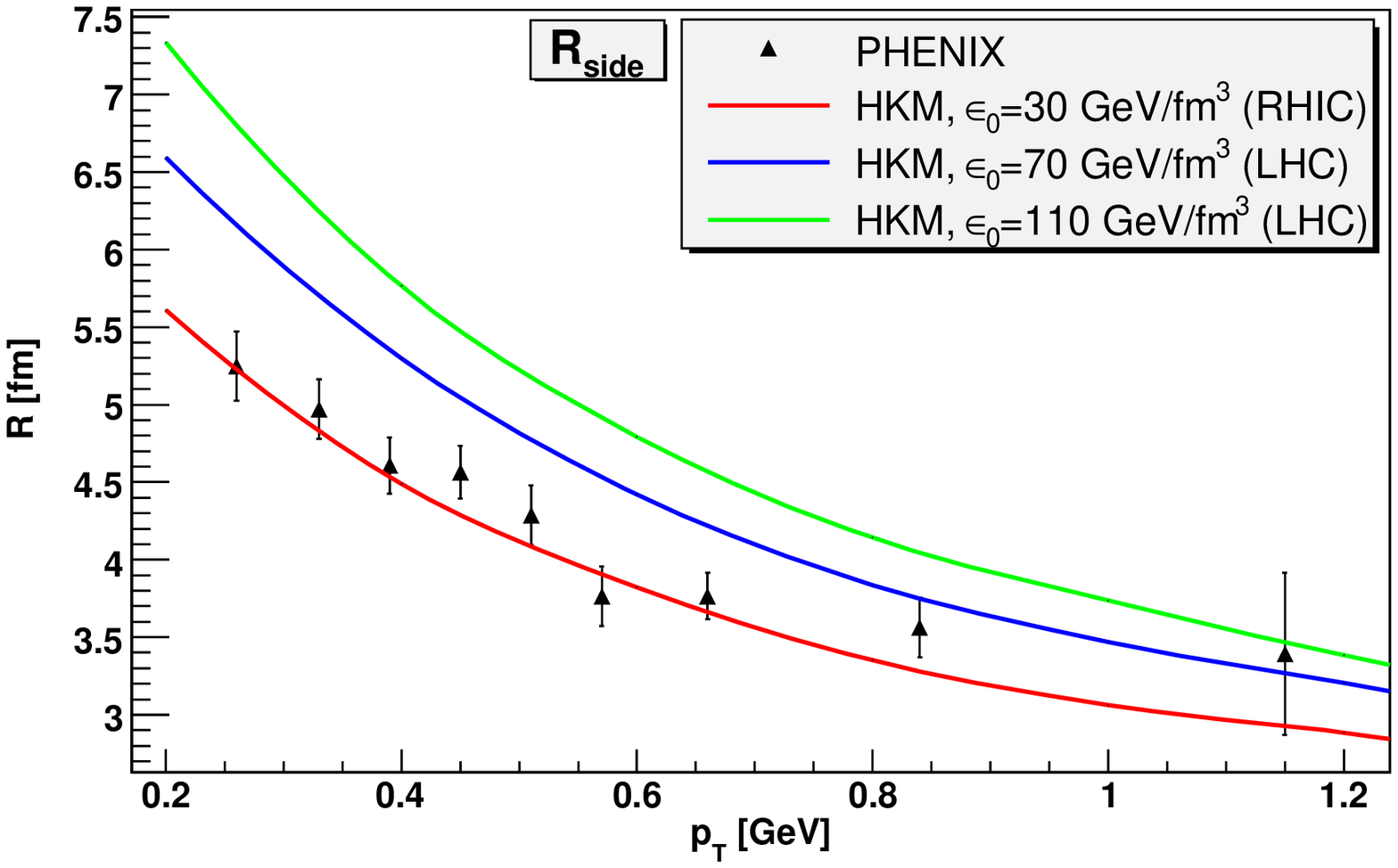}
\hspace{-0.3in}
\includegraphics[width=3.2in]{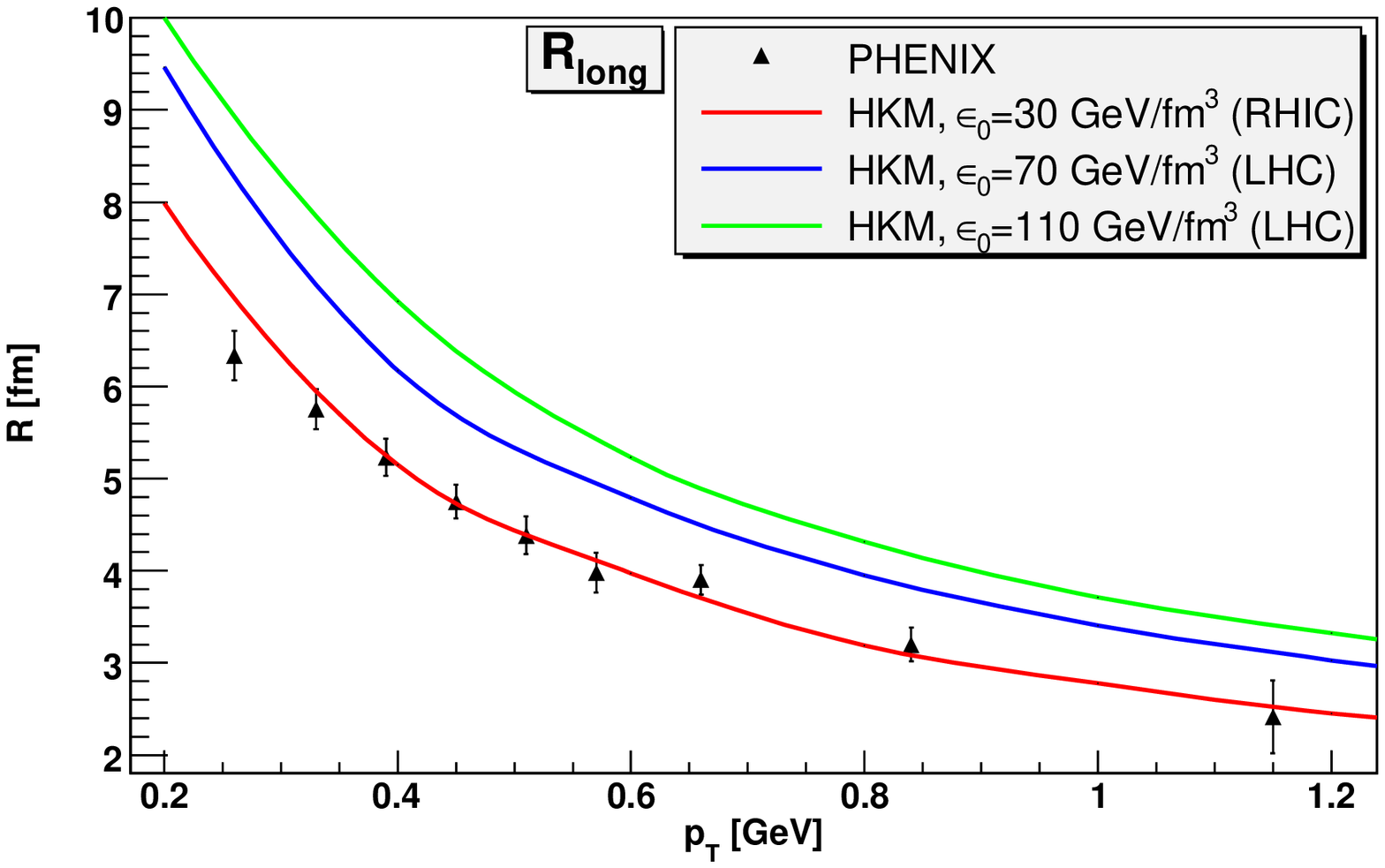}
\vskip -0.3cm
\renewcommand{\captionfont}{\footnotesize \rmfamily}
\renewcommand{\captionlabelfont}{}
\setcaptionwidth{6in}
 \caption{Comparison of the single-particle
momentum spectra of pions  and pion $R_{out}$, $R_{side}$,
$R_{long}$ radii measured by the PHENIX Collaboration for Au+Au
central collisions (HBT radii data were recalculated for $0-5\%$
centrality) at RHIC with the HKM calculations, and HKM predictions
 for Pb+Pb central collisions at LHC. For the sake of
convenience the calculated one-particle spectra  are enhanced in
$1.4$ times.}
\label{figsinyukov}
\end{minipage}
\end{narrow}
\vskip -0.8cm
\end{figure}

%\section{Spectra and correlations of pions at RHIC and LHC
%energies}
 For simulations we utilize ideal fluid  model
\cite{Hirano:2001eu,Hirano:2002ds,Hirano:priv}
and realistic equation of state (EoS) that combines
high temperature  EoS with crossover transition 
\cite{Laine:2006cp} adjusted to the QCD lattice
data and EoS of hadron resonance gas with partial chemical
equilibrium \cite{Hirano:2001eu,Hirano:2002ds,Hirano:priv}.
The gradual disappearance of pions
during the crossover transition to deconfinement and different
intensity of interactions of pions in pure hadronic and "mixed"
phases are taken into account in the  hydro-kinetic model (HKM),
but resonance contribution to pion spectra and interferometry
radii is not taken into account in the present  version of the HKM. We
assume the following initial conditions at  proper time
$\tau_{0}=1$ fm/c for HKM calculations: boost-invariance of a
system in longitudinal direction and cylindrical symmetry with
Gaussian energy density profile in transverse plane. The maximal
energy densities at RHIC, $\epsilon_{0}= 30$ GeV/fm$^3$ and at
LHC, $\epsilon_{0}=70$ GeV/fm$^3$, were calculated from Ref.
\cite{Lappi:2006hq}
in approximation of Bjorken expansion of free
ultrarelativistic partons till $\tau_{0}$ and adjusted for
transverse Gaussian density profile. The (pre-equilibrium) initial
transverse flows at $\tau_{0}$ were estimated assuming again a
free-streaming of partons, with transverse modes distributed
according to CGC picture, from proper time $\approx$ 0.1 fm/c till
$\tau_{0}=1$ fm/c. Finally, we approximate the transverse velocity
profile by $ v_T=\tanh (\alpha \cdot{r_T\over R_T})$ where $\alpha
=0.2$ both for RHIC and LHC energies and we suppose the fitting
Gaussian radius for RHIC top energy, $R_{T}=4.3$, to be the same
for LHC energy. Our results for RHIC and predictions for LHC are
presented in Fig. \ref{figsinyukov}. The relatively small increase of the
interferometry  radii with energy in HKM calculations  is
determined by early (as compare to sharp freeze-out prescription)
emission of hadrons, and also by increase of transverse flow at
LHC caused by longer time of expansion. It is noteworthy that in
the case of EoS related to first order phase transition, 
the satisfactory fitting of the RHIC
HBT data requires non-realistic high initial
transverse flows at $\tau_{0}=1$ fm/c: $\alpha =0.3$.

\subsection{Mach Cones at central LHC
Collisions via MACE}
\label{bauchle}

{\it B. B\"auchle,
H. St\"ocker and
L. P. Csernai}

{\small
The shape of Mach Cones in central lead on lead collisions at
$\sqrt{s_{NN}} = 5.5$~TeV are calculated and discussed using MACE.
}

\subsubsection{Introduction}

After the discovery of ``non-trivial parts'' in three-particle correlations
at RHIC \cite{Ulery:2007zb}, which are compatible with the existence of Mach
cones \cite{Satarov:2005mv}, it is interesting to see how the signal for
Mach cones will look like under the influence of a medium created at the
LHC in PbPb-Collisions. 

Mach cones caused by ultrarelativistic jets going in midrapidity will create
a double-peaked two-particle correlation function ${\rm d}N / {\rm d}(\Delta
\varphi)$. Those peaks are located at $\Delta \varphi = \pi \pm \cos^{-1}
c_{\rm S}$, where $c_{\rm S}$ is the speed of sound as obtained by the
equation of state. The model MACE (``Mach Cones Evolution'') has been
introduced to simulate the propagation of sound waves through a medium and
recognize and evaluate mach cones \cite{Baeuchle:2007qw}.

The medium is calculated without influence of a jet using the hydrodynamical
Particle-in-Cell-method (PIC) \cite{Clare:1986qj}. For the equation of
state, a massless ideal gas is assumed, so that $c_{\rm S} = 1/\sqrt{3}$ and
$\cos^{-1} c_{\rm S} = 0.96$.  The sound waves are propagated independently
of the propagation of the medium and without solving hydrodynamical
equations. Only the velocity field created by PIC is used.  To recognize
collective phenomena, the shape of the region affected by sound waves is
evaluated.

\subsubsection{Correlation functions}

\begin{figure}\begin{center} \includegraphics[width=.8\textwidth]{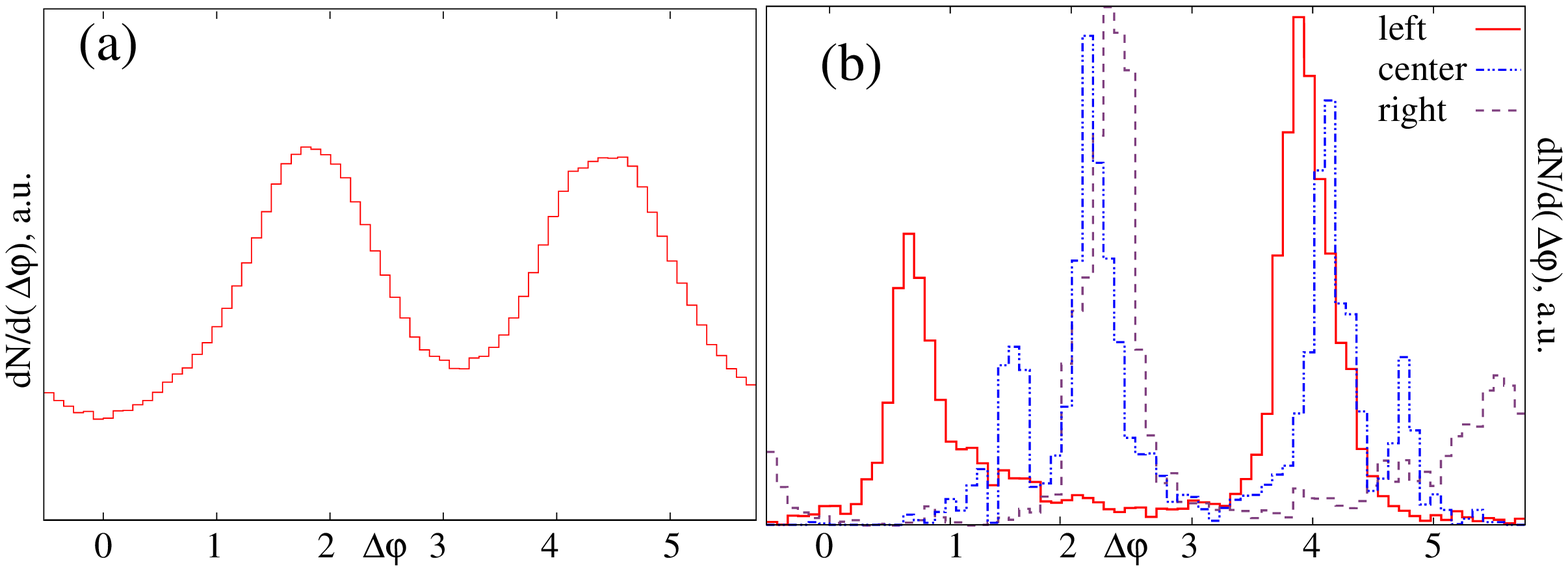}%
\caption{Two-particle-correlation function (away-side-part) for central
PbPb-Collisions at $\sqrt{s_{\rm NN}} = 5.5$~TeV. The peak created by the
forward jet is not calculated. (a): minimum jet bias (see text) with peaks
at $\Delta \varphi \approx \pi \pm 1.2$. (b): Midrapidity jets starting from
a position 70~\% on the way outside left and right of as well as in the
middle.} \label{fig} \end{center} \end{figure}

The correlation functions from the backward peak show a clear double-peaked
structure. The data for arbitrary jet origin and jet direction (minimum jet
bias) is shown in \fref{fig} (a). Here, the peaks are visible at $\Delta
\varphi \approx \pi \pm 1.2$. This corresponds to a speed of sound of
$c_{\rm S} \approx 0.36$. Note that the contributions from the forward
jet are not shown.
Deeper insight into different jet directions do not show a qualitatively
different picture.

\begin{figure}[b]\begin{center}
\includegraphics[width=.5\textwidth]{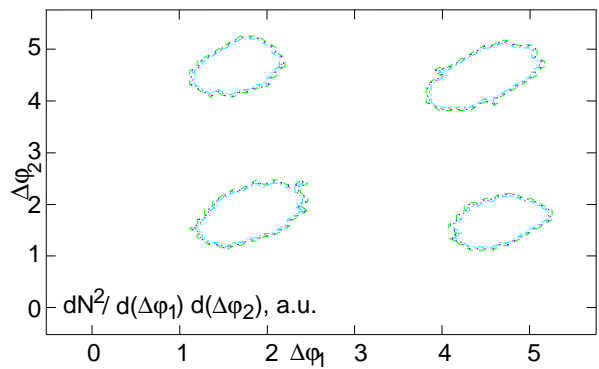}%
\caption{Three-particle-correlation function for the same data as in
\fref{fig}.} \label{3pc} \end{center} \end{figure}

Triggers on the origin of the jet, though, show the dependence of the
correlation function on the position where the jet was created (see
\fref{fig} (b)).
It shows that only the jet coming from the middle of the reaction results in
a symmetric correlation function with peaks at the mach angle $\Delta
\varphi = \pi \pm 0.96$. All other jets result in correlations that have
peaks at different angles, with the deviation getting bigger when going away
from the middle. Therefore, the speed of sound will always appear to be
smaller than it actually is.

\subsubsection{Conclusions}

If sound waves are produced from jet quenching in LHC-Collisions, the
two-particle correlation function will show the expected double-humped
structure in the backward region. The peaks will, though, be further apart
than $\delta(\Delta \varphi) = 2 \cos^{-1} c_{\rm S}$, thus alluding to a
speed of sound smaller than is actually present in the medium. 

The only case in which the true speed of sound can be measured is a
midrapidity jet that creates a symmetric correlation function.

\subsection{Study of Mach Cones in (3+1)d Ideal Hydrodynamics at LHC Energies}
\label{betz}

{\it B. Betz, P. Rau, G. Torrieri, D. Rischke and 
H. St{\"o}cker}

{\small
The energy loss of jets created in heavy--ion collisions shows an
anomalous behaviour of the angular distribution of particles created by
the away-side jet due to the interaction of the jet with
the medium \cite{Gyulassy:2000er,jetquen}. Recent three--particle correlations \cite{Ulery:2007zb,ulery,Adams:2005ph} confirm that a
Mach cone is created. Ideal (3+1)d hydrodynamics \cite{shasta} is used to
study the creation and propagation of such Mach cones under LHC
conditions.
}
\vskip 0.5cm

Jets are one possible probe to study the medium created in a heavy--ion collision. 
They are assumed to be formed in an early stage of the collision and to interact with the hot and dense nuclear matter. 

Experimental results from the Relativistic Heavy Ion Collider (RHIC) show a suppression
of the away--side jet in Au+Au collisions for  high-$p_\bot$ particles 
as compared to the away--side jet in p+p collisions. This effect is commonly
interpreted as jet energy loss or jet quenching \cite{Gyulassy:2000er,jetquen}. 
However, studies including low-$p_\bot$ particles \cite{Ulery:2007zb,ulery,Adams:2005ph} exhibit a double peaked away--side jet. 
Recent three--particle correlations confirm that this pattern is due to a creation of a Mach cone \cite{Ulery:2007zb,ulery,Adams:2005ph}.

The interaction of a jet with the medium is theoretically not well enough understood. Therefore, 
we compare two models of energy loss under LHC conditions. 
We consider a medium with an initial radius of $3.5$~fm and an initial energy density of 
$e_0 = 1.7$~GeV/fm$^3$ that undergoes a Bjorken--like expansion according to a bag--model
equation of state (EoS) with a first--order phase transition from a hadron gas to the
quark--gluon plasma (QGP) with a mixed phase between $e_H = 0.1$~GeV/fm$^3$ and $e_Q = 1.69$~GeV/fm$^3$.

In the first scenario, we implement a jet that completely deposits its energy and momentum 
during a very short time in a $0.25$~fm$^3$ spatial volume. Initially, the jet is located between
$-3.5 \,{\rm fm}<x<-2.5\,{\rm fm},\, |y|<0.25\,{\rm fm},\, |z|<0.25\,{\rm fm}$, has a velocity
of $v_x=0.99$~c and traverses the medium along the x-axis. 
Totally, it deposits an energy of $15$~GeV, no rapidity cut is applied.

In a second step, we study a $15$~GeV jet that gradually deposits its energy and momentum in equal time steps of
$\Delta t = 0.8$~fm/c. As in the first scenario, the jet traverses the medium with a velocity of $v_x=0.99$~c 
along the x--axis.

The hydrodynamic evolution is stopped after a time of $7.2$~fm/c. Using 
a Monte Carlo simulation based on the SHARE program \cite{Torrieri:2004zz}, 
an isochronous freezeout according to the Cooper--Frye formula is performed, 
considering a gas of rhos, pions and etas in the pseudorapidity interval of [-2.3,2.3].

\begin{figure*}[b]
\begin{minipage}[t]{7.75cm}
\includegraphics[width=7.75cm]{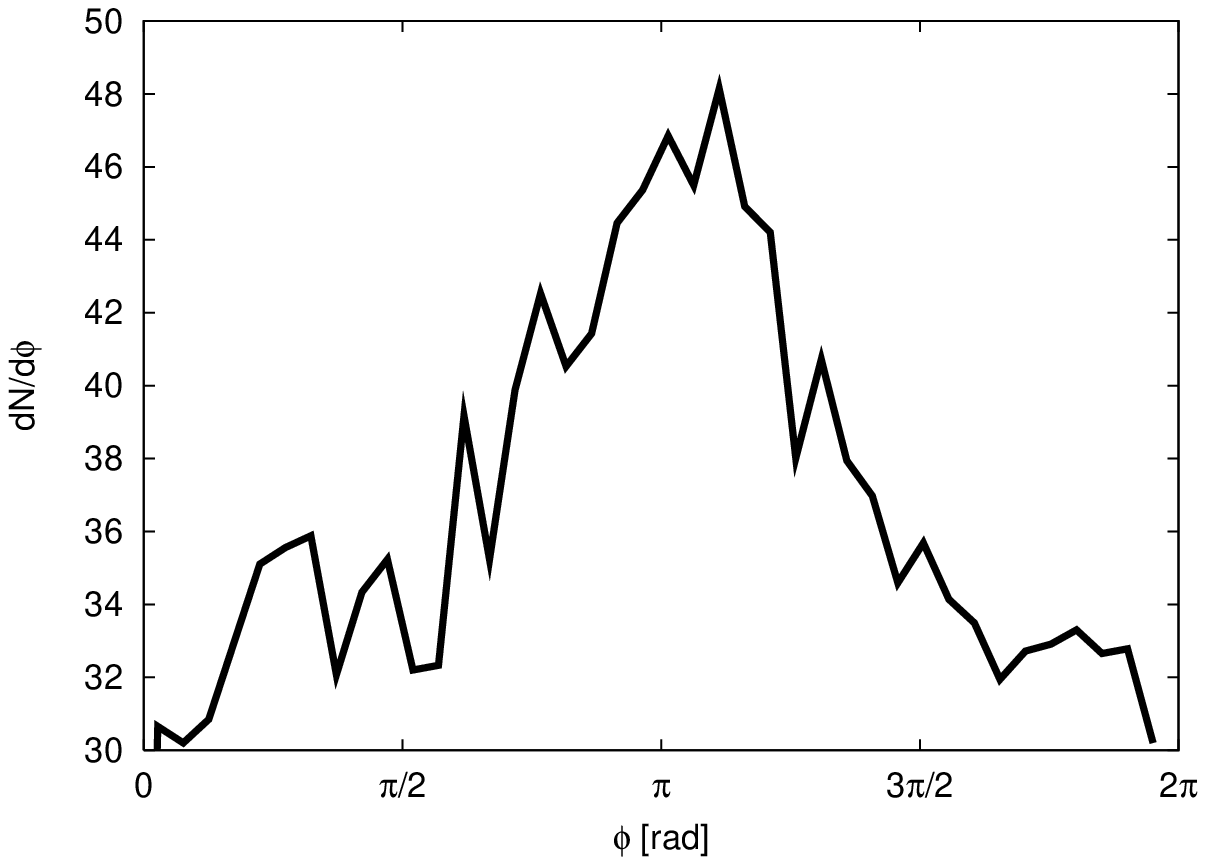}
\end{minipage}
\begin{minipage}[t]{7.75cm}
\includegraphics[width=7.75cm]{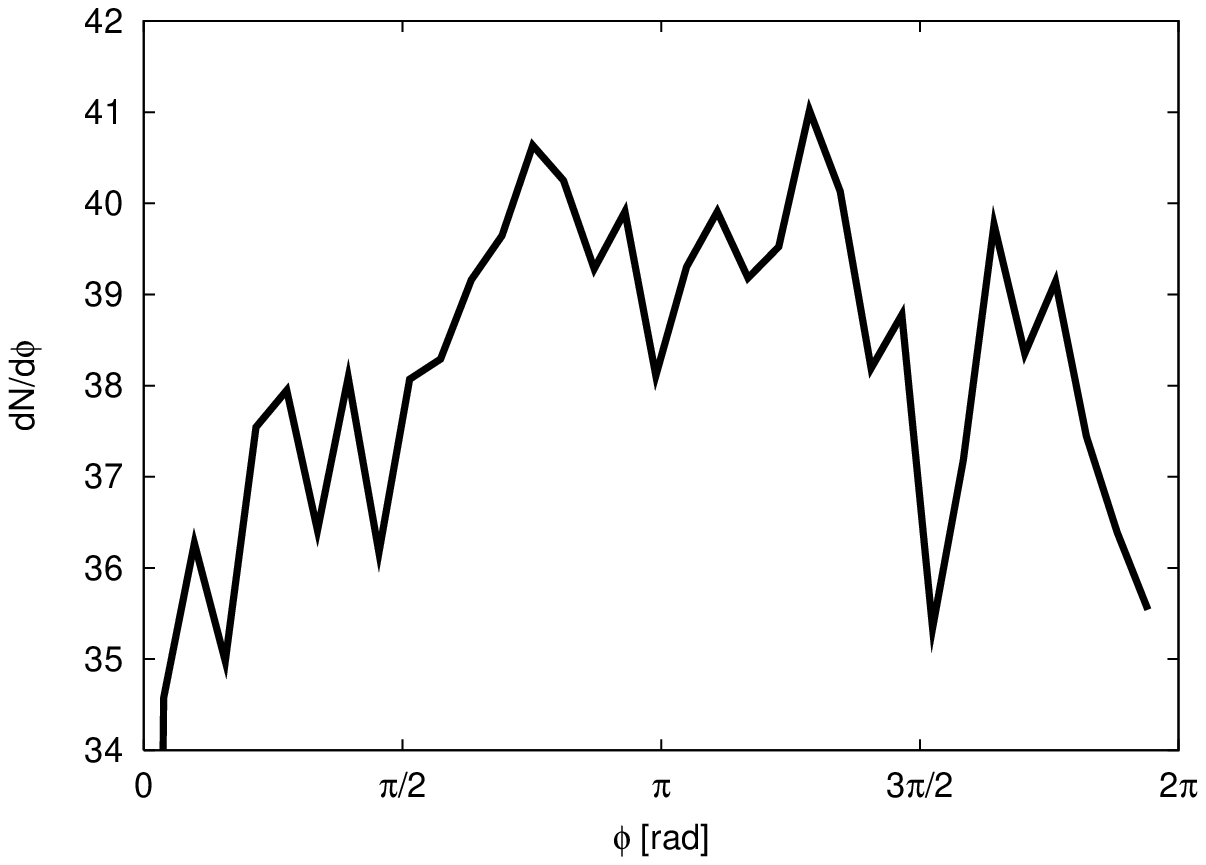}
\end{minipage}
\hspace*{-6cm}
\caption{Angular distribution of particles after isochronous freeze--out of a (3+1)d ideal hydrodynamical
evolution for a jet that deposits its energy and momentum
a) completely within a very short time (left panel) b) in equal timesteps (right panel)
in a medium that undergoes a Bjorken--like expansion according to a bag--model eos.}
\label{fig1betz}
\end{figure*}

Figure \ref{fig1betz} shows the angular distribution of particles for the first (left panel) and second (right panel)
scenario, without any background subtraction. The omitted near--side jet would appear at $\phi = 0$. 

In case of a short--time energy and momentum deposition, a broad away--side distribution 
(left panel) occurs, due to the deposition and dissipation of kinetic energy caused by the jet. 
However, if the jet gradually dispenses its energy and momentum (right panel), 
two maxima appear. This Mach cone--like structure agrees with the recent STAR and PHENIX data \cite{Ulery:2007zb,ulery,Adams:2005ph}.

\subsection{Forward-Backward (F-B) rapidity correlations in a two step scenario}
\label{s:DiasdeDeus}

{\em J. Dias de Deus and J. G. Milhano}

{\small 
We argue that in models where particles are produced in two steps, formation 
first of longitudinal sources (glasma and string models), followed by local 
emission, the Forward-Backward correlation parameter $b$ must have the structure
$b=(\langle n_B\rangle /\langle n_F\rangle) / (1+K/\langle n_F\rangle)$ where 
$\langle n_B\rangle (\langle n_F\rangle )$ is the multiplicity in the backward 
(forward) rapidity window and $1/K$ is the (centrality and energy dependent) 
normalized variance of the number of sources.
}
\vskip 0.5cm

Two-step scenario models  for particle production are based on: 
\begin{enumerate}
\item creation of extended objects in rapidity (glasma longitudinal colour 
  fields or coloured strings); followed by 
\item local emission of particles.
\end{enumerate}
The first step guarantees the presence of F-B correlations due to fluctuations 
in the colour/number of sources, while the second step accounts for local 
effects such as resonances.

The F-B correlation parameter $b$ is defined via
\begin{equation}
\label{eq:DiasdeDeus1}
\langle n_B\rangle_F = a+b n_F \, , \quad b\equiv D^2_{FB} / D^2_{FF} \, ,
\end{equation}
where $D^2$ is  the variance. 
In general, correlations are measured in two rapidity windows separated by a 
rapidity gap so that F-B short range correlations are eliminated. 
In the two-step scenario models we write \cite{DiasdeDeus:1997di,Braun:2000cc,%
  Brogueira:2006yk,Brogueira:2007ub},
\begin{eqnarray}
\label{eq:DiasdeDeus2}
D^2_{FB} \equiv \langle n_F n_B \rangle - \langle n_F \rangle \langle n_B\rangle = {\langle n_F \rangle \langle n_B\rangle \over K} \, ,  \\
\label{eq:DiasdeDeus3}
D^2_{FF} \equiv \langle n^2_F\rangle -\langle  n_F \rangle^2 =  {\langle n_F \rangle^2  \over K} + \langle n_F \rangle \, ,   
\end{eqnarray}
where $1/K$ is the normalized --- e.g., in the  number of elementary collisions 
---  long range fluctuation and depends on centrality, energy and rapidity 
length of the windows. 
We have assumed, for simplicity, that local emission is of Poisson type.

{}From equations~(\ref{eq:DiasdeDeus1}, \ref{eq:DiasdeDeus2}, 
\ref{eq:DiasdeDeus3}) we obtain 
\begin{equation}
\label{eq:DiasdeDeus4}
b= {\langle n_B\rangle /\langle n_F\rangle \over 1+ K/\langle n_F\rangle } \, .
\end{equation}
It should be noticed that $b$ may be larger than 1, and that a Colour Glass Condensate (CGC) model calculation \cite{Armesto:2007ia} shows a structure similar to (\ref{eq:DiasdeDeus4}): $b= A [1+B]^{-1}$ (for a discussion on general properties of (\ref{eq:DiasdeDeus4}) and on the CGC model, see \cite{Brogueira:2007ub}).

\begin{figure}[h] 
   \centerline{\includegraphics*[angle=0,width=10cm]{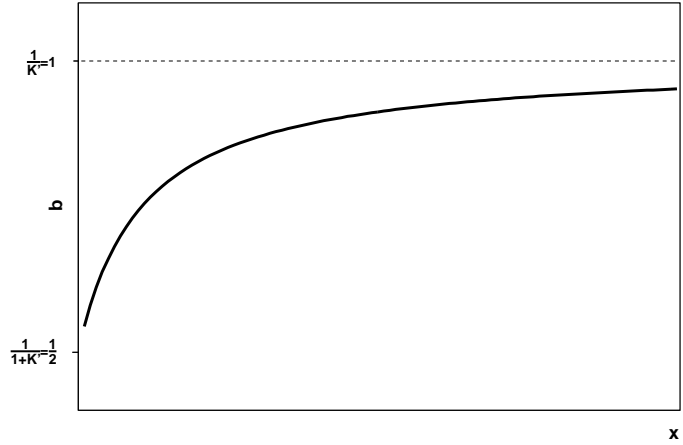}}
   \caption{F-B correlation parameter $b$ (\ref{eq:DiasdeDeus5}) with $K' =1$.}
   \label{fig:DiasdeDeus-fig1}
\end{figure}

A simple way of testing (\ref{eq:DiasdeDeus4}) is by fixing the backward 
rapidity window, or $\langle n_B\rangle$, in the region of high particle density
and move the forward window along the rapidity axis. 
We can rewrite equation (\ref{eq:DiasdeDeus4}) in the form
\begin{equation}
\label{eq:DiasdeDeus5}
b={x\over 1+K'x} \, , 
\end{equation}
where $K' \equiv K/\langle n_B\rangle$ is a constant and 
$x\equiv \langle n_B\rangle /\langle n_F\rangle$. 
In (\ref{eq:DiasdeDeus5}), one has $1< x < \infty$ with the limiting behaviour:
\begin{equation}
x \to 1\, , \quad b\to \frac{1}{1+K'} \, ;\qquad  
x \to \infty \, , \ b\to \frac{1}{K'} \, .
\end{equation}
The behaviour of (\ref{eq:DiasdeDeus5}) is shown in 
figure~\ref{fig:DiasdeDeus-fig1} (drawn for $K' =1$).

A similar curve is obtained for B-F correlations in the backward region of 
rapidity.
Note that in $aA$ collisions, $a\leq A$, the centrality and energy dependence 
of $K'$ is given by~\cite{Brogueira:2006yk,DiasdeDeus:2007wb},
\[
K'\sim a^{1/2} A^{-1/6} {\rm e}^{\lambda Y} \, , 
\]
where $Y$ is the beam rapidity and $\lambda$ a positive parameter. 
In the symmetric situation, $a=A$ and $K'$ \textbf{increases} with centrality 
(and the curve of the figure moves down) while in the asymmetric situation, 
$a=1,2\ll A$ and $K'$ \textbf{decreases} with centrality (and the curve in the 
figure moves up). 
As the energy increases $K'$ increases (and the curve moves down).

\subsection{Cherenkov rings of hadrons}
\label{dremin1}

{\it I. M. Dremin}

{\small
The ring-like structure of inelastic events in heavy ion collisions becomes
pronounced when the condition for the emission of Cherenkov gluons
is fulfilled.}
\vskip 0.5cm

In heavy ion collisions any parton can emit a gluon. On its way through the
nuclear medium the gluon collides with some internal modes. Therefore it
affects the medium as an ``effective'' wave which accounts also for the waves
emitted by other scattering centers. Beside incoherent scattering, there
are processes which can be described as the refraction of the initial wave
along the path of the coherent wave. The Cherenkov effect is the induced
coherent radiation by a set of scattering centers placed on the way of
propagation of the gluon. Considered first for events at very high energies
\cite{Dremin:1979yg,Dremin:1980wx},
the idea about Cherenkov gluons was extended to resonance production
\cite{Dremin:2005an,Dremin:2006vm}.
The refractive index and the forward scattering amplitude 
$F(E, 0^o)$ at energy $E = \sqrt{s}$ are related as

\begin{equation}
\Delta n = {\rm Re}n - 1 = {8\pi N_s {\rm Re}F(E, 0^o)\over E^2}.
\label{eq:dremin_eq1}
\end{equation}
$N_s$ is the density of the scattering centers in the medium.

The necessary condition for Cherenkov radiation is
\begin{equation}
\Delta n > 0\ \ \ {\rm or}\ \ \  {\rm Re}F(E, 0^o) > 0. 
\label{eq:dremin_eq2}
\end{equation}
If these inequalities are satisfied,
Cherenkov gluons are emitted along the cone with half-angle 
$\theta_c$ in the rest
system of the medium determined by $n$:
\begin{equation}
\cos \theta_c = {1\over n}
\label{eq:dremin_eq3}
\end{equation}
{\bf Prediction} The rings of hadrons similar to usual
Cherenkov rings of photons can be observed in the plane perpendicular to
the cone (jet) axis if $n > 1$. 

\noindent
{\bf Proposal} Plot the one-dimensional
pseudorapidity $(\eta = - \ln\tan \theta/2)$
hadron distribution with trigger jet
momentum as $z$-axis. It should have maximum at (\ref{eq:dremin_eq3}). 

This is the best possible one-dimensional projection of the ring. To define
the refractive index in the absence of the theory of nuclear media (for a
simplified approach see \cite{Koch:2005sx})
I prefer to rely on our knowledge of hadronic
reactions. From experiments at comparatively low energies we learn that the
resonances are abundantly produced. They are described by the Breit-Wigner
amplitudes which have a common feature of the positive real part in the
low-mass wing (e.g., see Feynman lectures). Therefore the hadronic
refractive index exceeds 1 in these energy regions. 

At high energies the experiment and dispersion relations indicate on positive
real parts of amplitudes for all hadronic reactions above a very high
threshold. Considering gluons as carriers of strong forces one can assume
that the similar features are typical for their amplitude as well. Then one
should await for two energy regions in which Cherenkov gluons play a role.
Those are either gluons with energies which fit the left wings of
resonances produced in their collisions with internal modes of the medium
or with very high energies over some threshold.

The indications on ``low'' energy effects come from RHIC \cite{Adams:2005ph}
where the
two-bump structure of the angular distribution of hadrons belonging to the
so-called companion (away-side) jet in central heavy-ion collisions has
been observed. It arises as the projection of a ring on its diameter and
provides important information on the properties of the nuclear medium 
\cite{Dremin:2005an,Dremin:2006vm}.
From the distance between peaks the cone half-angle is found to be about
$60^o-70^o$ in the c.m.s. which is equivalent to the target rest system for the
trigger at central rapidities. Derived from it and Eq.(\ref{eq:dremin_eq3})
are the large
refractive index $(n \approx 3)$ and parton density ($\nu \approx 20$
within a nucleon volume) that favor the state of a liquid. The energy loss
$(dE/dx \approx 1 {\rm GeV/fm})$
is moderate and the free path length is of a nuclear size. The
three-particle correlations also favor the ring-like structure.

The indications on high energy effects came from the cosmic ray event
\cite{Apanasenko:1979yh}
at energy about $10^{16}$eV (LHC!) with two ring-like regions. They are formed
at such angles in the target rest system which are equivalent to $60^o-70^o$
and $110^o-120^o$
in c.m.s. It corresponds to the refractive index close to 1
that well fits results of dispersion relations and experiment at these
energies. Such dependence on parton energy shows that the same medium could
be seen as a liquid by rather slow partons and as a gas by very fast ones.

It is crucial for applicability of Eq.(\ref{eq:dremin_eq3}) to define properly the target
rest system. In RHIC experiments the parton-trigger moves in the transverse
direction to the collision axis and, on the average, ``sees'' the target (the
primary fireball) at rest in c.m.s. dealing with rather low $x$ and $Q^2$.
In
the cosmic event the narrow forward ring is produced by fast forward moving
partons (large $x$) which ``see'' the target at rest in the lab. system. At LHC
one can await for both types of Cherenkov gluons produced. Thus, the
hadronic Cherenkov effect can be used as a tool to scan 
$(1/x, Q^2)$-plane and
plot on it the parton densities (see Eq.(\ref{eq:dremin_eq1})) corresponding to its
different regions.

To conclude, the ring-like structure of inelastic processes must be
observed if the gluonic Cherenkov effects are strong enough. The ring
parameters reveal the properties of the nuclear medium and their energy
dependence.

\subsection{Evolution of pion HBT radii 
  from RHIC to LHC -- predictions from ideal hydrodynamics}
\label{heinzhbt}

{\it E. Frodermann, R. Chatterjee and U. Heinz}
\vskip 0.5cm

%\section[Pion HBT radii from RHIC to LHC]{Evolution of pion HBT radii 
%  from RHIC to LHC -- predictions from ideal 
%  hydrodynamics\footnote[7]{$\!\!\!$Authors: E. Frodermann, R. Chatterjee, 
%  U. Heinz, Ohio State University, Columbus, Ohio, USA. Work supported
%  by U.S. DOE, grant DE-FG02-01ER41190 (UH), and an Ohio State University 
%  University Presidential Fellowship (EF).}}

We use the longitudinally boost-invariant relativistic ideal hydrodynamic
code AZHYDRO \cite{Kolb:2000sdd} to predict the expected trends for the 
evolution from RHIC to LHC of the HBT radii at mid-rapidity in central 
$(A{\approx}200){+}(A{\approx}200)$ collisions, as well as that of their 
normalized oscillation amplitudes in non-central collisions. We believe
that these trends may be trustworthy, in spite of the model's failure to 
correctly predict the HBT radii at RHIC \cite{Lisa:2005dd}. The results
shown here are selected from Ref.\,\cite{FCH07}.

Hydrodynamics can not predict the $\sqrt{s}$-dependence
of its own initial conditions, but it relates uniquely
the initial entropy density to the final hadron multiplicity. 
We compute hadron spectra and HBT radii as functions
of final multiplicity, parametrized by the initial peak 
entropy density $s_0$ at thermalization time $\tau_0$ in $b{\,=\,}0$ 
collisions. We 
%%%%%%%%%%%%%%%%%%%%%%%%%%%%%%% Fig. 1 %%%%%%%%%%%%%%%%%%%%%%%%%%%%%%%
\begin{figure}[h]
  \begin{center}
  \includegraphics[bb=7 2 789 607,width=\linewidth]{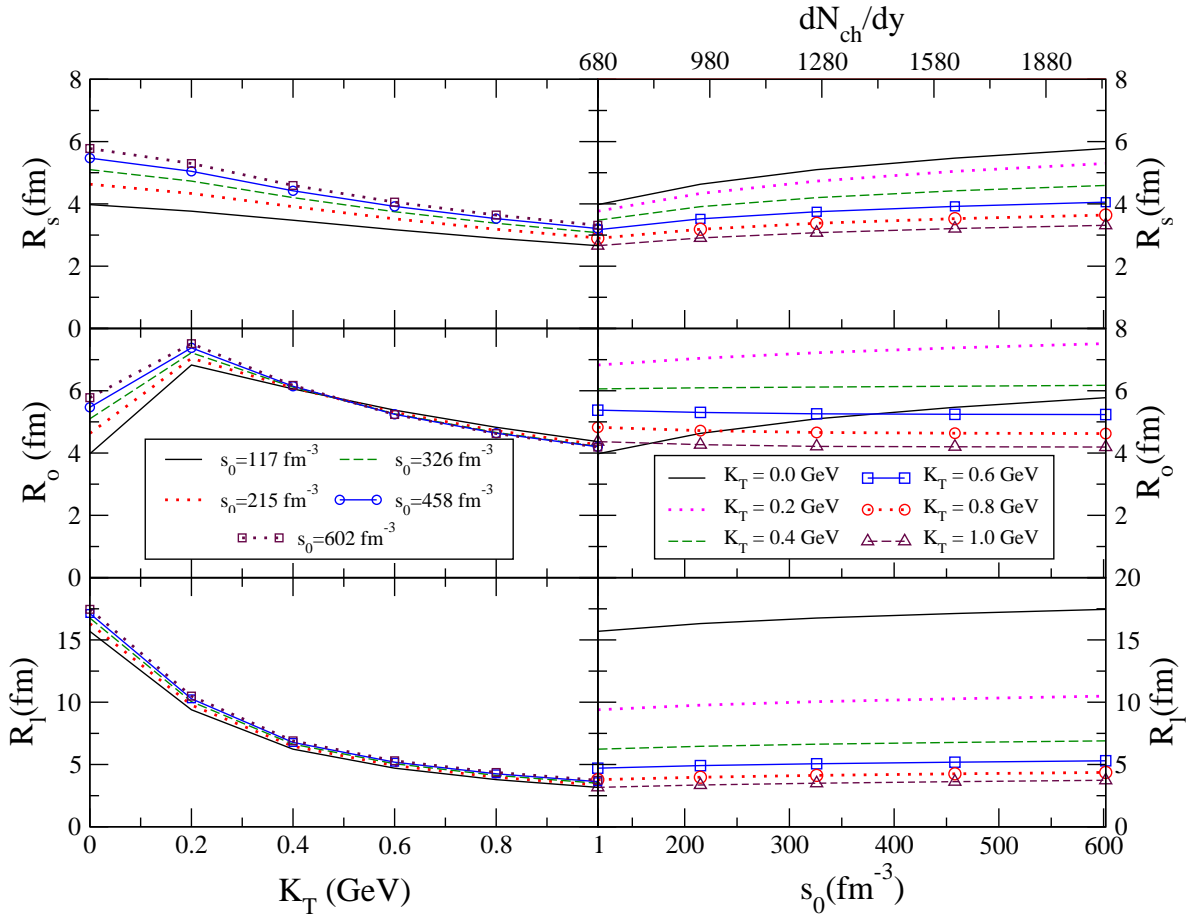}
  \end{center}
   \caption{\label{radiiKT}(Color online)
   Pion HBT radii for central ($b{=}0$) Au+Au collisions as a function of 
   transverse pair momentum $K_T$ (left) and of initial entropy density 
   $s_0$ or final charged multiplicity $\frac{dN_{\mathrm{ch}}}{dy}$ (right).
   For details see \cite{FCH07}. 
   } 
\end{figure}
%%%%%%%%%%%%%%%%%%%%%%%%%%%%%%%%%%%%%%%%%%%%%%%%%%%%%%%%%%%%%%%%%%%%%%
%
hold $T_0\tau_0$ constant (where $T_0 \sim s_0^{1/3}$ is the initial
peak temperature). Our results cover a range from 
$\frac{dN_{\mathrm{ch}}}{dy}{\,=\,}680$ (``RHIC initial conditions'': 
$s_0{\,=\,}117$\,fm$^{-3}$ at $\tau_0=0.6$\,fm/$c$) to 
$\frac{dN_{\mathrm{ch}}}{dy}{\,=\,}2040$ (``LHC initial conditions'':
$s_0{\,=\,}602$\,fm$^{-3}$ at $\tau_0{\,=\,}0.35$\,fm/$c$).

%We use an equation of state with a quark-gluon plasma to hadron resonance
%gas phase transition at $T_\mathrm{c}{\,=\,}164$\,MeV and assume chemical
%decoupling at $T_\mathrm{c}$. Hadron momentum distributions and HBT 
%correlations are assumed to decouple at $T_\mathrm{dec}{\,=\,}100$\,MeV.
%For simplicity, we compute the HBT radii from the space-time variances of 
%the emission function (computed via Cooper-Frye along the decoupling surface)
%instead of doing a Gaussian fit to the two-pion correlation function.
%Since we know that this overestimates the longitudinal radius $R_l$ by 
%20-25\% \cite{Frodermann:2006sp}, all $R_l$ values shown below should be 
%corrected downward by a corresponding factor. 

%%%%%%%%%%%%%%%%%%%%%%%%%%%%%%%%%%%%%%%%%%%%%%%%%%%%%%%%%%%%%%%%%%%%%%%%
\noindent\textbf{\emph{1.~Central collisions:}}
%%%%%%%%%%%%%%%%%%%%%%%%%%%%%%%%%%%%%%%%%%%%%%%%%%%%%%%%%%%%%%%%%%%%%%%%
%
\Fref{radiiKT} shows the pion HBT radii for central Au+Au (Pb+Pb) collisions 
in the $(osl)$ coordinate system \cite{Lisa:2005dd}. Since we computed the 
HBT radii from the space-time variances of the emission function instead of 
doing a Gaussian fit to the two-pion correlation function, all $R_l$ values 
should be corrected downward by about 20\% \cite{Frodermann:2006sp}. We see
no dramatic changes, neither in magnitude nor in $K_T$-dependence, of the 
HBT radii as we increase the multiplicity by up to a factor 3. The largest 
increase (by $\sim30\%$ at low $K_T$) is seen for $R_s$,
while $R_o$
%(which follows $R_s$ at $K_T{\,=\,}0$ by symmetry) 
even slightly decreases at large $K_T$. $R_l$ changes hardly at all. 
The main deficiency 
of hydrodynamic predictions for the HBT radii at RHIC (too weak 
$K_T$-dependence of $R_s$ and $R_o$ and a ratio $R_o/R_s$ much larger 
than 1) is not likely to be resolved at the LHC unless future LHC data 
completely break with the systematic tendencies observed so far 
\cite{Lisa:2005dd}.\\[-3mm]

\begin{figure}[thb]
  \begin{center}
  \includegraphics[width=0.93\linewidth]{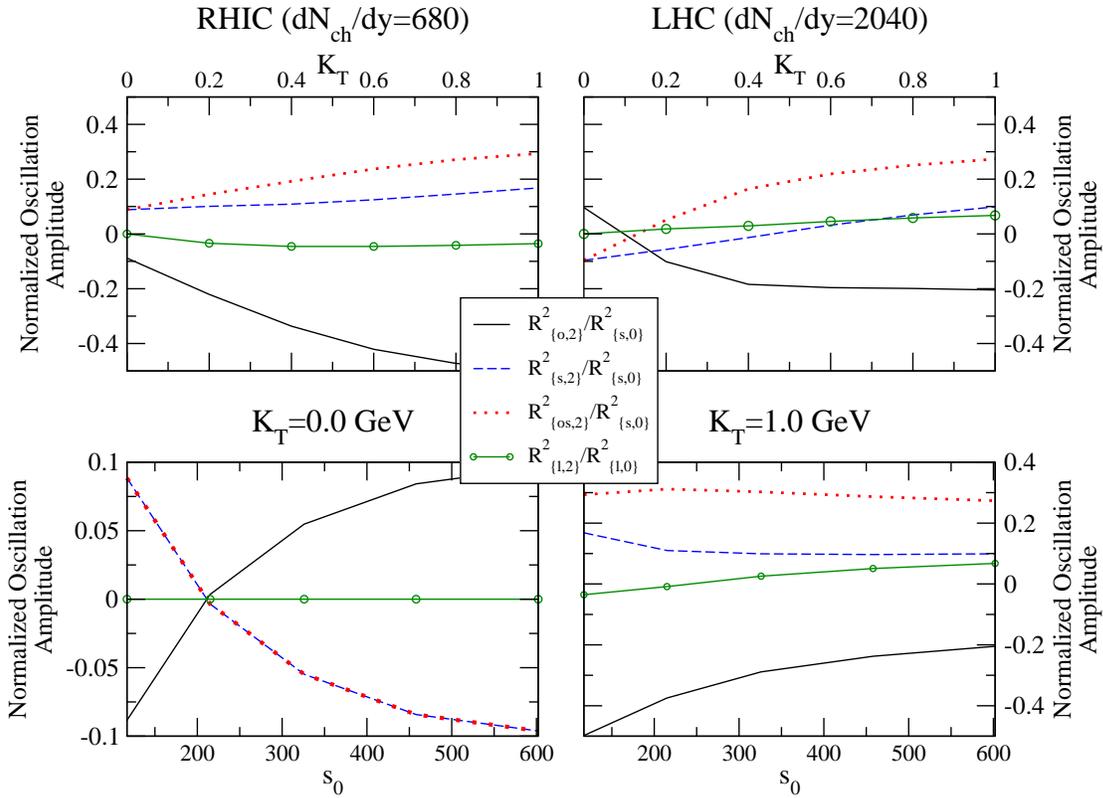}
  \vspace*{-4mm}
  \end{center}
  \caption{\label{osciAmp}(Color online)
   Normalized HBT oscillation amplitudes as a function of $K_T$ at RHIC 
   and LHC (top) and as function of $s_0$ for two values of $K_T$ 
   (bottom).
  }
\end{figure}

%%%%%%%%%%%%%%%%%%%%%%%%%%%%%%%%%%%%%%%%%%%%%%%%%%%%%%%%%%%%%%%%%%%%%%%%
\noindent\textbf{\emph{2.~Non-central collisions:}}
%%%%%%%%%%%%%%%%%%%%%%%%%%%%%%%%%%%%%%%%%%%%%%%%%%%%%%%%%%%%%%%%%%%%%%%%
%
%For RHIC initial conditions, although the magnitudes of the 
%HBT radii in central collisions were not predicted accurately, their 
%normalized oscillation amplitudes at small $K_T$ \cite{Adams:2003zg}
%(which measure the source eccentricity at freeze-out \cite{Retiere:2003kf}) 
%were correctly reproduced \cite{concepts}. Their extrapolation to LHC initial 
%conditions may therefore have predictive power.
\Fref{osciAmp} shows the normalized azimuthal oscillation amplitudes 
\cite{Retiere:2003kf} of the HBT radii for $b{\,=\,}7$\,fm Au+Au
%
%%%%%%%%%%%%%%%%%%%%%%%%%% Fig.2 %%%%%%%%%%%%%%%%%%%%%%%%%%%%%%%%%%%%%%%%%% 
%%%%%%%%%%%%%%%%%%%%%%%%%%%%%%%%%%%%%%%%%%%%%%%%%%%%%%%%%%%%%%%%%%%%%%%%%%%
%
collisions. The dashed line in the lower left panel gives the 
spatial eccentricity of the source at freeze-out \cite{Retiere:2003kf}: 
$\epsilon_x^\mathrm{f.o.} \approx 2\lim_{K_T\to0}
\left(R^2_{s,2}/R^2_{s,0}\right)$. 
The freeze-out eccentricity is seen to flip sign 
between RHIC and LHC: at the LHC the freeze-out source is elongated 
{\em in the reaction plane direction} by almost as much as it was 
still {\em out-of-plane elongated} at RHIC. 
%This is different in 
%runs with an ideal massless gas EoS where even with ``LHC initial 
%conditions'' the final freeze-out source is found to be still out-of-plane 
%elongated (although just barely so).

%\section{Conclusions}
%By varying the initial entropy density to control the final charged 
%multiplicity, we used the hydrodynamic model to predict trends for the
%pion HBT radii from Au+Au or Pb+Pb collisions as one moves from RHIC to 
%LHC energies. In spite of a documented failure of the hydrodynamic model
%to reproduce the HBT radii measured at RHIC, the model has had great success 
%for most other soft-hadron observables, so the predicted trends may still be 
%trustworthy. We find very little variation in the HBT radii for central
%collisions, and whatever small differences we see depends sensitively
%on details of the equation of state, in particular whether or not it 
%embodies a quark-hadron phase transition. Clear and characteristic changes
%are predicted for the normalized azimuthal oscillation amplitudes of
%the HBT radii from non-central collisions, indicative of a qualitative
%change of the shape of the source at freeze-out which evolves from 
%an out-of-plane elongated freeze-out configuration at RHIC to an in-plane
%elongated shape at the LHC.\\[1ex]

%%%%%%%%%%%%%%%%%%%%%%%%%%%%%%%%%%%%%%%%%%%%%%%%%%%%%%%%%%%%%%%%%%%%%%%%%%

\subsection{Correlation
radii by FAST HADRON FREEZE-OUT GENERATOR}
\label{lokhtinfast}

{\it Iu.~A.~Karpenko, R.~Lednicky, I.~P.~Lokhtin,
L.~V.~Malinina,
Yu.~M.~Sinyukov and A.~M.~Snigirev}

{\small
The predictions for correlation radii in the central Pb+Pb
collisions for LHC $\sqrt{s}_{NN}=5500$~GeV are given in the frame of
FAST HADRON FREEZE-OUT GENERATOR (FASTMC).
}
\vskip 0.5cm

\begin{figure}[htbp]
\begin{minipage}{18pc}
\includegraphics[width=18pc]{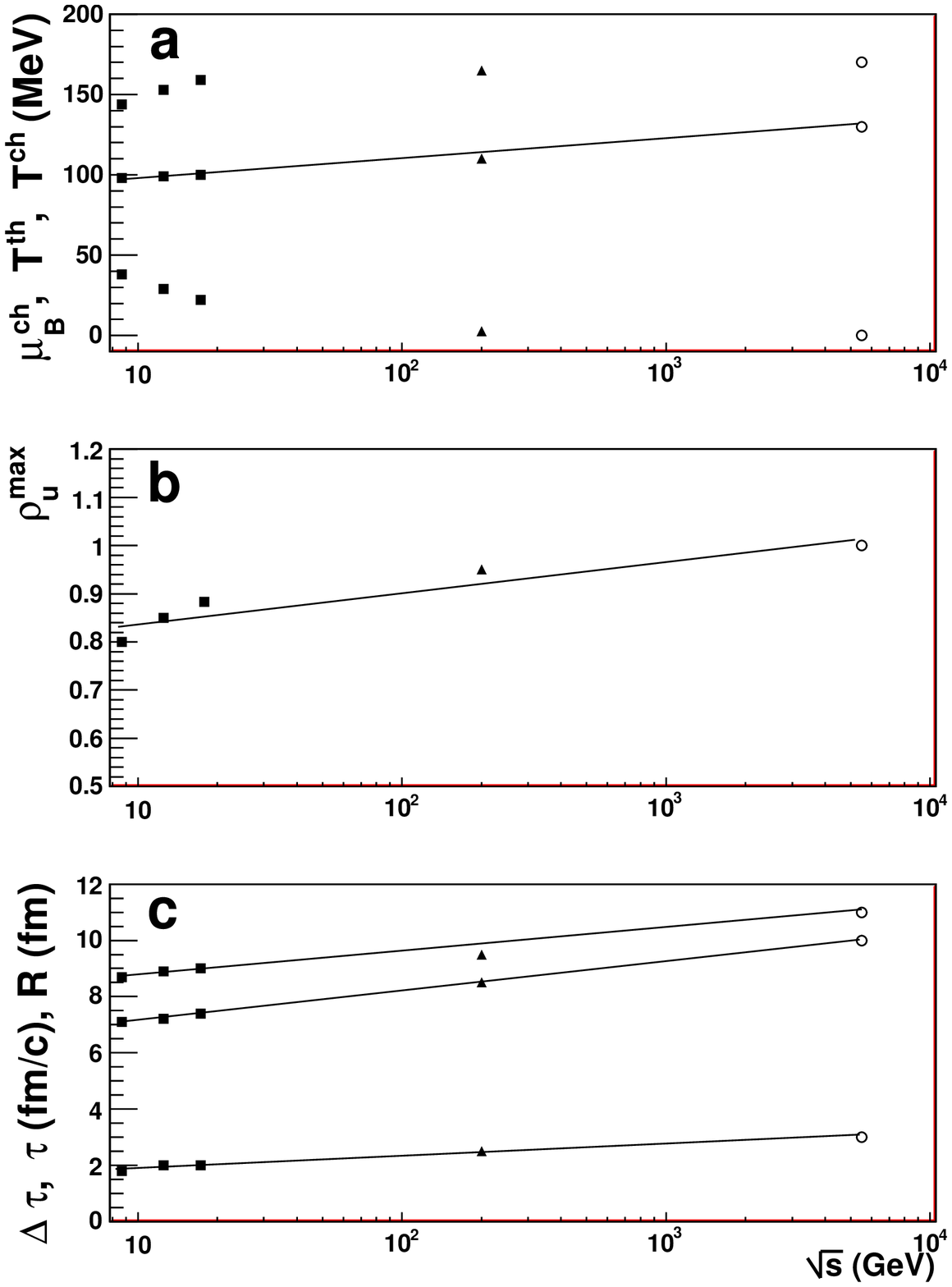}
\caption{\small{
FASTMC parameters versus $\log(\sqrt {s})$ for SPS $\sqrt{s}=8.7 - 17.3$~GeV
(black squares), RHIC $\sqrt{s}=200$~GeV (black triangles) and
LHC $\sqrt{s}=5500$~GeV(open circles): ({\bf a}) $T^{\rm ch}$, $T^{\rm th}$, $\mu^{B}$,
({\bf b}) $\rho_{u}^{max}$, ({\bf c}) $\tau$, $R$ and $\Delta \tau$.}
 \label{fig:Parameters}}
\end{minipage}
\hspace{\fill}%
\begin{minipage}{18pc}
\includegraphics[width=18pc]{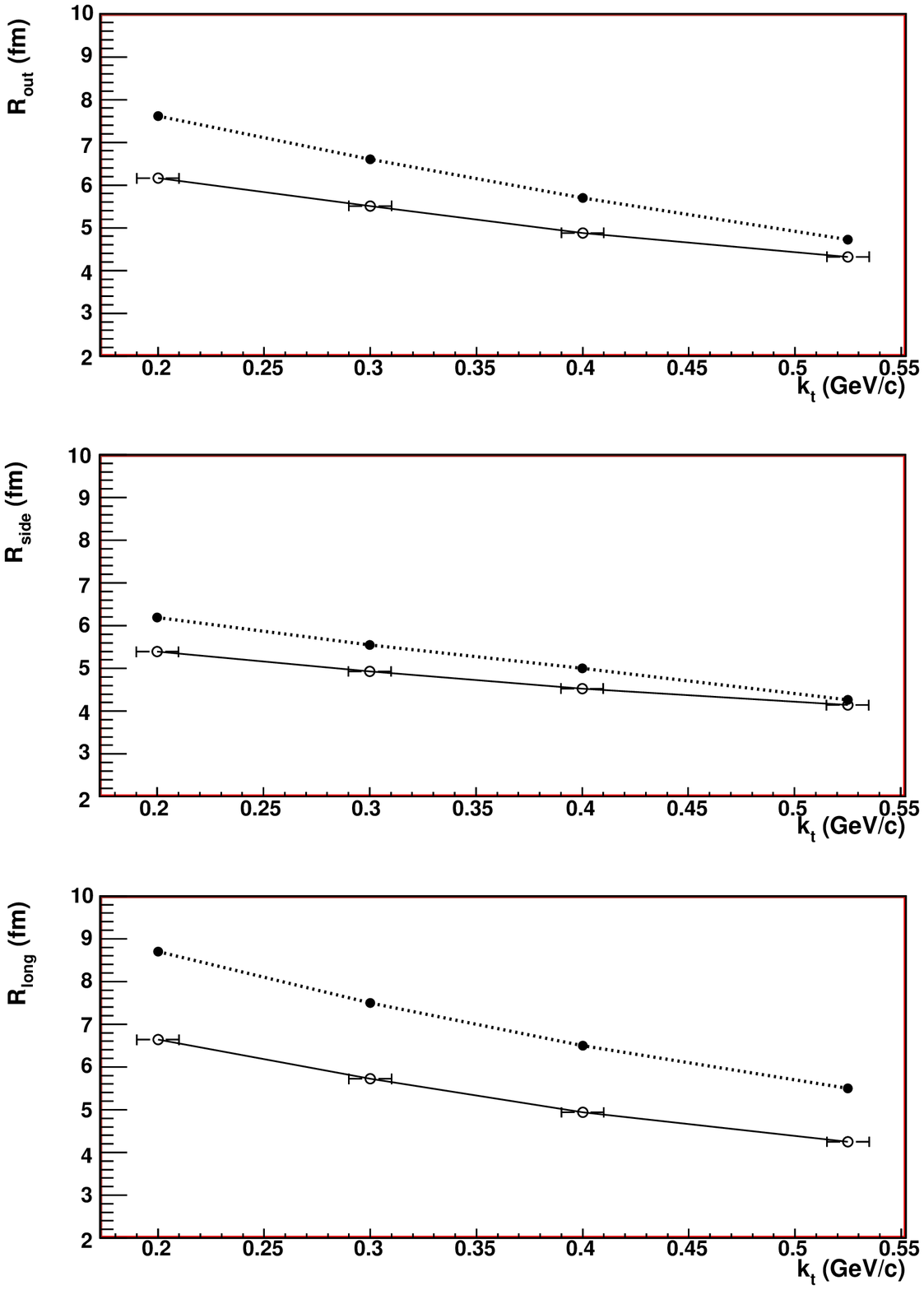}
\caption{\small{
The $\pi^{+} \pi^{+}$ correlation radii in longitudinally comoving
system
 at mid-rapidity in central Au+Au
collisions at $\sqrt s_{NN} = 200$~GeV from the STAR experiment \cite{Adams:2004yc}
(open circles) and the FASTMC calculations for LHC $\sqrt{s}=5500$~GeV
(black squares).}
\label{fig:CFsLHC} }
\end{minipage}
\end{figure}

One of the most  spectacular features of the RHIC data,
refereed as ``RHIC puzzle'', is the impossibility
to describe simultaneously  momentum-space measurements and
the freeze-out coordinate-space ones (femtoscopy)
by the existing hydrodynamic and cascade models or their hybrids.
However, a good description of SPS and RHIC data have been obtained
in various models based on
hydro-inspired parametrizations of freeze-out hypersurface.
Thus, we have achieved this goal
within our fast hadron freeze-out MC generator (FASTMC)
%within FAST HADRON FREEZE-OUT MC GENERATOR (FASTMC)
\cite{Amelin:2006qe}. In FASTMC, particle multiplicities
are determined based on the concept of chemical freeze-out.
Particles and hadronic resonances are
generated on the thermal freeze-out hypersurface,
the hadronic composition at this stage is defined by the parameters of the
system at chemical freeze-out \cite{Amelin:2006qe}.
%The hadrons consisting of light $u$,$d$ and $s$ quarks are only considered.
The input parameters which control the
execution of our MC hadron generator
in the case of Bjorken-like parameterization
of the thermal freeze-out hypersurface
(similar to the well known ``Blast-Wave'' parametrization
with the transverse flow)
for central collisions are the following:
temperature
 $T^{\rm ch}$ and chemical potentials per a unit charge
$\widetilde{\mu}_B, \widetilde{\mu}_S, \widetilde{\mu}_Q$ at chemical freeze-out,
temperature $T^{\rm th}$ at thermal freeze-out,
the fireball transverse radius $R$,
the mean freeze-out proper time $\tau$
and
its standard deviation $\Delta \tau$ (emission duration),
the maximal transverse flow rapidity $\rho_u^{\max}$.
We considered here the naive ``scaling'' of the existing
physical picture of heavy ion interactions over two order of
magnitude in $\sqrt{s}$ to
the maximal LHC energy $\sqrt{s}_{NN}=5500$~GeV.
%The same type of the freeze-out hypersurface as at lower energies
%is supposed.
The model parameters obtained by 
the fitting within FASTMC generator of the
existing experimental data on
$m_t$-spectra, particle ratios, rapidity density
$dN/dy$, $k_t$-dependence of the correlation radii
$R_{\rm out}, R_{\rm side}, R_{\rm long}$
from SPS ($\sqrt{s}_{NN}= 8.7 - 17.3$~GeV)
to RHIC ($\sqrt{s}_{NN} = 200$~GeV)
are shown in Fig.~\ref{fig:Parameters}.
For LHC energies we have fixed the thermodynamic parameters at chemical freeze-out
as the asymptotic ones: $T^{\rm ch}=170$~MeV,  $\widetilde{\mu}_B=
\widetilde{\mu}_S = \widetilde{\mu}_Q$=0~MeV.
The linear extrapolation of the model parameters
in $\log(\sqrt{s})$ to LHC ($\sqrt{s}_{NN} = 5500$~GeV)
is shown in Fig.~\ref{fig:Parameters} by open symbols.
The extrapolated values are the
following: $R \sim 11$~fm, $\tau \sim 10$~fm/c, $\Delta \tau \sim 3.0$~fm/c,
$\rho_u^{\max} \sim 1.0$, $T^{\rm th} \sim 130$~MeV. The density
of charged particles at mid-rapidity obtained with these parameters is
$dN/dy=1400$, i.e. twice larger than at RHIC $\sqrt{s}_{NN}=200$~GeV
in coincidence with the naive extrapolation of $dN/dy$.
These parameters yield only a small increase of the
correlation radii $R_{\rm out}, R_{\rm side}
, R_{\rm long}$
(Fig.~\ref{fig:CFsLHC}).

\subsection{Exciting the quark-gluon plasma with a relativistic jet}
\label{s:Manuel}

{\em M. Mannarelli and C. Manuel}

{\small 
We discuss the properties of a system composed by a static plasma traversed by a
jet of particles.
Assuming that both the jet and the plasma can be described using a 
hydrodynamical approach, and in the conformal limit, we find that unstable modes
arise when the velocity of the jet is larger than the speed of the sound of the 
plasma and only modes with momenta smaller than a certain values are unstable. 
Moreover, for ultrarelativistic velocities of the jet the most unstable modes 
correspond to relative angles between the velocity of the jet and momentum of 
the collective mode $\sim \pi/4$.
Our results suggest an alternative mechanism for the description of the jet 
quenching phenomenon, where the jet crossing the plasma loses energy exciting 
colored unstable modes. 
In LHC this effect should be seen with an enhanced production of hadrons for 
some specific values of their momenta and in certain directions of momenta 
space.
}
\vskip 0.5cm

It has been suggested that a high $p_T$ jet crossing the medium produced  after 
a relativistic heavy ion collision, and travelling at a velocity higher than 
the speed of sound should form shock waves with a Mach cone 
structure~\cite{Casalderrey-Solana:2004qm,Casalderrey-Solana:2006sq}. 
Such shock waves should be detectable in the low $p_T$ parton distributions at 
angles $\pi \pm 1.2$ with respect to the direction of the trigger particle.
A preliminary analysis of the azimuthal dihadron correlation performed by the 
PHENIX Collaboration~\cite{Adler:2005ee} seems to suggest the formation of such 
a conical flow.

We propose a novel possible collective process to describe the jet quenching 
phenomenon. 
In our approach a neutral beam of colored particles crossing an equilibrated 
quark-gluon plasma induces plasma instabilities~\cite{Mannarelli:2007gi}. 
Such instabilities represent a very efficient mechanism for converting the  
energy and momenta stored in the total system (composed by the plasma and the 
jet) into (growing) energy and momenta of gauge fields, which are initially 
absent.  
To the best of our knowledge, only reference~\cite{Pavlenko:1991ih} considers 
the possibility of the appearance of filamentation instabilities produced by 
hard jets in heavy-ion collisions.  

We have studied this phenomenon using the chromohydrodynamical approach 
developed in \cite{Manuel:2006hg}, assuming the conformal limit for the plasma. 
Since we are describing the system employing ideal fluid-like equations, our 
results are valid at  time scales shorter than the average time for collisions. 
A similar analysis using kinetic theory, and reaching to similar results, will 
soon be reported.

\begin{figure}[!th]
\label{plots}
\includegraphics[width=3in,angle=-0]{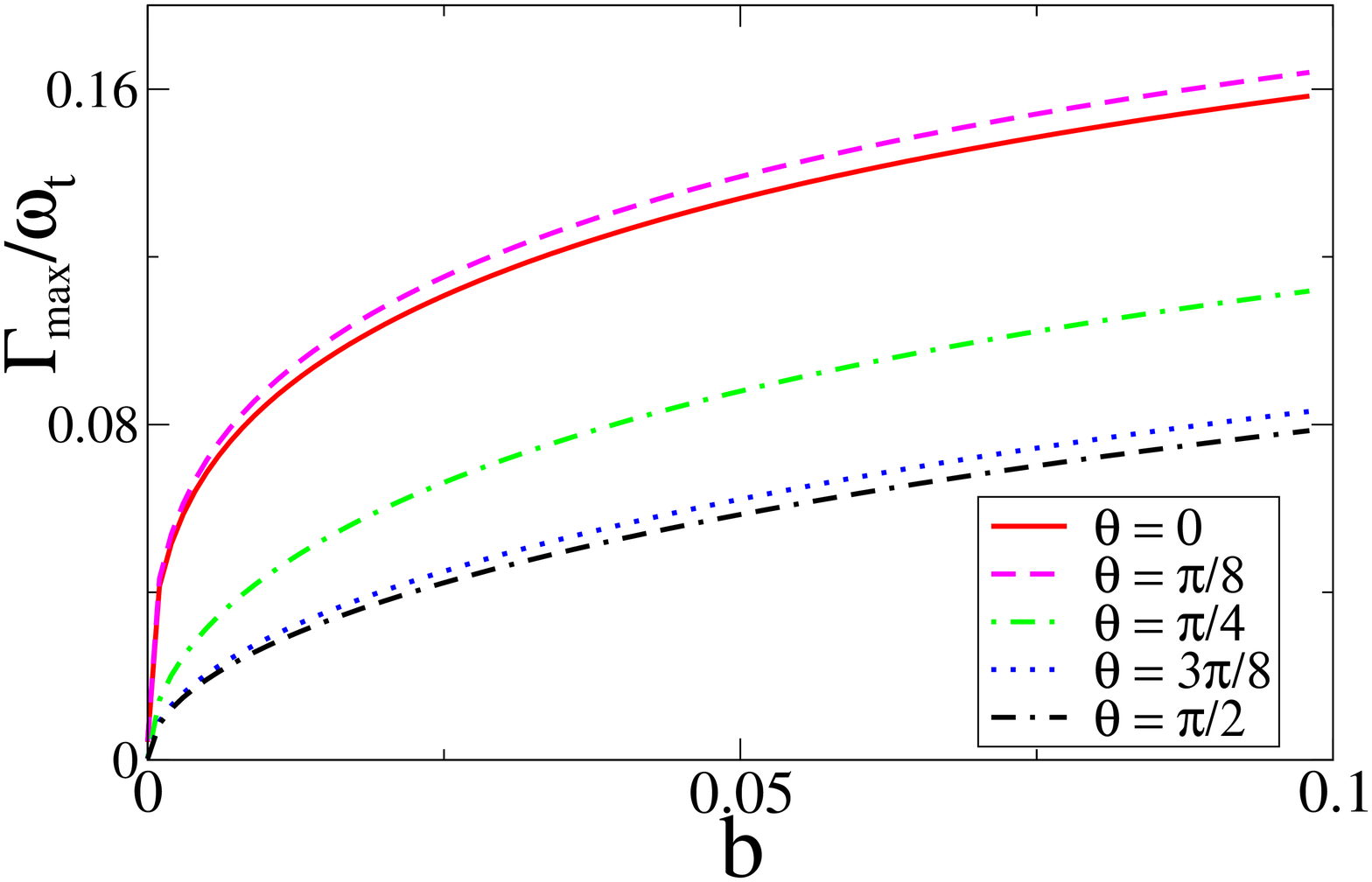}
\includegraphics[width=3in,angle=-0]{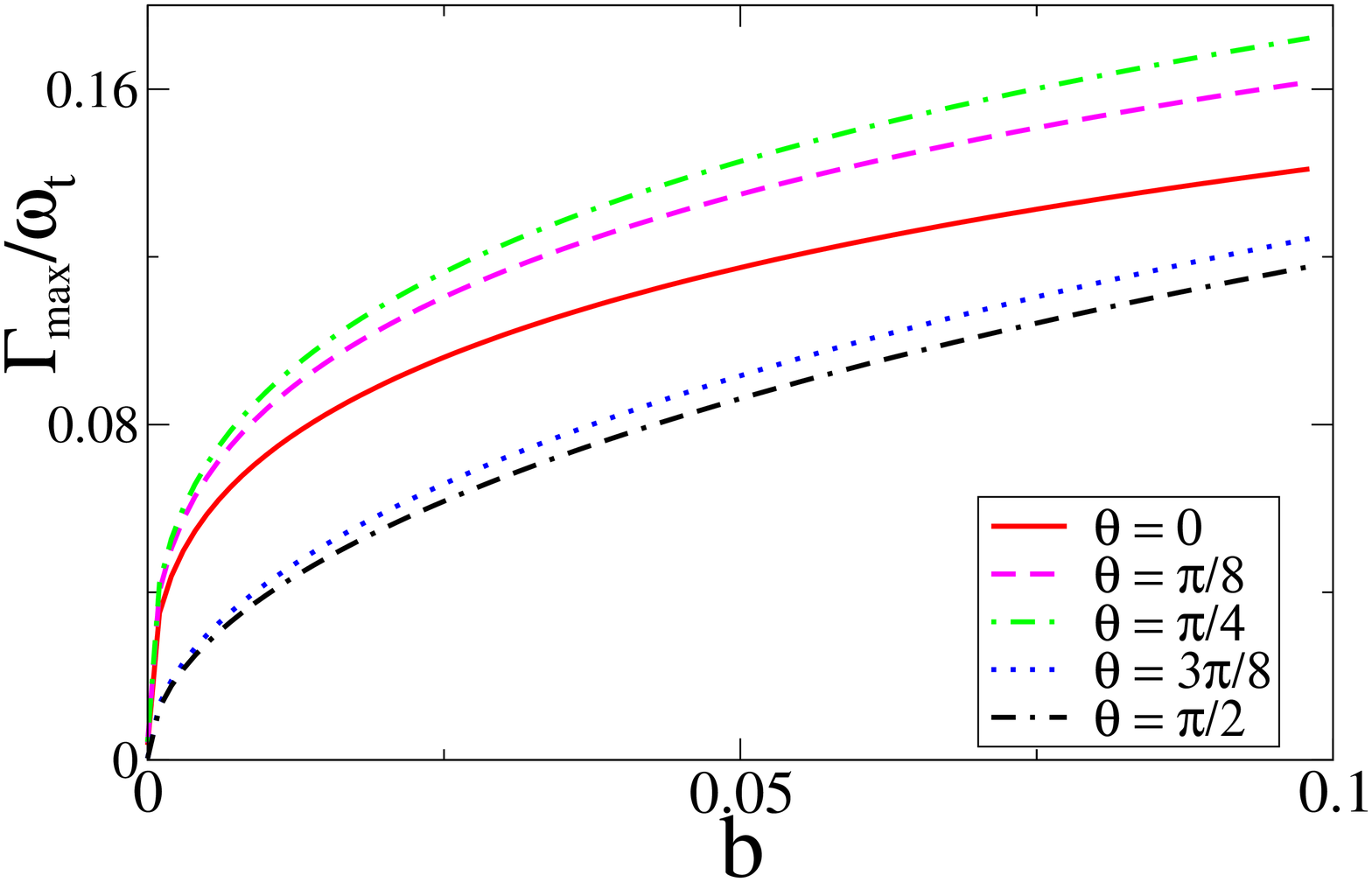}
\caption{Largest value of the imaginary part of the dispersion law for the 
  unstable mode as a function of $b=\omega^2_{\rm jet}/\omega_p^2$ for two 
  different values of the velocity of the jet $v$ and five different angles 
  between $\bf k$ and $\bf v$. The left/right panels correspond to $v=0.8/0.9$,
  respectively.}
\label{fig:Manuel-fig1}
\end{figure}

We have studied the dispersion laws of the gauge collective modes and their
dependence on the velocity of the jet $v$, the magnitude of the momentum of the 
collective mode ${\bf k}$, the angle $\theta$ between these quantities, and of 
the plasma frequencies of both the plasma $\omega_p$ and the jet 
$\omega_{\rm jet}$.
We find that there is always one unstable mode if the velocity of the jet is 
larger than the speed of sound $c_s = 1/\sqrt{3}$, and if the momentum of the 
collective mode is in modulus smaller than a threshold value.
Quite interestingly we find that the unstable modes with momentum parallel to 
the velocity of the jet is the dominant one for velocity of the jet 
$v \lesssim 0.8$.
For larger values of the jet velocity only the modes with angles larger than 
$\sim \pi/8$ are significant and the dominant unstable modes correspond to  
angles $\sim \pi/4$ (see figure~\ref{fig:Manuel-fig1}).

Our numerical results imply that both in RHIC and in the LHC these instabilities
develop very fast, faster in the case of the LHC as there one assumes that 
$\omega_p$ will attain larger values. 
Further, the soft gauge fields will eventually decay into soft hadrons, and may 
affect the hydrodynamical simulations of shock waves mentioned in 
reference~\cite{Casalderrey-Solana:2004qm,Casalderrey-Solana:2006sq}.

\section{Fluctuations}
\label{sec:fluct}

\subsection{Fluctuations and the clustering
of color sources}
\label{ferreirofluc}

{\it L. Cunqueiro, E. G. Ferreiro and C. Pajares}

{\small
We present our results on multiplicity and $p_T$ fluctuations at LHC energies
in the framework of the clustering of color sources.
In this approach, elementary color sources -strings- overlap forming
clusters, 
so the number of effective sources is modified.
We find that the fluctuations are proportional to the 
number of those clusters. 
%These clusters decay into particles with mean transverse momentum and
%mean multiplicity that
%depend on the
%number of elementary sources that conform each cluster, and the area
%occupied by the cluster.
}
\vskip 0.5cm

%\section{Transverse momentum fluctuations}
Non-statistical event-by-event fluctuations in relativistic
heavy ion collisions have been proposed as a probe of phase instabilities
near de QCD phase transition.
%Event-by-event fluctuations of %$p_T$
The transverse momentum and the multiplicity fluctuations
have been measured at SPS
and RHIC energies.
These
fluctuations show a non-monotonic behavior with the centrality of the
collision: they grow as the centrality increases, showing a maximum
at mid centralities, followed by a decrease at larger centralities.
Different mechanisms
have been proposed in order to explain those data. Here, we will apply the clustering of color sources.
In this approach, color strings are
stretched between 
the colliding partons. Those strings act as color sources of particles
which
are
successively
broken by creation of $q {\bar q}$ pairs from the sea.
The color strings correspond to small areas
 in the transverse space filled
with color field created by the
colliding partons. 
If the density of strings increases, they overlap in the transverse 
space, giving rise to a phenomenon of string fusion and 
percolation \cite{Armesto:1996kt}. 
Percolation indicates that the cluster size diverges, 
reaching the size of the system. 
Thus, variations of the initial state can lead to a
transition from
disconnected to connected color clusters. The percolation point signals the 
onset of color deconfinement.

These clusters decay into particles with mean transverse momentum and
mean multiplicity that
depend on the
number of elementary sources that conform each cluster, and the area
occupied by the cluster.
In this approach, the behavior of the $p_T$ \cite{Ferreiro:2003dw}
and multiplicity \cite{Cunqueiro:2005hx}
fluctuations can be
understood as
follows:
at low density, most of the particles are
produced by individual strings with
the same transverse momentum $<p_T>_1$ and the same multiplicity
$<\mu_1>$, so fluctuations are small.
At large density,
above the critical point of percolation, we have only one cluster, so
fluctuations are not expected either.
Just below the percolation critical density, we have
a large number of clusters formed by different number of strings $n$,
with
different size and thus different $<p_T>_n$
and different $<\mu>_n$
so the fluctuations are maximal. 

The variables to measure event-by-event $p_T$ fluctuations are $\phi$ and
$F_{p_T}$, 
that quantify the deviation of the observed fluctuations from
statistically independent particle emission:
\beq
\phi=\sqrt{\frac{<Z^2>}{<\mu>}}-\sqrt{<z^2>}\ ,
\eeq
where $z_i={p_T}_i - <p_T>$ is defined for each particle and
$Z_i=\sum_{j=1}^{N_i} z_j$ is defined for each event,
and
\beq
F_{p_T} = \frac{\omega_{data} - \omega_{random}}{\omega_{random}},\, \ \ \
\omega= \frac{\sqrt{<p_T^2>-<p_T>^2}}{<p_T>}\ .
%=\frac{\phi}{\sqrt{<z^2>}}=\frac{1}{\sqrt{<z^2>}}
%\sqrt{\frac{<Z^2>}{<\mu>}} -1\ .
\eeq
Moreover, 
in order to measure the multiplicity fluctuations, the variance of the multiplicity distribution scaled to the mean value of
the multiplicity has been used.
Its behavior is similar to the one obtained 
for $\Phi (p_T)$, used to quantify the $p_T$-fluctuations,
suggesting that they are related to each other. 
The $\Phi$-measure is independent of the distribution of number 
of particle sources
if the sources are identical and independent from each other. That is,
$\Phi$ should be
independent of the impact parameter if the nucleus-nucleus collision is
a simple superposition of nucleon-nucleon interactions.

\begin{center}
\begin{figure*}
\begin{minipage}[t]{80mm}
\epsfxsize=8.0cm
\epsfysize=5.5cm
\centerline{\epsfbox{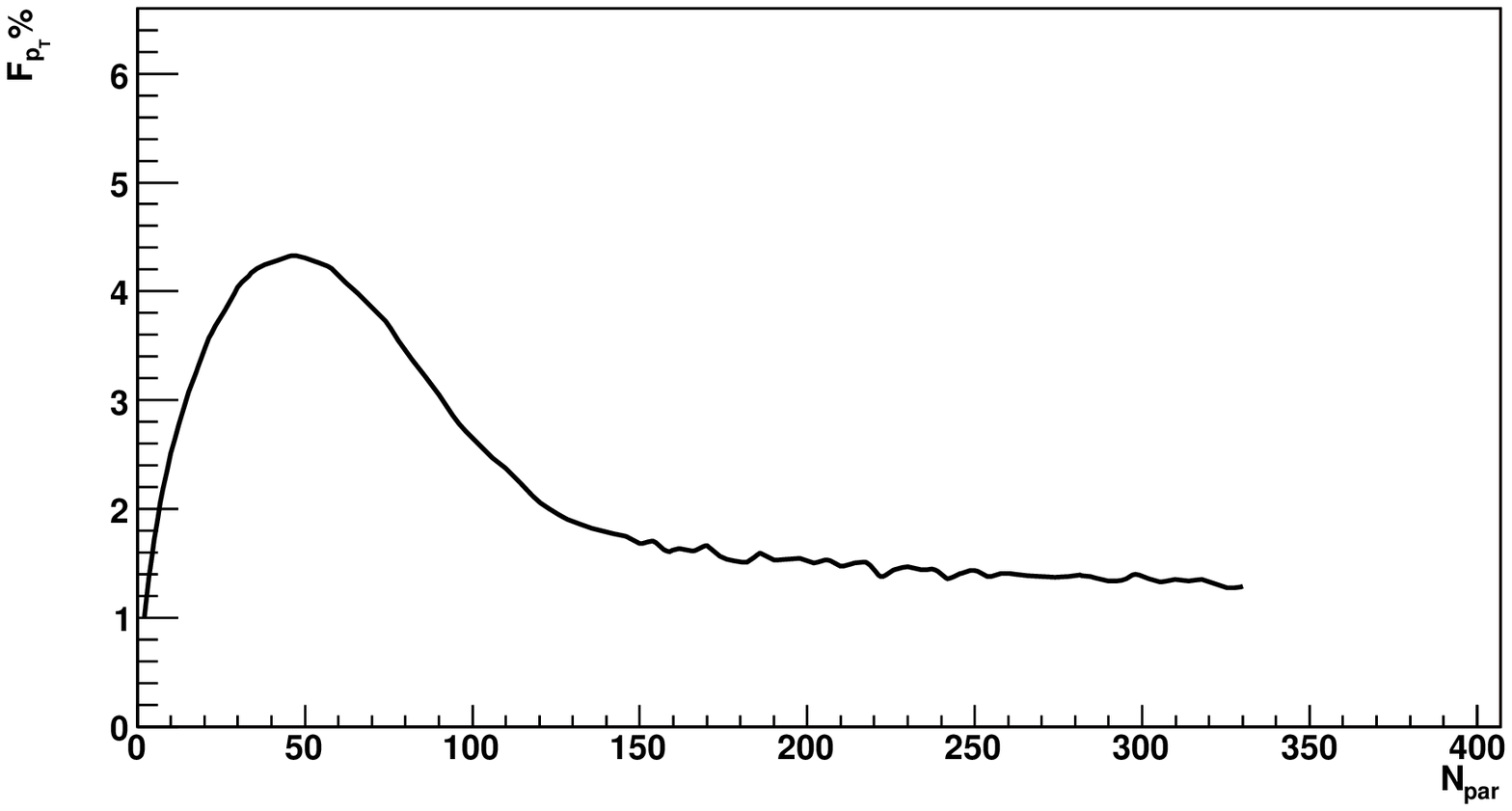}}
\vskip -0.25cm
\caption{$F_{p_T}$ at LHC.}
\label{ferreiroflucf1}
%RHIC (lower curve)
%and LHC (upper curve) energies.}
\end{minipage}
\hspace{\fill}
\begin{minipage}[t]{80mm}
%\vskip -5.8cm
%\epsfxsize=8.5cm
%\epsfysize=6.0cm
\vskip -5.8cm
\epsfxsize=8.5cm
\epsfysize=6.0cm
\centerline{\epsfbox{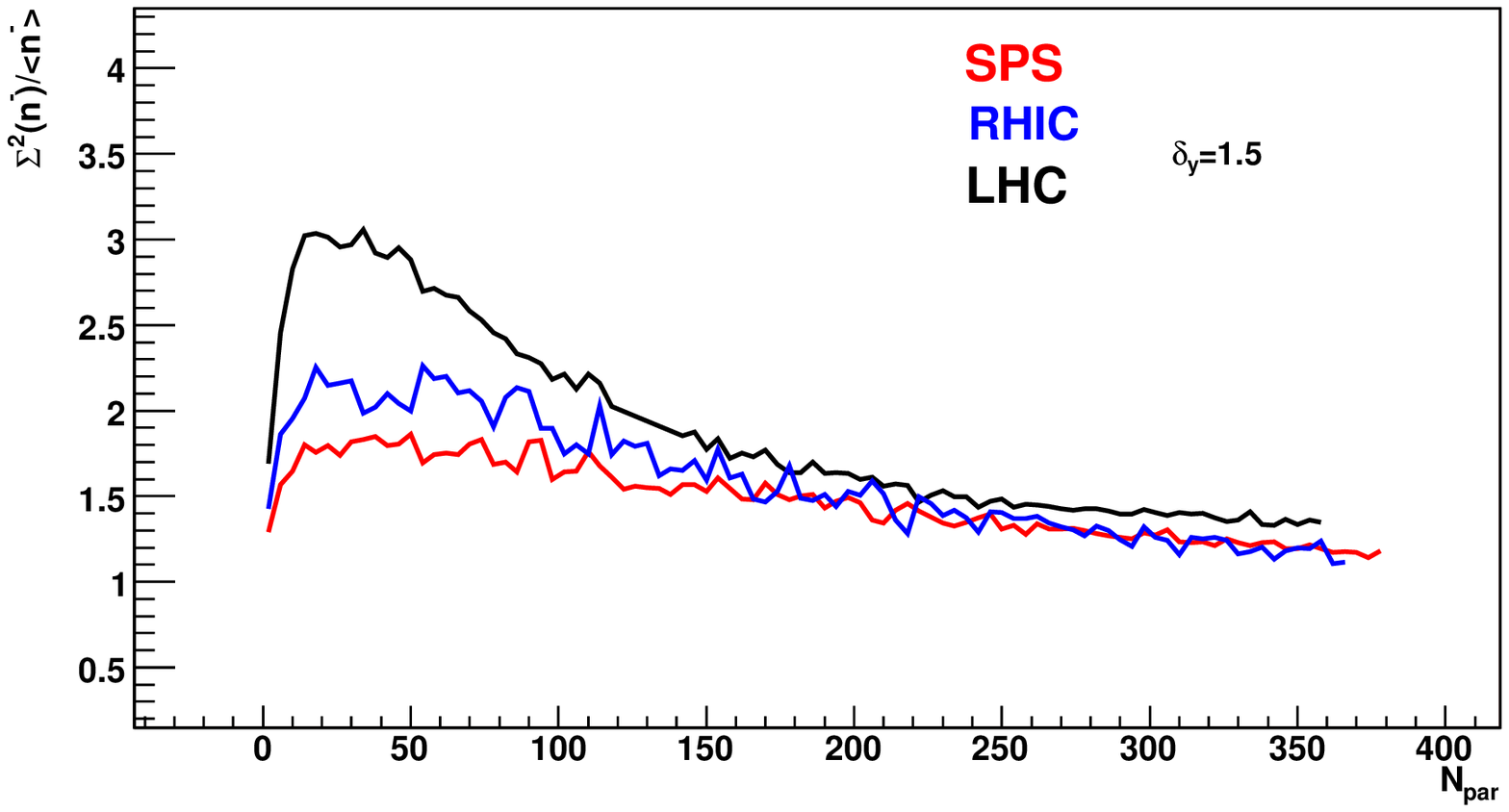}}
\vskip -0.25cm
\caption{Scaled variance on negatively charged particles at, from up to down,
LHC, RHIC and SPS.} 
\label{ferreiroflucf2}
\end{minipage}
\vskip -0.5cm
\end{figure*}
\end{center}

In Fig. \ref{ferreiroflucf1} we present our results on $p_T$ fluctuations at LHC. 
Note that the increase of the energy essentially 
shifts the maximum position to a lower number of participants 
\cite{Ferreiro:2003dw}.
%Note that we have an uncertainty on the absolute value of this quantity at 
%LHC, but the 
%behavior vs. centrality should not be affected.
In Fig. \ref{ferreiroflucf2} we show our values for the scaled variance of negatively charged particles
%and total charged particles 
at SPS, RHIC and LHC energies.

Summarizing:
the $p_T$ and multiplicity
fluctuations are due in our approach to
the
different mean $<p_T>$
and mean multiplicities of the clusters,
and they depend essentially on the number
of clusters.
In other words, 
a decrease in the number of effective sources
leads to a decrease of the 
fluctuations.

\subsection {Fluctuations of particle multiplicities from RHIC to LHC}
%{\it Giorgio Torrieri}
{\it G.~Torrieri}

{\small
We define an observable capable of determining which statistical model, if any, governs freeze-out in very high energy heavy ion collisions such as RHIC and LHC.
We calculate this observable for $K/\pi$ fluctuations, and show that it should be the same for RHIC and LHC, as well as independent of centrality, if the Grand-Canonical statistical model is appropriate and chemical equilibrium applies.  We describe variations of this scaling for deviations from this scenario, such as light quark chemical non-equilibrium, strange quark over-saturation and local (canonical) equilibrium for strange quarks.
}
\vskip 0.5cm

Particle yield fluctuations are a promising observable to falsify the statistical model and to constrain its parameters (choice of ensemble, strangeness/light quark chemical equilibrium) \cite{Torrieri:2005va}.
The uncertainities associated with fluctuations, however, warrant that care be taken to choose a fluctuation observable.

For instance, volume fluctuations could be originating from both initial state effects and dynamical processes, and are not well understood.  Their effect has to be factored out from multiplicity fluctuations data.
One way to do this is to concentrate on fluctuations of particle ratios, where volume factors out event by event \cite{Jeon:1999gr}
\begin{equation}
\label{fluctratio}
\sigma_{N_1/N_2}^2
= \frac{\ave{(\Delta N_1)^2}}{\ave{N_1}^2}
+ \frac{\ave{(\Delta N_2)^2}}{\ave{N_2}^2}
- 2 \frac{\ave{\Delta N_1 \Delta N_2}}{\ave{N_1}\ave{ N_2}}.
\end{equation}
This, however, introduces an {\em average} hadronization volume dependence through the $\ave{N_{1,2}}$ terms (two in the denominator, one in the numerator of Eq. \ref{fluctratio}).

This feature allows us to perform an invaluable consistency chech for the statistical model, since the volume going into the ratio fluctuations must, for consistency, be the same as the volume going into the yields.  Thus, observables such as $\frac{d \ave{N_1}}{d y} \sigma_{N_1/N_2}^2$ should be  strictly independent of multiplicity and centrality, as long as the statistical model holds and the physically appropriate ensemble is Grand Canonical.

\begin{figure*}[h]
\begin{center}
\epsfig{width=14cm,figure=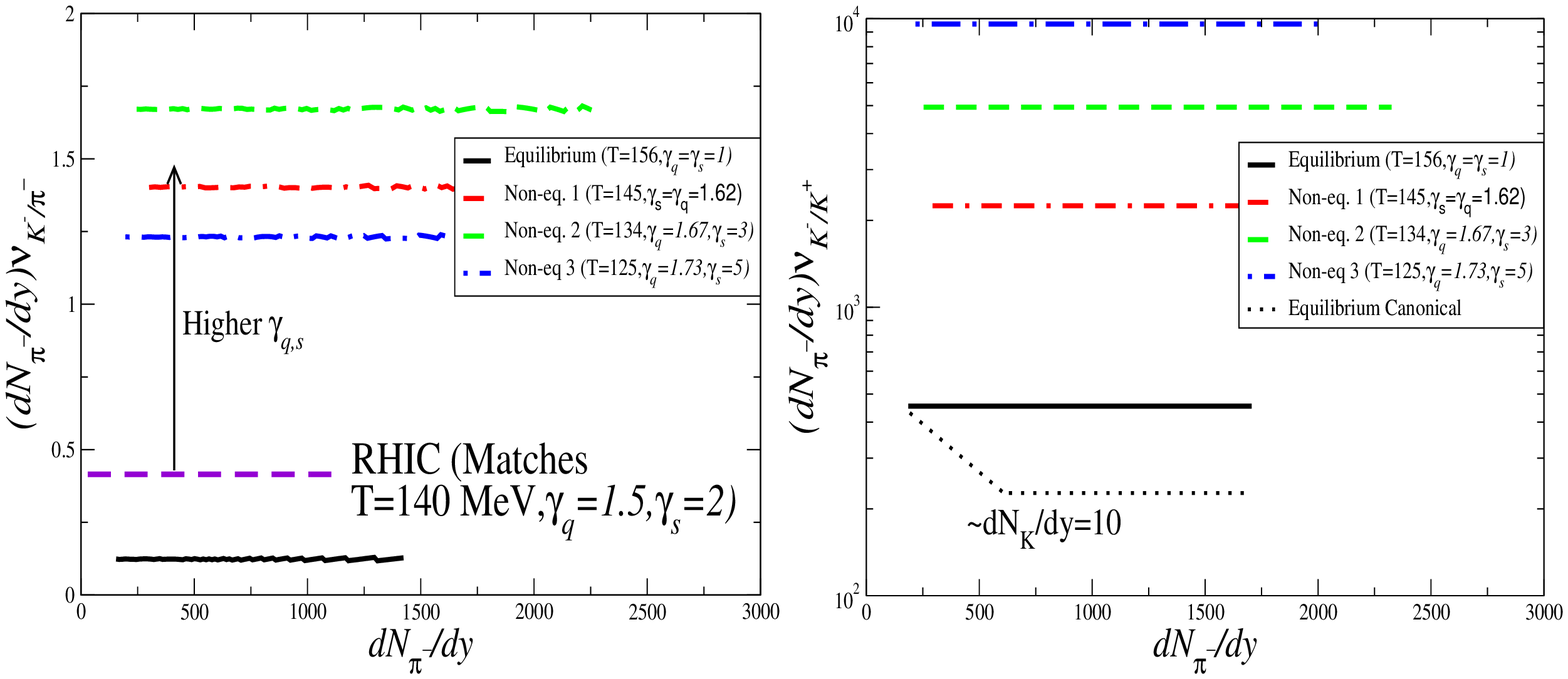}
\end{center}
\caption{\label{1fig}  }
\end{figure*}

We propose doing this test, at both RHIC and LHC, using the  corrected variance
\begin{equation}
\Psi^{N_1}_{N_1/N_2} = \frac{d N_1}{d y} \nu_{N_1/N_2}^{dyn}
\end{equation}
where $\nu_{N_1/N_2}^{dyn}$ is theoretically equal to the corrected mixed
variance \cite{Pruneau:2002yf}
\begin{eqnarray}
\nu^{dyn}_{N_1/N_2} = (\sigma^{dyn}_{N_1/N_2})^2 =  \sigma_{N_1/N_2}^2 - (\sigma^{Poisson}_{N_1/N_2})^2= \nonumber \\ = \frac{\ave{N_1 (N_1 - 1)}}{\ave{N_1}^2}+  \frac{\ave{N_2 (N_2 - 1)}}{\ave{N_2}^2} - 2 \frac{\ave{N_1 N_2}}{\ave{N_1}\ave{N_2}}
\end{eqnarray}
SHAREv2.X \cite{Torrieri:2006xi}
provides the possibility of calculating all ingredients of $\Psi^{N_1}_{N_1/N_2}$ for any hadrons, incorporating the effect of all resonance decays, as well as chemical (non)equilibrium.
The calculation for $\Psi^{\pi^-}_{K^-/\pi^-}$, as well as $\Psi^{\pi^-}_{K^-/\pi^-}$ is shown in Fig. \ref{1fig}.  These species were chosen because their correlations (from resonance decays, $N^* \rightarrow N_1 N_2$), which would need corrections for limited experimental acceptance, are small.

Equilibrium thermal and chemical parameters are very similar at RHIC and the LHC(the baryo-chemical potential will be lower at the LHC, but it is so low at RHIC that the difference is not experimentally detectable).
Thus, $\Psi^{N_1}_{N_1/N_2}$ should be identical, to within experimental error, for both the LHC and RHIC, over all multiplicities were the statistical model is thought to apply.

According to \cite{Rafelski:2005jc},
chemical conditions at freeze-out deviate from equilibrium, and reflect the higher entropy contect and strangeness per entropy content of the early deconfined phase through an over-saturated phase space occupancy for the light and strange quarks ($\gamma_{s}>\gamma_q>1$).
If this is true, than $\Psi^{N_1}_{N_1/N_2}$ should still be independent of
centrality for a given energy range, but should go markedly up for the LHC
from RHIC, because of the increase in $\gamma_q$ and $\gamma_s$.   Fig.
\ref{1fig} shows what effect three different sets of $\gamma_{q,s}$ inferred
in \cite{Rafelski:2005jc} would have on  $\Psi^{\pi^-}_{K^-/\pi^-}$ and $\Psi^{\pi^-}_{K^-/K^+}$

If non-statistical processes (minijets, string breaking etc.) dominate event-by-event physics, the flat $\Psi^{N_1}_{N_1/N_2}$ scaling on centrality/multiplicity should be broken, and $\Psi^{N_1}_{N_1/N_2}$ would exhibit a non-trivial dependence on $N_{part}$ or $dN/dy$.

This is also true if global correlations persist, such as is the case in Canonical and micro-canonical models \cite{Begun:2004gs}
If global correlations persist for particle $N_2$ and/or $N_1$, than $\Psi^{N_1}_{N_1/N_2}$ becomes reduced, and starts strongly varying with centrality in lower multiplicity events.   Thus, if strangeness at RHIC/the LHC is created and maintained locally,  $\Psi^{N_1}_{N_1/N_2}$ should develop a ``wiggle'' at low centrality, and be considerably lower than Grand Canonical expectation. 
For  $\Psi^{\pi^-}_{K^+/K^-}$ it should be lower by a factor of two.

In conclusion, measuring  $\Psi^{\pi^-}_{K^-/\pi^-}$ and $\Psi^{\pi^-}_{K^+/K^-}$, at comparing the results between the LHC and RHIC can provide an invaluable falsification of the statistical model, as well as constraints as to {\em which} statistical model applies in these regimes.

%GT thanks the Alexander Von Humboldt
%foundation, the Frankfurt Institute for Theoretical Physics and FIAS 
%for continued support, and CERN theory division for providing local 
%support necessary for attending the workshop where this work is presented.   
%We would also like to thank Sangyong Jeon, Marek Gazdzicki, 
%Mike Hauer, Johann Rafelski and Mark Gorenstein for useful and 
%productive discussions.

%%%%%%%%%%%%%%%%%%%%%%%%%%%%%%%%%%
%%%%%%%%%%%%%%%%%%%%%%%%%%%%%%%

\section{High transverse momentum observables and jets}
\label{sec:highpt}

\subsection{Jet quenching parameter $\hat q$ from Wilson loops in a thermal environment}
\label{antonov}

%{\it D. Antonov, D. D. Dietrich and H. J. Pirner}
{\it D. Antonov and H. J. Pirner}

{\small
The gluon jet quenching parameter is calculated in SU(3) quenched QCD within the stochastic vacuum model. At the LHC-relevant temperatures, it is defined by the gluon condensate and the vacuum correlation length. Numerically, when the temperature varies from $T_c=270{\,}{\rm MeV}$ to the inverse vacuum correlation length $\mu=894{\,}{\rm MeV}$, the jet quenching parameter rises from zero to $1.1{\,}{\rm GeV^2}/{\rm fm}$.
}
\vskip 0.5cm

\begin{figure}
\epsfig{file=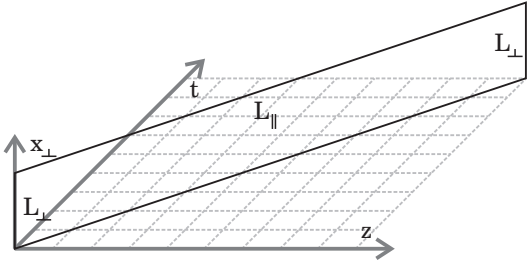, width=70mm}
\caption{\label{one} The contour of the Wilson loop of a gluon.}
\end{figure}

At LHC energies, radiative energy loss is
the dominant mechanism of jet energy loss in the quark-gluon plasma.  The expectation value of a light-like adjoint Wilson loop provides an estimate for the radiative energy loss of a gluon~\cite{Liu:2006he}:
\begin{equation}
\label{Wadj}
\left<W^{\rm adj.}_{L_{\parallel}\times L_{\perp}}\right>=\exp\left(-\frac{\hat q}{4\sqrt{2}}L_\parallel L_\perp^2\right).
\end{equation}
The contour of the loop at zero temperature is depicted in Fig.~\ref{one}. We have calculated the {\it jet quenching 
parameter} $\hat q$ in the SU(3) quenched theory through the evaluation of the Wilson loop~(\ref{Wadj}). To this end, we have used 
the stochastic vacuum model~\cite{svm} at $T>T_c$, where $T_c=270{\,}{\rm MeV}$ is the deconfinement temperature. 
This model incorporates the gluon condensate which, together with the vacuum correlation length, defines the jet quenching 
parameter. This is different from the results obtained within perturbative QCD~\cite{pq} and conformal field theories~\cite{Liu:2006he}, where $\hat q\propto T^3$.

\begin{figure}
\psfrag{a1}{$\hat q{\,}[{\rm GeV}^2/{\rm fm}]$}
\epsfig{file=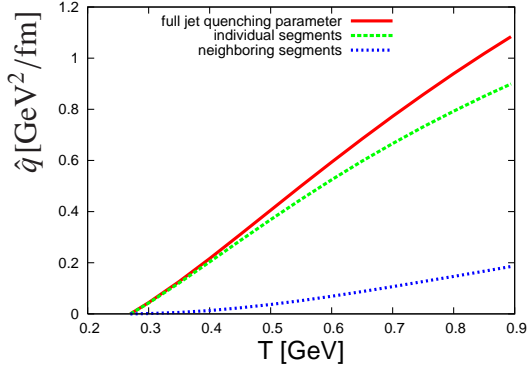, width=70mm}
\caption{\label{ra3} The full jet quenching parameter and relative contributions to it.}
\end{figure}

The hierarchy of scales in our problem is 
$\mu^{-1}\ll L_\perp\ll \beta\ll L_\parallel$,
where $\beta$ is the inverse temperature, and $\mu=894{\,}{\rm MeV}$ is the inverse vacuum correlation length.
Due to the $x_4$-periodicity at finite temperature, the contour depicted in Fig.~\ref{one} effectively splits into segments whose extensions along the 3rd and the 4th axes are $\beta$. Furthermore, due to the short-rangeness of gluonic correlations, which fall off at the vacuum correlation length, the dominant contribution to $\hat q$ stems from self-interactions of individual segments.
We have also calculated the contribution stemming from the correlations of neighboring segments, which turns out to be parametrically (and numerically) suppressed by the factor ${\rm e}^{-\mu/T}$. For this reason, the even smaller contributions from the next-to-nearest neighboring segments on are disregarded. 
The contributions of individual and neighboring segments read
$$
\hat q=\frac{g^2\left<(F_{\mu\nu}^a)^2\right>_{T=0}}{16\mu}\left[\sqrt{2}-\frac{T}{\mu}
\left(1-{\rm e}^{-\sqrt{2}\mu/T}\right)\right]
\left[\coth\left(\frac{\mu}{2T}\right)-\coth\left(\frac{\mu}{2T_c}\right)\right]~ {\rm and}
$$
$$
\Delta\hat q=\frac{g^2\left<(F_{\mu\nu}^a)^2\right>_{T=0}}{16\mu}{\rm e}^{-\mu/T}
\left[1-\frac{T}{\mu}
\left(1-{\rm e}^{-\mu/T}\right)\right]
\left[\coth\left(\frac{\mu}{2T}\right)-\coth\left(\frac{\mu}{2T_c}\right)\right],
$$
respectively. The right most brackets in these equations define the temperature dependence of the gluon condensate, 
corresponding to the exponential fall-off of its nonlocal counterpart~\cite{a}. As for the zero-temperature value of the 
gluon condensate, it can be expressed through the vacuum correlation length and the string tension in the fundamental representation of SU(3), $\sigma=(440{\,}{\rm MeV})^2$, and reads~\cite{dp}
$g^2\left<(F_{\mu\nu}^a)^2\right>_{T=0}=(72/\pi)\sigma\mu^2=3.55{\,}{\rm GeV}^4$. The above contributions 
together with their sum are plotted in Fig.~\ref{ra3}. Note finally that, in the 
large-$N_c$ limit, our full result for the jet quenching parameter behaves as $N_c^0$, i.e. it does not scale with $N_c$.
This behavior is similar to those of other models~\cite{Liu:2006he, pq}.

\subsection{Particle Ratios at High $p_T$ at LHC Energies}
\label{barnafoldi}

{\it G. G. Barnaf\"oldi, P. L\'evai,
B. A. Cole,
G. Fai and
G. Papp}

{\small
Hadron production has been calculated in a pQCD improved parton model for
 $pp$, $dA$ and heavy ion collisions. We applied KKP and AKK fragmentation
functions. %Hadron ratios -- measurable by the ALICE experiment -- have
%been investigated.  
Our jet fragmentation study shows, that hadron ratios
at high $p_T$ depend on quark contribution mostly and less on the gluonic 
one. This finding can be seen in jet-energy loss calculations, also. We 
display the suppression pattern on different hadron ratios in $PbPb$ 
collisions at LHC energies. 
}
\vskip 0.5cm

The precision of pQCD based parton model calculations was enhanced during 
the last decade. The calculated spectra allow to make predictions not 
only for the hadron yields, but for sensitive particle ratios and nuclear 
modifications. For the calculation of particle ratios new fragmentation 
functions are needed not only for the most produced light mesons, but for 
protons also. From the experimental point of view one requires identified 
particle spectra by RHIC and LHC. Especially the ALICE detector has a 
unique capability to measure identified particles at highest transverse 
momenta via \v{C}herenkov detectors. The $\pi^{\pm}/K^{\pm}$ and 
$K^{\pm}/p(\bar{p})$ ratios can be measured up to $3$ GeV/c and $5$ GeV/c 
respectively.     

\begin{figure}
\begin{center}
\includegraphics[width=7.5truecm,height=8.0truecm]{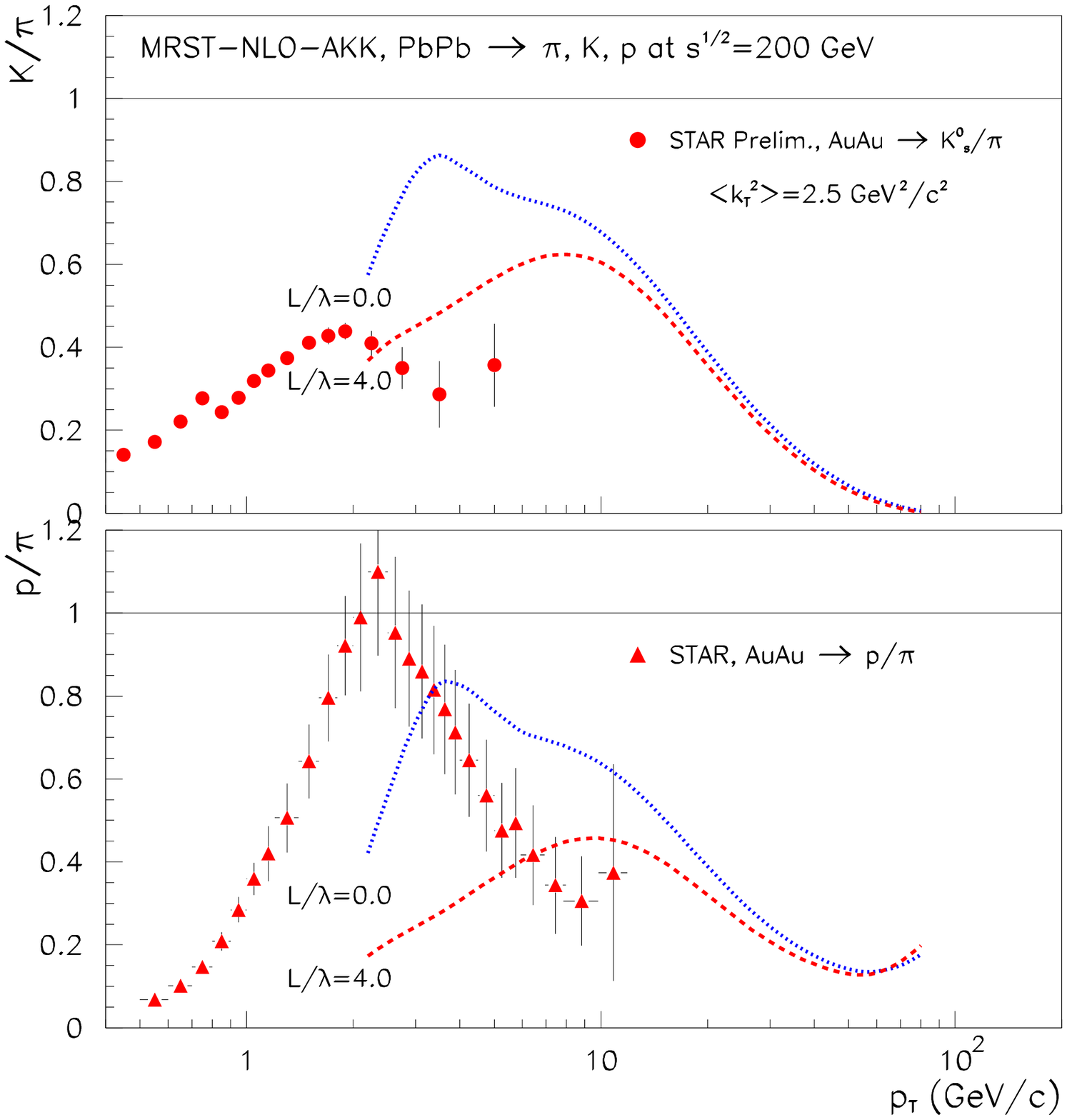}
\includegraphics[width=7.5truecm,height=8.0truecm]{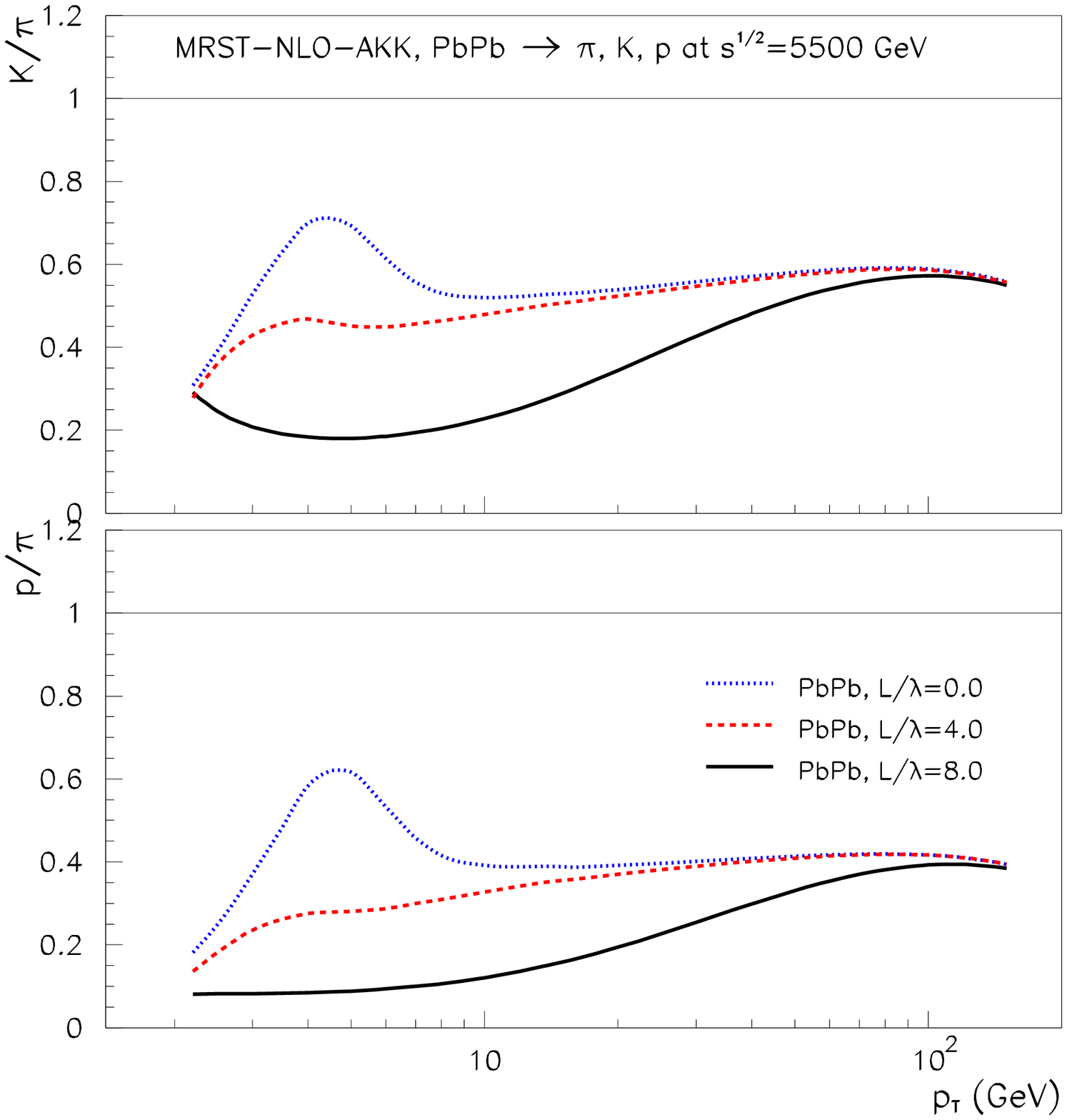}
\end{center}
\vspace*{-1.0truecm}
\caption{Calculated charge-averaged $K/\pi$ and $p/\pi$ ratios in  
$AA$ collisions at RHIC and LHC energies. RHIC curves are compared to 
STAR\cite{STAR_data,Abelev:2006jr} data at $\sqrt{s}=$ 200 $A$GeV.}
\label{fig1}
\end{figure}

Here we calculate hadron ratios in our next-to-leading order pQCD 
improved parton model based on Ref.\cite{bgg_qm06} %~\cite{Yi02} 
with intrinsic transverse momenta, determined by the expected c.m. 
energy evolution along the lines of Ref.\cite{bgg_qm06}. The presented ratios 
are based on $\pi$, $K$ and $p$ spectra which were calculated by AKK 
fragmentation functions~\cite{AKK1}. First we compare calculated particle ratios to the 
data of the STAR collaboration measured in $AuAu$ collisions at 
$\sqrt{s}= 200$ $A$GeV RHIC energy \cite{STAR_data,Abelev:2006jr}. Predictions 
for high-$p_T$ hadron ratios at RHIC and at LHC energies in most central 
($0-10\%$) $PbPb$ collisions are also shown in Fig. \ref{fig1}.   

%\newpage

On the {\sl left panel} of Fig. \ref{fig1}, particle ratios are compared 
to $AuAu$ 
collisions at $\sqrt{s}=200$ $A$GeV STAR $K/\pi$ ({\sl dots}) and $p/\pi$ 
({\sl triangles}) data. The agreement between the RHIC 
data and the calculations at RHIC energy can be considered acceptable 
at $p_T \gtrsim 5$ 
GeV/c, with an opacity of $L/\lambda=4$. However, at lower momenta, where 
pQCD is no longer reliable, the ratios differ from the calculated curves.  
 
The {\sl right panel} shows calculations for $PbPb$ collisions for 
$\sqrt{s}= 5.5$ $A$TeV energy. Using a simple 
$\dd N / \dd y \sim 1500-3000$ estimation, we expect a 
$L/\lambda \approx 8$ opacity in most central $PbPb$ collisions. 
For comparison, we plotted the $L/\lambda = 0$ and $4$ values 
also. The lower- and intermediate-$p_T$ variation of the hadron ratios 
arise from the different strengths of the jet quenching for quark and gluon 
contributions\cite{KaonJet}. Due to the quark dominated fragmentation, the 
difference disappears at high-$p_T$ in the ratios.

%%%%%%%%%%%%%%%%%%%%%%%%%%

\subsection{$\pi^0$
fixed p$_{\bot}$ suppression and elliptic flow at LHC}

{\it A. Capella, E. G. Ferreiro, A. Kaidalov and K. Tywoniuk}

{\small
Using a final state interaction model which describes the data on these two observables, at RHIC, we make predictions at the LHC -- using the same cross-section and $p_{\bot}$-shift. The increase in the medium density between these two energies (by a factor close to three) produces an increase of the fixed $p_{\bot}$ $\pi^0$ suppression by a factor 2 at large $p_{\bot}$ and of $v_2$ by a factor 1.5.
}

\subsubsection{$\pi^0$ fixed $p_{\bot}$ suppression} 
Final state interaction (FSI) effects have been observed in $AA$ collisions. They are responsible of strangeness enhancement, $J/\psi$ supression, fixed $p_{\bot}$ supression, azimuthal asymmetry, ... Is it the manifestation of the formation of a new state of matter or can it be described in a FSI model with no reference to an equation of state, thermalization, hydrodynamics, ...~? We take the latter view and try to describe all these obseervables within a unique formalism~: the well known gain and loss differential equations. We assume \cite{Capella:2004xj} that, at least for particles with $p_{\bot}$ larger than $<p_{\bot} >$, the interaction with the hot medium produces a $p_{\bot}$-shift $\delta p_{\bot}$ towards lower values and thus the yield at a given $p_{\bot}$ is reduced. There is also a gain term due to particles produced at $p_{\bot} + \delta p_{\bot}$. Due to the strong decrease of the $p_{\bot}$-distributions with increasing $p_{\bot}$, the loss is much larger than the gain. Asuming boost invariance and dilution of the densities in $1/\tau$ due to longitudinal expansion, we obtain 
\beq
\label{1e}
{\tau dN_{\pi^0} (b, s, p_{\bot}) \over d\tau}= - \sigma N(b,s) \left [ N_{\pi^0}(b,s,p_{\bot} ) - N_{\pi^0}(b,s,p_{\bot} + \delta p_{\bot})\right ]
\eeq

\begin{center}
\begin{figure*}
\begin{minipage}[t]{78mm}
\epsfxsize=7.5cm
\epsfysize=7.45cm
\centerline{\epsfbox{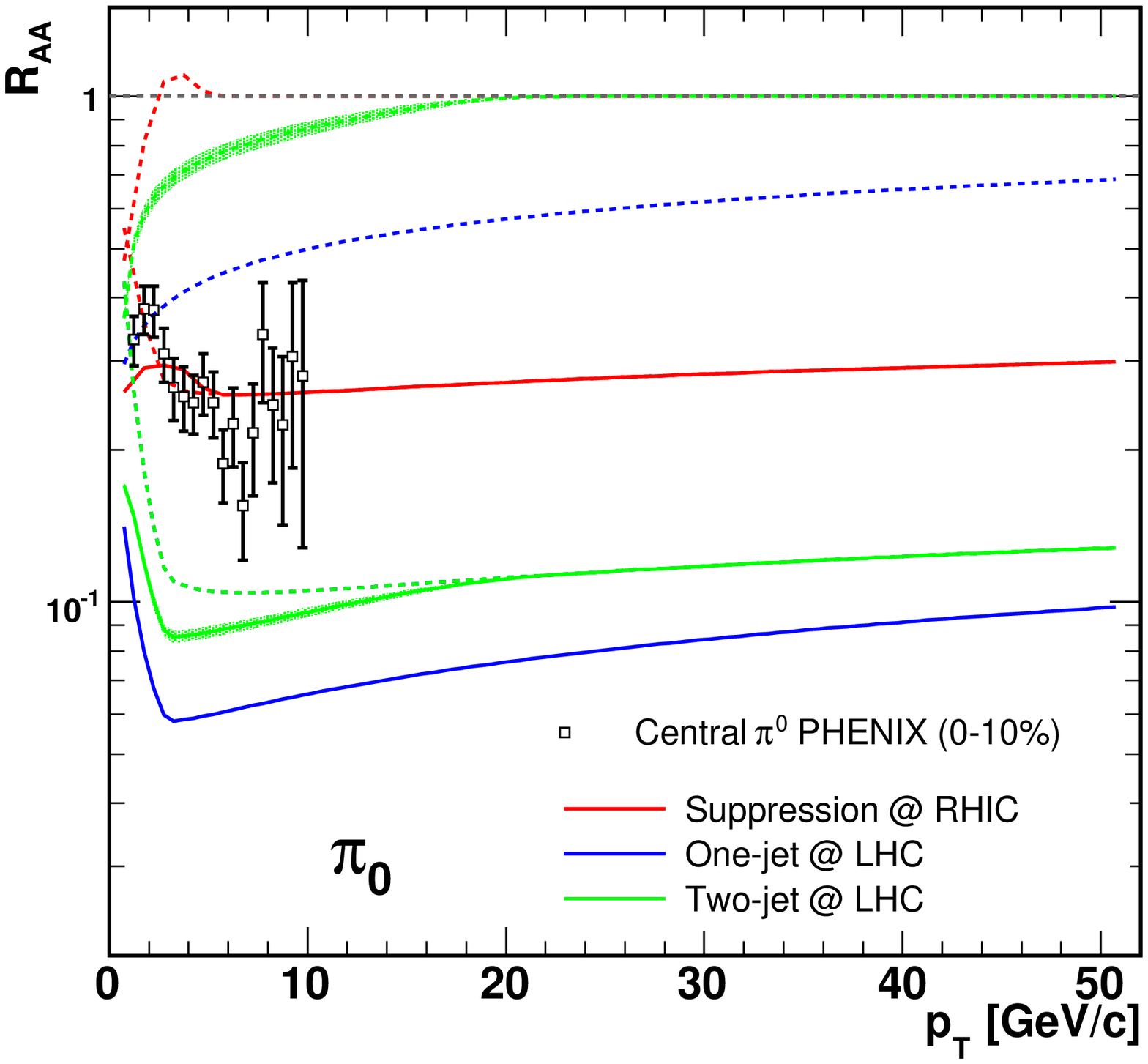}}
\vskip -0.25cm
\caption{
From up to down: RHIC initial, 2 LHC initial, 
RHIC final,
LHC FSI, LHC FSI+shadowing.} 
\label{ferreiroacf1}
\end{minipage}
\hspace{\fill}
\begin{minipage}[t]{78mm}
\vskip -7.45cm
\epsfxsize=7.5cm
\epsfysize=7.45cm
\centerline{\epsfbox{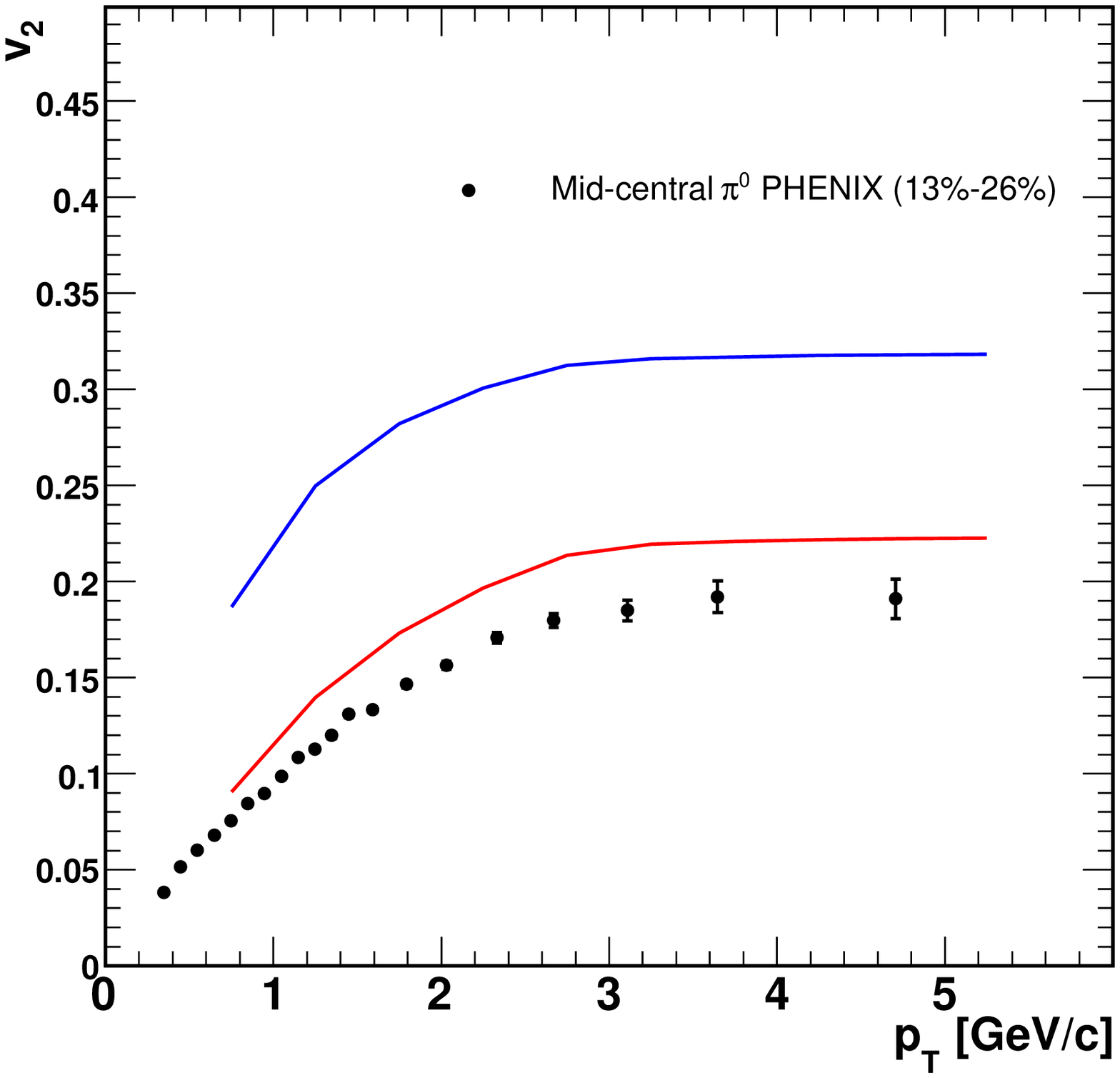}}
\vskip -0.25cm
\caption{$v_2$ 
for $\pi^0$ %in Au+Au collisions
%pions in mid-central Au+Au collisions 
%13 \%-26 \%, 
at 
%$\sqrt{s} = 200$~GeV 
RHIC (lower curve) and LHC (upper curve).}
%. The same quantity is shown for LHC (upper curve) for the same centrality class.}
\label{ferreiroacf2}
\end{minipage}
\vskip -0.3cm
\end{figure*}
\end{center}

Here $N \equiv dN/dy d^2s$ is the transverse density of the medium and $N_{\pi^0}$ the corresponding one of the $\pi^0$ \cite{Capella:2007bw}. This has to be integrated between initial time $\tau_0$ and freeze-out time $\tau_f$. The solution deepnds only on $\tau_f/\tau_0$. We use $\sigma = 1.4$~mb at both energies and $\delta p_{\bot} = p_{\bot}^{1.5}/20$ for $p_{\bot} < 2.9$~GeV and $\delta p_{\bot} = p_{\bot}^{0.8}/9.5$ for $p_{\bot} > 2.9$~GeV \cite{Capella:2006fw}. Eq. (\ref{1e}) at small $\tau$ describes an interaction at the partonic level. Indeed, here the densities are very large and the hadrons not yet formed. At later times the interaction is hadronic. Most of the effect takes place in the partonic phase. We use a single (effective) value of $\sigma$ for all values of the proper time $\tau$. 
The results at RHIC and LHC are given in Fig.~\ref{ferreiroacf1}. At LHC only shadowing \cite{Capella:2007bw} has been included in the initial state. The suppression is given by the dashed line. It coincides with $R_{AA}$ for $p_{\bot}$ large enough -- when shadowing and Cronin efffects are no longer present. The LHC suppression is thus a factor of two larger than at RHIC.
%\begin{figure}%[htb]
%\begin{center}
%\epsfig{file=figAC1.eps, width=6cm}
%\end{center}
%\caption{$\pi^0$ fixed $p_{\bot}$ suppression for central (0-10 \%) collisions. 
%The dashed (upper) and solid (medium) lines are for $AuAu$ at RHIC. The other five lines are for $PbPb$ collisions at LHC. The lower dashed line is for the FSI alone (i.e. initial $R_{AA} = 1$). The other lines correspond to initial and final $R_{AA}$ with shadowing of the $\pi^0$ in two kinematical configurations -- one-jet (lower) and two-jets (upper).}
%\end{figure}

\subsubsection{Elliptic flow} 
Final state interaction in our approach gives rise to a positive $v_2$ \cite{Capella:2006fw} (no need for an equation of state or hydro). Indeed, when the $\pi^0$ is emitted at $\theta_R = 90^{\circ}$ its path length is maximal (maximal absorption). In order to compute it we assume that the density of the hot medium is proportional to the path length $R_{\theta_R}(b,s)$ of the $\pi^0$ inside the interaction region determined by its transverse position $s$ and its azimuthal angle $\theta_R$. Hence, we replace $N(b,s)$ by $N(b,s) R_{\theta_R}(b,s)/$\break\noindent $<R_{\theta_R}(b,s)>$ where $R_{\theta_R}$ is the $\pi^0$ path length and $<>$ denotes its average over $\theta_R$. (In this way the averaged transverse density $N(b,s)$ is unchanged). The suppression $S_{\pi^0}(b,s)$ depends now on $\theta_R$ and $v_2$ is given by
\beq
\label{2e}
v_2(b, p_{\bot}) = 
{\displaystyle{\int} d\theta_R S_{\pi^0} (b, p_{\bot}, \theta_R) 
\cos 2 \theta_R \over \displaystyle{\int} d\theta_R S_{\pi^0} (b, p_{\bot}, \theta_R)}
\eeq
%
%\noi 
The results at RHIC and LHC are presented in Fig. \ref{ferreiroacf2}.
%\begin{figure}%[htb]
%\begin{center}
%\epsfig{file=figAC2.eps, width=6cm}
%\end{center}
%\caption{$v_2$ for pions in mid-central $Au+Au$ collisions. 13 \%-26 \%, at $\sqrt{s} = 200$~GeV (lower curves). The same quantity is shown for LHC (upper curve) for the same centrality class.}
%\end{figure}
%\includegraphics{figAC2.eps}

\subsection{Energy dependence of jet transport parameter}
\label{casalderrey}

{\it J. Casalderrey-Solana and X. N. Wang}

{\small
We study the evolution and saturation of the gluon distribution
function in the quark-gluon plasma as probed by a propagating parton and its effect
on the computation of the jet quenching or transport parameter $\hat{q}$.
For hard probes, this evolution at
small $x=Q^2_s/6ET$ leads to a jet energy dependence of $\hat{q}$
}
\vskip 0.5cm

Within the picture of multiple parton scattering in QCD, the
energy loss for an energetic parton propagating in a dense medium
is dominated by induced gluon bremsstrahlung. Taking into account
of the non-Abelian Landau-Pomeranchuck-Midgal (LPM) interference,
the radiative parton energy loss \cite{bdmps},
\begin{equation}
\Delta E= \frac{\alpha_s N_c}{4} \hat q_R L^2,
\label{bdmps}
\end{equation}
is found to depend
% quadratically on the medium length $L$ and a
on the
jet transport or energy loss parameter $\hat{q}$ which
 describes the averaged transverse momentum transfer squared
per unit distance (or mean-free-path). Here $R$ is the color representation
of the propagating parton in $SU(3)$.
% and $\rho$ is the color charge density of
%the medium.

The transport parameter $\hat q_R$
experienced by a propagating parton
can be defined in terms of the unintegrated gluon distributions
$\phi_k(x,q_T^2)$ of the color sources in the quark-gluon plasma,
\begin{eqnarray}
\hat q_R&=&\frac{4\pi^2C_R}{N_c^2-1} \rho \int_0^{\mu^2}
\frac{d^2q_T}{(2\pi)^2}\int dx
% \nonumber \\
%&&\hspace{0.2in} \times
 \delta(x-\frac{q_T^2}{2p^-\langle k^+\rangle})
\alpha_s(q_T^2)\phi(x,q_T^2),
\label{eq:xg1}
\end{eqnarray}
where $\langle k^+\rangle$ is the average energy of the color sources
and $\phi(x,q_T^2)$ is the corresponding average unintegrated gluon
distribution function per color source.
The integrated gluon distribution is
\begin{equation}
xG(x,\mu^2)=\int_0^{\mu^2} \frac{d^2q_T}{(2\pi)^2}  \phi(x,q_T).
\end{equation}

Since we are interested in the determination of $\hat{q}_R$ at
large jet energies, we need to know the unintegrated parton
distribution $\phi(x,q^2_T)$ in \Eq{eq:xg1} at small $x\sim
\left<q^2_T\right>/6 E T$. For a large path length, the typical
total momentum transfer, $\hat{q} L$, which will set the scale of
the process, is also large.
These scales lead to the evolution of the gluon distribution function.
In the medium, this evolution may be modified due to the interaction
of the radiated gluons with thermal partons. 
However, since the medium effects
are of the order of $\mu_D<< T$, we neglect those at hard scales. 
Given that both the scale and the rapidity are large, we describe the
the (linear) vacuum evolution in the double
logarithmic approximation (DLA) \cite{Gribov:1981ac}. 
The thermal gluon distribution function at a scale $\mu^2=T^2$ is 
determined via the hard thermal loop approximation and it is used 
as an initial condition for the evolution. As in vacuum, the growth of
the gluon distribution function leads to saturation which tame this growth for 
scales $\mu^2<Q_s^2$. The saturation scale is estimated from the linearly evolved
distribution. The details of the computation can be found in 
\cite{Casalderrey-Solana:2007sw}.

\begin{figure}
\begin{center}
%  \hspace{1cm}
 \includegraphics[width=7cm]{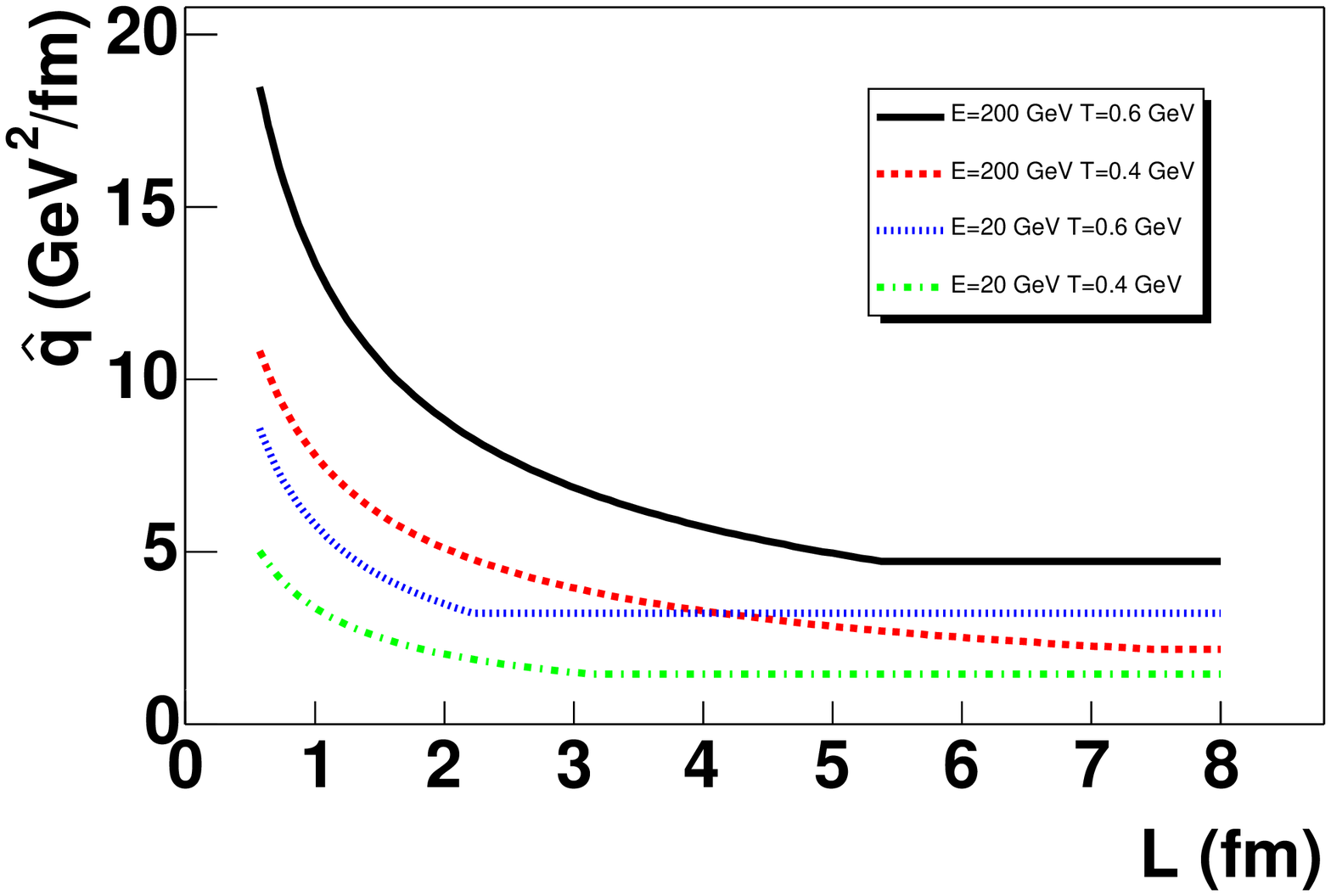}
 \includegraphics[width=7cm]{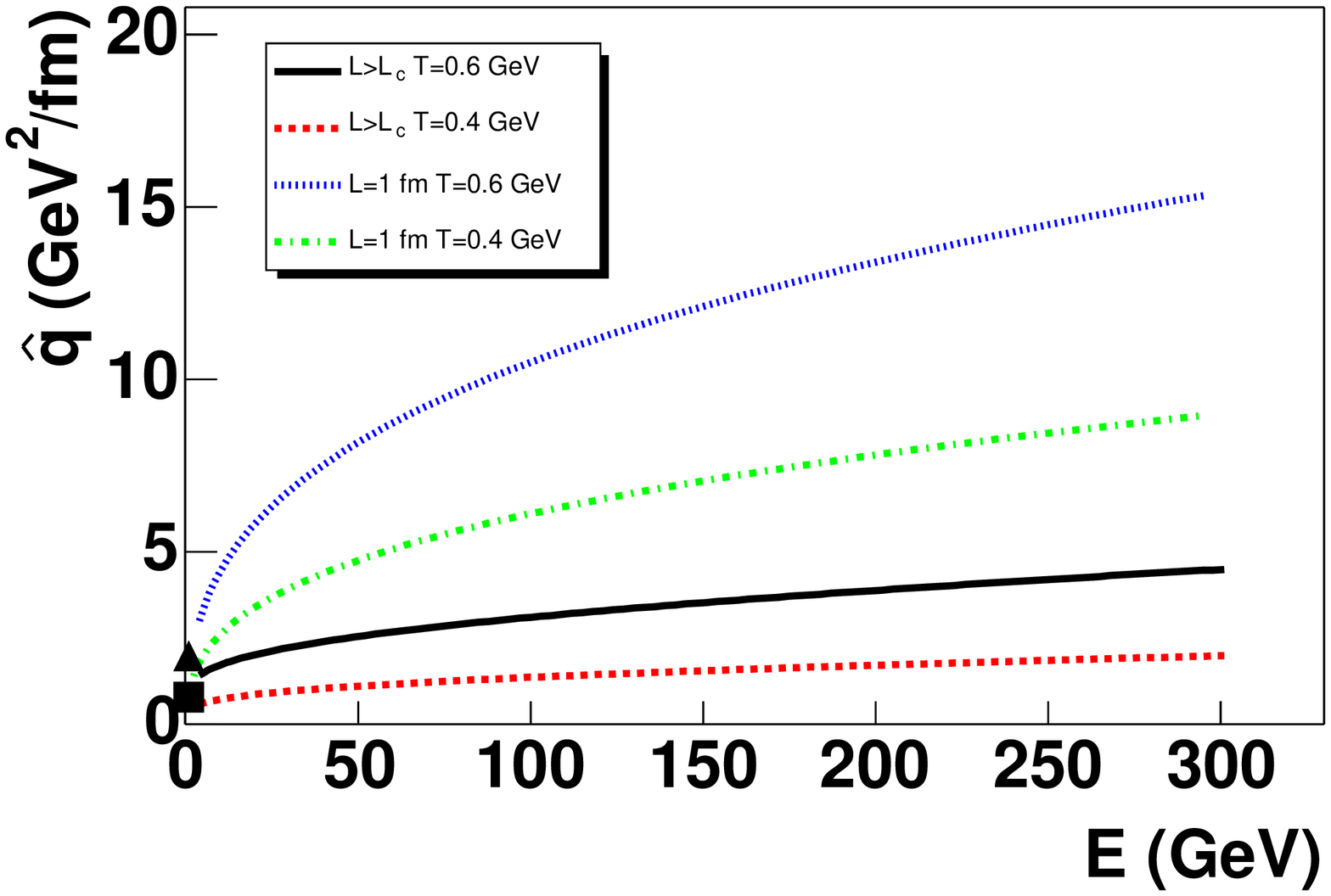}
 \caption{\label{qhatdep}
Jet quenching parameter $\hat{q}$ as a
function of the path length (left) 
and jet energy (right).
The square (triangle) marks the value of $\hat{q}$ for thermal particle 
at $T=0.4 $ GeV ($T=0.6 $ GeV).
Significant corrections to the energy dependence are expected at low energy 
which should approach their thermal value at $E=3\,T$.
}
\end{center}
\end{figure}

The evolution leads to a jet
energy dependence of the transport parameter that is stronger than
any power of logarithmic dependence. The saturation effect also
gives rise to a non-trivial length dependence of the jet transport
parameter. These two features are shown in Figure \ref{qhatdep}, 
where we compute the transport parameter for $T=0.4$ GeV (RHIC) and 
$T=0.6$ GeV (LHC). In both cases, the energy dependence of $\hat{q}$ is 
significant, leading to a factor of 2 difference between jets of 20 and 200 GeV. 
This difference is larger for small jet path lengths. The computation also
shows that $\hat{q}$ grows as the path length decreases. Both dependences 
translate into different amount of radiative energy loss \Eq{bdmps}. 
Let us note, however, that the derivation of \Eq{bdmps} assumes a constant
$\hat{q}$; thus, the relation between the radiative energy loss and the transport
parameter should be revisited for an energy/length dependent $\hat{q}$.

\subsection{PQM prediction of $\RAA($\pt$)$ and $\RCP($\pt$)$ at midrapidity in Pb--Pb collisions at the LHC}
\label{dainese}

{\it A.~Dainese,
        C.~Loizides and
	G.~Pai\'c}

{\small
The Parton Quenching Model (PQM) couples the BDMPS-SW quenching weights
for radiative energy loss with a realistic description of the nucleus--nucleus
collision geometry, based on the Glauber model.
We present the predictions for the nuclear modification factors, 
in Pb--Pb relative to pp collisions ($\RAA$) 
and in central relative to peripheral 
Pb--Pb collisions ($\RCP$), 
of the transverse momentum distributions of light-flavour hadrons 
at midrapidity.  
}
\vskip 0.5cm

The Parton Quenching Model (PQM)~\cite{Dainese:2004te}, which combines
the pQCD \mbox{BDMPS-SW} framework 
for the probabilistic calculation of parton energy loss 
in extended partonic matter 
of given size and density~\cite{Salgado:2003gb} with a realistic description of the 
collision overlap geometry (Glauber model) in a static medium, was shown to 
describe the transverse momentum and centrality dependence of 
the leading particle suppression in Au--Au collisions at top RHIC energy.
The model has one single parameter that sets the scale of the BDMPS
transport coefficient $\hat{q}$, hence of the medium density.
The parameter has been tuned~\cite{Dainese:2004te} on the basis of the $\RAA$ data at 
$\sqrtsNN=200~\gev$, that indicate a transport coefficient in the range 
4--14~$\gev^2/\fm$.
We scale the model parameter to LHC energy assuming its proportionality to the 
expected volume-density of gluons $n^{\rm g}$. Using the value of $n^{\rm g}$ 
predicted for the LHC 
by the EKRT saturation model~\cite{Eskola:1999fc} 
(which gives $\dNdy\simeq 3000$), 
we obtain $\hat{q}\simeq 25$--$100~\gev^2/\fm$.

In PQM we obtain the leading-particle suppression in nucleus--nucleus 
collisions by calculating the hadron-level 
transverse momentum distributions in a Monte Carlo approach.
The `event loop' that we iterate is the following:
1) Generation of a parton, quark or gluon, with \pt$>5~\gev$, 
      using the PYTHIA event generator in pp mode with 
      CTEQ\,4L parton distribution functions; nuclear shadowing is neglected,
      since it's effect is expected to be small above 5--10~GeV in \pt;
      the \pt-dependence of the quarks-to-gluons ratio is taken 
      from PYTHIA.
2) Sampling of a parton production point and propagation direction 
      in the transverse plane,
      according to the density of binary collisions, and 
      determination of the in-medium path length and of the 
      path-averaged $\hat{q}$, the inputs  
      for the calculation of the quenching weights, i.e. the energy-loss
      probability distribution $P(\Delta E)$.
3) Sampling of an energy loss $\Delta E$ according to $P(\Delta E)$
   (non-reweighted case~\cite{Dainese:2004te})
      and definition of the new parton
      transverse momentum, \pt$-\Delta E$;
4) Fragmentation of the parton to a hadron using the 
      leading-order Kniehl-Kramer-P\"otter (KKP) fragmentation 
      functions. 
Quenched and unquenched \pt distributions are obtained including or 
excluding the third step of the chain. The nuclear modification 
factor $R_{\rm AA}($\pt$)$ is given by their ratio.

%\section{Nuclear modification factors at the LHC}

\begin{figure}[!t]
  \begin{center}
    \includegraphics[width=0.49\textwidth]{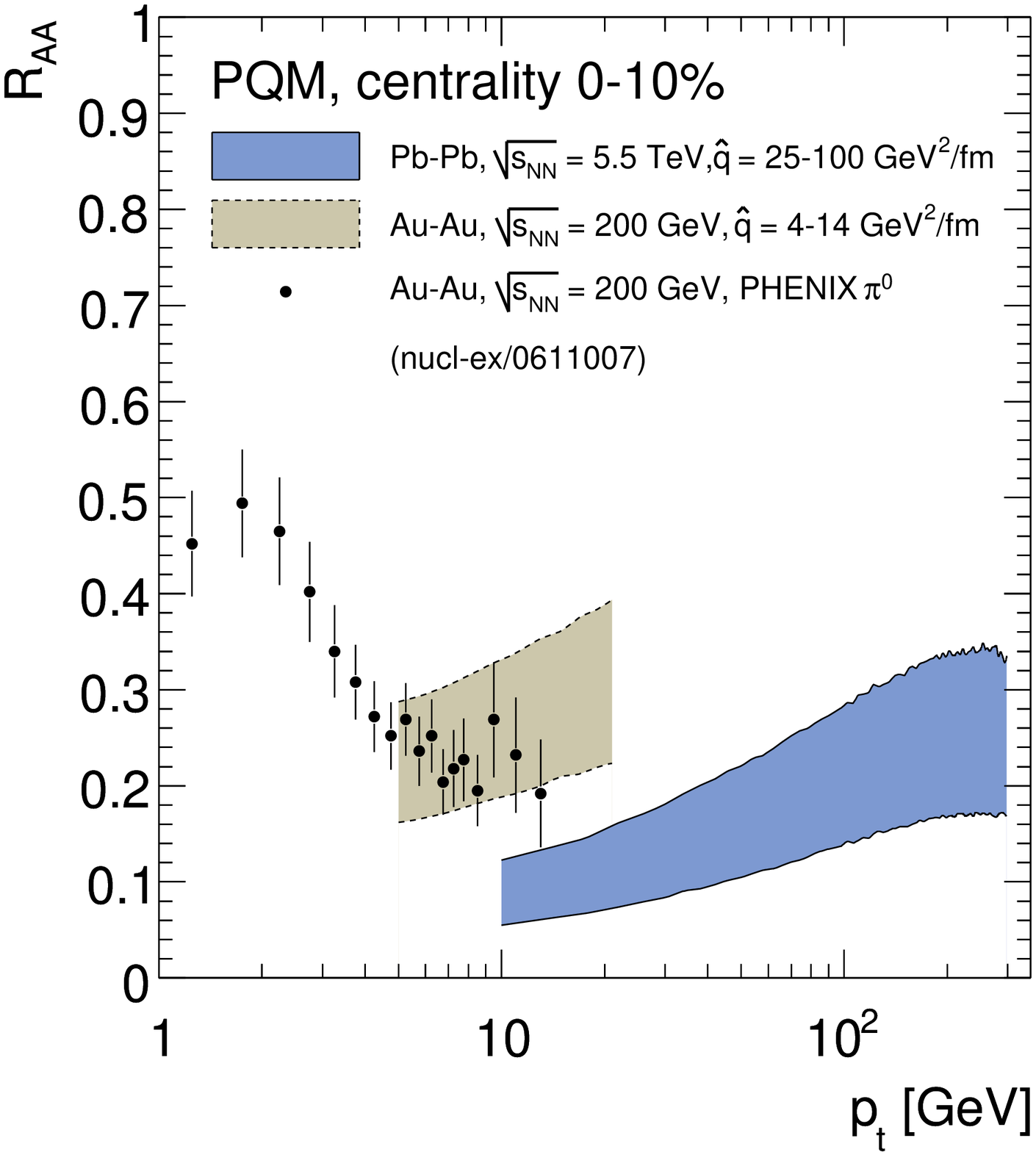}
    \includegraphics[width=0.49\textwidth]{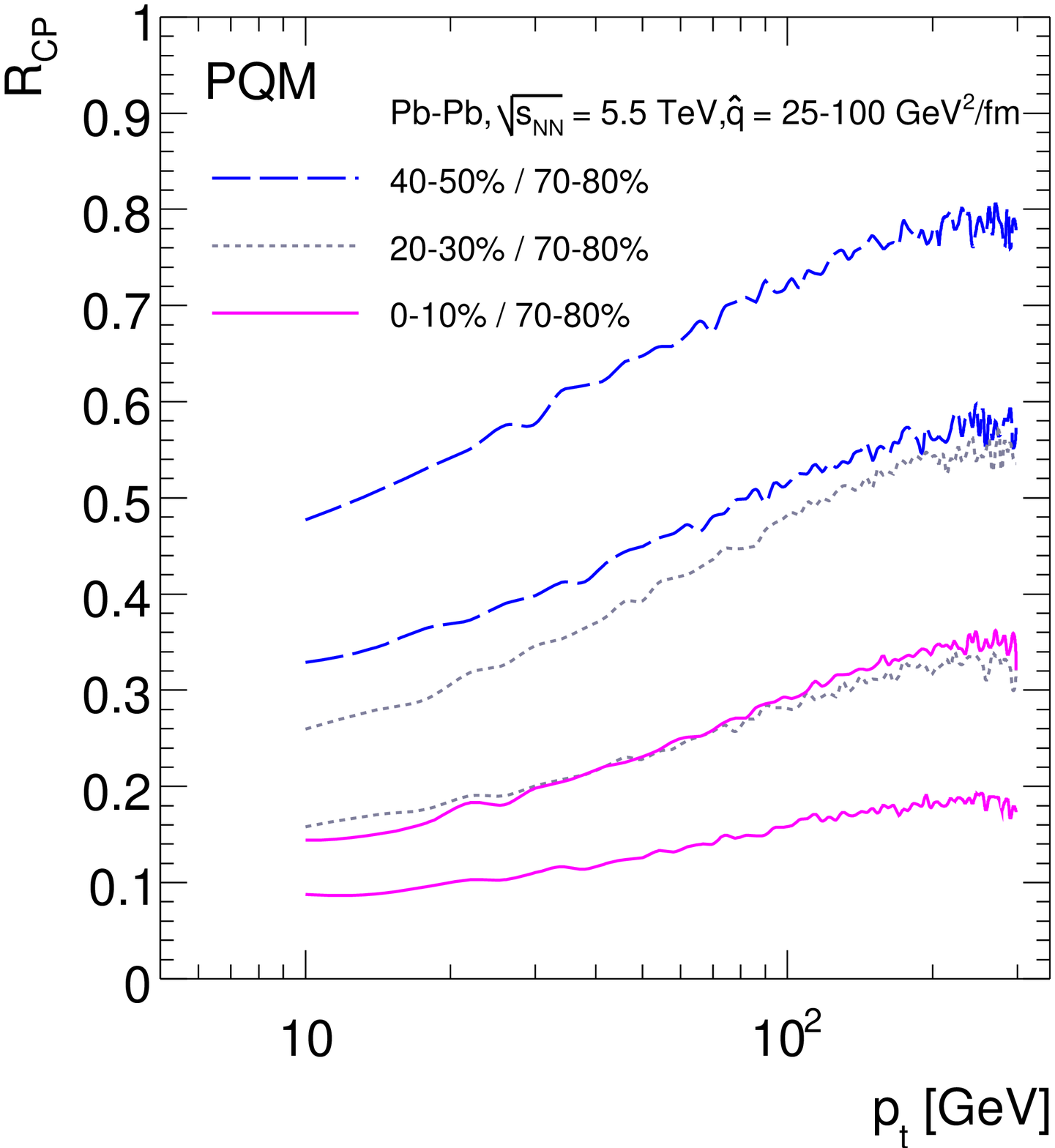}
    \caption{Left: $\RAA($\pt$)$ for central
             Pb--Pb collisions at $\sqrtsNN=5.5~\tev$ and central Au--Au 
             collisions at $\sqrtsNN=200~\gev$. The PHENIX $\pi^0$ data 
             are shown with statistical errors only and they have a 
             10\% normalization systematic error~\cite{Adler:2006bw}.
             Right: $\RCP($\pt$)$ for
             Pb--Pb collisions at $\sqrtsNN=5.5~\tev$.} 
    \label{fig:PQMRAARCPlhc}
  \end{center}
\end{figure}

The left-hand panel of Fig.~\ref{fig:PQMRAARCPlhc} shows 
the \pt-dependence of the 
$\RAA$ nuclear modification factor in 0--10\% central Pb--Pb at 
$\sqrtsNN=5.5~\tev$ relative to pp. 
The $\RAA$ for central Au--Au collisions at top RHIC energy is also shown 
and compared to $\pi^0$ data from the PHENIX experiment~\cite{Adler:2006bw}. 
PQM predicts for central Pb--Pb at the LHC a very slow increase of $\RAA$
with \pt, from about $0.1$ at $10~\gev$ to about $0.2$ at $100~\gev$.  
The right-hand panel of the figure shows the 
$\RCP$ central-to-peripheral nuclear
modification factor for different centrality classes relative to the
peripheral class 70--80\%.

\subsection{Effect of dynamical QCD medium on radiative heavy quark energy loss}
\label{djordjevic}
{\it M. Djordjevic and U. Heinz}

{\small
The computation of radiative energy loss in a dynamically screened QCD medium 
is a key ingredient for obtaining reliable predictions for jet quenching in 
ultra-relativistic heavy ion collisions. We calculate, to first order in the 
number of scattering centers, the energy loss of a heavy quark traveling 
through an infinite and time-independent QCD medium consisting of dynamical 
constituents. We show that the result for a dynamical medium is almost twice 
that obtained previously for a medium consisting of randomly distributed 
static scattering centers. A quantitative description of jet suppression in 
RHIC and LHC experiments thus must correctly account for the dynamics of the 
medium's constituents. 
}
\vskip 0.5cm

Heavy flavor suppression is considered to be a powerful tool to study the 
properties of a QCD medium created in ultra-relativistic heavy ion 
collisions~\cite{Brambilla}. The suppression results from the energy loss of 
high energy partons moving through the plasma~\cite{MVWZ:2004}. Therefore, the 
reliable computations of heavy quark (collisional and radiative) energy loss 
mechanisms are essential for the reliable predictions of jet suppression. 

However, currently available heavy quark radiative energy loss studies suffer 
from one crucial drawback: The medium induced radiative energy loss is 
computed in a QCD medium consisting of randomly distributed but static 
scattering centers (``static QCD medium''). Within such approximation, the 
collisional energy loss is exactly zero, which is contrary to the recent 
calculations~\cite{MD_Coll} that showed that the collisional 
contribution is important and comparable to the radiative energy loss.
Due to this, it became necessary to obtain the heavy quark radiative 
energy loss in a dynamical QCD medium, and to test how good is the static 
approximation in these calculations.
 
In this proceeding, we report on a first important step, the calculation of
heavy quark radiative energy loss in an infinite and time-independent 
QCD medium consisting of dynamical constituents. By comparing with the
static medium calculation this permits us to qualitatively assess the 
importance of dynamical effects on radiative energy loss.

We compute the medium induced radiative energy loss for a heavy quark to first 
(lowest) order in number of scattering centers. To compute this process, we 
consider the radiation of one gluon induced by one collisional interaction 
with the medium. In distinction to the static case, we take into account that 
the collisional interactions are exhibited with dynamical (moving) medium 
partons. To simplify the calculations, we consider an infinite QCD medium and 
assume that the on-shell heavy quark is produced at time $x_0=-\infty$, i.e. 
we consider the Bethe-Heitler limit. The calculations were performed by using 
two Hard-Thermal Loop approach, and are presented in~\cite{DH_Dyn}. As the
end result, we obtained a closed expression for the radiative energy loss in 
dynamical QCD medium. This result allows us to compare the radiative energy 
loss in dynamical and 
static QCD medium, from which we can observe two main differences. First, there
is an $\mathcal{O}(15\%)$ decrease in the mean free path which increases the 
energy loss rate in the dynamical medium by $\mathcal{O}(20\%)$. Second, there
is a change in the shape and normalization of the emitted gluon spectrum. This 
second difference leads to an additional significant increase of the heavy 
quark energy loss rate and of the emitted gluon radiation spectrum by about 
$50\%$ for the dynamical QCD medium. The numerical results are briefly 
discussed below.

\begin{figure}[ht]
\vspace*{4cm} 
\includegraphics{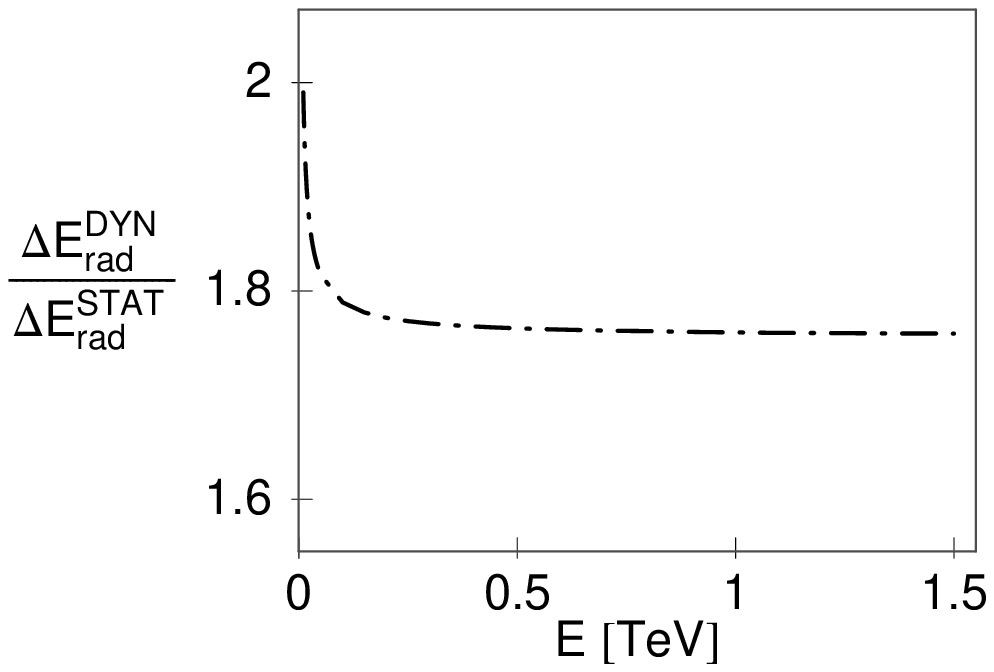}
\includegraphics{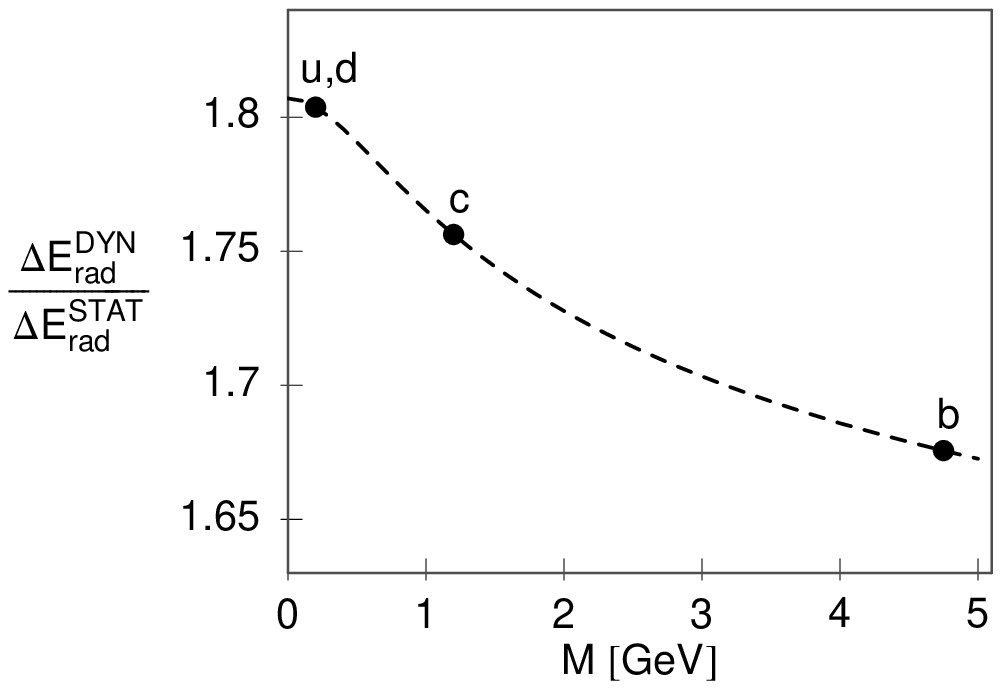}
\caption{ 
{\sl Left panel:} Ratio of the fractional radiative 
energy loss in dynamical and static media for charm quarks as a function 
of initial quark energy $E$. 
{\sl Right panel:} Asymptotic value of the radiative energy loss ratio for 
high energy quarks as a function of their mass, with marks indicating the 
light, charm and bottom quarks. For the parameter values, see~\cite{DH_Dyn}. }
\label{ratioLHC}
\end{figure}

Left panel of the Fig.~\ref{ratioLHC} shows the energy loss ratio between 
dynamical and static media for charm quark under the LHC conditions. We see 
that the ratio is almost independent of the momentum $p$ of the fast charm 
quark, saturating at $\simeq 1.75$ above $p\gtrsim 100$\,GeV and being even 
somewhat larger at smaller momenta. The dynamical enhancement persists at 
constant level to the largest possible charm quark energies. Therefore, we can 
conclude that there is no quark energy domain where the assumption of static 
scatterers in the medium becomes a valid approximation. Further, the mass of 
the fast quark plays only a minor role for its energy loss. The right panel in 
Fig.~\ref{ratioLHC} shows  the asymptotic energy loss ratio for very high 
energy quarks as a function of the quark mass. While the dynamical enhancement 
is largest for light quarks, the difference between light and bottom quarks is 
only about $15\%$, and $b$ quarks still suffer about 70\% more energy loss in
a dynamical medium than in one with static scattering centers.

In summary, we obtained an important qualitative conclusion that the 
constituents of QCD medium can {\it not} be approximated as static scattering 
centers in the energy loss computations. Therefore, the dynamical effects 
have to be included for the reliable prediction of radiative energy loss and 
heavy flavor suppression in the upcoming high luminosity RHIC and LHC 
experiments.

\subsection{Charged hadron $R_{\A\A}$ as a function of $p_T$ at LHC}
\label{s:Renk}

{\em T. Renk and K. J. Eskola}

{\small
We compute the nuclear suppression factor $R_{\A\A}$ for charged hadrons within 
a radiative energy loss picture using a hydrodynamical evolution to describe 
the soft medium inducing energy loss. 
A minijet + saturation picture provides initial conditions for LHC energies and 
leading order perturbative QCD (LO pQCD) is used to compute the parton spectrum 
before distortion by energy loss.
}
\vskip 0.5cm

We calculate the suppression of hard hadrons induced by the presence of a soft 
medium produced in central Pb-Pb collisions at $\sqrtsnn=5.5$~TeV at the LHC. 
Note that this prediction depends on knowledge of the medium. 
In the present calculation, the medium evolution is likewise predicted and has 
to be confirmed before the suppression can be tested. 
Note further that the calculation is only valid where hadron production is 
dominated by fragmentation and that it cannot be generalized to the suppression 
of jets since the requirement of observing a hard hadron leads to showers in 
which the momentum flow is predominantly through a single parton. 
This is not so for jets in which the momentum flow is shared on average among 
several partons (which requires a different framework).

We describe the soft medium evolution by the boost-invariant hydrodynamical 
model discussed in~\cite{Eskola:2005ue} where the initial conditions for LHC 
are computed from perturbative QCD+saturation~\cite{Eskola:1999fc}. 
Our calculation for the propagation of partons through the medium follows the 
BDMPS formalism for radiative energy loss using quenching weights~\cite{%
  Salgado:2003gb}.
Details of the implementation can be found in~\cite{Renk:2006pk}. 

The probability density $P(x_0, y_0)$ for finding a hard vertex at the 
transverse position ${\bf r_0} = (x_0,y_0)$ and impact parameter ${\bf b}$ is 
given by the normalized product of the nuclear profile functions.
We compute the energy loss probability $P(\Delta E)_{\rm path}$ for any given 
path from a vertex through the medium by evaluating the line integrals
\[
\omega_c({\bf r_0}, \phi) = \int_0^\infty\!{\rm d}\xi\,\xi\hat{q}(\xi) \quad  
{\rm and} \quad 
\langle\hat{q}L\rangle({\bf r_0},\phi)=\int_0^\infty\!{\rm d}\xi\,\hat{q}(\xi).
\]
Along the path where we assume the relation
\[
\hat{q}(\xi) = K \cdot 2 \cdot\epsilon^{3/4}(\xi) (\cosh\rho-\sinh\rho\cos\alpha)
\]
between the local transport coefficient $\hat{q}(\xi)$, the energy density 
$\epsilon$ and the local flow rapidity $\rho$ as given in the hydrodynamical 
model. 
The angle $\alpha$ is between flow and parton trajectory. 
We view  the constant $K$ as a tool to account for the uncertainty in the 
selection of $\alpha_s$ and possible non-perturbative effects increasing the 
quenching power of the medium (see~\cite{Renk:2006pk}) and adjust it such that 
pionic $R_{\A\A}$ for central Au-Au collisions at RHIC is described. 
The result for LHC is then an extrapolation with $K$ fixed.

Using the numerical results of~\cite{Salgado:2003gb}, we obtain 
$P(\Delta E; \omega_c, R)_{\rm path}$ for $\omega_c$ and 
$R=2\omega_c^2/\langle\hat{q}L\rangle$.
From this distribution given a single path, we can define the averaged energy 
loss probability distribution $P(\Delta E)\rangle_{T_{\A\A}}$ by averaging over 
all possible paths, weighted with the probability density $P(x_0, y_0)$ for 
finding a hard vertex in the transverse plane.

We consider all partons as absorbed whose energy loss is formally larger than 
their initial energy.
The momentum spectrum of produced partons is calculated in LO pQCD. 
The medium-modified perturbative production of hadrons is obtained from the 
convolution
\[
{\rm d}\sigma_{\rm med}^{\A\A\to h+X} = 
\sum_f{\rm d}\sigma_{\rm vac}^{\A\A\to f +X} \otimes 
\langle P(\Delta E)\rangle_{T_{\A\A}} \otimes D_{f\to h}^{\rm vac}(z, \mu_F^2)
\]
with $D_{f\to h}^{\rm vac}(z, \mu_F^2)$ the fragmentation function. 
From this we compute the nuclear modification factor $R_{\A\A}$ as
\[
R_{\A\A}(p_T,y) = 
\frac{{\rm d}N^h_{\A\A}/{\rm d}p_T{\rm d}y}{T_{\A\A}({\bf b})
{\rm d}\sigma^{pp}/{\rm d}p_T{\rm d}y}.
\]

\begin{figure}[htb]
\centerline{\includegraphics[width=7.2cm]{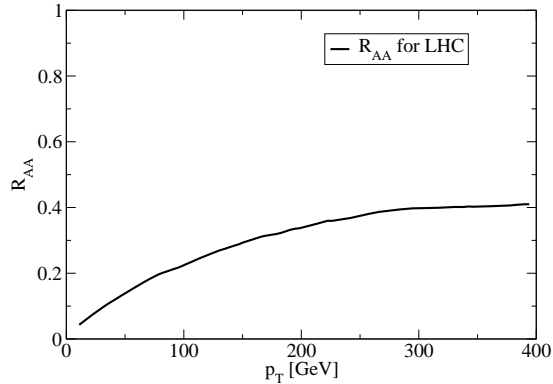}} 
\caption{\label{fig:Renk-fig1}Expectation for the $p_T$ dependence of the 
  nuclear suppression factor $R_{\A\A}$ for charged hadrons in central Pb-Pb 
  collisions at midrapidity at the LHC.}
\end{figure}

Figure~\ref{fig:Renk-fig1} shows the expected behaviour of $R_{\A\A}$ with 
hadronic transverse momentum $p_T$ at midrapidity. 
On quite general grounds, we expect a rise of $R_{\A\A}$ with $p_T$ for any 
energy loss model in which the energy loss probability does not strongly depend 
on the initial parton energy as more of the shift in energy becomes accessible 
(see~\cite{Renk:2006pk}). 
The detailed form of the rise is then sensitive to the form of 
$P(\Delta E)\rangle_{T_{\A\A}}$.

\subsection{Nuclear suppression of jets and $R_{\A\A}$  at the LHC}
\label{s:Qin}

{\em G. Y. Qin, J. Ruppert, S. Turbide, C. Gale and S. Jeon}

{\small 
The nuclear modification factor $R_{\A\A}$ for charged hadron production at the 
LHC is predicted from jet energy loss induced by gluon bremsstrahlung. 
The Arnold, Moore, and Yaffe~\cite{Arnold:2001ba,Arnold:2001ms,Arnold:2002ja}  
formalism is used, together with an ideal hydrodynamical 
model~\cite{Eskola:2005ue}.
}
\vskip 0.5cm

We present a calculation of the nuclear modification factor $R_{\A\A}$ for 
charged hadron production as a function of $p_T$ in Pb+Pb collisions at 
$\sqrtsnn=5.5$~TeV in central collisions at mid-rapidity at the LHC. 
The net-energy loss of the partonic jets is calculated by applying the Arnold, 
Moore, and Yaffe (AMY) formalism to calculate gluon bremsstrahlung~\cite{%
  Arnold:2001ba,Arnold:2001ms,Arnold:2002ja}.
The details of jet suppression relies on an understanding of the nuclear 
medium, namely the temperatures and flow profiles that are experienced by 
partonic jets while they interact with partonic matter at $T \geq T_c$. 
Our predictions use a boost-invariant ideal hydrodynamic model with initial 
conditions calculated from perturbative QCD + saturation~\cite{Eskola:2005ue,%
  Eskola:1999fc}. 
It is emphasized that the reliability of this work hinges on the validity of 
hydrodynamics at the LHC.  
It has been verified that $R_{\A\A}$ for $\pi_0$ production as a function of 
$p_T$ as obtained in the same boost-invariant ideal hydrodynamical model 
adjusted to Au+Au collisions at $\sqrtsnn=0.2$~TeV~\cite{Eskola:2005ue} is in
agreement with preliminary data from PHENIX in central collisions at RHIC (and 
the result is very close to the one obtained in 3D hydrodynamics presented 
in~\cite{Qin:2007ys}). 
In AMY the strong coupling constant $\alpha_s$ is a direct measure of the 
interaction strength between the jet and the thermalized soft medium and is the 
only quantity not uniquely determined in the model, once the temperature and 
flow evolution is fixed by the initial conditions and subsequent hydrodynamical 
expansion. 
We found that assuming a constant $\alpha_s=0.33$ describes the experimental 
data in most central collisions at RHIC. 
It is conjectured that $\alpha_s$ should not be changed very much at the LHC 
since the initial temperature is about twice larger than the one at RHIC 
whereas $\alpha_s$ is only logarithmically dependent on temperature. 
We present results for $\alpha_s=0.33$ and 0.25.

For details of the calculation of nuclear suppression, we refer the reader 
to~\cite{Qin:2007ys}. 
The extension to the LHC once the medium evolution and $\alpha_s$ are fixed is 
straightforward. 
The initial jets are produced with an initial momentum distribution of jets 
computed from pQCD in the factorization formalism including nuclear shadowing 
effects. 
The probability density $\mathcal{P}_{\A\B}(\vec{r}_\bot)$ of finding a hard jet 
at the transverse position $\vec{r}_\bot$ in central $\A$+$\B$ collisions is 
given by the normalized product of the nuclear thickness functions, 
$\mathcal{P}_{\A\B}(\vec{r}_\bot) = 
  {T_{\A}(\vec{r}_\bot)T_{\B}(\vec{r}_\bot})/{T_{\A\B}}$ and is calculated for Pb+Pb 
collisions. 
The evolution of the jet momentum distribution 
$P_j(p,t)={\rm d}N_j(p,t)/{\rm d}p{\rm d}y$ in the medium is calculated by 
solving a set of coupled rate equations with the following generic form,
\[
\frac{{\rm d}P_j(p,t)}{{\rm d}t} = \sum_{ab} \int\!{\rm d}k 
  \left[P_a(p+k,t) \frac{{\rm d}\Gamma^a_{jb}(p+k,p)}{{\rm d}k{\rm d}t} -
  P_j(p,t)\frac{{\rm d}\Gamma^j_{ab}(p,k)}{{\rm d}k{\rm d}t}\right],
\]
where ${\rm d}\Gamma^j_{ab}(p,k)/{\rm d}{\rm d}t$ is the transition rate for the 
partonic process $j\to a+b$ which depends on the temperature and flow profiles 
experienced by the jets traversing the medium. 
The hadron spectrum ${\rm d}N^h_{\A\B}/{\rm d}^2p_T{\rm d}y$ is obtained by the 
fragmentation of jets after their passing through the medium. 
The nuclear modification factor $R_{\A\A}$ is  computed as
\[
R^h_{\A\A}(\vec{p}_T,y) = \frac{1}{N_{\rm coll}} 
  \frac{{\rm d}N^h_{\A\A}/{{\rm d}^2p_T{\rm d}y}}
    {{\rm d}N^h_{pp}/{\rm d}^2p_T{\rm d}y}.
\]
\begin{figure}[htb]
\centerline{\includegraphics[width=8cm]{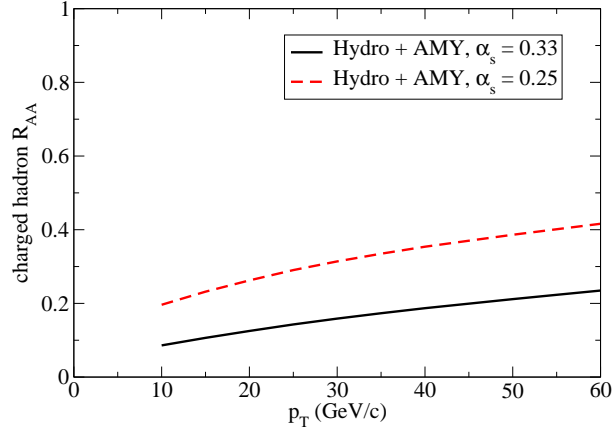}}
\caption{\label{fig:Qin-fig1} The $p_T$ dependence of the nuclear modification 
  factor $R_{\A\A}$ for charged hadrons in central Pb+Pb collisions at 
  mid-rapidity at the LHC.}
\end{figure}

In figure~\ref{fig:Qin-fig1} we present a prediction for charged hadron 
$R_{\A\A}$ as a function of $p_T$ at mid-rapidity for central collisions at the 
LHC. 
We consider that these two values of $\alpha_s$ define a sensible band of 
physical parameters.

\subsection
{Perturbative jet energy loss mechanisms: learning from RHIC, extrapolating to LHC}

{\it S. Wicks and M. Gyulassy}
 
 {\small
 In many recent papers, collisional energy loss has been found to be of the same order as radiative energy loss for parameters applicable to the QGP at RHIC. As the temperature and jet energy dependence of collisional energy loss differs from that of radiative loss, the interpretation of the results at RHIC affects our extrapolation to predictions for the LHC. We present results from a hybrid collisional plus radiative model, combining DGLV radiative loss with HTL-modified collisional loss, including the fluctuation spectrum for small numbers of collisions and gluons emitted.
}
\vskip 0.5cm

%\subsubsection{Introduction}
Collisional energy loss is an essential component of the physics of high momentum partonic jets traversing the quark-gluon plasma~\cite{Mustafa:2004dr,Wicks:2005gt}. If we do not properly understand the energy loss mechanisms that are important at RHIC, then we cannot accurately extrapolate in medium density and jet energy to make predictions for the LHC. 

WHDG~\cite{Wicks:2005gt} made a first attempt at including both collisional and DGLV radiative energy loss processes. A simple model of the collisional energy loss was used: leading log average loss with a Gaussian distribution around this average, the width given by the fluctuation-dissipation theorem. For the short lengths of interest in the QGP fireball ($\approx 0-6$fm), we expect a jet to undergo only a small number of significant collisions. But the fluctuation spectrum for this will be different than that implemented in the WHDG model: instead, the distinctly non-Gaussian fluctuation in energy loss in 0,1,2,3 collisions is necessary. We present here results and predictions from an improved hybrid radiative plus collisional energy loss model which include a full evaluation of these fluctuations.

\begin{figure}[!ht] %[eRAA]
\centerline{\hbox{
$\begin{array}{ccc}
\epsfig{file=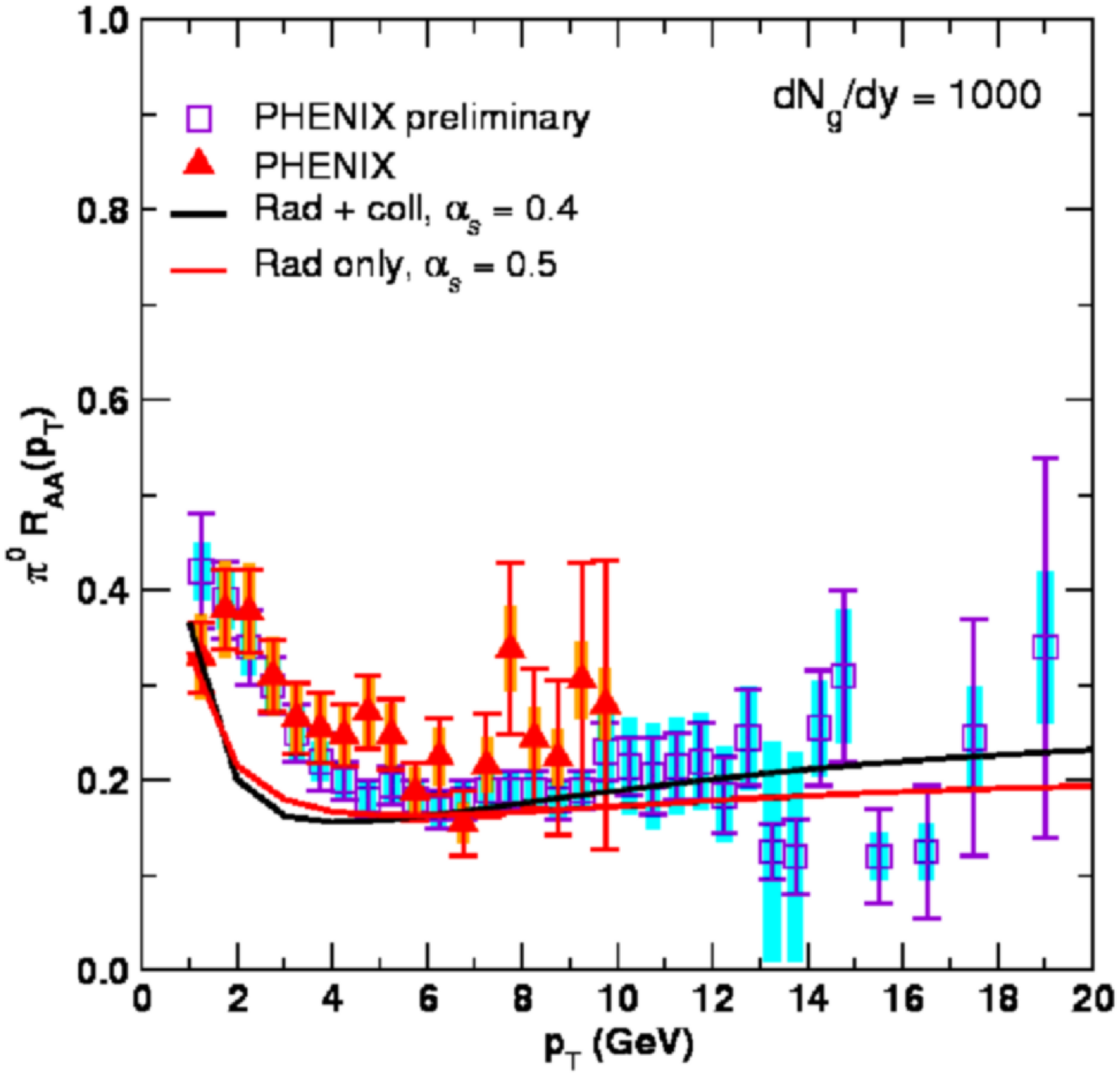,height=2.0in,,clip=,angle=0} &
\epsfig{file=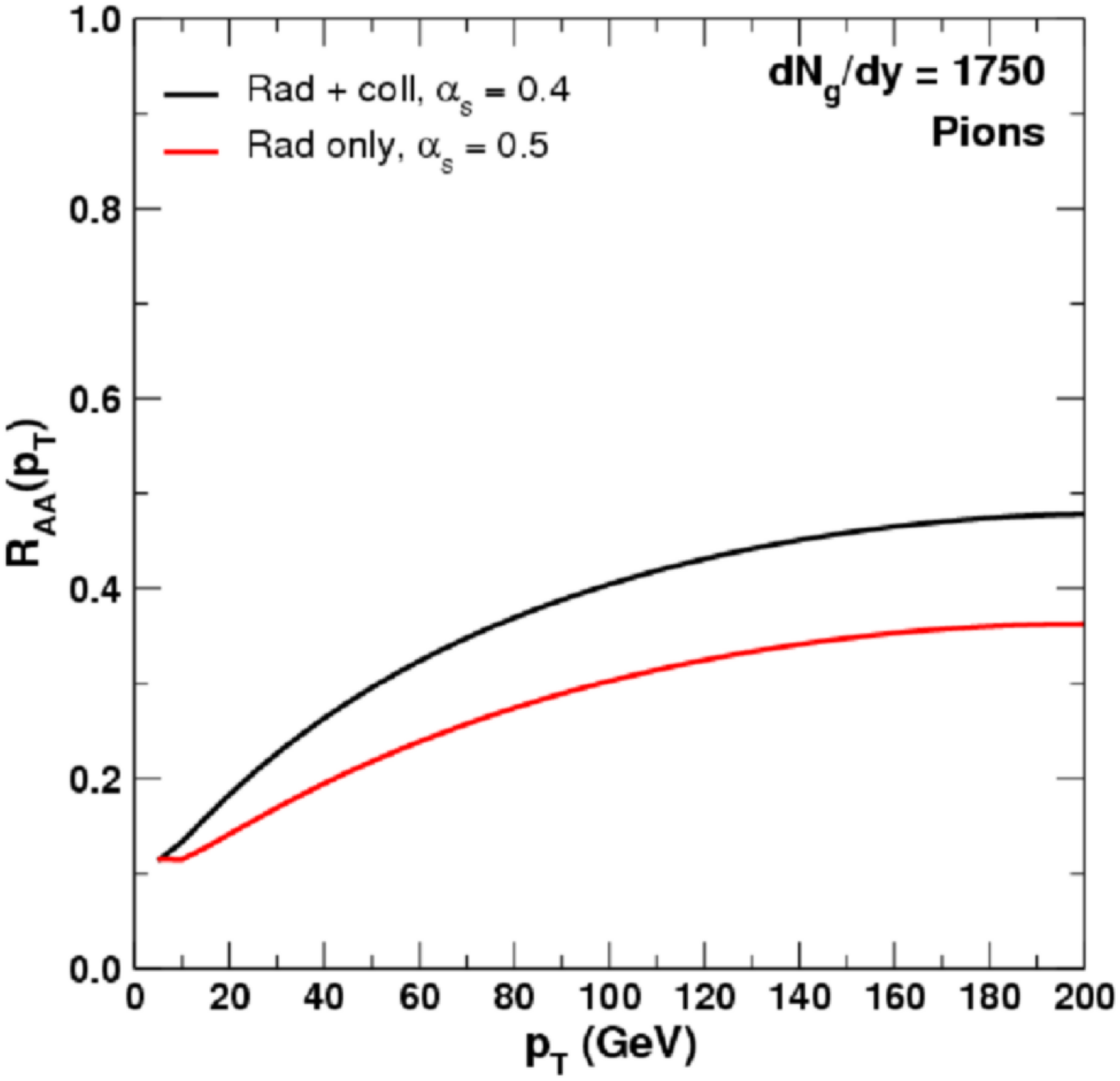,height=2.0in,,clip=,angle=0} &
\epsfig{file=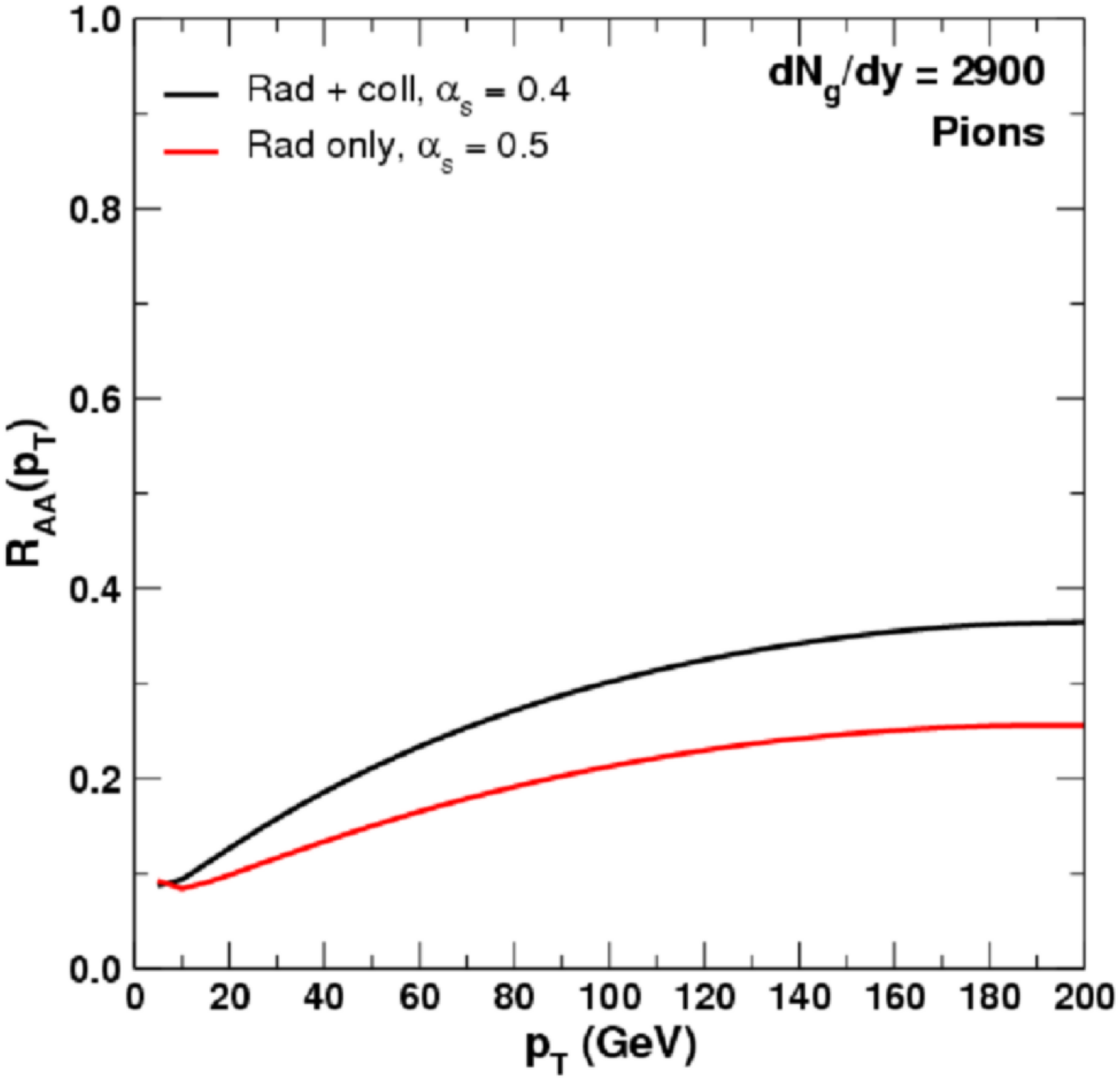,height=2.0in,,clip=,angle=0}
\end{array}$
}}
\caption{ 
$R_{AA}$ for pions for RHIC (left) and two possible densities at LHC (central and right). The main result, the hybrid radiative plus collisional energy loss model for $\alpha_s = 0.4$, is compared to a radiative energy loss alone model for an increase value of the strong coupling. The increased range in momentum available at the LHC enables the different slopes of the two models to be seen.
}
\label{plot:lhcpions2}
\end{figure}

A significant uncertainty in the model is the use of a fixed strong coupling constant. In WHDG, a canonical value $\alpha_s = 0.3$ was used, validated by the fitting of the pion $R_{AA}(p_T)$ at RHIC. Here, for a fixed density $dN_g/dy = 1000$, an increased coupling $\alpha_s=0.4$ is necessary to do the same. In fact, if the collisional component of the energy loss is neglected completely, a further increased coupling of $\alpha_s = 0.5$ would be necessary, as shown in the left-hand side of \fig{plot:lhcpions2}. Both values, while large, are still in a possible perturbative kinematical region, and are evaluated with medium densities constrained by the total entropy and multiplicity of the collision.

\begin{figure}[!ht] %[eRAA]
\centerline{\hbox{
$\begin{array}{cc}
\epsfig{file=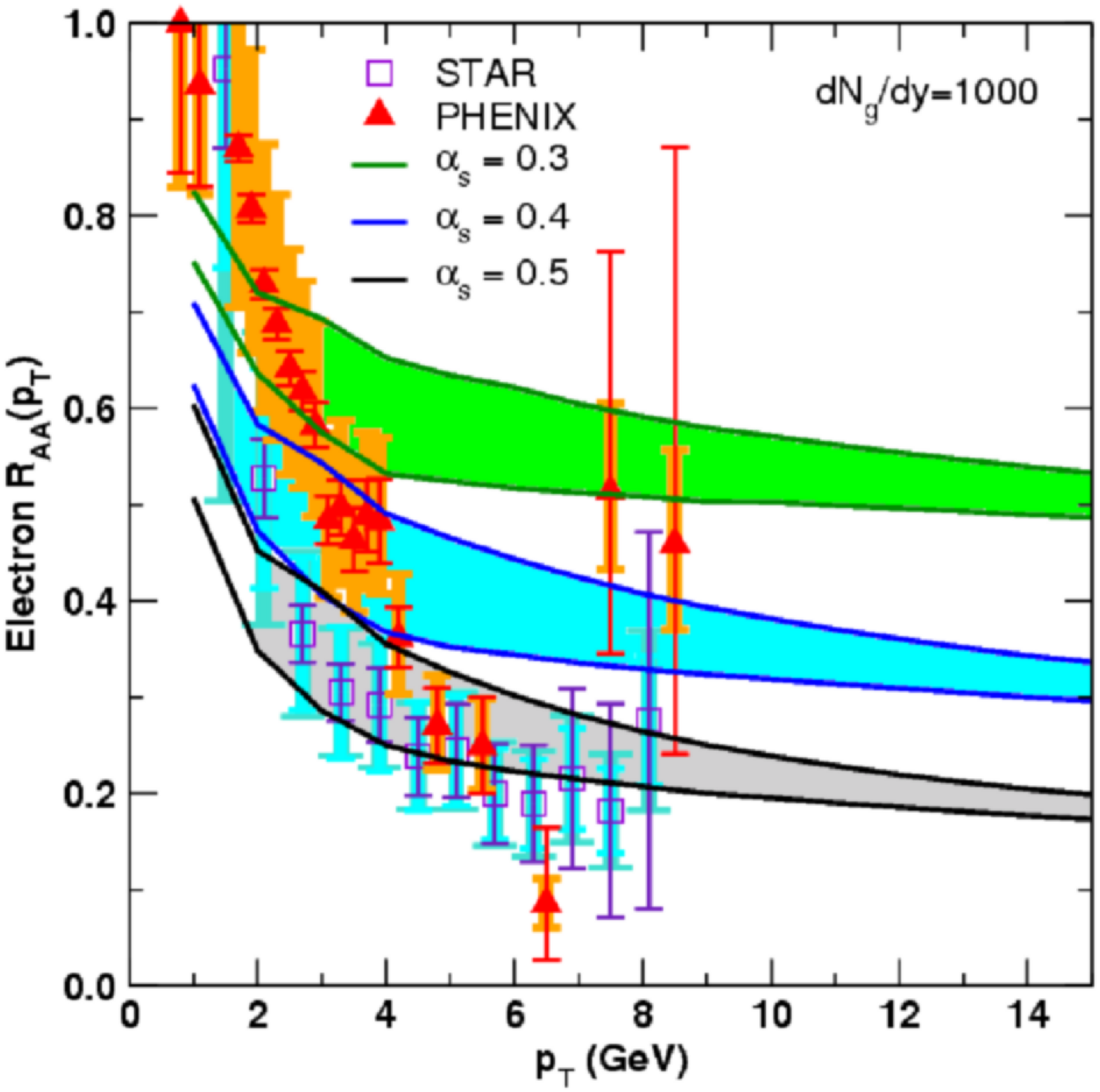,height=2.0in,,clip=,angle=0} &
\epsfig{file=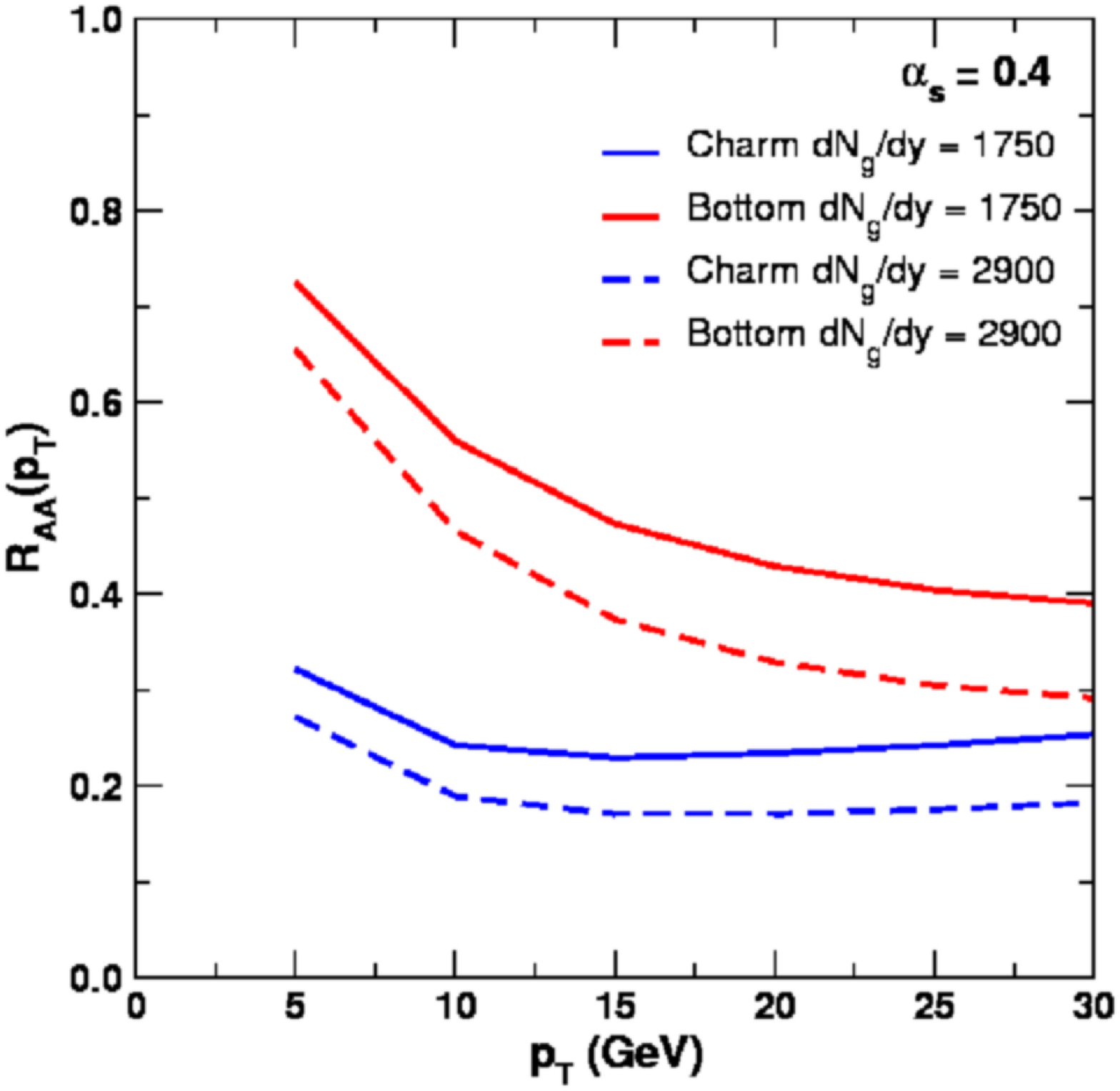,height=2.0in,,clip=,angle=0}
\end{array}$
}}
\caption{ 
$R_{AA}$ for observable products of heavy quark jets at RHIC (electrons - left) and two possible densities at the LHC (D and B mesons - right). There is considerable uncertainty in the perturbative production of c and b jets. This shows up in the results for electrons at RHIC in the large uncertainty band, $\pm 0.1$ or greater - as the ratio of c to b jets is very uncertain. However, the uncertainty in D and B meson $R_{AA}$s is small (approximately $\pm 0.02$) - the different slopes on the individual spectra have very little effect on the meson $R_{AA}$ results.
}
\label{plot:lhcheavies}
\end{figure}

Is it possible to differentiate between these two scenarios: one including collisional loss, the other neglecting it but increasing the coupling to compensate? Staying with the most simple observables, single particle inclusives in central collisions, we have three dependences to test: the dependence on medium density, jet energy and jet mass. The first is tested by the predicted increased density of the medium to be produced at the LHC (consistency between the left plot and either the central or right plot in \fig{plot:lhcpions2}). The very high momentum reach available for measurements involving gluon and light quark jets is valuable for the second (radiative versus radiative plus collisional in the central and right hand plots of \fig{plot:lhcpions2}), and the separate detection of D and B mesons gives us the third (as in \fig{plot:lhcheavies}). All these together will provide very strong constraints on the energy loss models, even before considering observables beyond the single-particle inclusives.

There are still significant uncertainties in the energy loss model. The most important kinematic region for evaluation of both the collisional and radiative energy losses are for energy and momentum transfers from the medium greater than $\mu_D$, the Debye mass. This is the region in which we know the least about the physics of the QGP: beyond the HTL region, but before a region of vacuum gluon exchange, especially if processes close to the light-cone of the exchanged gluon are important (as it is for collisional energy loss). This can produce an uncertainty of $\approx 50\%$ for the average collisional loss, which may not be correlated with an uncertainty in the radiative loss. Such large uncertainties affect both the explanation of RHIC data and the extrapolation to the LHC.

\subsection{Jet evolution in the Quark Gluon Plasma}
\label{s:Pirner}

{\em H. J. Pirner, K. Zapp, J. Stachel, G. Ingelman and J. Rathsman}

{\small
Jet evolution is calculated in the leading log approximation. 
We solve the evolution equation for the branching of gluons in vacuum, using a 
triple differential fragmentation function $D(x,Q^2,p_\perp^2)$. 
Adding an extra scattering term for evolution in the quark gluon plasma we 
investigate the influence of the temperature of the plasma on the differential 
cross section of partons ${\rm d}N/{\rm d}\ln(1/x)$ in a jet of virtuality 
$Q^2=(90\mbox{ GeV})^2$. 
Due to scattering on the gluons in the plasma the multiplicity increases, the 
centroid of the distribution shifts to smaller $x$ values and the width narrows.
}
\vskip 0.5cm

The evolution equation for the transition of a parton $i$ with virtuality $Q^2$ 
and momentum $(1,k_\perp)$ into a parton $j$ with momentum $(z,p_\perp)$ can be 
constructed in leading logarithmic approximation~\cite{Bassetto:1979nt}.  
In a dense medium they are modified due to the possibility that the parton is 
scattered. 
The scatterings change the transverse momentum of the leading fast parton by 
giving it $\vec q_\perp$ kicks, but they do not change the mass scale or 
virtuality of the fast parton. 
The lifetime of a virtual parton can be estimated as 
${\rm d}\tau= E/Q_0^2 ({\rm d}Q^2/Q^2)$ using the uncertainty principle ($E$ is 
the parton energy and $Q_0^2$ is the infrared scale). 
Evolving along a straight line path in a homogeneous plasma with a density of 
gluons $n_g$ we obtain a modified evolution equation
\begin{eqnarray*}
\fl\lefteqn{Q^2\frac{\partial D_i^j(z,Q^2,\vec p_\perp)}{\partial Q^2} = }\\ 
\fl & & \quad
  \frac{\alpha_s(Q^2)}{2\pi}\!\int_z^1\!\frac{{\rm d}u}{u} P_i^r(u,\alpha_s(Q^2))
  \frac{{\rm d}^2\vec q_\perp}{\pi}\, 
  \delta\!\left(\!u(1-u)Q^2-\frac{Q_0^2}{4}-q_\perp^2\right) 
  D_r^j\!\left(\frac{z}{u},Q^2,\vec p_\perp-\frac{z}{u}\vec q_\perp\right) \\
\fl & & \qquad + S(z,Q^2,\vec p_\perp)
\end{eqnarray*}
with the scattering term $S(z,Q^2,p_\perp)$
\begin{eqnarray*}
\fl\lefteqn{S(Q^2,\vec p_\perp) = } \\
\fl & & \quad
  \frac{z E n_g}{Q_0^2}\!\int_z^1\! {\rm d}w \int\!{\rm d}^2\vec q_\perp 
  \frac{{\rm d}\sigma_i^r}{{\rm d}^2\vec q_\perp} 
  \left[D_r^j(w,Q^2,\vec p_\perp-w \vec q_\perp) - D_r^j(z,Q^2,\vec p_\perp)\right]
  \delta\!\left(\!w- z-\frac{q_\perp^2}{2 m_g E}\right).
\end{eqnarray*}
The scattering term includes the probability for scattering into and out of the 
$p_\perp$ bin as well as the energy loss of the parton. 
The gluon mass in the plasma $m_g$ is related~\cite{Braun:2006vd} to the Debye 
mass $m_g=1/2 m_{\rm D}$.

There is an analytic solution for the $p_\perp$ integrated equation restricted 
to gluons, which give the dominant contribution to the multiplicity. 
The solution can be found via Mellin transformation in a similar fashion as in 
vacuum~\cite{Ellis:QCDbook}, the running of the coupling is taken into account.
\begin{figure}[ht]
\centering
\input{Pirner-fig1.pstex_t}
\caption{LHS: Multiplicity of two jets with invariant mass $Q^2$ in vacuum 
  (dashed line), at $T=0.8$~GeV (dotted line) and at $T=1.0$~GeV (full line) \\
  RHS: Differential multiplicity ${\rm d}N/{\rm d}\ln(1/x)$ of jet particles 
  inside a jet with invariant mass $Q^2=(90\mbox{ GeV})^2$ in vacuum (dashed 
  line) and at $T=1.0$~GeV (full line)}
 \label{fig:Pirner-fig1}
\end{figure}

One finds an increase of the
multiplicity with temperature and a shift of the centroid of the $\ln(1/x)$ distribution towards smaller $x$,
%One finds a behaviour which is close to the expectations, namely an increase of
%the multiplicity with temperature and a shift of the centroid of the $\ln(1/x)$
%distribution towards smaller $x$,
see figure~\ref{fig:Pirner-fig1}.
The width of the distribution, however, becomes smaller. 
It remains to be studied in vacuum how the choice of the parameters can be
optimized to the LEP data, for simplicity the above curves are calculated for 
$\Lambda_{\rm QCD}=250$~MeV. 
It is well known that in the evolution equation the QCD scale parameter may 
well be adjusted. 
Concerning the effects of the plasma, the form of the cross section and its 
dependence on $\alpha_s(Q^2)$ has to be further investigated. 
The results look encouraging and serve as an analytical model with which 
numerical Monte Carlo calculations can be compared. 

There has been a calculation of jet evolution in the modified leading log 
approximation~\cite{Borghini:2005mp} which has produced similar shapes for the 
differential multiplicity distribution. 
The advantage of our calculation is that it takes into account the scattering 
term explicitly and therefore gives results which depend on the plasma 
properties. 
The equation can also be used to investigate the $p_\perp$ broadening of the 
parton in the medium, since our input function contains the transverse momentum 
as an extra variable explicitly.

\noindent {\it Note added in proof:} The calculation described in the text has been undergoing several changes
during the last months. Therefore  we refer to a
forthcoming publication where these improvements are included.

\subsection{Pion and Photon Spectra at LHC}

{\it S. Jeon, I. Sarcevic and J. Jalilian-Marian}

{\small 
Using simple modification of jet fragmentation function that is tuned to
reproduce the RHIC $\pi^0$ data, we had previously predicted photon production 
at RHIC which is confirmed by recent PHENIX data. Using the same parameter set,
we predict hight $p_T$ pion and prompt photon spectra in Pb-Pb collisions at 
LHC.
}
\vskip 0.5cm

In perturbative QCD, the inclusive cross section
for pion production in a hadronic collision is given by: 
\[
\fl
E_\pi \frac{{\rm d}^3\sigma}{{\rm d}^3p_\pi}(\sqrt{s},p_\pi) =
\int {\rm d}x_{a}{\rm d}x_{b}{\rm d}z \sum_{i,j}F_{i}(x_{a},Q^{2})
F_{j}(x_{b},Q^{2}) D_{c/\pi}(z,Q^2_f) E_\pi 
\frac{{\rm d}^3\hat{\sigma}_{ij\to c X}}{{\rm d}^3p_\pi}
\]
where $i$ and $j$ label hadrons or nuclei and $a,b,c$ label partons.

In heavy-ion collisions, one needs to include nuclear effects.
In our model, we take the parton distribution function for a nucleus to be 
\[
F_{a/A}(x,Q^2,b_t)=T_A(b_t)\,S_{a/A}(x,Q^2)\,F_{a/N}(x,Q^2)
\]
where $T_A$ is the nuclear thickness function and $S_{a/A}$ is the shadowing
function (we use EKS98 parametrization).

Unfortunately, the interaction of parton-medium cannot be calculated within
perturbative QCD, but need to be modeled. 
The purpose of our model~\cite{Jeon:2002mf,Jeon:2002dv,Jeon:2002ca} is to be as 
simplistic as possible so that the essential nature of the energy loss process 
can manifest. To achieve this goal, we modify the fragmentation function in the 
following way~\cite{Wang:1996yh}
\[
z D_{c/\pi}(z,\Delta L,Q^2) =
 \sum_{n=0}^N P_a(n) z^a_n D^0_{c/\pi}(z^a_n,Q^2)
+\langle n_a\rangle z'_aD^0_{g/\pi}(z'_a,Q_0^2), 
\]
where $z^a_n=z/(1-n\epsilon_a/E_T)$, $z'_a=zE_T/\epsilon_a$, $N$ is the maximum 
number of collisions for which $z_n^a \le 1$ and $D^0_{c/\pi}$ is the hadronic 
fragmentation function. The second term comes from the emitted gluons each 
having energy $\epsilon_a$ on the average. The average number of scatterings 
within a distance $\Delta L$ is $\langle n_a\rangle = \Delta L/\lambda_a$. We 
take $\lambda_a=1$~fm and $\Delta L=R_A$. $P_a(n)$ is the Poisson distribution
function with $\langle n\rangle = (\Delta L/\lambda_a)$.

The three energy loss models we use are $\Delta E = 1.0\mbox{ GeV}$ (Const)
$\Delta E = \sqrt{E_{\rm LPM}E}$ (LPM) and $\Delta E = \kappa E$ (BH) per 
collision.  For RHIC, BH (Bethe-Heitler) gives best description of $\pi^0$ data,
and predictions for direct photons using the same energy loss is recently found 
to be in agreement with PHENIX data~\cite{Frantz:2007zz}. Within the same 
framework we present our predictions for the LHC.

\begin{figure}[htb]
\center{
\includegraphics[width=0.40\textwidth]{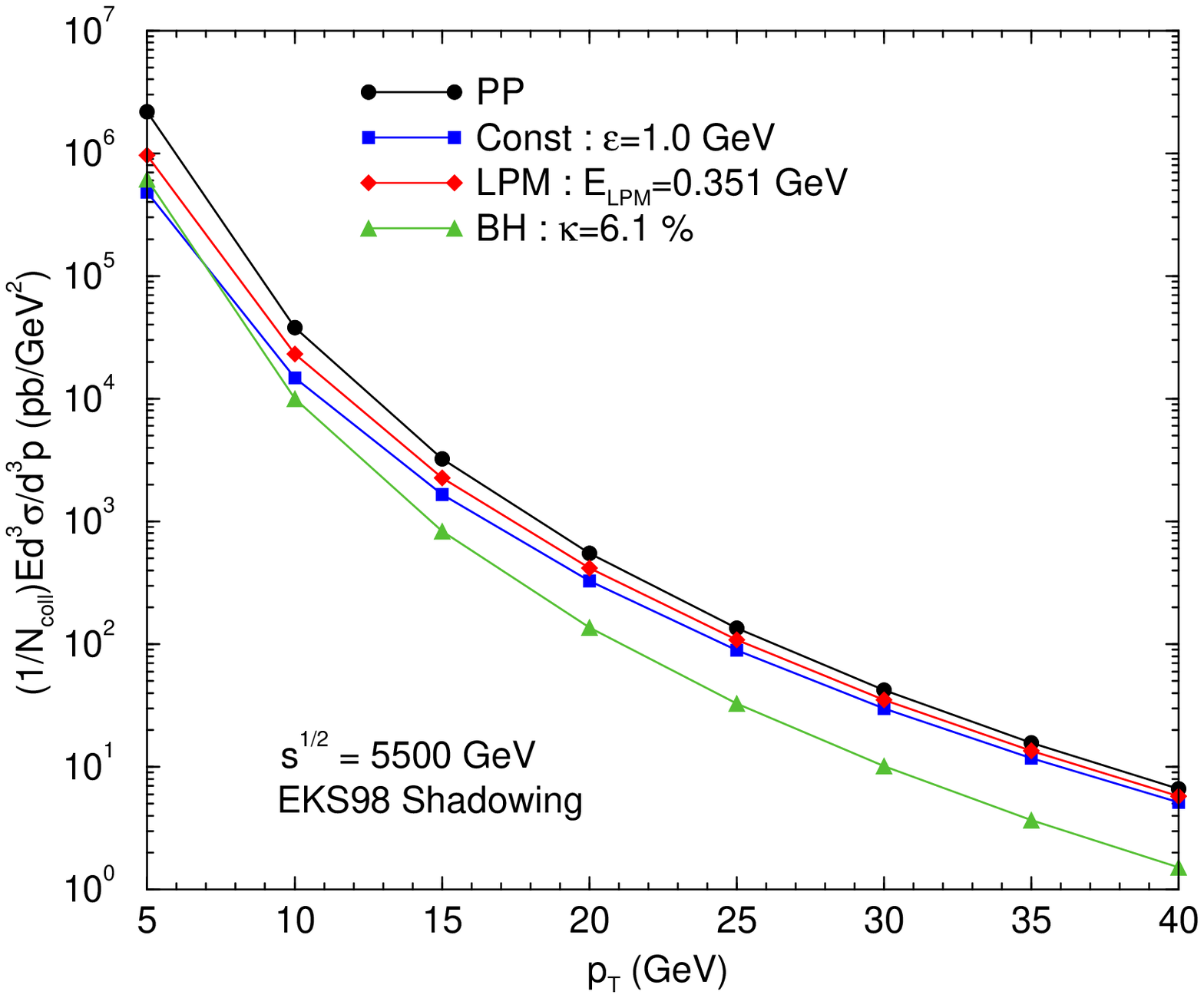}
\includegraphics[width=0.40\textwidth]{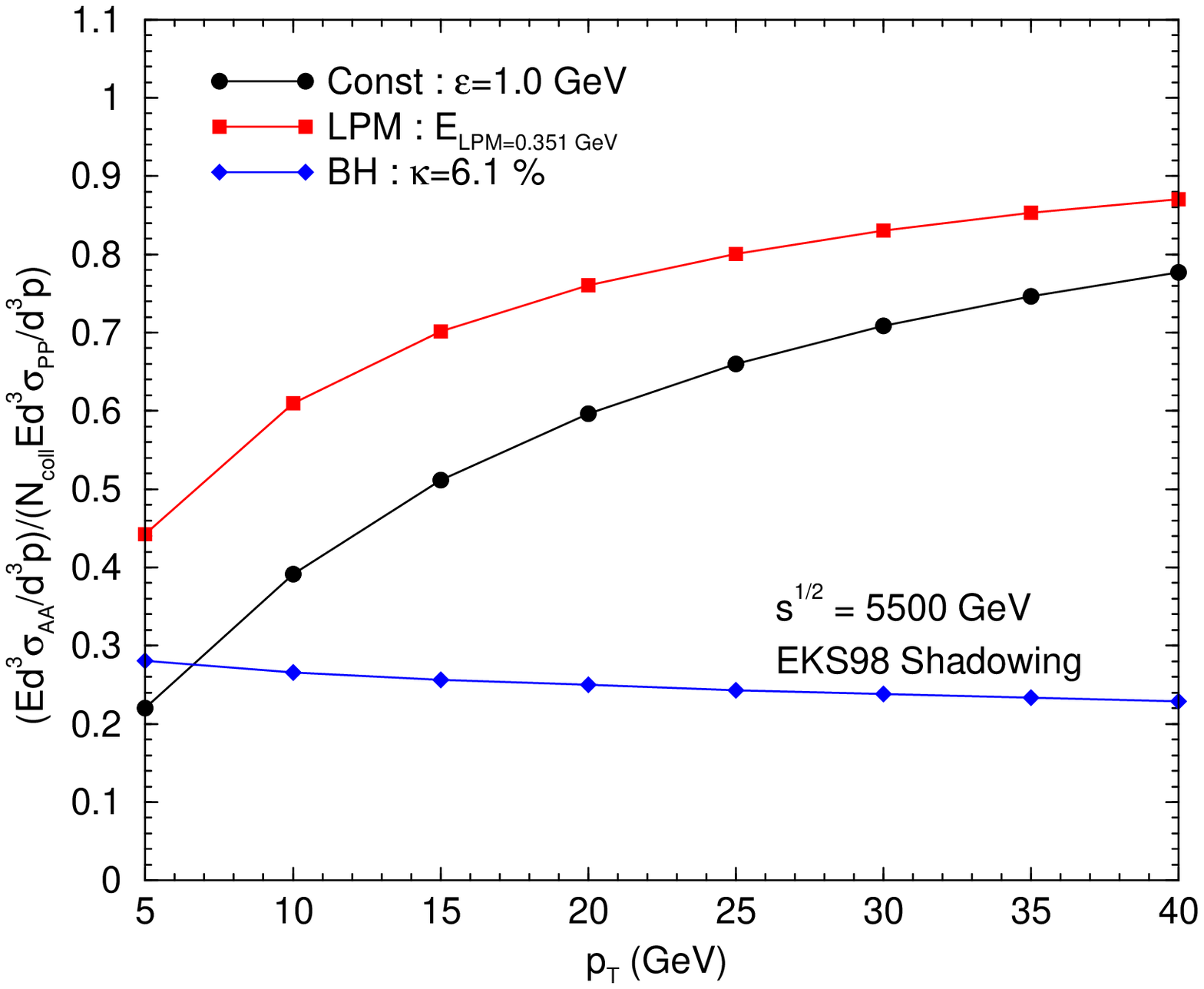}
}
\label{fig:Jeon2-fig1}
\caption{Neutral pion spectrum and $\raa$ at LHC. The energy loss parameter 
$\kappa$ is fixed by fitting the RHIC data.}
\end{figure}

\begin{figure}[htb]
\center{
\includegraphics[width=0.40\textwidth]{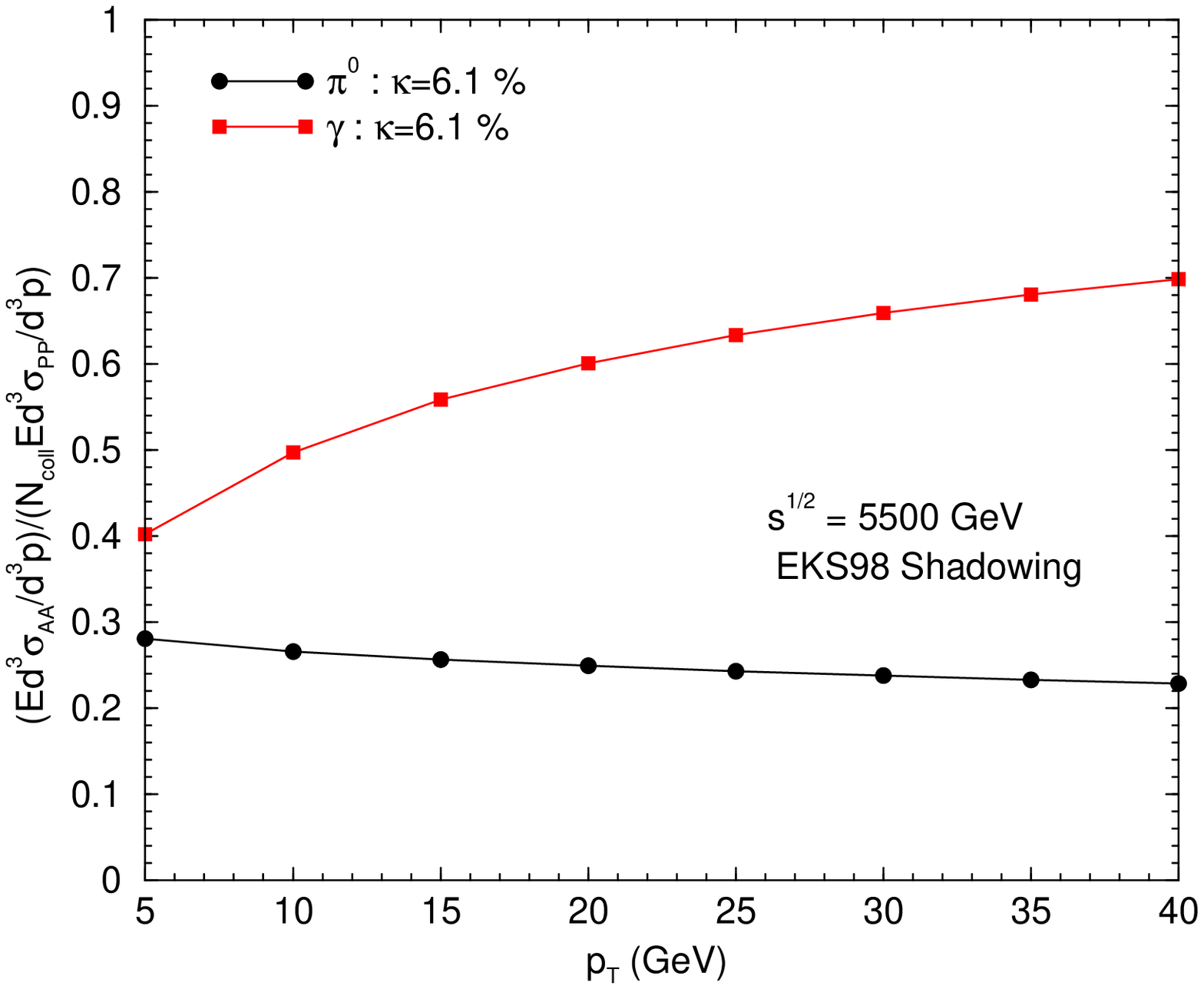}
\includegraphics[width=0.40\textwidth]{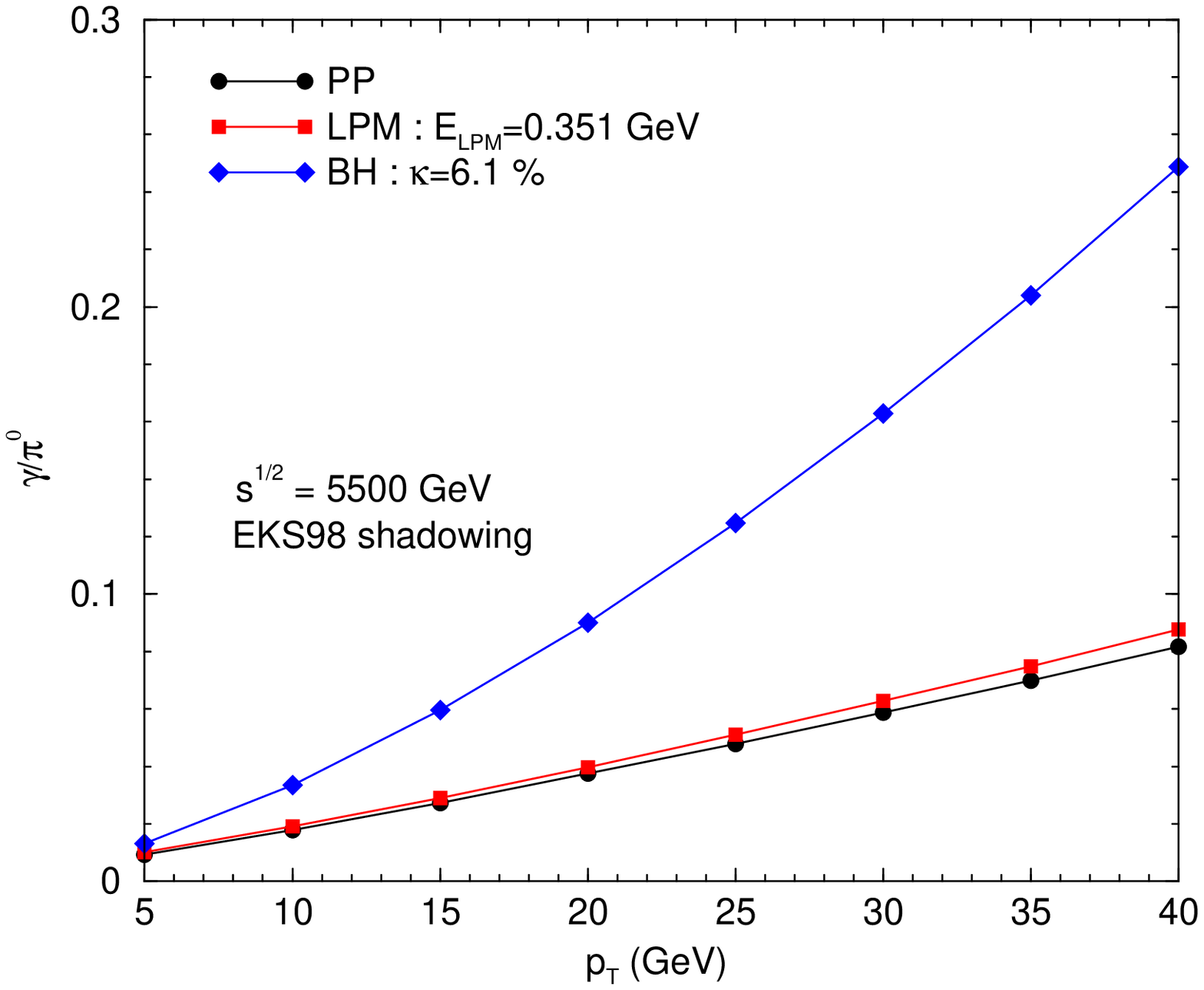}
}
\label{fig:Jeon2-fig2}
\caption{Direct photon $\raa$ and $\gamma/\pi^0$ ratio at LHC.} 
\end{figure}

Photons can be either produced during the primary collision or via 
fragmentation. The reason that the photon $\raa$ behaves qualitatively 
differently than that of $\pi^0$ is because in this energy range, the direct 
photons that come out of the primary collisions dominate over the fragmentation 
photons. Therefore the effect of energy loss is substantially reduced compared 
to the pion case.

\subsection{Transverse momentum broadening of vector bosons in heavy ion collisions at the LHC}
{\em Z.-B. Kang and J.-W. Qiu}
	
{\small 
We calculate in perturbative QCD the transverse momentum broadening of vector 
bosons in heavy ion collisions at the Large Hadron Collider (LHC).  
We predict transverse momentum broadening of $W/Z$ bosons constructed from 
their leptonic decay channels, which should be a clean probe of initial-state 
medium effect.  
We also predict the upper limit of transverse momentum broadening of J/$\psi$ 
and $\Upsilon$ production as a function of $N_{\rm part}$ at the LHC energy.
}
\vskip 0.5cm

Nuclear transverse momentum broadening of heavy vector bosons
($\gamma^*$, $W/Z$, and heavy quarkonia) is defined 
as a difference between the averaged transverse momentum square
measured in nuclear collisions and that measured in collisions of 
free nucleons,
\begin{equation}
\Delta \langle q_T^2\rangle_{AB} \equiv 
\langle q_T^2\rangle_{AB}- \langle q_T^2\rangle_{NN} \approx 
\int dq_T^2\, q_T^2\, \frac{d\sigma^{(D)}_{AB}}{dq_T^2} \left/ 
\int dq_T^2\, \frac{d\sigma_{NN}}{dq_T^2} \right. \, .
\label{eq:Kang1}
\end{equation}
Since single scattering is localized in space, the broadening is
a result of multiple parton scattering, and is a good
probe for nuclear medium properties.  Because the mass scale of 
the vector bosons is much larger than the characteristic momentum
scale of the hot medium, the broadening is likely dominated 
by double partonic scattering as indicated in equation~(\ref{eq:Kang1}).  
The broadening caused by the double scattering can be 
systematically calculated in terms of high twist formalism in 
QCD factorization~\cite{Luo:1993ui,Guo:1998rd}. 

\begin{figure}[hbt]
\centerline{
\psfig{file=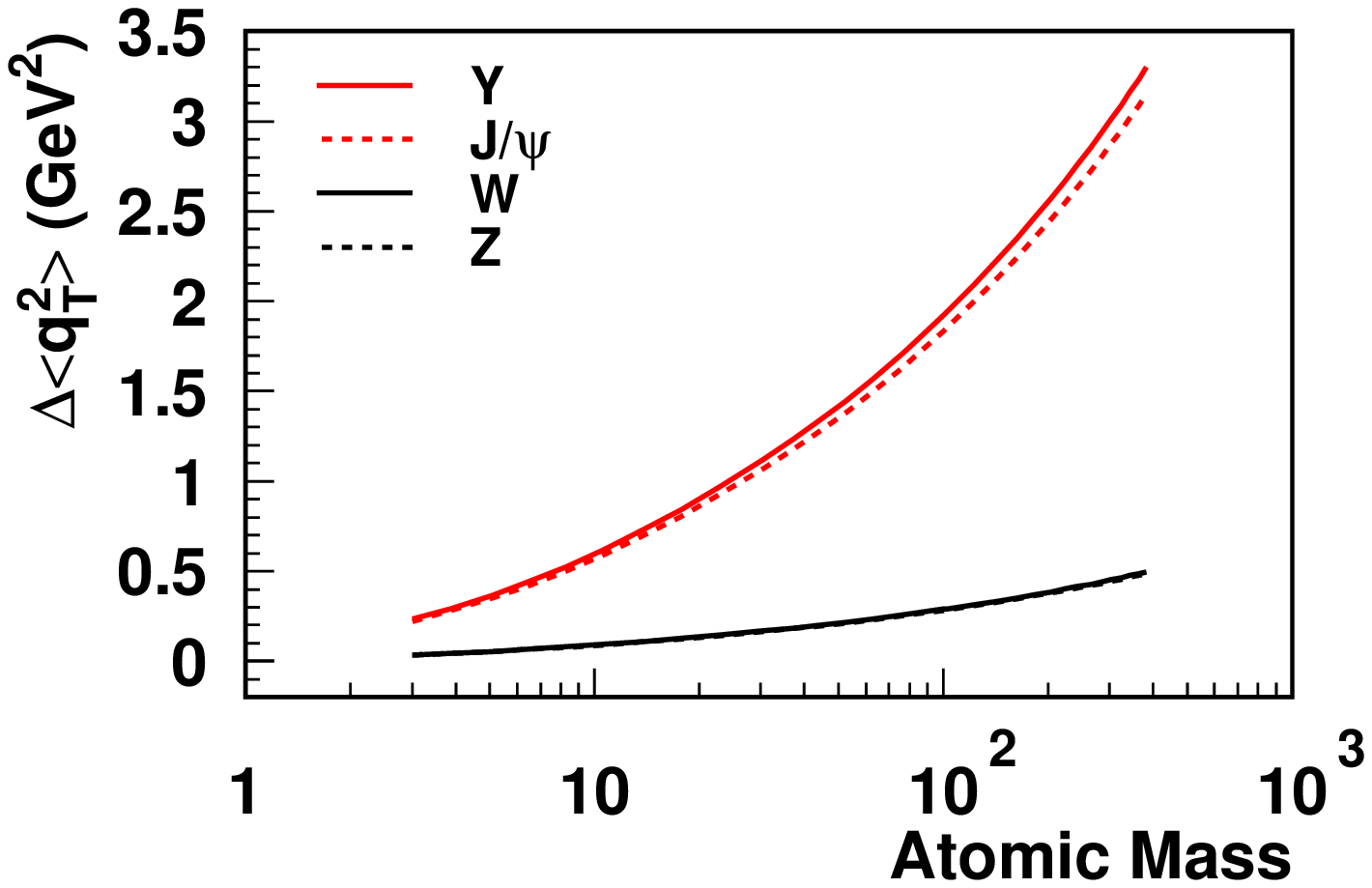,height=2in}
\psfig{file=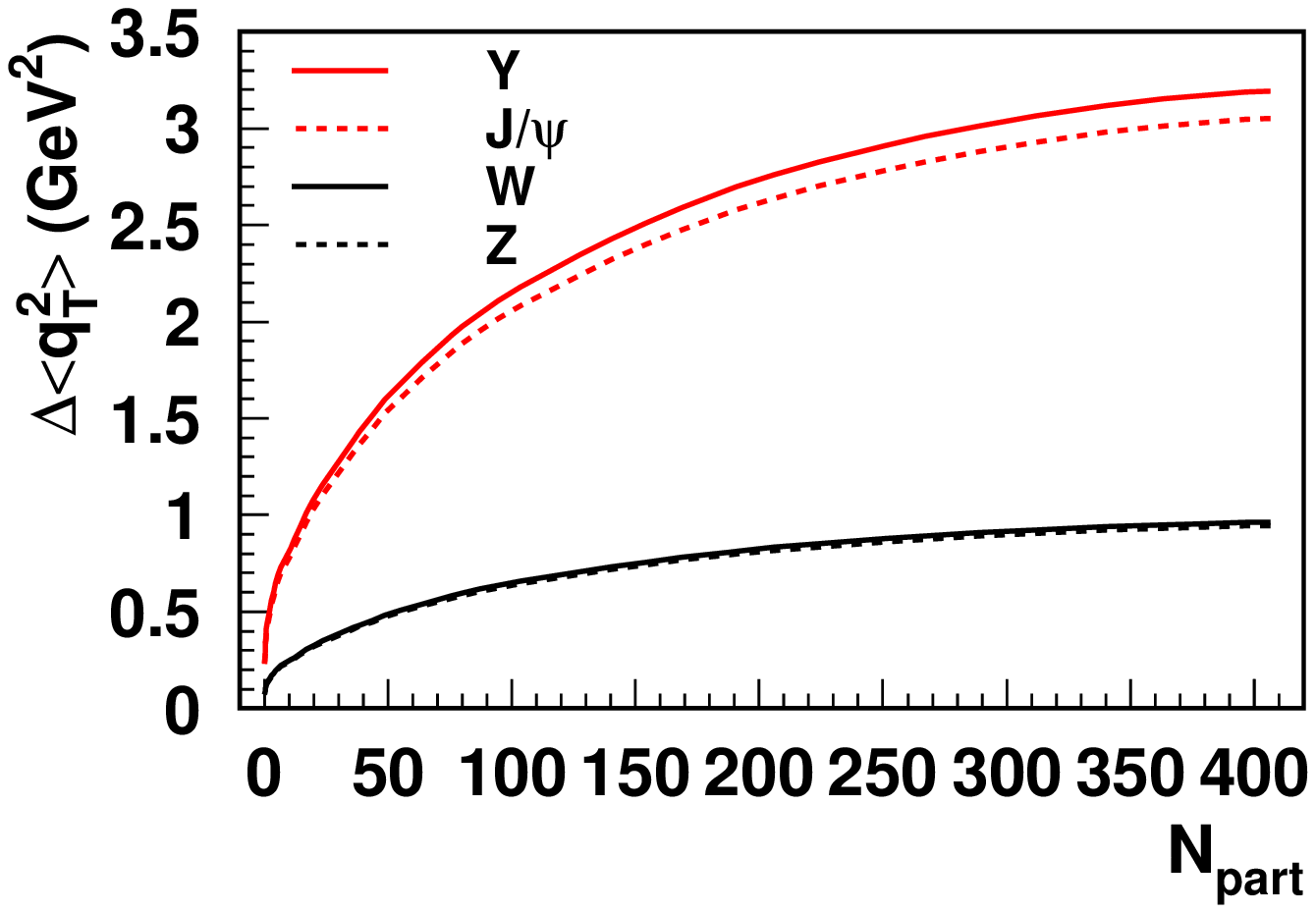,height=2in}}
\caption{Predicted broadening (maximum broadening) for 
         $W$ and $Z$ (J/$\psi$ and $\Upsilon$) production 
         in $p\A$ (left) and Pb-Pb (right) collisions 
         at $\sqrtsnn=5500$~GeV.}
\label{fig:Kang-fig}
\end{figure}

At the LHC energies, a lot $W$ and $Z$, and 
J/$\psi$ and $\Upsilon$ will be produced.  Most reconstructed 
$W/Z$ bosons will come from their leptonic decays.  
Their transverse momentum broadening is a result of purely 
initial-state multiple scattering.  By calculating the double 
scattering effect, we obtain~\cite{Guo:1998rd,Kang:2007xx}
\begin{equation}
\Delta\langle q_T^2\rangle_{pA}^W =
\frac{4\pi^2\alpha_s(M_W)}{3} \lambda^2_W A^{1/3} \, ,
\quad
\Delta\langle q_T^2\rangle_{pA}^Z
=\frac{4\pi^2\alpha_s(M_Z)}{3} \lambda^2_Z A^{1/3}
\label{eq:Kang2}
\end{equation}
for hadron-nucleus collisions.  The $\lambda^2 A^{1/3}$ 
in equation~(\ref{eq:Kang2}) was introduced 
in~\cite{Luo:1993ui} as a ratio of nuclear four parton correlation 
function over normal parton distribution.  The $\lambda$ 
is proportional to the virtuality or transverse momentum
of soft gluons participating in the coherent double scattering.  
For collisions with a large momentum transfer, $Q$, 
the $\lambda^2$ should be proportional to $\ln(Q^2)$~\cite{Kang:2007xx} and 
the saturation scale $Q_s^2$ if the active parton $x$ is small.
By fitting Fermilab E772 Drell-Yan data, it was found that 
$\lambda^2_{\rm DY}\approx 0.01$~GeV$^2$ at $\sqrt{s}=38.8$~GeV 
\cite{Guo:1998rd}.  From the $\lambda^2_{\rm DY}$, we
estimate the value of $\lambda^2$ for production of a vector boson 
of mass $M_V$ at the LHC energy as 
\begin{equation}
\lambda^2_V({\rm LHC})\approx
\lambda^2_{\rm DY}\
\frac{\ln(M_V^2)}{\ln(Q_{\rm DY}^2)}\,
\left(\frac{M_V/5500}{Q_{\rm DY}/38.8}\right)^{-0.3}\, ,
\label{eq:Kang3}
\end{equation}
where we used $Q_s^2 \propto 1/x^\delta$ with $\delta\approx 0.3$~\cite{%
  Golec-Biernat:1998js} and $\sqrtsnn=5500$~GeV for the LHC heavy ion 
collisions.  For an averaged $Q_{\rm DY}\sim 6$~GeV, we obtain
$\lambda^2_{W/Z}\approx 0.05$ at the LHC energy.  
We can also apply our formula in equation~(\ref{eq:Kang2}) 
to the broadening in nucleus-nucleus collisions by replacing 
$A^{1/3}$ by an effective medium length $L_{\rm eff}$. 
We calculate $L_{\rm eff}$ in Glauber model with inelastic 
nucleon-nucleon cross section $\sigma_{NN}^{in}=70$~mb at the 
LHC energy.  We plot our predictions (lower set curves) 
for the broadening of $W/Z$ bosons in figure~\ref{fig:Kang-fig}.  

Heavy quark pairs are produced at a distance scale much less than
the physical size of heavy quarkonia in high energy 
collisions.  The pairs produced in heavy ion collisions
can have final-state interactions before bound quarkonia could
be formed. We found~\cite{Kang:2007xx} that with both initial- and 
final-state double scattering, the broadening of heavy quarkonia 
is close to $2C_A/C_F$ times the Drell-Yan broadening
in proton-nucleus collision, which is consistent with existing 
data~\cite{Peng:1999gx}.  If all soft gluons of heavy ion beams are stopped to 
form the hot dense medium in nucleus-nucleus collisions, final-state
interaction between the almost stationary medium and the fast moving 
heavy quarks (or quarkonia) of transverse momentum $q_T$ is 
unlikely to broaden the $q_T$ spectrum, instead, it is likely to 
slow down the heavy quarks (or quarkonia) \cite{Kang:2007xx}.  
From equation~(\ref{eq:Kang3}). we obtain
$\lambda^2_{{\rm J/}\psi}\approx 0.035$, and 
$\lambda^2_{\Upsilon}\approx 0.049$ at the LHC energy; and we 
predict the maximum broadening for J/$\psi$ and $\Upsilon$ production
(upper set curves) in figure~\ref{fig:Kang-fig}.

\subsection{Nuclear modification factors for high transverse momentum pions and protons at LHC}
\label{s:Ko2}

{\em W. Liu, B.-W. Zhang and C. M. Ko}
\vskip 0.5cm

The inclusion of conversions between quark and gluon jets in a quark-gluon 
plasma (QGP) via both elastic $qg\leftrightarrow gq$ and inelastic 
$q\bar q\leftrightarrow gg$ reactions~\cite{Liu:2006sf} has recently been shown 
to give a plausible explanation for the observed similar $p/\pi^+$ and 
$\bar p/\pi^-$ ratios at large transverse momenta in both central Au+Au and $pp$
collisions at $\sqrtsnn=200$~GeV~\cite{Abelev:2006jr}. 
Extending this study to LHC, we predict the nuclear modification factor for 
both protons and pions as well as their ratios at large transverse momenta in 
central Pb+Pb collisions at $\sqrtsnn=5.5$~TeV.

For the dynamics of formed QGP at LHC, we assume that it evolves boost 
invariantly in the longitudinal direction but with an accelerated transverse 
expansion. 
Specifically, its volume expands in the proper time $\tau$ according to 
$V(\tau)=\pi R^2(\tau)\tau c$, where $R(\tau) = R_0+ a(\tau-\tau_0)^2/2$ is the 
transverse radius with an initial value $R_0=7$ fm, the QGP formation time
$\tau_0=0.5$~fm/$c$, and the transverse acceleration $a=0.1~c^2/$fm.
Starting with an initial temperature $T_0=700$~MeV, the time dependence of the 
temperature is obtained from entropy conservation, leading to the critical 
temperature $T_{\rm C}=170$~MeV at proper time $\tau_{\rm C}=8.4$~fm/$c$.  
For a quark or gluon jet moving through the QGP, the rate for the change in its 
mean transverse momentum $\langle p_T\rangle$ is given by 
${\rm d}\langle p_T \rangle /{\rm d}\tau\approx 
  \gamma(\langle p_T\rangle,T)\langle p_T\rangle$. 
The drag coefficient $\gamma(\langle p_T\rangle,T)$ is calculated from two-body 
scattering with thermal quark and gluon masses and the strong QCD coupling 
$\alpha_s(T) = g^2(T)/4\pi \approx 2.1\alpha_{\rm pert}(T)$ from lattice 
calculations~\cite{Kaczmarek:2004gv}. 
To take into account the contribution from two-body radiative scattering, we 
multiply the calculated drag coefficient by a factor $K_E\sim 2$, which is 
determined from fitting the light meson nuclear modification factor at RHIC. 
Because of conversion scatterings, the quark or gluon jet can also be converted 
to a gluon or quark jet with a rate given by corresponding collisional widths, 
which are also calculated by using the strong QCD coupling constant and 
multiplying with $K_C=K_E\sim 2$.
\begin{figure}[ht]
\centerline{
\includegraphics[width=2.4in,height=2.4in,angle=0.0]{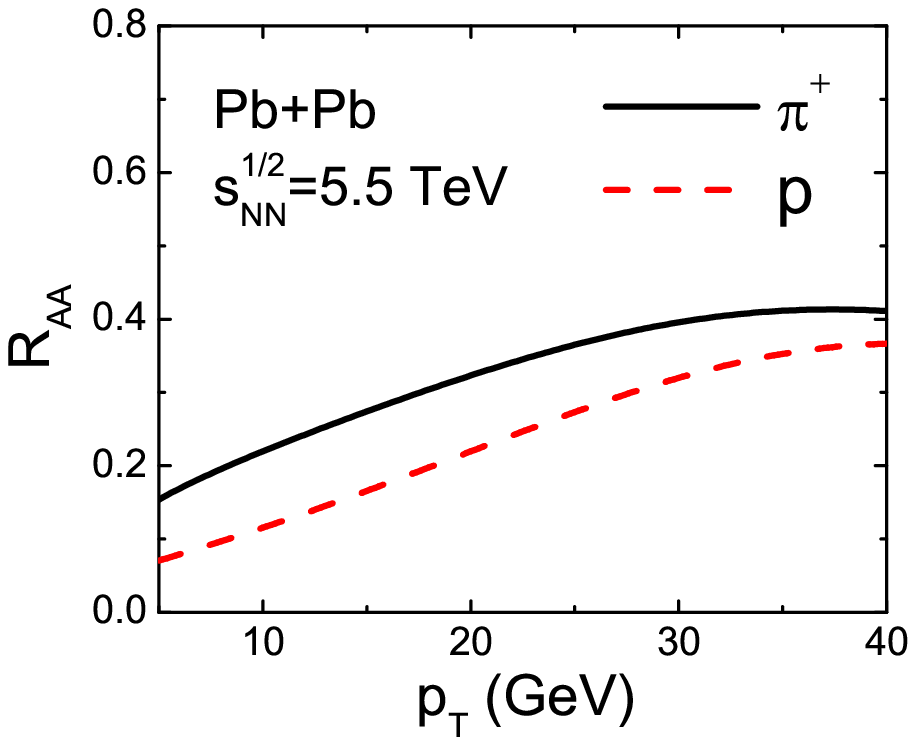}
\includegraphics[width=2.4in,height=2.4in,angle=0.0]{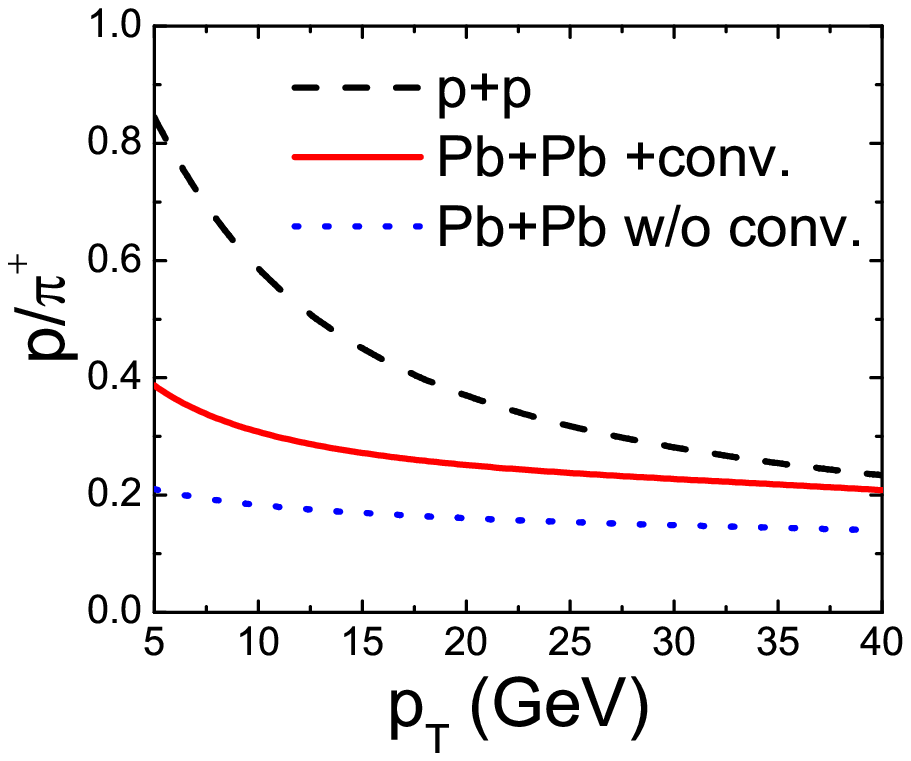}}
\caption{(Color online) Left window: Nuclear modification factor $\raa$ for 
  $\pi^+$ (solid line) and proton (dashed line) in central Pb+Pb collisions at 
  $\sqrtsnn=5.5$~TeV. Right window: $p/\pi^+$ ratio without (dotted lines) or 
  with jet conversions (solid lines). Dashed lines correspond to p+p collisions 
  at same energy.} 
\label{fig:Ko2-fig1}
\end{figure}

Using initial transverse momentum spectra of minijet gluons, quarks, and 
anti-quarks obtained by multiplying those from the PYTHIA for $pp$ collisions 
at same energy with the number of binary collisions, we simulate the 
propagation of jets in the QGP using the Monte Carlo method with test 
particles~\cite{Liu:2006sf}. 
Resulting charged pion and proton spectra from freeze-out quark and gluon jets 
are obtained via the AKK fragmentation functions~\cite{AKK1}. 
In the left window of figure~\ref{fig:Ko2-fig1}, we show predicted nuclear
modification factor $\raa$ for $\pi^+$ and $p$ at large transverse momenta in 
central Pb+Pb collisions at $\sqrtsnn=5.5$~TeV at LHC. 
It is seen that the $\raa$ of pions increases from 0.18 at $p_T=5$~GeV/$c$ to 
0.4 at $p_T=40$~GeV/$c$ due to a smaller drag coefficient at large transverse 
momenta. 
The $\raa$ of protons has a similar behavior, but its value is smaller because 
of stronger suppression of gluon than quark jets. 
The resulting $p/\pi^+$ ratio, shown by the solid line in the right window of 
figure~\ref{fig:Ko2-fig1}, approaches that in $pp$ collisions at same energy 
when the transverse momenta become very large. 
At lower transverse momenta, the $p/\pi^+$ ratio in Pb+Pb collisions remains, 
however, smaller than that in $pp$ collisions, which is different from that in 
heavy ion collisions at RHIC as a result of the larger ratio of gluon to quark
jets at LHC. 
Without conversions between quark and gluon jets, the $p/\pi^+$ ratio decreases 
by a factor of two as shown by the dotted line.

\subsection{Quenching of high-\boldmath$p_T$ hadrons: Alternative scenario}
\label{kopeliovichjets}

{\it B .Z. Kopeliovich, I. K. Potashnikova and
I. Schmidt}

 {\small
 A new scenario, alternative to energy loss, for the observed suppression
of high-$p_T$ hadrons observed at RHIC is proposed. In the limit of a very
dense medium crated in nuclear collisions the mean free-path of the
produced (pre)hadron vanishes, and and the nuclear suppression, $R_{AA}$
is completely controlled by the production length. The RHIC data are well
explained in a parameter free way, and predictions for LHC are provided.
}
\vskip 0.5cm

%\maketitle

The key assumption of the energy loss scenario for the observed
suppression of high-$p_T$ hadrons in nuclear collisions is a long
length of the quark hadronization which ends up in the medium. This has
got no justification so far and was challenged in \cite{within}.

\begin{figure}[htbp]
\centerline{
  \scalebox{0.32}{\includegraphics{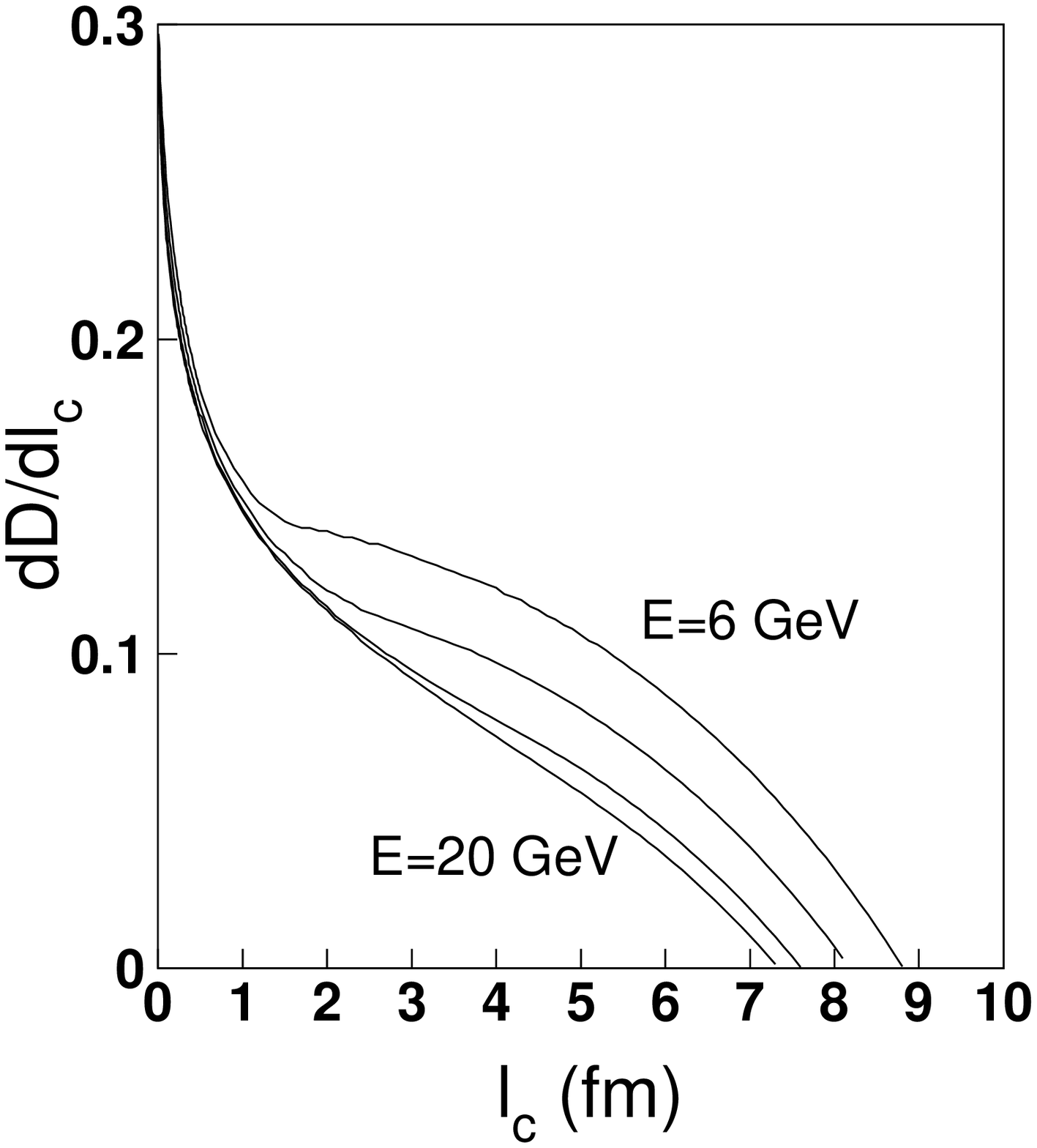}} %\hspace*{1mm}
  \scalebox{0.61}{\includegraphics{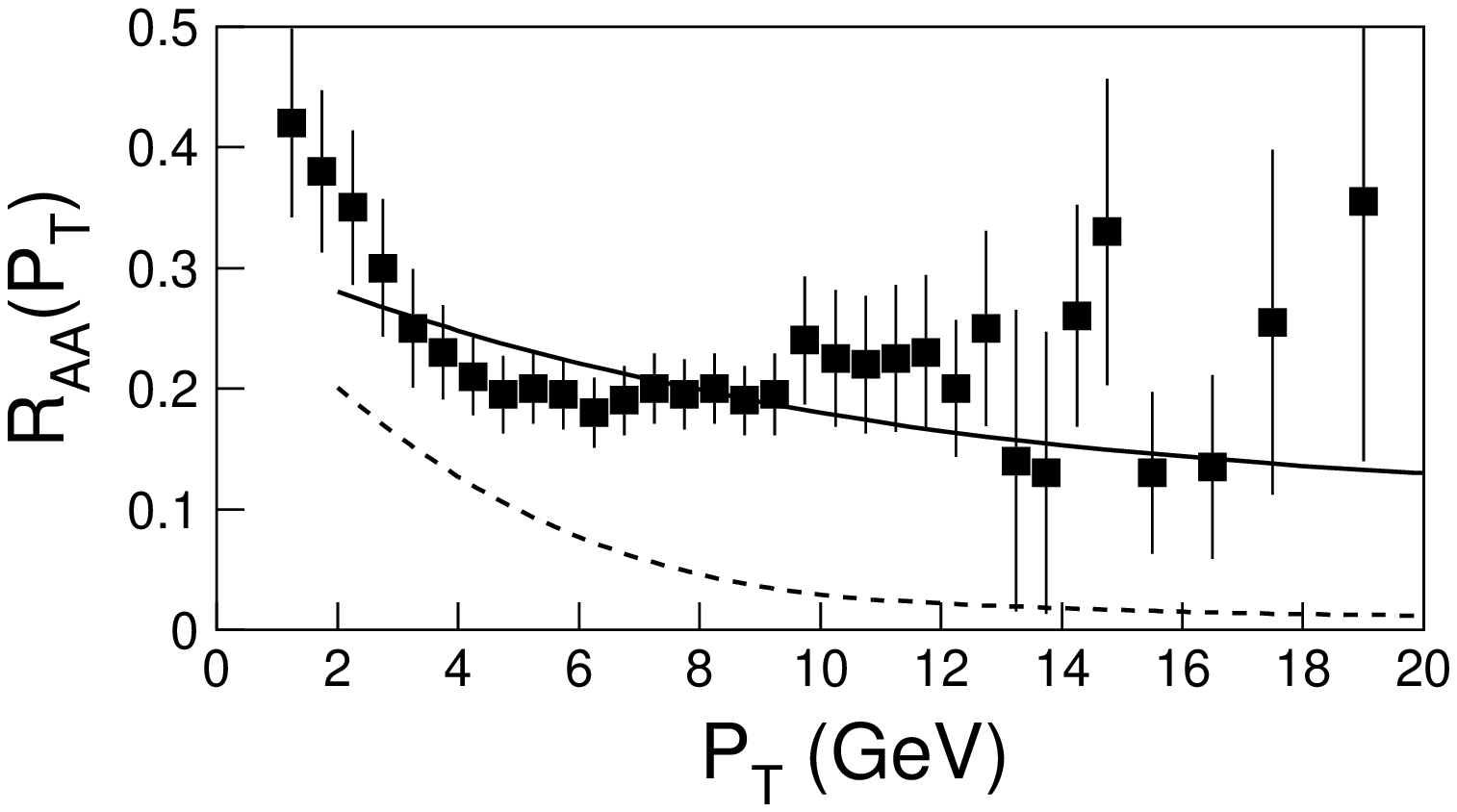}}
 }
 \caption{{\it Left:} $\partial D(z)/\partial l_c$
(in arbitrary units) at jet energies $6,\ 10,\ 16,\ 20\GeV$ and
$z=0.7$.  {\it Right:} Pion suppression in central $AA$ collisions  
($A\sim 200$) at $\sqrt{s}=200\GeV$ (solid) and $\sqrt{s}=5500\GeV$
(dashed). Data are from the PHENIX experiment.}
\label{fig:kopjets}
 \end{figure}

The quark fragmentation function (FF) was calculated in Born 
approximation in \cite{berger}: 
 \beq
\frac{\partial D^{Born}_{\pi/q}(z)}{\partial k^2}\propto
{1\over k^4}\,(1-z)^2\,,
\label{10}
 \eeq
 where $k$ and $z$ are the transverse and fractional longitudinal momenta
of the pion. One can rewrite this in terms of the coherence length
$l_c=z(1-z)E/k^2$, where $E$ is the jet energy. Then, $\partial
D^{Born}_{\pi/q}(z)/\partial l_c\propto (1-z)$, is $l_c$ independent.  
Inclusion of gluon radiation leads to the jet lag effect \cite{jetlag}
which brings $l_c$ dependence,
 \beq
\frac{\partial D_{\pi/q}(z)}{\partial l_c}\propto
(1-\tilde z)\,S(l_c,z)\ .
\label{20}
 \eeq
 Here $\tilde z=z[1+\Delta E(l_c)/E]$ accounts for the higher Fock
components of the quark, which are incorporated via the
vacuum energy loss $\Delta E(l_c)$ calculated perturbatively with a
running coupling. The induced energy loss playing a minor role is added as 
well. $S(l_c,z)$ is the Sudakov suppression caused by energy
conservation. Fig.~\ref{fig:kopjets} shows an example for the $l_c$-distributions
calculated for $z=0.7$ and different jet energies at $\sqrt{s}=200\GeV$.

The pre-hadron, a $\bar qq$ dipole, may be produced with a rather large
initial separation $\la r_0^2\rangle\approx 2l_c/E+1/E^2$ and it keeps
expanding.

To keep calculations analytic we consider a central, $b=0$, collision of
identical heavy nuclei with nuclear density
$\rho_A(r)=\rho_A\Theta(R_A-r)$.  Then we find,
 \beq
R_{AA}=\frac{\la l_c^2\rangle}{R_A^2}\left[1-
A\,\frac{L}{\la l_c\rangle} +
B\,\frac{L^2}{\la l_c^2\rangle}
\right]\,,
\label{30}
 \eeq
 where the effective absorption length has the form,
$L^3=3p_T/(8\rho_A^2R_A\,X)$, and $X$ 
includes the unknown density of the medium and is to be
fitted to data on $R_{AA}$. However. if the medium is very dense,
i.e. $X$ is large, the last two terms in (\ref{30}) can be neglected, and
we can {\it predict} $R_{AA}$,
 \beq
R^h_{AA}=\frac{\la l_c^2\rangle}{R_A^2}.
\label{40}
 \eeq
 With this expression we calculated $R_{AA}$ at the energies of RHIC and
LHC and in fig.~\ref{fig:kopjets} (right). This parameter free result well agrees with the
data supporting the assumption that the medium is very dense. Summarizing:
 \begin{itemize}
 \item
 The $A$-dependence, eq.~(\ref{40}), predicts $R_{AA}\approx0.42$
for $Cu-Cu$ confirmed by data.
 \item Vacuum radiation which
depends only on the current trajectory should be flavor independent. This
fact and the above consideration explains the strong suppression for heavy
flavors observed at RHIC.
 \item
 Since the strength of absorption does not affect $R_{AA}$, 
eq.~(\ref{40}), a single hadron and a pair of hadrons should be suppressed 
equally.
 \item The observed suppression $R_{AA}$ may not contain much information 
about the medium properties, except it is very dense.

\end{itemize}

\subsection{Expectations from AdS/CFT for Heavy Ion Collisions at the LHC}

%{\it Hong Liu, Krishna Rajagopal and Urs Achim Wiedemann}
{\it H.~Liu, K.~Rajagopal and U.~A.~Wiedemann}

{
\small
% The AdS/CFT correspondences provides a technique for doing strong coupling calculations 
% in non-abelian thermal quantum field theories. It is of interest for understanding heavy ion collisions 
% at the LHC, since the matter produced at the LHC is expected to be strongly coupled. Here, 
We summarize results obtained by use of the AdS/CFT correspondence for jet quenching 
and quarkonium dissociation, and we discuss the resulting expectations for heavy ion collisions 
at the LHC.}
\vskip 0.5cm

%%% INTRODUCTION AND MODEL DESCRIPTION %%%%%%%%%%%%%%%%%%%%%%%%%%%%%%%%%%%%%%%%%
The AdS/CFT correspondence maps nonperturbative problems in a large class of strongly coupled
non-abelian gauge theories onto calculable problems in dual gravity theories. The gravity dual of Quantum Chromodynamics is not known. However, one 
finds many commonalities amongst the quark-gluon plasmas
in large classes of strongly coupled
non-abelian thermal gauge theories, independent of their significantly differing microscopic degrees of freedom and interactions. Since 
these results are generic and do not seem to depend on 
microscopic features of the
theory such as its particle content at weak coupling, one may 
expect that they are shared by QCD. Where this
can be tested against QCD lattice results, the qualitative agreement
is fair (see Ref.~\cite{Liu:2006he}). However, many measurements 
in heavy ion collisions involve strong coupling and real-time dynamics, 
where lattice QCD results are not available or in their infancy. 
The practitioner faces the uncomfortable choice of calculating either with inappropriate (e.g. perturbative) techniques in 
QCD, or using appropriate strong coupling techniques but working
in a class of gauge theories that may not include QCD itself and
seeking universal commonalities.  
We report on two results from the latter approach.

\subsubsection{Jet quenching}
In QCD itself, the jet quenching parameter $\hat{q}$ 
has not been calculated in the strong coupling
regime. For the ${\cal N}=4$ SYM theory, it 
has been calculated for large t'Hooft coupling $\lambda = g^2\, N_c$
by use of the AdS/CFT correspondence\cite{Liu:2006ug}:
\begin{equation}
 \hat{q}_{SYM} = \frac{\pi^{3/2}\Gamma(3/4)}{\Gamma(5/4)} \sqrt{\lambda} T^3\, .
 \label{eq1.wiedemann}
\end{equation} 
If one relates this to QCD by fixing $N_c =3$ and $\alpha_{SYM} = .5$, then
$\hat{q}_{SYM} = 32.7\, T^3 = 4.5$~GeV$^2$/fm at $T=300$~MeV. 
This shows that a medium characterized 
by a momentum scale $T$ can give rise to
an apparently large quenching parameter, significantly larger than $T^3$. 
For a certain infinite class of
theories with gravity dual, one finds
that the quenching parameter scales with 
the square root of the entropy density~\cite{Liu:2006ug}. 
Assuming that QCD follows this systematic, 
%upon comparing 
%the number of degrees of freedom 
%of N=4 SYM and QCD, 
one finds
\begin{equation}
	\hat{q}_{QCD} = \sqrt{\frac{s_{QCD}}{s_{N=4}}}\, \hat{q}_{N=4} 
	= \sqrt{\frac{47.5}{120}}\, \hat{q}_{N=4} \simeq 0.63\, \hat{q}_{N=4}\, .
	\label{eq2}
\end{equation}

In extrapolating from RHIC to LHC, we assume that the change in $\hat{q}$ 
is dominated by 
the change in $T^3$, see eq. (\ref{eq1.wiedemann}). In the presence of 
expansion, the relevant 
temperature $T$ at RHIC and at the LHC must be compared at 
the {\it same} 
time $\tau$. This can be seen, e.g., in a Bjorken expansion scenario
in which $T(\tau)= T_0\, \left(\tau_0/\tau\right)^{1/3}$. The time-averaged
$\overline{\hat q} = (2/L^2)\, \int_0^L d\tau\, \tau\, \hat{q}(\tau)$, 
which determines parton 
energy loss and which is the quantity that has been extracted
by comparison with RHIC data, is then
$\overline{\hat q} \propto (2\tau_0/L) T_0^3 = (2\tau/L) T(\tau)^3$,
independent of the reference time $\tau_0$. Since the volume of 
the collision region at early times 
depends only on the nuclear overlap and is energy independent,
we can assume that at any particular $\tau$, 
$T^3_{\rm LHC}/T^3_{\rm RHIC} 
=(dN_{\rm ch}^{\rm LHC}/d\eta)/ (dN_{\rm ch}^{\rm RHIC}/d\eta)$
%so this temperature will grow only 
%mildly from RHIC to the LHC, 
and hence make the prediction
$\hat{q}_{\rm LHC} = \hat{q}_{\rm RHIC}
(dN_{\rm ch}^{\rm LHC}/d\eta)/ (dN_{\rm ch}^{\rm RHIC}/d\eta)$.
 
\subsubsection{Quarkonium suppression}
In lattice QCD, the temperature dependent 
potential between a heavy quark and anti-quark has been calculated as a function of their
separation $L$. At finite temperature, this potential is screened above a length 
$L_s \sim 0.5/T$ (see references in~\cite{Liu:2006nn}). These studies
indicate that the $J/\Psi$ dissociates 
at a temperature between $1.5 \, T_c$ and $2.5 \, T_c$.
For $N=4$ SYM theory, one finds $L_s \sim 0.277/T$. In contrast to QCD, 
the calculation
in theories with gravity dual can be done also for heavy quark-antiquark pairs which are
moving with a velocity $v$ through the heat bath. 
One finds that the screening length
decreases with increasing $\gamma = \sqrt{1/(1-v^2)}$~\cite{Liu:2006nn}:
\begin{equation}
	L_s(v,T) \simeq L_s(0,T)/\sqrt{\gamma} %\left( 1-v^2\right)^{1/4}
		\longrightarrow
	T_{\rm diss}(v) \simeq T_{\rm diss}(0)  / \sqrt{\gamma}\, .
	\label{eq3}
\end{equation}
So, bound states with a dissociation temperature $T_{\rm diss}(v)$ will 
survive if at rest
in a medium at temperature $T$ if $T_{\rm diss}(0)>T$. Yet, they will dissociate if they 
move sufficiently fast through the medium, such that $T_{\rm diss}(v)<T$. 
LHC data may test this prediction, depending on the quarkonium formation
mechanism.
Let us
consider three possibilities for the latter: i) A parent quark ($c$ or $b$) 
propagates through the medium but the quarkonium forms later, outside 
the medium. ii) As in (i) but with a parent gluon.
iii) A quarkonium bound state forms (from either
a parent quark or gluon) and propagates through the medium. 
These three scenarios can be
discriminated as follows:  i) The nuclear modification 
factor of quarkonium is the same
as that of open charm or beauty, which are known 
to be dominated by quark parents. It is
the same for all quarkonium bound states.
ii) The nuclear modification factor of quarkonium 
is the same as that of light hadrons,
which at the LHC are dominated by gluon parents. Again, 
all bound states are equally 
suppressed. iii) The nuclear modification factor will differ  for 
different bound states, since they will dissociate for different 
values of the transverse 
momentum $p_T$. The hierarchy in 
the $p_T$-dependence of the quarkonium suppression 
pattern would test (\ref{eq3}). For example, $\Upsilon$ (or $J/\Psi$)
suppression
could set in only above some $p_T$ while $\Upsilon'$ (or $\Psi'$) are 
suppressed even at low $p_T$.  Details of the formation mechanism
cancel in ratios like 
$\Upsilon/\Upsilon'$, making the $p_T$-dependent
pattern predicted by (\ref{eq3}) visible
as long as the quarkonia form in the medium.

\subsection{High-$p_T$ observables in PYQUEN model}
\label{lokhtinhighpt}

{\it I. P. Lokhtin, A. M. Snigirev and C. Yu. Teplov}

{\small
Predictions of PYQUEN energy loss model for high-$p_T$ observables 
at the LHC are discussed. Nuclear modification factors and elliptic flow for 
hard jets and high-$p_T$ hadrons, medium-modified jet fragmentation function, 
$p_T$-imbalance for dimuon tagged jets, high-mass dimuon and secondary 
$J/\psi$ spectra are calculated for PbPb collisions.
}
\vskip 0.5cm

In this paper, the various high-p$_T$ observables in PbPb collisions at 
$\sqrt{s_{\rm NN}}=5.5 A$ TeV are analyzed in the frame 
of PYQUEN partonic energy loss model~\cite{Lokhtin:2005px}. The pseudorapidity 
cuts for jets $|\eta^{\rm jet}|<3$, charged hadrons $|\eta^{\rm h^{\pm}}|<2.5$ 
and muons $|\eta^{\rm \mu}|<2.5$ were applied. The jet energy was determined
here as the total transverse energy of the final particles around the direction of a 
leading particle inside a cone 
$R\,=\,\sqrt{\Delta \eta^2+\Delta \varphi ^2}=0.5$ ($\varphi$ is the azimuthal
angle).

\subsubsection{Nuclear modification factors for jet and high-p$_T$ hadrons}

\begin{figure}[htbp]
\begin{minipage}{18pc}
\includegraphics[width=18pc]{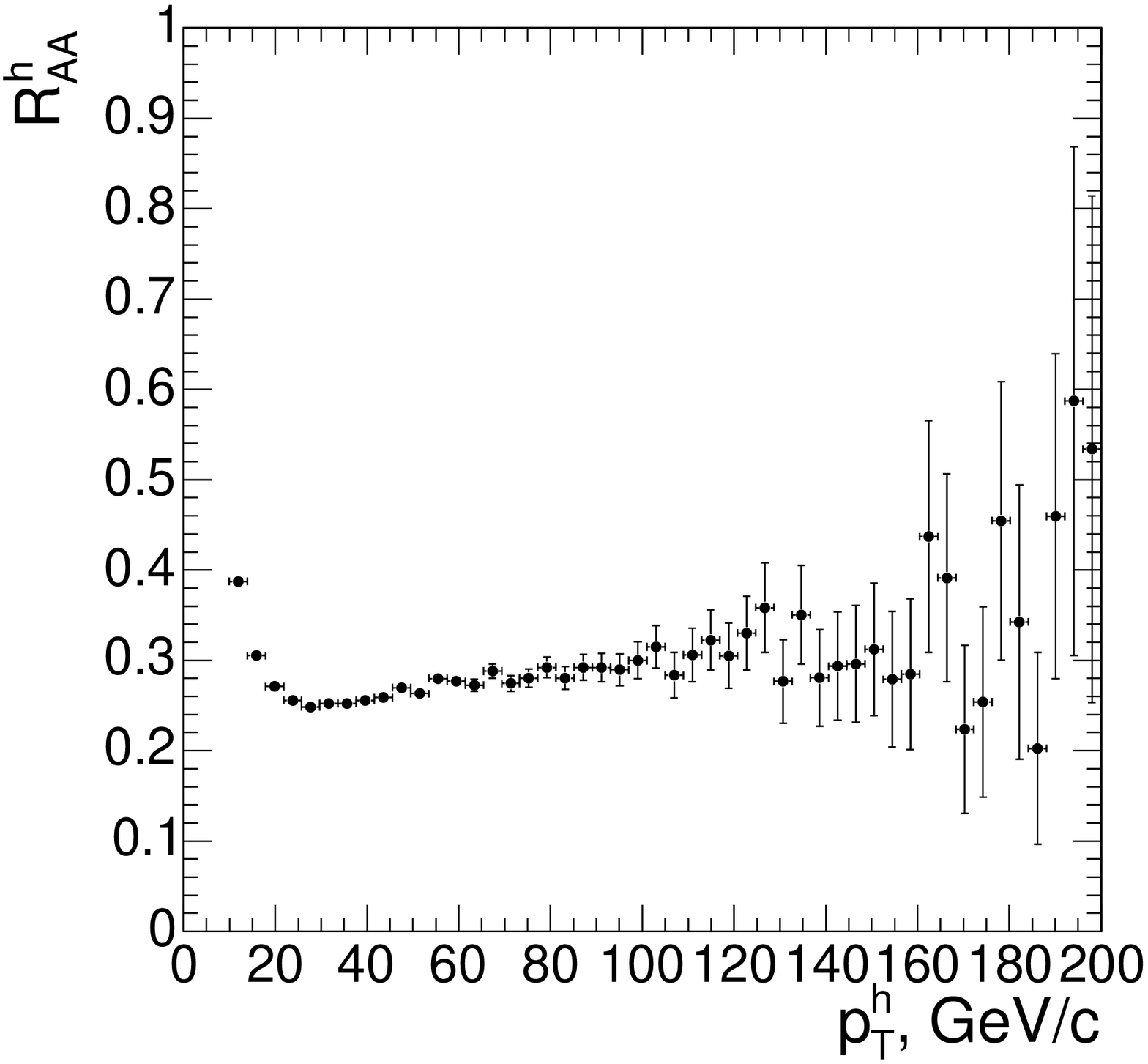}
\caption{The nuclear modification factor for charged hadrons in central PbPb 
collisions triggered on jets with $E_T^{\rm jet}>100$ GeV.  
\label{fig:raa_had}}
\end{minipage}
\hspace{\fill}%
\begin{minipage}{18pc}
\includegraphics[width=18pc]{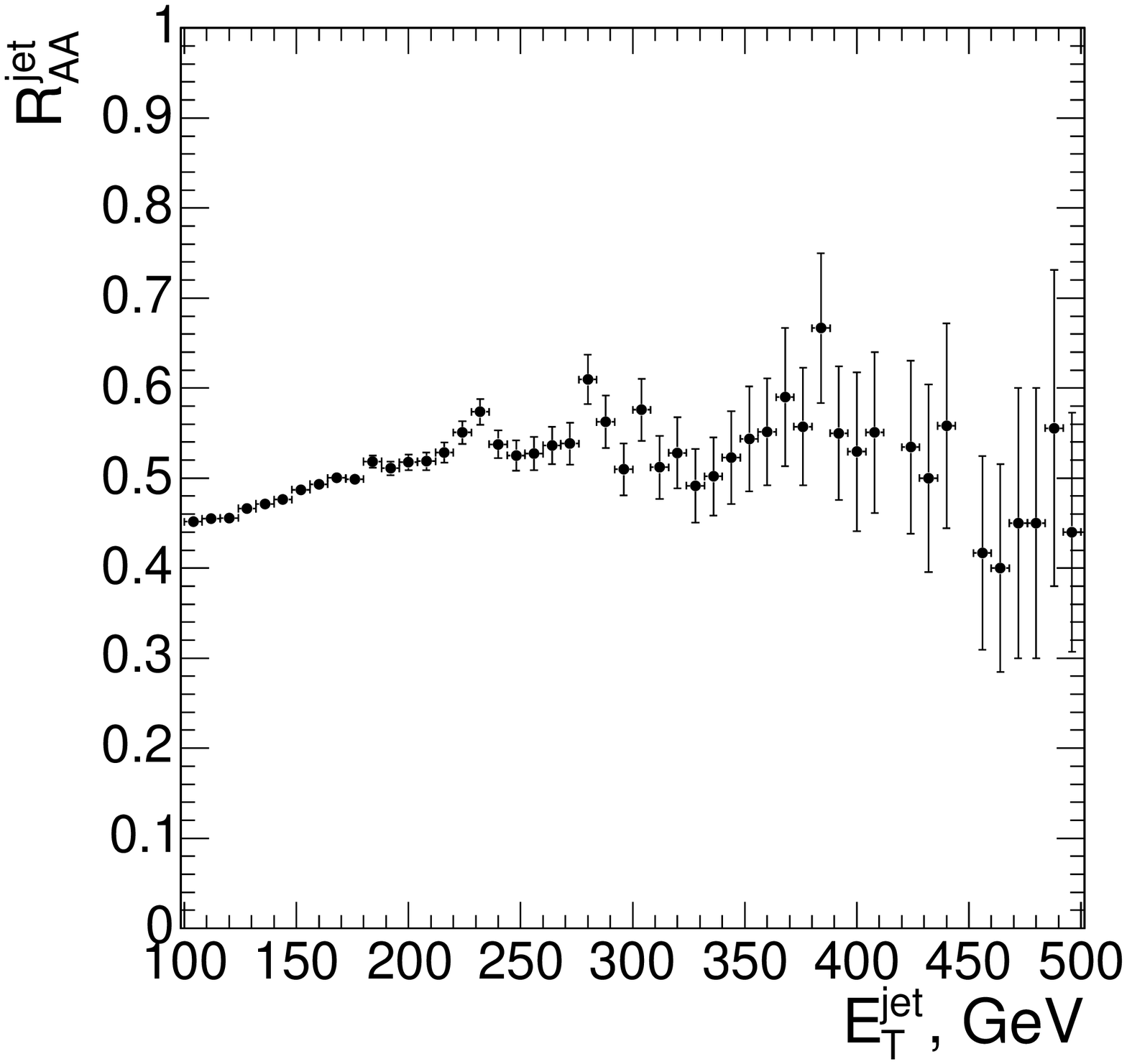}
\caption{The nuclear modification factor for jets in central PbPb collisions. 
\label{fig:raa_jet} }
\end{minipage} 
\end{figure}

The nuclear modification factor is defined as a ratio of particle yields in 
$AA$ and pp collisions normalized on the number of binary nucleon-nucleon 
collisions. Figures \ref{fig:raa_had} and \ref{fig:raa_jet} show 
$p_T$-dependences of nuclear modification factors for inclusive charged hadrons 
(in central PbPb events triggered on jets with $E_T^{\rm jet}>100$ GeV) and for 
jets respectively. The number of entries and the statistical 
errors correspond to the estimated event rate for one month of LHC run and a 
nominal integrated luminosity of 0.5 nb$^{-1}$~\cite{Virdee:1019832}. The 
predicted hadron suppression factor slightly increases with
$p_T$($>20$ GeV), from $\sim 0.25$ at $p_T \sim 20$ GeV to $\sim 0.4$ at $p_T \sim 
200$ GeV. This behaviour manifests the specific implementation of partonic energy 
loss in the model, rather weak energy dependence of loss and the shape of initial 
parton spectra. Without event triggering on high-$E_T$ jet(s), the suppression factor 
is stronger ($\sim 0.15$ at 20 GeV and slightly increasing with $p_T$ up to 
$\sim 0.3$ at 200 GeV). The predicted jet suppression factor (due to partial 
gluon bremsstrahlung out of jet cone and collisional loss) is about $2$ and 
almost independent on jet energy. It is clear that the measured jet nuclear 
modification factor will be very sensitive to the fraction of partonic energy 
loss carried out of the jet cone. 

\subsubsection{Medium-modified jet fragmentation function}

\begin{figure}[htbp]
\begin{minipage}{18pc}
\includegraphics[width=18pc]{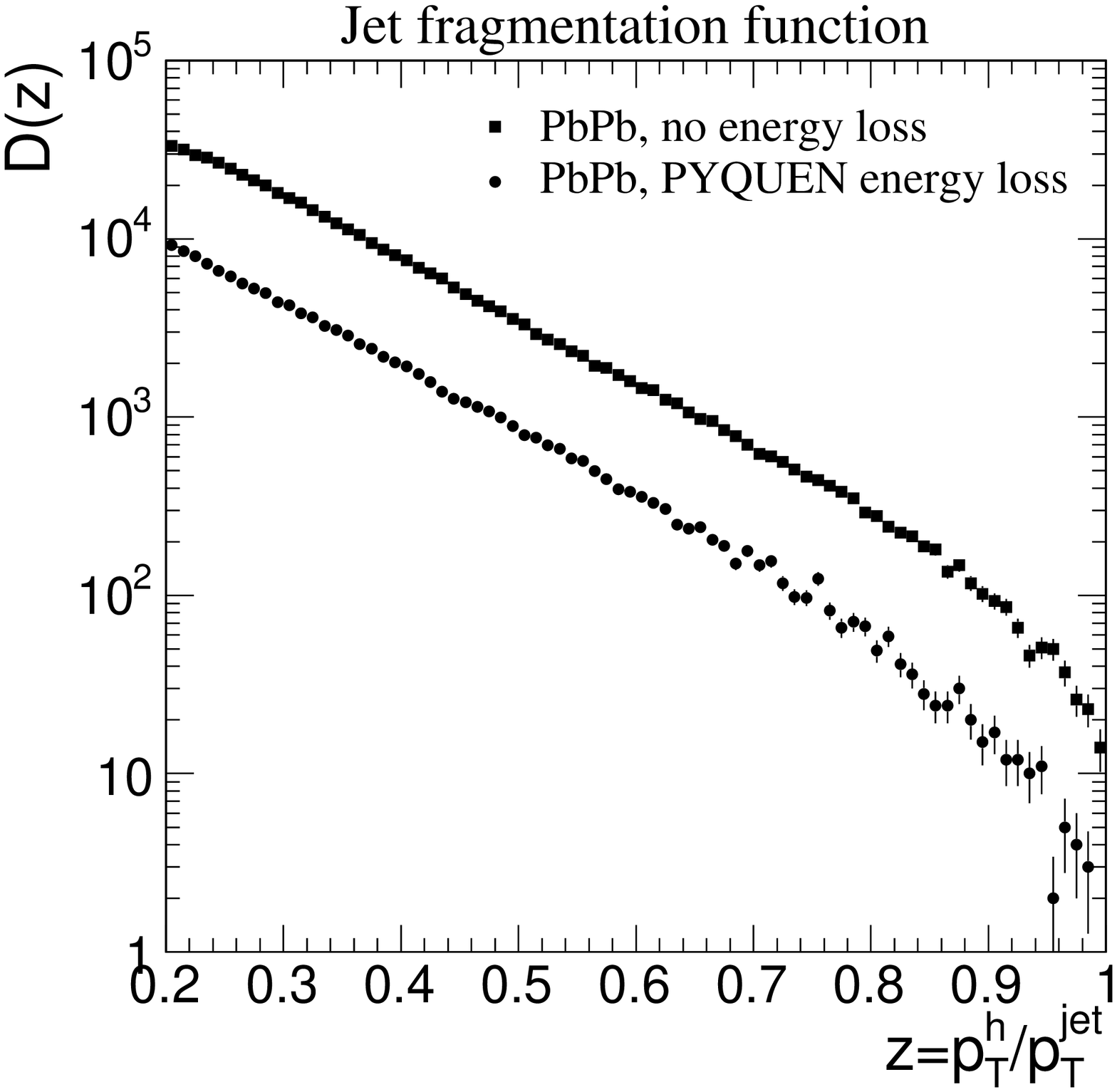}
\caption{Jet fragmentation function for leading hadrons in central PbPb 
collisions triggered on jets with $E_T^{\rm jet}>100$ GeV without (squares) 
and with (circles) partonic energy loss.
 \label{fig:jff}}
\end{minipage}
\hspace{\fill}%
\begin{minipage}{18pc}

\vskip -0.4 cm

\includegraphics[width=18pc]{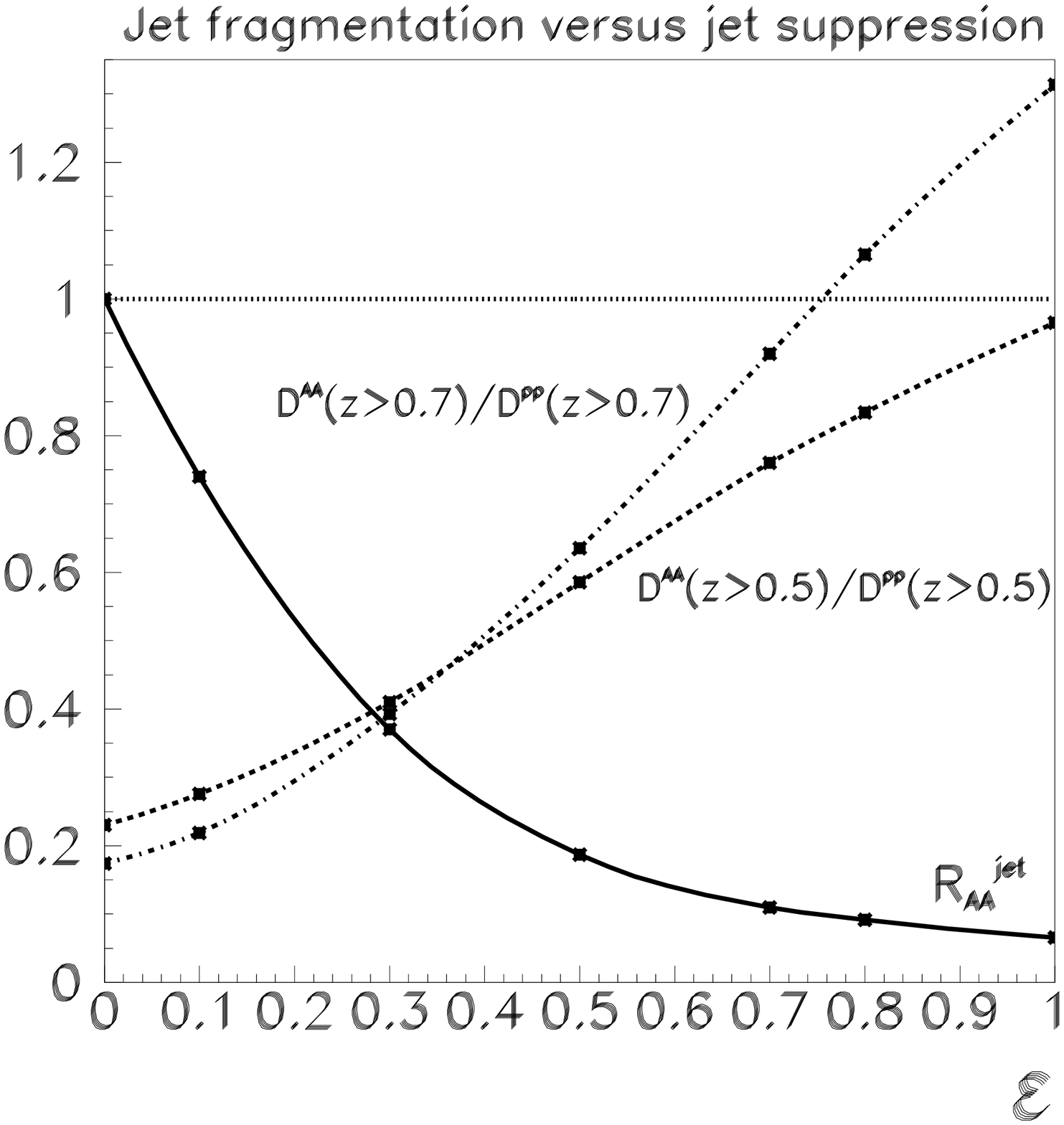}
\caption{Jet nuclear modification factor (solid curve) and 
ratio of JFF with loss to JFF without loss (dashed, dash-dotted curves) as 
a function of $\varepsilon$ (see text).  
\label{fig:jff_vs_quen} }
\end{minipage} 
\end{figure}

The ``jet fragmentation function'' (JFF), $D(z)$, is defined as the probability 
for a given product of the jet fragmentation to carry a fraction $z$ of the 
jet transverse energy. Figure~\ref{fig:jff} shows JFF's in central PbPb 
collisions with and without partonic energy loss. The number of entries and the 
statistical errors correspond again to the estimated event rate for one month 
of LHC run. Significant softening of the JFF (by a factor of $\sim 4$ and 
slightly increasing with $z$) is predicted. 

The medium-modified JFF is sensitive to a fraction $\varepsilon$ of partonic 
energy loss carried out of the jet cone. 
Figure~\ref{fig:jff_vs_quen} shows the 
$\varepsilon$-dependences of jet nuclear modification factor $R_{\rm AA}^{\rm 
jet}$ and ratio of JFF with energy loss to JFF without loss, 
$D^{\rm AA} (z>z_0) / D^{\rm pp} (z>z_0)$, for $z_0=0.5$ and $0.7$. 
If $\varepsilon$ close to 0, then $R_{\rm AA}^{\rm jet} \sim 1$ (there is  
no jet rate suppression), and JFF softening is maximal. 
Increasing $\varepsilon$ results in stronger jet rate suppression, but effect 
on JFF softening becomes smaller. Indeed, final jet transverse momentum
(which is the denominator in definition of $z$) decreases in this case 
without an influence on the numerator of $z$ and, as a consequence, the effect 
on JFF softening reduces, while the integral jet suppression factor becomes 
larger. Thus a novel study of the softening of the 
JFF and suppression of the absolute jet rates can be carried out in order to 
differentiate between various energy loss mechanisms (``small-angular''
radiative loss versus ``wide angular'' and collisional 
loss)~\cite{Lokhtin:2003yq}. 

Other correlation measurements which also can be useful extracting information 
about medium-modified jets are jet shape broadening and jet quenching versus
rapidity~\cite{Lokhtin:2006dp} and monojet-to-dijet ratio versus 
dijet acoplanarity~\cite{Lokhtin:1999mf}. 

\subsubsection{Azimuthal anisotropy of jet quenching}

\begin{figure}[htbp]
\begin{minipage}{18pc}
\includegraphics[width=18pc]{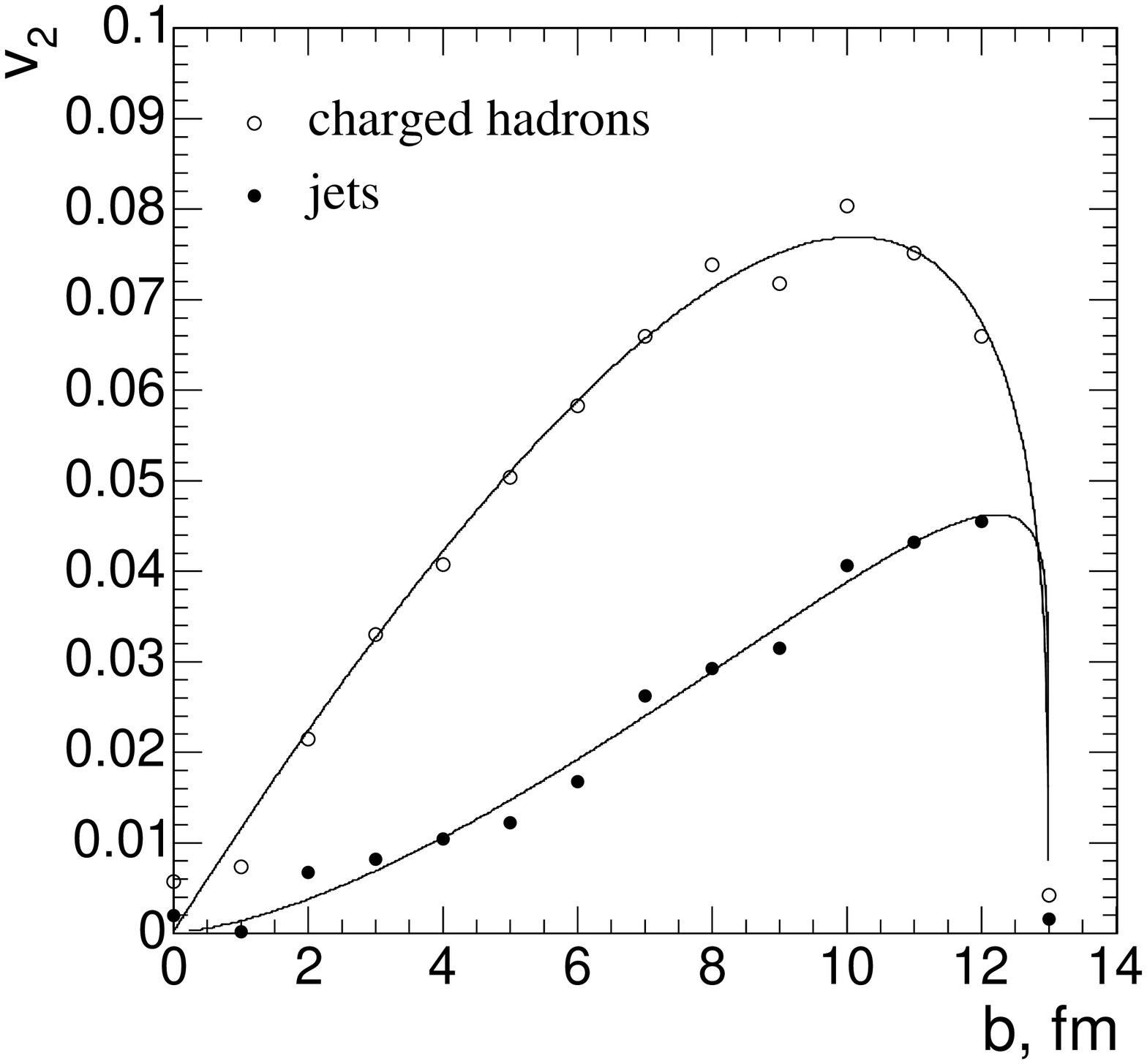}
\caption{The impact parameter dependence of elliptic flow coefficients 
$v^{\rm jet}_2$ for jets with $E_T^{\rm jet}>100$ GeV (black circles) and 
$v^{\rm h}_2$ for inclusive charged hadrons with $p_T>20$ GeV$/c$ (open 
circles) in PbPb events triggered on jets.}
\label{fig:v2_jet}
\end{minipage}
\hspace{\fill}%
\begin{minipage}{18pc}
\includegraphics[width=18pc]{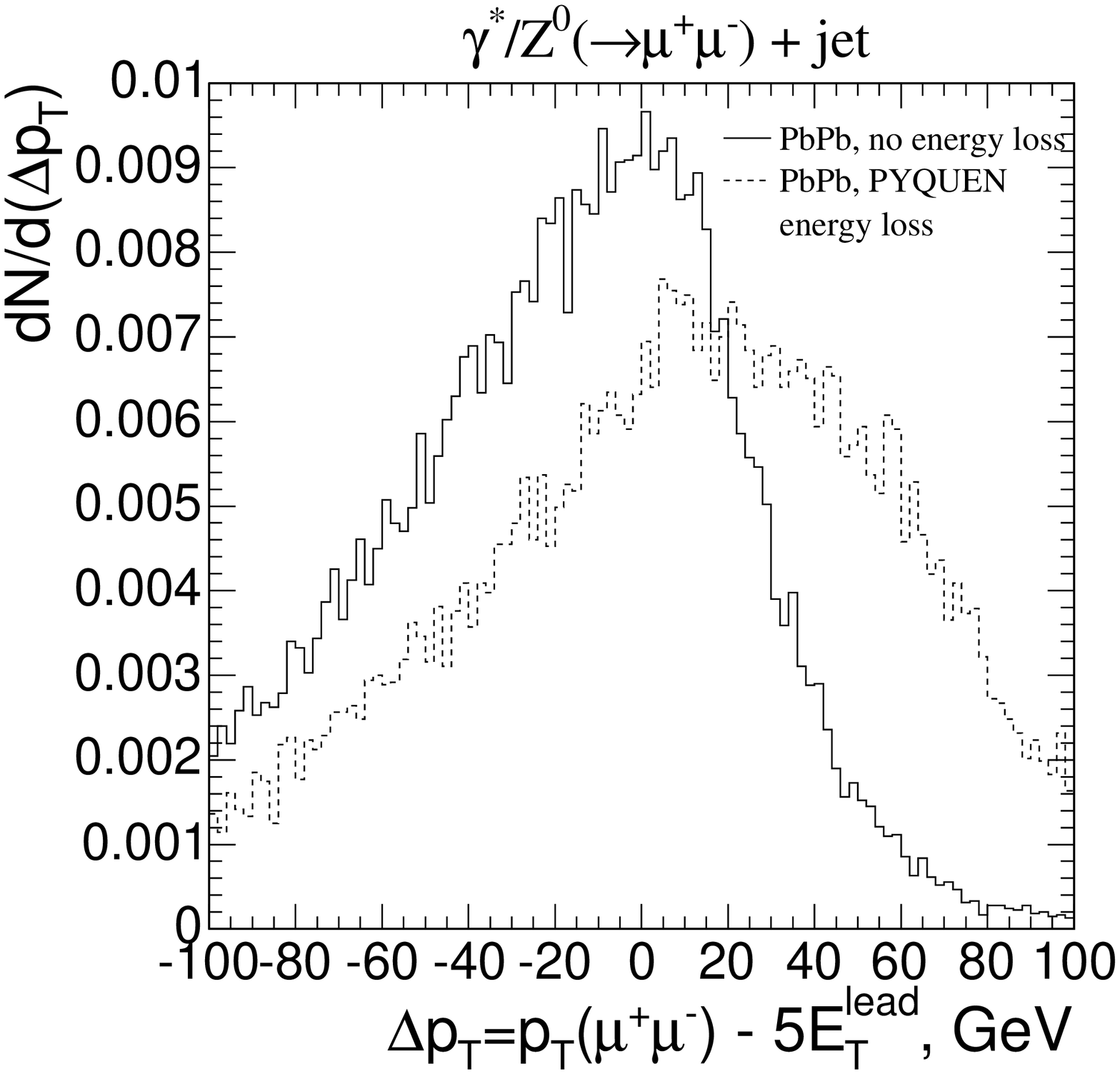}
\caption{The distribution of the difference between the transverse
momentum of a $\mu ^+\mu ^-$ pair and five times the transverse energy of the 
leading particle in a jet in PbPb collisions with (dashed histogram) and 
without (solid histogram) energy loss.}
\label{fig:mujet_elos}
\end{minipage} 
\end{figure}

The azimuthal anisotropy of particle spectrum is characterized by the second 
coefficient of the Fourier expansion of particle azimuthal distribution, 
elliptic flow coefficient, $v_2$. The non-uniform dependence of medium-induced 
partonic energy loss in non-central heavy ion collisions on the parton 
azimuthal angle $\varphi$ (with respect to the reaction plane) is mapped 
onto the final hadron spectra~\cite{Lokhtin:2001kb,Lokhtin:2002jd}. 
Figure~\ref{fig:v2_jet} shows the calculated impact parameter dependence of 
$v_2$ coefficient for jets with $E_T^{\rm jet}>100$ GeV and for inclusive 
charged hadrons with $p_T>20$ GeV$/c$ in PbPb events triggered on jets. 
The absolute values of $v_2$ for high-$p_T$ hadrons is larger that one's for 
jets by a factor of $\sim 2-3$. However, the shape of $b$-dependence of 
$v^{\rm h}_2$ and $v^{\rm jet}_2$ is similar: it increases almost 
linearly with the growth of $b$ and becomes a maximum at 
$b \sim 1.6 R_A$ (where $R_A$ is the nucleus radius). After that, the $v_2$ 
coefficients drop rapidly with increasing $b$. 

\subsubsection{$P_T$-imbalance in dimuon tagged jet events}

An important probe of medium-induced partonic energy loss in ultrarelativistic 
heavy ion collisions is production of a single jet opposite to a gauge boson 
such as $\gamma^\star$/$Z^0$ decaying into dileptons. The advantage of such 
processes is that the mean initial transverse momentum of the hard jet equal to 
the mean initial/final transverse momentum of boson, and the energy lost by the 
parton can be estimated from the observed $p_T$-imbalance between the leading 
particle in a jet and the lepton pair. Figure~\ref{fig:mujet_elos} shows the 
difference between the transverse momentum of a $\mu ^+\mu ^-$ pair from 
$\gamma^\star$/$Z^0$ decay, 
$p_{\rm T}^{\mu ^+\mu ^-}$, and five times the transverse energy of the 
leading particle in a jet (since the average fraction of the parent parton 
energy carried by a leading hadron at these energies is 
$z\approx 0.2$) for minimum bias PbPb collisions~\cite{Lokhtin:2004zb}. 
The cuts, $p_{\rm T}^{\mu^+\mu^-}, E_{\rm T}^{\rm jet} >50$ GeV$/c$ were 
applied. Despite the fact that the initial distribution is smeared and 
asymmetric due to initial-state gluon radiation, hadronization effects, etc., 
one can clearly see the additional smearing and the displaced mean and maximum 
values of the $p_{\rm T}$-imbalance due to partonic energy loss. The 
$p_{\rm T}$-imbalance between the $\mu ^+\mu ^-$ pair and a leading particle in 
a jet is directly related to the absolute value of partonic energy loss, and 
almost insensitive to the form of the angular spectrum of the emitted 
gluons and to the experimental jet energy resolution~\cite{Lokhtin:2004zb}.

\subsubsection{High-mass dimuon and secondary $J/\psi$ spectra}

\begin{figure}[htbp]
\begin{minipage}{18pc}
\includegraphics[width=18pc]{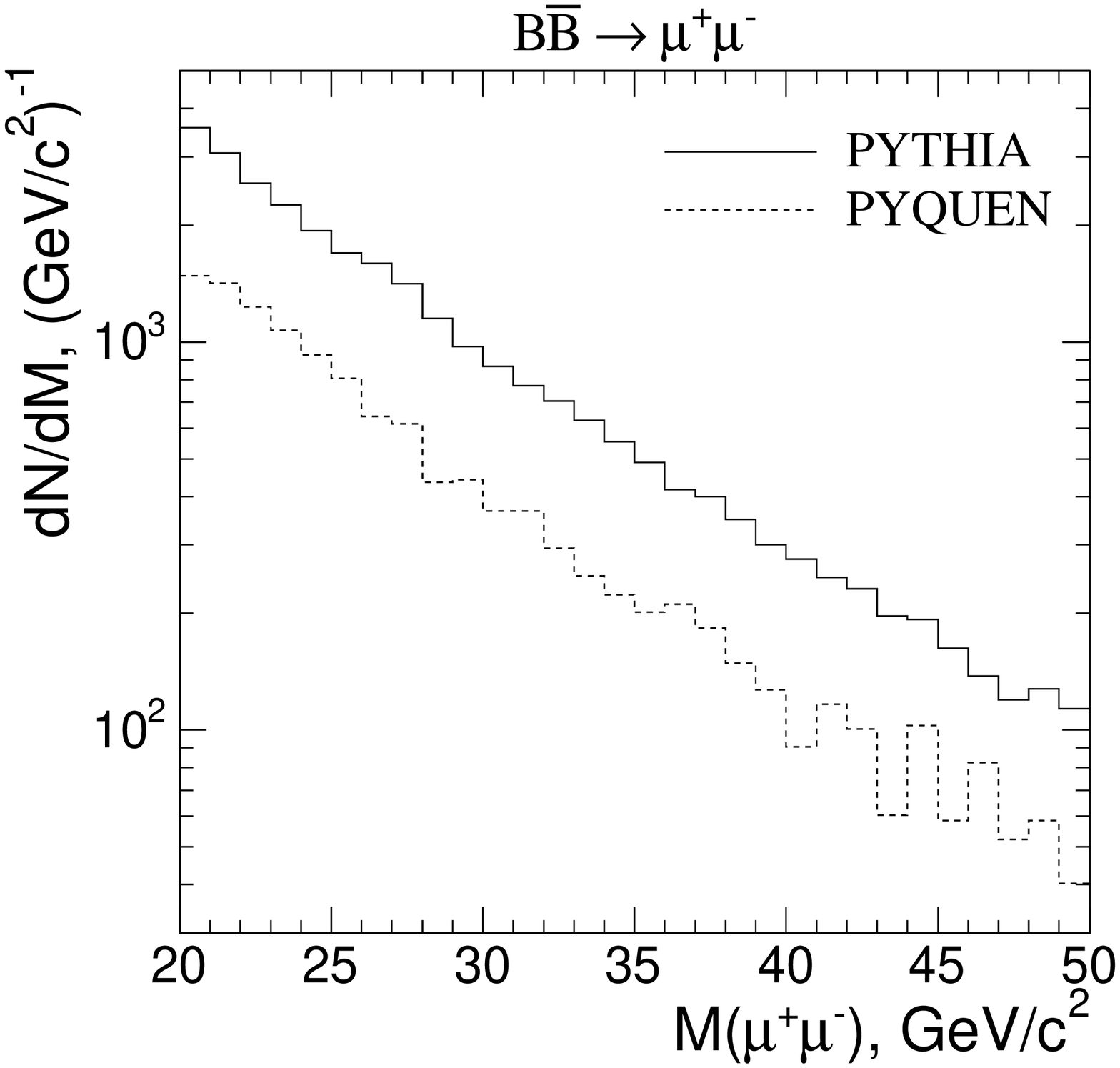}
\caption{Invariant mass distribution of $\mu^+\mu^-$ pairs from
$b \bar{b}$ decays in minimum bias PbPb collisions, with (dashed histogram) and
without (solid histogram) bottom quark energy loss. \label{fig:b-bbar}}
\end{minipage}
\hspace{\fill}%
\begin{minipage}{18pc}
\includegraphics[width=18pc]{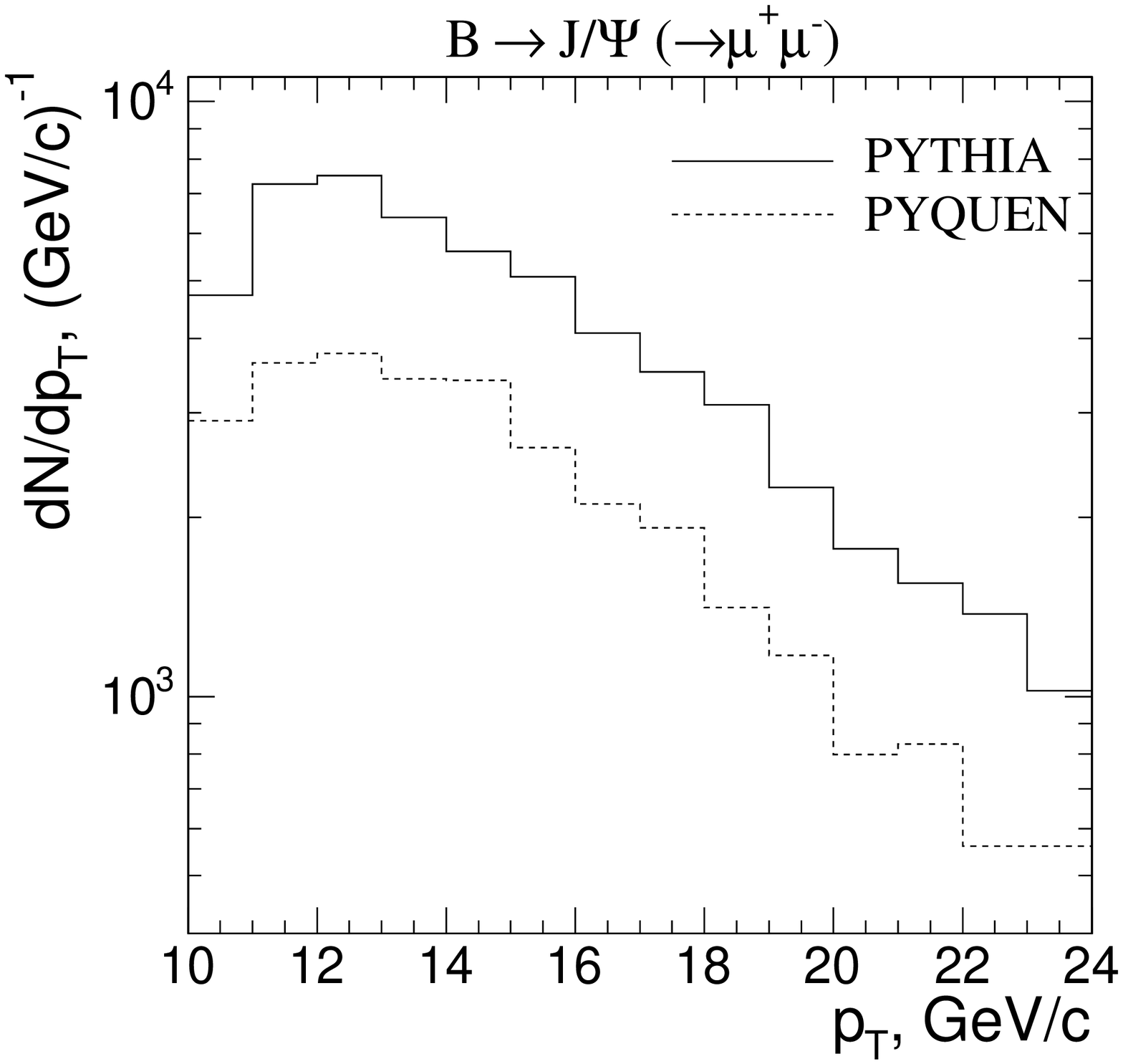}
\caption{Transverse momentum spectrum of secondary $J/\psi$ in 
minimum bias PbPb 
collisions, with (dashed histogram) and without (solid histogram) bottom quark  
energy loss. \label{fig:jpsi-b-bbar} }
\end{minipage} 
\end{figure}

While the study of inclusive high-$p_{\rm T}$ jet production in heavy ion
collisions provides information on the response of created  
medium to gluons and light quarks, the study of open heavy flavour production 
gives corresponding information on massive colour charges. 
The open charm and bottom semileptonic decays are the main sources of muon 
pairs in the resonance-free high invariant mass region, $10<M_{\mu^+ \mu^-}<70$ 
GeV$/c^2$~\cite{Virdee:1019832}. 
Other processes which also carry information about medium-induced bottom 
rescattering are secondary $J/\psi$ production from the B meson 
decay~\cite{Lokhtin:2001ay,Lokhtin:2002wu} and muon tagged 
$b$-jets~\cite{Lokhtin:2004rg}. 
Figures~\ref{fig:b-bbar} and \ref{fig:jpsi-b-bbar} show the spectra of 
high-mass $\mu^+\mu^-$ pairs and the $p_{\rm T}$-distributions of the 
secondary $J/\psi$'s respectively, for minimum bias PbPb collisions with and 
without energy loss of bottom quarks ($p_{\rm T}^{\mu} > 5$ GeV/c). A factor of 
around $2.5$ suppression for $b \bar{b} \rightarrow \mu^+\mu^-$ and $2$ for 
secondary $J/\psi$  would be clearly observed over the initial state nuclear 
shadowing expected in this kinematic region~\cite{Lokhtin:2001ay,Lokhtin:2002wu}. 

\subsection{Predictions for LHC heavy ion program within finite sQGP formation time}
\label{s:Pantuev}

{\em V. S. Pantuev}

{\small
Predictions for some experimental physical observables in nucleus-nucleus 
collisions at LHC energies are presented. 
I extend the previous suggestion that the retarded jet absorption, at RHIC by 
time about 2.3~fm/$c$, in opaque core is a natural explanation of many 
experimental data. 
At LHC this time should be inversely proportional to the square root of parton 
hard scattering density, thus about 2 times shorter than at RHIC, or 1.2~fm/$c$.
Predictions were done for hadrons, including charm hadrons, with transverse 
momentum above 5~GeV/$c$. 
I calculate nuclear modification factor $R_{\A\A}$, azimuthal anisotropy 
parameter $v_2$, jet suppression $I_{\A\A}$ for the away side jet and its
dependence versus the reaction plane orientation. 
The system under consideration is Au+Au at central rapidities. 
}
\vskip 0.5cm

In previous paper~\cite{Pantuev:2005jt} I propose a simple model, driven by 
experimental data, to explain the angular dependence of the nuclear modification
factor $R_{\A\A}$ at high transverse momentum in and out the reaction plane. 
I introduce one free parameter $L\simeq 2.3$~fm to describe the the thickness 
of the corona area with no absorption wich was adjusted to fit the experimental 
data of Au-Au collisions at centrality 50-60\%. 
The model uses realistic Woods-Saxon nuclear density distribution and nicely 
describes the $R_{\A\A}$ dependence for all centrality classes. 
I extract the second Fourier component amplitude, $v_2$, for high $p_T$ particle
azimuthal distribution and found $v_2$ should be at the level of 11-12\% purely 
from the geometry of the collision with particle absorption in the core. 
At that time I made a prediction for $R_{\A\A}$ in Cu+Cu collisions at 200~GeV 
which, as later was found, is in very good agreement with experimental data. 
Physical interpretation of the parameter $L$ could be that it is actually 
retarded jet absorption caused by {\bf the plasma formation time} 
$T=L/c\simeq 2.3$~fm/$c$ at RHIC, or at least non-trivial response of strongly 
interacting plasma  to fast moving color charge.

From experimental data at 62~GeV center-of-mass beam energy I found that this 
time should be about 3.5~fm/$c$.  
This follows the expectation on the significance of mean distance between the 
centers with mini jet production (hard scatterings) at particular beam energy. 
At LHC energy of about 5~TeV we expect $\simeq 1.2$~fm/$c$ formation 
time~\cite{Pantuev:2006iu}. 

\begin{figure}[htb]
\begin{minipage}[t]{0.48\linewidth}
\resizebox{\columnwidth}{!}{\includegraphics{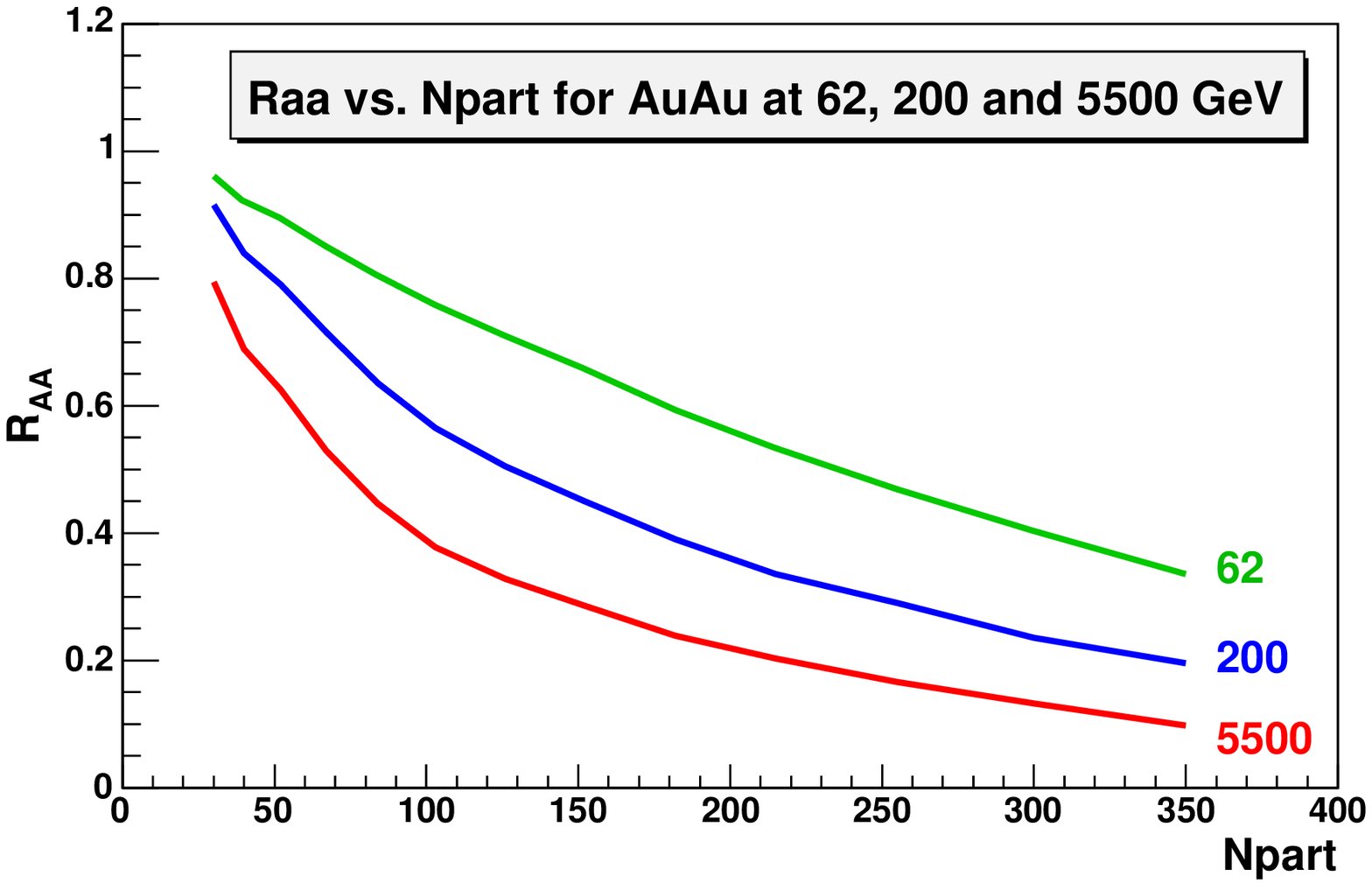}}\\
\end{minipage}
\hfill
\begin{minipage}[t]{0.45\linewidth}
\resizebox{\columnwidth}{!}{\includegraphics{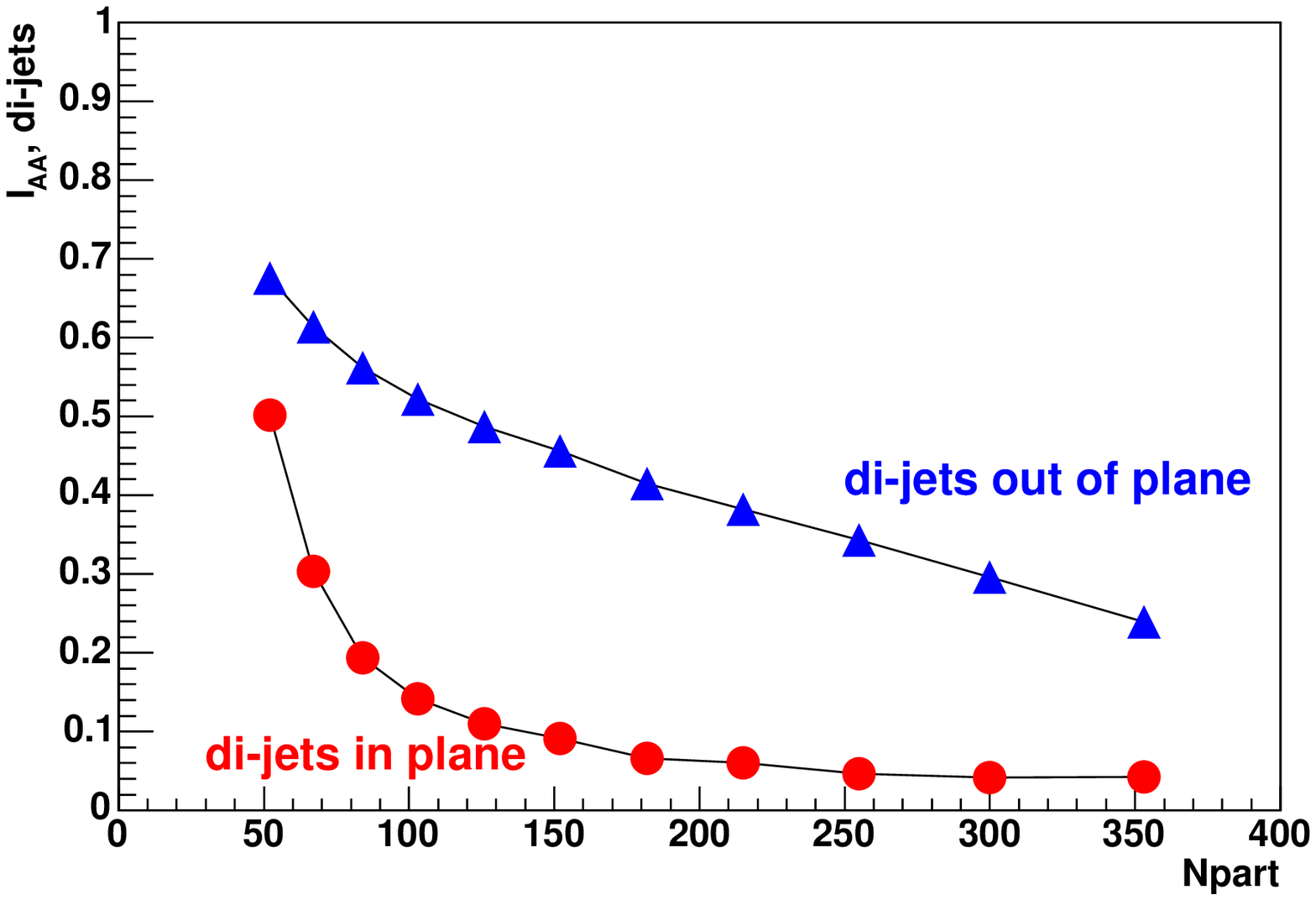}}\\
\end{minipage}
\caption{\label{fig:Pantuev-fig1} $R_{\A\A}$, nuclear modification factor, (left)
  and $I_{\A\A}$, suppression of away-side jet compared to $pp$ data versus 
  number of participants, right.
  Hadrons are at transverse momentum 5 to 20~GeV/$c$. 
  Width of the away side jet was assumed to be $\sigma=0.22$ radians.}
\end{figure}

In figure~\ref{fig:Pantuev-fig1} I show predictions for $R_{\A\A}$ and $I_{\A\A}$ 
at central rapidities. 
As usual, nuclear modification factor $R_{\A\A}$ is defined as:
\[
R_{\A\A}(p_T) = \frac{(1/N_{\rm evt}) \;{\rm d}^2N^{\A\A}/{\rm d}p_T{\rm d}\eta}
{(\langle N_{\rm binary} \rangle/\sigma^{NN}_{\rm inel}) \;
  {\rm d}^2\sigma^{NN}/{\rm d}p_T{\rm d}\eta},
\]
where $\langle N_{\rm binary} \rangle$ is a number of binary nucleon-nucleon 
collisions at particular centrality class. 

In all cases I consider hadrons (mesons and baryons, including charm) at $p_T$ 
above 4-5~GeV/$c$. 
$R_{\A\A}(p_T)$ at a such momentum should be independent on $p_T$, flat 
distribution at least to 20~GeV/$c$.

\begin{figure}[htb]
\begin{center}
\includegraphics[width=0.4\linewidth]{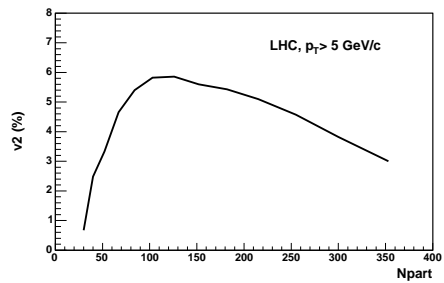}
\caption{\label{fig:Pantuev-fig2} Azimuthal assimetry parameter $v_2$ for mesons
  and baryons at transverse momentum between 5 to 20~GeV/$c$.} 
\end{center}
\end{figure}

$I_{\A\A}$ is defined as a ratio of away-side yield per trigger high $p_T$ 
particle to the similar value from $pp$ collisions. 
The major feature of this model is the dominant tangential back to back di-jet 
production from the surface region. 
Because of that we may expect {\bf significantly larger di-jet production out 
  of the reaction plane}, figure~\ref{fig:Pantuev-fig1}, in contrast to punch 
through jet scenario. 
Predictions for azimuthal asymmetry parameter $v_2$ are shown in 
figure~\ref{fig:Pantuev-fig2}.

\subsection{Hadrochemistry of jet quenching at the LHC}

%{\it Sebastian Sapeta and Urs Achim Wiedemann}
{\it S.~Sapeta and U.~A.~Wiedemann}

{\small
We point out that jet quenching can leave signatures not only in the longitudinal and transverse multiplicity distributions, but also in the hadrochemical composition of the jet fragments. As a theoretical framework, we use the MLLA+LPHD formalism, supplemented by medium-modified splitting functions. 
}
\vskip 0.5cm


%%% INTRODUCTION AND MODEL DESCRIPTION %%%%%%%%%%%%%%%%%%%%%%%%%%%%%%%%%%%%%%%%%

In heavy ion collisions at the LHC, the higher energies of produced jets will facilitate their separation
from the soft background. The interactions of jets with the matter produced in these collisions is 
expected to modify  both the longitudinal and transverse jet distributions.  In addition, we expect that
these interactions affect also  the hadrochemical composition of jets.

Within current models of jet quenching, this may be expected, since color is transfered between
the projectile and the medium - and a changed color flow in the parton shower can be expected 
to change the hadronization. More generally, one may imagine that partonic fragments of the
jet participate in hadronization mechanisms not available in the vacuum (such as a recombination
mechanism, which depends on the density of recombination partners), or that recoil effects 
kick components of the medium into the jet cone. Also, any exchange of quantum numbers
between medium and jet (e.g. baryon number or strangeness) may be reflected in the
hadrochemical composition. In the following, we consider a model which does not implement
such mechanisms, but considers solely the enhanced parton splitting due to medium 
effects~\cite{sebastianurs07}.

To calculate multiplicities of the identified hadrons we use the framework
of Modified Leading Logarithmic Approximation (MLLA) \cite{Dokshitzer:1988bq}.
This perturbative approach supplemented by the hypothesis of Local Parton-Hadron Duality (LPHD) was shown to reproduce correctly the single inclusive hadron spectra in jets both in $e^+e^-$ and $pp/p\bar p$ collisions. It provides good description not only for the distributions of all charged particles but also for the spectra of identified hadrons such as pions, kaons and protons 
\cite{Azimov:1984np, Azimov:1985by}. Moreover, the dependence on jet opening angle can be implemented. 
The general form of the multiplicity of hadrons of mass $M_h$ in the jet of energy $E_{\rm jet}$ and opening angle $\theta_c$  is given by
\begin{equation}
\frac{dN^h}{d\xi} = 
K_{\scriptscriptstyle \rm LPHD}\, D(\xi, E_{\rm jet}, \theta_c, M_h, \Lambda)\, ,
\label{eq:spec}
\end{equation}
where $\xi = \ln 1/x$ and $x = p/E_{\rm jet}$ is the fraction of the jet energy carried by the hadron~$h$. The regularization scale $\Lambda$ is a parameter of the model.

The medium-modification of jets is formulated within the MLLA formalism~ \cite{Borghini:2005em} by enhancing the singular parts of the LO splitting functions by a factor $1+f_{\rm med}$. This accounts for the nuclear modification factor at RHIC when $f_{\rm med}$ is of the order of 1, and it provides a model for the distribution of subleading jet fragments.  

\begin{figure}[t]
\begin{center}
\includegraphics[width=11.0cm]{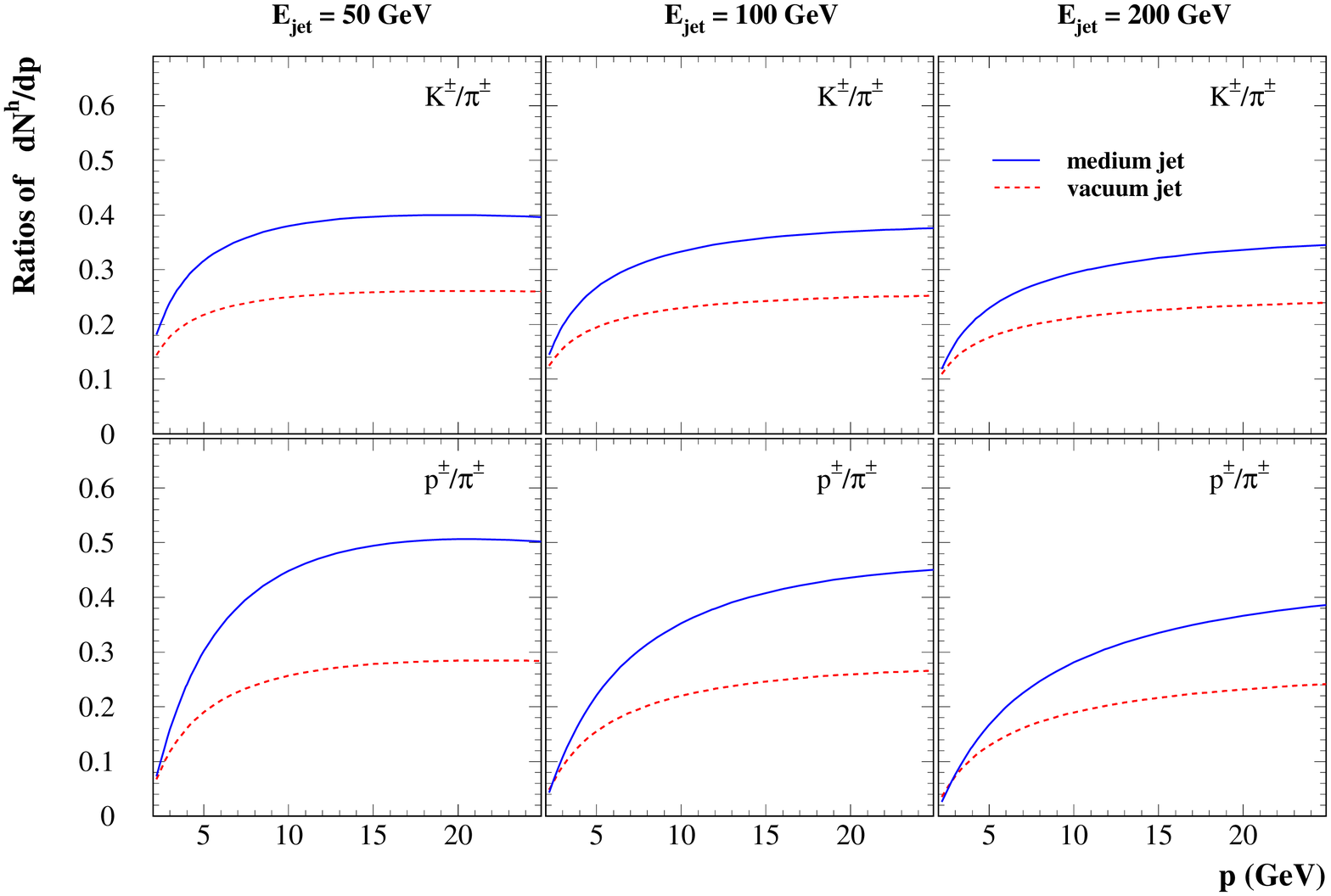}
\caption{Ratios of kaons and protons to pions in the jets with or without medium modification for
jet opening angle $\theta_c = 0.28\ $rad. The $\theta_c$-dependence is weak. The kaon multiplicity 
was in adjusted by a strangeness suppression factor 0.73 as in \cite{Azimov:1985by}. }
\label{figure1}
\end{center}
\end{figure}

One result of our studies is shown in Figure~\ref{figure1}. We observe a significant difference of the $K^{\pm}/\pi^{\pm}$ and $p^{\pm}/\pi^{\pm}$ ratios of medium-modified (with $f_{\rm med} = 1$) and 
'standard' vacuum fragmenting jets.  We have also shown, that Figure~\ref{figure1} 
remains largely unchanged if the soft background is included in forming the ratio~\cite{sebastianurs07}.

The precise numerical change of the hadrochemical composition, shown in  Figure~\ref{figure1},
is model-depement of course. We emphasize, however, that in our model, medium effects are implemented  on the partonic level only, and the hadronization mechanism remains unchanged. 
Nevertheless, the observed change is significant. Thus, our model provides a first example for our expectation, that the hadrochemical composition of jets may be
very fragile to medium effects, and provides additional information about the microscopic
mechanism underlying jet quenching.

%%% REFERENCES %%%%%%%%%%%%%%%%%%%%%%%%%%%%%%%%%%%%%%%%%%%%%%%%%%%%%%%%%%%%%%%%%

\subsection{GLV predictions for light hadron production 
and suppression at the LHC}

{\it I. Vitev}

{\small
Simulations of neutral pion quenching in Pb+Pb reactions at 
$s^{1/2} = 5.5$~A.TeV at the LHC  are presented to high transverse  
momentum $p_T$. At low and moderate  $p_T$, we study the contribution 
of medium-induced gluon bremsstrahlung to single inclusive hadron  
production. At the LHC, the redistribution of the lost energy is 
shown to play a critical role in yielding nuclear suppression 
that does not violate the participant scaling limit. Energy loss 
in cold nuclear matter prior to the formation of the QGP is also 
investigated and shown to have effect on particle 
suppression as large as doubling the parton rapidity density.
}
\vskip 0.5cm

\begin{figure}[b!]
\includegraphics[width=2.5in,height=2.5in,angle=0]{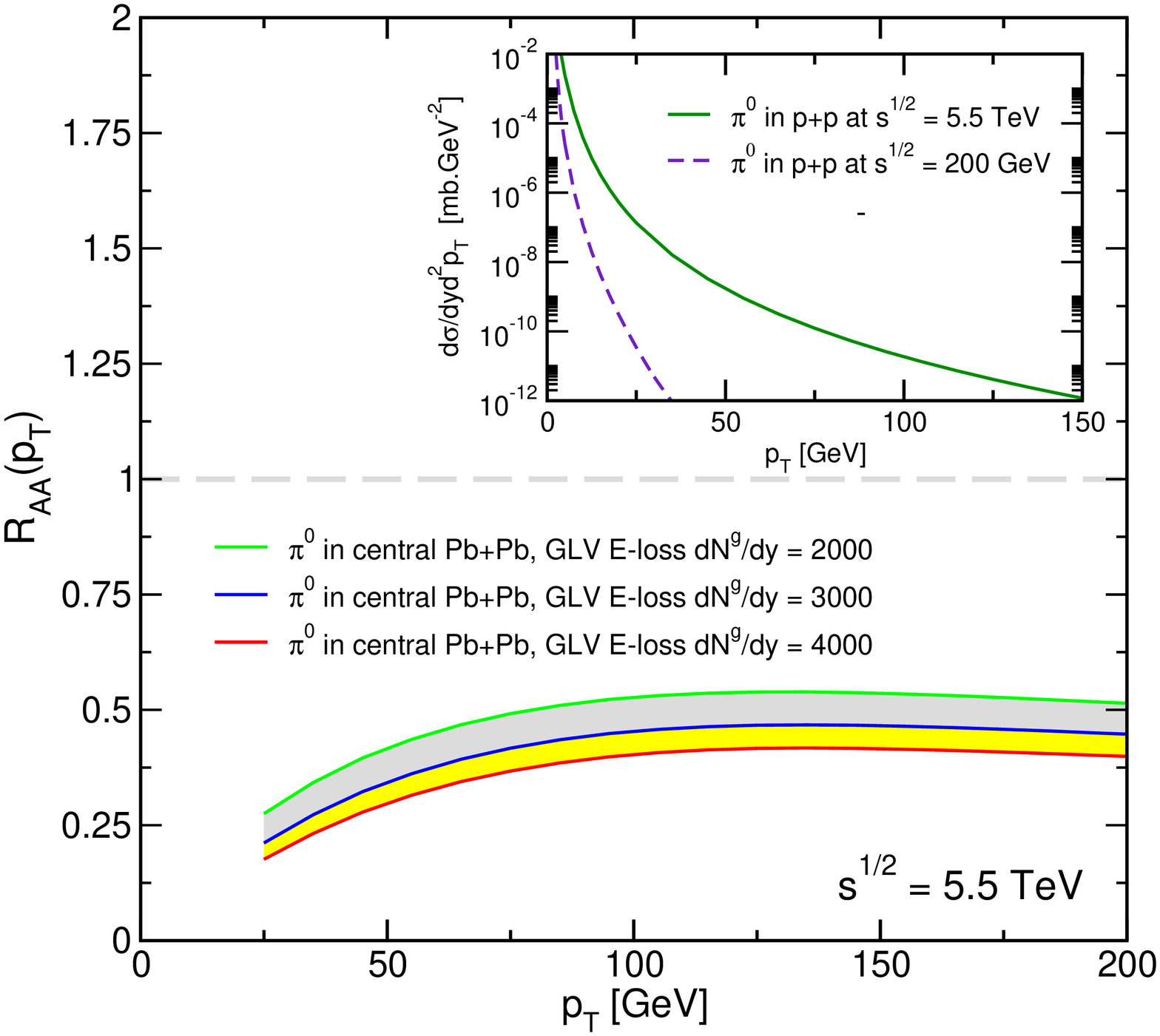}
\hspace*{.5cm}
\includegraphics[width=2.5in,height=2.5in,angle=0]{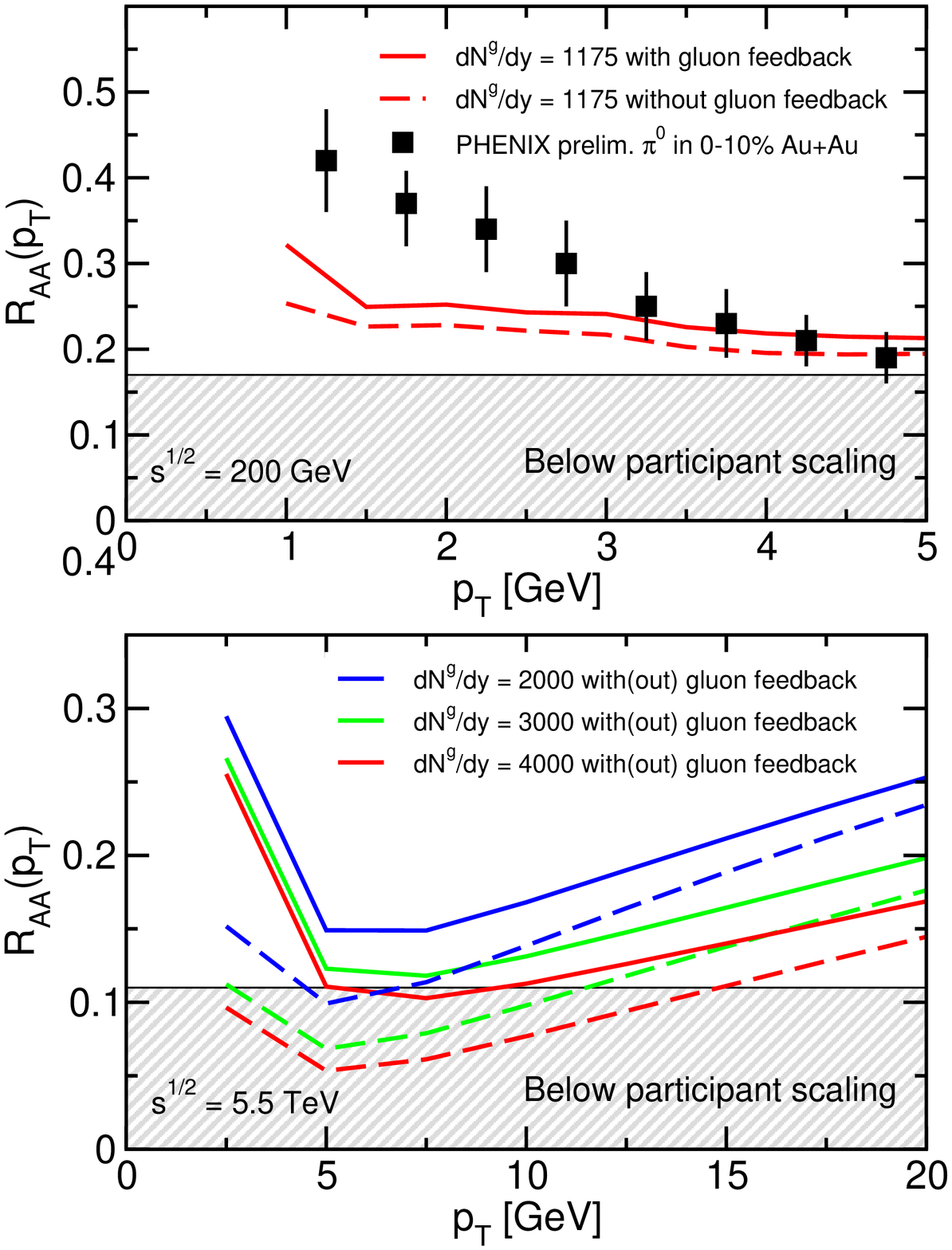}
\caption{ Left panel: Suppression of $\pi^0$ production in central Pb+Pb 
collisions at the LHC as a function of the parton rapidity 
density. Insert shows the baseline p+p $\pi^0$ cross section 
at $\sqrt{s}=200$~GeV and 
$\sqrt{s} = 5.5$~TeV~\cite{Vitev:2002pf,Vitev:2003xu}. 
Right panel: nuclear modification factor $R_{AA}$ in 
central Au+Au collisions at moderate $p_T$ with (solid line) and without 
(dashed line) gluon feedback, $dN^g/dy = 1175$. Central Pb+Pb collisions
with (solid line) and without (dashed line) gluon feedback are shown,  
$dN^g/dy \simeq 2000, 3000, 4000$~\cite{Vitev:2003xu}.}
\label{cros-LHC}
\end{figure}

Pb+Pb collisions at the LHC represent the future energy 
frontier of QGP studies in heavy ion reactions. Energy loss
of jets in the final state is calculated in the GLV 
formalism~\cite{Gyulassy:2000er}. Numerical simulations
follow the technique outlined in~\cite{Vitev:2002pf}  and
incorporate the Cronin effect~\cite{Vitev:2003xu}.
We have explored the sensitivity of $R_{AA}(p_T)$ to 
the parton rapidity density in central nuclear reactions with 
$dN^g/dy \simeq 2000, 3000$ and $4000$.  In this work we 
adhere to a more modest  two- to  four-fold increase 
of the soft hadron rapidity density and emphasize that future 
measurements of jet quenching must be  
correlated to  $dN^g/dy  \approx (3/2) 
dN^{ch}/dy$~\cite{Vitev:2002pf,Vitev:2003xu} to verify the 
consistency of the phenomenological results. See left panel
of Fig.~\ref{cros-LHC}. The contribution of the bremsstrahlung gluons to 
low- and moderate-$p_T$ inclusive particle production at the 
LHC is shown to be significant. See right panel of Fig. \ref{cros-LHC}.

\begin{figure}[b!]
\vskip 0.3cm
\includegraphics[width=2.5in,height=2.5in,angle=0]{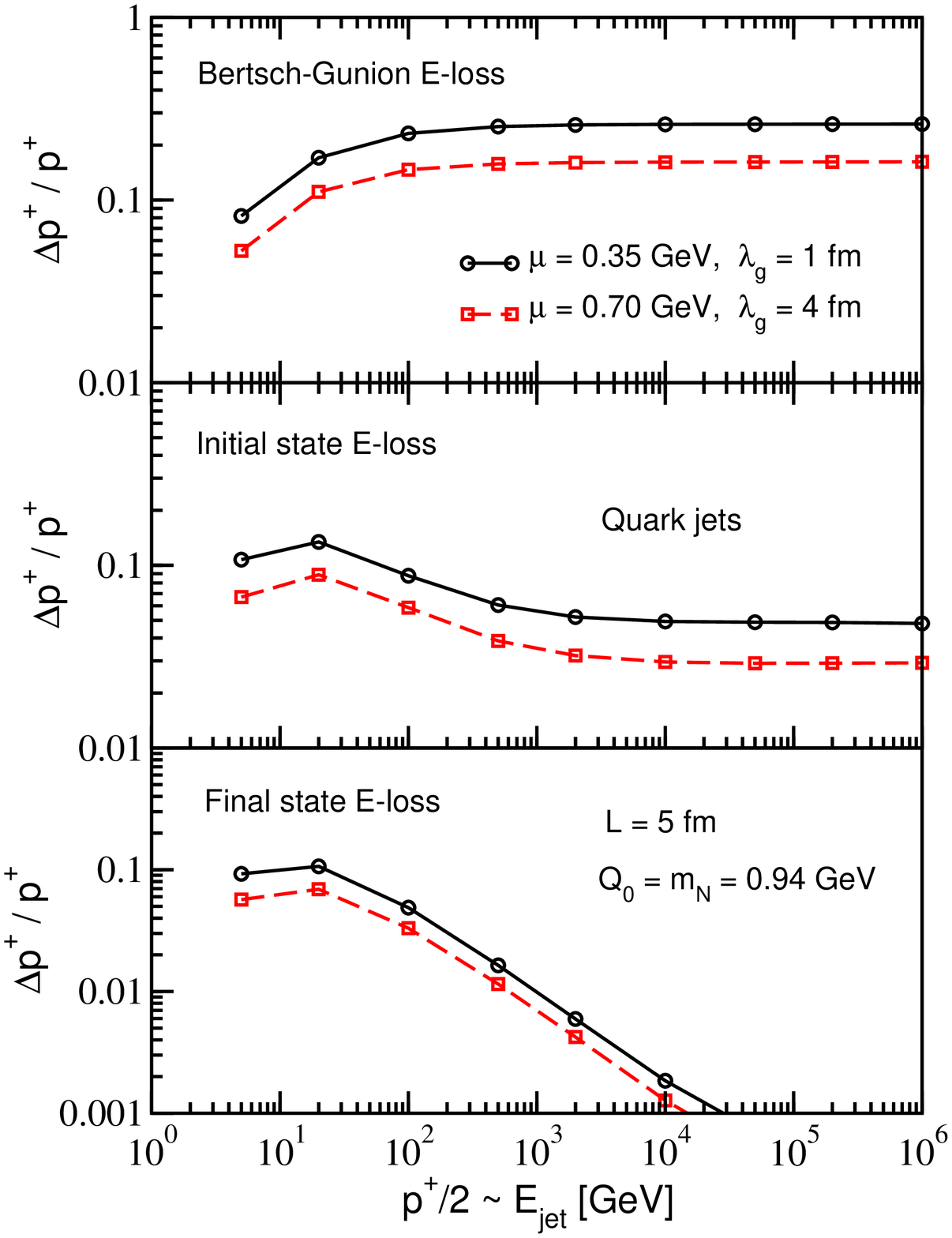}
\hspace*{0.5cm}
\includegraphics[width=2.5in,height=2.5in,angle=0]{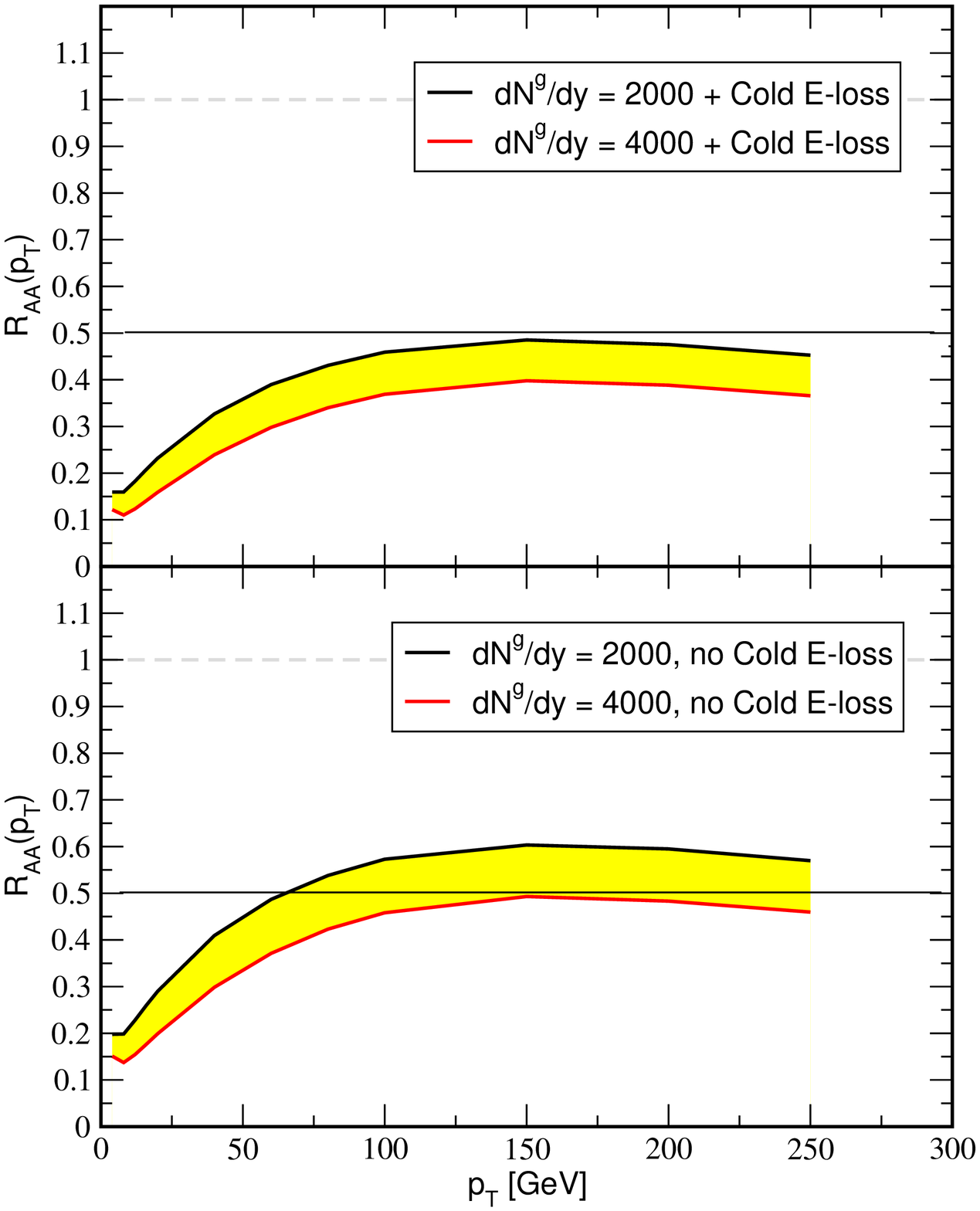}
\caption{ Left panel: Comparison of Bertsch-Gunion, initial-state
and final-state quark energy loss in a large nucleus, such as Au or Pb. 
The cancellation of initial-state energy loss is finite and cannot be 
neglected even at high parton energies~\cite{Vitev:2007ve}.
Right panel: Effects of could nuclear matter energy loss on 
suppressed $\pi^0$ production in central Pb+Pb collisions at the LHC. 
Two parton rapidity densities  $dN^g/dy \simeq 2000,  4000$. 
are shown; cold nuclear matter energy loss effects can be as large 
as the effect of doubling the parton rapidity 
density~\cite{Vitev:2003xu,Vitev:2007ve}.}
\label{cros-LHC1}
\end{figure}

Energy loss of jets in cold nuclear matter has not been 
considered before. Recent calculations in the GLV approach
show that, in contract to final-state energy loss, the 
cancellation of the bremsstrahlung in the initial-state 
is finite~\cite{Vitev:2007ve}. With $\Delta E / E \sim $few 
\%, the observable effect of the bremsstrahlung associated with 
the multiple soft scattering in nuclei is non-negligible 
even for very energetic partons in the nuclear rest 
frame. See left panel of Fig. \ref{cros-LHC1}. At the LHC, in central 
Pb+Pb collisions, the effect of cold nuclear matter energy 
loss can be as large as doubling the parton rapidity 
density $dN^g/dy$ mainly due to reduced sensitivity in the
final state. See right panel of Fig. \ref{cros-LHC1}.

\subsection{NLO Predictions for Single and Dihadron Suppression in
Heavy-ion Collisions at LHC}

%{\it Enke Wang, Xin-Nian Wang and Hanzhong Zhang}
{\it E.~Wang, X.-N.~Wang and H.~Zhang}

{\small
Suppresions of high transverse momentum
single and dihadron spectra at LHC are calculated within a next-to-leading
order perturbative QCD model with energy parton energy loss.
}
\vskip 0.5cm

The predictions presented here 
are calculated within a NLO pQCD Monte 
Carlo based program \cite{Owens:2001rr}. For the study of large $p_T$ 
single and dihadron production in $A+A$
collisions, we assume that the initial hard scattering cross sections are
factorized as in $p+p$ collisions. We further assume that the effect 
of final-state interaction between produced parton and the bulk medium 
can be described by the effective medium-modified FF's.
%\begin{eqnarray}
%D_{h/c}(z_c,\Delta E_c,\mu^2) &=&(1-e^{-\langle \frac{L}
%{\lambda}\rangle}) \left[ \frac{z_c^\prime}{z_c}
%D^0_{h/c}(z_c^\prime,\mu^2)
%\right.
% \nonumber \\
%& &\hspace{-0.9in} \left.
% + \langle \frac{L}{\lambda}\rangle
%\frac{z_g^\prime}{z_c} D^0_{h/g}(z_g^\prime,\mu^2)\right] +
%e^{-\langle\frac{L}{\lambda}\rangle} D^0_{h/c}(z_c,\mu^2),
%\label{eq:modfrag}
%\end{eqnarray}
%where $z_c^\prime=p_T/(p_{Tc}-\Delta E_c)$,
%$z_g^\prime=\langle L/\lambda\rangle p_T/\Delta E_c$ are the rescaled
%momentum fractions, $\Delta E_c$ is the average radiative parton
%energy loss and $\langle L/\lambda\rangle$ is the number of
%scatterings. The FF's in vacuum $D^0_{h/c}(z_c,\mu^2)$ are given by the
%KKP parameterization \cite{KKP}.
The total parton energy loss in a finite and expanding medium is
approximated as a path integral,
\begin{equation}
\Delta E \approx \langle \frac{dE}{dL}\rangle_{1d}
\int_{\tau_0}^{\infty} d\tau \frac{\tau-\tau_0}{\tau_0\rho_0}
\rho_g(\tau,{\bf b},{\bf r}+{\bf n}\tau),
\end{equation}
for a parton produced at a transverse position ${\bf r}$ and 
traveling along the direction ${\bf n}$.
$\langle dE/dL\rangle_{1d}$ is the average parton energy loss per
unit length in a 1-d expanding medium with an initial uniform gluon
density $\rho_0$ at a formation time $\tau_0$ for the medium gluons. 
The energy dependence of the energy loss is
parameterized as
\begin{equation}
 \langle\frac{dE}{dL}\rangle_{1d}=\epsilon_0 (E/\mu_0-1.6)^{1.2}
 /(7.5+E/\mu_0),
\label{eq:loss}
\end{equation}
from the numerical results in Ref.~\cite{Wang:2001cs,Wang:2002ri}.
The parameter $\epsilon_0$ should be proportional
to the initial gluon density $\rho_0$.
The gluon density distribution
in a 1-d expanding medium in $A+A$ collisions at impact-parameter
${\bf b}$ is assumed to be proportional to the transverse profile of
participant nucleons ,
\begin{equation}
\rho_g(\tau,{\bf b},{\bf r})=\frac{\tau_0\rho_0}{\tau}
        \frac{\pi R^2_A}{2A}[t_A({\bf r})+t_A(|{\bf b}-{\bf r}|)].
\end{equation}
In fitting the RHIC data \cite{Zhang:2007ja} we have chosen
the parameters as $\mu_0=1.5$ GeV,
$\tau_0=0.2$ fm/$c$ and $\epsilon_0=1.68$ GeV/fm. We assume $\epsilon_0$
is proportional to the final multiplicity density and 
$\epsilon_0=5.6$ GeV/fm in the central $Pb+Pb$ collisions $\sqrt{s}=5.5$ TeV.
%The parton
%distributions per nucleon $f_{a/A}(x_a,{\bf r})$ inside the
%nucleus are assumed to be factorizable into the parton
%distributions in a nucleon given by CTEQ6M parameterization
%\cite{CTEQ} and different parameterizations of the
%impact-parameter dependent nuclear modification factor, including
%the isospin dependence.

\begin{figure}[htb]
\begin{minipage}[t]{75mm}
\includegraphics[width=80mm]{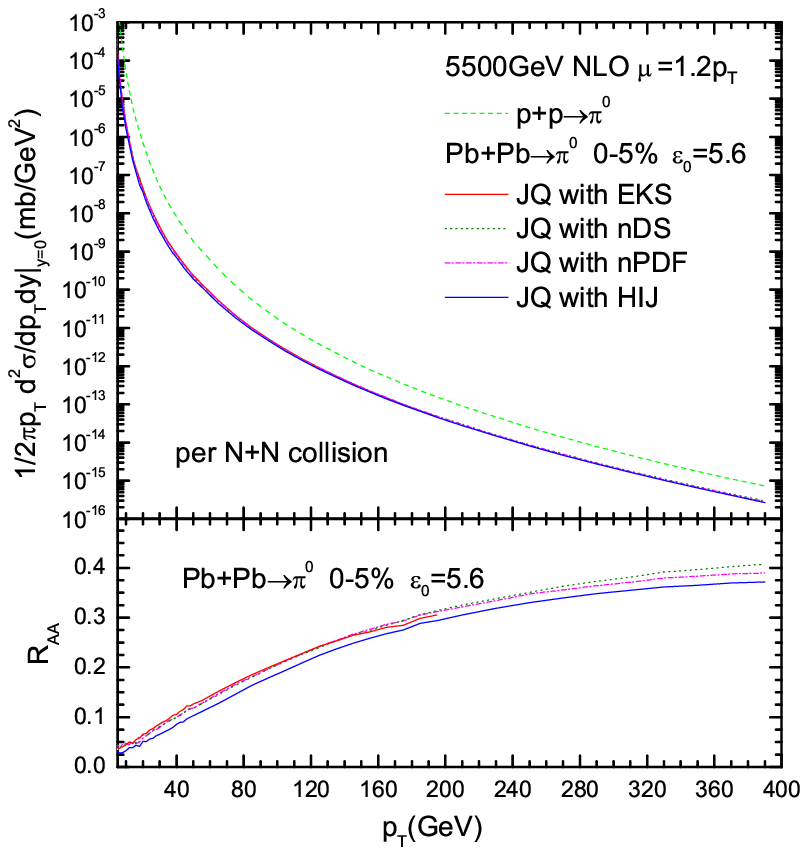}
\end{minipage}
\hspace{\fill}
\begin{minipage}[t]{75mm}
\includegraphics[width=80mm]{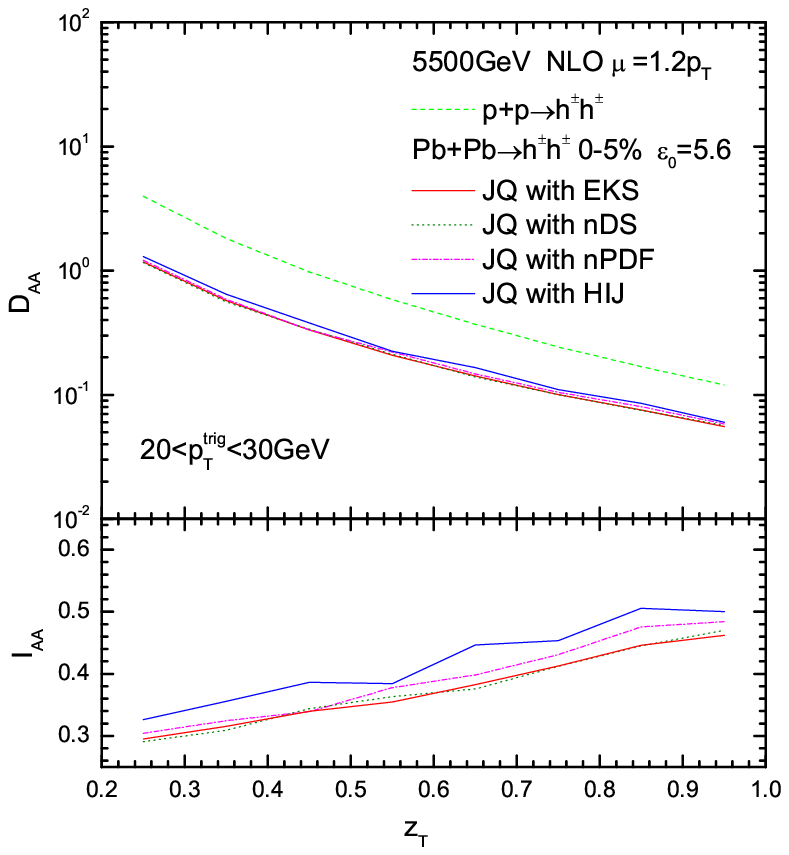}
\end{minipage}
\caption{(a) $\pi^0$ spectra and suppression factor in $Pb+Pb(0-5\%)$ 
collisions at $\sqrt{s}=5.5$ TeV. (b)Hadron-triggered fragmentation 
functions $D_{AA}(z_T)$ and the medium modification factors $I_{AA}(z_T)$ 
in NLO pQCD in central $Pb+Pb$ collisions at $\sqrt{s}=5.5$ TeV.}
\label{fig:rau0-10}
\end{figure}

We use the factorization scale $\mu=1.2 p_T$ in both $p+p$ and $A+A$ 
collisions in our calculation. Shown in Fig.~\ref{fig:rau0-10}(a) are the 
single $\pi^0$ spectra in both $p+p$ and $Pb+Pb$ collisions
at $\sqrt{s}=5.5$ TeV and the corresponding nuclear
modification factor,
$R_{AA}=d\sigma_{AA}/dp_T^2dy
[\int d^2b\, T_{AA}({\bf b})d\sigma_{NN}/dp_T^2dy]^{-1}$.

The hadron-triggered fragmentation function,\\
$D_{AA}(z_T,p_T^{\rm trig}) \equiv p_T^{\rm trig}
  d\sigma^{h_1h_2}_{AA}/dy^{\rm trig}
dp^{\rm trig}_T dy^{\rm asso}dp_T^{\rm asso}
[d\sigma^{h_1}_{AA}/dy^{\rm trig}dp^{\rm trig}_T]^{-1}$
as a function
of $z_T=p^{\rm asso}_T/p^{\rm trig}_T$ is
essentially the away-side hadron spectrum associated with a triggered
hadron within
$|y^{\rm trig, asso}|<0.5$ and the azimuthal angle relative to the
triggered hadron is integrated over $|\Delta\phi|>2.5$.

The factorization scale in the NLO calculation of dihadron
spectra is chosen to be  $\mu=1.2 M$, where $M$ is 
the invariant mass of the dihadron $M^2=(p_1+p_2)^2$.  
The associated hadron spectra $D_{pp}(z_T,p_T^{\rm trig})$ 
in $p+p$ and central $Pb+Pb$ collisions at $\sqrt{s}=5.5$ TeV
and the suppression factor
$I_{AA}=D_{AA}(z_T,p_T^{\rm trig})[D_{pp}(z_T,p_T^{\rm trig})]^{-1}$
for central $Pb+Pb$ collision at LHC are shown in 
Fig.~\ref{fig:rau0-10}(b).

%%%%%%%%%%%%%%%%%%%%%%%%%%%%%%%%%%%%%%%%%%%%%%%%%%%%%%%%%%%%%%%%%%%%%%%%%%%%
%\myack
%This work was supported 
%by DOE under contracts No. DE-AC02-05CH11231, 
%by NSFC under Project No. 10440420018, No. 10475031
%and No. 10635020, and by MOE of China under projects No. NCET-04-0744,
%No. SRFDP-20040511005 and No. IRT0624.

\section{Heavy quarks and quarkonium}
\label{sec:quarkonium}

\subsection{Statistical hadronization model predictions for charmed hadrons}
\label{andronic_charm}

{\it A. Andronic, P. Braun-Munzinger, K. Redlich and J. Stachel}

{\small
We present predictions of the statistical hadronization model for charmed 
hadrons production in Pb+Pb collisions at LHC.
}
\vskip 0.5cm

The results presented below are discussed in detail in our recent 
publication \cite{Andronic:2006ky}. 
We summarize here the values of the model parameters:
i) characteristics at chemical freeze-out: temperature, $T$=161$\pm$4 MeV;
baryochemical potential, $\mu_b$=0.8$^{+1.2}_{-0.6}$ MeV; volume 
corresponding to one unit of rapidity $V$=6200 fm$^3$;
ii) charm production cross section:
$\ud \sigma_{c\bar{c}}^{pp}/\ud y=0.64^{+0.64}_{-0.32}$ mb.

\begin{figure}[htb]
\begin{tabular}{lr}
\begin{minipage}{.49\textwidth}
\vspace{-1cm}
\centering\includegraphics[width=1.15\textwidth]{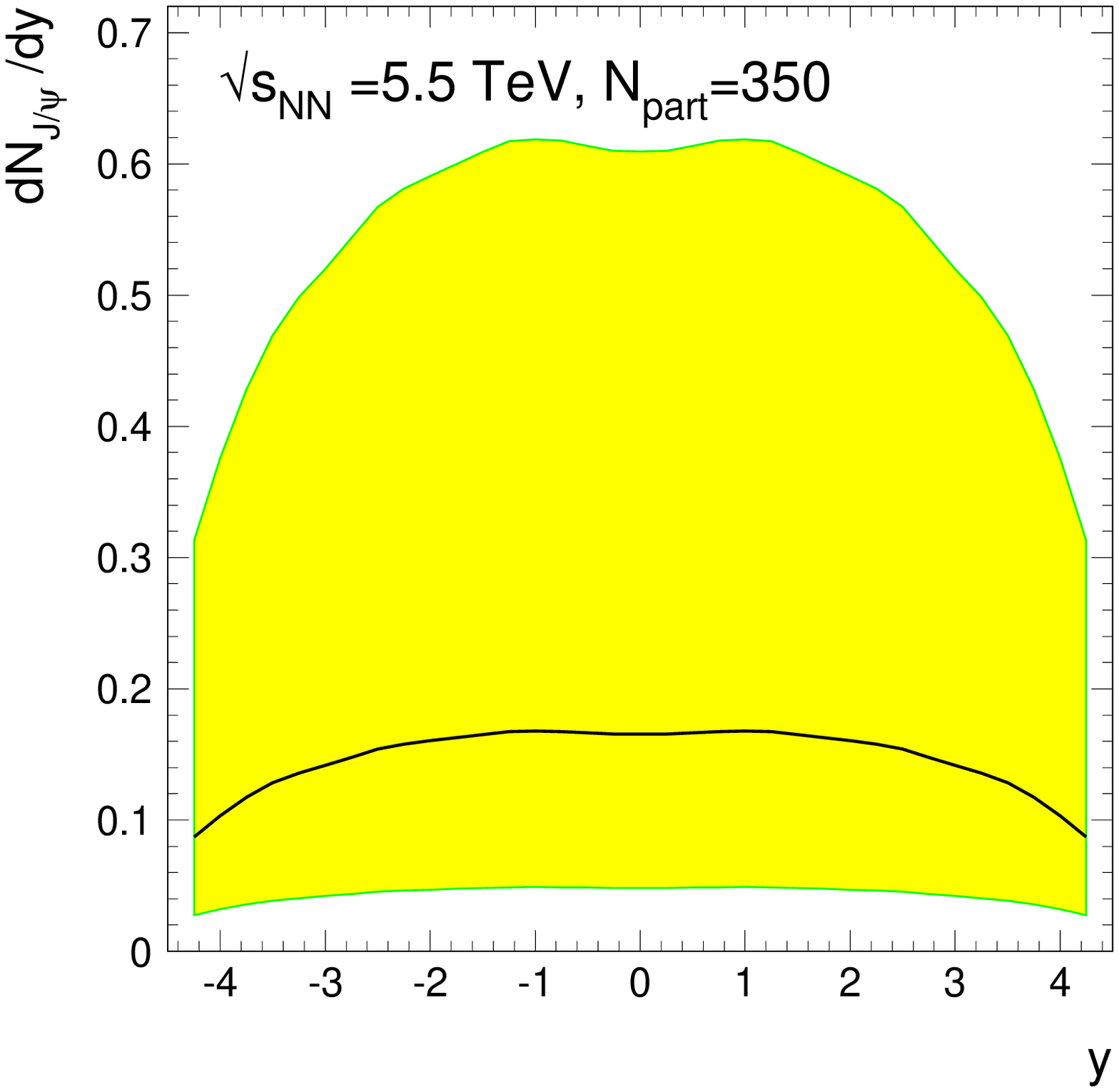}
\end{minipage}  & \begin{minipage}{.49\textwidth}
\vspace{-1cm}
\centering\includegraphics[width=1.15\textwidth]{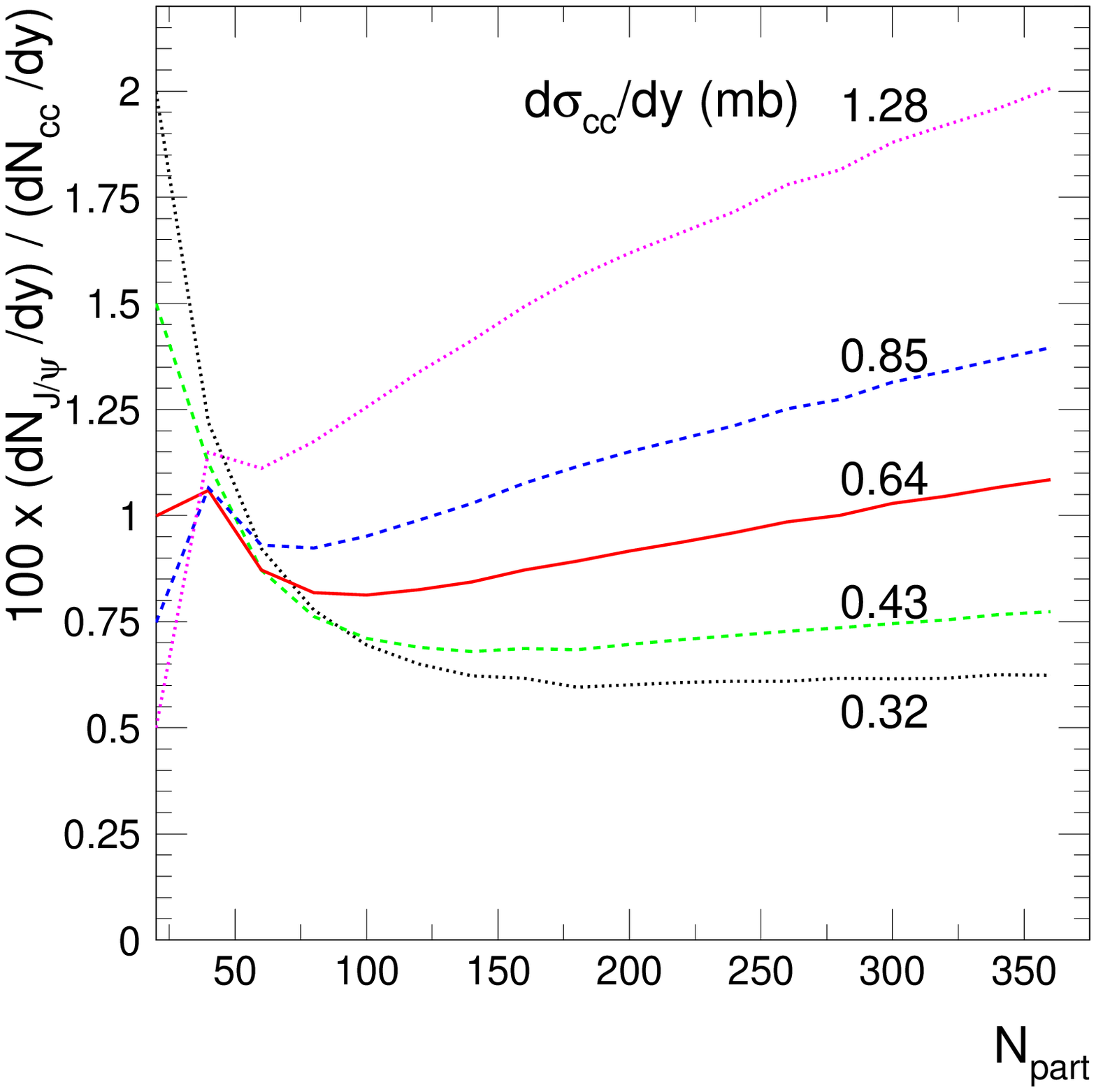}
\end{minipage} \end{tabular}
\vspace{-.8cm}
\caption{Predictions for $J/\psi$ yield: rapidity distribution for 
central collisions (left panel) and centrality dependence 
of the yield relative to the charm production yield for different
values of the charm cross section indicated on the curves (right panel).}
\label{aa_fig1c}
\end{figure} 

In Fig.~\ref{aa_fig1c} we present predictions for the yield of $J/\psi$.
The left panel shows the rapidity distribution with the band reflecting
the uncertainty in the charm production cross section.
The right panel shows the centrality dependence of the yield relative 
to the charm production yield for five values of the input charm cross section.

The statistical hadronization model predictions for charmed hadron yield 
ratios in central Pb+Pb collisions at LHC are shown in Table~\ref{tab_aa_t1}.
We expect that these ratios are independent of centrality down to values
of $N_{part}\simeq$100.

\begin{table}[hbt]
\caption{Predictions of the statistical hadronization model for charmed 
hadron ratios for Pb+Pb collisions at LHC. The numbers in parantheses 
represent the error in the last digit(s) due to the uncertainty of $T$.\\}
\label{tab_aa_t1}
\begin{tabular}{ccccccc}
$D^-/D^+$ & $\bar{D_0}/D_0$ & $D^{*-}/D^{*+}$ & $D_s^-/D_s^+$ & 
$\bar{\Lambda_c}/\Lambda_c$ & $D^+/D_0$ & $D^{*+}/D_0$\\ \br
1.00(0) & 1.01(0) & 1.01(
0) & 1.00(1) & 1.00(1) & 0.425(18) & 0.387(15) \\ \br \\
$D_s^+/D_0$ & $\Lambda_c/D_0$ & $\psi'/\psi$ & $\eta_c/\psi$ & 
$\chi_{c1}/\psi$ & $\chi_{c2}/\psi$ &\\ \br
0.349(14) & 0.163(16) & 0.031(3) & 0.617(14) & 0.086(5) & 0.110(8) & \\ \br 

\end{tabular}
\end{table}

%\vspace{-1.5cm}
\begin{figure}[htb]
\begin{tabular}{lr} \begin{minipage}{.49\textwidth}
\centering\includegraphics[width=1.22\textwidth]{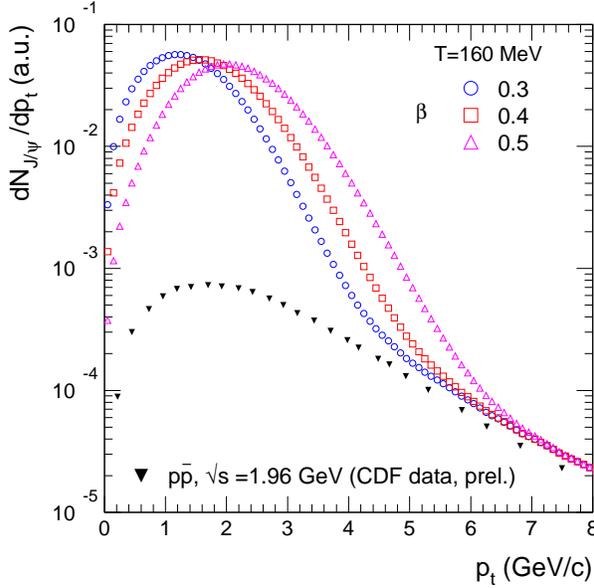}
\end{minipage}  & \begin{minipage}{.45\textwidth}
\vspace{-.8cm}
\caption{Predictions for momentum spectrum of $J/\psi$ meson
for different values of the average expansion velocity, 
$\beta$, for central Pb+Pb collisions ($N_{part}$=350). 
Also included is the measured spectrum in p$\bar{\mathrm{p}}$ collisions at 
Tevatron \cite{Pauletta:2005nt}, which is used to calculate the contribution 
from the corona (see ref. \cite{Andronic:2006ky}).}
\label{aa_fig2c}
\end{minipage} \end{tabular}
\end{figure}

Following from our model assumption of charm quark thermalization and 
assuming decoupling of charm at hadronization, the transverse momentum 
spectra of charmed hadrons can be calculated \cite{Andronic:2006ky}. 
As seen in Fig.~\ref{aa_fig2c}, a precision measurement of the spectrum 
of $J/\psi$ meson will allow the determination of the expansion velocity in QGP.

\subsection{Nuclear suppression for heavy flavors in PbPb collisions at the LHC}
\label{salgado1}

%{\it N\'estor Armesto, Matteo Cacciari, Andrea Dainese,
%Carlos A.~Salgado and Urs Achim Wiedemann}

{\it N.~Armesto, M.~Cacciari, A.~Dainese, C.~A.~Salgado 
and U.~A.~Wiedemann}

{\small
We predict the nuclear suppression factors for D and B mesons, and for electrons from their semi-leptonic decays, in PbPb collisions at the LHC. The results are obtained supplementing a perturbative
  next-to-leading order + next-to-leading log (FONLL) calculation with
  appropriate non-perturbative fragmentation functions and radiative
  energy loss. 
}
\vskip 0.5cm

\bigskip

\noindent Medium-induced gluon radiation is usually identified as the dominant source of energy loss of high-$p_T$ particles traversing a hot medium. Different models which use different approximations to this physical mechanism of energy loss, provide a successful phenomenological description of available experimental data for light hadron suppression. Most of these calculations assume independent multiple gluon emission to model the exclusive distributions essential to compute the suppression which convolutes the fragmentation functions with a steeply falling perturbative spectrum. This convolution biases the observed particle yields to small in-medium energy losses and surface emission which, on the other hand, leads to a lack of precision in the determination of the medium parameters \cite{Dainese:2004te, Eskola:2004cr}. The value of the transport coefficient obtained in these approaches, by using the multiple soft scattering approximation is  \cite{Dainese:2004te, Eskola:2004cr}
\begin{equation}
\hat q=5\div 15 \, {\rm GeV}^2/{\rm fm}.
\label{eq:qhat}
\end{equation}
One proposal to increase the sensitivity to the value of the transport coefficient is to measure the corresponding effects on heavy mesons as the formalism predicts a calculable hierarchy of energy losses $\Delta E^g>\Delta E^q_{\rm m_q=0}>\Delta E^Q_{\rm m_Q\neq 0}$, due color factors for the first and mass terms for the second inequality \cite{Armesto:2003jh}. The implementation of mass effects does not add any new parameter to the calculation once the transport coefficient $\hat q$ is fixed by e.g. light hadrons (\ref{eq:qhat}). The description of non-photonic electrons data from RHIC given by this formalism is reasonable \cite{Armesto:2005mz} although the uncertainties in the benchmark relative contribution from beauty and charm quarks are still large.

%figure---------------------------------------------
 \begin{figure}[h]
\begin{minipage}{0.5\textwidth}
\begin{center}
\includegraphics[width=7.cm,angle=-90]{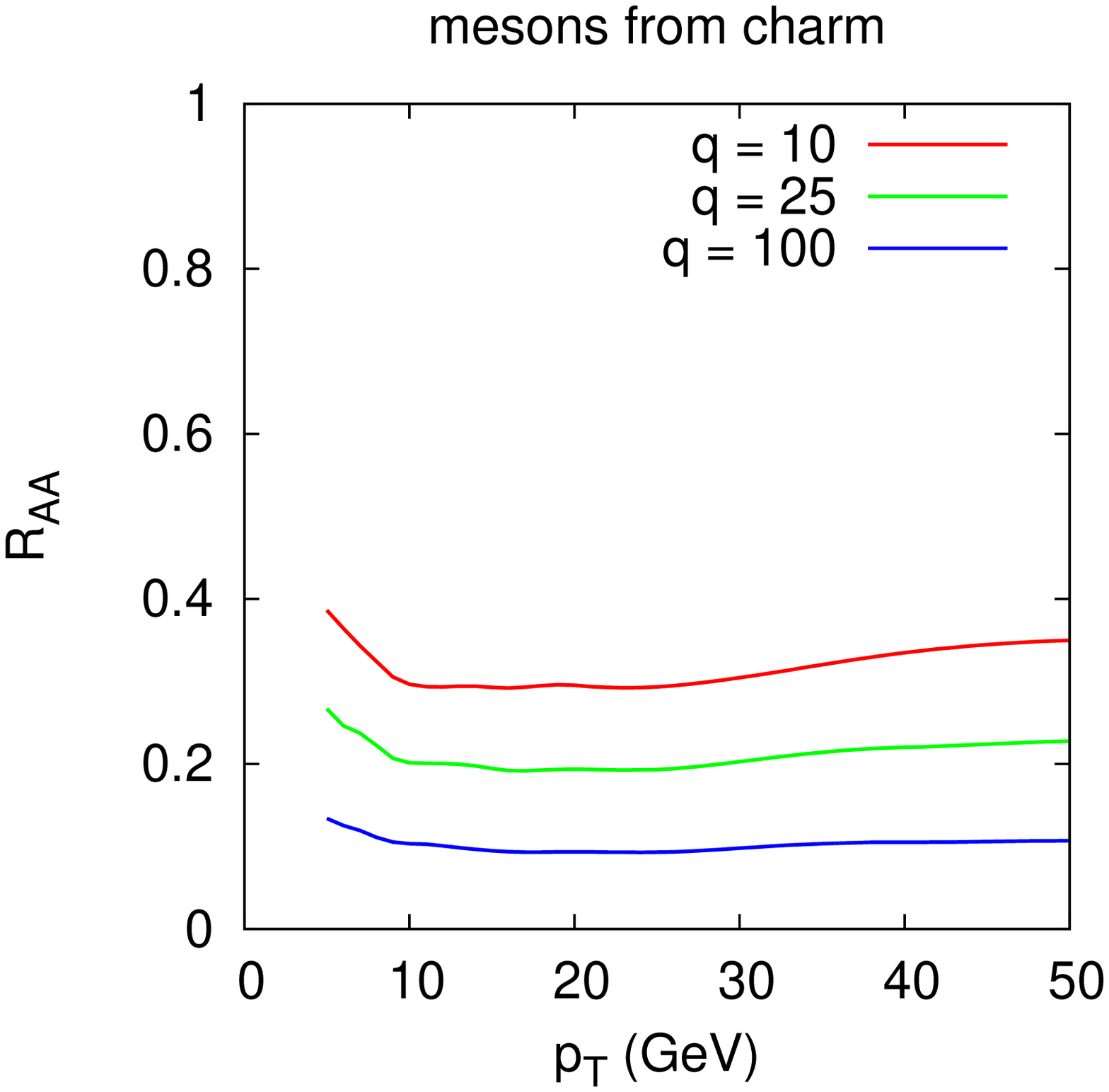}
\end{center}
\end{minipage}
\hskip -0.5cm
\begin{minipage}{0.5\textwidth}
\begin{center}
\includegraphics[width=7.cm,angle=-90]{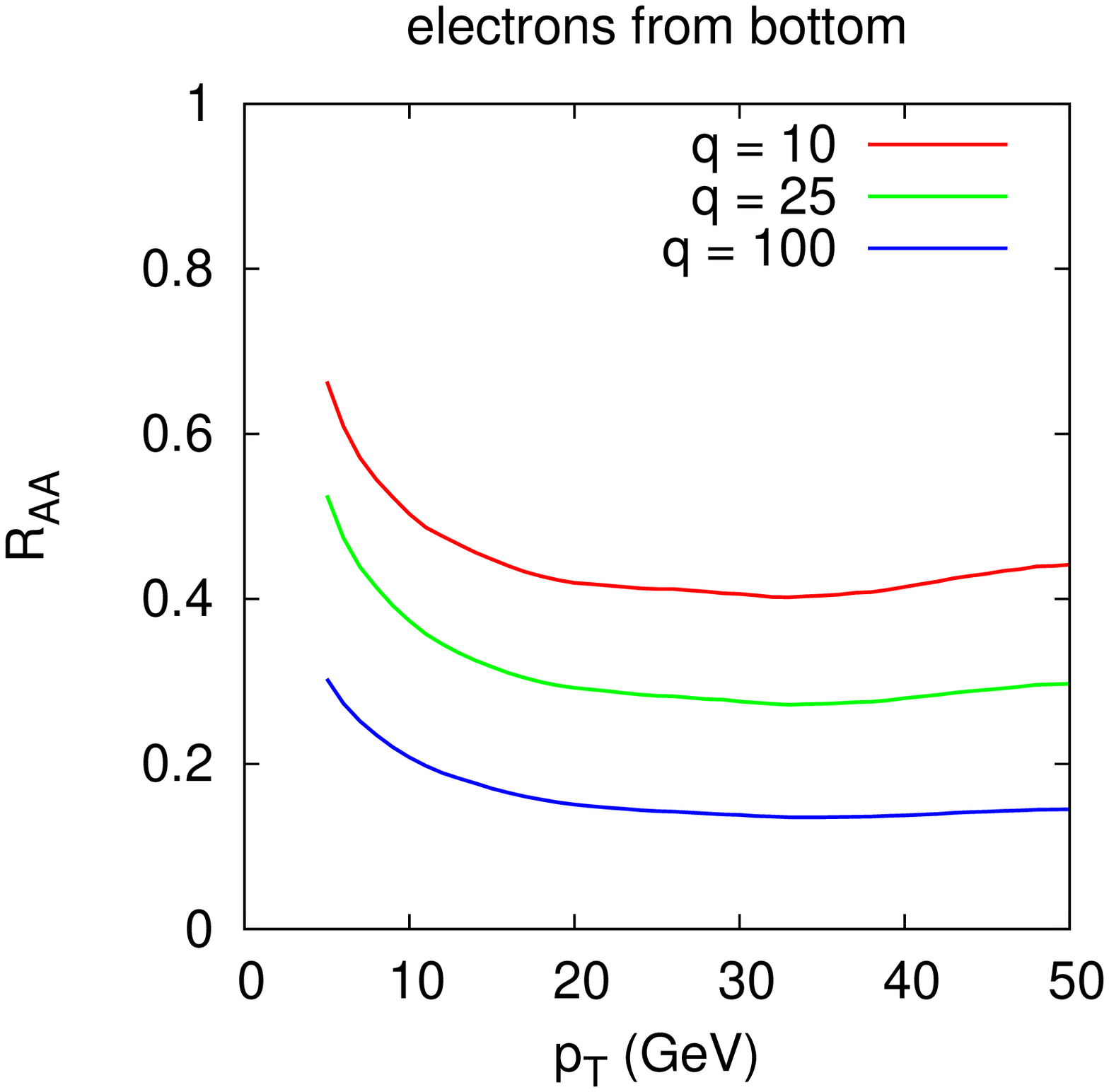}
\end{center}
\end{minipage}
\caption{$R_{AA}$ for D's (left) and for electrons coming from bottom decays (right) at $y=0$ for 10\% PbPb collisions at $\sqrt{s}=5.5$ TeV/A, for different $\hat{q}$ (in GeV$^2$/fm).}
\label{fig:raa1}
\end{figure}
%figure---------------------------------------------

At the LHC, where charm and beauty suppression will be measured separately, the situation will be improved and a definite check on the influence of mass terms in the medium-induced gluon radiation and the corresponding energy loss will be done. We here present predictions based on the formalism developed in references \cite{Dainese:2004te,Eskola:2004cr, Armesto:2003jh,Armesto:2005mz, Armesto:2005iq}, for $0-10$\% and $30-60$\% PbPb collisions at LHC energy. This also updates the calculations in  \cite{Armesto:2005iq} by taking into account the FONLL baseline for the perturbative calculation (see \cite{Armesto:2005mz} and references therein).

\begin{figure}
 \begin{minipage}{0.5\textwidth}
\begin{center}
\includegraphics[width=0.7\textwidth]{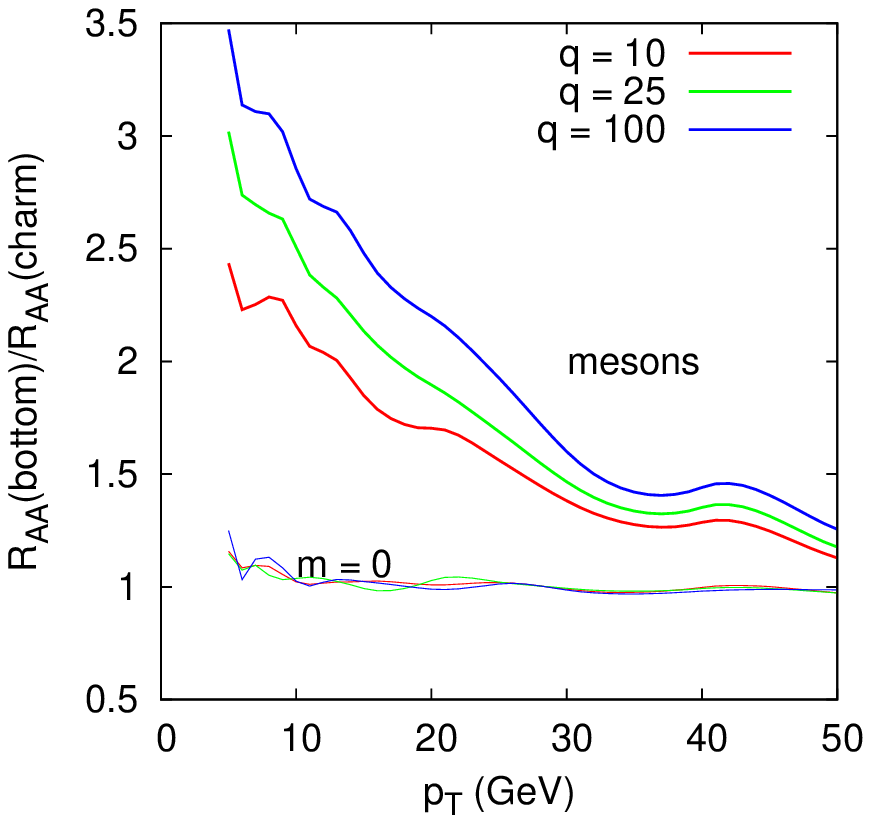}
\includegraphics[width=0.7\textwidth,height=0.64\textwidth]{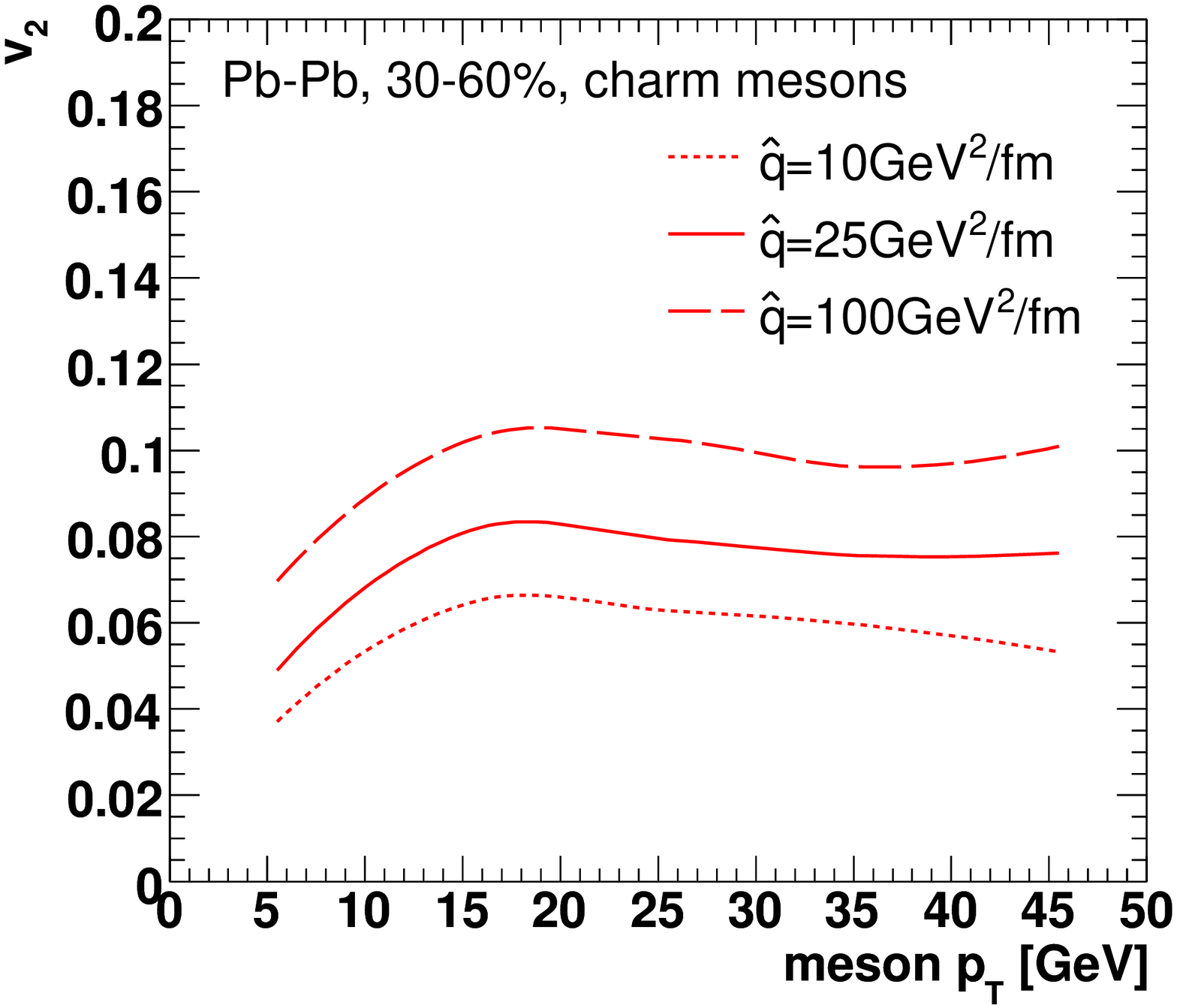}
\end{center}
\end{minipage}
\hskip -1.5cm
\begin{minipage}{0.5\textwidth}
\begin{center}
\includegraphics[width=0.7\textwidth]{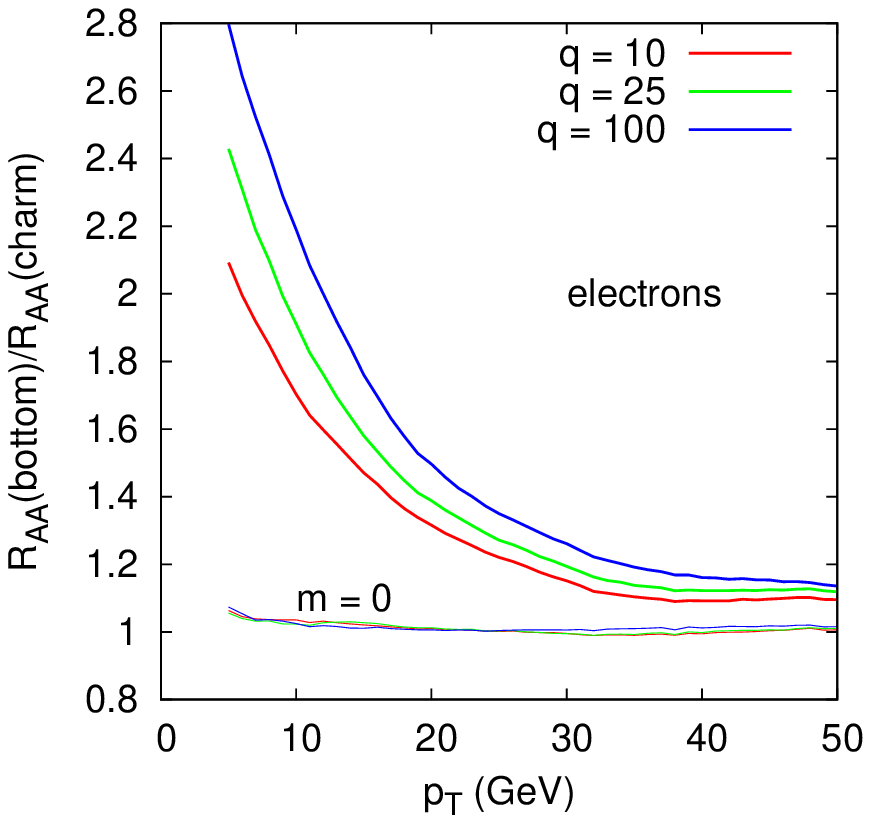}
\includegraphics[width=0.7\textwidth,height=0.64\textwidth]{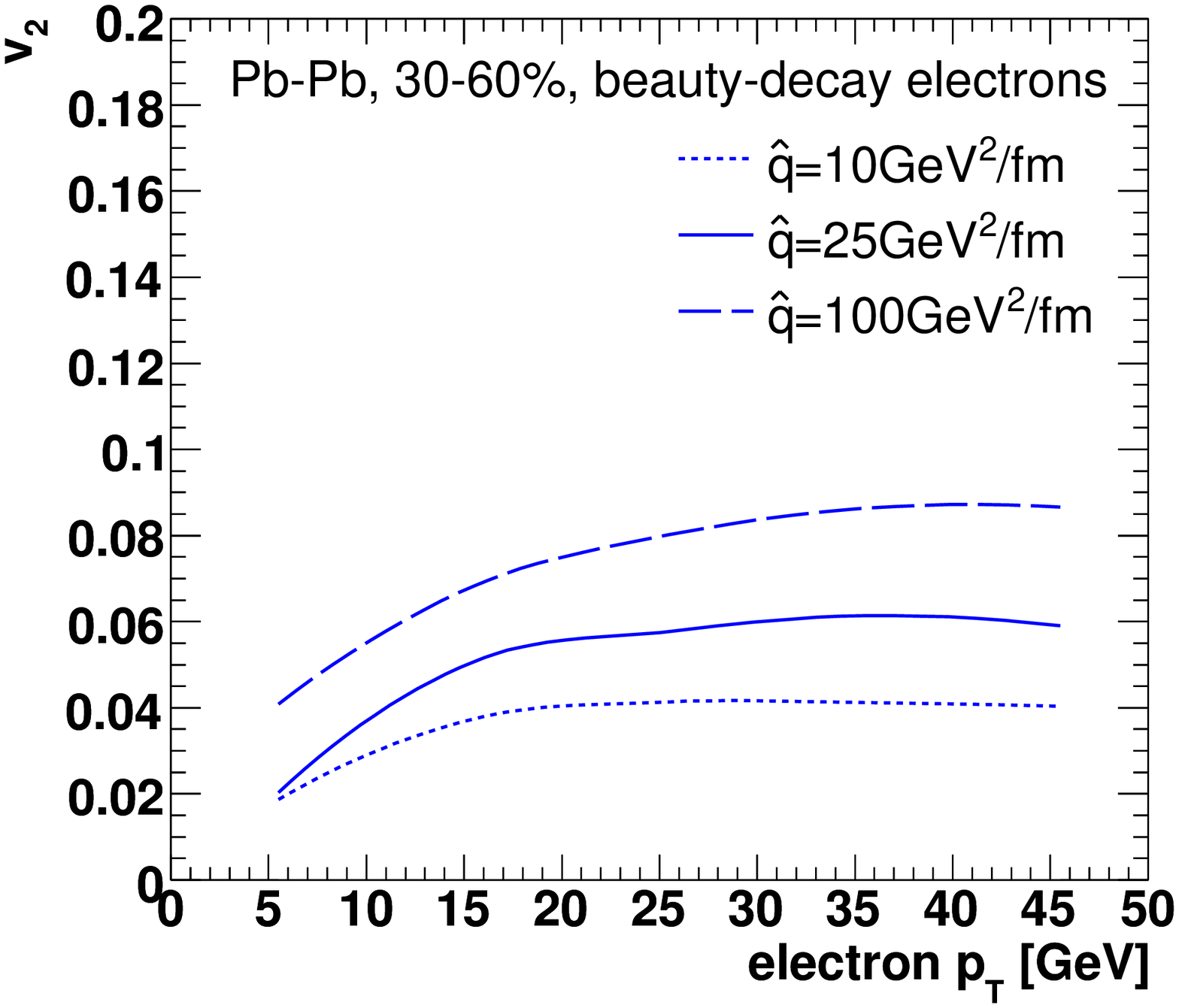}
\end{center}
\end{minipage}
\caption{Upper plots: double ratio for mesons (left) and decay electrons (right) for 10\% PbPb collisions at $\sqrt{s}=5.5$ TeV/A for $y=0$, for different $\hat{q}$ (in GeV$^2$/fm). Lower plots: $v_2$ for D's (left) and from electrons coming from bottom decays (right) at $y=0$ for $30-60$\% PbPb collisions at $\sqrt{s}=5.5$ TeV/A, for different $\hat{q}$. }
\label{fig:kinem1}
\end{figure}

In Figs. \ref{fig:raa1} and \ref{fig:kinem1}  we present our predictions for $R_{AA}$, double ratios and $v_2$,  for mesons and/or decay electrons at $y=0$. While the mass effects in charm are very modest, they are clearly visible for bottom quarks at $p_T \lesssim 20$ GeV. At larger $p_T$ they tend to disappear and the typical suppression is that of massless particles \cite{Dainese:2004te,Eskola:2004cr}. We have used several values of $\hat q$ ranging from 10 GeV$^2$/fm, which is the lowest one still compatible with RHIC non-photonic electrons data, to 100 GeV$^2$/fm, which is our estimated upper limit, from the most extreme extrapolation of the multiplicities at the LHC.

%figure---------------------------------------------
%figure---------------------------------------------

\subsection{Heavy-quark production from Glauber-Gribov theory at LHC}
\label{bravina}

{\it I. C. Arsene, L. Bravina, A. B. Kaidalov, K.
  Tywoniuk and E. Zabrodin}

{\small
We present predictions for heavy-quark production for proton-lead
collisions at LHC energy 5.5 TeV from Glauber-Gribov theory of
nuclear shadowing. We
have also made predictions for baseline cold-matter (in other words
inital-state) nuclear effects in
lead-lead collisions at the same energy that has to be taken into
account to understand properly final-state effects.
}

\subsubsection{Introduction}
In the Glauber-Gribov theory \cite{Gribov:1968jf} nuclear shadowing at low-$x$ is
related to diffractive structure functions of the nucleon, which are
studied experimentally at HERA. The space-time picture of the
interaction for production of a heavy-quark state on nuclei changes
from longitudinally ordered rescatterings at energies below the
critical energy, corresponding to $x_2$ of an active parton from a
nucleus becoming smaller than $1/m_N R_A$, to the coherent interaction
of constituents of the projectile with a target nucleus at energies
higher thant the critical one \cite{Bor93}. For production of $J/\psi$ and
$\Upsilon$ in the central rapidity region the transition happens at
RHIC energies. In this kinematical region the contribution of
Glauber-type diagrams is small and it is necessary to calculate
diagrams with interactions between pomerons, which, in our approach,
are accomodated in the gluon shadowing. A similar model for
$J/\psi$-suppression in d+Au collisions at RHIC has been considered in
Ref.~\cite{Capella:2006mb}.

Calculation of gluon shadowing was performed in our recent paper
\cite{Tywoniuk:2007xy}, where Gribov approach for the calculation of nuclear
structure functions was used. The gluon diffractive distributions were
taken from the most recent experimental parameterizations of HERA data
\cite{Aktas:2006hx}. The Schwimmer model was used to account for higher-order
rescatterings. 

\subsubsection{Heavy-quark production at the LHC}
\begin{figure}[t!]
  \begin{minipage}[t]{1.\linewidth}
    \begin{center}
      \includegraphics[scale=.5]{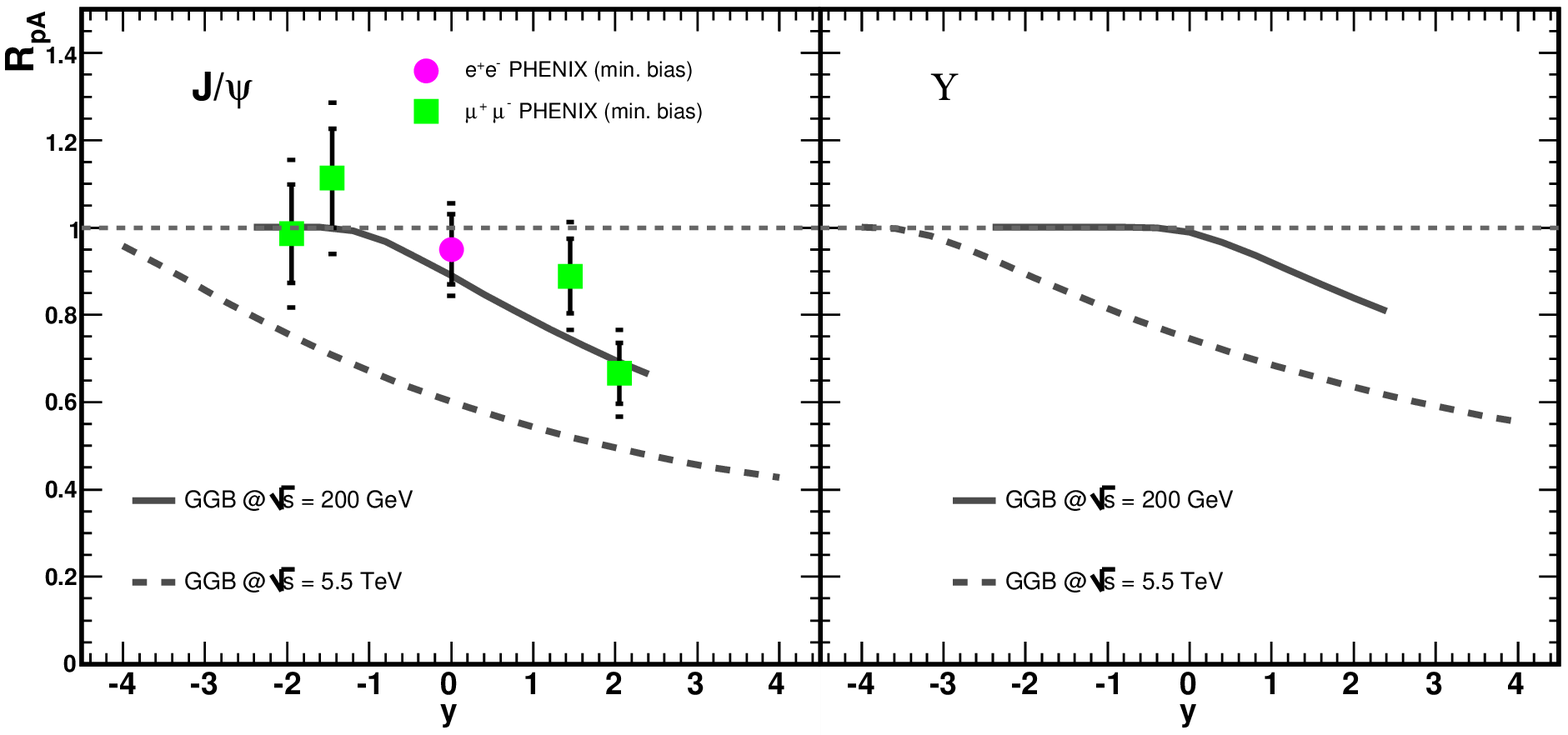}\\
      \includegraphics[scale=.5]{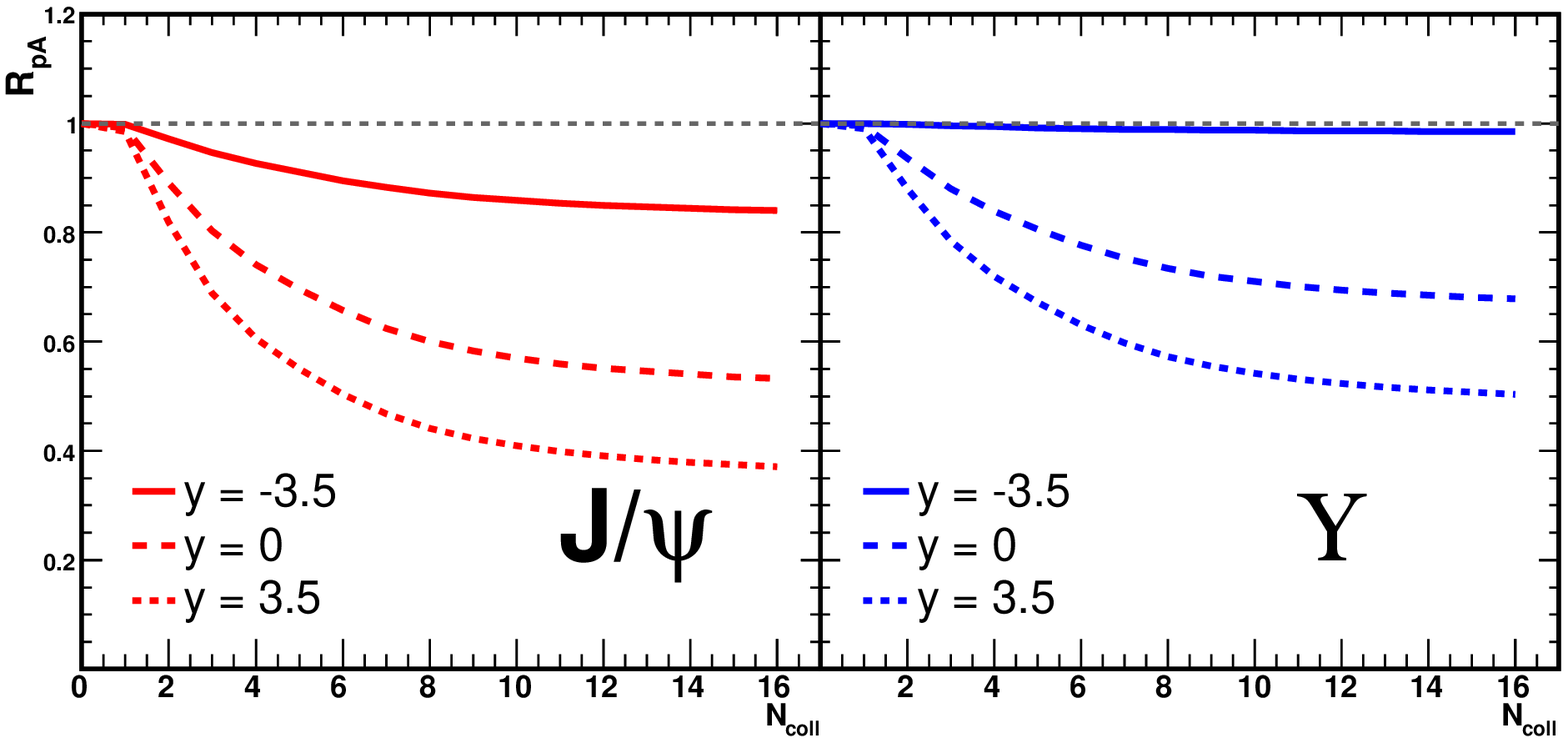}
    \end{center}
  \end{minipage}
  \caption{Rapidity (top) and centrality (bottom) dependence of the
    nuclear modification factor for 
    $J/\psi$ (left) and $\Upsilon$ (right) production in p+Pb (d+Au)
    collisions at $\sqrt{s} =$ 5500 (200) GeV. Experimental data are
    from \cite{PHE03}.}
  \label{fig:RapDep}
\end{figure}

We present predictions for the rapidity and centrality dependence of
the nuclear modification factor in proton-lead (p+Pb) collisions for
both $J/\psi$ and $\Upsilon$ in 
Fig.~\ref{fig:RapDep} (the data on $J/\psi$ suppression at $\sqrt{s} =
200$ GeV is taken from \cite{PHE03}, where also a 
definition of the nuclear modification factor can be found). We
predict a similar suppression for open charm, $c\bar{c}$, and bottom,
$b\bar{b}$, as for the hidden-flavour particles.
The observed $x_F$ scaling at low energies of the parameter $\alpha$
(from $\sigma_{pA} = \sigma_{pp} A^\alpha$) for $J/\psi$ production,
which is broken already at RHIC, will go to a scaling in $x_2$ at
higher energies. This will also be the case for $\Upsilon$ and open
charm and bottom. 

\begin{figure}[t]
  \begin{center}
    \includegraphics[scale=.4]{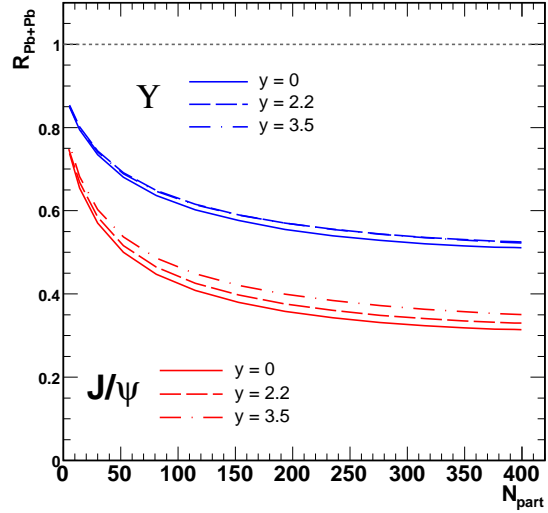}
  \end{center}
  \caption{Baseline cold-nuclear matter effects in Pb+Pb collisions at
    5.5 TeV for $J/\psi$ and $\Upsilon$ production.}
  \label{fig:ColdNucl}
\end{figure}

In Fig.~\ref{fig:ColdNucl} we present predictions for cold-nuclear
matter effects due to gluon shadowing in lead-lead (Pb+Pb) collisions
at LHC energy $\sqrt{s} = 5.5$ TeV for the production of $J/\psi$ and
$\Upsilon$. The suppression is given as a function of rapidity and
centrality. . 

\subsection{$R_{\rm AA}(p_{\rm t})$ and $R_{\rm CP}(p_{\rm t})$ of single muons from heavy quark and 
vector boson decays at the LHC}
\label{dainese2}

{\it Z. Conesa Del Valle, A. Dainese, H.-T. Ding, G. Mart\'inez and D. Zhou}

{\small
We study the effect of heavy-quark energy loss on the nuclear modification 
factors $R_{\rm AA}$ and $R_{\rm CP}$ of the high-$p_{\rm t}$ distribution of
single muons 
%at large pseudorapidities, $2.5<\eta<4$, 
in Pb--Pb collisions at $\sqrt{s_\mathrm{NN}}$=5.5~TeV. 
The energy loss of
heavy quarks is calculated using the mass-dependent BDMPS quenching weights
and taking into account the decrease of medium density at large rapidity.
Muons from W and Z decays, that dominate the yield at high $p_{\rm t}$,
can be used as a medium-blind reference that scales with the number of 
binary collisions.
}
\vskip 0.5cm

The PHENIX and STAR experiments at RHIC have measured
a suppression, in central Au--Au relative to pp collisions, 
of the high-$p_{\rm t}$ yield of non-photonic electrons,
which are assumed to come from semi-electronic 
decays of charm and beauty particles. 
This suppression is interpreted as an indication of a strong energy loss
of c and b quarks in the medium formed in Au--Au collisions. 
At the LHC, the nuclear modification factors $R_{\rm AA}$ and $R_{\rm CP}$
of the single-muon inclusive $p_{\rm t}$ distribution will be among the first 
measurements sensitive to heavy-quark energy loss. 
Moreover, the very high $p_{\rm t}$ domain ($p_{\rm t}>30~$GeV$/c$) 
of the muon spectrum will be dominated
by muonic decays of electroweak boson W (mainly) and Z,
that should be medium-insensitive and follow binary scaling, thus 
making of the nuclear modification factor a self-normalized observable.
%The single muon $p_{\rm t}$ 
%distribution will be measured by ALICE in $2.5<\eta<4$, 
%and by ATLAS and CMS in $|\eta|<2.5$. 
%For the present prediction we consider the ALICE acceptance for muons, 
%$p>1~$GeV and $2.5<\eta<4$.

We obtain the charm and beauty contributions to the muon spectrum 
from the NLO pQCD calculation (MNR~\cite{Mangano:1991jk}) supplemented with the  
mass-dependent BDMPS quenching weights for radiative energy 
loss~\cite{Armesto:2005iq},
quark fragmentation \`a la Peterson 
and semi-muonic decay with the spectator model. We account for the medium 
density decrease at large rapidity by assuming the transport coefficient
to scale as $\hat{q}(\eta)\propto dN_{ch}/d\eta$. 
We use PYTHIA to calculate the W and Z decay contribution~\cite{zaida}.
More details can be found in Ref.~\cite{hengtongQM06}.

\begin{figure}[htb]
\begin{center}
     \includegraphics[width=\textwidth]{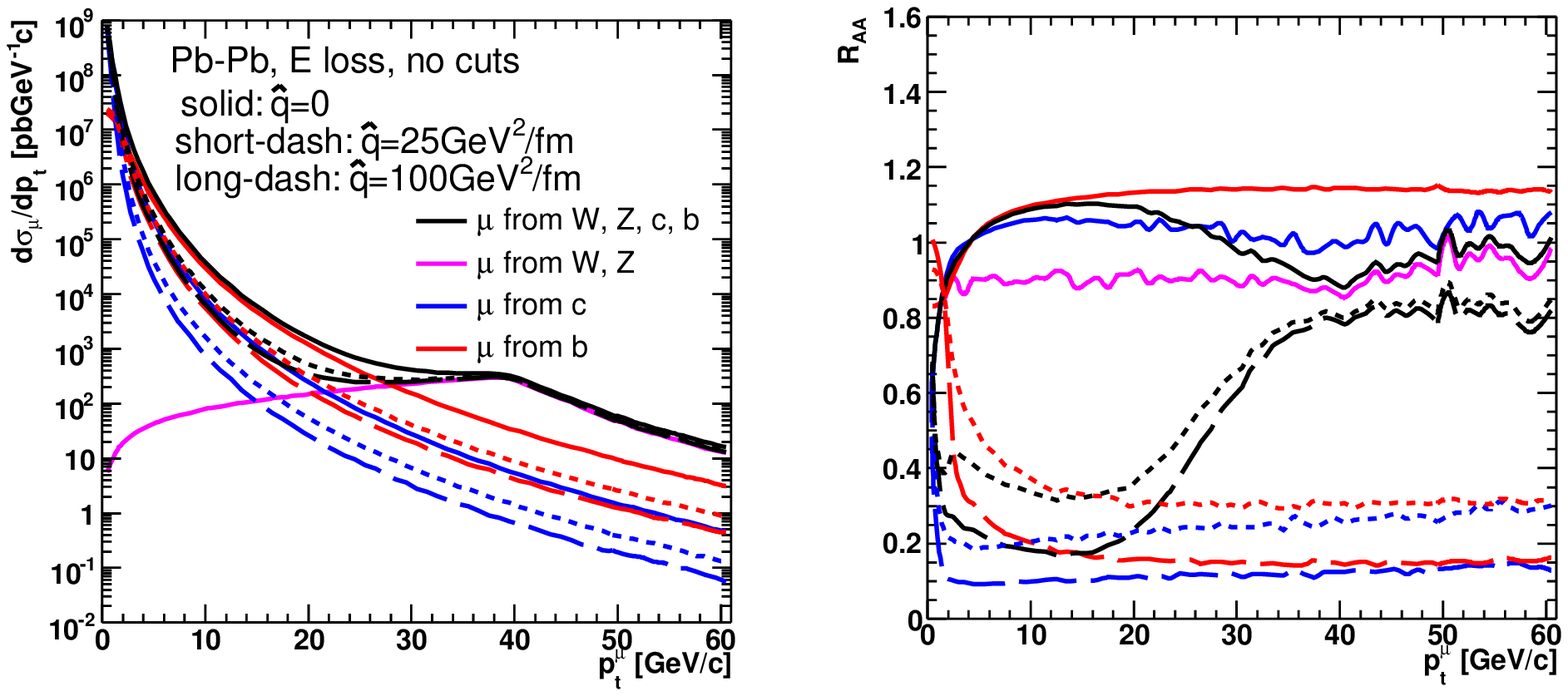}
     \caption{$p_{\rm t}$ distribution (left) and
              $R_{\rm AA}$ (right) of single muons in central (0--10\%) Pb--Pb
              collisions at $\sqrt{s_{\mathrm{NN}}}=5.5~$TeV.}
        \label{fig:raamu}
%\end{center}
%\end{figure} 
%\begin{figure}[htb]
%\begin{center}
   \includegraphics[width=0.48\textwidth]{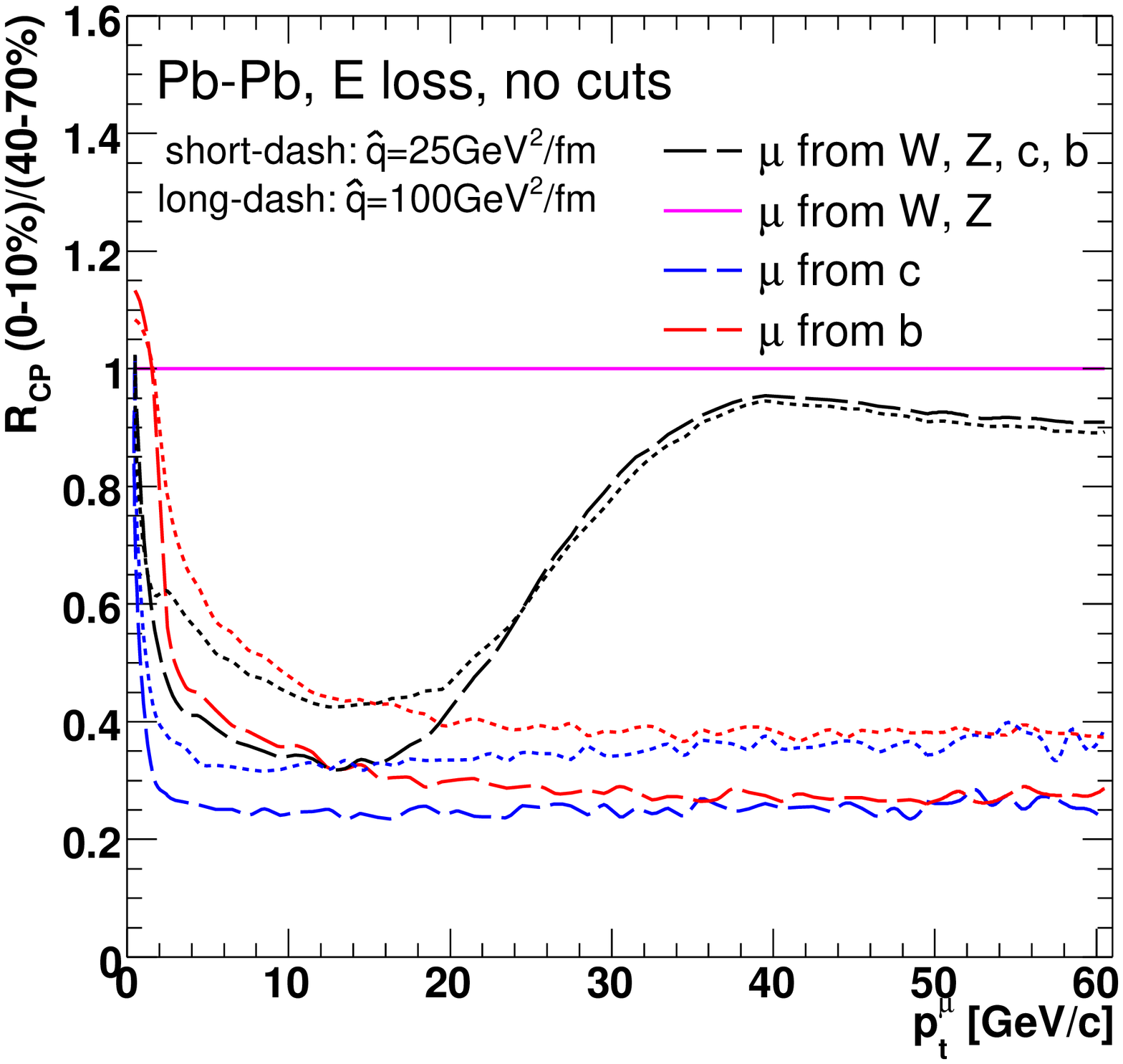}
   \includegraphics[width=0.49\textwidth]{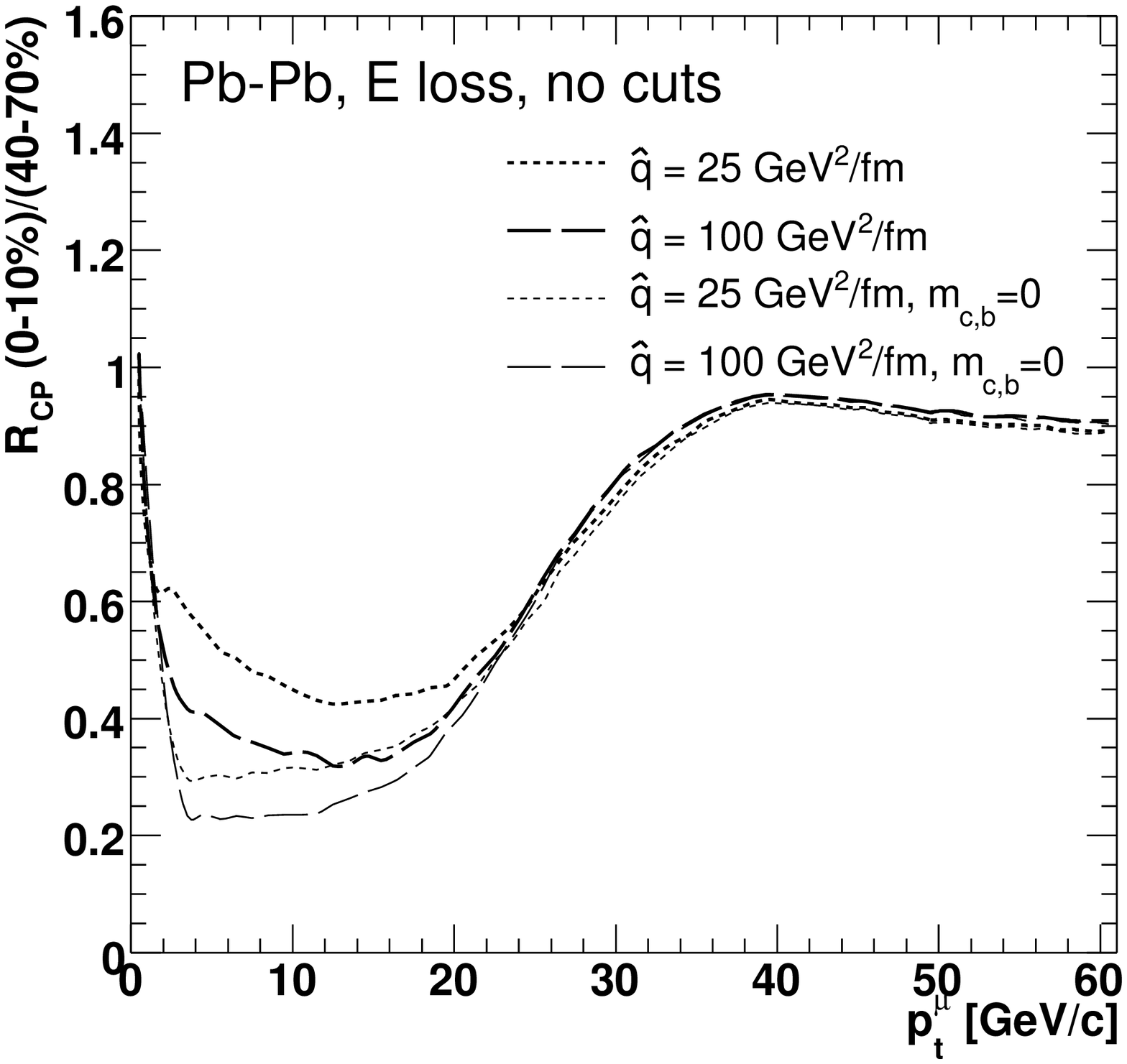}
    \caption{$R_{\rm CP}$ (0--10\%)/(40--70\%) of single muons in Pb--Pb
              collisions at $\sqrt{s_{\mathrm{NN}}}=5.5~$TeV.}
    \label{fig:rcpmu}
\end{center}
\end{figure}

Fig.~\ref{fig:raamu} shows the $p_{\rm t}$
spectrum and $R_{\rm AA}(p_{\rm t})$ of the single muons from heavy quark
and W/Z bosons in the central Pb--Pb collisions at
$\sqrt{s_{\mathrm{NN}}}=5.5~$TeV, with the transport coefficient values
$\hat{q}=0,25,100~$GeV$^{2}$/fm.
The crossing point of b and W decay muons shifts down by 5--7~GeV$/c$. 
$R_{\rm AA}$ rapidly increases
from 0.3 to 0.8 between 20 (b-dominated) and 40~GeV$/c$
(W-dominated), as does $R_{\rm CP}$ (0--10\%)/(40--70\%), shown in 
Fig.~\ref{fig:rcpmu}. The effect of the heavy-quark mass on the medium-induced
suppression of $R_{\rm CP}$ is shown in the left-hand panel of 
Fig.~\ref{fig:rcpmu}.

\subsection{Quarkonium production in coherent $pp/\A\A$  collisions and small-$x$ physics}

{\em V. P. Gon\c{c}alves and M. V. T. Machado}

{\small 
We study the photoproduction of quarkonium in coherent proton-proton and 
nucleus-nucleus interactions at the LHC.
The integrated cross sections and rapidity distributions are estimated using 
the Color Glass Condensate (CGC) formalism, which takes into account the parton 
saturation effects at high energies. 
Nuclear shadowing effects are also taken into account.
}
\vskip 0.5cm

In this contribution we study the photoproduction of vector mesons in the 
coherent $pp/\A\A$ interactions at the LHC energies. 
The main advantage of using colliding hadrons and nuclear beams for studying 
photon induced interactions is the high equivalent photon energies and 
luminosities that can be achieved at existing and future accelerators (for a 
review see reference~\cite{Goncalves:2005sn}).
Consequently, studies of $\gamma p$ interactions at LHC could provide valuable 
information on the QCD dynamics at high energies. 
The basic idea in coherent hadron collisions is that the total cross section 
for a given process can be factorized in terms of the equivalent flux of 
photons of the hadron projectile and the photon-photon or photon-target 
production cross section. 
In exclusive processes, a certain particle is produced while the target remains 
in the ground state (or is only internally excited). 
The typical examples of these processes are the exclusive vector meson 
production, described by the process $\gamma h \to V h$ ($V=\rho$, $J/\Psi$, 
$\Upsilon$). 
In the last years we have discussed this process in detail considering 
$pp$~\cite{Goncalves:2005yr}, $p\A$~\cite{Goncalves:2005ge} and 
$\A\A$~\cite{Goncalves:2005yr} collisions as an alternative to investigate the 
QCD dynamics at high energies. 
Here, we revised these results and present for the first time our predictions 
for the $\Upsilon$ production.

The cross section for the  photoproduction of a vector meson $X$ in an 
ultra-peripheral hadron-hadron collision is given by
\[
\sigma (h_1 h_2 \to h_1 h_2 X) = \int_{\omega_{min}}^{\infty} {\rm d}\omega\,
\frac{{\rm d}N_{\gamma}(\omega)}{{\rm d}\omega}\,\sigma_{\gamma h \to X h} 
(W_{\gamma h}^2),
\]
where $\omega$ is the photon energy and ${\rm d}N_{\gamma}(\omega)/{\rm d}\omega$ 
is the equivalent flux of photons from a charged hadron. 
The total cross section for vector meson photoproduction is calculated 
considering the color dipole approach, which is directly related with the 
dipole-target forward amplitude ${\cal N}$. 
In the Color Glass Condensate (CGC) formalism (see 
e.g.~\cite{Jalilian-Marian:2005jf}), ${\cal{N}}$ encodes all the information
about the hadronic scattering, and thus about the non-linear and quantum 
effects in the hadron wave function. 
In our analyzes we have used the phenomenological saturation model proposed in 
references~\cite{Golec-Biernat:1998js,Iancu:2003ge}. 
Nuclear effects are also properly taken into account.

\begin{figure}
\begin{tabular}{cc}
\psfig{file=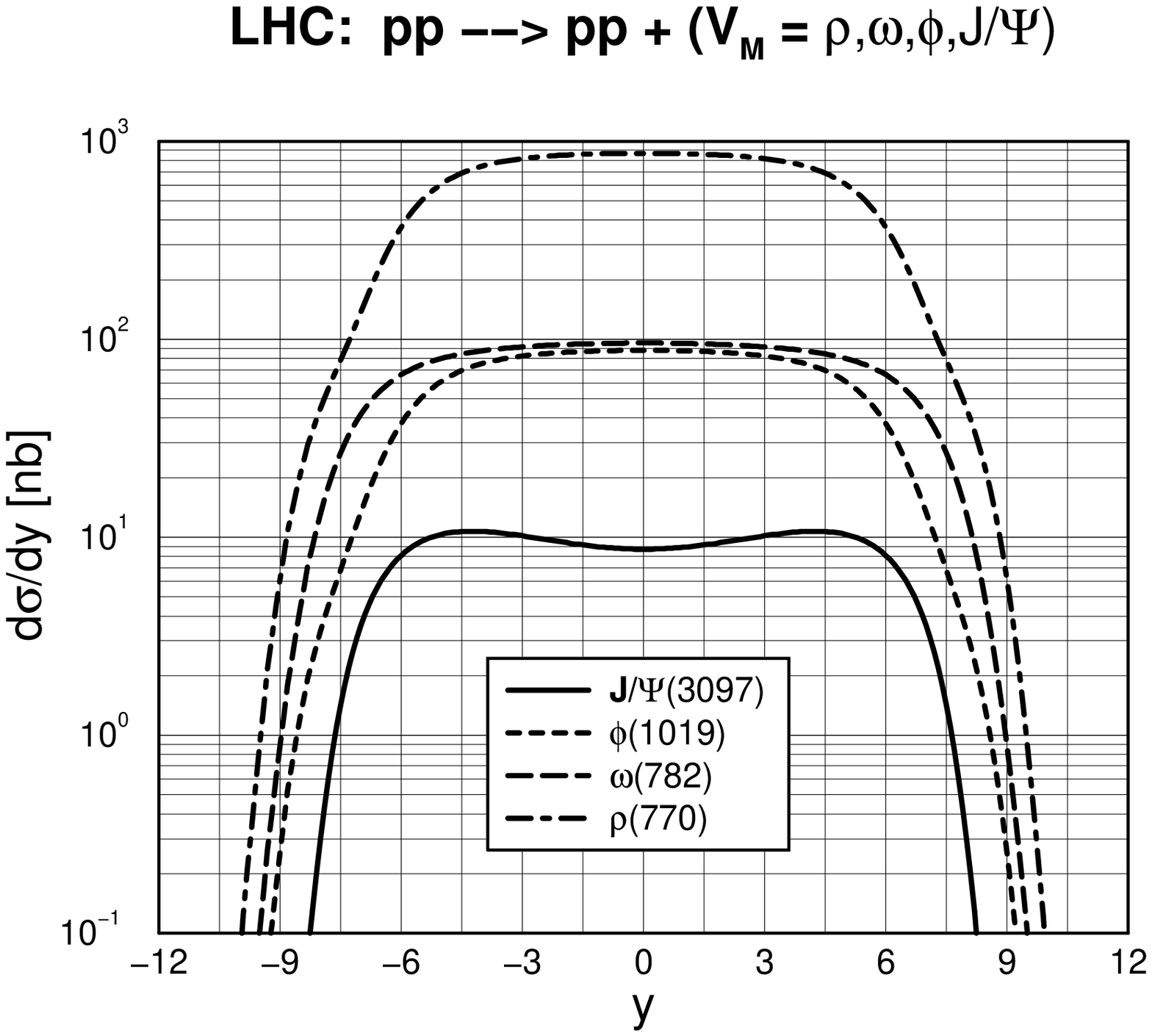,width=80mm} &
\psfig{file=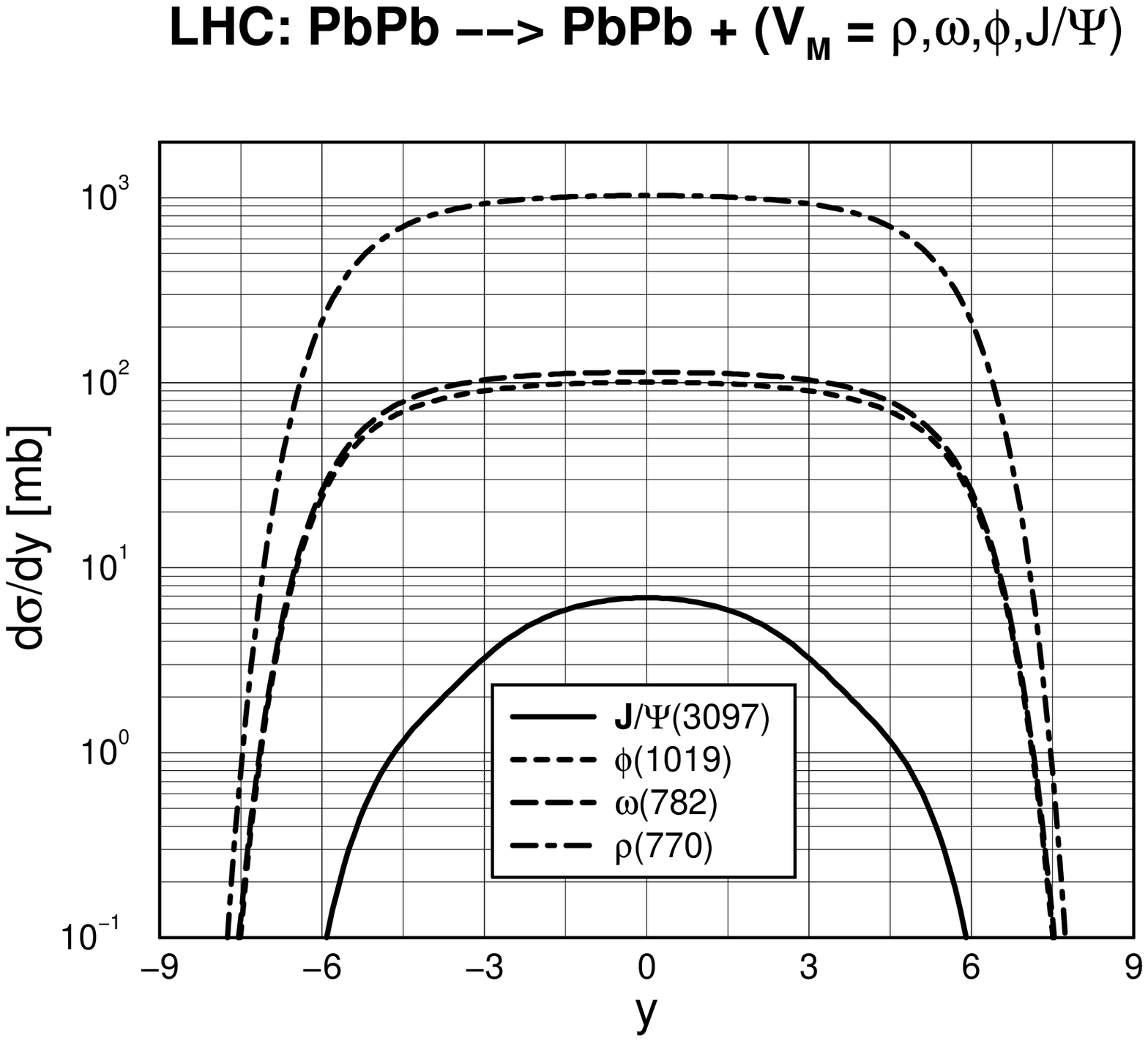,width=80mm}
\end{tabular}
\caption{\it The rapidity distribution for nuclear vector meson photoproduction
  on coherent $pp$ (left panel) and $\A\A$ (right panel) reactions at the LHC.}
\label{fig:Machado-fig1}
\end{figure}
Our predictions for the rapidity distributions are presented in figure~\ref{%
  fig:Machado-fig1} and for the total cross section in table~\ref{%
  tab:Machado-tab1}. 
The main uncertainties are the photon flux, the quark mass and the size of 
nuclear effects for the photonuclear case. 
In addition, specific predictions for $\rho$ and $J/\Psi$ phoproduction in 
$p\A$ collisions can be found in reference~\cite{Goncalves:2005ge}. 
The rates are very high, mostly for light mesons. 
Although the rates are lower than hadroproduction, the coherent photoproduction 
signal would be clearly separated by applying a transverse momentum cut
$p_T<1$~GeV and two rapidity gaps in the final state. 
\begin{table}
\begin{center}
\caption{\it The integrated cross section for nuclear vector mesons
photoproduction in coherent $pp$ and $\A\A$ collisions at the LHC.}
\label{tab:Machado-tab1}
\begin{tabular} {||c|c|c|c|c|c||}
\hline \hline
    & $\Upsilon\,(9460)$ & $J/\Psi\,(3097)$ & $\phi\,(1019)$ & $\omega\,(782)$ & $\rho\,(770)$  \\
\hline \hline
 pp & 0.8 nb  & 132 nb &  980 nb & 1.24 $\mu$b & 9.75 $\mu$b \\
\hline
 Ca-Ca & 9.7 $\mu$b &436 $\mu$b & 12 mb & 14 mb &  128 mb \\
\hline
  Pb-Pb & 96 $\mu$b & 41.5 mb &  998 mb & 1131 mb & 10069 mb \\
\hline \hline
\end{tabular}
\end{center}
\end{table}

\subsection{Heavy-Quark Kinetics in the QGP at LHC}

{\it H.~van Hees, V.~Greco and R.~Rapp} 

{\small
  We present predictions for the nuclear modification factor and
  elliptic flow of $D$ and $B$ mesons, as well as of their decay
  electrons, in semicentral Pb-Pb collisions at the LHC.  Heavy quarks
  are propagated in a Quark-Gluon Plasma using a relativistic Langevin
  simulation with drag and diffusion coefficients from elastic
  interactions with light anti-/quarks and gluons, including
  non-perturbative resonance scattering.  Hadronization at $T_c$ is
  performed within a combined coalescence-fragmentation scheme.
}
\vskip 0.5cm

In Au-Au collisions at the Relativistic Heavy Ion Collider (RHIC) a
surprisingly large suppression and elliptic flow of ``non-photonic''
single electrons ($e^\pm$, originating from semileptonic decays of $D$
and $B$ mesons) has been found, indicating a strong coupling of charm
($c$) and bottom ($b$) quarks in the Quark-Gluon Plasma (QGP).

We employ a Fokker-Planck approach to evaluate drag and diffusion
coefficients for $c$ and $b$ quarks in the QGP based on elastic
scattering with light quarks and antiquarks via $D$- and $B$-meson
resonances (supplemented by perturbative interactions in color
non-singlet channels)~\cite{vanHees:2004gq}. This picture is motivated
by lattice QCD computations which suggest a survival of mesonic states
above the critical temperature, $T_c$.  Heavy-quark (HQ) kinetics in the
QGP is simulated with a relativistic Langevin
process~\cite{vanHees:2005wb}.  Since the initial temperatures at the
LHC are expected to exceed the resonance dissociation temperatures, we
implement a ``melting'' of $D$- and $B$-mesons above
$T_{\mathrm{diss}}$=$2 T_c$=360~MeV by a factor
$(1+\exp[(T-T_{\mathrm{diss}})/\Delta])^{-1}$ ($\Delta$=50~MeV) in the
transport coefficients.

The medium in a heavy-ion reaction is modeled by a spatially homogeneous
elliptic thermal fireball which expands isentropically.  The temperature
is inferred from an ideal gas QGP equation of state with $N_f$=2.5
massless quark flavors, with the total entropy fixed by the number of
charged hadrons which we extrapolate to $\dd N_{\mathrm{ch}}/\dd
y$$\simeq$1400 for central $\sqrt{s_{NN}}$=5.5~TeV Pb-Pb collisions.
The expansion parameters are adjusted to hydrodynamic simulations,
resulting in a total lifetime of $\tau_{\mathrm{fb}}$$\simeq$6~fm/c at
the end of a hadron-gas QGP mixed phase and an inclusive light-quark
elliptic flow of $\langle v_2 \rangle$=7.5\%.  The QGP formation time,
$\tau_0$, is estimated using $\tau_0 T_0$=const ($T_0$: initial
temperature), which for semicentral collisions (impact parameter
$b$$\simeq$7~fm) yields $T_0$$\simeq$520~MeV.

Initial HQ $p_T$ spectra are computed using PYTHIA with parameters 
as used by the ALICE Collaboration. $c$ and $b$ quarks are hadronized
into $D$ and $B$ mesons at $T_c$ by coalescence with light 
quarks~\cite{Greco:2003vf}; ``left over'' heavy quarks are 
hadronized with $\delta$-function fragmentation. For semileptonic 
electron decays we assume 3-body kinematics~\cite{vanHees:2005wb}. 
\begin{figure}[t]
\begin{center}
\begin{minipage}{0.345\textwidth}
\includegraphics[width=\textwidth]{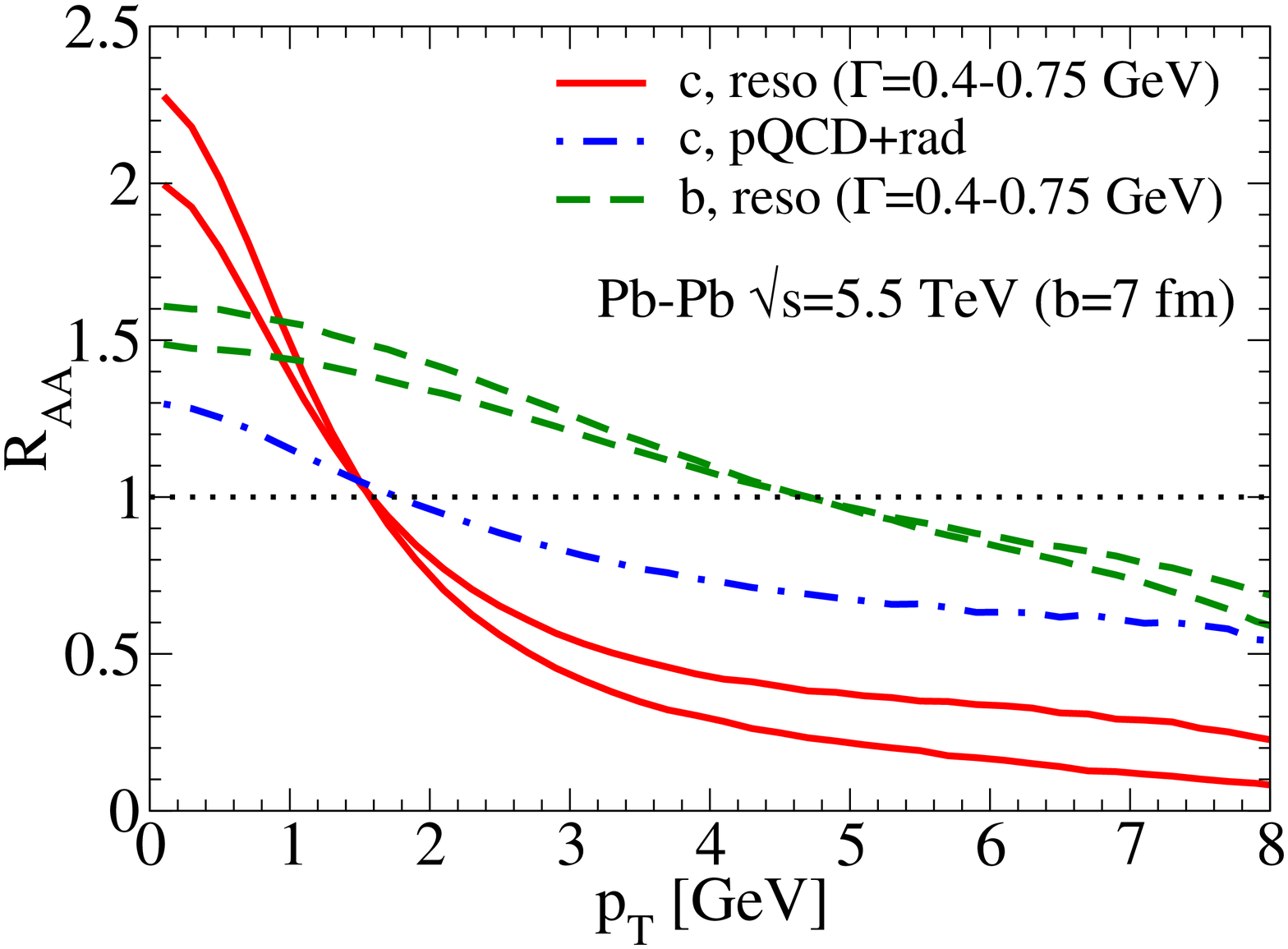}
\end{minipage}\hspace*{0.7cm}
\begin{minipage}{0.345\textwidth}
\includegraphics[width=\textwidth]{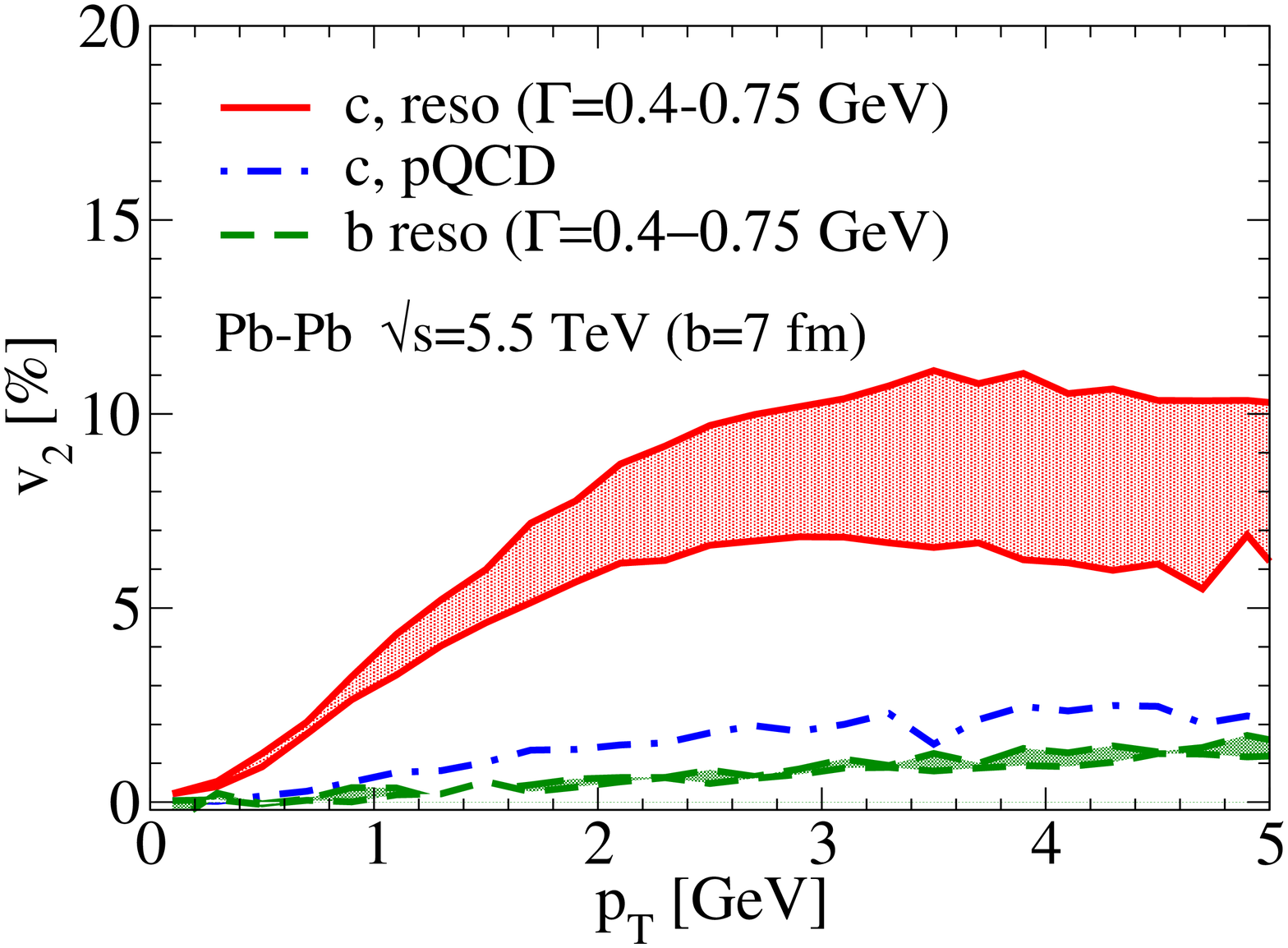}
\end{minipage}

\begin{minipage}{0.355\textwidth}
\includegraphics[width=\textwidth]{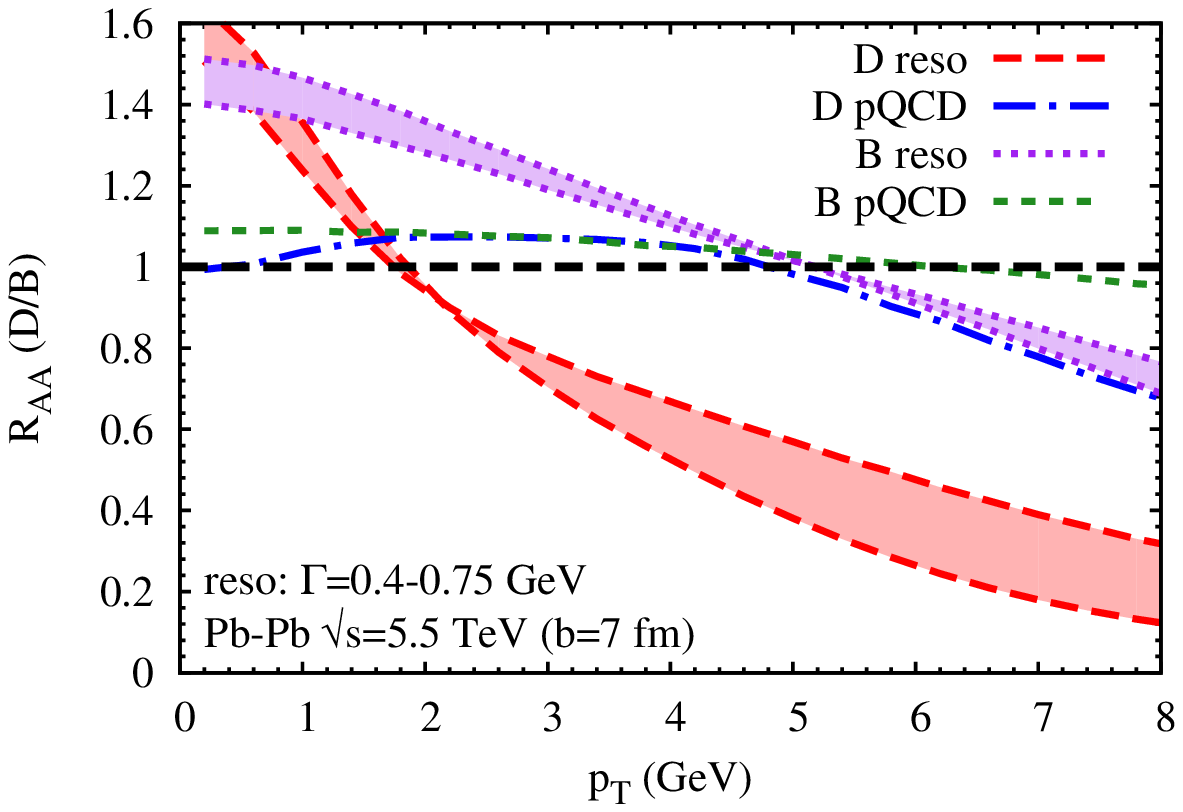}
\end{minipage}\hspace*{0.7cm}
\begin{minipage}{0.355\textwidth}
\includegraphics[width=\textwidth]{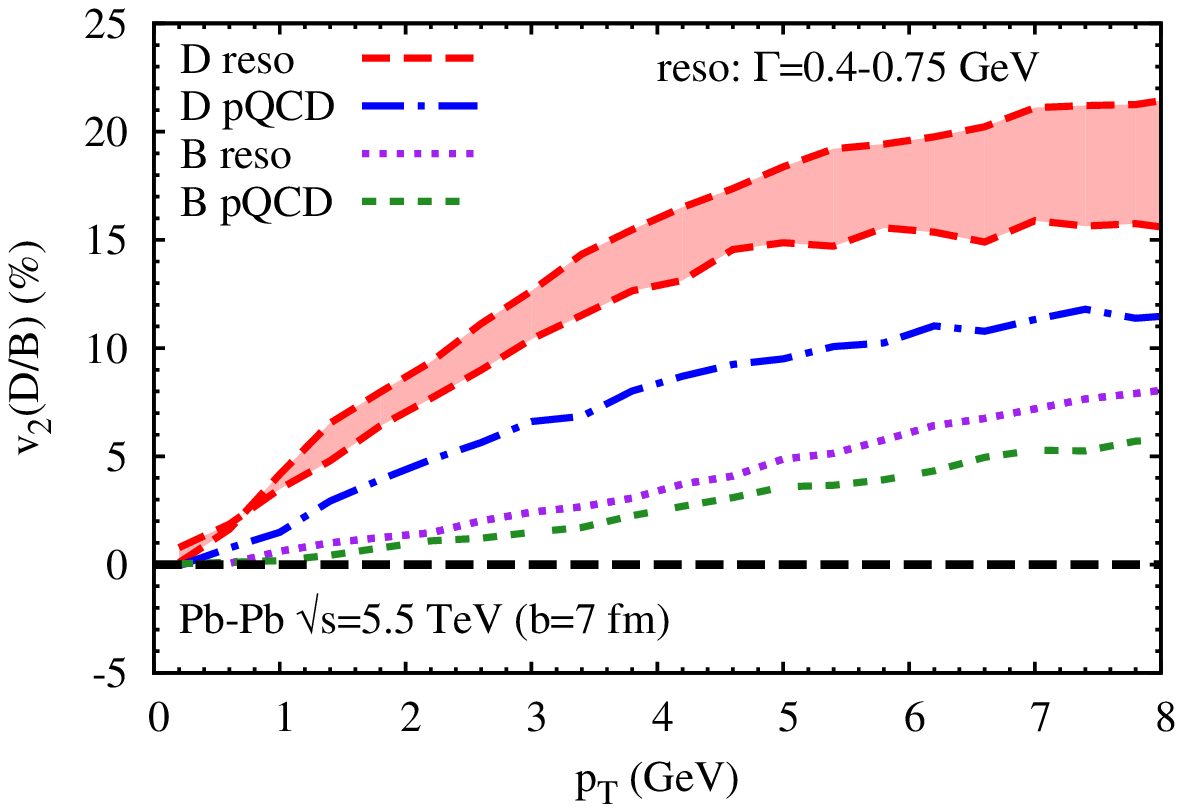}
\end{minipage}

\begin{minipage}{0.355\textwidth}
\includegraphics[width=\textwidth]{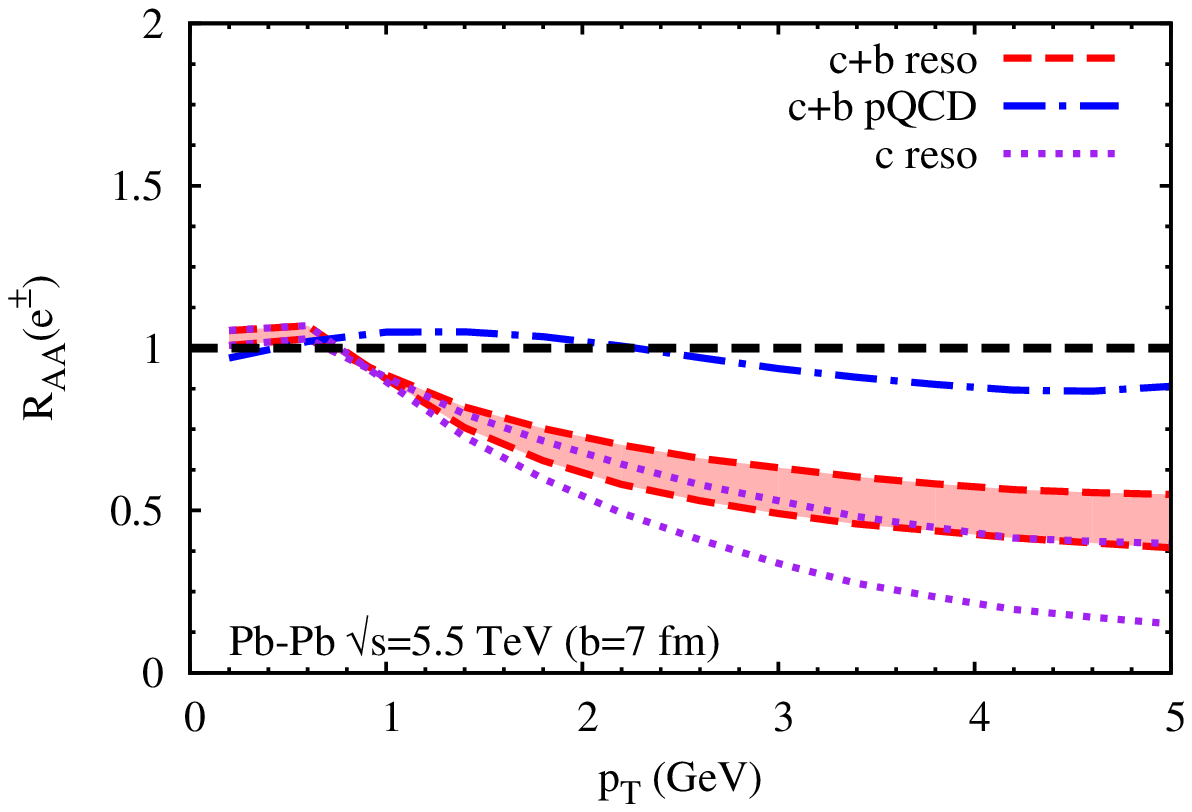}
\end{minipage}\hspace*{0.7cm}
\begin{minipage}{0.355\textwidth}
\includegraphics[width=\textwidth]{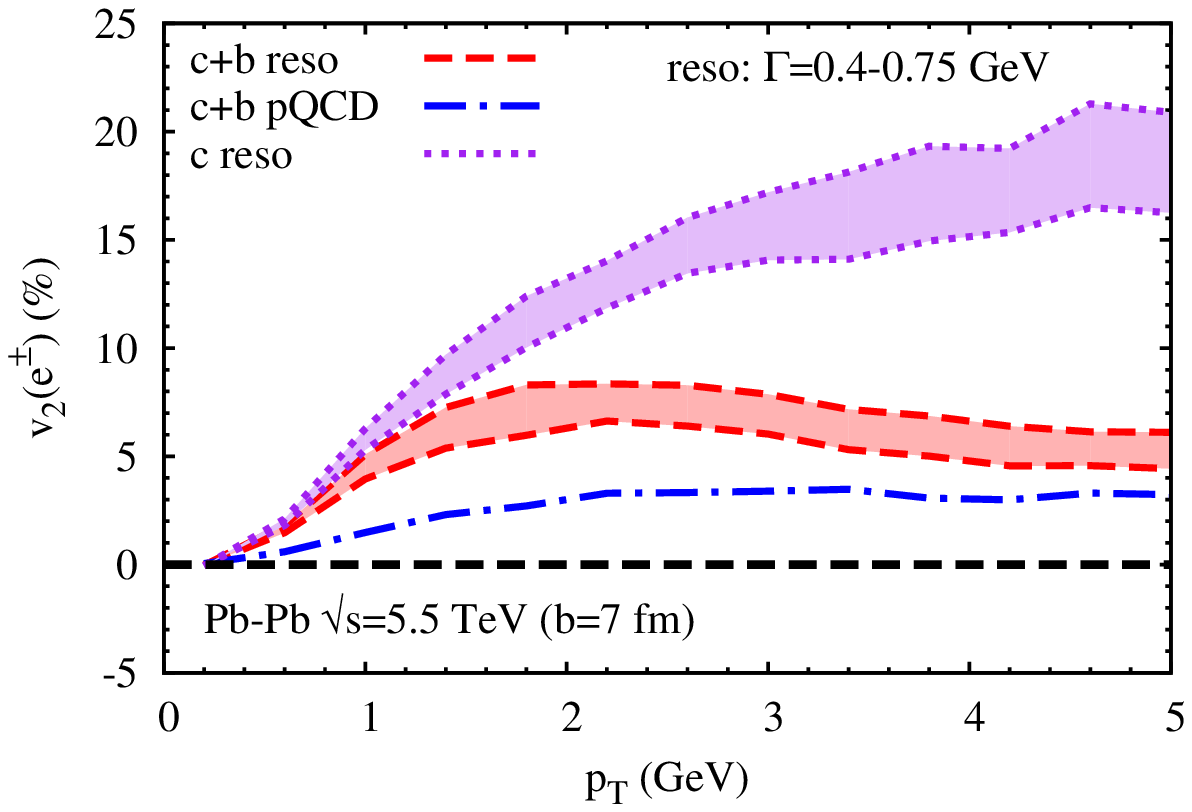}
\end{minipage}

\end{center}
\vspace*{-0.5cm}

\caption{(Color online) Predictions of relativistic Langevin simulations
  for heavy quarks in a sQGP for $b$=7~fm $\sqrt{s_{NN}}$=5.5~TeV Pb-Pb
  collisions: $R_{AA}$ (left column) and $v_2$ (right column) for heavy
  quarks (1$^\mathrm{st}$ row), $D$ and $B$ mesons (2$^\mathrm{nd}$ row)
  and decay-$e^\pm$ (3$^\mathrm{rd}$ row).}
\label{fig.1.vanhees}
\end{figure}

Fig.~\ref{fig.1.vanhees} summarizes our results for HQ diffusion in a QGP in
terms of $R_{AA}(p_T)$ and $v_2(p_T)$ at the quark, meson and
$e^\pm$ level for $b$=7~fm Pb-Pb collisions at the LHC
(approximately representing minimum-bias conditions). Our most
important findings are: (a) resonance interactions substantially
increase (decrease) $v_2$ ($R_{AA}$) compared to perturbative
interactions; (b) $b$ quarks are much less affected than $c$ quarks,
reducing the effects in the $e^\pm$ spectra; (c) there is a strong
correlation between a large $v_2$ and a small $R_{AA}$ at the quark
level, which, however, is partially reversed by coalescence
contributions which increase \emph{both} $v_2$ and $R_{AA}$ at the meson
(and $e^\pm$) level. This feature turned out to be important in the
prediction of $e^\pm$ spectra at RHIC; (d) the predictions for LHC are
quantitatively rather similar to our RHIC
results~\cite{vanHees:2005wb,Rapp:2006ta}, due to a combination
of harder initial HQ-$p_T$ spectra with a moderate increase in
interaction strength in the early phases where non-perturbative
resonance scattering is inoperative. 

%\myack
%{This work is supported by a U.S. NSF CAREER Award, grant no. PHY-0449489.}

\subsection{Ratio of charm to bottom \raah as a test of pQCD vs.~\ads energy loss}
\label{horowitz}

{\it W. A. Horowitz}
\vskip 0.5cm

The theoretical framework of a \weaklycoupled QGP used in pQCD models that quantitatively describe the \highpt \pizerocomma, $\eta$ suppression at RHIC is challenged by several experimental observables, not limited to \highpt only, suggesting the possibility that a \stronglycoupled picture might be more accurate.  One seeks a measurement that may clearly falsify one or both approaches; heavy quark jet suppression is one possibility.  Strongly-coupled calculations, utilizing the \ads correspondence, have been applied to \highpt jets in three ways \cite{Liu:2006ug,Casalderrey-Solana:2006rq,Herzog:2006gh}%: to compute \qhat \cite{Liu:2006ug}, the input parameter in a radiative energy loss model \cite{Armesto:2003jh}; to compute $D$ \cite{Casalderrey-Solana:2006rq}, the diffusion coefficient that is an input parameter in a relativistic Langevin model for heavy quarks \cite{Moore:2004tg}; and a direct computation of the drag felt by a heavy quark in a \stronglycoupled SYM background \cite{Herzog:2006gh}.  We will concentrate on the latter as it has not until now been used to calculate experimental observables; we will compare these values to pQCD predictions from the WHDG model of convolved radiative and elastic energy loss \cite{Wicks:2005gt}.
.  We will focus on predictions from the \ads heavy quark drag model % as it has not until now been used to calculate experimental observables; we will 
and compare them to pQCD predictions from the full radiative and elastic loss WHDG model and radiative alone WHDG model \cite{Wicks:2005gt}.  
Comparisons between \ads models and data are difficult.  First, one must accept the double conjecture of QCD$\leftrightarrow$SYM$\leftrightarrow$\adscomma.  Second, to make contact with experiment, one must make further assumptions to map quantities such as the coupling and temperature in QCD into the SUGRA calculation.  For example, the \ads prediction for the heavy quark diffusion coefficient is $D = 4/\sqrt{\lambda}(/2\pi T)$ \cite{Casalderrey-Solana:2006rq}, where $\lambda=g_{SYM}^2 N_c$ is the \thooft coupling.  The ``obvious'' first such mapping \cite{Gubser:2006qh} simply equates constant couplings, $g_s=g_{SYM}$, and temperatures, $T_{SYM}=T_{QCD}$.  Using this prescription with the canonical $N_c=3$ and $\eqnalphas=.3$ yields $D\approx1.2(/2\pi T)$.  It was claimed in \cite{Casalderrey-Solana:2006rq} that $D = 3(/2\pi T)$ agrees better with data; this requires $\eqnalphas\approx.05$.  An ``alternative'' mapping \cite{Gubser:2006qh} equates the quark-antiquark force found on the lattice to that computed using \adscomma, giving $\lambda\approx5.5$, and the QCD and SYM energy densities, yielding $T_{SYM}=T_{QCD}/3^{1/4}$.% (due to the approximately factor of 3 difference in the number of degrees of freedom).  
The medium density to be created at LHC is unknown%There is an additional uncertainty in the medium density to be used for LHC
; we will take the PHOBOS extrapolation of $\eqndngslashdy=1750$ and the KLN model of the CGC, $\eqndngslashdy=2900$, as two sample values.%predictions
  We will search for general trends associated with \ads drag (denoted hereafter simply as \adscomma) or pQCD as these uncertainties mean little constrains the possible normalizations of \ads \raaQ predictions for LHC.%As a result of these inherent uncertainties there is little constraining the possible normalizations of \ads \raaQ predictions for LHC.  We will thus search for general trends that might allow experimental falsification of \ads or usual pQCD ideas.

\begin{figure}[htb!]
%\vspace{.15in}
\begin{center}
$\begin{array}{c@{\hspace{.00in}}c}
\epsfxsize=.53\columnwidth
\epsffile{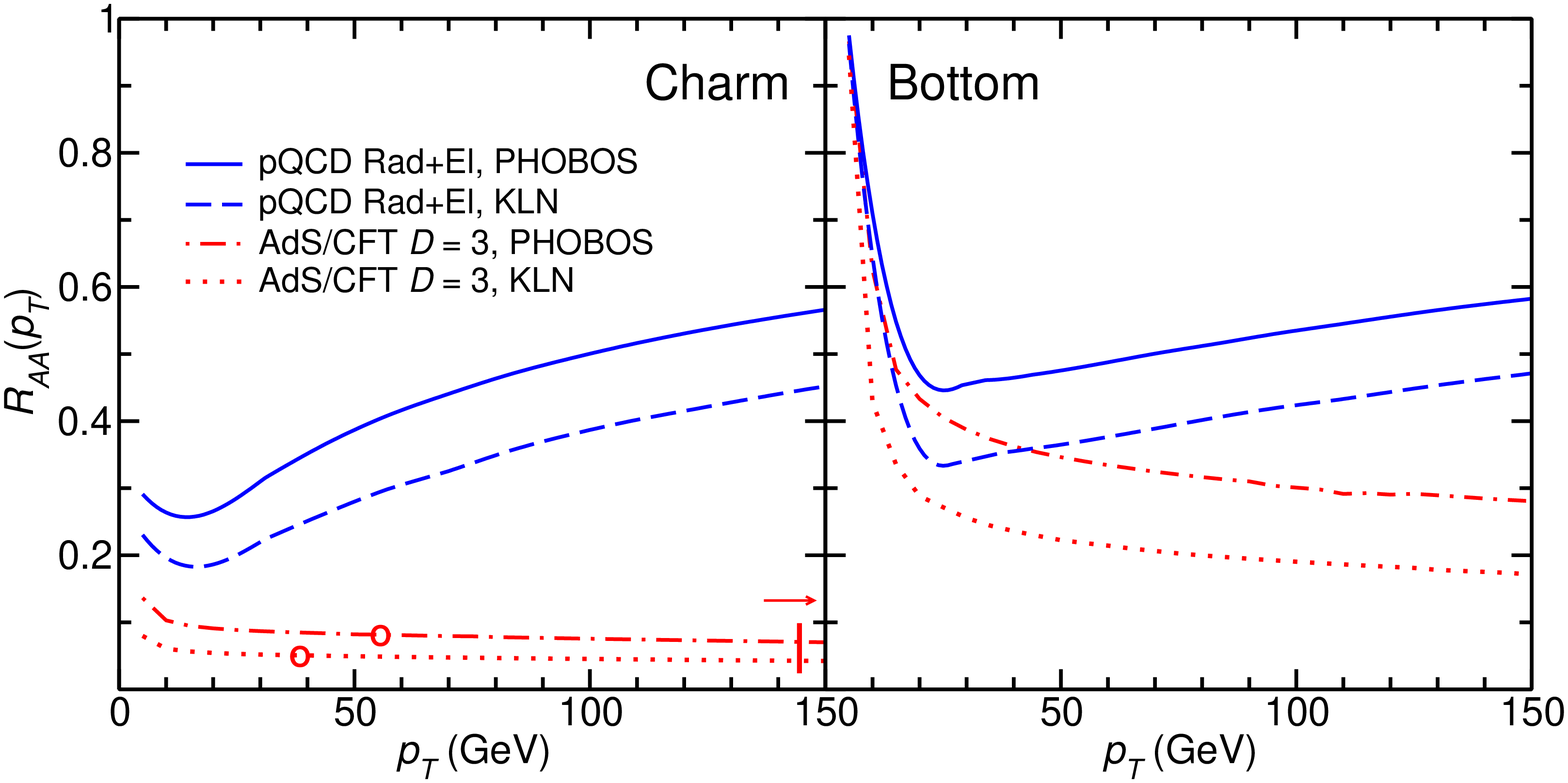 } & \epsfxsize=.47\columnwidth
\epsffile{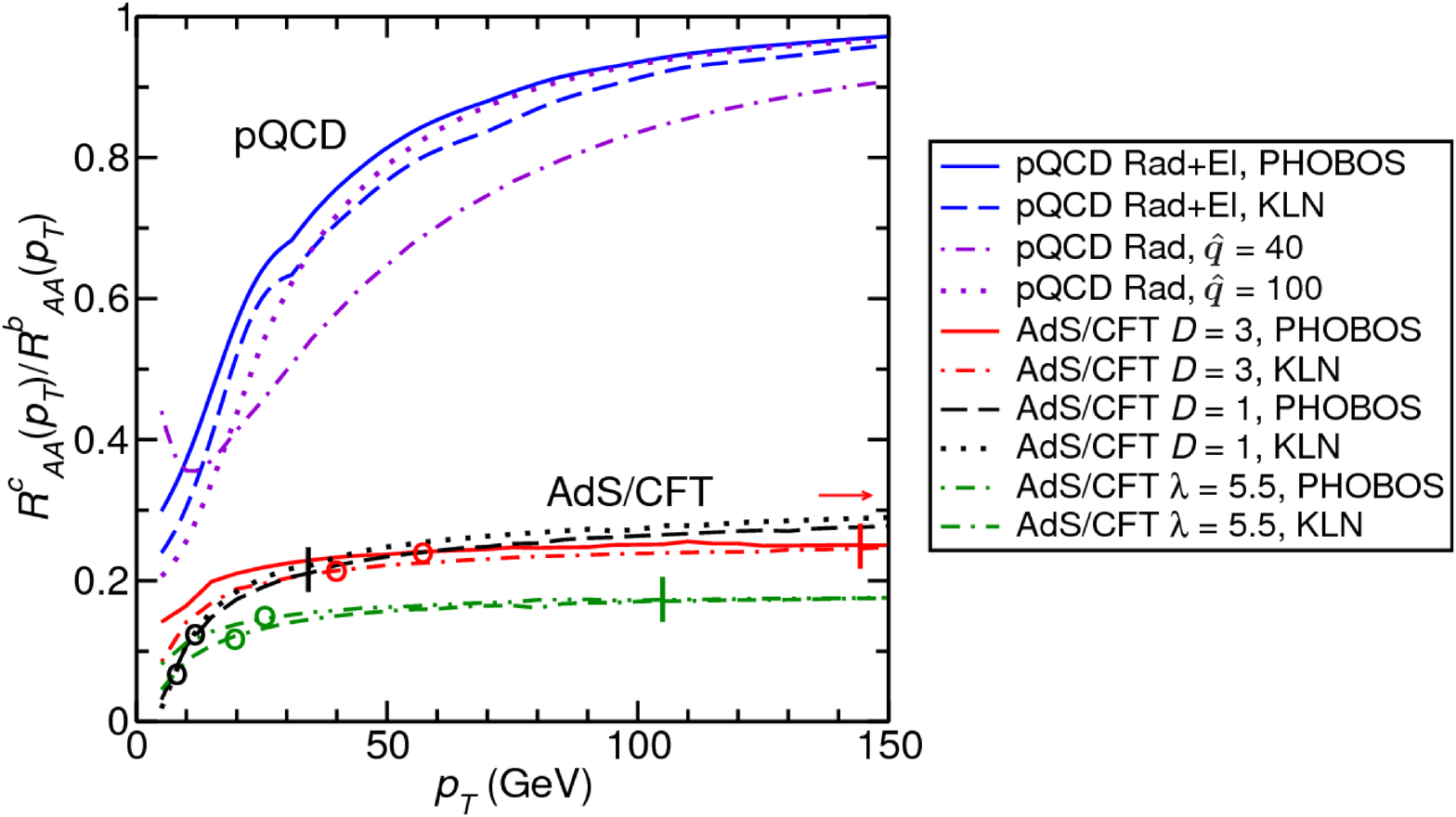 } \\ [-.1in]
{\mbox {\scriptsize {\bf (a)}}} & {\mbox {\scriptsize {\bf (b)}}}
\end{array}$
\end{center}
\vspace{-.2in}
\caption{(a) Charm and bottom \raapt predictions with representative input parameters for LHC.  The generic trend of pQCD curves increasing with \pt while \ads curves decrease is seen for representative input parameters; similar trends occurred for the other input possibilities considered.  (b) Ratio of charm to bottom \raapt bunches the two models for a wide range of input parameters; the LHC should easily distinguish between the two trends.}
\label{fig1horo}
\vspace{-.2in}
\end{figure}%Comparisons between \ads and data are difficult.  First, one must accept the {\em double} conjecture of QCD$\leftrightarrow$SYM$\leftrightarrow$\adscomma.  Second, to make contact with experiment, one must make further assumptions to map quantities such as the coupling and temperature in QCD into the SUGRA calculation.  For this reason, as well as on purely theoretical grounds, there is much debate over the magnitude of \ads predictions for the \qhat and $D$ input parameters.  Nevertheless, pQCD predictions for these quantities do not agree with the RHIC nonphotonic electron spectra while values in the (large) range of \ads predictions are bettert.  While the third model has not yet been compared with data, its results should be similar to those of the diffusion approach at the relatively low momenta, compared to the mass the bottom quark, probed at RHIC.  Thus one would like to find a more discerning measurement, and the identified charm and bottom spectra along with the large \pt reach of the LHC will facilitate just such a differentiation.

%The first two approaches have been compared to data from RHIC; there is a debate over the magnitude of \ads predictions for the input parameters, but pQCD predictions are not in quantitative agreement with the data-determined input parameters.  While published comparisons between the last model and nonphotonic electron \raah data has not been made, one finds that within the wide range of possible input, it is generally consistent.  Thus one would like to find a discerning (distinguishing) measurement, and the identified charm and bottom spectra along with the large \pt reach of the LHC will facilitate a differentiation.%We note that the first two approaches use AdS/CFT results onto models inspired by perturbative thinking while the latter is carried out fully from a string theoretic perspective.

\ads calculations of the drag on a heavy quark yield
$d\eqnpt/dt = -\eqnmuQ\eqnpt = -(\pi\sqrt{\lambda}T_{SYM}^{2}/2\eqnmQ)\eqnpt$ \cite{Herzog:2006gh}, 
%\be
%\frac{d\eqnpt}{dt} = -\mu\eqnpt = -\frac{\pi\sqrt{\lambda}(T^*)^2}{2\eqnmq}\eqnpt,
%\ee
%where $\lambda=N_c^2g_s$ is the \thooft coupling, and $T^*$ is the SYM plasma temperature.  
giving an average fractional energy loss of $\bar{\epsilon}=1-\exp(-\int \! dt\mu_Q)$.  Asymptotic pQCD energy loss for heavy quarks in a static medium goes as
$\bar{\epsilon}\approx\kappa L^2 \eqnqhat \log(\eqnpt/\eqnmQ)/\eqnpt$,
%\be
%\epsilon = \frac{\Delta\eqnpt}{\eqnpt} \sim \alphas\eqnqhat L^2 \log(\eqnpt/\eqnmq)/\eqnpt,
%\ee
where $\kappa$ is a proportionality constant and $L$ is the pathlength traversed by the heavy quark.  Note that \ads fractional momentum loss is independent of momentum while pQCD loss decreases with jet energy.  %The key difference between the two calculations is that the \ads drag is inversely dependent on quark mass and independent of momenta; on the other hand, pQCD energy loss decreases with increasing momenta.  
The heavy quark production spectrum may be approximated by a slowly varying power law of index $n_Q(\eqnpt)+1$% with $n_Q+1=-\frac{d}{d\log \eqnpt}\log\left(\frac{d\sigma_Q}{dyd\eqnpt}\right)$
, then $\eqnraaQ\approx(1-\bar{\epsilon})^{n_Q(\eqnpt)}$.  Since $n_Q(\eqnpt)$ is a slowly increasing function of momentum, we expect $R_{AA,SYM}^Q(\eqnpt)$ to decrease while $R_{AA,pQCD}^Q(\eqnpt)$ to increase as momentum increases.  This behavior is reflected in the full numerical calculations shown in Fig.~\ref{fig1horo} (a); details of the model %, in which partons are created and propagated through a realistic, Bjorken-expanding medium, 
can be found in \cite{Horowitz:2007su}.  

For high suppression pQCD predicts nearly flat \raaQcomma, masking the difference between \ads and pQCD.  One can see in Fig.~\ref{fig1horo} (b) that the separation of AdS/CFT and pQCD predictions is enhanced when the double ratio of charm to bottom nuclear modification, $R^{cb}(\eqnpt)=\eqncbratio$, is considered.  Asymptotic pQCD energy loss goes as $\log(\eqnmQ/\eqnpt)/\eqnpt$, becoming insensitive to quark mass for $\eqnpt\gg\eqnmQ$; hence $R^{cb}_{pQCD}\rightarrow1$.  Expanding the \raah formula for small $\epsilon$ yields $R^{cb}_{pQCD}(\eqnpt)\approx1-p_{cb}/\eqnpt$, where $p_{cb}=\kappa n(\eqnpt)L^2 \log(m_b/m_c) \eqnqhat$ and $n_c\approx n_b=n$.  Therefore the ratio approaches unity more slowly for larger suppression.  This behavior is reflected in the full numerical results for the moderately quenched pQCD curves, but is violated by the highly oversuppressed $\eqnqhat=100$ curve.  The \ads drag, however, is independent of \ptcomma.  A back of the envelope approximation gives $\eqnraaQ\approx\int_0^L\! d\ell \exp(-n_Q\eqnmuq\ell)\approx1/n_Q\eqnmuq$ which yields $R^{cb}(\eqnpt)\approx n_b(\eqnpt)m_c/n_c(\eqnpt)m_b\approx m_c/m_b\approx.27$.
This behavior is also reflected in the full numerical results shown in Fig.~\ref{fig1horo} (b), and so, remarkably, the pQCD and \ads curves fall into easily distinguishable bunches, robust to changes in input parameters.
%Although poorly understood, one probably cannot apply the \ads drag calculation to all momenta.  
An estimate for the momentum after which corrections to the above \ads drag formula are needed, $\gamma>\gamma_c$, found in the static string geometry is $\gamma_c=1/1+(2\eqnmQ/T\sqrt{\lambda})$ \cite{Gubser:2006nz}.  Since temperature is not constant we show the smallest speed limit, using $T(\tau_0,\vec{x}=\vec{0})$, and largest, from $T_c$, represented by ``O'' and ``$|$,'' respectively.  A deviation of $R^{cb}$ away from unity at LHC in year 1 would pose a serious challenge to the usual pQCD paradigm.  An observation of a significant increase in $R^{cb}$ with jet momenta would imply that the current \ads picture is only applicable at low momenta, if at all.  For a definitive statement to be made a $p+Pb$ control run will be crucial.

\subsection{Thermal charm production at LHC}
\label{s:Ko5}

{\em B.-W. Zhang, C. M. Ko and W. Liu}
\vspace{0.5cm}

Charm production from an equilibrated quark-gluon plasma (QGP)
produced in heavy ion collisions at LHC is studied to the
next-to-leading order in perturbative QCD
\cite{Zhang:2007dm}.
Specifically, we consider
the process $q(g)+\bar{q}(g)\to c+ \bar{c}$ and its virtual
correction as well as the processes $q(g)+\bar{q}(g)\to
c+\bar{c}+g$, and $g+q (\bar{q})\to c+ \bar{c}+q(\bar{q})$. The
amplitudes for these processes are taken from
Refs. \cite{Beenakker:1988bq,Beenakker:1990ma,Nason:1987xz,Nason:1989zy}
using massless
quarks and gluons, the QCD coupling constant $\alpha_s(m_c)\approx
0.37$, and a charm quark mass $m_c=1.3$ GeV. The charm quark
production rate in the QGP is then evaluated by integrating over the
thermal quark and gluon distributions in the QGP. Both thermal
quarks and gluons are taken to have thermal masses given by
$m_q=m_g=gT/\sqrt{6}$, where $T$ is the temperature of the QGP and
$g$ is related to the thermal QCD coupling constant $\alpha_s(2\pi
T)=g^2/4\pi$, which has values ranging from $\sim 0.23$ for $T=700$
MeV to $\sim 0.42$ for $T=170$ MeV.

For the dynamics of formed QGP in central Pb+Pb collisions at
$\sqrt{s_{NN}}=5.5$ TeV at LHC, we assume that it evolves boost
invariantly in the longitudinal direction but with an accelerated
transverse expansion. Specifically, its volume expands in the proper
time $\tau$ according to $V(\tau)=\pi R^2(\tau)\tau c$, where
$R(\tau) = R_0+ a(\tau-\tau_0)^2/2$ is the transverse radius with an
initial value $R_0=7$ fm, the QGP formation time $\tau_0$=0.2
fm/$c$, and the transverse acceleration $a=0.1~c^2$/fm. Starting
with an initial temperature $T_0=700$ MeV, which gives an initial
energy density of about 50\% higher than that predicted by the AMPT
model \cite{Lin:2004en} or the Color Glass Condensate
\cite{Lappi:2006hq}, the time dependence of the temperature is
obtained from entropy conservation, leading to the critical
temperature $T_{\rm C}=170$ MeV at proper time $\tau_{\rm C}=6.4$
fm/$c$. The initial number of charm pairs is taken to be
$dN_{c\bar{c}}/dy = 20 $ at midrapidity, which is of similar
magnitude as that estimated from initial hard nucleon-nucleon
collisions based on the next-to-leading order pQCD calculations.

\begin{figure}[htb]
\begin{center}
\vskip 1cm
\includegraphics[width=2.7in,height=2in,angle=0]{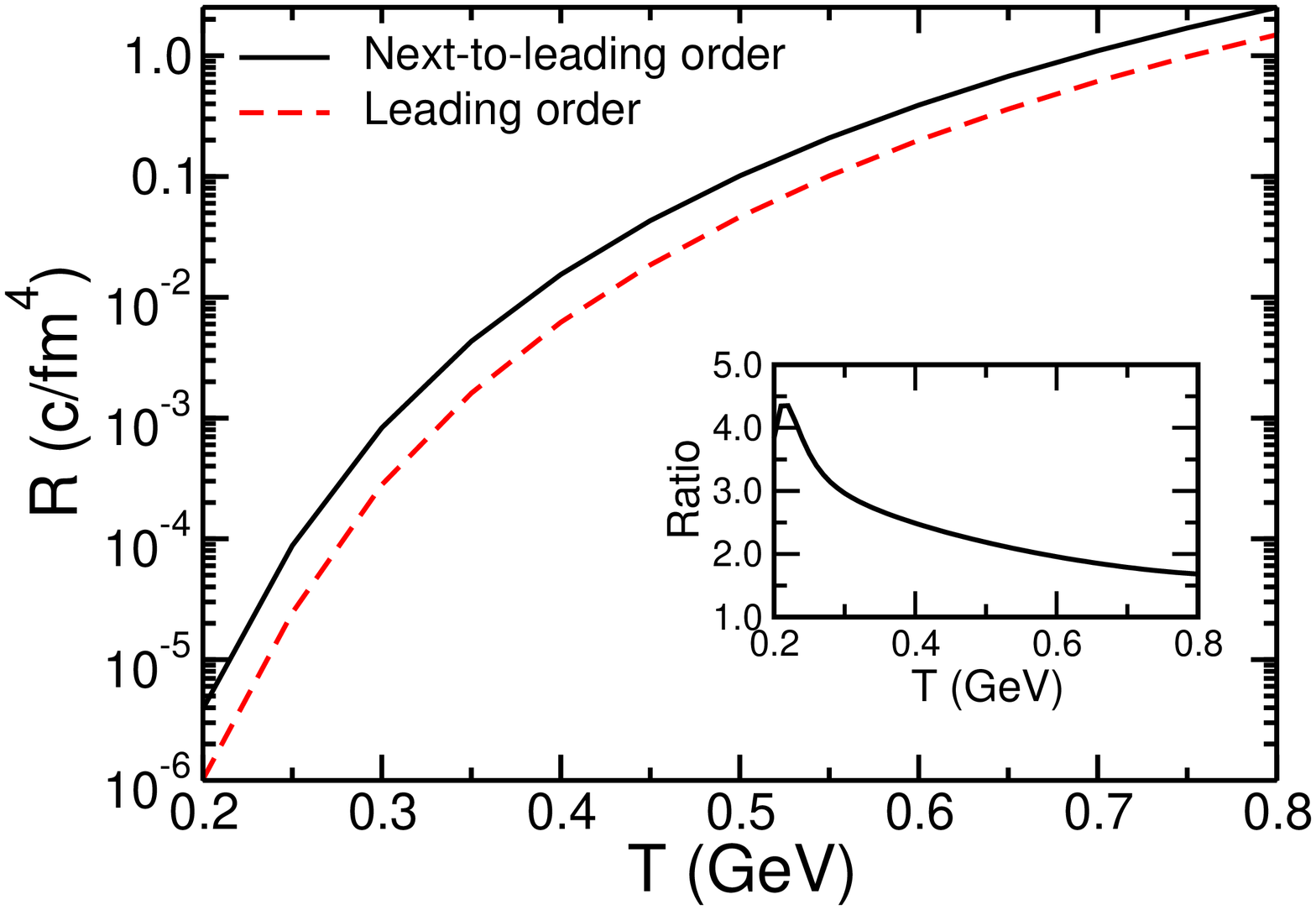}
\hspace{0.5cm}
\includegraphics[width=2.5in,height=2in,angle=0]{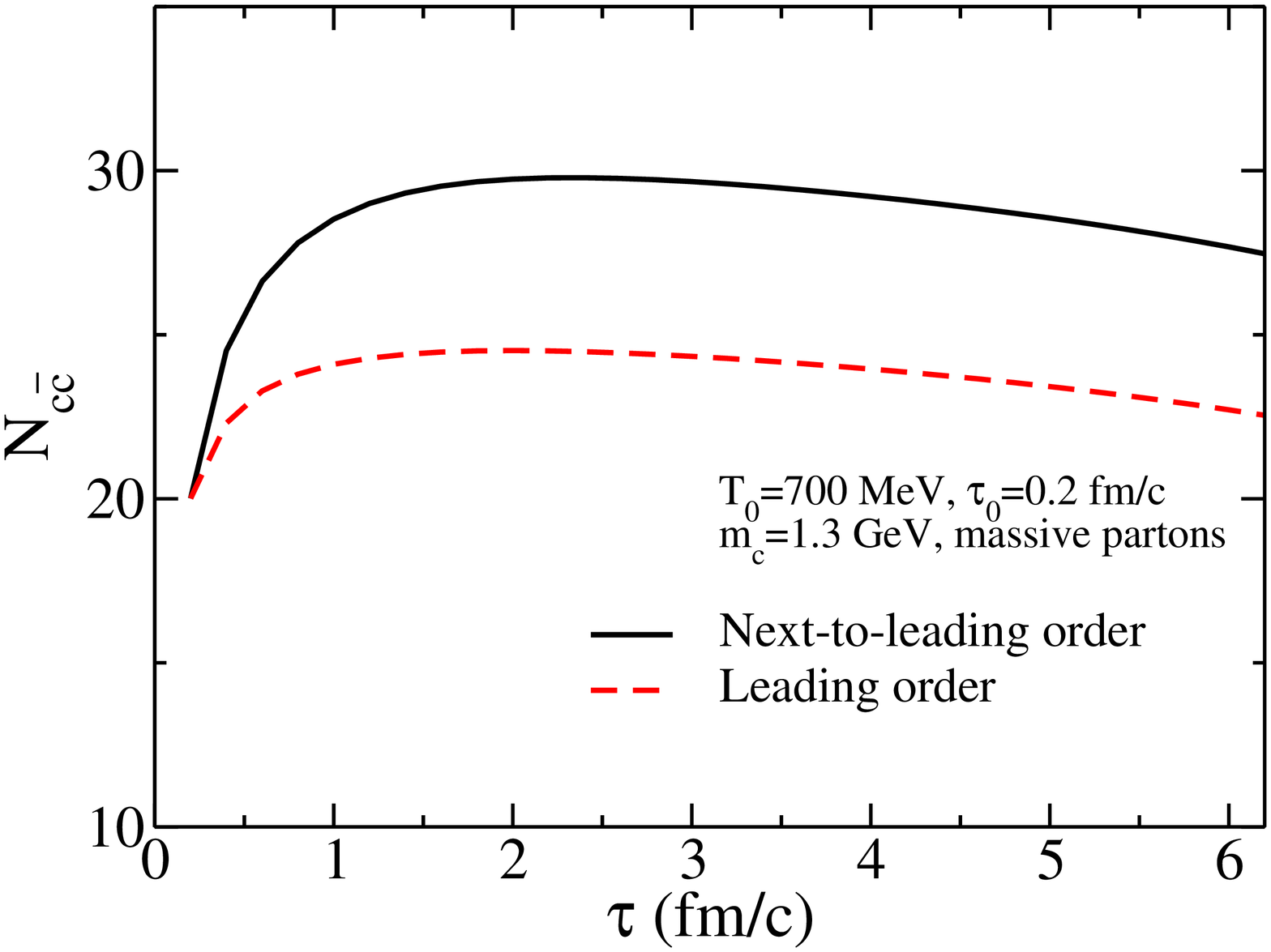}
\end{center}
\caption{Time evolution of charm pair production rate (left window)
and number (right window) in central Pb+Pb collisions at
$\sqrt{s_{NN}}=5.5$ TeV for an initial QGP temperature of 700 MeV.
Dashed and solid lines are results from the leading order and
next-to-leading order calculations, respectively. The inset in left
window gives the ratio of charm production rate in the
next-to-leading order to that in the leading order.}
\label{fig:kocharm}
\end{figure}

In the left window of Fig.~\ref{fig:kocharm}, we show the temperature dependence of
the charm quark pair production rates from the leading order (dashed
line) and the next-to-leading order (solid line) with their ratio
shown in the inset. The contributions from the leading order and
next-leading order are of similar magnitude and both are appreciable
at high temperatures. The total number of charm pairs as a function
of the proper time $\tau$ in an expanding QGP produced at LHC is
shown in the right window of Fig.~\ref{fig:kocharm}. As shown by the dashed line,
including only the leading-order contribution from two-body
processes increases the number of charm pairs by about 10\% during
the evolution of the QGP. Adding the next-leading-order contribution
through virtual corrections to two-body precesses as well as the
$2\to 3$ processes further increases the charm quark pair number by
about 25\% as shown by the solid line. The charm quark pair number
reaches its peak value at $\tau\sim 2$ fm/$c$ and then deceases with
the proper time as a result of larger charm annihilation than
creation rates when the temperature of the QGP drops. At the end of
the QGP phase, it remains greater than both its initial value and
the chemically equilibrium value of about 5 at $T_C=170$ MeV. The
number of charm quark pairs produced from the QGP would be reduced
by a factor of about 3 if a larger charm quark mass of 1.5 GeV or a
lower initial temperature of $T_0=630~{\rm MeV}$ is used. It is,
however, not much affected by using massless gluons due to increase
in the gluon density. On the other hand, increasing the initial
temperature to 750 MeV would enhance the thermally produced charm
quark pairs by about a factor of 2.

\subsection{Charm production in nuclear collisions}
\label{s:Kopeliovich5}

{\em B. Z. Kopeliovich and I.~Schmidt }

{\small
Nuclear suppression of heavy flavor inclusive production in hard partonic 
collisions has a leading twist component related to gluon shadowing, as well as 
a higher twist contribution related to the nonzero separation of the produced 
$\bar QQ$ pair. 
Both terms are evaluated and suppression for charm production in heavy ion
collisions at LHC is predicted.
}

\subsubsection{Higher twist shadowing}

Heavy flavors are produced via gluon fusion, therefore they serve as a good 
probe for the gluon distribution function in nuclei. 
The light-cone dipole approach is an effective tool for the calculation of 
nuclear effects in these processes, since the phenomenological dipole cross 
section includes by default all higher order and higher twist terms.

\begin{figure}[htp]
\centerline{
  \scalebox{0.50}{\includegraphics{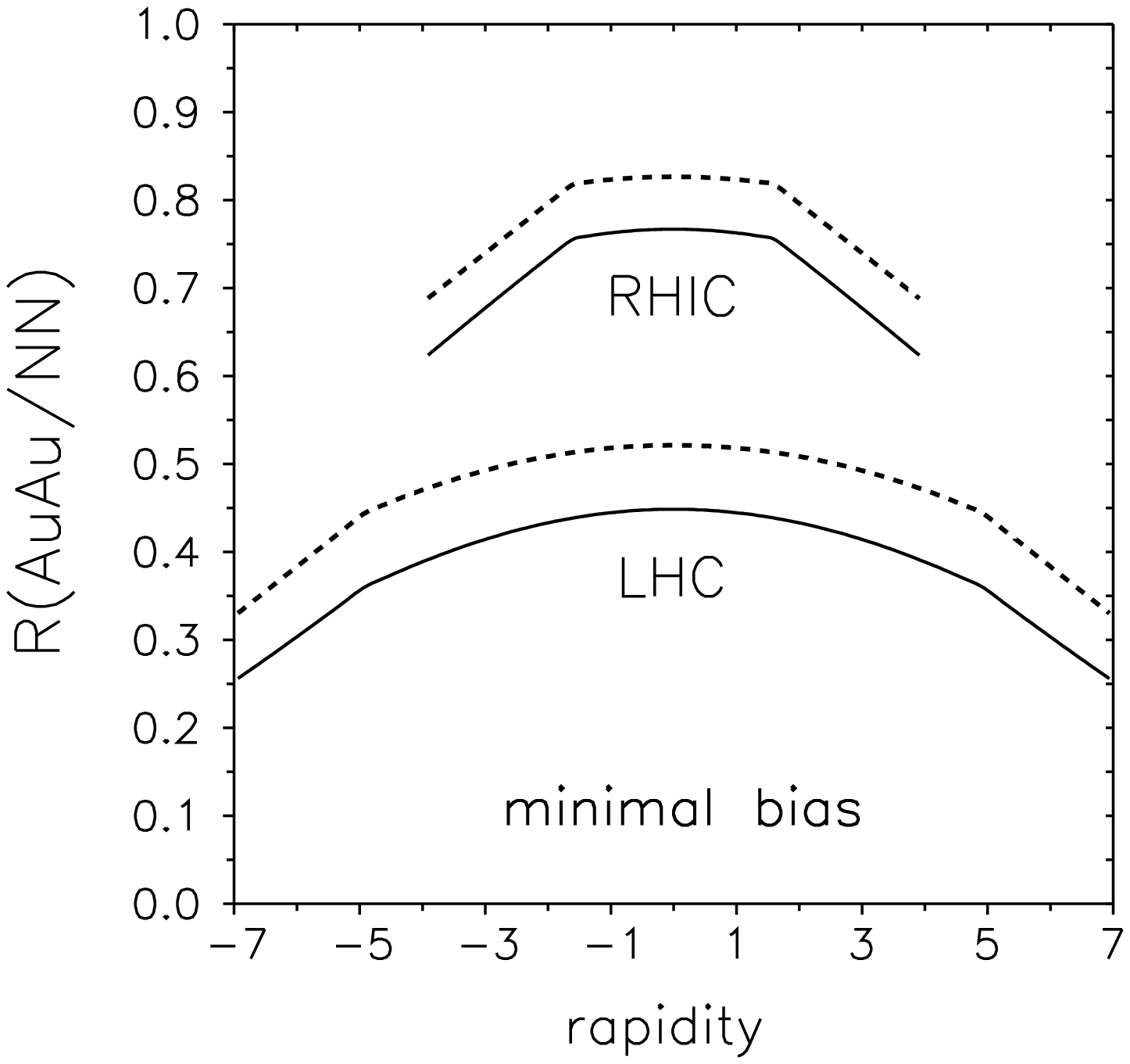}}
  \scalebox{0.50}{\includegraphics{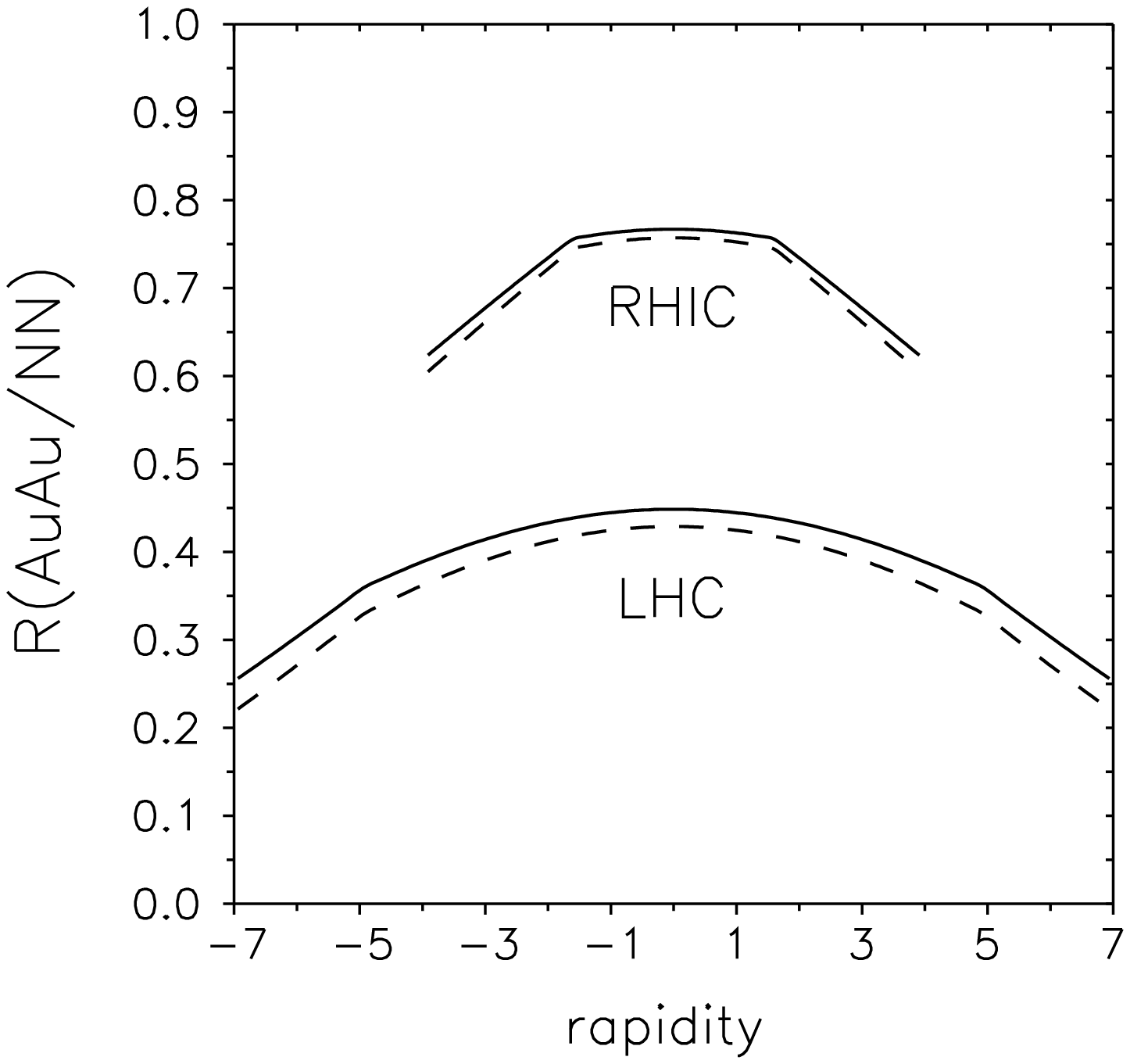}}
 }
\caption{\label{fig:Kopeliovich5-fig1} Shadowing for DY reaction in $p\A$ 
  (upper curves) and $d\A$ (lower curves) collisions at the energies of RHIC 
  ($\sqrtsnn=200$~GeV) and LHC ($\sqrtsnn=5500$~GeV), as function of $x_F$ and 
  dilepton mass $M^2$.
  The left and right figures are calculated at $M=4.5$~GeV and $x_F=0.5$ 
  respectively.}
\end{figure}

The production of heavy flavors can be treated as freeing of a $\bar QQ$ 
fluctuation in the incoming hadron, in which the interaction with such a small 
dipole (actually, with a three-body $\bar QQg$ dipole) results in nuclear 
shadowing, which is a higher twist, $1/m_Q^2$, effect. 
Although very small, it steeply rises with energy and reaches sizable magnitude 
at the energy of LHC.
The effect of this higher twist shadowing on charm production in minimal bias 
and central collisions of heavy ions at the energies of RHIC and LHC is shown 
in figure~\ref{fig:Kopeliovich5-fig1} as the difference between solid and 
dashed curves.

\subsubsection{Process dependent leading twist gluon shadowing}

The projectile fluctuations containing, besides the $\bar QQ$, also gluons, are 
responsible for gluon shadowing, which is a leading twist effect. 
Indeed, the aligned jet configurations, i.e. the fluctuation in which the 
$\bar QQ$ pair carries the main fraction of the momentum, have a large and 
scale independent, transverse size. 
Gluon shadowing is expected to be a rather weak effect~\cite{Kopeliovich:1999am}
due to the localization of the glue inside small spots in the proton~\cite{%
  Kopeliovich:2006bm}. 
This is confirmed by the latest NLO analysis~\cite{deFlorian:2003qf} of data on 
DIS on nuclei.

Unlike for the DIS case, where the produced $\bar qq$ is predominantly in a 
color octet state, in the case of hadroproduction the $\bar QQ$ may be either 
colorless or a color octet. 
Moreover, in the latter case it may have different symmetries~\cite{%
  Kopeliovich:2002yv,Kopeliovich:2001ee}. 
Nonperturbative effects, which cause a contraction of the gluon cloud, may be 
absent for a colorless $\bar QQ$, leading to a much stronger shadowing compared 
to DIS.
This possibility was taken into account predicting the rather strong nuclear 
effects depicted in figure~\ref{fig:Kopeliovich5-fig1}. 
This part of the prediction should be taken with precaution, since it has never
been tested by data.

Figure~\ref{fig:Kopeliovich5-fig1} shows our results for $R_{\A\A/NN}$ as function
of rapidity for minimal bias and central collisions. 
These calculations do not include the suppression caused by energy conservation 
at the ends of the rapidity interval~\cite{Kopeliovich:2005ym}.

The rather strong suppression of charm production that we found should be taken 
into account as part of the strong suppression of high $p_T$ charm production 
observed in central nuclear collisions at RHIC. 
At high $p_T$ this effect should fade away because of the rise of $x_2$, 
although at the LHC this may be a considerable correction.

\subsection{Charm and Beauty  Hadrons from Strangeness-rich QGP at LHC}

{\it I. Kuznetsova  and J. Rafelski}

{\small
The yields of heavy flavored hadrons emitted by strangeness rich QGP are
evaluated within chemical non-equilibrium statistical hadronization
model,  conserving strangeness, heavy flavor, and entropy yields at hadronization.
}
\vskip 0.5cm

%%%%%%%%%%%%%%%%%%%%%%%%%%%%%%%
A relatively large number of hadrons containing charm ($dN_c/dy\simeq 10$) and bottom
 ($dN_b/dy\simeq 1$) quarks are expected to be produced at central rapidity in heavy ion (Pb--Pb) collisions at
the Large Hadrons Collider (LHC). This report summarized results of our
more extensive recent report~\cite{Kuznetsova:2006bh}, and amplifies its findings with
reference to the `first day' LHC-ion results. Differing from  other recent studies   which assume that
the hadron yields after hadronization are in chemical
equilibrium~\cite{Becattini:2005hb},
we form the charm hadron yields in the statistical
hadronization approach based on an
abundance of $u,d,s$ quark pairs fixed by the
bulk properties of a practically chemically equilibrated QGP phase.

In proceeding in this fashion we are  respecting the constraints of
the recombinant dynamic model~\cite{Thews:2000rj}. The absolute
yields (absolute chemical equilibrium) depend in addition to
recombination  on absolute heavy quark yield $dN_{b,c}/dy$. We are
fully implementing the relative chemical equilibrium, that is the
formation of heavy (charmed) hadrons according the the relative
phase space, thus ratios of yields presented here are a complete and
reliable prediction characterized by QGP entropy and strangeness content.
%%%%%%%%%%%%%%%%%%%%%%%%%%%%%%%%%%%%%%%%%%%%%%%%%%%%%%%%%%%%%%%

It is energetically more effective for strange quarks to emerge bound to heavy quarks.
Said differently, the reaction ${\rm K}+D\to \pi+D_s$ is strongly
exothermic, with $\Delta Q\simeq 240$\,MeV, and similarly for the bottom quark.
Considering that the phase space for hadronization
is characterized by a domain temperature $T=160\pm20$ MeV, in presence of strangeness  the  yield  tilts in favor  $D_s$ over $D$,
and  $B_s$ over $B$.The  variability in the light and strange quark content at given hadronization
temperature $T$ is accomplished introducing the  phase space occupancy
$\gamma^{\rm H}_s>1, \, \gamma^{\rm H}_q>1$ of strange, and, respectively,  light constituent
quarks in the hadron phase. In chemical equilibrium $\gamma^{\rm H}_s = \gamma^{\rm H}_q = 1$.
%%%%%%%%%%%%%%%%%%%%%%%%%%%%%%%%%%%%%%%%%%%%%%%Fig 1
\begin{figure}
\centering\hspace*{0.41cm}
\includegraphics[width=7.8cm,height=7.8cm]{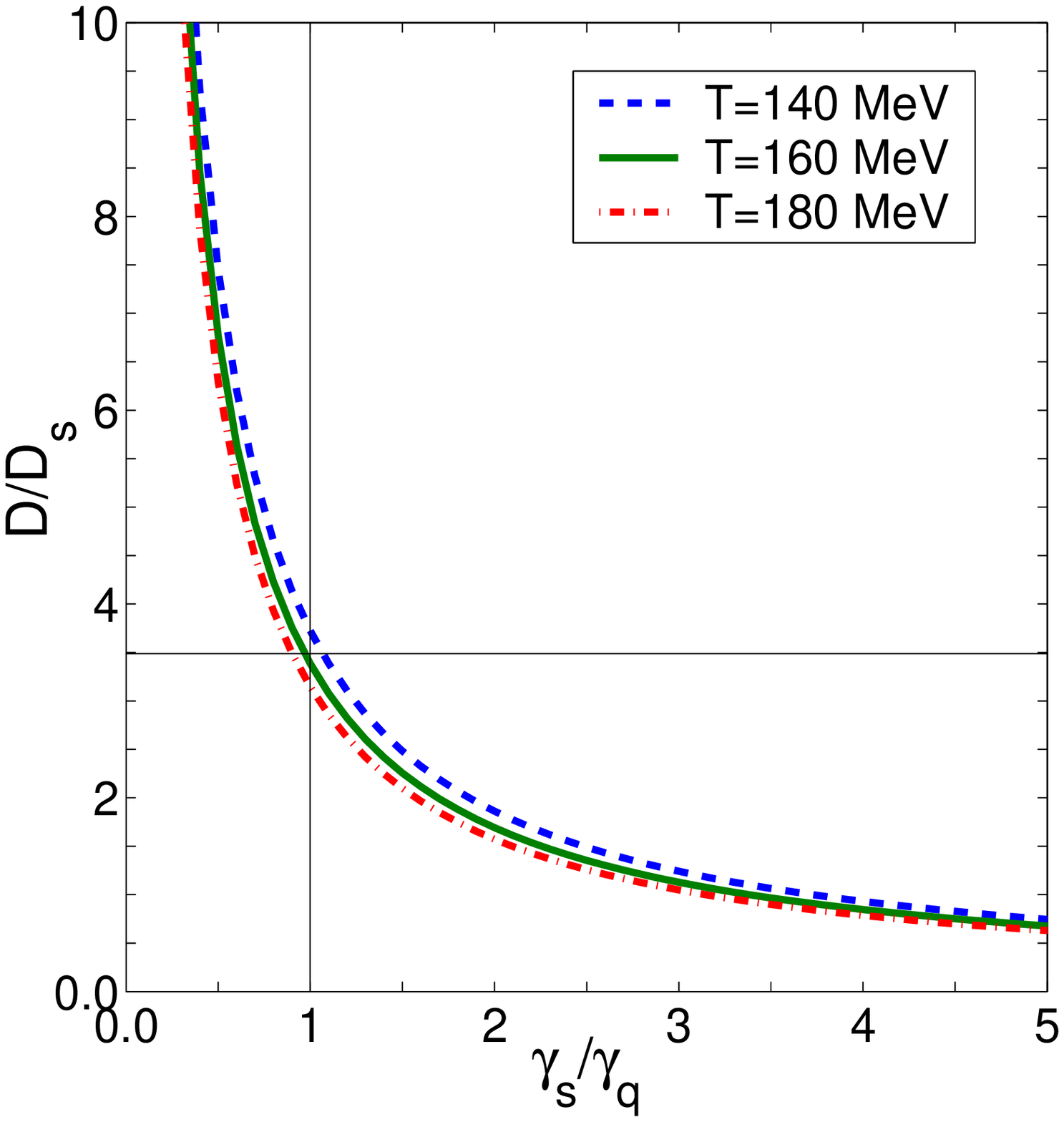}\hspace*{-0.41cm}
\includegraphics[width=7.8cm,height=7.8cm]{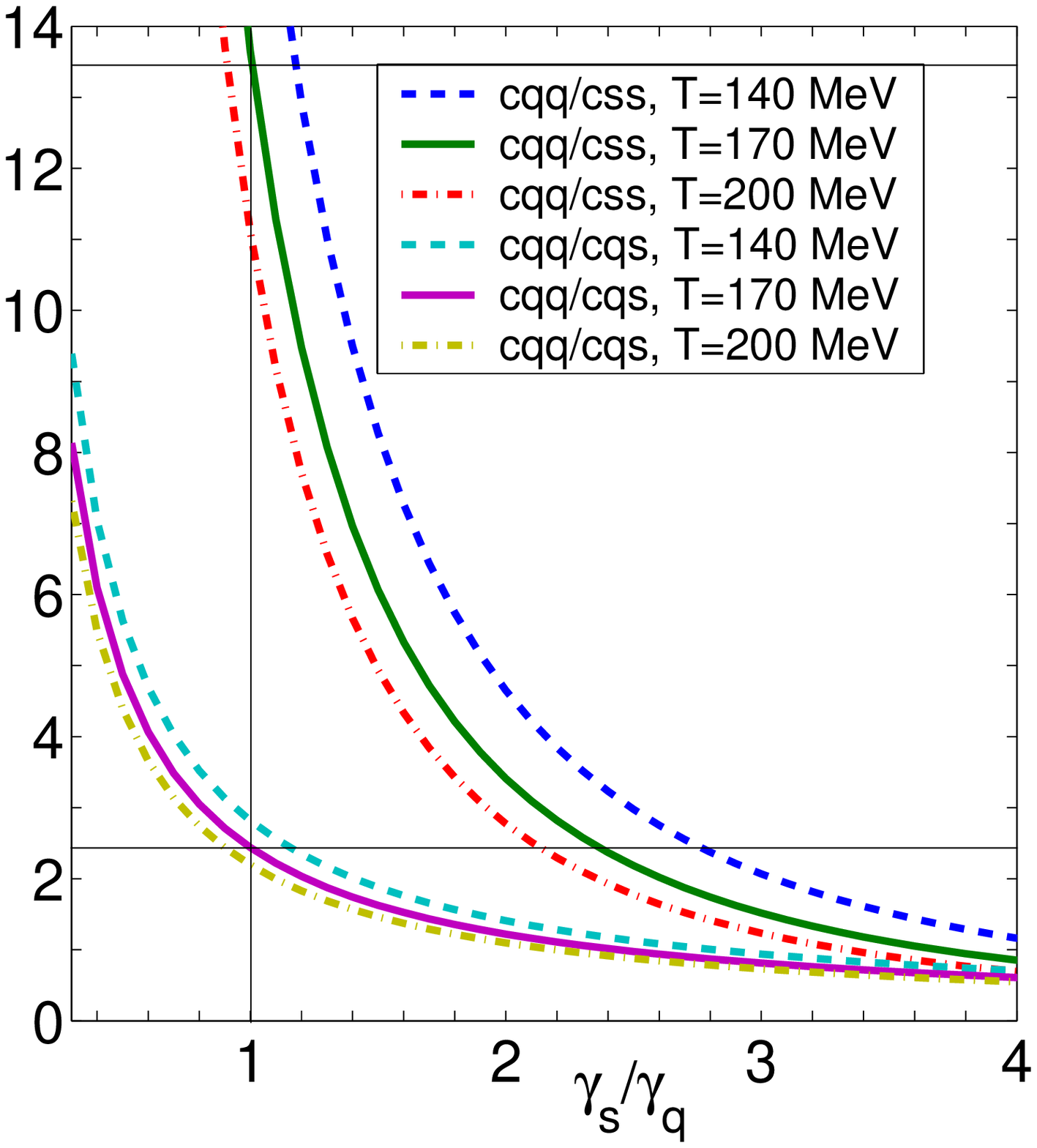}
\caption{ \small As a function of
$\gamma^\mathrm{H}_s/\gamma^\mathrm{H}_q$ on
left: $D/D_s$ ratio  and on right:  $cqq/css=(\Lambda_c+\Sigma_c)/\Omega_c$ (upper lines)
and  $cqq/cqs=(\Lambda_c+\Sigma_c)/\Xi_c$ (lower lines) ratios.
} \label{rDsDg}\label{barratio}
\end{figure}
%%%%%%%%%%%%%%%%%%%%%%%%

A phase space evaluation of the relative yields leads to the results
presented in figure~\ref{rDsDg}, where we show ratio of open charm
strange meson and baryons with the corresponding `less' strange open
charmed (strange) meson  and baryons, as a function of $\gamma^{\rm
H}_s/\gamma^{\rm H}_q$, which is the controlling variable for three
values  $T=200$ MeV, $T=180$--$160$ MeV  and $T=140$ MeV. The
corresponding chemical reference results are indicated by the
crossing vertical and  horizontal lines. For $B$, $B_s$ mesons the
results are the same as for $D$, $D_s$ mesons,
see~\cite{Kuznetsova:2006bh} for details. 

The challenge is to understand
what values of $\gamma^{\rm H}_s/\gamma^{\rm H}_q$ a fast hadronizing QGP implies.
We obtain these by requiring that the
hadronization of QGP proceeds  conserving the entropy $dS/dy$ and strangeness $ds/dy=d\bar s /dy$
content of QGP.  For LHC the expected ratio $s/S=0.038$~\cite{Letessier:2006wn} at $T=140$--180 MeV which implies in the
hadron phase ${\gamma_s}/{\gamma_q}=1.8$--2~\cite{Rafelski:2005jc}. This entails a considerable shift of open charm
hadrons  away from hadron
chemical equilibrium yield towards states containing strangeness in all cases considered
in figure~\ref{rDsDg} (and similarly for the bottom flavor). The hadronization process, as expected,
favors formation of strange charmed meson and baryons, once the
actual QGP  strangeness yield near/above-chemical  equilibrium is
allowed for.

\subsection{Charmonium Suppression in Strangeness-rich QGP}

{\it I. Kuznetsova  and J. Rafelski}

{\small
The yields of $c\bar c$ mesons formed in presence of  
entropy and strangeness rich QGP are evaluated within chemical non-equilibrium statistical hadronization
model,  conserving strangeness and entropy yields at hadronization.
We find that for a given $dN_c/dy$ charm  yield, the abundant presence of light an strange quarks favors  
 formation of $D,\,D_s$ mesons and  to suppression of charmonium.
}
\vskip 0.5cm

There is considerable energetic advantage for a charm quark to bind with 
a strange quark  -- most, if not all, charmonium--strange meson/baryon 
reactions of the type
$c\bar c+sX\to cX+\bar c s$, where $X\equiv \bar q=\bar u, \bar d$ or
 $X\equiv qq, qs, ss $ are strongly exothermic. 
In statistical hadronization this phase space effect
favors formation of $D_s$ over $c\bar c$. Seen from the kinetic model
perspective~\cite{Thews:2000rj}, this observation shows a strong channel of charmonium destruction.
Thus presence of strangeness facilitates  a novel charmonium 
suppression mechanism~\cite{Kuznetsova:2006hx,Kuznetsova:2006bh}.
To implement this effect hadronization of QGP must  conserve strangeness
and entropy and thus cannot be ad-hoc associated with chemical equilibrium. 

In the non-equilibrium statistical hadronization model we   balance total yield
of charmed particles  within a given volume $dV/dy$ to  the level available in
the QGP phase $dN_c/dy\propto dV/dy  ( \gamma^{\rm H}_{c} \gamma^H_i +\ldots)$, where a
few percent of the yield is in multi-charm baryons and charmonium
 involving higher powers of  $   \gamma^{\rm H}_{c}$. 
This constraint determines a value of  $\gamma^H_{c}$, 
which   for the case of LHC can be considerably above unity. 
 Therefore, the hadronization yields we  compute for hidden charm mesons:
$dN_{c\bar c}/dy\propto dV/dy \gamma^{H\,2}_c\propto (dN_c/dy)^2/(\gamma^{H\,2}_idV/dy)$.
depends  on the
inverse of the model dependent reaction volume,  and scales with the square of the
 total  charm yields~\cite{Thews:2000rj}. We also show above that for the case that 
$\gamma^H_i>1$ a hereto unexpected  suppression of 'onium yield is expected.

\begin{figure}[!t]
%\centering%\hspace*{2cm}
\centerline{\hspace*{1cm}\includegraphics[width=12.0cm,height=7.5cm]{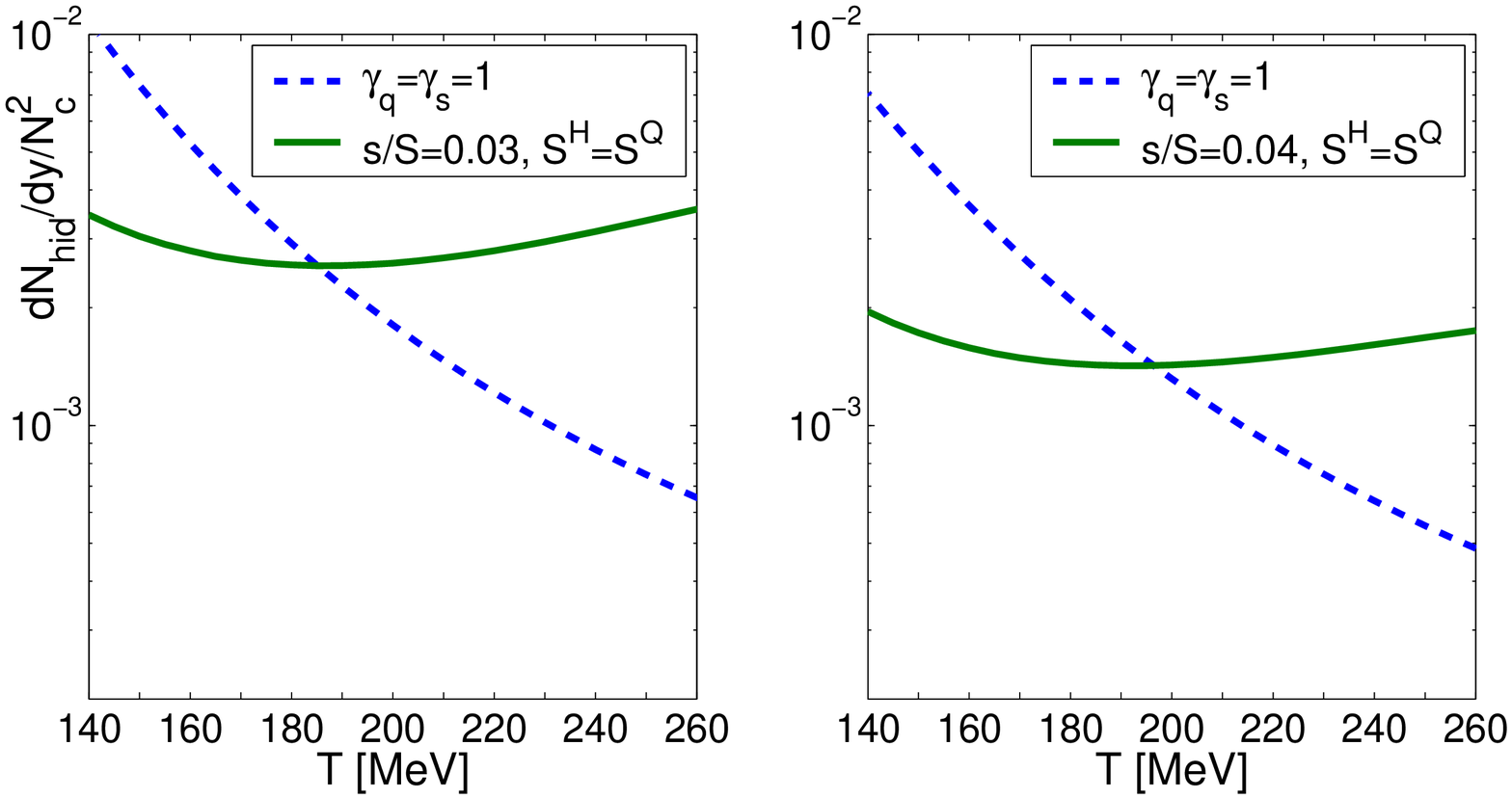}
\hspace*{-2.5cm}\includegraphics[width=6cm,height=6.5cm]{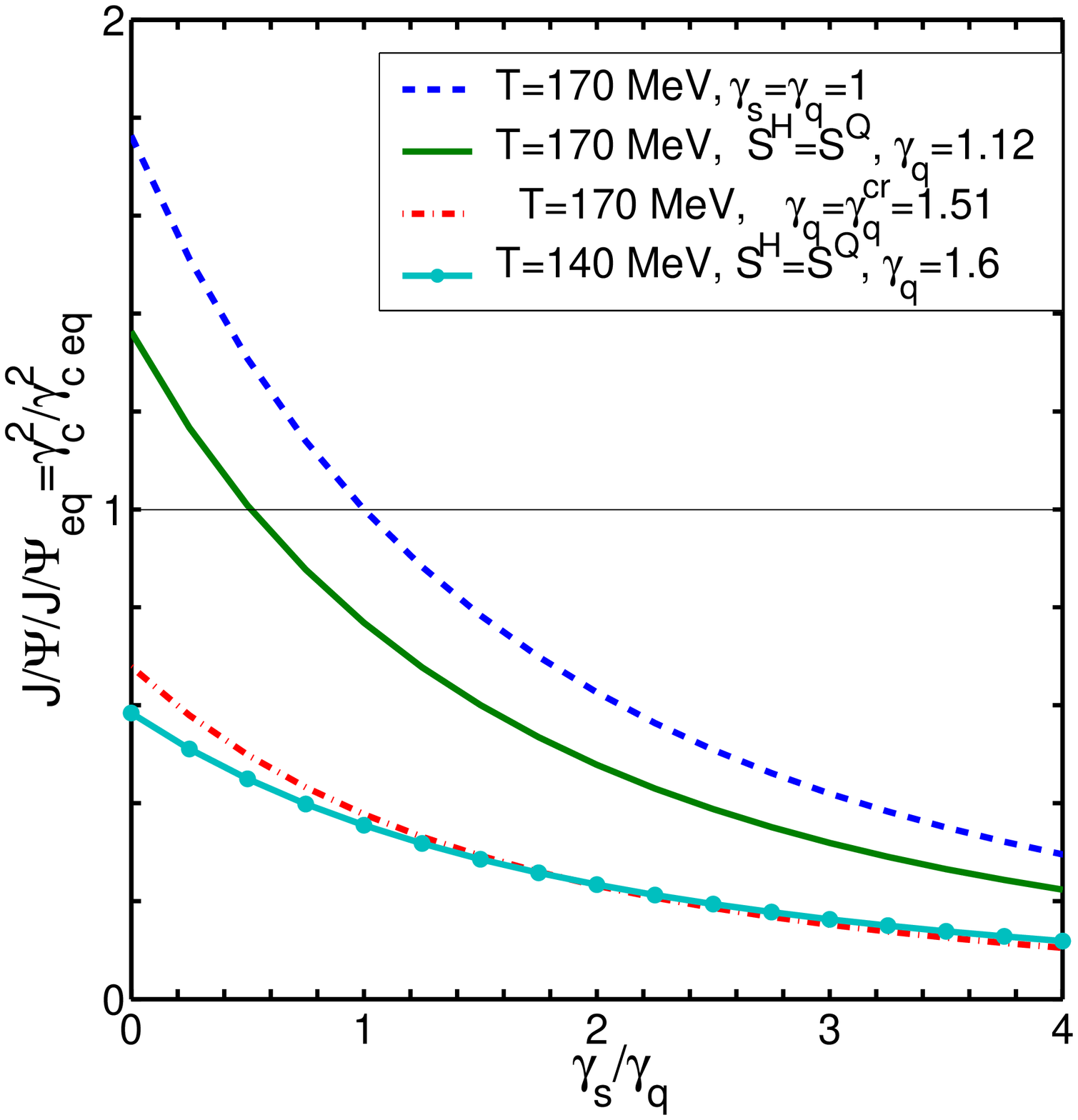}}
\caption{Left two panels: $c\bar{c}/N^2_c$ relative yields as a function of
hadronization temperature T, right panel ratio $J\!/\!{\Psi}/J\!/\!\Psi_{eq}$ as a function
of $\gamma^{\rm H}_s/\gamma^{\rm H}_q$, see tex.} 
\label{cc}  \label{jpsrg} 
\end{figure}

In figure~\ref{cc} the yield of all hidden charm $c\bar{c}$ (sum
over all  $c\bar{c}$ mesons)  is shown, normalized by the square of
 $dN_c/dy=10$ (middle panel for LHC environment) and $dN_c/dy=3$ (left panel, RHIC environment), as a function of
hadronization temperature $T$. We show result for $s/S=0.03$ with
$dV/dy=600$\,fm$^3$, $T = 200$ MeV (solid line, left
panel) and   for $s/S=0.04$ with  $dV/dy=800$\,fm$^3$, $T =
200$ MeV  (solid line, middle panel).   Results shown 
for chemical equilibrium case (dashed lines) are for the values $\gamma_s=\gamma_q=1$.
For the chemical non-equilibrium hadronization (solid lines $\gamma^H_i>1, i=q,s$), 
the QGP and hadron phase space is evaluated conserving  entropy    $S^Q=S^H$  
and strangeness    $s^Q=s^H$  between phases. 

%%%%%%%%%%%%%%%%%%%%%%%%%
%%%%%%%%%%%%%%%%%%%%%%%%%%%%%%
 
We see, comparing the left and middle panel 
that the yield of $c\bar{c}$ mesons decreases with increasing specific
strangeness content  (note logarithmic scale).   The chemical suppression effect is further quantified  in third, right
panel in  figure~\ref{jpsrg}, where  we show the ratio $J\!/\!{\Psi}/J\!/\!\Psi_{eq}=\gamma^2_c/\gamma^2_{c\,eq}$ as a function
of $\gamma^{\rm H}_s/\gamma^{\rm H}_q$ at fixed value of $\gamma^{\rm H}_q$ and, as required, entropy
conservation  for $T=140, 170$ MeV.
For $T=140$ MeV we show result with $\gamma_q = 1.6$ (solid dotted
line) which corresponds entropy conservation between QGP and
hadronic phase for this hadronization temperature. For $T=170$ MeV we
show results with $\gamma_q \approx \gamma_q^{cr} = 1.51\equiv e^{m_\pi/2T}$ (dash-dot
line), $\gamma_q=1.12$ (solid line) and $\gamma_q=1$ (dashed line). For  $\gamma_q=1.12$  
entropy  is conserved in hadronization at $T=170$ MeV .

The formation of the $B_c(\bar b c)$ proposed as another QGP signature~\cite{Schroedter:2000ek}
 has not been evaluated in the present work,  since this  particle yield
suffers from additional (canonical) suppression. Kinetic 
formation models suggest significant enhancement of this double exotic meson, as compared to 
a cascade of NN reactions.

\subsection{ $J/\psi$   $p_T$ spectra from in-medium recombination }

{\it R.~L.~Thews and M.~L.~Mangano}

{\small
We consider production of $J/\psi$ by recombination of c, $\bar{c}$
quarks produced in separate N-N interactions during Pb-Pb collisions.
Inputs for the calculation include the NLO pQCD spectra of charm
quarks, plus a range of nuclear parameters taken from extrapolation of
results at RHIC energy.
}
\vskip 0.5cm

\begin{figure}[htb]
\vskip 0.5 truecm
\begin{center}
\epsfig{file = 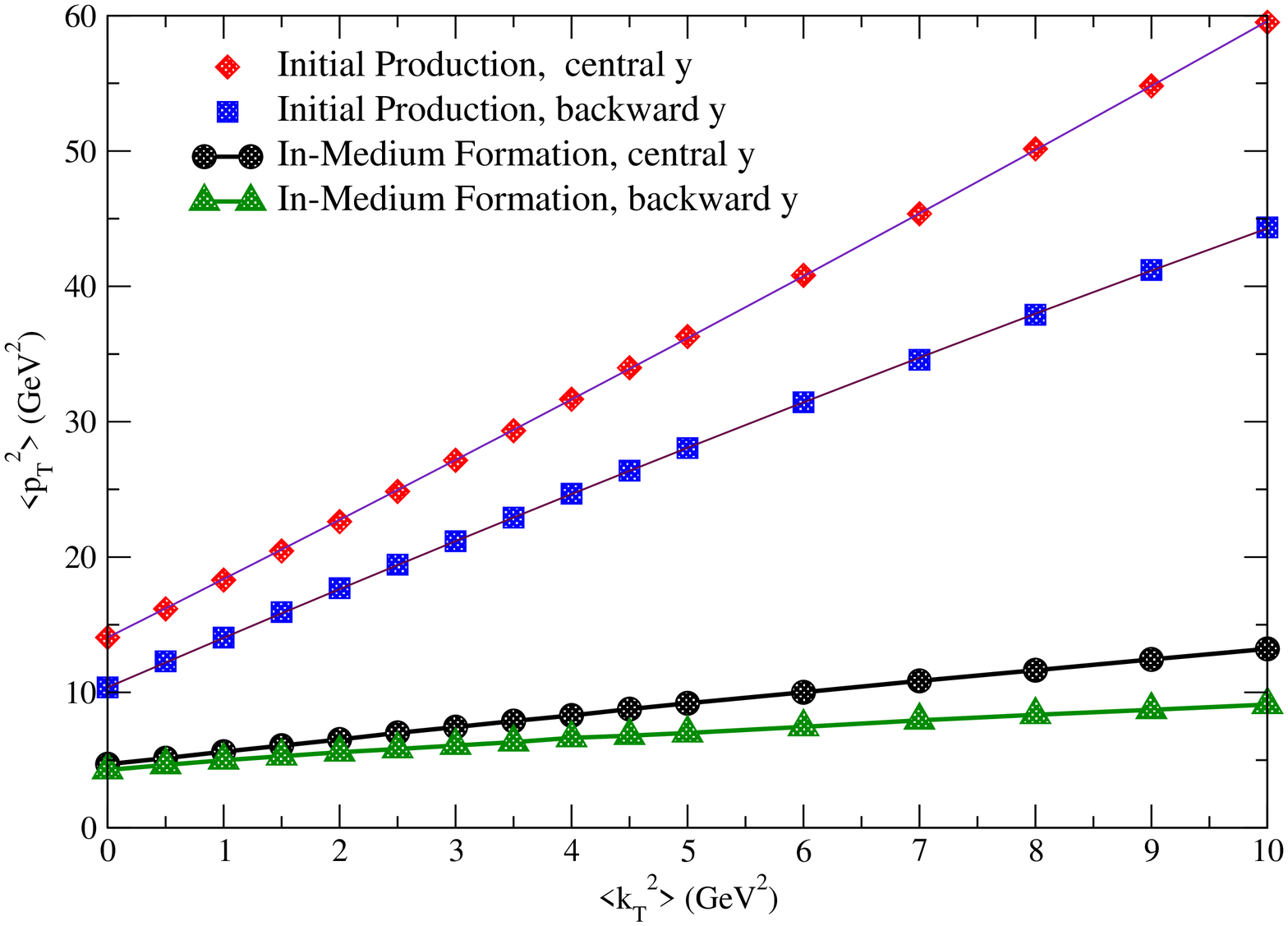, width=7.5cm} \hfill \\
\epsfig{clip=,width=7.5cm,figure=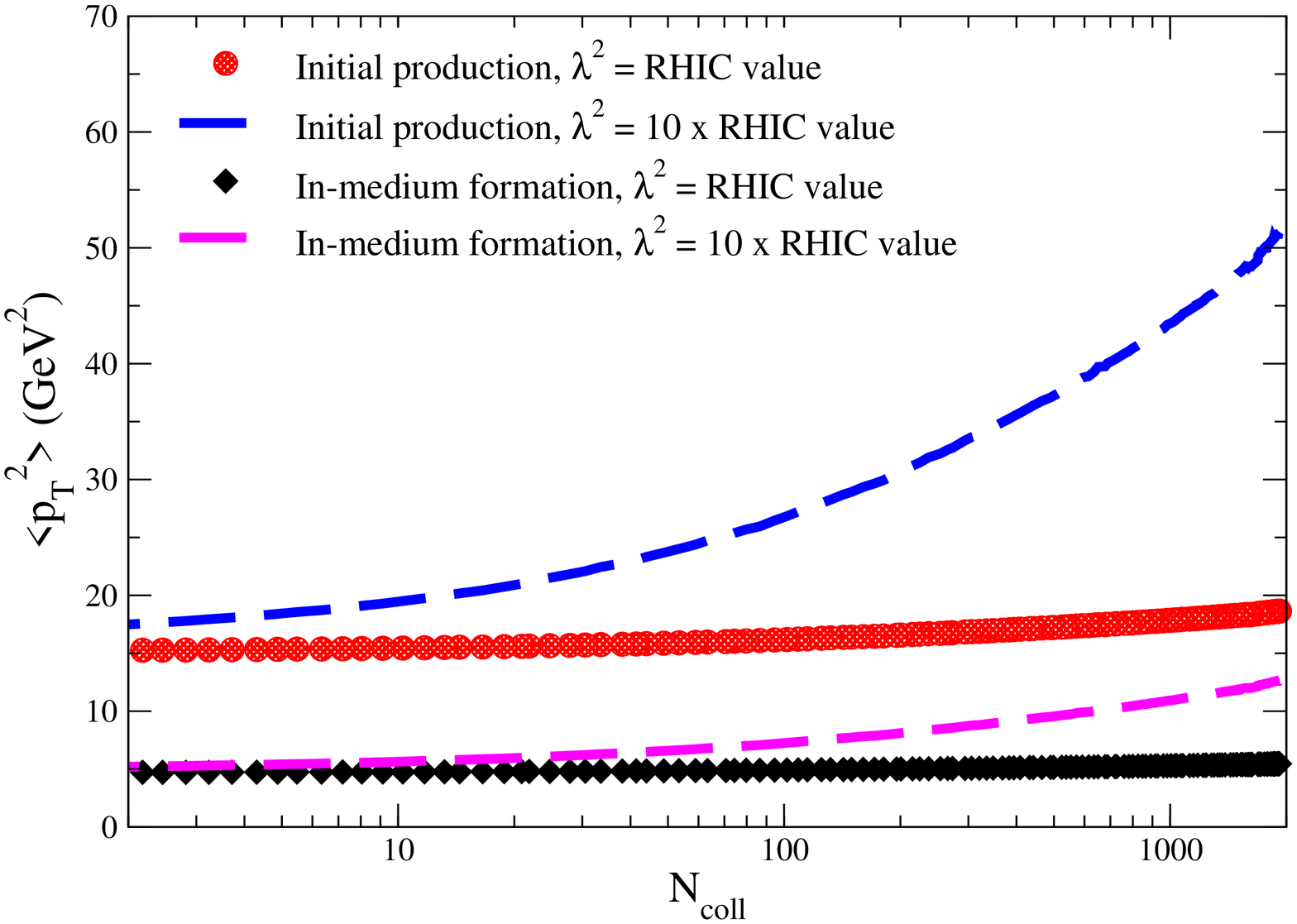} 
\epsfig{clip=,width=7.5cm,figure=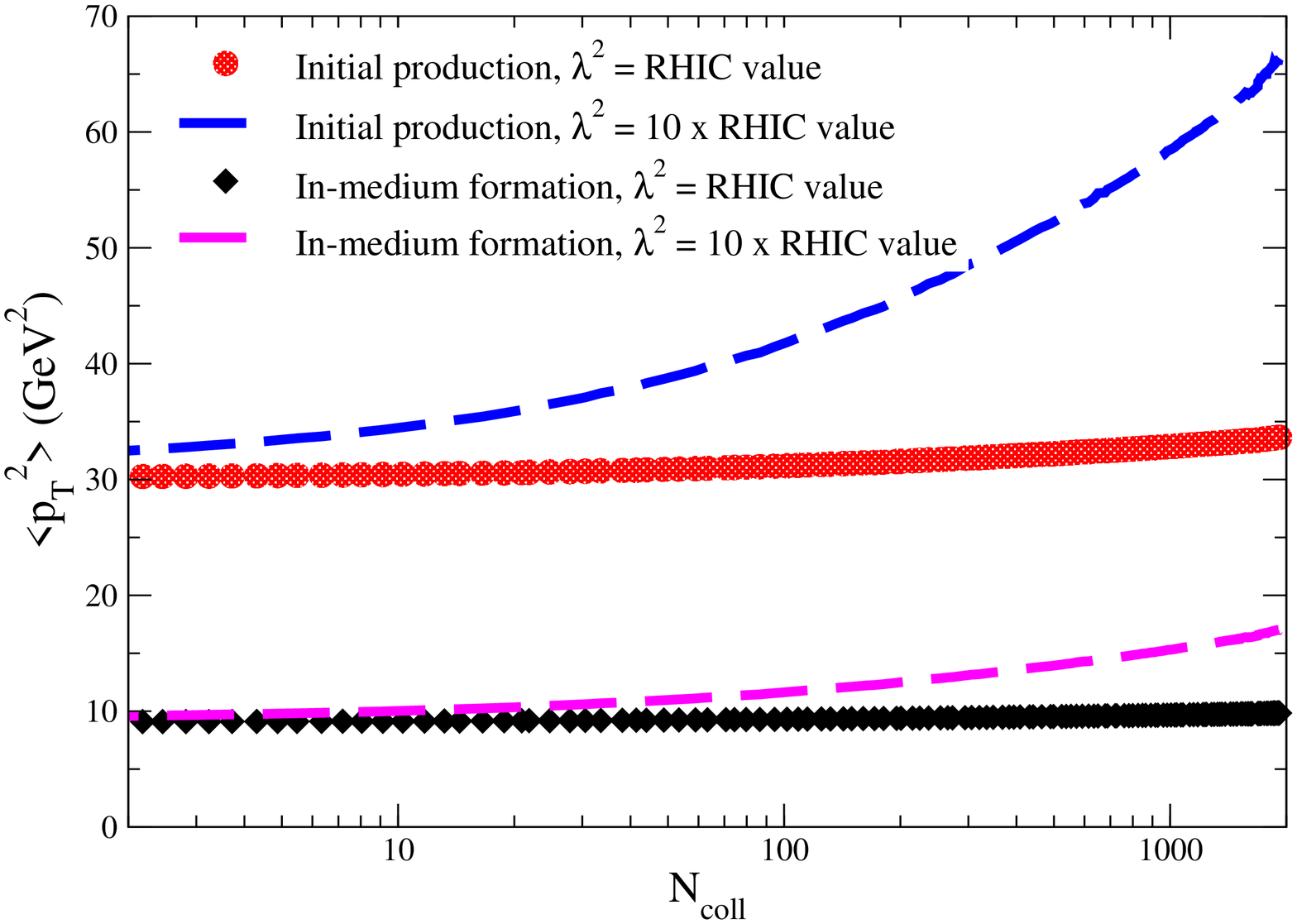}
\end{center}
\caption{\label{F1} Upper: Variation of $J/\psi$ $\pts$ with the
  nuclear smearing parameter. Lower: dependence on the intrinsic
  $\kts_{pp}$, with $\kts_{pp}$ = 0.0 (left) and 5.0 (right). }
\vskip -0.5 truecm
\end{figure}

The possibility that $J/\psi$ could be formed in AA collisions 
by recombination in a region of color deconfinement was first
developed in Ref.\cite{Thews:2000rj}.  It was motivated by the
realization that the total formation probability would be proportional
to the square of the total number of $c\bar{c}$ pairs, which
at RHIC and especially LHC provide a large enhancement factor.
One can calculate the $p_T$ and $y$ spectra of $J/\psi$ formed 
either through recombination or direct initial production,
using the corresponding quark spectra from a pQCD NLO
calculation~\cite{Mangano:1991jk} in individual 
nucleon-nucleon interactions. The method involves generating
a sample of these initially-produced $c\bar{c}$ pairs, 
smearing the transverse momentum with a gaussian distribution
of width $\kts$ to simulate nuclear broadening and confinement
effects, 
and weighting each pair with a formation cross section.
This procedure naturally divides the total pair sample 
into two categories:  the so-called 
``diagonal" sample, which pairs
the c and $\bar{c}$ 
from the same nucleon-nucleon interaction and  the
``off-diagonal" sample, where  c and
$\bar{c}$ come from different nucleon-nucleon interactions.
The spectra of the resulting $J/\psi$ will retain some 
memory of the charm quark spectra 
and provide signatures of the two different origins.
For example, one expects the $p_T$ spectrum of non-diagonal pairs to
be softer, since it is less likely for high-$p_T$ $c$ and $\bar{c}$ quarks from
independent scatterings to be close enough in phase-space to coalesce
into a $J/\psi$.
Results for RHIC were presented
in Ref. \cite{Thews:2005vj}, where the primary signal
was found to be a narrowing of the non-diagonal $y$ and $p_T$ spectra,
relative  to the diagonal ones.
We show in Fig.~\ref{F1} the calculated $J/\psi$ width $\pts$
as a function of $\kts$, for central and forward production in ALICE. 
$\pts$ grows with  $\kts$ for both the direct initial
production and the in-medium formation, but  the latter widths are 
always smaller than the former. 
Widths at small $y$ are  also  greater than at large $y$, reflecting
the underlying pQCD distributions.  
To fix the nuclear smearing parameter values, we use a relation between
measurable  $\pts$ in pp and pA interactions, 
\begin{equation}
\pts_{pA} - \pts_{pp}\; = \lambda^2\; [\bar{n}_A - 1],
\label{pApt}
\end{equation}
where $\bar{n}_A$ is the impact-averaged number of inelastic interactions
of the proton projectile in nucleus A, and $\lambda^2$ is 
proportional to the square of the
transverse momentum transfer per initial state collision. 
We use a Glauber model to calculate the centrality
dependence of the $\bar{n}_A$,  and parameterize the
centrality by the total number of collisions, $N_{coll}$.
Thus with measurements of $\pts_{pp}$ and $\pts_{pA}$ one can
extract $\lambda^2$ and calculate the corresponding
nuclear broadening for AA interactions.
The lower plots of Fig.~\ref{F1} show the results for
Pb-Pb at 5.5 TeV, with 
$\kts_{pp}$=0 and 5 $GeV^2$.   For both cases
the $J/\psi$ widths will provide a clear discrimination between
direct initial production and in-medium formation.
In general, one would expect some combination of initial production 
and in-medium formation, so the prediction is bounded from above
and below.
%\begin{figure}[htb]
%\begin{center}
%\epsfig{file = thews/thewsfig3.eps, width=12cm}
%\end{center}
%\caption{\label{F2} Variation of $J/\psi$ transverse momentum widths with 
%the collision centrality, for $\kts_{pp}$ = 0.}
%\end{figure}
%
%\begin{figure}[htb]
%\begin{center}
%\epsfig{file = thews/thewsfig4.eps, width=12cm}
%\end{center}
%\caption{\label{F3} Variation of $J/\psi$ transverse momentum widths with 
%the collision centrality, for $\kts_{pp}$ = 5 GeV$^2$.}
%
%\end{figure}
There is almost no change in the $p_T(c)$ spectra between 5.5 and 
14 TeV. Thus we can use the 14 TeV pp data 
to determine $\kts_{pp}$ at 5.5 TeV.
One can then expect that the absence of energy dependence will
also hold for p-Pb results, allowing us to also determine $\lambda^2$
at 5.5 TeV from a measurement at any LHC energy, thus fixing the
prediction for curves such as those in Fig.~\ref{F1}.
\vskip -0.5 truecm

\subsection{Predictions for quarkonia dissociation}
\label{s:Mocsy}

{\em \'A. M\'ocsy and P. Petreczky}

{\small
We predict the upper bound on the dissociation temperatures of different
quarkonium states.
}
\vskip 0.5cm

In a recent paper~\cite{Mocsy:2007yj} we analyzed in detail the quarkonium 
spectral functions. 
This analysis has shown that spectral functions calculated using potential 
model for the non-relativistic Green's function combined with perturbative QCD 
can describe the available lattice data on quarkonium correlators both at zero 
and finite temperature in QCD with no light quarks~\cite{Mocsy:2007yj}. 
Charmonia, however, were found to be dissolved  at temperatures significantly 
lower than quoted in lattice QCD studies, and in contradiction with other 
claims made in recent years from different potential model studies. 
In~\cite{Mocsy:2007jz} we extended the analysis to real QCD with one strange 
quark and two light quarks using new lattice QCD data on quark anti-quark free
energy obtained with small quark masses~\cite{Petrov:2006pf}.  

Here we briefly outline the main results of the analysis of~\cite{Mocsy:2007jz},
in particular the estimate for the upper limit on the dissociation temperatures.
There is an uncertainty in choosing the quark-antiquark potential at finite 
temperature. 
In~\cite{Mocsy:2007jz} we considered two choices of the potential, both 
consistent with the lattice data~\cite{Petrov:2006pf}.  
The more extreme choice, still compatible with lattice data, leads to the 
largest possible binding energy.  
In this most binding potential some of the quarkonium states survive above 
deconfinement, but their strongly temperature-dependent binding energy is
significantly reduced. 
This is shown in figure~\ref{fig:Mocsy-fig1}. 
Due to the reduced binding energy thermal activation can lead to the 
dissociation of quarkonia, even when the corresponding peak is present in the 
spectral function. 
Knowing the binding energy we estimate the thermal width using the analysis 
of~\cite{Kharzeev:1995ju}. 
The expression of the rate of thermal excitation has particularly simple form 
in the two limiting cases:
\begin{figure}
\begin{center}
\includegraphics[width=6.5cm]{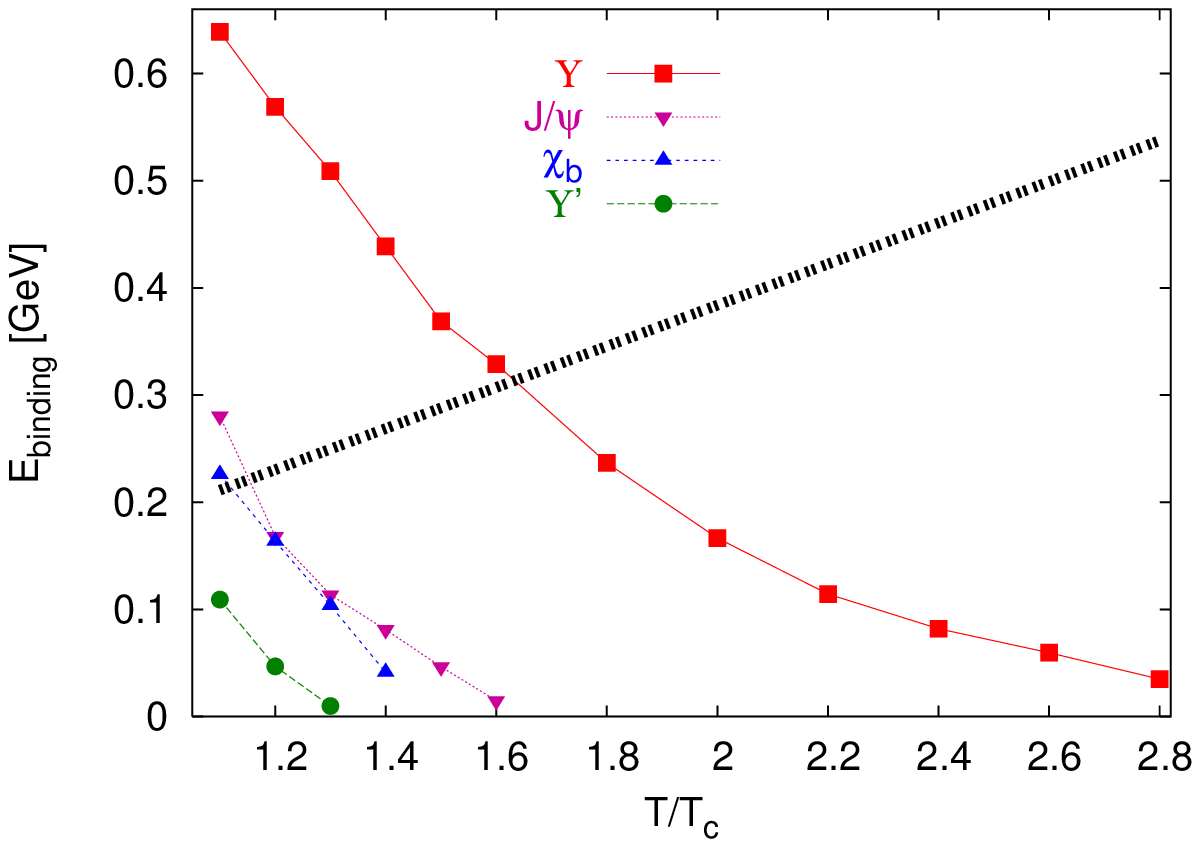}
\includegraphics[width=6.5cm]{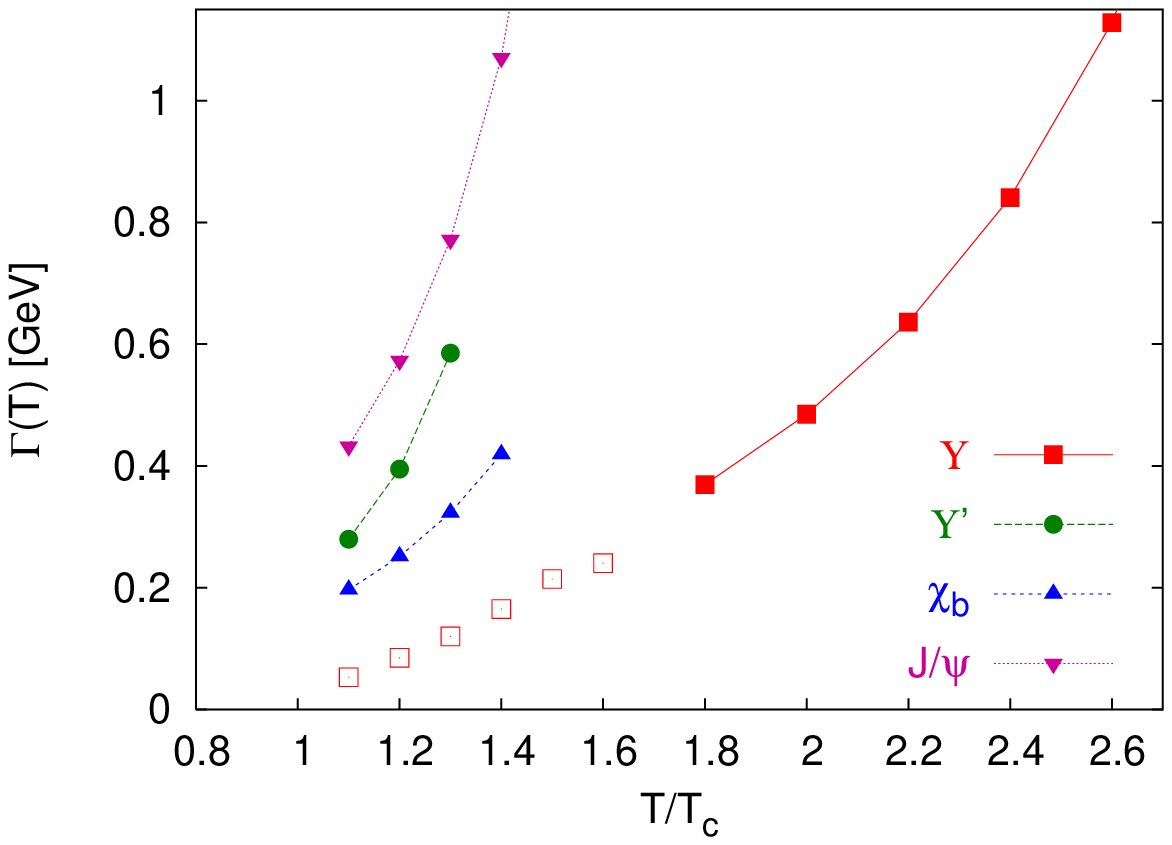}
\caption{Upper limit of the binding energy (left) and the width (right) of 
  quarkonium states. 
  For better visibility, in the limit of small binding, the open squares show 
  the width of the 1S bottomonium state multiplied by six.}
\label{fig:Mocsy-fig1}
\end{center}
\end{figure}
\[
\Gamma(T)=\frac{(LT)^2}{3\pi}M{\rm e}^{-E_{\rm bin}/T}\, ,~E_{\rm bin}\gg T\, \quad 
\Gamma(T)=\frac{4}{L}\sqrt{\frac{T}{2\pi M}}\, ,~ E_{\rm bin}\ll T \, .
\]
Here $M$ is the quarkonium mass, $L$ is the size of the spatial region of the 
potential, given by the distance from the average quarkonium radius to the top 
of the potential, i.e. $L=r_{\rm med}-\langle r^2\rangle^{1/2}$, $r_{\rm med}$ being 
the effective range of the potential~\cite{Mocsy:2007jz}. 
Using the above formulas we estimate the thermal width of charmonium and 
bottomonium states.
Since in the deconfined phase $E_{\rm bin}<T$ the $1S$ charmonium and $2S$ and 
$1P$ bottomonium states are in the regime of weak binding, and their width is 
large, as shown in figure~\ref{fig:Mocsy-fig1}. 
The $1S$ bottomonium is strongly bound for $T<1.6T_c$ and its thermal width is 
smaller than $40$~MeV. 
For $T>1.6T_c$, however, even the 1S bottomonium states is in the weak binding 
regime resulting in the large increase of the width, see 
figure~\ref{fig:Mocsy-fig1}.
When the thermal width is significantly larger than the binding energy no peak 
structure will be present in the spectral functions, even though the simple 
potential model calculation predicts a peak.
Therefore, we define a conservative dissociation temperature by the condition
$\Gamma>2 E_{\rm bin}$. 
The obtained dissociation temperatures are summarized in 
table~\ref{tab:Mocsy-tab1}. 
\begin{table}[h]
\renewcommand{\arraystretch}{0.81}
\begin{center}
\caption{Upper bound on quarkonium dissociation temperatures.}
\begin{minipage}{8.5cm} \tabcolsep 5pt
\begin{tabular}{|ccccccc|}
 state&$\chi_c$&$\psi'$&$J/\psi$&$\Upsilon'$&$\chi_b$&$\Upsilon$\\ \hline 
 $T_{\rm dis}$&$\le T_c$& $\le T_c$ &$1.2T_c$&$1.2T_c$&$1.3T_c$&$2T_c$\\ 
\end{tabular}
\end{minipage}
\end{center}
\label{tab:Mocsy-tab1}
\end{table}

From the table it is clear that all quarkonium states, except the $1S$ 
bottomonium, will melt at temperatures considerably smaller than previous 
estimates, and will for certain be dissolved in the matter produced in heavy 
ion collision at LHC. 
Furthermore, it is likely that at energy densities reached at the LHC a large 
fraction of the $1S$ bottomonium states will also dissolve. 
It has to be seen to what extent these findings will result in large $R_{AA}$ 
suppression at LHC. 
For this more information about initial state effects is needed. 
Moreover, the spectral functions are strongly enhanced over the free case even 
when quarkonium states are dissolved~\cite{Mocsy:2007yj,Mocsy:2007jz} indicating
significant correlations between the heavy quark and antiquark. 
Therefore, one should take into account also the possibility of quarkonium 
regeneration from correlated initial quark-antiquark pairs.

\subsection{Heavy flavor production and suppression at the LHC}

{\it I. Vitev}

{\small
Predictions for the baseline $D$- and $B$-mesons production cross
sections at $s^{1/2} = 5.5$~TeV at the LHC in p+p collisions are given 
for $p_T > M_{c,b}$, respectively. New measurements that allow to 
identify the underlying hard partonic processes in heavy flavor
production are discussed. Based on the short $D$- and $B$-mesons  
formation times, medium-induced dissociation is proposed as
a mechanism of heavy flavor suppression in the QGP at intermediate $p_T$.
In contrast to previous results on heavy quark modification, 
this approach predicts suppression of $B$-mesons comparable to 
that of $D$-mesons at transverse momenta as low as $p_T \sim 10$~GeV. 
Suppression of non-photonic electrons form the primary semi-leptonic
decays of charm and beauty hadrons is calculated in the $p_T$ region 
where collisional dissociation is expected to be relevant.
}
\vskip 0.5cm

\begin{figure}[t]
\begin{center} 
%\vspace*{-1.in}
\includegraphics[width=2.4in,height=2.9in,angle=0]{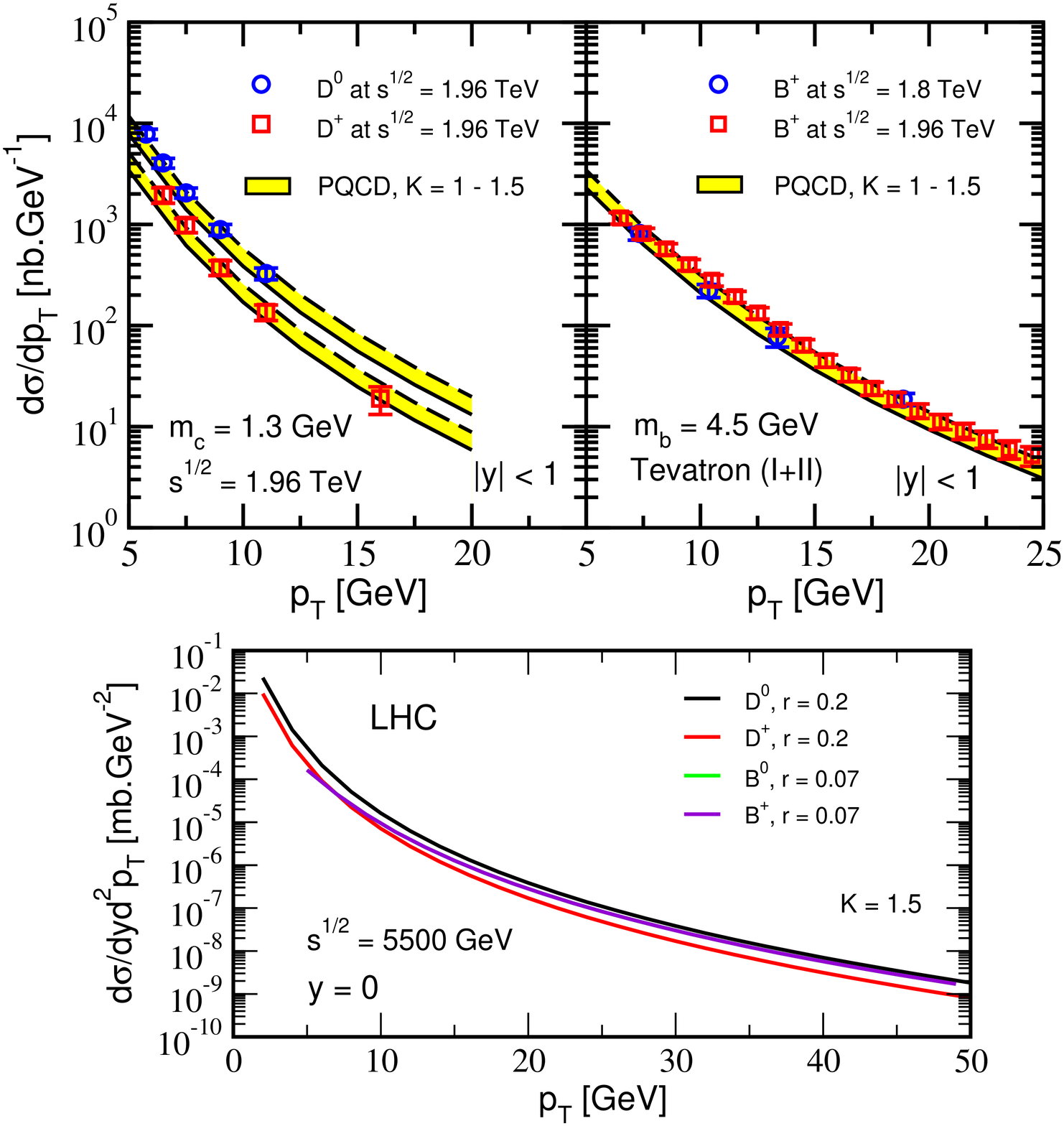}
\hspace*{.5cm}
\includegraphics[width=2.4in,height=2.5in,angle=0]{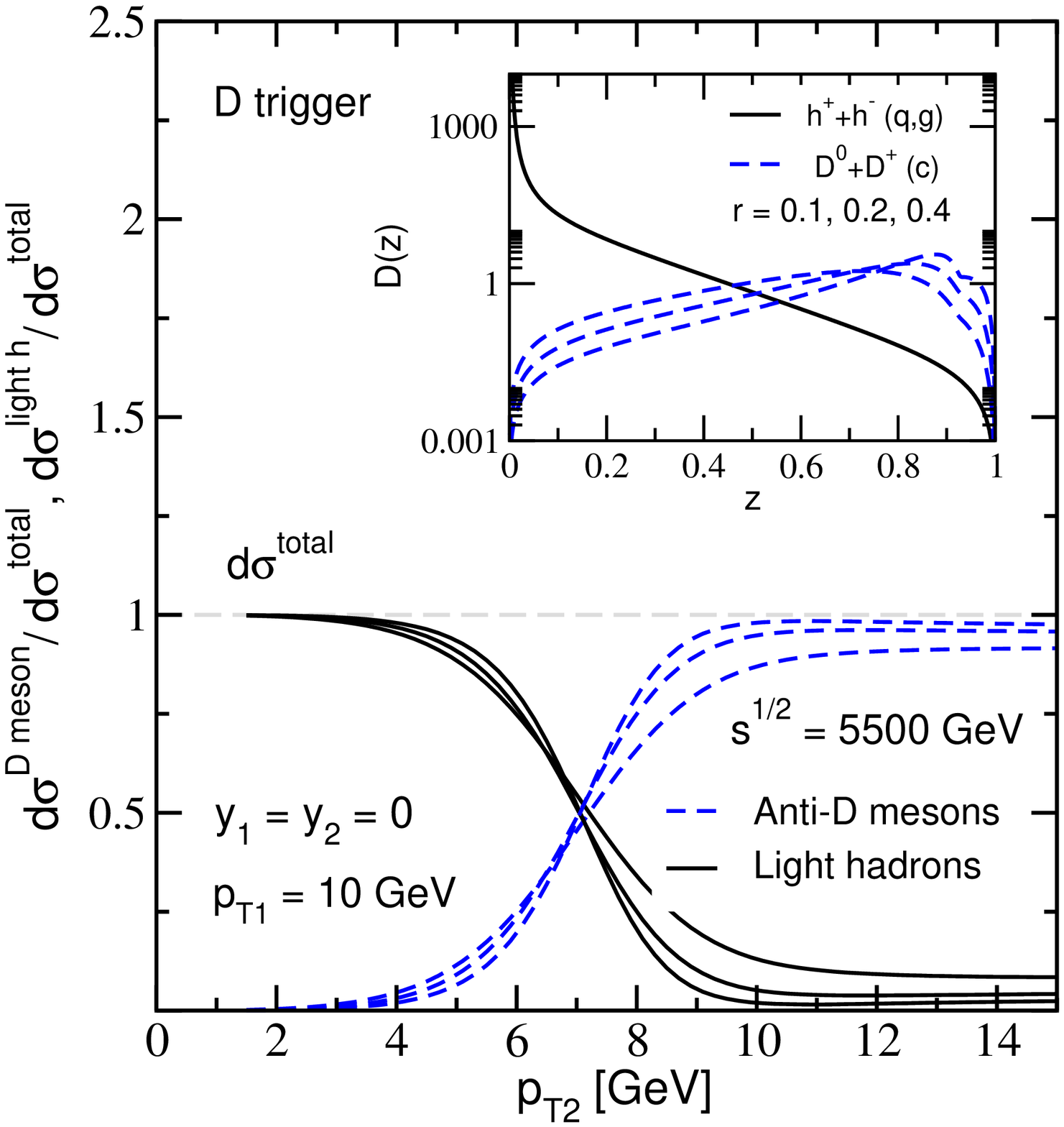}
\caption{ Left panel: $D$- and $B$-meson production cross sections
st $s^{1/2}=5.5$~TeV~\cite{Vitev:2006bi}. Comparison to available 
data at Tevatron is
also shown. Away-side hadron composition of $p_T = 10$~GeV 
$D$-meson triggered jet~\cite{Vitev:2006bi}. 
Right panel: Hadron composition of the away-side $D$-meson 
triggered jet at LHC energies as a function of the  hardness of 
the heavy quark fragmentation function.} 
\label{tevbase}
\end{center} 
\end{figure}

Predictions for the baseline $D^0, D+, B^0, B+$ cross sections in p+p 
collisions at the LHC at $s^{1/2}=5.5$~TeV are given in the left 
panel of Fig.~\ref{tevbase}~\cite{Vitev:2006bi}. At lowest order we also 
include $Q+g \rightarrow Q+g$, $Q+q(\bar{q}) \rightarrow Q+q(\bar{q})$ 
and  processes that give a dominant  contribution to the single 
inclusive $D$- and 
$B$-mesons~\cite{Vitev:2006bi}. The right panel of Fig.~\ref{tevbase} 
illustrates a  method to determine the underlying heavy
flavor production mechanism through the away-side hadron composition  
of $D-$ and $B-$meson triggered jets~\cite{Vitev:2006bi}.  
 
\begin{figure}[t!]
\begin{center} 
\vskip 0.2cm
\includegraphics[width=2.4in,height=2.7in,angle=0]{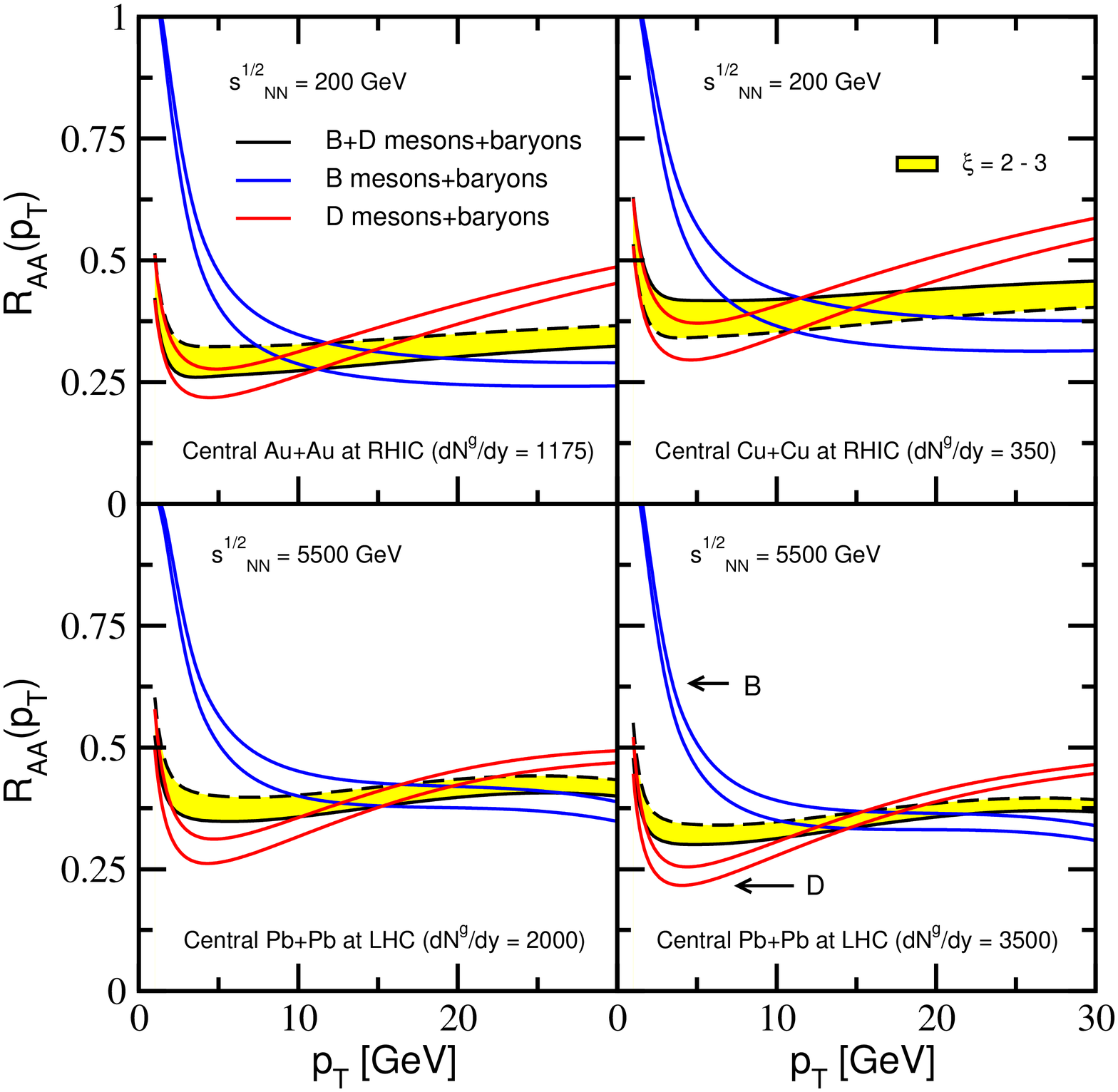}
\hspace*{.5cm}
\includegraphics[width=2.4in,height=2.9in,angle=0]{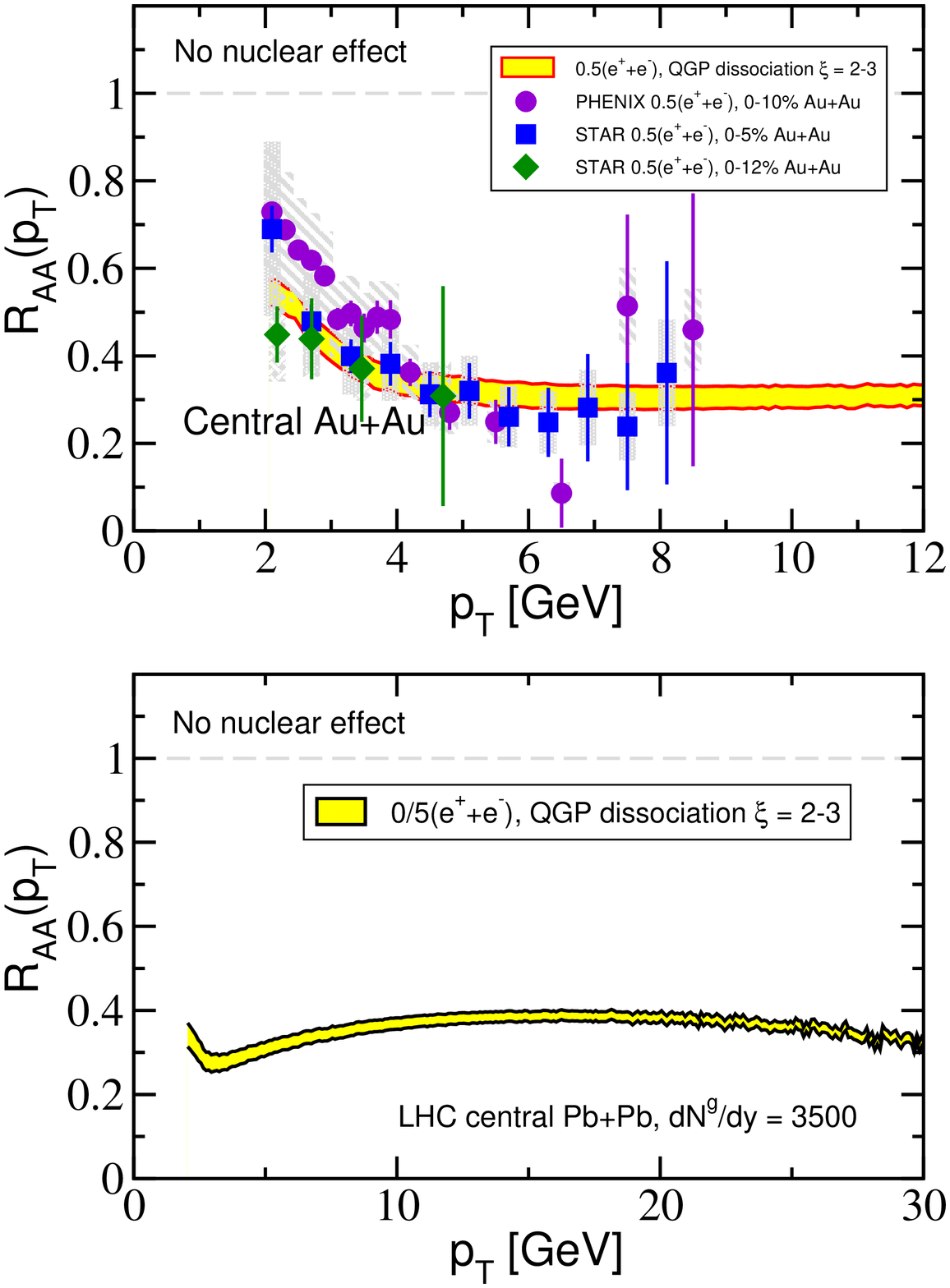}
\caption{ Left panel: Suppression of $D$- and $B$-meson production 
via collisional dissociation in the QGP. Results on $R_{AA}(p_T)$ in 
central Pb+Pb collisions at the LHC are compared to central Au+Au and 
Cu+Cu collisions at RHIC\cite{Adil:2006ra,Vitev:2007jj}. 
Right panel: Suppression of inclusive non-photonic electrons  
from $D$- and $B$-meson spectra softened 
by collisional dissociation  in central  
Au+Au collisions at RHIC compared to data  and Pb+Pb collisions 
at the LHC. }
\label{hadquench}
\end{center} 
\end{figure}

The GLV approach is to multiple parton scattering~\cite{Gyulassy:2002yv}
can be easily generalized to various compelling high energy nuclear 
physics problems, such as meson dissociation in dense nuclear 
matter~\cite{Adil:2006ra}. $R_{AA}(p_T)$ results for charm and beauty
from this novel suppression mechanism at RHIC and LHC are shown in
the left panel of Fig. \ref{hadquench}. Attenuation rate similar  to the 
light hadron quenching from radiative energy loss~\cite{Gyulassy:2002yv}
is achieved. The right panel of Fig.
\ref{hadquench} shows the suppression of the
single non-photonic $0.5(e^+ + e^-)$ in central Au+Au and Pb+Pb 
collisions at RHIC and LHC respectively~\cite{Adil:2006ra}.
The separate measurement of intermediate $p_T$ $D-$ and $B-$meson 
quenching will allow to experimentally determine the correct 
physics mechanism of heavy flavor suppression~\cite{Vitev:2007jj}.

\bigskip

\subsection{Quarkonium shadowing in $p$Pb and Pb+Pb collisions}
{\it R. Vogt}
\vskip 0.5cm

The d+Au data from RHIC, including the $pA$ results from the
fixed-target CERN SPS $pA$ data, suggest increased importance of initial-state
shadowing and decreasing nuclear absorption with increasing energy 
\cite{LLVqm06}.  This is not surprising since smaller $x$ is
probed at higher energy while absorption due to multiple scattering is
predicted to decrease with energy \cite{Braun:1997qw}.  The CERN SPS data
suggest a $J/\psi$ absorption cross section of about 4 mb without 
shadowing, and a larger absorption cross section if it is included since
the SPS $x$ range is in the antishadowing region.  The d+Au RHIC 
data support smaller absorption, $\sigma_{\rm abs}^{J/\psi} \sim 0-2$ mb.  
Thus our predictions for $J/\psi$
and $\Upsilon$ production in $p$Pb and Pb+Pb interactions at the LHC are 
shown for initial-state shadowing alone with no absorption or dense matter
effects.  We note that
including absorption would only move the calculated ratios down in proportion
to the absorption survival probability since, at LHC energies, any rapidity
dependence of absorption is at very large $|y|$ \cite{Vogt:2004dh}, outside 
the detector acceptance.

We present $R_{p{\rm Pb}}(y) = p{\rm Pb}/pp$ and $R_{\rm PbPb}(y) = 
{\rm PbPb}/pp$ for $J/\psi$ and $\Upsilon$.
Since the $pp$, $p$Pb and Pb+Pb data are likely to be taken at different
energies (14 TeV, 8.8 TeV and 5.5 TeV respectively), to make the calculations 
as realistic as possible we show several different scenarios for $R_{p{\rm 
Pb}}(y)$ and $R_{\rm PbPb}(y)$.  The lead nucleus is assumed to
come from the right in $p$Pb.  
All the $pA$ calculations employ the EKS98
shadowing parameterization \cite{Eskola:1998iy,Eskola:1998df}. 
The difference in the $J/\psi$ and $\Upsilon$ results is primarily due to the
larger $\Upsilon$ mass which increases the $x$ values by 
about a factor of three.  In addition, the higher $Q^2$ 
reduces the overall shadowing effect.

\begin{figure}[htbp]
\setlength{\epsfxsize=0.6\textwidth}
\setlength{\epsfysize=0.3\textheight}
\centerline{\epsffile{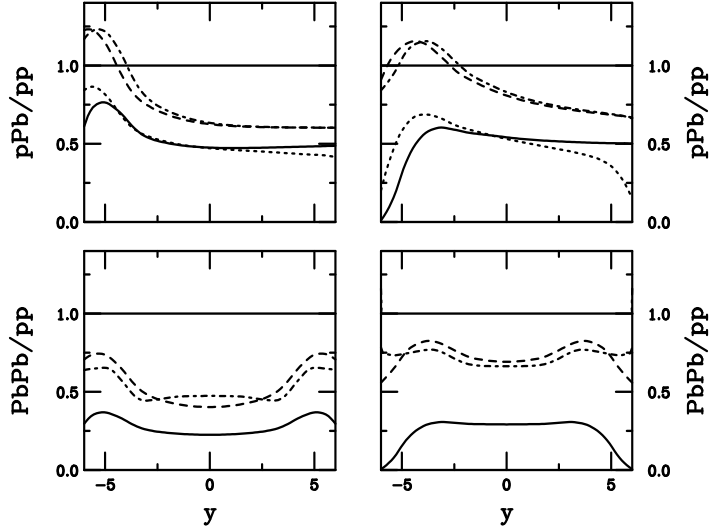}}
\caption[]{The $J/\psi$ (left) and $\Upsilon$ (right) $p$Pb/$pp$ (top)
and PbPb/$pp$ (bottom) ratios as a function of rapidity.  The $p$Pb/$pp$
ratios are given for 8.8 (dashed) and 5.5 (dot-dashed) TeV collisions in 
both cases and 8.8 TeV $p$Pb to 14 TeV $pp$ without (dotted) and with (solid)
the beam rapidity shift taken into account.  The Pb beam comes from the right.
The PbPb/$pp$ ratios are shown for 5.5 TeV in both cases with EKS98 (dashed)
and nDSg (dot-dashed) shadowing and also for 5.5 TeV Pb+Pb and 14 TeV $pp$
(solid).
}
\label{rparaa}
\end{figure}

The top of Fig.~\ref{rparaa} shows $R_{p {\rm Pb}}(y)$ for 
$p$Pb/$pp$ with both systems at $\sqrt{S_{NN}}= 8.8$ and 5.5 TeV
(dashed and dot-dashed curves respectively),
ignoring the $\Delta y = 0.46$ rapidity shift at 8.8 TeV.  For the
$J/\psi$, these ratios are relatively flat at forward rapidity where the
$x$ in the lead is small.  The larger $x$ and greater $Q^2$ for the
$\Upsilon$ brings the onset of antishadowing closer to midrapidity,
within the range of the ALICE dimuon spectrometer. 
At far backward rapidity, a rise due to
the antishadowing region is seen.  The
lower energy moves the antishadowing peak to the right for both
quarkonia states.  We show $R_{p{\rm Pb}}(y)$ with $p$Pb at 8.8 TeV and 
$pp$ at 14 TeV with $\Delta y = 0$ in the dotted curves.
The effect on the $J/\psi$ is an apparent lowering of the dashed curve.
Since the $\Upsilon$ rapidity distribution is narrower at 8.8 TeV than at
14 TeV in the rapidity range shown here, the $\Upsilon$ curve turns over at 
large $|y|$.  (This effect occurs at $|y| > 6$ for the $J/\psi$.)  
The solid curves show $R_{p{\rm Pb}}(y)$ for 8.8 TeV $p$Pb and 
14 TeV $pp$ with the rapidity shift.
Both the $J/\psi$ and $\Upsilon$ ratios are essentially constant for
$y > -2.5$.  Thus relying on ratios of $pA$ to $pp$ collisions at different
energies to study shadowing (or other small $x$ effects) may be difficult
because the shadowing function is hard to unfold when accounting for the 
$pA$ $\Delta y$ as well as the difference in
$x$.  If d+Pb collisions were used, $\Delta y$ would be
significantly reduced \cite{Klein:2003dj}.

The lower part of Fig.~\ref{rparaa} shows $R_{\rm PbPb}(y)$ for the $J/\psi$
and $\Upsilon$ at 5.5 TeV for both systems.  No additional dense matter effects
such as $Q \overline Q$ coalescence or plasma screening are included.
The EKS98 (dashed) and nDSg \cite{deFlorian:2003qf} 
(dot-dashed) shadowing parameterizations are compared.  The 
results are very similar over the entire rapidity range.  (Other shadowing 
parameterizations,which do not agree with the RHIC d+Au data, 
give different $R_{\rm PbPb}(y)$.)  There are antishadowing peaks at 
far forward and backward rapidity.  As at RHIC, including 
shadowing on both nuclei lowers the overall ratio relative to 
$R_{p {\rm Pb}}(y)$ as well as making $R_{\rm PbPb}(y>2)$ similar to or larger
than $R_{\rm PbPb}(y=0)$ because, without any other effects,
$R_{\rm PbPb}(y) \sim R_{p{\rm Pb}}(y) R_{p{\rm Pb}}(-y)$ when all systems
are compared at the same $\sqrt{S_{NN}}$.  The solid
curves show the ratios for Pb+Pb at 5.5 TeV relative to $pp$ at 14 TeV with
the EKS98 parameterization.  The
trends are similar but the magnitude is lower.  

Since these calculations reflect what should be seen if nothing
else occurs, $R_{\rm PbPb}(y)$ is expected to differ 
significantly due to dense
matter effects.  If the initial $J/\psi$ production is strongly suppressed
by plasma screening, then the only observed $J/\psi$'s would be from $c
\overline c$ coalesence \cite{Thews:2005vj}
or $B$ meson decays.  It should be possible to experimentally
distinguish secondary production from the primordial distributions
by displaced vertex cuts.  Secondary $J/\psi$ production should
have a narrower rapidity distribution and a lower average $p_T$. Both are 
indicated in central Au+Au collisions at $\sqrt{S_{NN}} = 200$ GeV
at RHIC \cite{Adare:2006ns}.  If $J/\psi$ production in central collisions is 
dominated by secondary $J/\psi$'s, peripheral collisions should still
reflect initial-state effects.  Predictions of the centrality dependence of
shadowing on $J/\psi$ production at RHIC agree with the most peripheral
Au+Au data.

Finally, the $J/\psi$ and $\Upsilon$ rapidity distributions
are likely to be inclusive, including feed down from higher quarkonium states.
Initial-state effects should be the same for all members of a quarkonium family
so that these ratios would be the same for direct and inclusive production.

%\myack
%{This work was 
%performed under the auspices of the U.S. Department of Energy by
%University of California, Lawrence Livermore National Laboratory under
%Contract W-7405-Eng-48 and supported in part by the National Science
%Foundation Grant NSF PHY-0555660.}

\subsection{Quarkonium suppression as a function of $p_T$}
{\it R. Vogt}
\vskip 0.5cm

\begin{figure}[htbp]
\setlength{\epsfxsize=0.6\textwidth}
\setlength{\epsfysize=0.3\textheight}
\centerline{\epsffile{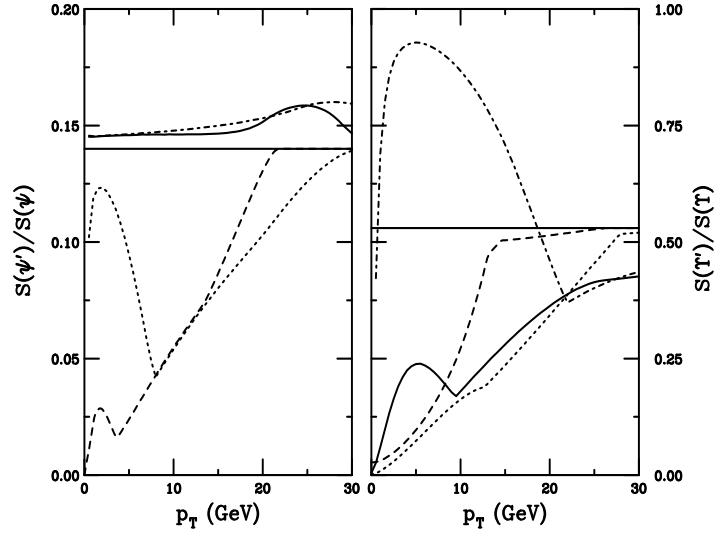}}
\caption[]{The indirect $\psi'/\psi$ (left) and $\Upsilon'/\Upsilon$ (right) 
ratios as a function of $p_T$ in Pb+Pb collisions at 5.5 TeV for $T_0 = 700$
MeV (solid and dashed) and 850 MeV (dot-dashed and dotted).  The $\psi$ 
($\Upsilon$) results are shown for assumed dissociation temperatures of 
$1.1T_c$ ($2.3T_c$) (solid
and dot-dashed) and $2.1T_c$ ($4.1T_c$) (dashed and dotted) respectively.
}
\label{ptrats}
\end{figure}

\begin{figure}[htbp]
\setlength{\epsfxsize=0.95\textwidth}
\setlength{\epsfysize=0.3\textheight}
\centerline{\epsffile{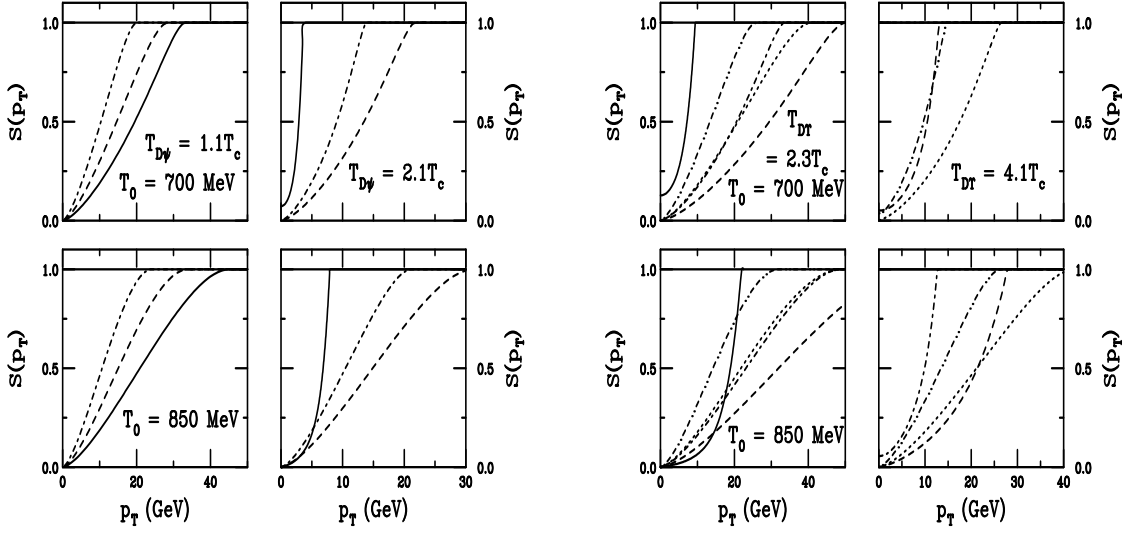}}
\caption[]{The survival probabilities as a function of $p_T$ for the charmonium
(left-hand side) and bottomonium (right-hand side) states for initial 
conditions at the LHC.
The charmonium survival probabilities are $J/\psi$ (solid), $\chi_c$ 
(dot-dashed) and $\psi'$ (dashed) respectively.  The bottomonium survival
probabilities are given for $\Upsilon$ (solid), $\chi_{1b}$ (dot-dashed), 
$\Upsilon'$ (dashed), $\chi_{2b}$ (dot-dot-dash-dashed) and $\Upsilon''$ 
(dotted) respectively.  The top plots are for $T_0 = 700$ MeV while the bottom
are for $T_0 = 850$ MeV.  The left-hand sides of the plots for each state
are for the lower dissociation temperatures, $1.1T_c$ for the $J/\psi$
and $2.3T_c$ for the $\Upsilon$ while the right-hand sides show the results for
the higher dissociation temperatures, $2.1T_c$ for the $J/\psi$ and $4.1T_c$
for the $\Upsilon$.
}
\label{sofpt}
\end{figure}

We present a revised look at the predictions of Ref.~\cite{Gunion:1996qc}, taking
into account newer calculations of the screening mass with temperature and
the quarkonia dissociation temperature based on both potential models and
calculations of quarkonium spectral functions.  The estimates of Digal {\it 
et al.} \cite{Digal:2001ue} predict lower quarkonium dissociation temperatures,
$1.1T_c$ for the $J/\psi$ and $2.3T_c$ for the $\Upsilon$, with $\mu = 1.15T$.
A later review by Satz \cite{Satz:2005hx}, predicts higher values, more in line with 
the recent calculations of quarkonium spectral functions, $2.1T_c$ for the 
$J/\psi$ and $4.1T_c$ for the $\Upsilon$, as well as $\mu \sim 1.45T$ for
$T>1.1T_c$.  We assume $700 < T_0 < 850$ MeV and
$\tau_0 = 0.2$ fm \cite{Vitev:2004bh}.  The $p_T$ dependence of the screening is
calculated as first discussed in Ref.~\cite{Karsch:1987zw}.  Since it may be unlikely
for feed down contributions to be separated from the inclusive $\psi$ and
$\Upsilon$ yields in $AA$ collisions, we present the indirect $\psi'/\psi$ and 
$\Upsilon'/\Upsilon$ ratios, with feed down included, in
Fig.~\ref{ptrats}.  While the individual suppression factors are smooth as a
function of $p_T$, as shown in Fig.~\ref{sofpt} for all four sets of initial
conditions and dissociation temperatures,
due to their different predicted dissociation temperatures
and formation times, they contribute differently to the ratios in 
Fig.~\ref{ptrats}.  

We have assumed that the $\psi'/\psi$ and $\Upsilon'/\Upsilon$ ratios are
independent of $p_T$, as predicted in the color evaporation model 
\cite{Gavai:1994in}.
However, if this is not the case, any slope of the $p_T$ ratios in $pp$
collisions can be calculated and/or evalulated experimentally and deconvoluted
from the data. Quarkonium regeneration by coalescence \cite{Thews:2005vj}
has not been included
here.  While it is unknown how coalescence production populates the quarkonium
levels, since the $p_T$ of quarkonium states produced by coalescence
is lower than those produced in the initial $NN$ collisions, higher $p_T$
quarkonia should have a smaller coalescence contribution.  The lower $B 
\overline B$ rates should reduce the coalescence probability of $\Upsilon$
production.  By taking the $\psi'/\psi$ and $\Upsilon'/\Upsilon$ ratios, we
reduce systematics and initial-state effects.

In the case where $T_D = 1.1T_c$ for the $J/\psi$, its
shorter formation time leads to suppression over a larger $p_T$ range than
that for the $\chi_c$ and $\psi'$, leading to a larger $\psi'/\psi$ ratio
than the $pp$ value over all $p_T$.  On the other hand, for the higher 
dissociation temperature, the $p_T$ range of $J/\psi$ suppression is shorter
than for the other charmonium states, giving a smaller ratio than in $pp$. 
The low $p_T$ behavior of the dashed and dotted curves in the left-hand side of
Fig.~\ref{ptrats} is due to the disappearance of $\chi_c$ suppression since
the $\chi_c$ is suppressed over a shorter $p_T$ range than the $\psi'$.

Since there are more states below the $B \overline B$ threshold for the
$\Upsilon$ family, the suppression is more complicated, in part because there
are also feed down contributions to the $\Upsilon'$, leading to more structure
in the $\Upsilon'/\Upsilon$ ratios on the right-hand side of Fig.~\ref{ptrats}.
For $\mu = 1.15 T$, the $\Upsilon$ itself is suppressed, albeit over a short
$p_T$ range.  The dips in the solid and dashed curves occur at the $p_T$
where direct $\Upsilon$ suppression ceases.  In the case where $T_D = 4.1T_c$
for the $\Upsilon$, the initial temperature is not large enough to suppress
direct $\Upsilon$ production so that $\Upsilon'/\Upsilon < 1$ for all $p_T$.
The $\chi_b$ contributions are responsible for the slopes of the ratios at
$p_T > 12$ GeV.

%\myack
%{This work was 
%performed under the auspices of the U.S. Department of Energy by
%University of California, Lawrence Livermore National Laboratory under
%Contract W-7405-Eng-48 and supported in part by the National Science
%Foundation Grant NSF PHY-0555660.} 

\section{Leptonic probes and photons}
\label{sec:photons}

\subsection{Thermal photons to dileptons ratio at LHC}

%{\it Jajati K. Nayak, Jan-e Alam, Sourav Sarkar and Bikash Sinha}
{\it J.~K. Nayak, J.~Alam, S.~Sarkar and B.~Sinha}
\vskip 0.5cm

Photons and dileptons are  considered to be efficient
probes of quark gluon plasma (QGP) expected to be created  
in heavy ion collisions at ultra-relativistic
energies.  However, the theoretical calculations of
the transverse momentum ($p_T$) spectra of photons 
($d^2N_\gamma/d^2p_Tdy_{y=0}$)
and dileptons ($d^2N_{\gamma^\ast}/d^2p_Tdy_{y=0}$)
depend on several parameters which are model dependent 
(see~\cite{Nayak:2007xv,Alam:2005za} and 
references therein). 
In the present work it is shown that the 
model dependences  involved in 
individual photon and dilepton 
spectra are canceled out in the ratio, $R_{em}$ defined as:
$R_{em} = (d^2N_\gamma/d^2p_Tdy)_{y=0}/(d^2N_{\gamma^\ast}/d^2p_Tdy)_{y=0}$.

The invariant yield of thermal photons can be written as 
${d^2N_\gamma}/{d^2p_{T}dy}=\sum_{i=Q,M,H}{\int_{i}{\left({d^2R_\gamma}/
{d^2p_{T}dy}\right)_i d^4x}}$,  
where $Q, M$ and $H$ represent QGP, mixed (coexisting
phase of QGP and hadrons)
and hadronic phases respectively.
$(d^2R/d^2p_{T}dy)_i$ is the static rate of photon
production from the phase $i$, which is convoluted over
the expansion dynamics
through the integration over $d^4x$.
The thermal photon rate from QGP up to 
$O(\alpha \alpha_{s})$ have been considered.  For photons from 
hadronic matter an exhaustive set of reactions
(including those involving strange 
mesons) and radiative decays of higher resonance states 
have been considered in which form factor effects have been
included.

Similar to photons, the $p_T$ distribution 
of thermal dileptons is given by,  
${d^2N_{\gamma^\ast}}/{d^2p_{T}dy}=\sum_{i=Q,M,H}{\int_{i}
{\left({d^2R_{\gamma^\ast}}/{d^2p_{T}dydM^2}\right)_i dM^2d^4x.}}$
The limits for the integration over $M$ are fixed  
from experimental measurements. Here  we consider
$2m_{\pi} < M $ $ <1.05$ GeV. 
Thermal dilepton rate from QGP 
up to $O(\alpha^2 \alpha_{s})$ has been considered. 
For the hadronic phase we include the dileptons 
from the decays of light vector mesons
~\cite{Nayak:2007xv}. The space time evolution 
of the system has been studied using 
(2+1) dimensional  relativistic hydrodynamics 
with longitudinal boost invariance and cylindrical 
symmetry. 
The calculations have been performed for the initial conditions
mentioned in table \ref{tabsinha} (see also~\cite{Nayak:2007xv}). 
The values of parameters shown in table \ref{tabsinha} reproduce
the various experimental data from SPS and RHIC. 
For LHC we have chosen two values of $T_i$
corresponding to two values of $dN/dy$.
We use the Bag   model EOS
for the QGP phase. For  EOS of the hadronic matter
all resonances with mass $\leq 2.5$ GeV have been considered

%%%%%%%%%%%%%%%%%%%%% FIGURE %%%%%%%%%%%%%%%%%%%%%%%%%%%%%%%%
\begin{figure}
%\hspace{-1.0cm}
%\vspace{-5.0mm}
\begin{flushleft}
%\vspace{-1.4cm}
\includegraphics[width=8.0cm]{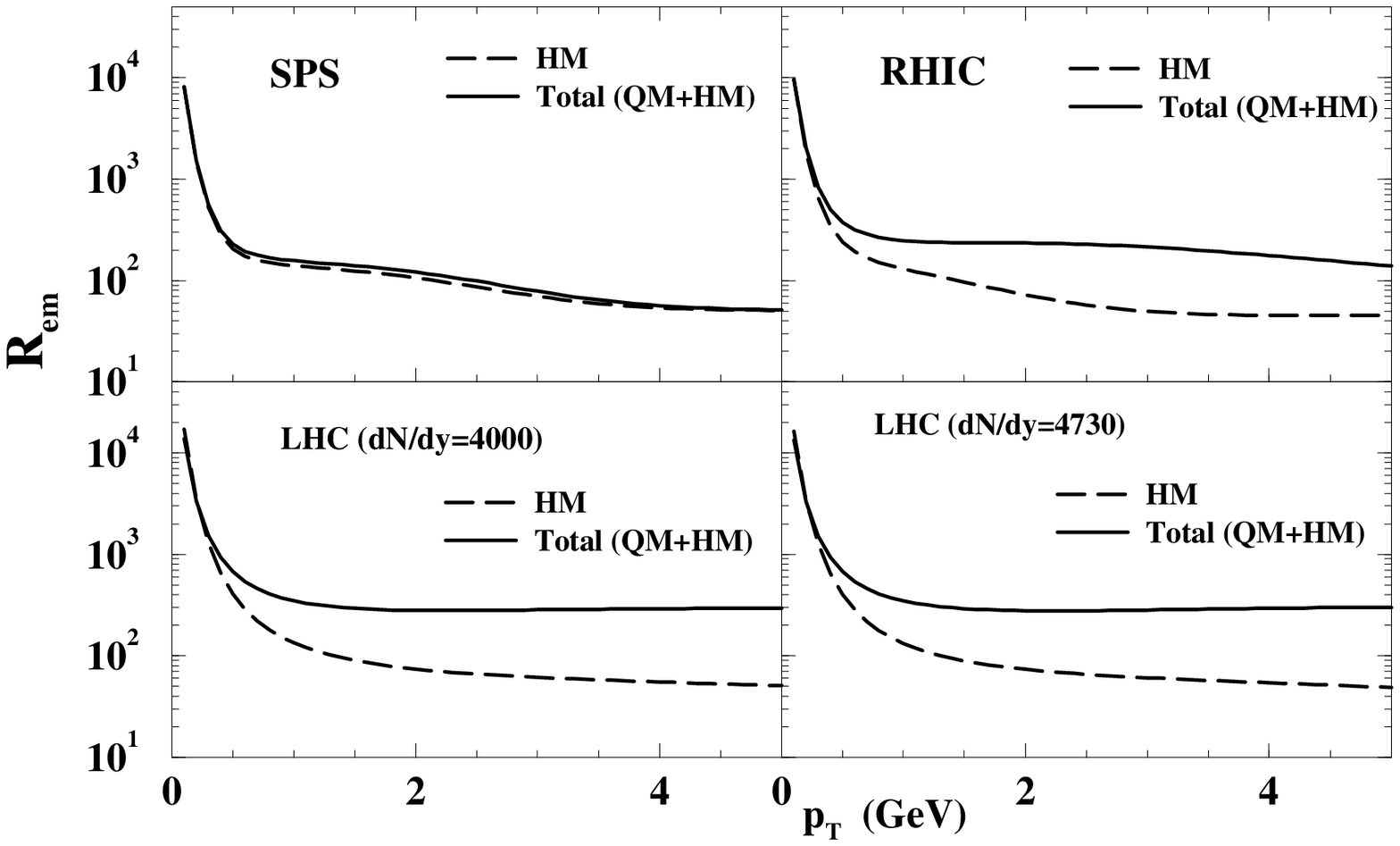}
%\label{fig1}
%\end{figure}
%\begin{figure}
\end{flushleft}
\begin{flushright}
\vspace{-5.5cm}
\includegraphics[width=6.3cm]{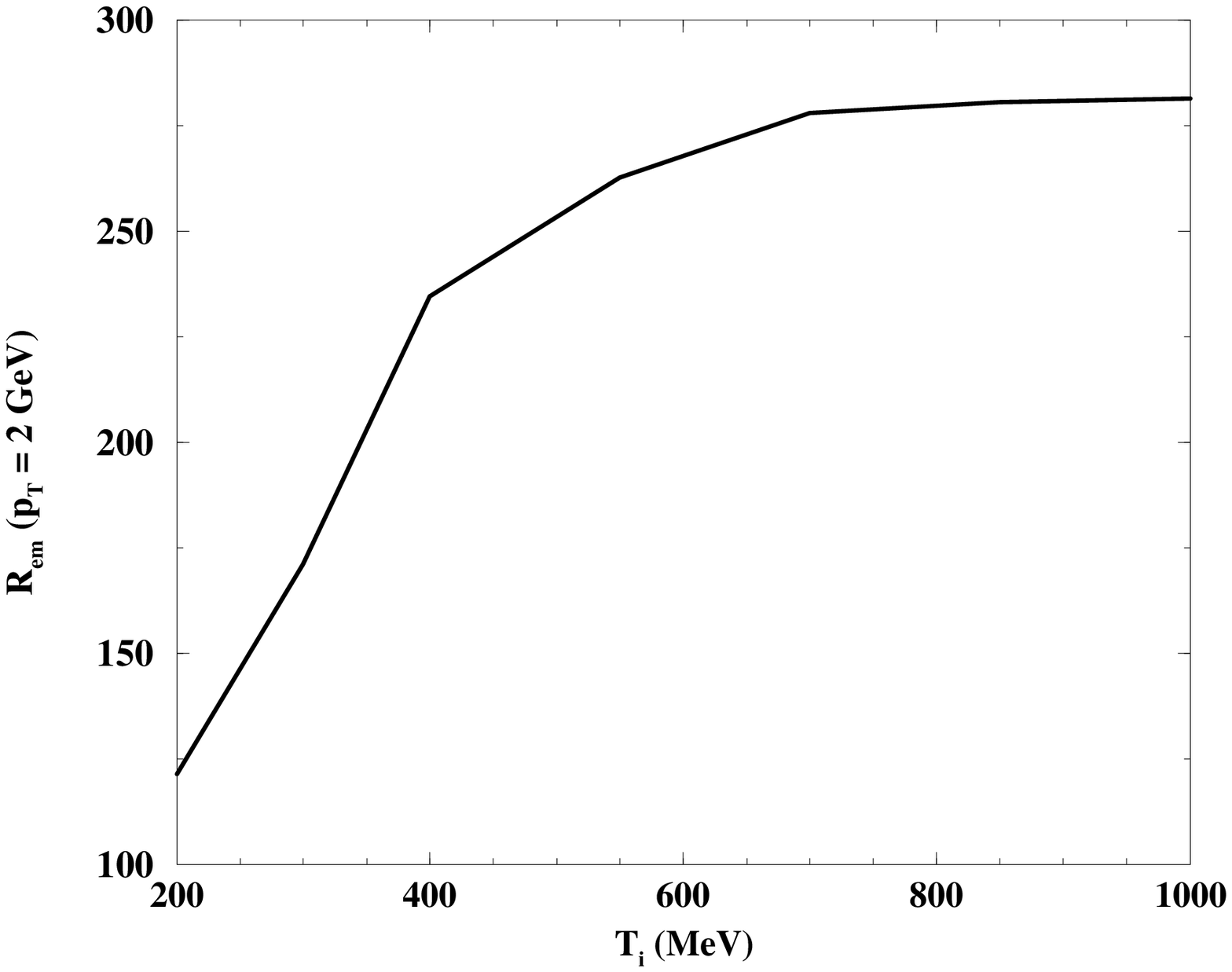}
%\vspace{-1.2cm}
\end{flushright}
\hspace{-12.0cm}{\caption{Left panel: Variation of $R_{em}$ with $p_T$,
right panel: variation of $R_{em} (p_T=2$GeV) with $T_i$.}
\label{fig1-sinha}}
%\vspace{+1.0cm}
%\hspace{-2.0cm}label{fig1}
%\hspace{12.0cm}{fig2}
\end{figure}
%%%%%%%%%%%%%%%%%%%%% FIGURE %%%%%%%%%%%%%%%%%%%%%%%%%%%%%%%%
%%%%%%%%%%%%%%%%%%%%%%%%%%%%%%%%%%%%%%%%%%%%%%%%%%%%%%%%%%%%%%%%%%%%
\begin{table}
\caption{The values of various parameters - thermalization
time ($\tau_i$), initial temperature ($T_i$), freeze-out temperature
($T_f$) and hadronic multiplicity $dN/dy$  - used
in the present calculations.}
\label{tabsinha}
\begin{tabular}{lcccr}
%\tableline
Accelerator&$\frac{dN}{dy}$&$\tau_i(fm)$&$T_i(GeV)$ &$T_f (MeV)$\\
%\tableline
SPS&700&1&0.2&120\\
RHIC&1100&0.2&0.4&120\\
LHC&4000&0.08&0.85&120\\
LHC&4730&0.08&0.905&120\\
%\tableline
\end{tabular}
\end{table}
%$R_{em}$ for SPS, RHIC and LHC energies have been calculated
%assuming a first order phase transition with transition temperature
%$T_{c}$=192 MeV~\cite{Nayak:2007xv}. 
The variation of $R_{em}$ with $p_T$ for different
initial conditions are  
depicted in Fig.~\ref{fig1-sinha} (left panel). 
At SPS, the contributions from hadronic matter (HM) coincides with
the total and hence it becomes difficult to make any
conclusion about the formation of QGP. However,
for RHIC and LHC the  contributions from HM are
less than the total indicating large contributions from
quark matter. The quantity, $R_{em}$, reaches a plateau 
beyond $p_T=1$ GeV
for all the three cases {\it i.e.} for SPS, RHIC and LHC.
However, it is very important to note that the values
of $R_{em}$ at the plateau region are different,
{\it e.g.} $R_{em}^{LHC}\,>\,R_{em}^{RHIC}\,>\, R_{em}^{SPS}$.
Now for all the three cases, SPS, RHIC and LHC, except
$T_i$  all other quantities {\it e.g.} $T_c$, $v_0$, $T_f$ and EOS
are same, indicating that  the difference in the value of $R_{em}$ 
in the plateau
region originates only due to different values of $T_i$ for the 
three cases (Fig. \ref{fig1-sinha}, right panel).
This, hence can be used as a measure of $T_i$.

We have observed that although the individual $p_T$ distribution 
of photons and lepton pairs are sensitive to different
EOS (lattice QCD, for example) the ratio $R_{em}$ is not.
It is also noticed that $R_{em}$ in the plateau 
region is not sensitive to the medium effects on hadrons, 
radial flow, $T_c$, $T_f$ and  other parameters.

It is interesting to note that the nature of variation of the
 quantity, $R_{em}^{pQCD}$, which is the corresponding ratio of
photons and lepton pairs from  hard processes only is
quite different from  $R_{em}^{thermal}$ for $p_T$ up to $\sim 3 $ GeV
indicating that the observed saturation is a thermal effect.

\subsection{Prompt photon in heavy ion collisions at the LHC: A ``multi-purpose'' observable}
\label{arleo}

{\it F. Arleo}

{\small
I emphasize in this contribution how prompt photons can be used to probe nuclear parton densities as well as medium-modified fragmentation functions in heavy ion collisions. Various predictions in $p$--A and A--A collisions at LHC energies are given.
}
\vskip 0.5cm

\begin{figure}[h]
%  \hspace{-1.2cm}
  \begin{minipage}[t]{7.0cm}
    \begin{center}
      \includegraphics[height=6.9cm]{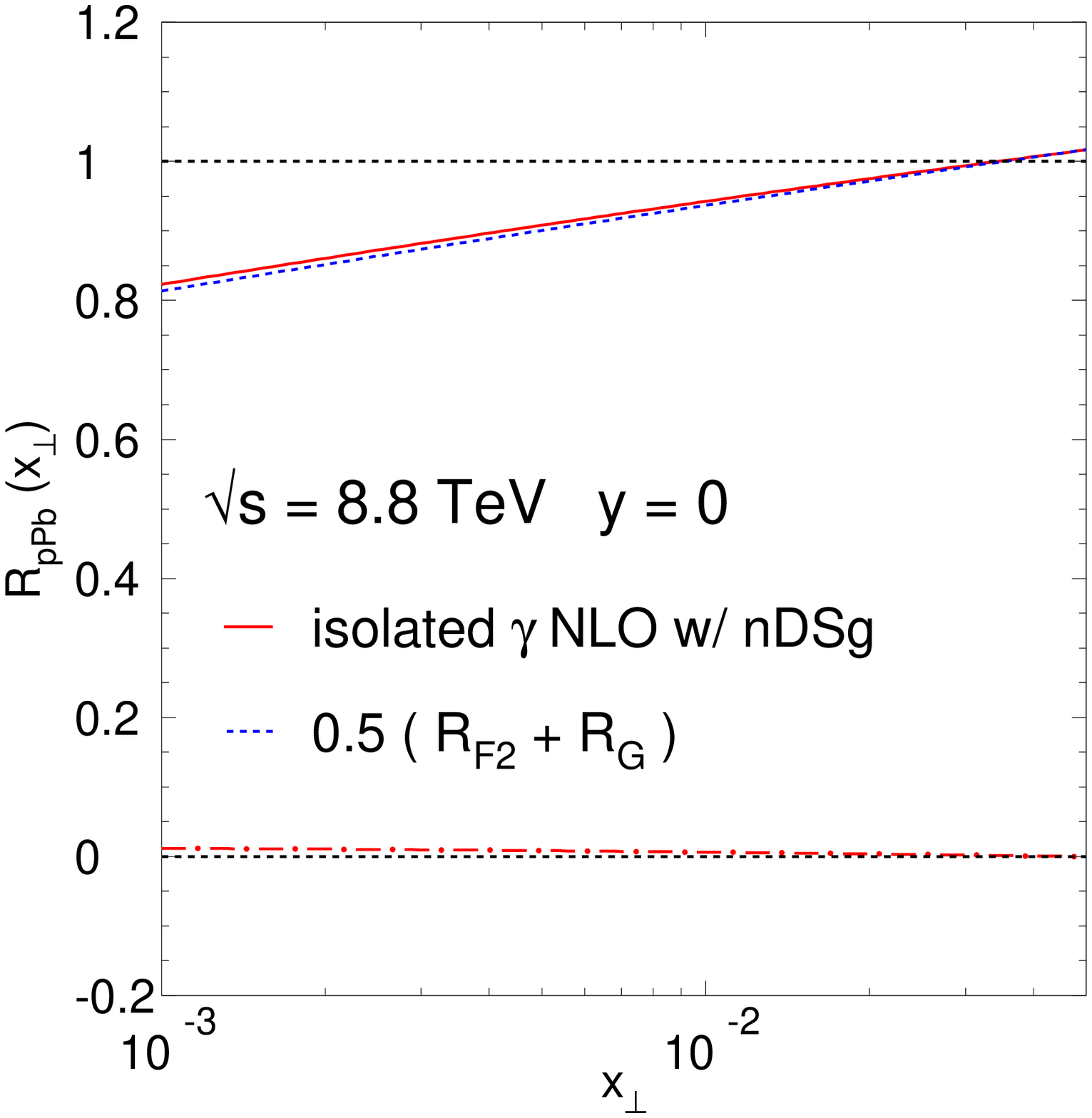}
    \end{center}
  \vspace{-0.5cm}
    \caption{$\rppb$ of $y=0$ isolated photons in $p$--Pb collisions at $\sqrtsnn=8.8$~TeV.}
    \label{fig:isoy0lhc}
  \end{minipage}
\hfill
%  \hspace{-1.8cm}
  \begin{minipage}[t]{7.0cm}
    \begin{center}
      \includegraphics[height=6.9cm]{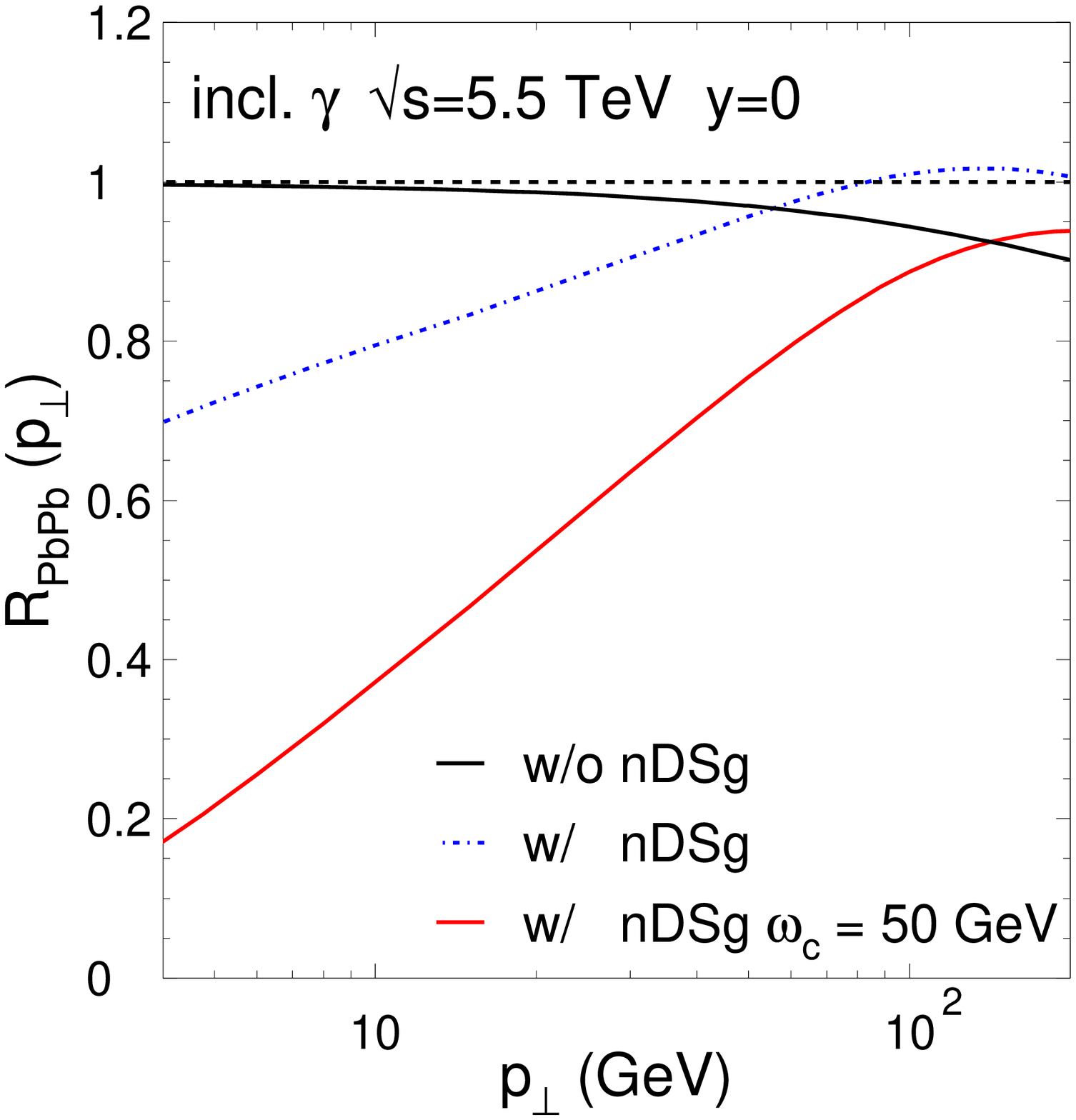}
    \end{center}
  \vspace{-0.5cm}
    \caption{$\rpbpb$ of $y=0$ inclusive photons in Pb--Pb collisions at $\sqrtsnn=5.5$~TeV.}
    \label{fig:incy00lhc}
  \end{minipage}
\end{figure}

Prompt photon production in hadronic collisions has been extensively studied, both experimentally and theoretically, over the past 25 years (see~\cite{Aurenche:2006vj} and references therein). As indicated in Ref.~\cite{Aurenche:2006vj}, it is remarkable that almost all existing data from fixed-target to collider energies can be very well understood within perturbative QCD at NLO. In these proceedings, I briefly discuss how prompt photons in nuclear collisions ($p$--A and A--A) may allow for a better understanding of interesting aspects discussed in heavy-ion collisions, namely the physics of nuclear parton distribution functions and medium-modified fragmentation functions. Parton distribution functions in nuclei are so far poorly constrained, especially in contrast with the high degree of accuracy currently reached in the proton channel, over a wide $x$ and $Q^2$ domain. In particular, only high-$x$ ($x\gtrsim 10^{-2}$) and low $Q^2$ ($Q^2\lesssim 100$~GeV$^2$) have been probed in fixed-target experiments. In order to predict hard processes in nuclear collisions at the LHC, a more accurate knowledge on a wider kinematic range is necessary. As stressed in~\cite{Arleo:2007pc}, the nuclear production ratio of isolated photons in $p$-A collisions,
\begin{equation*}
\label{eq:ratio}
  \rpa(\xt, y) = \frac{1}{A} \ \
  \frac{\dd^3\sigma}{\dd{y}\ \dd^2\ptrans}(p+\A\to\gamma+\X)
  \Big/ \frac{\dd^3\sigma}{\dd{y}\ \dd^2\ptrans}(p+p\to\gamma+\X)
\end{equation*} 
can be related to a good accuracy (say, less than 5\%) to the parton density ratios
\begin{equation*}
  \label{eq:rpa_approx}
\hspace{-2.cm}\rapprox(\xt,y=0) \simeq 0.5\ \rafd(\xt)+ 0.5\ \rag(\xt)\ ;\ \rapprox(\xt,y=3) \simeq \rag(\xt e^{-y}),
\end{equation*}
with $\xt=2\ptrans/\sqrtsnn$. To illustrate this, the ratio $\rpa$ is computed for isolated photons produced at mid-rapidity in $p$--Pb collisions at $\sqrtsnn=8.8$~TeV in Fig.~\ref{fig:isoy0lhc} (solid line), assuming the de Florian and Sassot (nDSg) nuclear parton distributions~\cite{deFlorian:2003qf}. The above analytic approximation $\rapprox_{y=0}$ (dotted line) demonstrates how well this observable is connected to the nuclear modifications of the gluon density and structure function; see also the agreement $(\rpa-\rapprox_{y=0})/\rpa$ as a dash-dotted line in Fig.~\ref{fig:isoy0lhc}. In nucleus-nucleus scattering, the energy loss of hard quarks and gluons in the dense medium presumably produced at LHC may lead to the suppression of prompt photons coming from the collinear fragmentation process. In Fig.~\ref{fig:incy00lhc}, the expected photon quenching in Pb--Pb collisions at $\sqrtsnn=5.5$~TeV is plotted. A significant suppression due to energy loss (taking $\omega_c=50$~GeV, see~\cite{Arleo:2007bg} for details) is observed, unlike what is expected when only nuclear effects in the parton densities are assumed in the calculation (dash-dotted line).

%%%%%%%%%%%%%%%%%%%%%%%%%%%%%%%%%%%%%%%%%%%%%%%%%%%%%%%%%%%%%%%%%%

%%%%%%%%%%%%%%%%%%%%%%%%%%%%%%%%%%%%%%%%%%%%%%%%%%%%%%%%%%%%%%%%%%
Finally performing momentum correlations between a prompt photon and a leading hadron in $p$--$p$ and A--A collisions, yet experimentally challenging, appears to be an interesting probe of vacuum and medium-modified fragmentation function, as discussed in detail in Refs.~\cite{Arleo:2004xj,Arleo:2006xb}. We refer in particular the interested reader to Fig.~10 of~\cite{Arleo:2004xj} for the predictions of $\gamma$--$\pi^0$ momentum-imbalance distributions at the LHC.

\subsection{Direct photon spectra in Pb-Pb at $\sqrtsnn$ = 5.5 TeV: hydrodynamics+pQCD predictions}
 
{\it F. Arleo, D. d'Enterria and D. Peressounko}

{\small
  The $p_T$-differential spectra for direct photons produced in Pb-Pb collisions at the LHC, 
  including thermal (hydrodynamics) and prompt (pQCD) emissions are presented.
}
\vskip 0.5cm

We present predictions for the transverse momentum distributions of direct-$\gamma$ (i.e. photons 
not coming from hadron decays) produced at mid-rapidity in Pb-Pb collisions at $\sqrtsnn$ = 5.5 TeV based 
on a combined hydrodynamics+pQCD approach. Thermal photon emission in Pb-Pb at the LHC is computed 
with a hydrodynamical model successfully used in nucleus-nucleus collisions at RHIC energies~\cite{d'Enterria:2005vz}. 
The initial entropy density of the produced system at LHC is obtained by extrapolating empirically 
the hadron multiplicities measured at RHIC~\cite{dde_peress_arleo}. Above 
$p_T\approx$~3~GeV/$c$, additional prompt-$\gamma$ production from parton-parton scatterings 
is computed perturbatively at next-to-leading-order (NLO) accuracy~\cite{Aurenche:1998gv}.
We use recent parton distribution functions (PDF)~\cite{Pumplin:2002vw} and parton-to-photon 
fragmentation functions (FF)~\cite{Bourhis:2000gs}, modified resp. to account for initial-state 
shadowing+isospin effects~\cite{deFlorian:2003qf} and final-state parton energy loss~\cite{Arleo:2007}.

%We use the same cylindrically-symmetric boost-invariant 2+1-D relativistic hydrodynamics ofour previous works. 
We follow the evolution of the hot and dense system produced in central Pb-Pb at LHC
by solving the equations of (ideal) relativistic 2D+1 hydrodynamics~\cite{d'Enterria:2005vz,dde_peress_arleo} 
starting at a time $\tau_0 = 1/Q_s\approx$ 0.1 fm/$c$. The system is assumed to have an 
initial entropy density of $s_0$~=~1120~fm$^{-3}$, which corresponds to a maximum temperature 
at the center of $T_0\approx$ 650~MeV ($\mean{T_0}\approx$ 470 MeV). 
We use a quark gluon plasma (QGP) and hadron resonance gas (HRG) equation of state
above and below $T_{\rm crit}\approx$ 170 MeV resp., connected by a 
standard Maxwell construction assuming a first-order phase transition at $T_{\rm crit}$. 
%we chemically freeze-out the system (i.e. fix the hadron ratios) at $T_{\rm crit}$.
%explicitly conserving particle numbers by introducing individual (temperature-dependent)
%chemical potentials for each hadron.
Thermal photon emission is computed using the most recent parametrizations 
of the QGP and HRG $\gamma$ rates. For the QGP phase we use the AMY complete leading-log 
emission rates including LPM suppression~\cite{Arnold:2001ms}. For the HRG phase, we employ the
improved parametrization from Turbide {\it et al.}~\cite{turbide}.

Our NLO pQCD predictions are obtained with the code of ref.~\cite{Aurenche:1998gv}
with all %factorization and fragmentation 
scales set to $\mu=p_{T}$. Pb-Pb yields are obtained scaling the NLO cross-sections 
by the number of incoherent nucleon-nucleon collisions: $N_{\rm coll}$ = 1670, 12.9 for 0-10\% central 
($\mean{b}$ = 3.2 fm) and 60-90\% peripheral ($\mean{b}$ = 13 fm). 
Nuclear (isospin and shadowing) corrections of the CTEQ6.5M PDFs~\cite{Pumplin:2002vw} 
are introduced using the NLO nDSg parametrization~\cite{deFlorian:2003qf}. At relatively low 
$p_T$, prompt photon yields have a large contribution from jet fragmentation processes.
%[The typical LO quark-gluon Compton and $q-\bar{q}$ annihilation photon processes are subdominant
%below $p_T\approx$ 20(?) GeV/$c$]. 
As a result, final-state parton energy loss in central Pb-Pb affects also the expected prompt 
$\gamma$ yields. We account for medium-effects on the $\gamma$-fragmentation component 
by modifying the BFG parton-to-photon FFs~\cite{Bourhis:2000gs} with 
BDMPS quenching weights. The effects of the energy loss are
encoded in a single parameter, $\omega_c=\mean{\hat{q}}\,L^2\approx$ 50 GeV,
extrapolated from RHIC. The combination of initial-state (shadowing) and final-state 
(energy loss) effects results in a quenching factor for prompt photons %at the LHC 
of $R_{PbPb}\approx$ 0.2 (0.8) at $p_T$ = 10 (100) GeV/$c$~\cite{Arleo:2007}.

\begin{figure}[htbp]
\begin{centering}
\hspace{6mm}
\includegraphics[width=0.45\textwidth,height=8.cm]{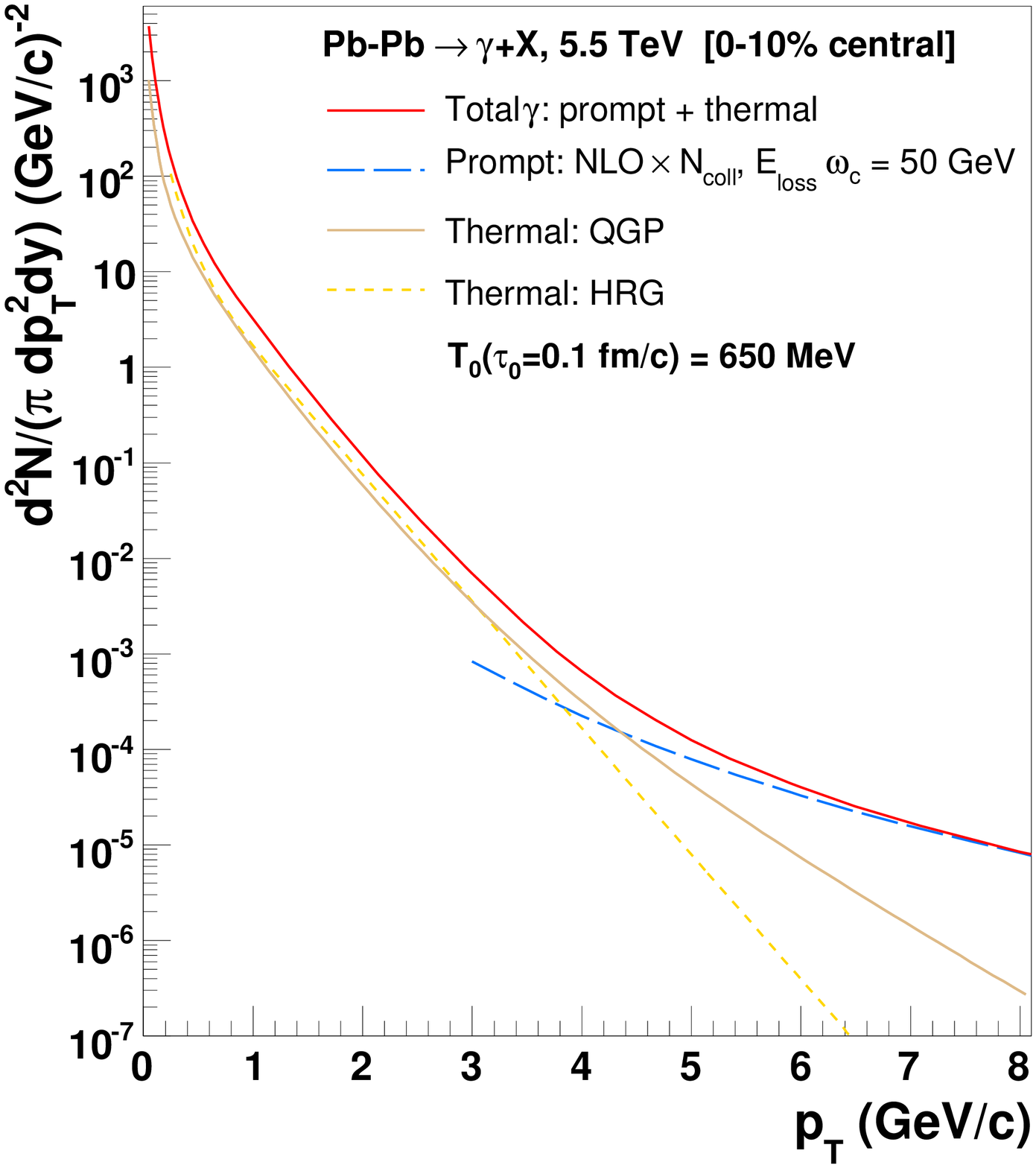}
\hspace{4mm}
\includegraphics[width=0.45\textwidth,height=8.cm]{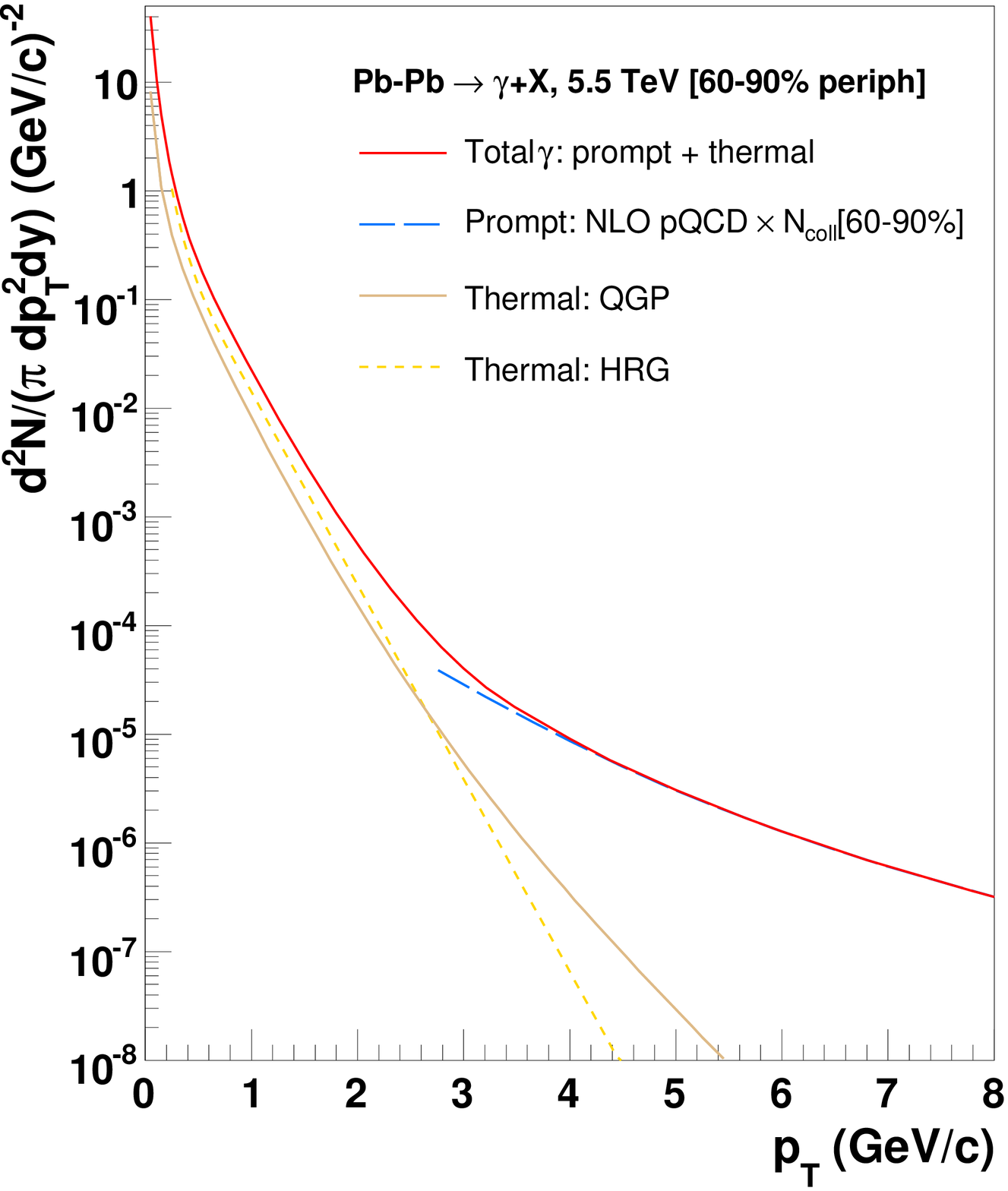}
\end{centering}
\vspace{-5mm}
\caption{Direct-$\gamma$ spectra in 0-10\% central (left) and 60-90\% peripheral (right) Pb-Pb at 
$\sqrtsnn$ = 5.5 TeV, with the thermal (QGP and HRG) and prompt (pQCD) contributions differentiated.}\label{fig:spectraph}
\end{figure}

Our predictions for the direct photon spectra at y=0 in Pb-Pb at 5.5 TeV are 
shown in Fig.~\ref{fig:spectraph}. The thermal contribution dominates over the (quenched) 
pQCD one up to $p_T\approx$ 4 (1.5) GeV/$c$ in central (peripheral) Pb-Pb. Two differences
are worth noting compared to RHIC results~\cite{d'Enterria:2005vz}: 
(i) the thermal-prompt crossing point moves up from $p_T\approx$ 2.5 GeV/$c$ to 
$p_T\approx$~4.5~GeV/$c$, and (ii) most of the thermal production in this transition
region comes solely from the QGP phase. Both characteristics make of semi-hard 
direct photons at LHC, a valuable probe of the thermodynamical properties of the system.

%%%%%%%%%%%%%%%%%%%%%%%%%%%%%%%%%%%%%%%%%%%%%%%%%%%%%%%%%%%%%%%
\vspace{-2mm}

\subsection{Elliptic flow of thermal photons 
       from RHIC to LHC}
\label{heinzflow}
{\it R. Chatterjee,  E. Frodermann, U. Heinz and 
  D.~K.~Srivastava}
\vskip 0.5cm

%\section[Elliptic flow of thermal photons]{Elliptic flow of thermal photons 
%         from RHIC to LHC\footnote[7]{$\!\!\!$Authors: R. Chatterjee,  
%         E. Frodermann, U. Heinz and D.~K.~Srivastava (Ohio State University
%         and VECC Kolkata).\\ 
%         Work supported by U.S. DOE, grant DE-FG02-01ER41190 (UH), and an 
%         Ohio State University University Presidential Fellowship (EF).}}

We use the longitudinally boost-invariant relativistic ideal hydrodynamic
code AZHYDRO \cite{Kolb:2000sdd} to predict the evolution from RHIC to LHC 
of the transverse momentum spectra and elliptic flow of thermal photons 
and dileptons at mid-rapidity in $(A{\approx}200){+}(A{\approx}200)$ 
collisions. Here we discuss only photons for Au+Au collisions at 
$b{\,=\,}7$\,fm; for other results and more details see 
Refs.~\cite{phot_flow}.

%%%%%%%%%%%%%%%%%%%%%%%%%%%%%%% Fig. 1 %%%%%%%%%%%%%%%%%%%%%%%%%%%%%%%
\begin{figure}[hb]
  \begin{center}
  \includegraphics[bb=7 40 710 540,width=\linewidth,clip=]{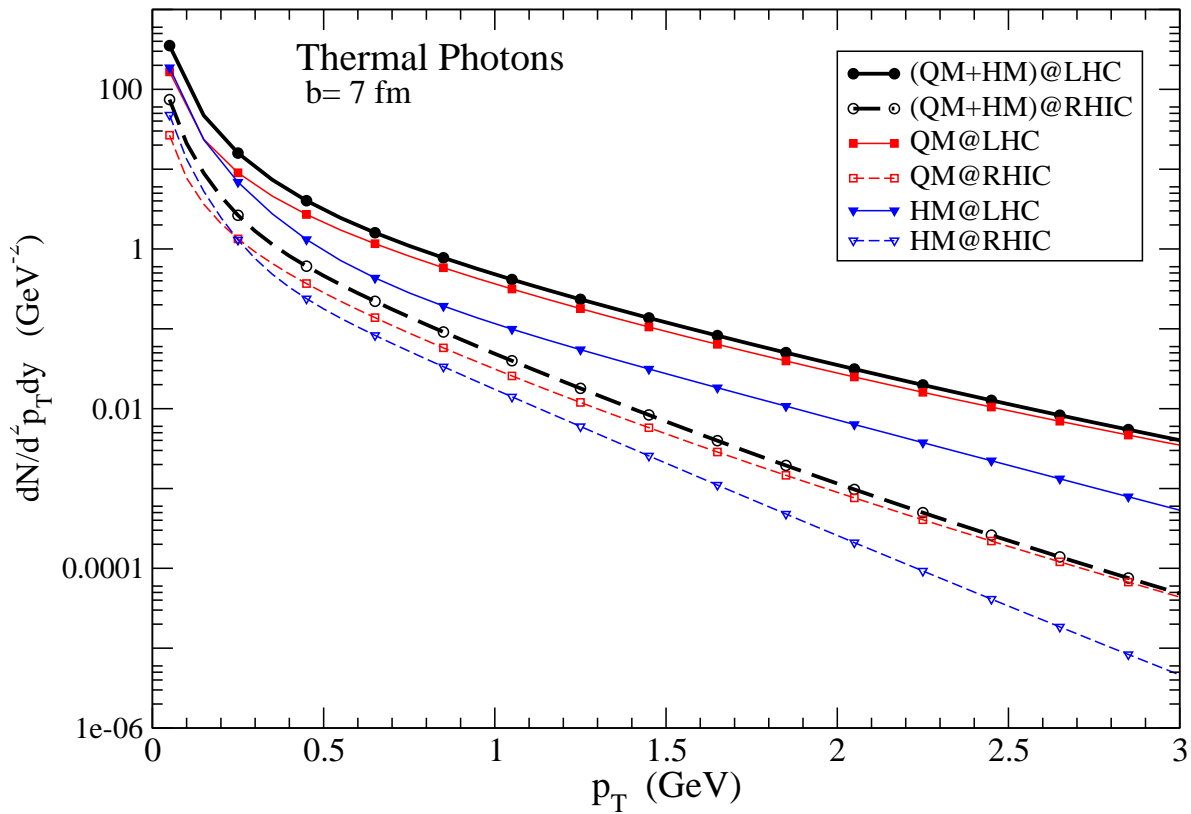}
  \end{center}
   \caption{\label{spectraphot}(Color online)
   Thermal photon spectra Au+Au collisions at RHIC and Pb+Pb collisions
   at LHC, both at $b{\,=\,}7$\,fm.}
\end{figure}
%%%%%%%%%%%%%%%%%%%%%%%%%%%%%%%%%%%%%%%%%%%%%%%%%%%%%%%%%%%%%%%%%%%%%%
%

The hydrodynamic initial conditions for RHIC collisions are described 
in \cite{phot_flow}. For the LHC simulations shown in comparison we 
assume a final charged hadron multiplicity near the upper end of the 
predicted range: $\frac{dN_{\mathrm{ch}}}{dy}(b{=}y{=}0){\,=\,}2350$ 
(680 at RHIC). Correspondingly we increase the initial peak entropy 
density in central Au+Au collisions from $s_0{\,=\,}351$\,fm$^{-3}$ at 
$\tau_0{\,=\,}0.2$\,fm/$c$ for RHIC to $s_0{\,=\,}2438$\,fm$^{-3}$ at 
$\tau_0=0.1$\,fm/$c$ for LHC. 

%%%%%%%%%%%%%%%%%%%%%%%%%%%%%%%%%%%%%%%%%%%%%%%%%%%%%%%%%%%%%%%%%%%%%%%%
\noindent\textbf{\emph{1.~Thermal photon spectra:}}
%%%%%%%%%%%%%%%%%%%%%%%%%%%%%%%%%%%%%%%%%%%%%%%%%%%%%%%%%%%%%%%%%%%%%%%%
%
\Fref{spectraphot} shows the thermal photon $p_T$-spectra (angle-integrated) 
for RHIC and LHC. At both collision energies the total spectrum is 
dominated by quark matter once $p_T$ exceeds a few hundred MeV. Its 
inverse slope (``effective temperature'') in the range 
$1.5{\,<\,}p_T{\,<\,}3$\,GeV/$c$ increases by almost 50\%, from $303$\,MeV 
at RHIC to $442$\,MeV at LHC, reflecting the higher initial temperature 
and significantly increased radial flow (visible in the HM contribution) 
at LHC.\\[-3mm]

%%%%%%%%%%%%%%%%%%%%%%%%%%%%%%%%%%%%%%%%%%%%%%%%%%%%%%%%%%%%%%%%%%%%%%%%
\noindent\textbf{\emph{2.~Thermal photon elliptic flow:}}
%%%%%%%%%%%%%%%%%%%%%%%%%%%%%%%%%%%%%%%%%%%%%%%%%%%%%%%%%%%%%%%%%%%%%%%%
%
%%%%%%%%%%%%%%%%%%%%%%%%% Fig. 2 %%%%%%%%%%%%%%%%%%%%%%%%%%%%%%%%%%%%%%%%%
\begin{figure}[t]
  \begin{center}
  \includegraphics[bb=15 40 710 528,width=0.95\linewidth,clip=]{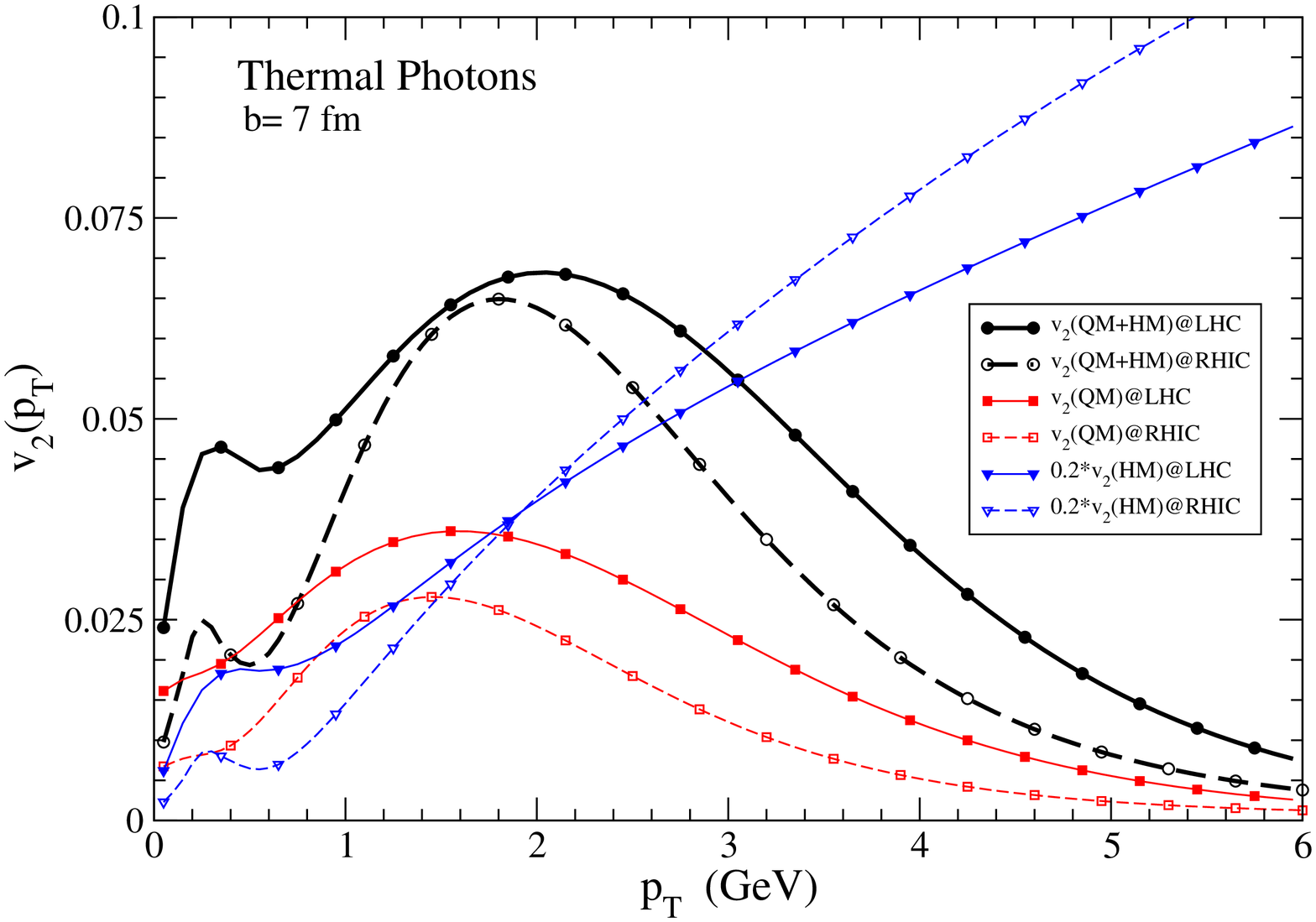}
  \end{center}
  \caption{\label{oscillationRHIC}(Color online)
  Thermal photon elliptic flow for Au+Au collisions at RHIC (dashed) and 
  Pb+Pb at LHC (solid lines), both at $b{\,=\,}$7\,fm. 
  } 
 \end{figure}
%%%%%%%%%%%%%%%%%%%%%%%%%%%%%%%%%%%%%%%%%%%%%%%%%%%%%%%%%%%%%%%%%%%%%%%%%%%
%
\Fref{oscillationRHIC} shows the differential elliptic flow of thermal
photons at RHIC and LHC, with quark matter (QM) and hadronic matter (HM)
radiation shown separately for comparison. The decrease at high $p_T$ of 
the QM and total photon $v_2$ reflects the dominance of QM radiation at 
high $p_T$ (emission from the early, hot stage when radial and elliptic 
flow are still small). At fixed $p_T$, the photon elliptic flow from QM 
radiation is larger at LHC than at RHIC since the LHC fireballs start hotter 
and fluid cells with a given temperature thus flow more rapidly. At low 
$p_T$, hadronic radiation dominates, and since it flows more rapidly
at LHC than at RHIC the corresponding photon elliptic is significantly 
larger at LHC than RHIC. This is different from hadrons whose elliptic
flow at low $p_T$ {\em decreases} from RHIC to LHC, reflecting a 
redistribution of the momentum anisotropy to higher $p_T$ by increased
radial flow \cite{Kestin}. For photons, the elliptic flow is not yet 
saturated at RHIC, and at low $p_T$ it keeps increasing towards LHC at 
a rate that overwhelms the loss of momentum anisotropy to the high-$p_T$
domain via radial flow. Contrary to pion $v_2$ \cite{Kestin}, the 
$p_T$-integrated photon elliptic flow roughly doubles (!) from RHIC to LHC.

%%%%%%%%%%%%%%%%%%%%%%%%%%%%%%%%%%%%%%%%%%%%%%%%%%%%%%%%%%%%%%%%%%%%%%%%%%

\subsection{Asymmetrical in-medium mesons}
\label{dremin2}

{\it I. M. Dremin}

{\small
Cherenkov gluons may be in charge of mass asymmetry of in-medium mesons
which reveals itself in the asymmetry of dilepton spectra.
}
\vskip 0.5cm

The hypothesis about the nuclear analogue of the well known Cherenkov
effect 
\cite{Dremin:1979yg,Dremin:1980wx,Dremin:2005an,Dremin:2006vm,Koch:2005sx}
is widely discussed now. The necessary condition for
Cherenkov effect in usual or hadronic media is the excess of the
corresponding refractive index $n$ over 1. There exists the general linear
relation between this excess $\Delta n = n - 1$ and the real part of the
forward scattering amplitude $F(E, 0^o)$. In electrodynamics, it is the
dipole excitation of atoms in the medium by light which results in the
Breit-Wigner shape of the photon amplitude. In a nuclear medium, this
should be the amplitude of gluon scattering on some internal modes of the
medium. In absence of the theory of such media I prefer to rely on our
knowledge of hadronic reactions. From experiments at comparatively low
energies we learn that the resonances are abundantly produced. They are
described by the Breit-Wigner amplitudes which have a common feature of the
positive real part in the low-mass wing (for the electrodynamic analogy
see, e.g., Feynman lectures). Therefore the hadronic refractive index
exceeds 1 in these energy regions. 

\noindent
{\bf Prediction} Masses of Cherenkov states
are less than in-vacuum meson masses. This leads to the asymmetry of decay
spectra of resonances with increased role of low masses. 

\noindent
{\bf Proposal} Plot the
mass distribution of $\pi^+\pi^-$, $\mu^+\mu^-$, $e^+e^-$-pairs 
near resonance peaks. Thus,
apart from the ordinary Breit-Wigner shape of the cross section for
resonance production, the dilepton mass spectrum would acquire the
additional term proportional to $\Delta n$
(that is typical for Cherenkov
effects) at masses below the resonance peak \cite{Dremin:2007sn}.
Therefore its excess
(e.g., near the $\rho$-meson) can be described by the following
formula\footnote{%
We consider only $\rho$-mesons here because the most precise experimental
data are available for them. To include other mesons, one should evaluate
the corresponding sum of similar expressions.}
\begin{equation}
{dN_{ll} \over dM}
=
{A\over  (m^2_\rho  - M^2)^2 + M^2\Gamma^2}
\left(1 + w {m^2_\rho - M^2\over M^2} \theta(m_\rho - M )\right)
\label{eq:dremin2_eq1}
\end{equation}
Here $M$ is the total c.m.s. energy of two colliding objects (the dilepton
mass), $m_\rho =775$ MeV is the in-vacuum $\rho$-meson mass. The first term
corresponds to the Breit-Wigner cross section. According to the optical
theorem it is proportional to the imaginary part of the forward scattering
amplitude. The second term is proportional to $\Delta n$
where the well known
ratio of real to imaginary parts of Breit-Wigner amplitudes has been used.
It vanishes for $M > m_\rho$
because only positive $\Delta n$ lead to the Cherenkov
effect. Namely it describes the distribution of masses of Cherenkov states.
In Eq.(\ref{eq:dremin2_eq1}) 
one should take into account the in-medium modification of the
height of the peak and its width. We just fit the parameters $A$ and
$\Gamma$
by describing the shape of the mass spectrum at $0.75 < M < 0.9$ GeV measured
in 
\cite{Arnaldi:2006jq,Damjanovic:2006bd,Damjanovic:2007qm}.
Let us note that $w$ is not used in this procedure. The values
$A=104\,{\rm GeV}^3$ and $\Gamma  = 0.354$ GeV were obtained. The width of the
in-medium peak is larger than the in-vacuum $\rho$-meson width equal to 
$150$ MeV.

Thus the low mass spectrum at $M < m_\rho$
depends only on a single parameter $w$
which is determined by the relative role of Cherenkov effects and ordinary
mechanism of resonance production. It is clearly seen from 
Eq.(\ref{eq:dremin2_eq1}) that the
role of the second term in the brackets increases for smaller masses $M$.
The excess spectrum
\cite{Arnaldi:2006jq,Damjanovic:2006bd,Damjanovic:2007qm}.
in the mass region from 0.4 GeV to 0.75 GeV has
been fitted by $w = 0.19$. The slight downward shift about 40 MeV of the peak
of the distribution compared with $m_\rho$ may be estimated from
Eq.(\ref{eq:dremin2_eq1}) at
these values of the parameters.

Whether the in-medium Cherenkov gluonic effect is strong can be verified by
measuring the angular distribution of the lepton pairs with different
masses. The trigger-jet experiments similar to that at RHIC are necessary to
check this prediction. One should measure the angles between the companion
jet axis and the total momentum of the lepton pair. The Cherenkov pairs
with masses between 0.4 GeV and 0.7 GeV should tend to fill in the rings
around the jet axis. The angular radius $\theta$
of the ring is determined by the usual condition
\begin{equation}
\cos\theta = {1\over n}
\label{eq:dremin2_eq2}
\end{equation}
Another way to demonstrate it is to measure the average
mass of lepton pairs as a function of their polar emission angle
(pseudorapidity) with the companion jet direction chosen as $z$-axis. Some
excess of low-mass pairs may be observed at the angle (\ref{eq:dremin2_eq2}).

The prediction of asymmetric in-medium widening of any resonance at its
low-mass side due to Cherenkov gluons is universal. This universality is
definitely supported by experiment. Very clear signals of the excess on the
low-mass sides of $\rho, \omega$ and $\phi$
mesons have been seen in KEK. This effect
for $\omega$-meson is also studied by CBELSA/TAPScollaboration. There are some
indications at RHIC on this effect for $J/\psi$-meson.

To conclude, the universal asymmetry of in-medium mesons with an excess
over the usual Breit-Wigner form at low masses is predicted as a signature
of Cherenkov gluons produced with energies which fit the left wings of
resonances where $n$ exceeds 1.

\subsection{Photons and Dileptons at LHC}
\label{fries}

{\it R. J. Fries, S. Turbide, C. Gale
and D. K. Srivastava}

{\small
We discuss real and virtual photon sources in heavy ion collisions and 
present results for dilepton yields in Pb+Pb collisions at the LHC
at intermediate and large transverse momentum $p_T$.
}
\vskip 0.5cm

Electromagnetic radiation provides a valuable tool to understand the dynamics
of heavy ion collisions. Due to their long mean free path real and virtual 
photons carry information from very early times and from deep inside the 
fireball. We discuss the sources of photons which will be important for
the upcoming heavy ion experiments at LHC. We focus on intermediate and large 
transverse momenta $p_T$ and masses $M$. We also present our numerical results 
for dilepton yields.

At asymptotically large $p_T$ the most important source of real and virtual 
photons is the direct hard production in primary parton-parton collisions
between the nuclei, via Compton scattering, annihilation, and the 
Drell-Yan processes. These photons do not carry any signature of the
fireball. They are augmented by photons fragmenting from hard jets also
created in primary parton-parton collisions. The emission of this vacuum 
bremsstrahlung is described by real and virtual photon fragmentation 
functions. Vacuum fragmentation is assumed to happen outside the fireball,
so the jets are subject to the full energy loss in the medium. This 
contribution to the photon and dilepton yield is therefore depleted in
heavy ion collisions analogous to the high-$p_T$ hadron yield.

At intermediate scales jet-induced photons from the medium become 
important. It has been shown that high-$p_T$ jets interacting with 
the medium can produce real and virtual photons by one of two processes: 
(i) by Compton scattering or annihilation with a thermal parton, leading 
to an effective conversion of the jet into a photon \cite{Fries:2002kt}; 
(ii) by medium induced Bremsstrahlung \cite{Zakharov:2004bi}. 
Jet-medium photons have a steeper spectrum than primary photons and 
carry information about the temperature of the medium. They are also 
sensitive to the partial energy loss that a jet suffers from its creation 
to the point of emission of the photon.
At even lower $p_T$ and $M$ thermal radiation from the quark gluon plasma
(and also the hadronic phase not considered here) has to be taken into
account.

\begin{figure}
\begin{center}
\epsfig{file=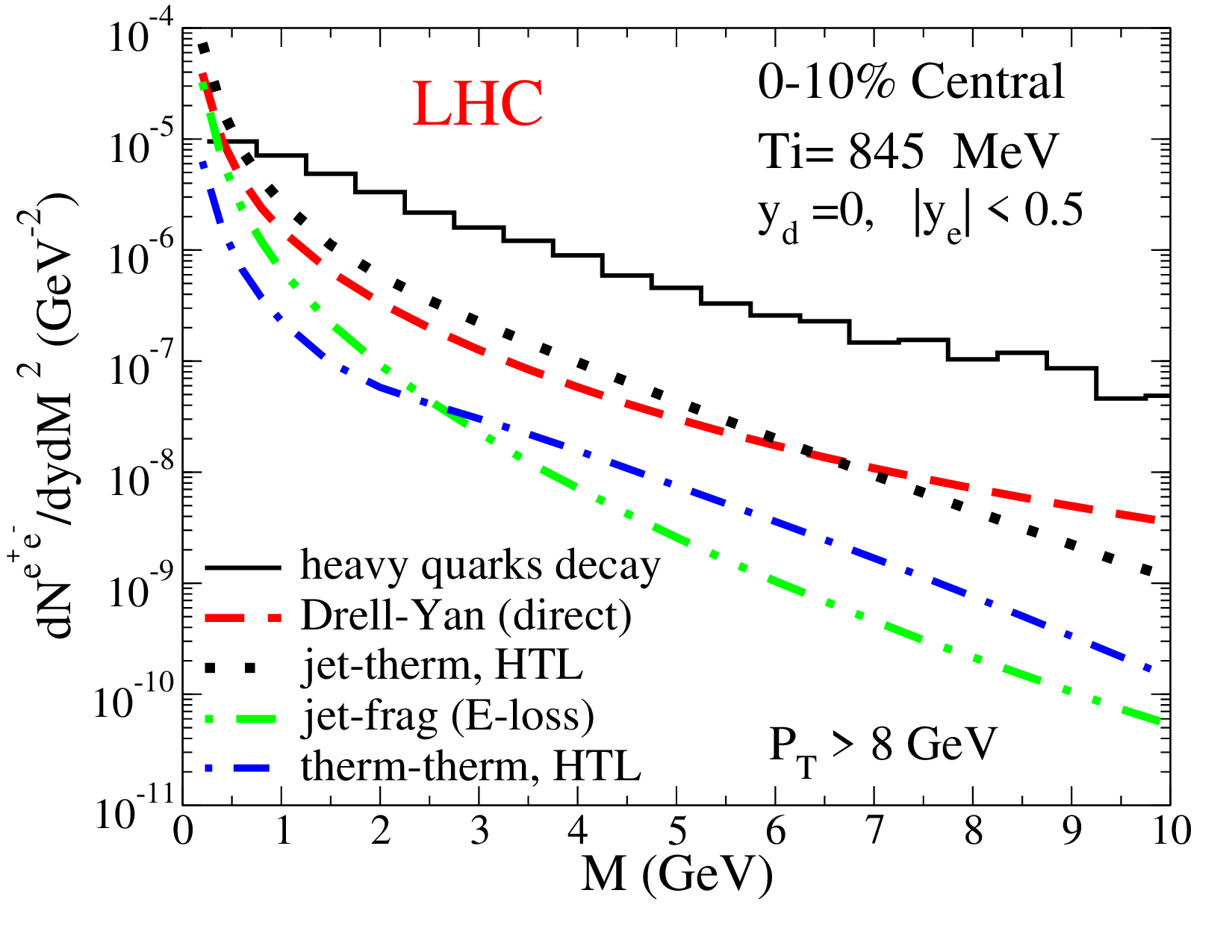,width=7.3cm} 
\epsfig{file=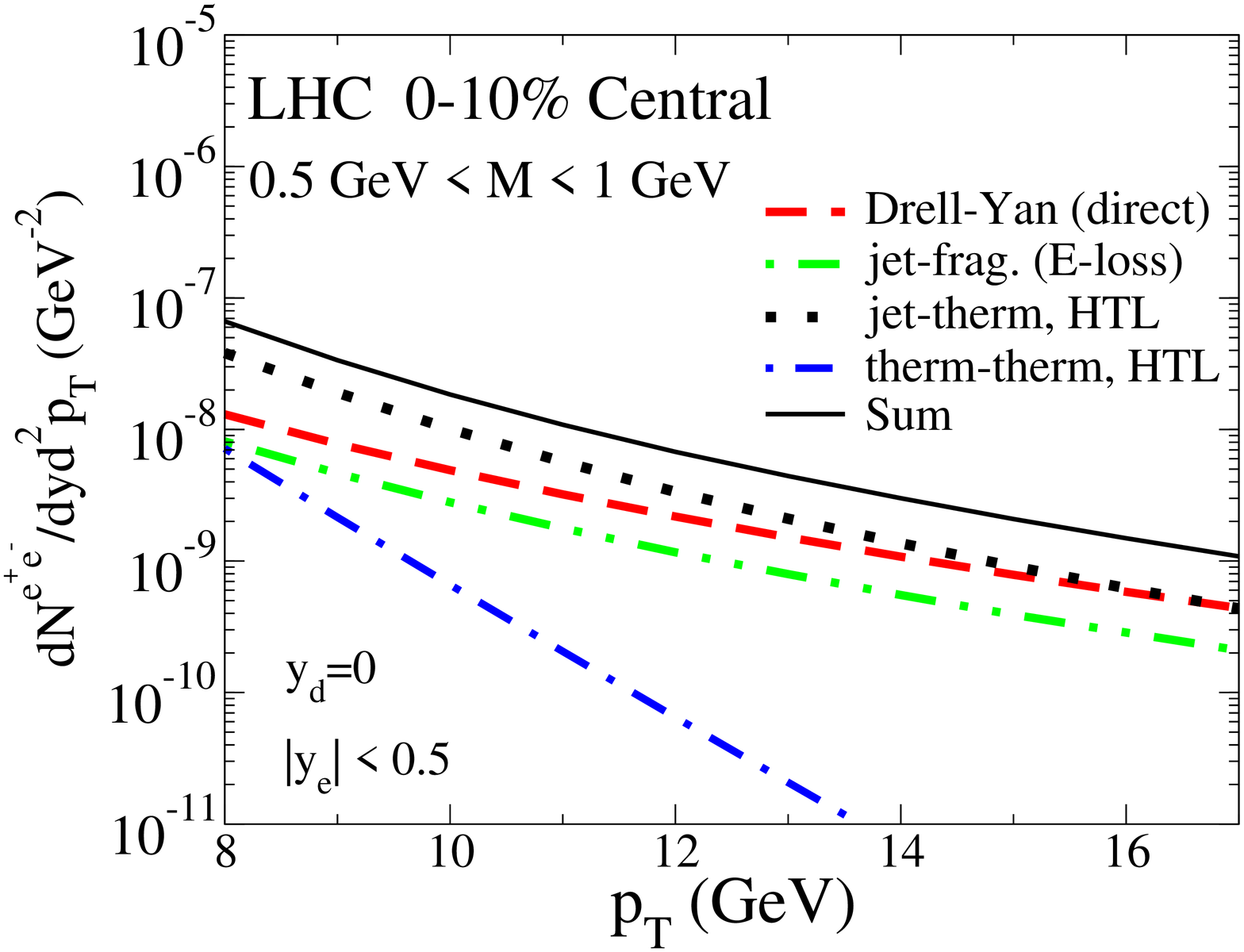,width=7.5cm}
\caption{The yield of e$^+$e$^-$ pairs in central Pb+Pb collisions at 
$\sqrt{s_{\mathrm{NN}}} =5.5$ TeV. 
{\it Left}: Mass spectrum $dN/(dy_d dM^2)$ integrated over the transverse 
momentum $p_T$ of the pair for $p_T > 8$ GeV/$c$.
{\it Right}: Transverse momentum spectrum $dN/(dy_d d^2 p_T)$ integrated over 
a mass range 0.5 GeV $< M<$ 1 GeV. 
Both panels show the case $y_d=0$ for the pair rapidity $y_d$ and a cut 
$|y_e| < 0.5$ for the single electron rapidity.}
\label{friesf1}
\end{center}
\end{figure}

Figure \ref{friesf1} shows numerical evaluations of the different contributions discussed
above to the e$^+$e$^-$ transverse momentum and mass spectrum for central 
Pb+Pb collisions at LHC. We use next-to-leading order pQCD calculations for 
Drell Yan and a leading order calculation for jet production. Energy 
loss of jets is computed with the AMY formalism \cite{Turbide:2005fk}. 
Jet-medium emission and thermal emission have been evaluated in the Hard 
Thermal Loop (HTL) resummation scheme. For the mass spectrum we also show 
the expected background from correlated heavy quark decays. The full 
calculation for dileptons with a more extended discussion is presented in 
\cite{Turbide:2006mc}. Predictions for direct photon yields including 
jet-medium photons can be found in \cite{Turbide:2005fk}.

Dileptons from jet-medium interactions will be more important at LHC than
at previous lower energy experiments. They will be as important or even 
exceeding the Drell-Yan yields at intermediate masses up to about 8 GeV.
They offer a new way to access information about the temperature and the
partonic nature of the fireball.

\subsection{Direct photons at LHC}
\label{rezaeian}

%{\it A. H. Rezaeian, B. Z. Kopeliovich, H. J. Pirner and Iv\'an Schmidt}
{\it A. H. Rezaeian, B. Z. Kopeliovich, H. J. Pirner and I. Schmidt}

{\small
 The DGLAP improved color dipole approach provides a good description of
data for inclusive direct photon spectra at the energies of RHIC and
Tevatron.  Within the same framework we predict the transverse momentum
distribution of direct photons at the CERN LHC energies. 
}

\subsubsection{Introduction} Direct photons, i.e. photons not from hadronic
decay, provide a powerful probe for the initial state of matter created in
heavy ion collisions, since they interact with the medium only
electromagnetically and therefore provide a baseline for the
interpretation of jet-quenching models. The primary motivation for
studying the direct photons has been to extract information about the
gluon density inside proton in conjunction with DIS data. However, this
task has yet to be fulfilled due to difference between the measurement and
perturbative QCD calculation which is difficult to explain by
altering the gluon density function (see Ref.~\cite{Kopeliovich:2007yv}
and references therein). We have recently shown that the color
dipole formalism coupled to DGLAP evolution is an viable alternative to
the parton model and provided a good description of inclusive photon and
dilepton pair production in hadron-hadron collisions
\cite{Kopeliovich:2007yv}. Here we
predict the transverse momentum spectra of direct photons at the LHC
energies $\sqrt{s}=5.5$ TeV and $14$ TeV within the same framework.

\subsubsection{Color dipole approach and predictions for LHC}
\begin{figure}[!t]
       \centerline{\includegraphics[width=11 cm] {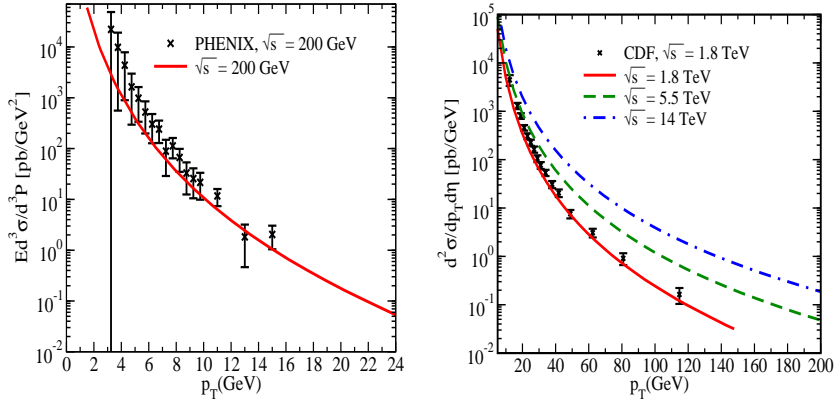}}
       \caption{ Direct photon spectra obtained from GBW-DGLAP dipole
       model at midrapidity for RHIC, CDF and LHC energies. Experimental data (right)
       are for inclusive isolated photon from CDF experiment for
       $|\eta|<0.9$ at $\sqrt{s}=1.8$ TeV \cite{Abe:1994rr} and (left)
       for direct photon at $\eta=0$ for RHIC energy $\sqrt{s}=200$ GeV 
       \cite{Adler:2006yt}.  The error bars are
       the linear sum of the statistical and systematic
       uncertainties. \label{fig-1}}
\end{figure}

 Although in the process of electromagnetic bremsstrahlung by a quark no
dipole participates, the cross section can be expressed via the more
elementary cross section $\sigma_{q\bar{q}}$ of interaction of a $\bar qq$
dipole. For the dipole cross section, we employ the saturation model of
Golec-Biernat and W\"usthoff coupled to DGLAP evolution (GBW-DGLAP)
\cite{Bartels:2002cj}
which is better suited at large transverse momenta. Without
inclusion of DGLAP evolution, the direct photon cross section is
overestimated \cite{Kopeliovich:2007yv}.
In Fig.~\ref{fig-1}, we show the GBW-DGLAP
dipole model predictions for inclusive direct photon production at
midrapidities for RHIC, CDF and LHC energies. We stress that the
theoretical curves in Fig.~\ref{fig-1}, are the results of a parameter
free calculation. Notice also that in contrast to the parton model,
neither $K$-factor (NLO corrections), nor higher twist corrections are
to be added. No quark-to-photon fragmentation function is needed
either.  Indeed, the phenomenological dipole cross section is fitted
to DIS data and incorporates all perturbative and non-perturbative
radiation contributions. For the same reason, in contrast to the
parton model, in the dipole approach there is no ambiguity in defining
the primordial transverse momentum of partons. Such a small purely
non-perturbative primordial momentum does not play a significant role
for direct photon production at the given range of $p_{T}$ in
Fig.~\ref{fig-1}. Notice that the color dipole picture accounts only
for Pomeron exchange from the target, while ignoring its valence
content. Therefore, Reggeons are not taken into account, and as a
consequence, the dipole is well suited mainly for high-energy
processes. As our result for RHIC and CDF energies indicate, we expect
that dipole prescription to be at work for the LHC energies. At the
Tevatron, in order to reject the overwhelming background of secondary
photons isolation cuts are imposed \cite{Abe:1994rr}.
Isolation conditions are
not imposed in our calculation. However, the cross section does not vary by more
than $10\%$ under CDF isolation conditions \cite{Kopeliovich:2007yv}.
One should also notice that the
parametrizations of the dipole cross section and proton structure
function employed in our computation have been fitted to data at
considerably lower $p_{T}$ values \cite{Kopeliovich:2007yv}.

%\myack
%This work was supported by Fondecyt grants 1070517 and 1050519 and 
%by DFG grant PI182/3-1.

\subsection{Thermal Dileptons at LHC}

{\it H.~van Hees and R.~Rapp} 

{\small
  We predict dilepton invariant-mass spectra for central 5.5~ATeV Pb-Pb
  collisions at LHC. Hadronic emission in the low-mass region is
  calculated using in-medium spectral functions of light vector mesons
  within hadronic many-body theory. In the intermediate-mass region
  thermal radiation from the Quark-Gluon Plasma, evaluated
  perturbatively with hard-thermal loop corrections, takes over. An
  important source over the entire mass range are decays of correlated
  open-charm hadrons, rendering the nuclear modification of charm and
  bottom spectra a critical ingredient.
}
\vskip 0.5cm

Due to their penetrating nature, electromagnetic probes (dileptons and
photons) are an invaluable tool to investigate direct radiation from the
hot/dense matter created in heavy-ion collisions.  At low
invariant mass, $M$$\leq$1~GeV, the main source of dileptons is the
decay of the light vector mesons, $\rho$, $\omega$ and $\phi$, giving
unique access to their in-medium spectral properties, most prominently
for the short-lived $\rho$ meson. If the chiral properties of the
$\rho$-meson can be understood theoretically, dilepton spectra can serve
as a signal for the restoration of chiral symmetry at high temperatures
and densities.

We employ medium-modified vector-meson spectral functions in hot/dense
matter following from hadronic many-body theory, phenomenologically
constrained by vacuum $\pi\pi$ scattering, decay branching ratios for
baryonic and mesonic resonances, photo-absorption cross sections on
nucleons and nuclei, etc.~\cite{Rapp:1999us}. The resulting spectral
functions, especially for the $\rho$ meson, exhibit large broadening
with little mass shift, with baryonic interactions as the prevalent
agent, especially in the mass region below the resonance peaks. Note
that $CP$ invariance of strong interactions implies equal interactions
with baryons and antibaryons. Thus, even in a net-baryon free
environment, the $\rho$ resonance essentially ``melts'' around the
expected phase transition temperature, $T_c$$\simeq$180~MeV.  Other
sources of thermal dileptons taken into account are (i) four-pion type
annihilation in the hadronic phase (augmented by chiral
vector-axialvector mixing)~\cite{vanHees:2006ng}, which takes over the
resonance contributions at intermediate mass, and (ii) radiation from
the Quark-Gluon Plasma (QGP), computed within hard-thermal loop improved
perturbation theory for in-medium $q$-$\bar{q}$ annihilation.

Thermal dilepton spectra are computed by evolving pertinent emission
rates over the time evolution of the medium in central 5.5~ATeV Pb-Pb
collisions.  To this end, we employ a cylindrical homogeneous thermal
fireball with isentropic expansion and a total entropy fixed by the
number of charged particles, which we estimate from a phenomenological
extrapolation to be $\dd N_{\mathrm{ch}}/\dd y$$\simeq$1400. We use an
ideal-gas equation of state (EoS) with massless gluons and $N_f$=2.5
quark flavors for the QGP, and a resonance gas for the hadronic EoS with
chemical freezeout at ($\mu_B^c$,$T_c$)=(2,180)~MeV (finite meson and
anti-/baryon chemical potentials are implemented to conserve the
particle ratios until thermal freezeout at
$T_{\mathrm{fo}}$$\simeq$100~MeV, with a mass-action law for short-lived
resonances).  We start the evolution in the QGP phase at initial time
$\tau_0$=0.17~fm/c, translating into $T_0$$\simeq$560~MeV.  The volume
expansion parameters are taken to resemble hydrodynamic simulations. A
standard mixed-phase construction connects QGP and hadronic phase at
$T_c$, and the total fireball lifetime is
$\tau_{\mathrm{fb}}$$\simeq$18~fm/c.
\begin{figure}
\begin{center}
\begin{minipage}{0.49\textwidth}
\includegraphics[width=\textwidth]{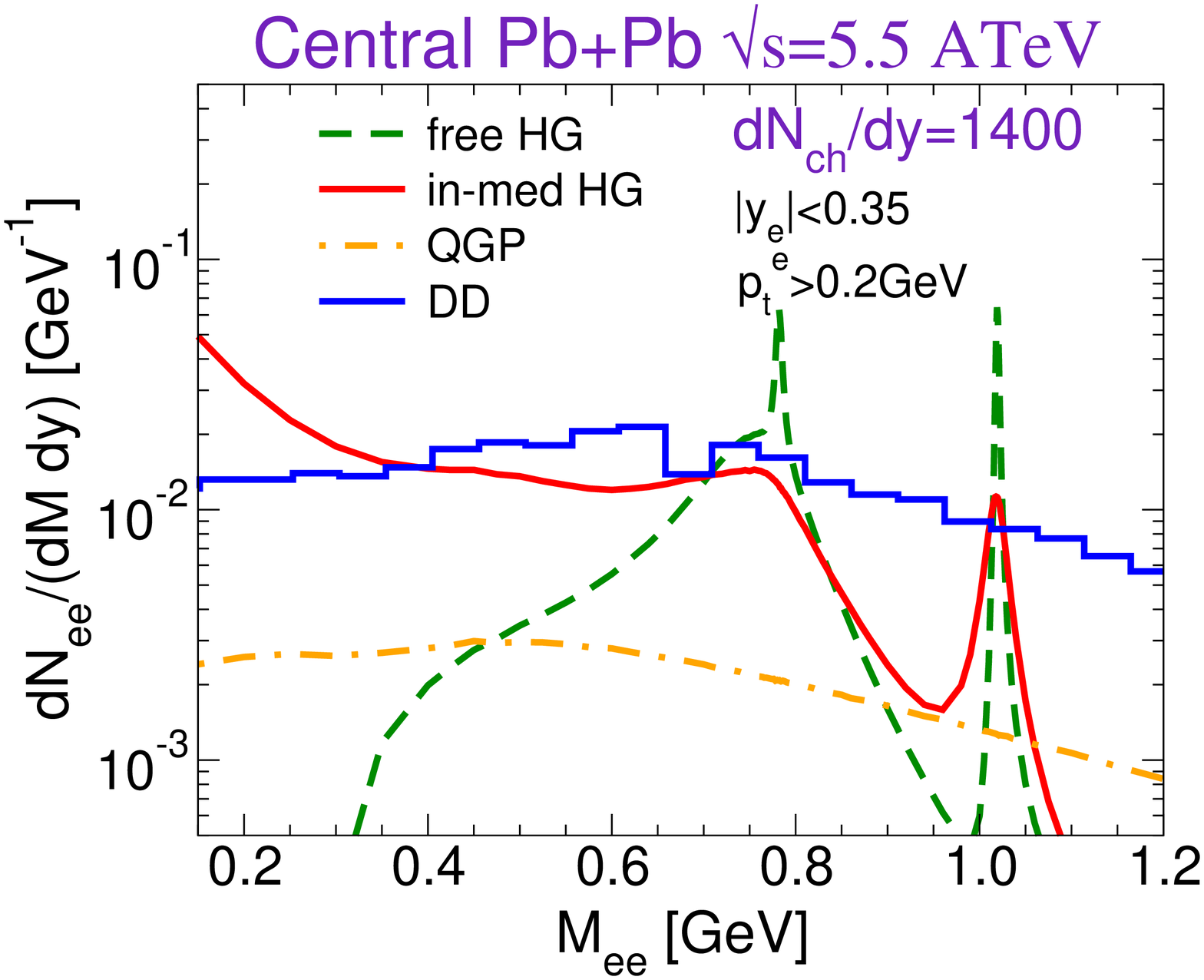}
\end{minipage}\hspace*{1mm}
\begin{minipage}{0.49\textwidth}
\includegraphics[width=\textwidth]{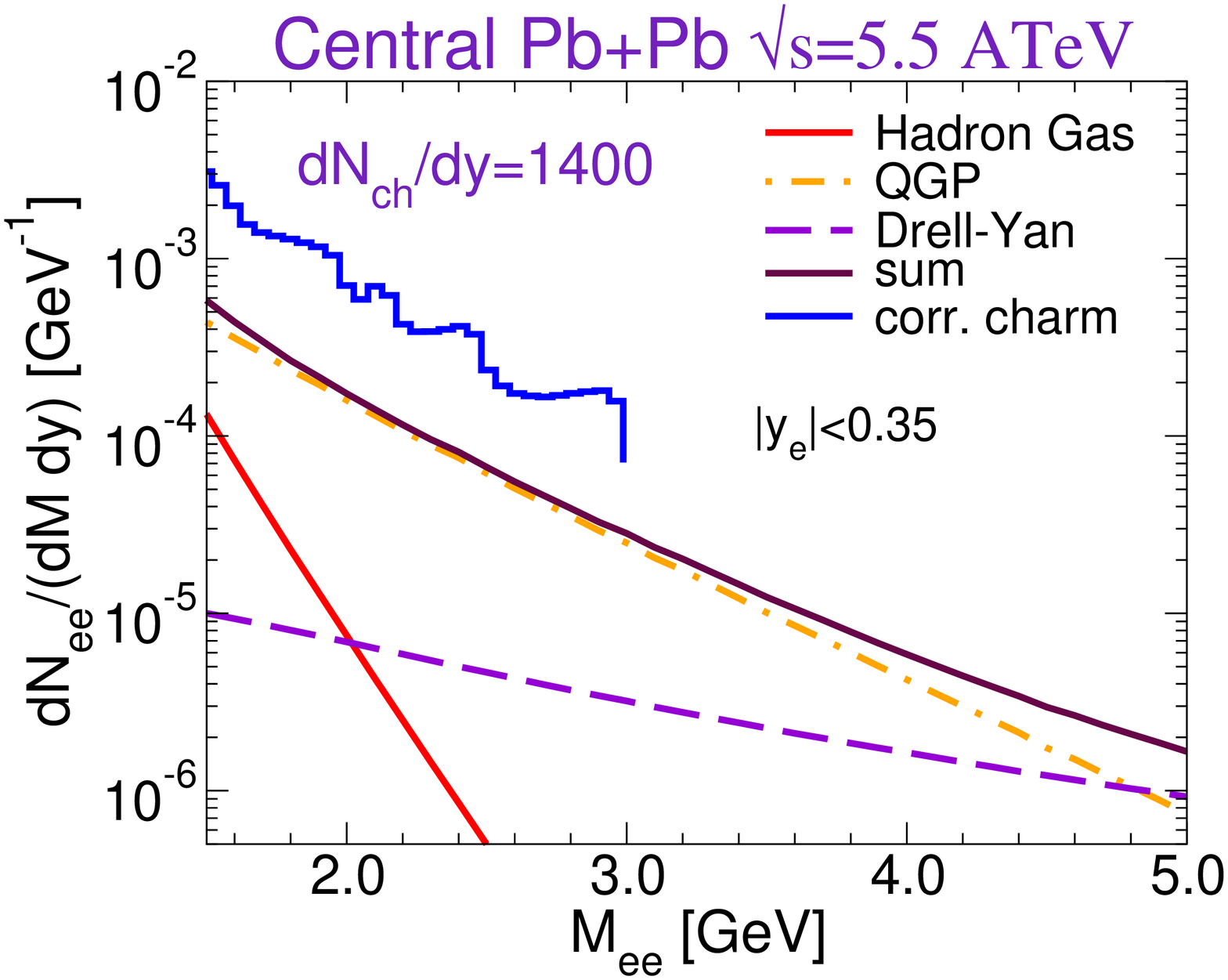}
\end{minipage}
\end{center}
\caption{(Color online) Predictions for dilepton spectra in 
central 5.5~ATeV Pb-Pb collisions at LHC in the 
low- (left panel) and intermediate-mass region (right panel).}
\label{fig.1}
\end{figure}

As for non-thermal sources, we include primordial Drell-Yan annihilation
and decays of correlated charm pairs.  The latter are estimated by
scaling the spectrum at RHIC with a charm-cross section anticipated at
LHC, which implies somewhat softer charm spectra than expected for
primordial N-N collisions (and thus softer invariant-mass spectra).
We neglect contributions from jet-plasma interactions.

Our predictions are summarized in Fig.~\ref{fig.1}. At low mass 
thermal dileptons are dominated by hadronic radiation, with large
modifications due to in-medium vector-meson spectral functions. The QGP
contribution takes over at around $M$$\gtrsim$1.1~GeV. The yield from
correlated open-charm decays is comparable to hadronic emission already
at low mass, and dominant at intermediate mass. However, this result
will have to be scrutinized by including the nuclear modification
of heavy-quark spectra in the QGP (as well as analogous
contributions from correlated bottom decays). Also, larger values of $\dd
N_{\mathrm{ch}}/\dd y$ would help to outshine correlated open-charm
decays, at least at low mass.

%\myack
%{This work is supported by a U.S. NSF CAREER Award, grant no. PHY-0449489.}

\subsection{Direct $\gamma$  production and 
modification at the LHC}

{\it I. Vitev}

{\small
Baseline direct photon production cross sections are studied in 
$\sqrt{s} = 5.5$~TeV p+p collisions at the LHC. The fraction of 
fragmentation photons, which suffer QGP effects, is shown to be 
non-negligible even at very high $p_T \sim 200$~GeV. We first 
examine important cold nuclear matter effects for direct photon
production, related to dynamical shadowing, isospin and initial state energy 
loss, in comparison to neutral pion production at $\sqrt{s} = 200$~GeV. 
Simulations of direct $\gamma$ suppression in Pb+Pb reactions at 
$s^{1/2} = 5.5$~A.TeV at the LHC are also presented to high 
transverse   momentum. Results are given in for central
nuclear collisions and energy loss in the QGP calculated in the 
GLV approach. Direct photon quenching is shown to strongly 
depend on the ratio $\gamma_{\rm prompt} / \gamma_{\rm fragmentation}$
At high $p_T> 100$~GeV cold nuclear matter attenuation can be 
as large as the QGP effects for the net suppression
of direct photons.  
}
\vskip 0.5cm

\begin{figure}[b!]
\includegraphics[width=2.5in,height=2.8in,angle=0]{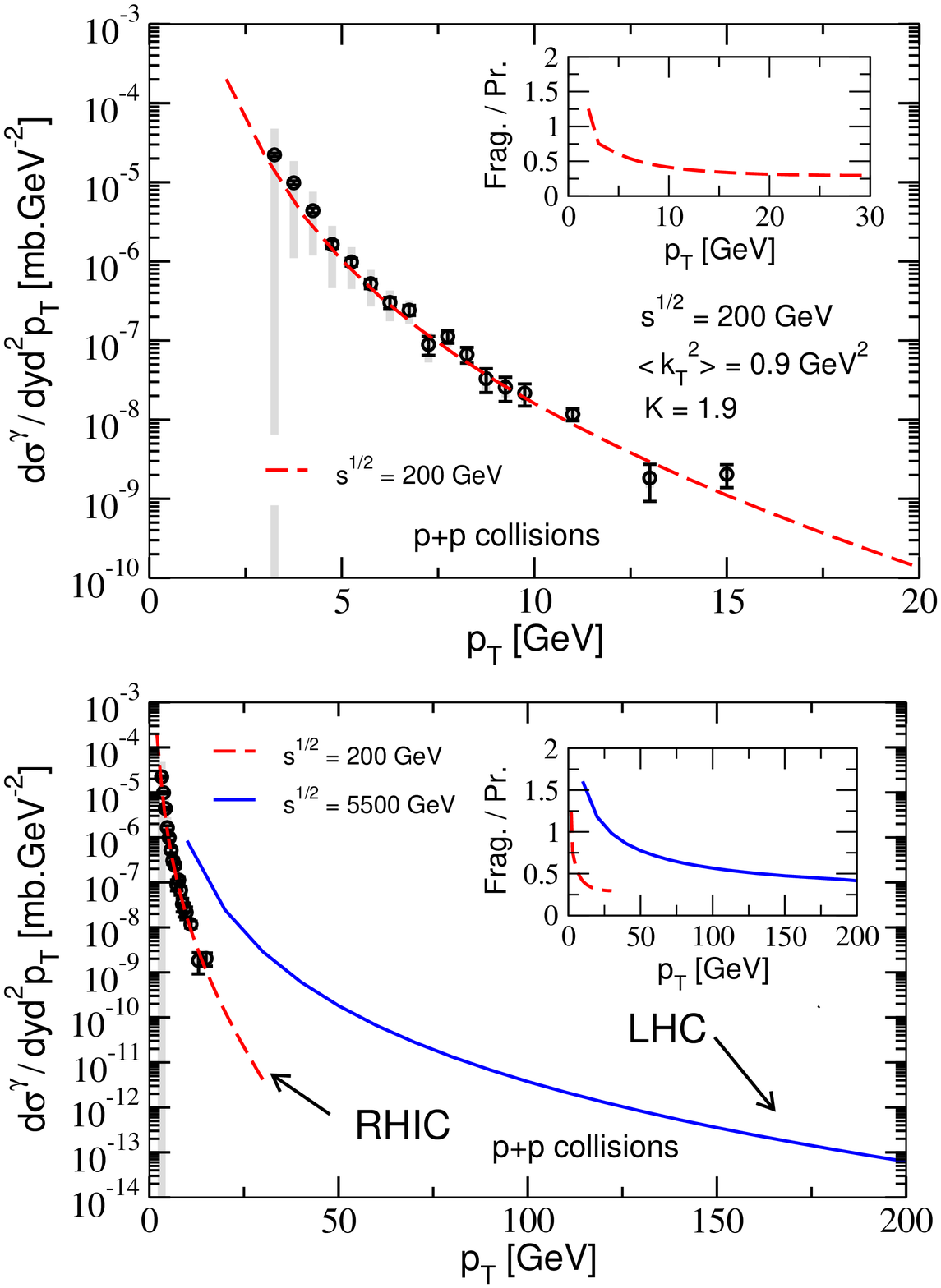}
\hspace*{.5cm}
\includegraphics[width=2.5in,height=2.8in,angle=0]{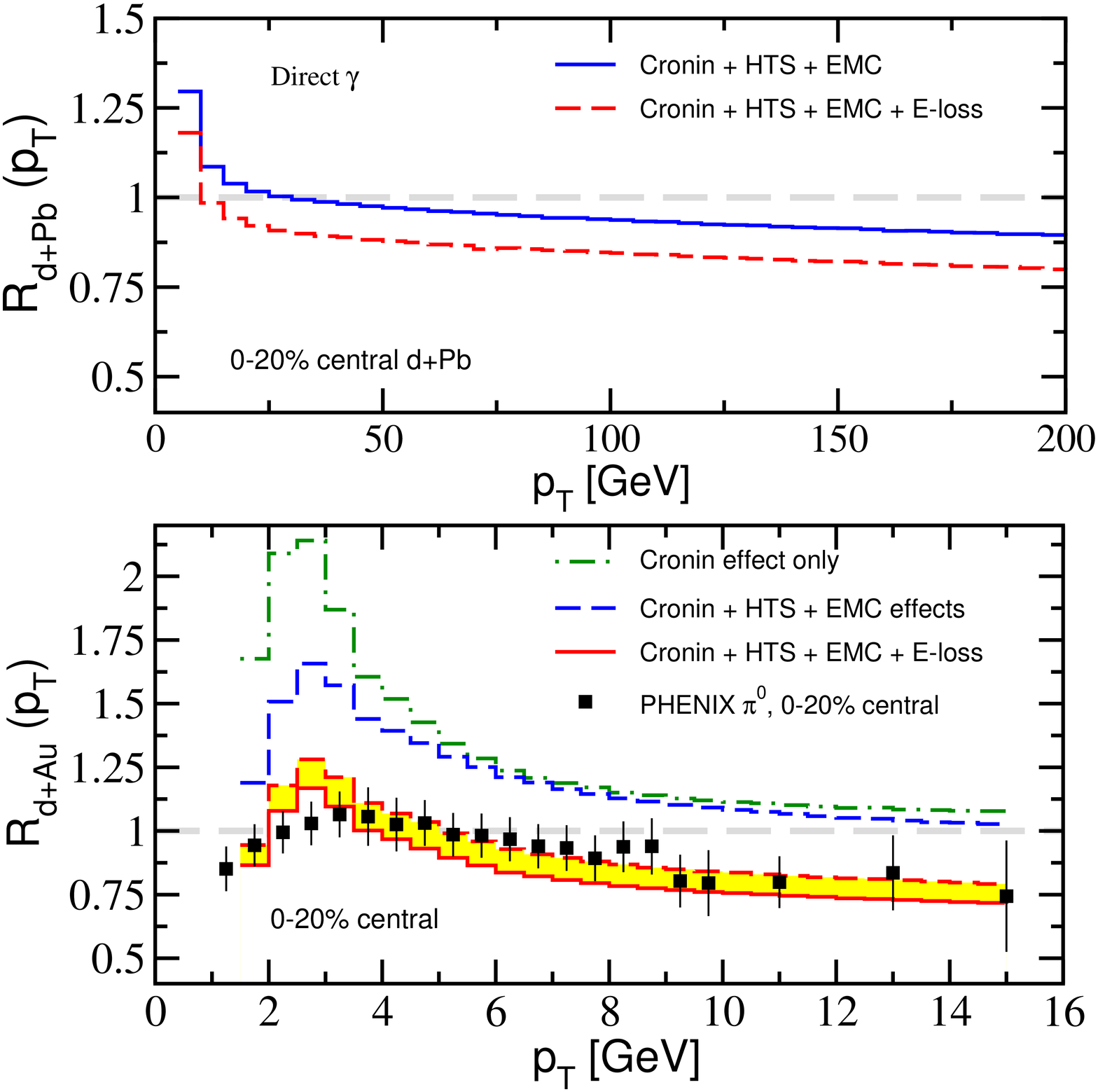}
\caption{ Left panel: Direct photon production cross section in p+p 
collisions at the LHC $\sqrt{s}=5.5$~TeV. Comparison to the 
same cross section calculation at RHIC at $\sqrt{s}=200$~GeV 
and to current high $p_T$ data is also shown. Insert illustrates 
the ratio of fragmentation  to  prompt photons vs  $p_T$ at LO. 
Right panel: 
Nuclear modification  factor $R_{dA}$ in central d+Au collisions 
at RHIC and central d+Pb
at the LHC, 0-20\%. The high $p_T$ behavior indicates the 
isospin (charge) effect and initial-state energy loss in cold 
nuclear matter. Comparison 
to similar effects on neutral pion production in d+Au collisions 
at RHIC, indicative for the first time for cold nuclear matter 
$-\Delta E_{\rm rad}$ effects at high $p_T$ is also shown.} 
\label{crosph-LHC}
\end{figure}

It has been argued that direct photon production and 
direct photon tagged jets provide error-free gauge for the 
quenching of quarks and gluons and for fixing their initial energy.  We show 
that quantitatively large nuclear corrections must be taken 
into account for direct $\gamma$ to become precision probes 
of the QGP. The left panel of Fig.~\ref{crosph-LHC} shows the  
direct photon production cross section
in p+p collisions at $\sqrt{s}= 5.5$~TeV the LHC compared to the 
corresponding cross section at RHIC  $\sqrt{s}= 200$~GeV to LO in 
perturbative QCD~\cite{Vitev:prep}. Insert shows the fraction of 
fragmentation to prompt photons versus $p_T$.  The right panel 
of Fig.~\ref{crosph-LHC} shows  cold nuclear effects, the Cronin~\cite{Vitev:2002pf}, 
dynamical 
shadowing~\cite{Qiu:2004da}  and cold nuclear matter 
energy loss~\cite{Vitev:2007ve}, in d+A  reactions at LHC energies. 
Comparison to data in 0-20\% central d+Au collisions at RHIC 
is also presented.

\begin{figure}[b!]
\includegraphics[width=2.5in,height=1.8in,angle=0]{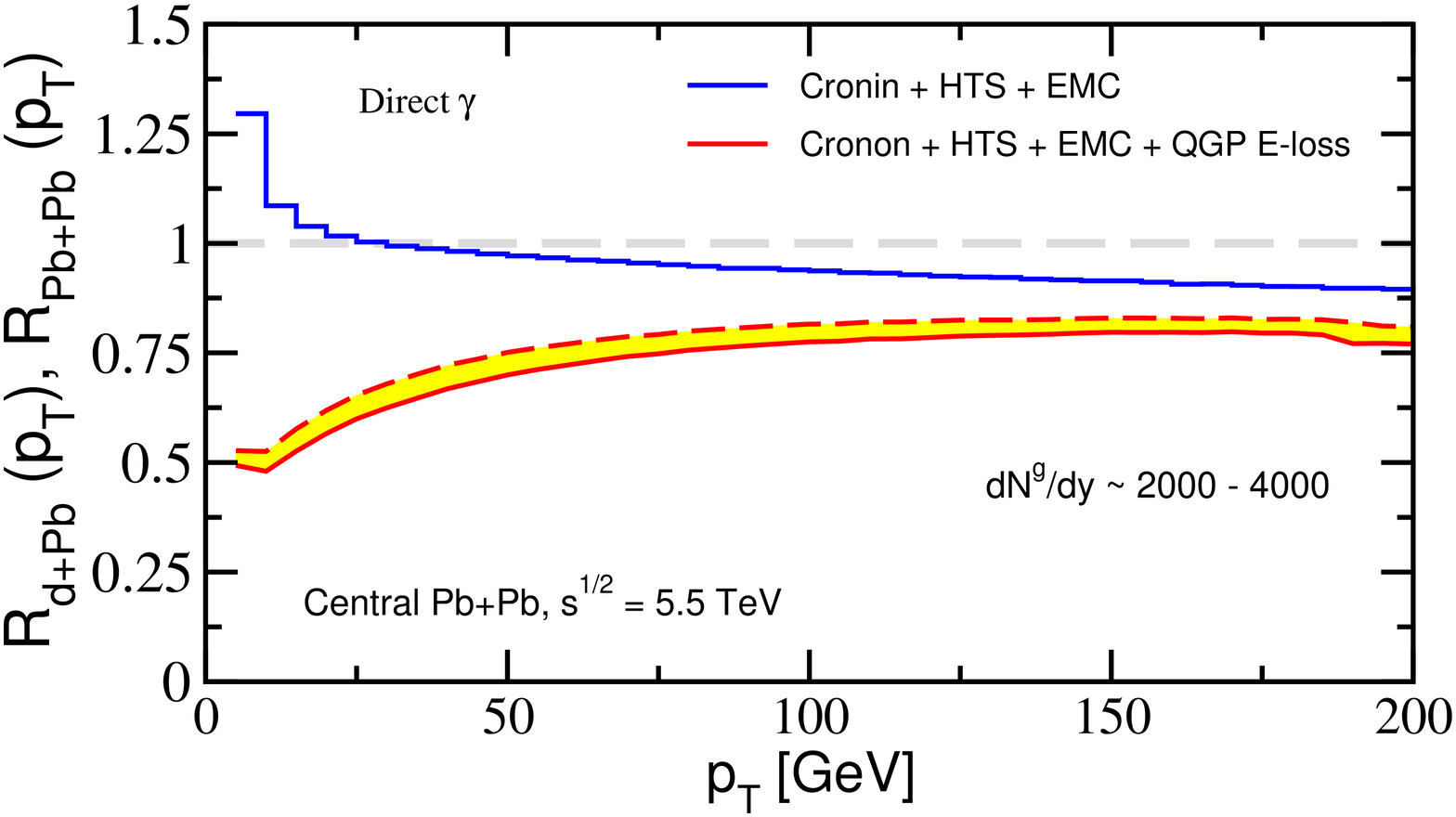}
\hspace*{0.5cm}
\includegraphics[width=2.5in,height=1.8in,angle=0]{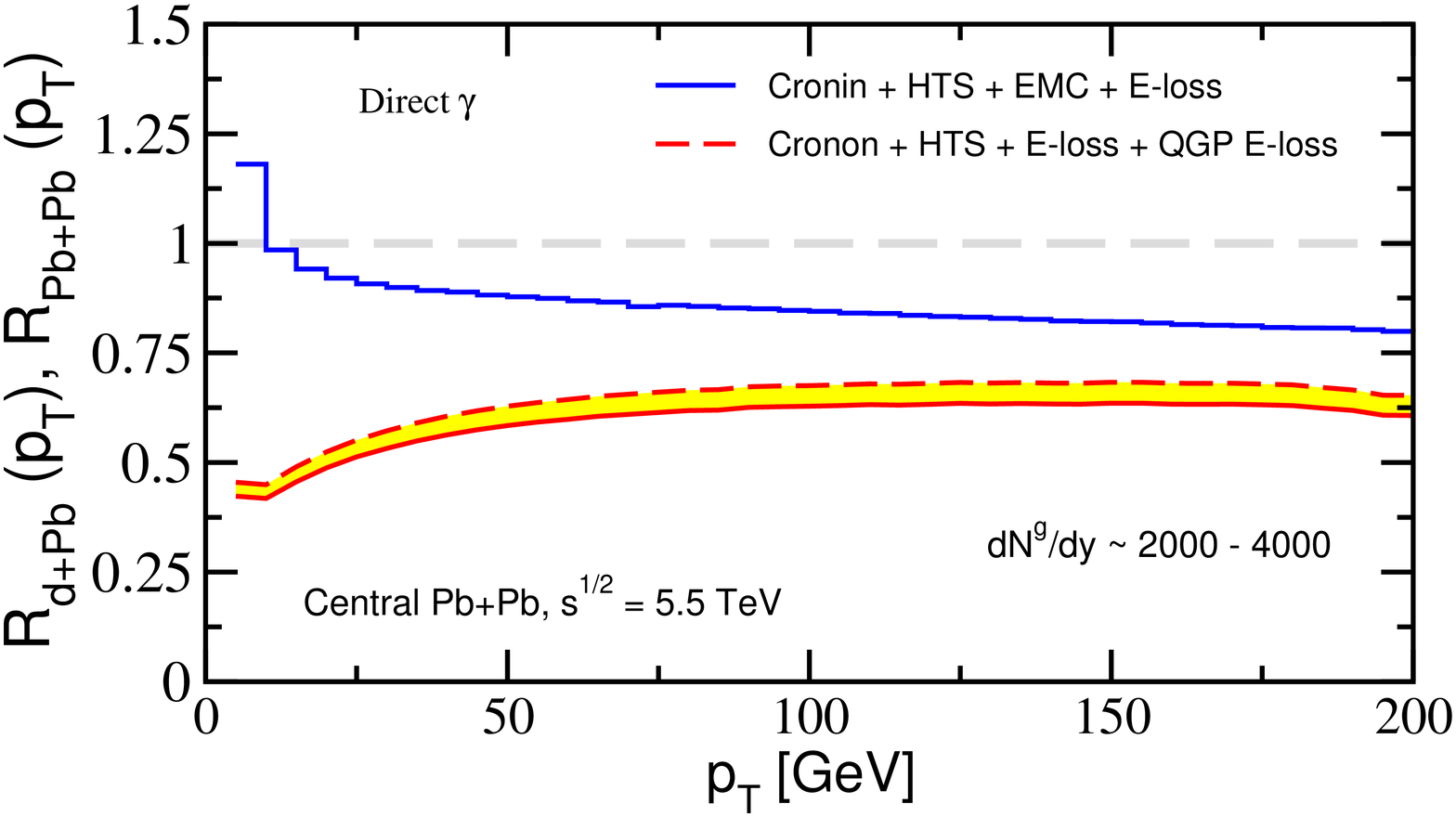}
\caption{ Left panel: Comparison of cold nuclear matter effects to 
QGP effects on direct photon production at the LHC. 
Central d+Pb and central Pb+Pb at $\sqrt{s}=5.5$~TeV are 
shown. Calculations do not include initial-state energy loss. 
QGP suppression trend with $dN^g/dy \sim  2000 - 4000$ follows 
the fragmentation/prompt ratio for direct $\gamma$. 
Right panel: Similar calculations including initial-state cold 
nuclear matter energy loss effects. Note that these can yield 
50\% larger suppression at high $p_T$. }
\label{crosph1-LHC}
\end{figure}

The left panel of Fig. \ref{crosph1-LHC} shows the QGP effect (final-state
interactions) in central Pb+Pb collisions at $\sqrt{s}=5.5$~TeV. 
Parton rapidity densities $dN^g/dy \sim 2000 - 4000$~\cite{Vitev:2002pf}, 
as for $\pi^0$ quenching and  heavy meson dissociation, are used. Direct 
photon quenching closely follows the ratio   $\gamma_{\rm prompt} / 
\gamma_{\rm fragmentation}$~\cite{Vitev:prep}. At low $p_T$
attenuation is QGP-dominated with significant and measurable 
suppression $R_{AA}(p_T) \sim 0.5$.  Nevertheless, such quenching 
is smaller than the one for $\pi^0$'s and reflects the $C_F/C_A$
average squared color charge difference for quark and gluon jets. 
The right panel of Fig. \ref{crosph1-LHC} 
includes the effect of initial-state cold nuclear matter energy 
loss. At high $p_T$ these can be comparable to the final-state 
quenching in the QGP~\cite{Vitev:prep,Qiu:2004da,Vitev:2007ve}.

\section{Others}
\label{sec:others}

\subsection{The effects of angular momentum conservation in relativistic heavy
ion collisions at the LHC}
\label{becattini}

{\it F. Becattini and F. Piccinini}

{\small
We argue that in peripheral heavy ion collisions at the LHC there
might be the formation of a spinning plasma with large intrinsic angular
momentum. If the angular momentum is sufficiently large, there could be
striking observable effects: a decrease of chemical freeze-out temperature
and an increase of transverse momentum spectra broadening (enhanced radial
flow) as a function of centrality; a large enhancement of elliptic flow;
a polarization of emitted particles along the direction of angular momentum.
The latter would be the cleanest signature of such effect.
}
\vskip 0.5cm

In peripheral relativistic heavy ion collisions colliding ions have a large
relative orbital angular momentum. While the fragments keep flying away from the
interaction region essentially unaffected, a fraction of the initial angular momentum 
is transferred to the interaction region. Much of it is probably spent into relative
orbital angular momentum of the newly formed fireballs at large rapidity, but it may 
happen that another significant fraction is given to the midrapidity region giving
rise to a spinning plasma with an intrinsic angular momentum $J$. If $J$ is 
sufficiently large, one has remarkable observable effects. It has been suggested
that such a phenomenon can produce an azimuthal anisotropy in the transverse plane
very similar to the well known elliptic flow \cite{Castorina:2007eb}. Also, a large $J$ may
result in a polarization of emitted particles \cite{wang}. We make a quantitative
determination of observable effects by assuming that the spinning system is at
statistical equilibrium, taking advantage of a recent calculation of the microcanonical
partition function of a relativistic quantum gas with fixed angular momentum 
\cite{micro2,prep} which allowed us to provide the expression of particle spin 
density matrix and polarization in a rotating thermodynamical system. Here, a 
possible scenario for the LHC energy is just sketched; a more detailed paper 
will appear \cite{prep}.

\begin{figure}[htb]
\begin{minipage}{17pc}
\includegraphics[width=17pc]{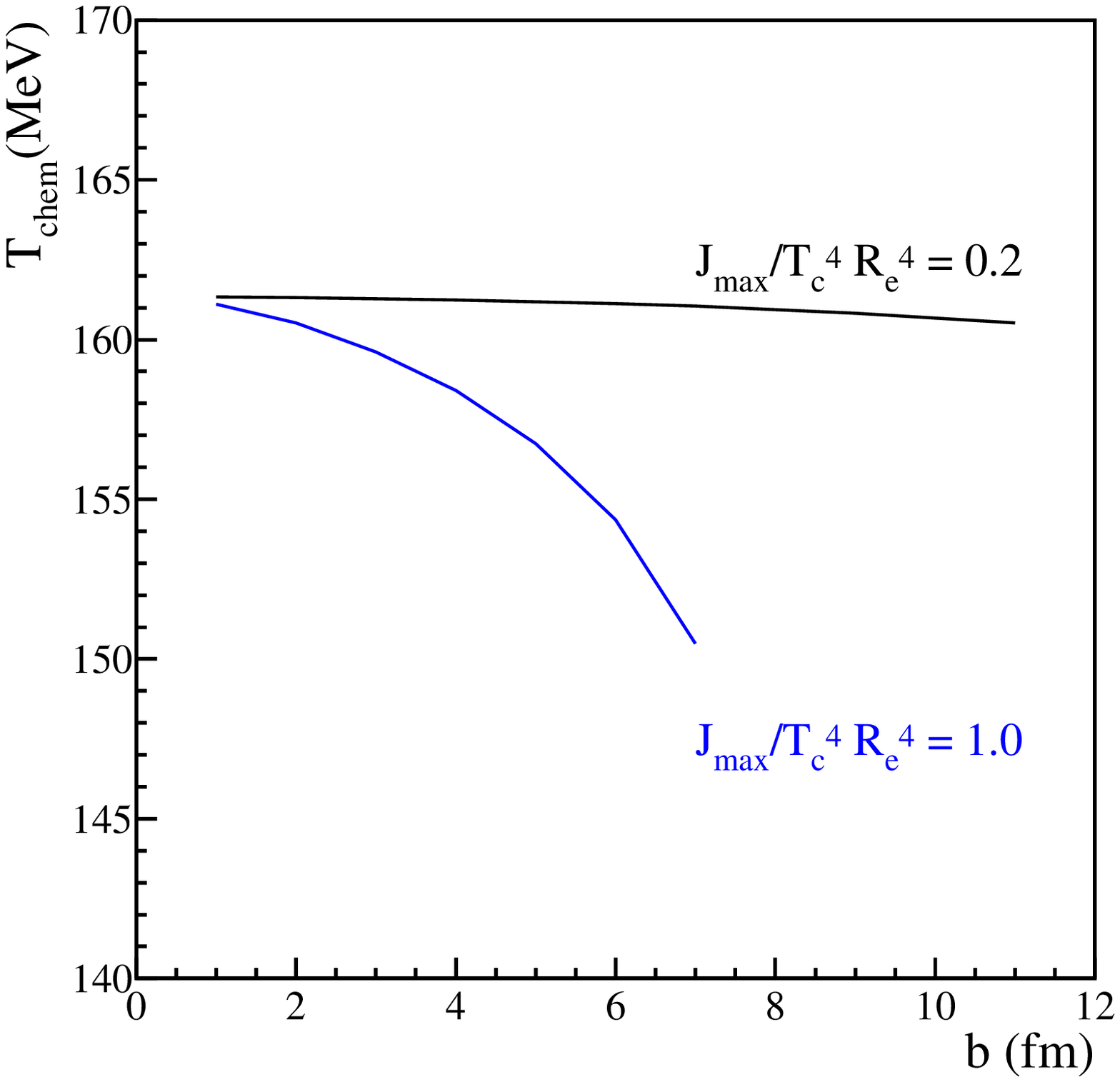}
\end{minipage}\hspace{2pc}%
\begin{minipage}{17pc}
\includegraphics[width=17pc]{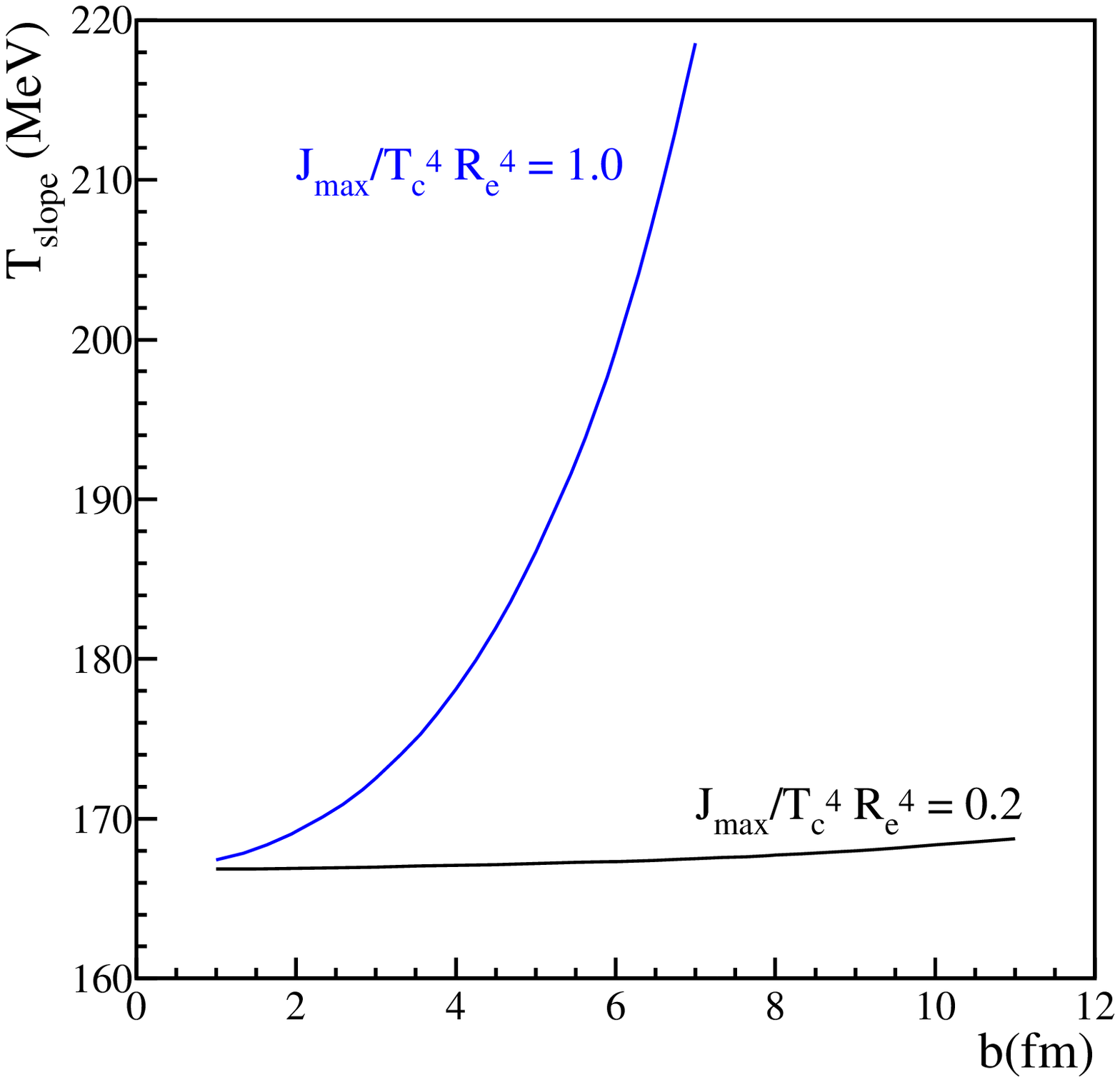}
\end{minipage} 
\hfill
\begin{minipage}{17pc}
\includegraphics[width=17pc]{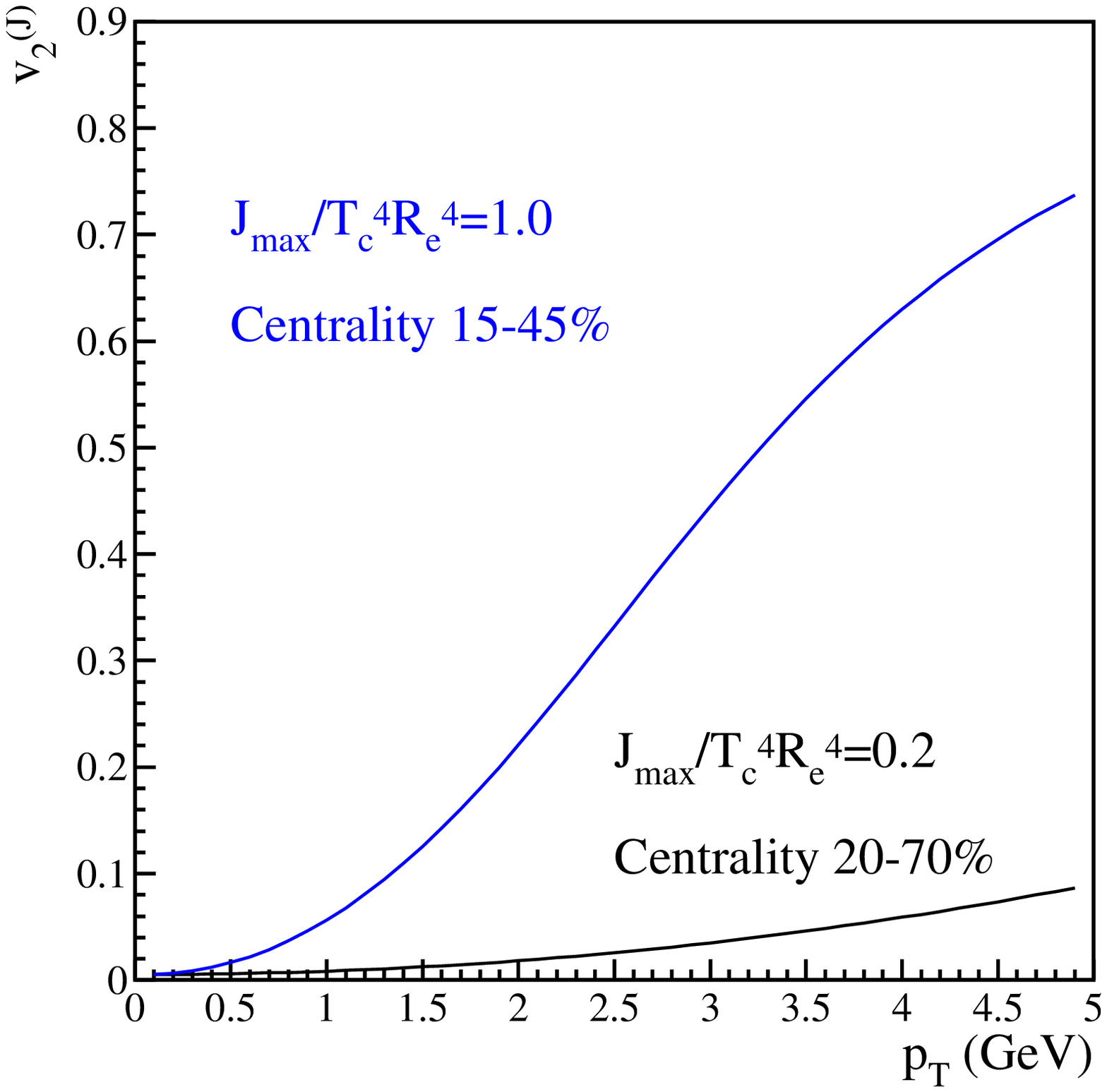}
\end{minipage}\hspace{2pc}%
\begin{minipage}{17pc}
\includegraphics[width=17pc]{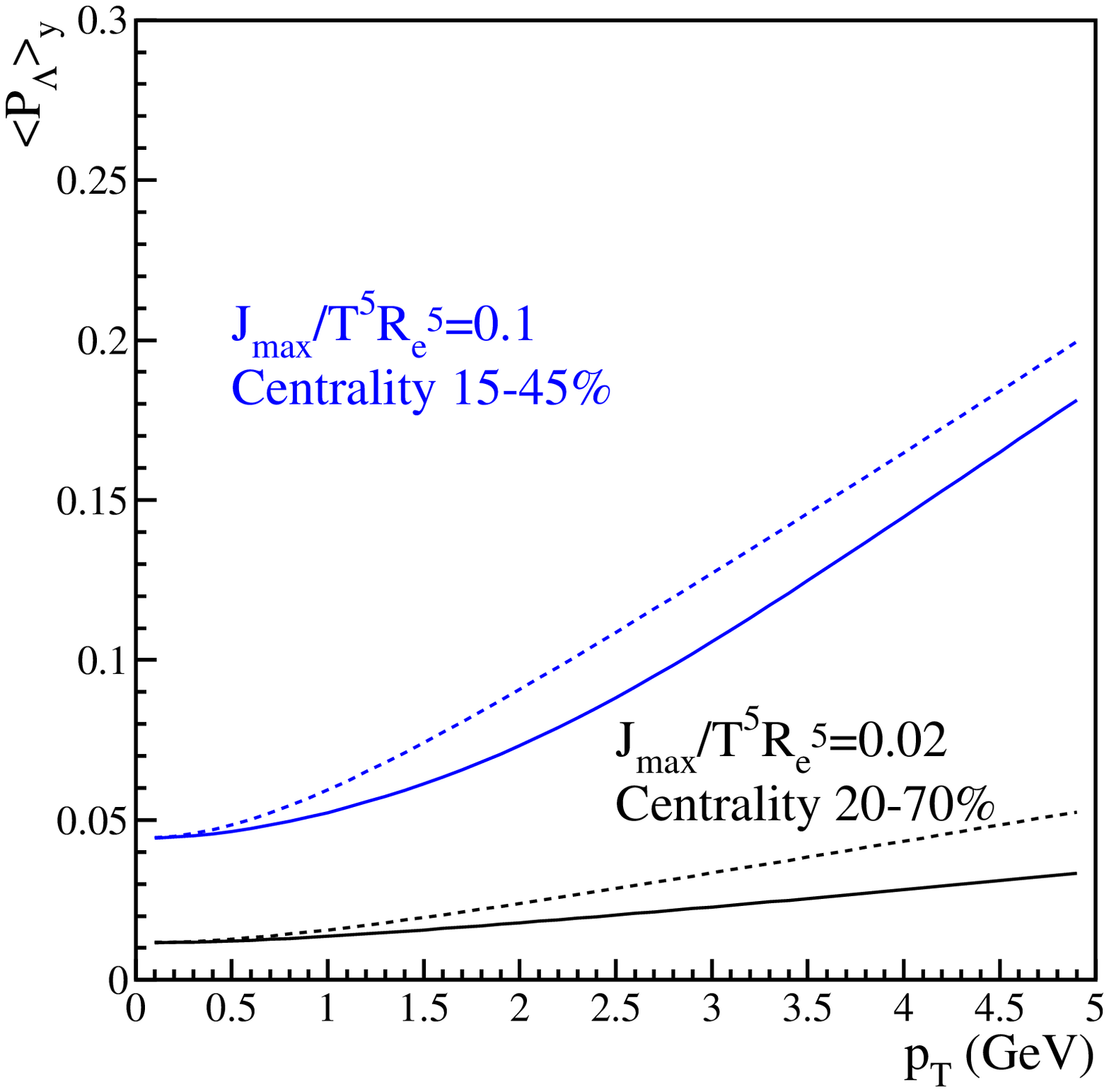}
\end{minipage} 
\caption{  }
\label{fig-beca}
\end{figure}

Under reasonable assumptions, the main observables which signal the presence of
an equilibrated spinning system are (see figure \ref{fig-beca}): a decrease of chemical freeze-out 
temperature and an increase of transverse momentum spectra broadening (enhanced radial
flow) as a function of centrality; a large enhancement of elliptic flow
and a polarization of emitted particles along the direction of angular momentum. 
The latter is the cleanest signature of a spinning system. These observables
scale with the parameter $J/T_c^4 R^4$, $T_c$ being the critical
temperature and $R$ the maximal transverse radius of the system. They are shown in 
figures below, as a function of the impact parameter or transverse momentum, for 
the upper bound of this parameter set by the RHIC $\Lambda$ polarization 
measurement (=0.2, blue line) and at LHC (=1.0, black line) under the assumption of a 
scaling of $J/T_c^4 R^4$ by $\sqrt{s}/\ln \sqrt{s}^{5/3}$.\\
{\em Caveat}: the calculations shown in the plots concern only primary hadrons
emitted from an equilibrated source. Dilution effects such as resonance decays, 
perturbative production at large $p_T$ and partial equilibration are not taken
into account.

%------------------------------------------------------------------------------
%-------------------------------------------------------------------------------

\subsection{Black hole predictions for LHC}

%{\it Horst St\"{o}cker and Ben Koch}
{\it H.~St\"{o}cker and B.~Koch}

{\small
The speculative prediction of
the production of microscopical
black holes, which would be possible
at the large hadron collider due to large extra dimensions,
is discussed. We review observables
for such black holes and for the
their possibly stable final state.
}

\subsubsection{From the hierarchy-problem
to black holes in large extra dimensions}

One of the problems
in the search for a unified description of gravity and
the forces of the standard model (SM),
is the fact that the Planck-scale $m_{Pl}\sim 10^{19}$~GeV
(derived from Newtons constant $G_N$)
is much bigger than the energy scales like the Z-mass $m_Z\sim 90$~GeV.
This huge difference
is the so-called hierarchy problem.
Several theories can explain this hierarchy
by the assumption of extra spatial dimensions 
\cite{ArkaniHamed:1998rs,Antoniadis:1998ig,Randall:1999vf}.
These theories assume a true fundamental scale
$M_f$ which is of the order of just a few TeV and they interpret
the Planck scale $m_{Pl}$ as an effective magnitude 
which comes into the game due
to unobservable and compactified extra spatial dimensions.
In the model  suggested 
by Arkani-Hamed, Dimopoulos and Dvali
\cite{ArkaniHamed:1998rs,Antoniadis:1998ig}
the $d$ extra space-like dimensions
are compactified on tori with radii $R$.
In this model the SM 
particles are confined to our 3+1-dimensional sub-manifold (brane)
and the
gravitons are allowed to propagate freely in the (3+d)+1-dimensional bulk.
Planck mass $m_{Pl}$ and the
fundamental mass $M_f$ are related by
\begin{eqnarray}
m_{Pl}^2 = M_f^{d+2} R^d \quad. \label{Master}
\end{eqnarray}

One exciting consequence of such models is that
 up to $10^9$ black holes (BH)
might be produced at the Large Hadron Collider (LHC)
\cite{Banks:1999gd}.
The intuitive approximation of 
the cross section for such events
can be made by using the Hoop conjecture 
%\cite{hoop}
and taking the classical area
of the (to be produced) BH with
radius $R_H$
\begin{eqnarray} \label{cross}
\sigma(M)\approx \pi R_H^2 \quad,
\end{eqnarray}
where $M$ is the BH mass.
The Scharzschild radius is given at 
distances smaller than the size of the extra
dimensions by 
%\cite{my}
%
\begin{equation} \label{ssradD}
R_H^{d+1}=
\frac{2}{d+1}\left(\frac{1}{M_{f}}\right)^{d+1} \; \frac{M}{M_{f}}
\quad .
\end{equation}
This radius is 
much larger than the Schwarzschild radius corresponding  
to the same BH mass in 3+1 dimensions, which translates
directly into a much larger cross section (\ref{cross}). 
This estimate seems to keep its validity also in
more elaborated picture 
%\cite{Voloshin:2001fe,Rychkov:2004sf,Jevicki:2002fq}
.\\

\subsubsection{From black hole evaporation to LHC observables}

Once a BH is produced it is
assumed to undergo a rapid evaporation
process. This happens first in the so called bolding phase
where angular momentum and internal degrees of
freedom are assumed to be radiated off.
%\cite{hair}.
For a BH mass
much bigger than the fundamental mass scale ($M\gg M_f$)
the following phase is the Hawking phase, 
%\cite{ehm}
where particles are thermally radiated off according to the Hawking
temperature
\footnote{
The process of Hawking radiation would in principle
allow to transform the BH mass into thermal energy
and was therefore subject to further speculations}
\cite{Stoecker:2006yz}
$T_H\approx M_f  (M_f/M)^{1/(d+1)}$.
As soon as the BH mass becomes comparable
to the fundamental mass scale, the underlying physics of
the BH is not understood and exact predictions
are hardly possible at the current state of knowledge.
Discussed scenarios reach from a sudden 
final explosion 
%\cite{Harris:2003db,Cavaglia:2006uk}
over a slowed down evaporation 
%\cite{Koch:2007um}
to the formation of
stable black hole remnant (BHR)
%\cite{39,Koch:2005ks}.
As most BHs would be produced close
to the production threshold the experimental outcome
will be influenced strongly by this final phase of BH evolution.

We analyzed the predictions for different scenarios.
It turned out that the, suppression of hard (TeV)
di-jets above the BH formation threshold 
%\cite{own2,Hofmann:2001pz,Casadio:2001wh}
would be the most scenario independent observable for the LHC.
Other observables such as event multiplicities or
$p_T$ distributions 
%\cite{Harris:2003db,Cavaglia:2006uk} 
should be definitely studied although
they are more model dependent.
Speculations about the formation of BHRs
can be tested experimentally at the LHC:
Charged stable BHRs would leave single
stiff tracks in the LHC detectors.
Uncharged BHRs with their very small reaction cross sections
could be observed by searching for events with $\sim 1$~TeV missing
energy and quenching of the high $p_T$ hadron spectra.
For further references on BHs, BHRs and their observables please see
\cite{Koch:2007um}.

We conclude that BHs at the LHC could provide a
unique experimental window to the understanding of quantum gravity.
As many principles of BH production and decay are not fully
understood, a large variability of experimental observables 
is absolutely essential to pin down the 
underlying physics.
%

%\myack {This work was supported by GSI and BMBF.}

\subsection{Charmed exotics from heavy ion collision}
\label{s:Lee}

{\em S. H. Lee, S. Yasui, W. Liu and C. M. Ko}

{\small
We discuss why charmed multiquark hadrons are likely to exist and explore the 
possibility of observing such states in heavy ion reactions at the LHC.
}
\vskip 0.5cm

Multiquark hadronic states are usually unstable as their quark configurations 
are energetically above those of combined meson and/or baryon states.  
However, constituent quark model calculations suggest that multiquark states 
might become stable when some of the light quarks are replaced by heavy quarks. 
Two possible states that could be realistically observed in heavy ion collisions
at LHC are the tetraquark $T_{cc} (ud\bar{c}\bar{c})$~\cite{Navarra:2007yw} and 
the pentaquark $\Theta_{cs} (ud us \bar{c})$~\cite{Sarac:2005fn}. 
The driving mechanism for the stability of these states can be traced to the
quark color-spin interaction, which can be effectively parameterized as 
$C_H \sum_{i>j} \vec{s}_i \cdot \vec{s_j} \frac{1}{m_i m_j}$.
Baryon mass splittings between states sensitive to the color-spin interaction 
are well explained with a single constant coefficient
$\frac{C_B}{m_u^2}=193$~MeV~\cite{Lee:2007tn}. 
Similarly, corresponding meson mass splittings are well reproduced with
$\frac{C_M}{m_u^2}=635$~MeV~\cite{Lee:2007tn}. 
Hence, the correlation energy in a quark-antiquark pair is about a factor 3 
larger than that in a quark-quark pair that is in the color antitriplet channel.
For heavy quarks, the size of the relative wave function decreases 
substantially, and the parameter $C_H$ extracted from the mass difference 
between $J/\psi$ and $\eta_c$ is $\frac{C_{c\bar{c}}}{m_c^2}=117$~MeV. 
As in the case of light quarks, we choose
$\frac{C_{cc}}{m_c^2}=\frac{1}{3}\frac{C_{c\bar{c}}}{m_c^2}=39$~MeV.
These numbers suggest that two quarks and two antiquarks would rather become 
two mesons than form a single tetraquark state.
However, when one or both of the antiquarks become heavy, the attractions to 
form mesons are relatively suppressed compared to the strong diquark 
correlation among light quarks, making multiquark states possibly stable. 
Using the constants $C_H$ discussed above, we find that the mass of $T_{cc}$ 
($\Theta_{cs}$) is -79~MeV below (8~MeV above) its hadronic decay threshold. 
These results are well reproduced by full constituent quark model calculations.
Although the binding becomes larger when the $c$ quark is replaced by a $b$
quark, the expected number of $b$ quarks produced in a heavy ion collision at 
the LHC is small for a realistic observation of such states. 
Therefore, we only give predictions for the multiquark states containing $c$ 
quarks.

Employing the coalescence model \cite{Chen:2007zp}, we have studied $T_{cc}$ and 
$\Theta_{cs}$ production in central Au+Au collisions at RHIC and Pb+Pb 
collisions at LHC. 
Using the $u$ (or $d$) quark numbers 245 and 662, the anti-strange quark numbers
150 and 405, and the charm quark numbers 3 and 20 based on initial hard 
collisions at RHIC and LHC, respectively all in one unit of midrapidity, we 
find that the numbers of $T_{cc}$ produced at RHIC and LHC are about 
$5.4 \times 10^{-6}$ and $8.9\times 10^{-5}$, respectively, while those of
$\Theta_{cs}$  are about $1.2\times 10^{-4}$ and $8.3\times 10^{-4}$, respectively.
Since these numbers are significantly smaller than $7.5\times 10^{-4}$ and
$8.6\times 10^{-3}$ for $T_{cc}$, and $4.5\times 10^{-3}$ and $2.7\times 10^{-2}$ 
for $\Theta_{cs}$ from the statistical hadronization model for RHIC and LHC, 
respectively, we expect additional production of these exotic charmed hadrons 
from the hadronic stage of the collisions. 
We note that these charmed hadrons would be more abundantly produced, 
particularly the $T_{cc}$, if charm quarks are produced from the QGP formed in 
these collisions.
\begin{table}[htdp]
\caption{Possible decay modes of $T_{cc}$. 
  Additional ($\pi^+ \pi^-$)'s  are possible in the bracket.}
\label{tab:Lee-tab1}
\begin{center}
\begin{tabular}{c|c|c}
\hline
  threshold & decay mode & life time \\
\hline
$M_{T_{cc}} > M_{D^{\ast}}+M_{D}$ & $D^{\ast -}\bar{D}^{0}$ & hadronic decay \\
$2M_{D}+M_{\pi} < M_{T_{cc}} < M_{D^{\ast}}+M_{D}$ &
$\bar{D}^{0}\bar{D}^{0}\pi^{-}$ & hadronic decay \\
$M_{T_{cc}} < 2M_{D}+M_{\pi}$ & $D^{\ast -} (K^{+} \pi^{-})$ & $0.41 \times 10^{-12}$~s \\
& $\bar{D}^0 (\pi^- K^{+} \pi^{-})$  & weak decay \\ \hline
\end{tabular}
\end{center}
\end{table}%
\begin{table}[htdp]
\caption{Possible decay modes of $\Theta_{cs}$.}
\label{tab:Lee-tab2}
\begin{center}
\begin{tabular}{c|c|c}
\hline
  threshold & decay mode & life time \\
\hline
$M_{\Theta_{cs}} > M_{N}+M_{D_{s}}$ & $pD_{s}^{-}$ & hadronic decay \\
$M_{\Lambda}+M_{D} < M_{\Theta_{cs}} < M_{N}+M_{D_{s}}$ &
$\Lambda \bar{D}^{0}$ & hadronic decay \\
   & $\Lambda D^{-}$ & hadronic decay \\
$M_{\Theta_{cs}} < M_{\Lambda}+M_{D}$ & $\Lambda K^{+} \pi^{-}$,
$\Lambda K^{+} \pi^{+} \pi^{-} \pi^{-}$ & $ 0.41 \times 10^{-12}$~s \\
   & $\Lambda K^{+} \pi^{-} \pi^{-}$ & $ 1.05 \times 10^{-12}$~s \\
\hline
\end{tabular}
\end{center}
\end{table}

To observe $T_{cc}$ and $\Theta_{cs}$ in experiments, we need to know their decay 
modes.  
While our analysis suggests that $T_{cc}$ is bound and $\Theta_{cs}$ is slightly
unbound with respect to their hadronic decays, we give predictions in 
tables~\ref{tab:Lee-tab1} and \ref{tab:Lee-tab2} for all possible $T_{cc}$ and 
$\Theta_{cs}$ masses. 
These exotic hadrons can then be observed through reconstructed final states if 
they decay hadronically or reconstructed final-state vertices if they decay 
weakly.

\subsection{Alignment as a result 
from QCD jet production or new still unknown physics at the LHC?}
\label{lokhtinalig}

{\it I. P. Lokhtin, A. M. Managadze, L. I. Sarycheva and A. M. Snigirev}

{\small
We would like to draw attention of the high-energy physics community to very
important experimental results indicating our lack of understanding of
features of hadron interactions at super-high energies and the necessity of
improving recent theories.
}
\vskip 0.5cm

The intriguing phenomenon of the strong collinearity of cores in emulsion
experiments, closely related to  coplanar scattering of 
secondary particles in the interaction, has been observed a long time ago. So 
far there is no simple satisfactory explanation of these cosmic ray 
observations in spite of numerous attempts
to find it (see, for instance,~\cite{Kopenkin:1994hu, Mukhamedshin:2005nr} 
and references therein).
Among them, the jet-like mechanism~\cite{Halzen:1989rg} looks 
very attractive and gives a natural explanation of alignment of three spots 
along a straight line which results from momentum 
conservation in a simple parton picture of
scattering.

In the Pamir experiment~\cite{Kopenkin:1994hu}
the families with the total energy of the $\gamma$-quanta larger 
than a certain threshold and at least one hadron present 
were selected and analyzed.
The alignment becomes apparent considerably at $\sum E_{\gamma} > 0.5$ PeV
(that corresponds to interaction energies  $\sqrt{s}\geq
4$ TeV).
The families are produced, mostly, by a  proton with  energy  $\geq 10^4$ TeV
interacting at a height $h$ of several hundred meters to several kilometers
in the atmosphere above the chamber~\cite{Kopenkin:1994hu}. 
The collision products are observed within a 
radial distance $r_{\rm max}$ up to several centimeters in the emulsion 
where the spot 
separation $r_{\rm min}$ is of the order of 1 mm.

Our analysis~\cite{Lokhtin:2005bb, Lokhtin:2006xv}
shows that
the jet-like mechanism can, in principle,  attempt to 
explain the results of  emulsion experiments. For such an explanation
it is necessary 
that particles from both hard jets (with rapidities 
close to zero in the center-of-mass system)  hit   the observation
region due to the large Lorentz factor under the transformation from the
center-of-mass system to the laboratory one.
This is possible when the combination of $h$,  $\sqrt{s}$ 
and $r_{\rm max}$  meets the following condition:
\begin{equation}
\label{r00} 
2h m_p/ \sqrt{s} \leq k r_{\rm max},
\end{equation} 
where $m_p$ is the proton mass.
$k \sim 1/2 $ is needed in order to have particles with adjoint positive
and negative rapidities in the center-of-mass system that
hit the detection region. 
At the height $h=1000$ m (mostly used in emulsion experiment estimations)
and $r_{\rm max}=15$ mm the condition (\ref{r00}) is fulfilled at the
energy $\sqrt{s}\geq 270$ TeV that is much higher than the LHC energies
$\sqrt{s}\simeq  5.5 \div 14$ TeV and the threshold
efficient interaction energies $\sqrt{s_{\rm eff}}\simeq 4$ 
TeV~\cite{Kopenkin:1994hu, Mukhamedshin:2005nr}, 
corresponding to the alignment phenomenon. Eq.~(\ref{r00}) can be fulfilled and
at the LHC energy (14 TeV) also, but at the considerably lesser height $h\leq 50$ m which is in
some contradiction with emulsion experiment vague estimations.

On the other hand
if  particles from the central rapidity region 
and the jet-like mechanism are insufficient
to describe the observed alignment, and there is another {\it still unknown}
mechanism of its appearance at the energy  $\sqrt{s}\sim 5.5 \div 14$ TeV and the
accepted height $h\sim1000$ m, then
in any case some sort  of alignment should arise at the LHC
too in the mid-forward rapidity region
(following from the laboratory acceptance 
criterion for, e.g., pp collisions)~\cite{Lokhtin:2005bb, Lokhtin:2006xv}: 
\begin{equation} 
\label{mini1}
 r_{\rm min}<r_i \Longrightarrow \eta_i < \eta_{\rm max} =
 \ln(r_0/r_{\rm min})\simeq 4.95,
\end{equation} 
\begin{equation} 
\label{max1}
 r_{i} < r_{\rm max} \Longrightarrow \eta_i > \eta_{\rm min} =
 \ln(r_0/r_{\rm max})\simeq 2.25,
\end{equation}
where
$r_0~=~2h/e^{\eta_o}$,
$\eta_0=9.55$ is the rapidity of center-of-mass system in the laboratory reference
frame, $\eta_i$ is the particle rapidity in the center-of-mass system,
$r_i$ is the radial particle spacing in the $x$-ray film.
Namely, at the LHC the {\it strong azimuthal anisotropy of energy flux
(almost all main energy deposition along a radial direction) will be 
observed}
for all events with the total energy deposition in the 
rapidity interval (\ref{mini1}, \ref{max1}) larger than some threshold $\sim 1$
TeV. Stress once more that at present there are no models or theories giving such azimuthal 
anisotropy following from the experimentally observed 
alignment phenomenon at $\sqrt{s} \geq\sqrt{s_{\rm eff}}\simeq 4$TeV
and $h\sim 1000$ m~\cite{Kopenkin:1994hu, Mukhamedshin:2005nr}.

This mid-forward rapidity region must be investigated more carefully
on the purpose to study the azimuthal anisotropy of energy flux in accordance
with the procedure applied in the emulsion and other experiments, i.e. one
should analyze the energy deposition in the cells of $\eta \times \phi$-space
in the rapidity interval (\ref{mini1}, \ref{max1}).  Note that the absolute 
rapidity interval can be shifted in  correspondence with the variation of the
height:
it is necessary only that the difference ($\eta_{\rm max}-\eta_{\rm min})$
is equal to $\simeq 2.7$ 
in accordance with the variation of radial distance by a factor of $\sim 15$  
($r_{\rm max}/r_{\rm min}= 15$ independently of $r_0(h)$) due to the relationship 
$r_i\simeq r_0/e^{\eta_i}$. 

Such an investigation  both in  pp and in heavy ion collisions (to differentiate
between
hadronic and nuclear interaction effects) at the LHC can clarify
the origin of the alignment phenomenon, give the new restrictions on the values
of height and energy,
and possibly discover new still unknown physics.

% Nicolas
%\input{Ivanov.tex}
%\input{Ko1.tex}
%\input{Ko3.tex}
%\input{Ko4.tex}
%\input{Kopeliovich2.tex}
%\input{Kopeliovich3.tex}
%\input{Kopeliovich4.tex}
%\input{Kopeliovich6.tex}

% Sangyong

\section*{Acknowledgements}
The research of J. L. Albacete is sponsored in part by
the U.S. Department of Energy under Grant No. DE-FG02-05ER41377.

The work of D. Antonov
has been supported  by the Marie-Curie fellowship through the
contract MEIF-CT-2005-024196.

Part of the work of F. Arleo has been done in collaboration with
T.~Gousset~\cite{Arleo:2007pc} and P.~Aurenche, Z.~Belghobsi, and
J.-P.~Guillet~\cite{Arleo:2004xj}.

N. Armesto acknowledges financial support by MEC of Spain under a contract
Ram\'on y Cajal.
J. G. Milhano acknowledges the financial support of  the Funda\c c\~ao para a 
Ci\^encia e a Tecnologia  of Portugal (contract SFRH/BPD/12112/2003).
C. A. Salgado is supported by the 6th Framework Programme of
the European Community under the Marie Curie contract MEIF-CT-2005-024624.
They and  L. Cunqueiro, J. Dias de Deus, E. G. Ferreiro and C.
Pajares acknowledge financial support by MEC under grant FPA2005-01963, and by
Xunta de Galicia (Conseller\'{\i}a de Educaci\'on).

The work of G. G. Barnaf\"oldi, P. Levai, B. A. Cole,
G. Fai and G. Papp was supported in part by
Hungarian OTKA T047050, NK62044 and  IN71374, by the U.S. Department
of Energy under grant U.S. DE-FG02-86ER40251, and jointly by the U.S.
and Hungary under MTA-NSF-OTKA OISE-0435701.
Special thanks to Prof. John J. Portman for computer
time at Kent State University.

The work of V. Topor Pop, J. Barrette, C. Gale and S. Jeon
was partly supported by the Natural Sciences and
Engineering Research Council of Canada and by the U. S. DOE
under Contract No. DE-AC03-76SF00098 and
DE-FG02-93ER-40764. M. Guylassy gratefully acknowledges partial
support also from FIAS and GSI, Germany.

D. Boer, A. Utermann and E. Wessels
thank Adrian Dumitru and Jamal Jalilian-Marian for helpful
discussions.

W. Busza wishes to acknowledge Alex Mott, Yen-Jie Lee and Andre Yoon for help
with many of the plots, and Yetkin Yilmaz for $N_{part}$ calculations.

The work of C. M. Ko was supported by the US National Science Foundation under
Grant PHY-0457265 and the Welch Foundation under Grant No. A-1358, that of B.
Zhang by NSF under Grant PHY-0554930, that of B.-A. Li by NSF under Grant
PHY-0652548 and  the Research Corporation, that of B.-W. Zhang by the NNSF of
China under Grant No. 10405011 and MOE of China under project IRT0624,
that of L.-W. Chen by the SRF for ROCS, SEM of
China, and their joint work by the NNSF of China under Grants Nos. 10575071 and
10675082, MOE of China under project NCET-05-0392 and Shanghai Rising-Star
Program under Grant No.  06QA14024.

The work of  G. Kestin and U. Heinz  was supported by NSF grant PHY-0354916,
and theirs and the one of M. Djordjevic by U.S. DOE grant DE-FG02-01ER41190.
The work of E. Frodermann was supported by an
Ohio State University Presidential Fellowship.

D. d'Enterria and D. Peressounko acknowledge respectively
support from 6th EU FP contract MEIF-CT-2005-025073 and
MPN Russian Federation grant NS-1885.2003.2.

K. J. Eskola, H. Niemi, P. V. Ruuskanen and S. S. R\"as\"anen
 thank the Academy of Finland, Projects 206024, 
115262, and GRASPANP for financial support.

The work of R. Fries, S. Turbide, C. Gale and D. K. Srivastava was supported in
parts by DOE grants DE-FG02-87ER40328, DE-AC02-98CH10886, RIKEN/BNL, the Texas
A\&M College of Science, and the Natural Sciences and Engineering Research
Council of Canada.

G. Y. Qin, J. Ruppert, S. Turbide, C. Gale and S. Jeon thank the authors 
of~\cite{Eskola:2005ue} for providing their hydrodynamical evolution calculation
at RHIC and LHC energies, T. Renk for discussions, and the Natural Sciences and 
Engineering Research Council of Canada for support.

The work of H. van Hees and R. Rapp
is supported by a U.S. NSF CAREER Award, grant no. PHY-0449489.

The work of Z.-B. Kang and J.-W. Qiu
is supported in part by the US Department of 
Energy under Grant No. DE-FG02-87ER40371 and contract number 
DE-AC02-98CH10886.

The work of D. Kharzeev was supported by the U.S. Department of Energy under 
Contract No. DE-AC02-98CH10886. 
The work of E. Levin was supported in part by the grant of Israeli Science 
Foundation founded by Israeli Academy of Science and Humanity. 

The work of H. St\"ocker and B. Koch was supported by GSI and BMBF.

The work of A. H. Rezaeian, B. Z. Kopeliovich, H. J. Pirner, I. K. Potashnikova
and I. Schmidt was supported in part by Fondecyt (Chile) grants 1070517and
1050519, and by DFG (Germany)  grant PI182/3-1.

The work of I. Kuznetsova, J. Letessier and J. Rafelski
has been
supported by a grant from the U.S. Department of Energy  DE-FG02-04ER4131.
LPTHE, Univ.\,Paris 6 et 7 is: Unit\'e mixte de Recherche du CNRS, UMR7589.

The work of M. Mannarelli and C. Manuel has been supported by the Ministerio de 
Educaci\'on y Ciencia (MEC) under grant AYA 2005-08013-C03-02.

D. Moln\'ar thanks RIKEN, Brookhaven National Laboratory and the US Department 
of Energy [DE-AC02-98CH10886] for providing facilities essential for the 
completion of his work.

G. Torrieri thanks the Alexander Von Humboldt
foundation, the Frankfurt Institute for Theoretical Physics and FIAS for
continued support, and CERN theory division for providing local support
necessary for attending the workshop where this work is presented.
He would also like to thank Sangyong Jeon, Marek Gazdzicki, Mike Hauer,
Johann Rafelski and Mark Gorenstein for useful and productive discussions.

K. Tuchin is grateful to Javier Albacete for showing me the results of his
calculations of open heavy quark production; his results are in a
qualitative agreement with \fig{fig2tuchin} and \fig{fig3tuchin}. He would like to thank
RIKEN, BNL and the U.S. Department of Energy (Contract No.
DE-AC02-98CH10886) for providing the facilities essential for the completion
of this work.

The work of R. Vogt was
performed under the auspices of the U.S. Department of Energy by
University of California, Lawrence Livermore National Laboratory under
Contract W-7405-Eng-48 and supported in part by the National Science
Foundation Grant NSF PHY-0555660.

The work of E. Wang, X. N. Wang and H. Zhang was supported
by DOE under contracts No. DE-AC02-05CH11231,
by NSFC under Project No. 10440420018, No. 10475031
and No. 10635020, and by MOE of China under projects No. NCET-04-0744,
No. SRFDP-20040511005 and No. IRT0624.

\end{document}